\documentclass[a4paper,11pt]{book}

\usepackage[latin1]{inputenc}
\usepackage[american]{babel}
\usepackage{amsmath,amssymb,epsfig}

\usepackage{euscript}
\usepackage{amssymb}
\usepackage{amsfonts}
\usepackage{amsbsy}
\usepackage{amsmath}
\usepackage{epsfig}
\usepackage{hyperref}

 \usepackage{a4wide}
 \usepackage{graphics}
 \usepackage{subfigure}
 \usepackage{fancyhdr}

\newcommand{\beq}{\begin{equation}}
\newcommand{\eeq}{\end{equation}}
\newcommand{\ba}{\begin{array}}
\newcommand{\ea}{\end{array}}
\newcommand{\bea}{\begin{eqnarray}}
\newcommand{\eea}{\end{eqnarray}}
\newcommand{\bean}{\begin{eqnarray*}}
\newcommand{\eean}{\end{eqnarray*}}
\newcommand{\eref}[1]{(\ref{#1})}
\newcommand{\sref}[1]{\S\ref{#1}}
\newcommand{\fref}[1]{Figure~\ref{#1}}

\newcommand{\tr}{\mathop{\rm Tr}}
\newcommand{\comment}[1]{}

\newcommand{\CM}{{\cal M}}
\newcommand{\CN}{{\cal N}}

\newcommand{\cO}{{\cal O}}
\newcommand{\cC}{{\cal C}}
\newcommand{\cX}{{\cal X}}
\newcommand{\IP}{\mathbb{P}}
\newcommand{\IC}{\mathbb{C}}
\newcommand{\IV}{\mathbb{V}}
\newcommand{\IZ}{\mathbb{Z}}
\newcommand{\f}{{\cal F}^{\flat}}
\newcommand{\firr}[1]{{}^{{\rm Irr}}\!{\cal F}^{\flat}_{#1}}

\newcommand{\vev}[1]{\langle #1 \rangle}
\newtheorem{theorem}{\bf THEOREM}

\newtheorem{conjecture}{\bf CONJECTURE}

\newtheorem{observation}{\bf OBSERVATION}

\newcommand{\setall}{\setcounter{equation}{0}
        \setcounter{theorem}{0}}

\newcommand{\tmat}[1]{{\tiny \left(\begin{matrix} #1 \end{matrix}\right)}}

\newcommand{\bee}{\begin{equation}}
\newcommand{\ee}{\end{equation}}
\newcommand{\baa}{\begin{array}}
\newcommand{\eaa}{\end{array}}

\begin{document}

\begin{flushright}
LPTENS 09/04\\
\end{flushright}
\vskip 0.25in

\renewcommand{\thefootnote}{\fnsymbol{footnote}}
\vskip 0.1in
\centerline{{\Huge Master Space and Hilbert Series}} 

\centerline{{\Huge for $\mathcal{N}=1$ Field Theories}}
~\\
\vskip 0.25in
\begin{center}
{\bf
Davide Forcella${}^{+}$\footnote{\tt forcella@lpt.ens.fr} 
}
\end{center}
~\\
~\\
{\hspace{-1in}
\scriptsize
%\begin{tabular}{ll}
  \begin{center}
${}^{+}${\it  CNRS et Laboratoire de Physique Théorique de l'École Normale Supérieure \\
24, rue Lhomond, 75321 Paris Cedex 05, France}
\end{center}
%\end{tabular}
}

\vskip 0.7in

\begin{center}
Abstract
\end{center}
Master Space and Hilbert Series are general tools to study any $\mathcal{N}=1$ supersymmetric field theory. We concentrate on the particular case of $\mathcal{N}=1$ super conformal field theories living on D3 branes at toric Calabi Yau singularities. We start reviewing the topic of branes at singularities, their algebraic geometric and field theory description. We then study the complete moduli spaces for N=1 and generic N number of branes, their different branches, their geometric properties and their symmetries. We study the spectrum of BPS operators both mesonic and baryonic, for N=1 and generic N number of branes, in the field theory and in the dual gravity side and the map between the two. More topics like the study of fermionic BPS operators, the modification of the moduli spaces and the chiral rings under marginal deformations and Seiberg Duality, are also briefly discussed. This paper is a revised version of the author's PhD thesis and it is aimed to give a pedagogical introduction to the use of Master Space and Hilbert Series to investigate supersymmetric field theories.

\setcounter{footnote}{0}
\renewcommand{\thefootnote}{\arabic{footnote}}

\newpage

%\begin{flushright}

%RUNHETC-2000-44
%26th May 1998
%\end{flushright}
%\begin{flushright}
%{\sf \today}
%\end{flushright}
%\begin{center}
%\huge{ \sc TITLE}\\
%\vskip 2cm

%\Large{\sc Davide Forcella$^a$ %\ft{valandro@sissa.it}
%}
%\\
%\vskip 15mm \normalsize
%$^a$  University \\
%\smallskip
%{\it e-mail}

%\end{center}
%\renewcommand{\abstractname}{\sc Abstract}
%\begin{Abstract}
%\vskip 15mm
%\begin{center}
%{\bf {\sc Abstract}}
%\end{center}
%\bigskip
%\normalsize

%This article is the author's PhD thesis.

%\newpage

\normalsize

\begin{titlepage}
 \vspace{3cm}
\center{\huge{\textsf{Davide Forcella}}}
%\begin{flushright}MÃ©tro Simplon
%  \huge{\textsl{Padova,  07 marzo 2003.}}
%\end{flushright}

\vspace{2cm}
\center{\LARGE{\textsl{PhD Thesis:}}}

\vspace{1cm}
% \center{\Huge{\textbf{Phenomenology from the }}}
% \vspace{5mm}
\center{\Huge{\textbf{Moduli Space and Chiral Ring}}}
\vspace{5mm}
\center{\Huge{\textbf{of}}}
\vspace{5mm}
\center{\Huge{\textbf{D3 Branes at Singularities}}}

\vspace{3cm}

\Large{Supervisors: }\\
\Large{\textsf{Prof. Alberto Zaffaroni}}\\
\Large{\textsf{Prof. Loriano Bonora}}\\
\Large{\textsf{Prof. Angel Uranga}}

\vspace{5cm}
\center{\LARGE{\textsl{September 2008}}}

\vspace{1cm}

\center{\Large{\it International School for Advanced Studies \\ (SISSA/ISAS)}}

\end{titlepage}

{\tiny .}
\\
\\
\\
\\
{\it This thesis is dedicated to my parents and to everyone who share with me the love for life }

\newpage

{\tiny .}

\vspace{3cm}

\begin{center}
\LARGE{\textsc{Acknowledgments}}
\end{center}

\vspace{10mm}

During the years of PhD I had the great fortune to meet very nice people, and I would like to acknowledge them with this thesis. 

It is a great pleasure to thank Alberto Zaffaroni for being a nice supervisor, an excellent collaborator and over all a good friend all along this thesis.

I would like to thank Loriano Bonora for being so kind and attentive with a ``not so easy'' PhD student like me.

A big ``GRACIAS !'' go to Angel Uranga that was my supervisor in this last year in CERN. GRACIAS for all the time spent together and for your kind presence! 

Many thanks to Amihay Hanany for good collaborations, great discussions and his nice help and presence in ``the moments of life''.

A special thank go to Luciano Girardello that was like a firm and quiet stone all along my PhD.
 
It was a great honor to share time, passion, knowledge and friendship with my collaborators Agostino Butti, Andrea Brini, 
Amihay Hanany, Alberto Zaffaroni, Angel Uranga, David Vegh, Luca Martucci, Ruben Minasian, Michela Petrini, 
Yang-Hui He, Antonio Amariti, Luciano Girardello, Alberto Mariotti, Inaki Garcia-Etxebarria.

It is very nice to have the possibility to thank, for all the enlightening discussions about life and physics 
and for the kind encouragement, Giuseppe Milanesi, Roberto Valandro, Francesco Benini, Stefano Cremonesi, Luca Ferretti,  Christiane Frigerio Martins, Diego Gallego, Luca Lepori, Andrea Prudenziati, Houman Safaai, Raffaele Savelli, Luca Vecchi, Giulio Bonelli, Marco Serone, Andrea Romanino, Serguey Petcov, Roberto Percacci, Matteo Bertolini, Barbara Fantechi , Alessandro Tanzini, Fabio Ferrari Ruffino, Annibale Magni, Luca Philippe Mertens, Luis Alvarez-Gaume, Guillermo Ballesteros, Massimo Bianchi, Nadejda Bogolioubova, Roberto Contino, Jan de Rydt, Cedric Delaunay, Elena Gianolio, Fernando Marchesano Buznego, Lotta Mether, Domenico Orlando, Mauro Papinutto, Anne-Marie Perrin, Nikolaos Prezas, Francesco Riva, Thomas Ryttov, Alireza Tavanfar, Marlene Weiss, Tziveloglou Pantelis, Wolfgang Bartmann, Balint Radics, Korinna Zapp, Mohamed Koujili, Mikail Acar, Alberto Romagnoni, Gabriele Tartagliono-Mazzuchelli, Roberto Casero, Marco Pirrone, Franceso Virotta, Claudio Destri, Giu!
 seppe Marchesini, Federico Rapuano, Carlo Oleari, Simone Alioli, Alessandro Tomasiello, Jose D. Edelstein, Bo Feng, Sergio Benvenuti, Constantin Bachas, Raphael Benichou, 
Umut Gursoy, Bernard Julia, Bruno Machet, Boris Pioline, Giuseppe Policastro, Jan Troost, Dario Martelli, Gianguido Dall'Agata, Riccardo Iancer.

I acknowledge my ``sponsors'': SISSA and the INFN of Trieste, the MIUR under  
contract 2005-024045-004, the European Community's Human Potential Program
MRTN-CT-2004-005104, the European Superstring Network MRTN-CT-2004-512194, the CNRS, and the Marie Curie
fellowship under the programme EUROTHEPHY-2007-1.

Thanks to the Laboratoire de Physique Th\'eorique de l'\'Ecole Normale
Sup\'erieure, Universit\'es Paris VI et VII, Jussieu, the physics theory group in CERN and Milano Bicocca 
for the very kind hospitality during part of this PhD.

\tableofcontents

\chapter*{Introduction}

String theory was born to describe strong interactions. 
The intuitive idea is still valid today. In the IR the 
hadrons in the nucleus get 
extremely coupled and if we try to separate two quarks a tube of color 
form between them. It has the tension proportional to its length and 
indeed it behaves essentially as a string. Because 
the elementary quarks are not observable at low energies, these stringy degrees of 
freedom would give the correct effective description of the strong interactions in the far IR. 
This approach reproduce some nice features of the hadronic spectrum, 
but was definitively surpassed by the advent of the QCD, that seems 
to describe, in a coherent picture, the strong interaction as a quantum 
field theory. 

After that first stop string theory left its original purpose and re-propose itself 
as the theory of everything. Namely as the only theory that 
could in principle unify all the interactions we know in a coherent picture, 
and explain every kind of phenomena as the result of some simple elegant postulates. 
Indeed at the base of string theory there is just the idea of giving a quantum mechanical 
description of the general covariant dynamics of extended objects. Starting 
with this logic, and some simple assumptions, it is possible to show that string theory is rich enough 
to contain the Standard Model and to unify it with an in principle consistent theory of
quantum gravity. Even more string theory seems to have not any free parameters and to be unitary 
and finite at least at the perturbative level. All these properties seem miraculous and gave the 
starting point to a full fledged investigation.

After a lot of years of very intensive research string theory reveals as a very hard beast. 
We do not even had a real definition of the theory and its celebrated uniqueness 
is basically lost in the very hard problem of 
finding the right vacuum describing our world. But we don't need to lose 
our enthusiasm, this theory is full of surprises and it is able to regenerate 
it self from its dust. Indeed, thanks to the discovery of new non perturbative degrees of
freedom, first of all D branes \cite{Polchinski:1995mt}, there were many new progresses. These news ingredients changed 
forever our point of view about string theory: it is no more a theory of one dimensional objects but 
contains in itself the spectra of all possible dimensional extended objects, like branes. 
A D brane in first approximation is a minkowski hyperplane embedded 
in the ten dimensional stringy space time. The interesting properties of D branes is that they 
are charged under the RR form fields of the perturbative strings and a supersymmetric 
gauge theory lives on its world volume. Using D branes it is possible to reproduce very interesting phenomenological 
models, study the microscopic origin of the black hole entropy, giving in this way a 
window over quantum gravity, study non perturbative corrections to string theory 
compactifications, or to gauge theory dynamics, and formulate the AdS/CFT correspondence \cite{Maldacena:1997re,Witten:1998qj,Gubser:1998bc}. 
Indeed regarding
to this last topic we can claim that sometime the history seems circular and looking into the future 
you can see your past. Today we believe that string theory is still able to teach 
us many things related to the strongly coupled regime of field theory that in many cases 
is not accessible in any other way! 

More than ten years ago the AdS/CFT 
conjecture was formulated in the context of the modern string theory \cite{Maldacena:1997re,Witten:1998qj,Gubser:1998bc}, and definitively 
realized the idea that the geometry, and in particular the string theory, could be the 
correct description for the strongly coupled regime of quantum field theories. The AdS/CFT 
correspondence tells that, in some way, the degrees of freedom of a flat non gravitational 
gauge theory, in the strong coupling IR regime, reorganize themselves into the degrees of freedom 
of a ten dimensional string theory on some specific background. This conjecture is formidable 
and very powerful. One can make prediction on the far IR quantum regime of a gauge theory 
doing essentially classical geometric computations. 
The actual formulation of the AdS/CFT makes use of D3 branes in flat space time, and in its generalizations, of D3 
branes placed on some particular singularities \cite{Morrison:1998cs,Acharya:1998db,Klebanov:1998hh}.
 
Along this thesis we will mainly focalize on the study of D3 branes in the kind of setup 
nearer to the AdS/CFT correspondence. Namely we will consider N D3 branes at the CY conical singularity $\cX=C(H)$. The low energy dynamics of D3 branes at $\cX$ 
is a quiver gauge theory \cite{Douglas:1996sw}: a UV free field theory that flows in the IR to a superconformal fixed point; 
it has gauge group $\prod_i^g SU(N_i)$, chiral bifoundamental matter fields, and a particular form for the superpotential. 
As we will see this setup is 
very rich and it has at least two main applications: local construction for model building, 
and study of strongly coupled field theories using the AdS/CFT correspondence. The first application is mainly due to the fact that
the structure of the field theory on D3 branes at singularities is rich enough to give a big variety of 
interesting gauge theory dynamics and naturally contains chiral fields fundamental for any phenomenological
applications, moreover it is flexible enough to skip some global consistency 
conditions required by effective compactifications. The second application is based on the fact that the AdS/CFT 
tells that the field theory living 
on $N$ D3 branes at the $\cX=C(H)$ singularity is dual ( in a strong/weak coupling duality sense ) 
to type IIB string theory propagating on the $AdS_5 \times H $ background 
with $\int_H F_5 = N $.

In both cases the central topic is the study of gauge theory phenomena.
In this thesis we will be humble and we will concentrate on the super conformal case, but 
it is possible to analyze also the non conformal 
and non supersymmetric case using a similar setup. 
As usual to go far we need first to learn how to move some easy steps, and we will 
see that there are many things to say and to learn even in this rather simpler SCFT case. 

In the recent years there were a lot of efforts to construct the dictionary between the geometrical singularity and the gauge theory. This dictionary is very well understood in the toric case \cite{Hanany:2005ve,Franco:2005rj,Feng:2005gw,Franco:2005sm,Hanany:2005ss,Martelli:2004wu,Benvenuti:2004dy,Benvenuti:2005ja,Butti:2005sw,Martelli:2005tp,Butti:2005vn}, namely when the singularity $\cX$ admits at least a $U(1)^3$ isometry.
We will not spend too much time to illustrate the
very interesting problem of constructing gauge theories at singularities. 
What is important for us is that, at the end of the day, we obtain 
a superconformal field theory that is described in the UV as a quiver gauge theory
with some specific form of the superpotential: every bifundamental field appears in the 
superpotential just two times: one time with the positive sign, the other time with the negative sign. 
Our main interest will be to 
study some supersymmetric properties of these theories. 

Once we have a supersymmetric field theory the first information we would like
to obtain is the structure of its vacua. Supersymmetric field theories indeed 
usually have a manifold of vacua called the moduli space $\CM$ and the study of these varieties 
is of fundamental importance to understand the dynamics of the theory. 
Another very important subject is the study of the degrees of freedom of a 
theory, and in particular of its BPS operators, namely operators that preserve half of the
supersymmetries of the theory. In this thesis we will be interested to study 
the moduli space, and the spectrum of BPS operators of gauge theories living of stacks of N
D3 branes at singularity $\cX$. We will learn a lot 
of properties about these two subjects and in particular their relations, namely 
how to get informations about one of the two once we have enough informations 
related to the other \cite{Forcella:2008bb,Forcella:2008eh,Martelli:2006yb,Benvenuti:2006qr,Butti:2006au,Hanany:2006uc,Feng:2007ur,Forcella:2007wk,Butti:2007jv,Forcella:2007ps,Butti:2007aq,Forcella:2008au}. 
To study these topics we will introduce the concepts of Master Space, Hilbert Series and Plethystic Exponential. We will learn soon to appreciate their properties, but let us now start with some basic facts.

There are a set of nice properties of this kind of setup that will make easier our analysis.
 A first crucial observation is that at least a part of the moduli space of the gauge theory is explicitly 
realized in the geometry. Indeed some of the scalar degrees of freedom parameterizing the moduli space $\CM$ must 
also parameterize the space transverse to the branes. In the abelian case we just have one D3 brane and it is just 
a free point in the transverse geometry, hence the moduli space of the theory must contain at least the six 
dimensional singularity $\cX$. In the generic non abelian case we have N D3 branes that, 
being mutually supersymmetric, will behave as a set on N non interacting particle on $\cX$. 
Because the ground state of the D3 branes is bosonic we have a system of N bosons 
and the non abelian moduli space 
must contain the N times symmetric product of the transverse singularity: Sym$^N(\cX)= \cX^N / \Sigma_N$, where 
$\Sigma_N$ is the group of permutations of N objects acting on the N factors $\cX$.
Another nice property is that the gauge theories on D branes at singularities
have in principle a natural gravity dual that describe the strong coupling regime 
of the field theory, and this could help to interpolate from 
strong to weak coupling. 

These nice properties are a good starting point to begin our study about the complete moduli space $\CM$ and the complete chiral ring of these theories.

The moduli space $\CM$ of a supersymmetric gauge theory is given by the
vanishing of the scalar potential as a function of the scalar components
of the superfields of the field theory. This, in turn, is the set of
zeros of so-called {\bf D-terms} and {\bf F-terms},
constituting a parameter, or moduli, space of solutions describing the
vacuum. It is a standard fact now that $\CM$ can be represented as the 
symplectic quotient of the space of
F-flatness, by the gauge symmetries provided by the D-flatness. We will
denote the space of F-flatness by $\f$ and symmetries prescribed by
D-flatness as $G_{D^{\flat}}$, then we have
\begin{equation}\label{M}
 \CM \simeq \f // G_{D^{\flat}} \ .
\end{equation}
$\f$ is called the master space and, as we will see all along this thesis, it is a parent 
space from which we can obtain a lot of informations regarding the moduli space $\CM$. 

We will start to fully characterize the moduli space of gauge theories for branes at singularities in the abelian case.
With just one brane we are left with a set of $SU(1)$ groups, 
hence the gauge dynamics of the  theory is trivial and $\CM=\f$. The master space $\f$ 
describe the complete moduli space of the theory and contains as a sub set ( the locus where 
all the baryonic operators have zero vacuum expectation value ) the geometrical six dimensional 
transverse singularity $\cX$.

The complete moduli space is some sort of $(\mathbb{C}^*)^{g-1}$ fibration over $\cX$ with the action induced by the 
baryonic symmetries:
\begin{itemize}
\item{$\CM=\f$ is generically a $g+2$ dimensional toric variety, and can be completely characterized in term of algebraic geometry. Indeed we will see that $\f$ is generically reducible in 
\begin{itemize}
\item{a $g+2$ dimensional component $\firr{~}$ that is a toric Calabi Yau cone,} 
\item{and a set of lower dimensional generically linear components $L_i$.}
\end{itemize}
}
\end{itemize} 
We will illustrate the properties of $\f$ with examples and with the use of some mathematical theorems. 

The natural and subsequent step is to analyze the 
non abelian case. In the non abelian case the equations describing the master space $\f$ get very quickly 
extremely involved, and on top of this fact, the gauge dynamics is highly non trivial. 
For these reasons we miss a clear direct description in term of algebraic equations and we need 
to look for other tools to get informations. It is easy to see that the moduli space for 
$N>1$ is in general no more toric and Calabi Yau. 
This two facts complicate enormously the direct analysis, and hence the idea 
will be to follow the algebraic geometric paradigm: the 
set of holomorphic functions over an algebraic variety completely 
characterize the variety itself. This means that if we are not able to analyze the variety directly 
we can extract all the informations we need studying the spectrum of holomorphic functions over it.

The set of holomorphic functions over the moduli space is exactly the set of all 
1/2 BPS operators of the gauge theory. We want to construct a function that encode all the 
informations of the chiral ring of the theory. We proceed in the following way. The gauge theory has 
generically $g+2$ global abelian symmetries: $U(1)^2 \times U(1)_R$ coming from the isometries of $\cX$, they are the 
flavor and R symmetries of the field theory, and they are generically called mesonic symmetries; $U(1)^{g-1}$ baryonic symmetries: anomalous and non anomalous, 
coming from the $U(1)$ factors inside the $U(N_i)$ of the UV description of the quiver gauge theory (the total 
diagonal $U(1)$ is always completely decoupled), whose gauge dynamics decouple in the IR superconformal fixed points. 
We want to realize a function that counts the BPS operators according
to their charges under these various $U(1)$. Let us introduce a set of $g+2$ chemical potentials $t_i$ with $i=1,...,g+2$. 
We would like to have a function that once expanded for small values of the chemical potentials 
\begin{equation}\label{HSintro}
H(t;\CM)=\sum_{i_1,...,i_{g+2}}c_{i_1,...i_{g+2}}t_1^{i_1}...t_{g+2}^{i_{g+2}}
\end{equation}
gives the numbers $c_{i_1,...,i_{g+2}}$ of BPS operators with charges $i_j$ under the 
corresponding global abelian symmetries. This function is called the 
Hilbert series of $\CM$. 

The holomorphic functions 
on the moduli space are the BPS operators of the gauge theory and we could try
 to construct these functions directly in field theory. 
We will give some examples of this construction, but it is worthwhile to say immediately that this 
way to proceed has two important 
drawbacks: the actual computational power is highly restricted by the increasing of the complexity 
of the operations with the increase of the number of colors N; and it gives back the classical, 
weak coupling spectra of BPS operators of the theory, while we know that 
the gauge theory is generically in strongly coupled fixed points.

To solve the computational problem and to try to obtain informations 
in the strongly coupled regime of the field theory we will analyze the dual stringy description 
following the AdS/CFT correspondence. The AdS/CFT 
proposes a one to one map between the operators 
in the gauge theory and the stringy states in the $AdS_5 \times H$ 
background. To get (\ref{HSintro}) in the strong regime we need to find all the states dual 
to the complete spectrum of BPS operators in gauge theory. It come out that to every BPS operators 
one can associate a quantum state of a D3 brane wrapped over a three cycle 
in H. If the three cycle is trivial the corresponding operator will be a mesonic 
operator, namely it will have charge zero under all the $U(1)^{g-1}$ baryonic symmetries; 
if the three cycle is non trivial the associated operator is a baryonic 
operator. There exist an infinite number of states of D3 branes wrapped in H. Indeed every section
 of every possible line bundle over $\cX$ individuates a 
suspersymmetric embedding of a D3 brane, while, through the geometric quantization of the system,
 the N times symmetric product of 
these sections, modulo the specification of a flat connection over the three cycle,
 is a state of D3 brane dual to a BPS operator. 
We will show how to explicitly construct all these states and how to obtain from them the partition 
function counting 
all the BPS operators of the gauge theory in the strongly coupled regime.

The basic idea is the following. It is possible to show that the knowledge of the abelian $N=1$ case is enough 
to obtain the Hilbert function for the generic non abelian $N>1$ case. In the $N=1$ case we need to count sections over 
the transverse space $\cX$. These are infinite in number, because the manifold is non compact, 
and are topologically distinguished by the baryonic numbers $\vec{B}$ in the field theory. 
Hence the $N=1$ partition function admits a decomposition in sectors having 
a different set of baryonic charges:
\begin{equation}\label{g1intro}
g_1(t;\cX)= \sum _{\vec{B}} m(\vec{B}) g_{1,\vec{B}}(t;\cX)
\end{equation}
$m(\vec{B})$ are the number of BPS operators associated to the same geometrical 
cycle: multiplicities. They are related to the possible set of flat connections that one can 
turn on the cycle. $g_{1,\vec{B}}(t;\cX)$ count the set of sections of the line bundle with topology given by $\vec{B}$ and weighted
by their charges under the isometries of $\cX$. We compute the multiplicities $m(\vec{B})$ with the use of some combinatorial techniques in the field theory, while $g_{1,\vec{B}}(t;\cX)$ is obtained applying some index theorems to the bundles on $\cX$. 

We will learn that these partition functions contain a lot of informations 
related to the CY singularity $\cX$: 
\begin{itemize}
\item{$g_{1,0}(t;\cX)$, the mesonic partition function, contains the structure of the algebraic equations 
defining the variety $\cX$;}
\item{from $g_{1,\vec{B}}(t; \cX)$, for generic $\vec{B}$, we can obtain the volume of H, and for $\vec{B}\ne 0$ 
the volumes of all the possible non 
trivial three cycles $C_3$ in H.}
\end{itemize}
They contain even important informations 
related to the gauge theory: 
\begin{itemize}
\item{$g_1(t;\cX)$, and in general $g_N(t;\cX)$, contain the structure of the generators and the relations 
in the chiral ring for different numbers of colors;}
\item{$g_{1,\vec{B}}(t; \cX)$ contain the value of the central charge and of the various R charges of the field theory at strong coupling.
These are related by AdS/CFT to the volume of H and to the volumes of the three cycles in H.}
\end{itemize}
 
The problem of counting chiral operators is influenced by the multiplicities problem: 
namely by the fact that to every geometrical D3 brane state in string theory 
correspond more than one operator in the gauge theory. We will need to put a 
lot of effort to solve this multiplicity problem. Along this way we will find a 
rather surprising relation between the baryonic numbers and the kahler moduli 
of the geometry. Pushing on this relation we will be able to give a precise 
prescription to compute complete BPS partition functions for a generic toric conical 
CY singularity. 

Afterward we want to pass to the generic $N>1$ non abelian case. This is in principle a non 
trivial step, because it introduces in the game non abelian dynamics, and non trivial algebraic relations
among the operators. The fundamental observation is that looking at the gravity side, 
or at the explicit construction of the BPS operators, it is
possible to see that to pass to the generic N case we just need to construct the partition function counting
all the possible N times symmetric products of the functions counted by $g_{1,\vec{B}}(t; \cX)$. 
This is essentially due to the fact that the operators for finite $N$ are symmetric functions 
of the operators for $N=1$. There exist a function, called the plethystic exponential PE, that exactly play this role: 
it takes a generating function for a set of operators and it gives back a function that counts all the possible symmetric products of the operators counted by the input function.

The Hilbert series counting the BPS operators for arbitrary number $N$ of branes is obtained from (\ref{g1intro}) applying the 
PE to every fixed $\vec{B}$ sector:
\begin{equation}
\sum_{\nu=0}^{\infty} \nu^N g_N(t;\cX)=\sum _{\vec{B}} m(\vec{B}) PE_{\nu}[ g_{1,\vec{B}}(t;\cX) ]
\end{equation} 
where $g_N(t;\cX)$ is the Hilbert series counting the BPS operators for N D3 branes.
We have in this way developed a procedure to construct 
partition functions counting BPS operators for arbitrary number $N$ of branes. This ability gives us 
new tools and a new point of view towards the study of the moduli space for generic N. 

The non abelian moduli space is indeed very complicated and a direct attack to the 
problem seems quite hard. All along the history of physics the search for symmetries 
was a guide line. Indeed the search for symmetries will be our guide line in the 
study of the moduli space $\CM$. It turn out that the Master Space $\f$ for 
just one D3 brane has some symmetries not manifest in the UV Lagrangian 
describing the field theory; we call them Hidden Symmetries. 
These symmetries are typically non abelian extensions 
of some anomalous or non anomalous baryonic $U(1)$ symmetries. The symmetries of the moduli space $\CM$ can be naturally study with the help of the Hilbert series that we have just described. 
Indeed we will show that It is possible to reorganize the spectrum of BPS operators 
in multiplets of the Hidden Symmetries in such a way that the Hilbert series for the $N=1$ moduli space will be explicitly 
written in term of characters $\chi_{\vec{l}}$ of these symmetries. This fact imply that all the 
BPS operators comes in representation of the symmetries. Now we know how to pass from the one brane case 
to the generic number of branes N. If we are able to reorganize the generating functions $g_N$ for the non abelian case 
in representations of the global symmetries guessed form the abelian case, we can show that the complete 
moduli space $\CM$ for generic $N$ has the Hidden Symmetries of the $N=1$ Master Space $\f$. 
Indeed in a case by case analysis it is possible to show that if we introduce a set of chemical potentials $q_i$, 
$i=1,...,g+2$ parameterizing the Cartan sub algebra of the Hidden 
Symmetries group, it is sometimes possible to rewrite the generating function $g_N$ in such a way that it will 
have a nice expansion in character of the Hidden Symmetries:
\begin{equation}
g_N(q)=\sum_{\vec{l}}\chi_{\vec{l}}(q)
\end{equation}
When this happens, it means that the non abelian moduli space is invariant under the Hidden Symmetries group. 

Let us try to summarize where we are. We tried an alternative algebraic geometrical approach to four dimensional $\mathcal{N}=1$ supersymmetric field theories. We make strong use of the concept of Master Space, Hilbert Series and Plethystic Exponential. For the specific setup of D3 branes at CY singularities we got as a result the complete control over the spectrum of BPS operators and its symmetry properties for generic N; the complete description for the moduli space for one brane, and interesting properties for generic N.

Due to the success of this approach it is natural to wonder about possible generalizations. We will describe only some generalizations while many other are present in literature, and we will comment about them in the conclusions. 
  
Till now we focalized our attention to the scalar sector of the BPS operators.
One possible extension is to try to construct partition functions counting 
all the supersymmetric degrees of freedom of the theory including the non bosonic ones. 
One of the reasons for this extension could be the recently proposed black holes solutions in $AdS_5$ space time \cite{Gutowski:2004ez,Gutowski:2004yv}. If we believe that string theory can give a microscopical explanation of the entropy of such black holes, It is natural to look for the counting of degrees of freedom of some kind of dual gauge theory. To take into account the entropy of such black holes we presumably need to count also non 1/2 BPS operators, but for the moment we will just try to count at least all the chiral operators. 
To get the partition function counting all the BPS operators we need to add the fermionic degrees of 
freedom to our previous counting, namely we need to make use of the superfield strength $W_{\alpha}$ 
associated to the gauge fields. 
We will see that this can be done in a rather systematic way, generalizing the formalism just proposed 
to some kind of superfield formalism, and taking into
account the mixed bosonic/fermionic statistic to go from $N=1$ to generic $N$. 

Another possible extension is to try to understand what happen to the moduli space of a gauge theory and to its chiral ring when the gauge theory is marginally deformed. 
Indeed all the theories living on branes at 
singularities have generically a manifold of fixed points. 
This means that the superconformality conditions imposes some constrains on the coupling 
constants that can be solved locally in term of a subset of all the coupling constants. The SCFT just 
described correspond to the more symmetric point in the conformal manifold. 
One can move along this manifold of conformal fixed points turning on some marginal 
operators in the superpotential. 
The marginal operators preserve the superconformality of the gauge theory, the values 
of the central charge and of all the R charges of the theory, but they generically 
break part or all the global symmetries, and turn on non trivial field configurations 
in the gravity dual. The result is the modification of both the moduli space and the chiral ring.
A marginal deformation common to all these toric gauge theories is the so called $\beta$-deformation. 
In this case the insertion of marginal operators in the gauge theory always preserve a $U(1)^3$ isometry 
and it is seen in the dual gravity setup as the presence of
non trivial fluxes. These fluxes act as a potential on the system of D3 branes and they drastically 
change the moduli space and the structure of the chiral ring of the gauge theory. Let us summarize what happen to 
the mesonic moduli space. In the undeformed case the $N=1$ mesonic moduli space is the transverse CY singularity $\cX$, 
while in the case in which $\beta$ is an irrational number the mesonic moduli space is just a set of complex lines. 
For rational values of $\beta \sim 1/n$ the $N=n$ mesonic moduli space becomes the $\mathbb{Z}_n\times\mathbb{Z}_n$ quotient of the original 
$\cX$ singularity. 
The set of BPS mesonic operators will be the set of holomorphic functions defined on these varieties, 
and it change discontinuously with $\beta$.

Along the thesis we will also make some comments about the use of the Master Space and the Hilbert Series to study seiberg or toric dualities.

There are of course a lot of other possible extensions to the concepts and techniques we introduce in this thesis and we will comment about some of them in the conclusions.

%\vskip 3cm
%\newpage
\section*{Structure of the Thesis}

The thesis is structured as follows.

\

In the {\it chapter 1} we give a short pedagogical introduction to the topic of branes at singularities. 
We will use as examples the easy case of $\mathbb{C}^3$ 
and the conifold. We will introduce some concepts of toric geometry, 
brane tiling, AdS/CFT, and we will give a quick overview of the mesonic moduli space and
the related counting problem of mesonic operators. 
We will finally define the PE function and explain some of its basic properties. 

In {\it chapter 2} we study the moduli space $\CM=\f$ for $N=1$ brane. In particular we will 
study its reducibility properties, its CY conditions and the structures of its Hilbert series. 
We will give different types of description of $\f$: explicit, toric, using dimer techniques, and linear sigma models.
We will give some orbifold and non orbiford examples and we will analyze some properties of their IR flows.

In {\it chapter 3} we study the problem of counting BPS operators looking to the dual stringy states. 
We realize a one to one map between the baryonic operators and the D3 brane states in the geometry. 
We learn how to construct the partition functions with fixed baryonic charges and how to obtain from them the 
volume of the horizon manifold H and of all the non trivial three cycles $C_3$ inside H, 
and hence the value of the central charge and of all the R charges in field theory. 

In {\it chapter 4} using the field theory intuition we will construct the first complete partition functions 
counting operators according to all their charges, baryonic and mesonic. 
Using the example of the conifold we will study some properties of the moduli spaces and the chiral rings for different values of N.

In {\it chapter 5} we will put together the abilities developed in the last two chapters and we will formulate 
a general recipe to obtain complete partition functions for every toric quiver gauge theory. 
This general proposal strongly relies on a correspondence between the Kahler moduli of the geometry 
and the baryonic numbers of the field theory. We give some explicit examples and we match the results 
between the gravity and the gauge theory computations.

In {\it chapter 6} we go back to the problem of studying the moduli space $\CM_N$ for generic number of branes N. 
In particular, using the Hilbert series associated to $\f$, we will be able to show that the fully non abelian 
moduli spaces have some non abelian hidden symmetries for generic N.

In {\it chapter 7} we explain how to add the fermionic BPS operators to the previously obtained Hilbert series, 
and we will give a recipe to construct the Hilbert series counting all the possible BPS operators of the field theory.

Finally in {\it chapter 8} we study how the mesonic moduli space and the mesonic chiral ring are modified 
if we marginally deform the gauge theory.
%\vskip 15mm

\

This thesis is based on the following papers:\\

%\begin{description}

  D.~Forcella, A.~Hanany, Y.~H.~He and A.~Zaffaroni,

  ``The Master Space of N=1 Gauge Theories,''

   JHEP {\bf 0808}, 012 (2008)
  [arXiv:0801.1585 [hep-th]].
  \\

  D.~Forcella, A.~Hanany, Y.~H.~He and A.~Zaffaroni,
  
 ``Mastering the Master Space,''
  
   Lett.\ Math.\ Phys.\  {\bf 85} (2008) 163
  [arXiv:0801.3477 [hep-th]].
\\

  A.~Butti, D.~Forcella and A.~Zaffaroni,

  ``Counting BPS baryonic operators in CFTs with Sasaki-Einstein duals,''

  JHEP {\bf 0706} (2007) 069
  [arXiv:hep-th/0611229].
\\
   
D.~Forcella, A.~Hanany and A.~Zaffaroni,

``Baryonic generating functions,''

  JHEP {\bf 0712} (2007) 022
  [arXiv:hep-th/0701236].
\\

A.~Butti, D.~Forcella, A.~Hanany, D.~Vegh and A.~Zaffaroni,
  
``Counting Chiral Operators in Quiver Gauge Theories,''

  JHEP {\bf 0711} (2007) 092
  [arXiv:0705.2771 [hep-th]].
\\  

 D.~Forcella,
  
``BPS Partition Functions for Quiver Gauge Theories: Counting Fermionic
  Operators,''
  
arXiv:0705.2989 [hep-th].
\\

A.~Butti, D.~Forcella, L.~Martucci, R.~Minasian, M.~Petrini and A.~Zaffaroni,
  
``On the geometry and the moduli space of beta-deformed quiver gauge
  theories,''

JHEP {\bf 0807}, 053 (2008)
  [arXiv:0712.1215 [hep-th]].

%\end{description}

%\part{Branes at Singularities}

\chapter{Branes at Singularities}\label{braneasing}

The discovery of D branes revolutionized our understanding of String Theory. 
D branes are promising tools 
for model building, holographic correspondences, geometry, gauge theories, and many other topics.

Local analysis, namely D branes in non compact space, is a good laboratory to study all these topics, ant it is also very well motivated by the AdS/CFT correspondence. 
Indeed the local setup can be seen as the first step to construct 
a global consistent model. Even more in the local setup, thanks to the AdS/CFT correspondence, we have some control on the strong dynamics of the system.

We are interested in the dynamics of the low energy degrees of freedom of N D3 branes in non-compact setup: 
namely extending in four flat minkowski space time directions and transverse to a non compact six dimensional variety $\cX$. Their low energy degrees of freedom is a special class of gauge theories called quiver gauge theories. 

\section{Generalities: $\mathbb{C}^3$ and the Conifold}

The easiest example of this kind of setup is a stack of N D3 branes in the flat space time i.e.: 
$M^{1,3} \times \mathbb{C}^3$.
It is well known that their low energy dynamics is described by the $\mathcal{N}=4$ supersymmetric $U(N)$ gauge theory. 
In the language of $\mathcal{N}=1$ supersymmetry this theory contains one $U(N)$ vector superfield and three 
chiral superfields $X$, $Y$, $Z$ in the adjoint representation of the gauge group, interacting through the superpotential:
\begin{equation}
W= \tr(XYZ-XZY)
\end{equation} 

\begin{figure}[t]
\begin{center}
\includegraphics[scale=0.6]{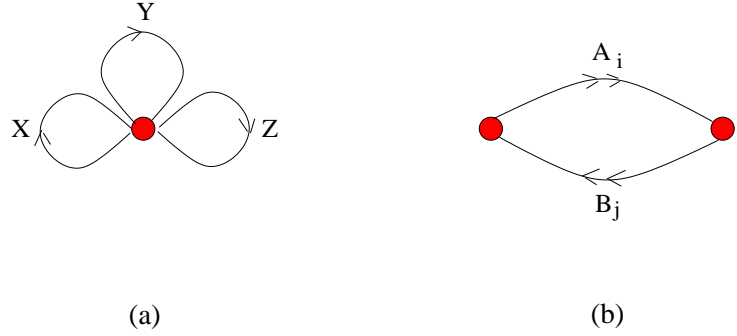} 
\caption{\small The quiver diagram a) for $\mathbb{C}^3$ and b) for the conifold.}
\label{Ccon}
\end{center}
\end{figure}

The D3 brane behaves like a point in the transverse space $\cX$. Being point like implies that the low energy space time 
dynamics of the brane is sensible only to the local geometry of the six dimensional space $\cX$. 
Indeed the low energy dynamics of a stack of N D3 branes in $M^{1,3} \times \cX $, when $\cX$ is a smooth manifold, 
is still $\mathcal{N}=4$ supersymmetric $U(N)$ gauge theory. To obtain more interesting dynamics in this setup, 
we are forced to consider singular varieties. We are indeed interested in studying the properties of a stack of 
N D3 branes at conical Calabi Yau ( CY from now on) singularities $\cX$. CY because we want to have at least $\mathcal{N}=1$ supesymmetry, 
and conical because we want a conformal field theory. A D3 brane at singularity fractionates in a system of 
$g$ fractional branes whose low energy dynamics is described by a $\prod_{i=1}^g U(N_i)$ gauge theory with chiral 
bifoundamental fields that are the zero mass modes of the strings stretching between the various fractional branes. 
These gauge theories are called quiver gauge theories and they have an easy graphical representation \cite{Douglas:1996sw}. 
For every gauge group factor draw a node while for every bifoundamental field draw an arrow with the tail 
on the node under which the field transform in the fundamental representation, and the tip on the node 
under which the field transform in the anti-fundamental representation, see Figure \ref{Ccon}. 
These theories are UV free and flow in the IR to a superconformal fixed manifold. The diagonal $U(1)$ gauge factor
is completely decoupled and there are no chiral fields charge under its action. Along this flow the other $g-1$ $U(1)$ 
factors of the gauge group decouple and remain as global baryonic symmetries, anomalous and non anomalous, in the IR, 
where the Super Conformal Field Theory ( SCFT from now on ) have gauge group $\prod_{i=1}^g SU(N_i)$.

One easy and very well known example is the conifold singularity $\mathcal{C}$ \cite{ossa,Klebanov:1998hh}. 
This is a six dimensional singularity described by the quadric $xy=wz$ in $\mathbb{C}^4 $. 
This space is singular and has an isolated singularity at the origin of $\mathbb{C}^4 $. 
A D3 brane at the conifold singularity fractionates into two fractional branes that forms 
the physical D3 bound state. Indeed the low energy dynamics of a stack of N D3 branes at $\mathcal{C}$ 
is given by an $\mathcal{N}=1$ supersymmetric gauge theory with gauge group $SU(N) \times SU(N)$, 
bifoundamental fields $A_i$ , $ B_j$ with $i,j=1,2$ 
transforming as shown in the quiver (Figure \ref{Ccon}), and superpotential:
\begin{equation}\label{consup}
W= \tr ( A_1 B_1 A_2 B_2 - A_1 B_2 A_2 B_1 )
\end{equation} 
The conifold is a real cone over the the coset space $T^{1,1}=\frac{SU(2)\times SU(2)}{U(1)_D}$, where the $U_D(1)$ is the
diagonal $U(1)$ contained into the two $SU(2)$: $\mathcal{C}=C(T^{1,1})$. The field theory has the global
symmetry $SU(2) \times SU(2) \times U(1)_R \times U(1)_B$, under which the fields $A_i$ transform as $(2,1)$ and have 
charges $1/2$, $+1$; while the fields $B_j$ transform as $(1,2)$ and have charges $1/2$, $-1$; and the superpotential (\ref{consup}) is invariant.
 
We refer to the literature the reader interested in the dictionary between the low energy dynamics of N D3 brane and the conical CY singularity $\cX$ \cite{Morrison:1998cs,Klebanov:1998hh,Hanany:2005ve,Franco:2005rj,Feng:2005gw,Franco:2005sm,Hanany:2005ss,Martelli:2004wu,Benvenuti:2004dy,Benvenuti:2005ja,Butti:2005sw,Franco:2005sm,Martelli:2005tp,Butti:2005ps}. It is enough to say that in the case in which the singularity $\cX$ is toric, meaning that it has at least 
$U(1)^3$ as isometry group, the dictionary to pass from the singularity to the gauge theory is well under control \cite{Hanany:2005ss,Feng:2005gw}.
The gauge theory is a quiver gauge theory of the type previously explained, and it has a specific form for the 
superpotential: all the bifundamental fields appear exactly two times, one time with the plus sign and one time with the minus sign. In what follows we will give more details about these types of gauge theories.

To give a feeling that the gauge theories proposed are reasonable, and to get familiar with these theories, 
there exist an intuitive test. As already explained a D3 brane is a point in $\cX$ and being a physical brane it can 
move all along $\cX$. Namely due to supersymmetry the D3 branes do not feel any forces along its transverse directions. 
On the other end the transverse degrees of freedom of a D3 brane are the scalar fields living on the brane constrained 
by their four dimensional interactions encoded into the action of the SCFT. 
The scalar fields are the lowest components of the chiral superfields in the theory, 
and their constraints are given by the superpotential and the gauge interactions, namely the F terms and the D terms flat 
conditions of the SCFT. Hence the vacua of the theory (moduli space), or at least a subset, 
must reproduce the transverse space $\cX$. The transverse space is a physical quantity and to describe 
it we need to fix the gauge redundancy i.e. we need to form gauge invariant combinations of the elementary 
chiral fields and then impose the superpotential constrains.

Let us see how it works in the two easy example just explained.
Let us consider the case with a single $N=1$ D3 brane: the gauge theory is abelian and 
its gauge group is the product of $U(1)$ factors. We want to show that the D3-brane
probes the Calabi-Yau singularity $\cX$.
For ${\cal N}=4$ SYM in the case $N=1$ the superpotential is trivial and the three fields $X$, $Y$, $Z$ are c-numbers 
and they are unconstrained by the supersymmetric conditions, giving a copy of
$\mathbb{C}^3$.
Also for the conifold the superpotential is trivial in the case $N=1$ and hence the four fields $A_i$, $ B_j$ 
are four unconstrained c-numbers: the coordinates of $\mathbb{C}^4$. 
We can now construct the four gauge invariant mesonic operators $ x=\tr( A_1B_1 )$, $y=\tr (A_2B_2 )$, 
$z=\tr ( A_1B_2 )$, $w=\tr (A_2B_1 )$. In the case $N=1$ the trace is trivial and the four operators are not 
independent but subject to the obvious relation $xy=wz$. 
This is indeed the equation of the conifold singularity where the D3 brane lives. 

For the generic number of colors N what happens is still intuitive. A single D3 brane is a free point 
probing the singularity $\cX$, N parallel D3 branes are mutually supersymmetric and do not feel any 
interactions among themselves, hence they behaves as $N$ free particles on $\cX$ describing $N$ copies of $\cX$. 
Because the ground state of the D3 is bosonic it implies that the system of $N$ free particles on $\cX$ is a bosonic system 
and its ground state is the N time symmetric product of $\cX$: Sym$^N(\cX)=\cX / \Sigma_N$, where $\Sigma_N$ is the group 
of permutation of N objects acting on the N copies of $\cX$. 

It is easy to see in the gauge theory that this is indeed the case. For ${\cal N}=4$ SYM the F-term equations read
\beq
\label{N4}
X Y =  Y X \hbox{ , } Y Z = Z Y \hbox{ , } Z X = X Z 
\eeq 
These are the equations for three $N\times N$ matrices, meaning that we can simultaneously diagonalize the fields 
$X$, $Y$, $Z$. Their eigenvalues are unconstrained and, taking the quotient by the remaining Weyl group, not fixed by the diagonalization, we obtain that the moduli space is Sym$^N(\mathbb{C}^3)$.

For the conifold the F-term equations are:
\bea\label{conFtermc}
& & B_1 A_1 B_2 =  B_2 A_1 B_1  \hbox{ , } B_1 A_2 B_2 = B_2 A_2 B_1\nonumber\\  
& & A_1 B_1 A_2 =  A_2 B_1 A_1 \hbox{ , } A_1 B_2 A_2  = A_2 B_2 A_1 .
\eea
we can define four composite mesonic fields which transform in the adjoint 
representation of one of the two gauge groups
\begin{equation}
x=(A_1B_1)_\alpha^\beta,\,\,\,\, y=(A_2B_2)_\alpha^\beta,\,\,\,\, z=(A_1B_2)_\alpha^\beta,\,\,\,\,  w=(A_2B_1)_\alpha^\beta
\end{equation}
and consider the four mesons $x,y,z,w$ as $N\times N$ matrices. We could use 
the second gauge group without changing the results. With a simple 
computation using the F-term conditions (\ref{conFtermc}) we derive the
matrix commutation equations
\begin{eqnarray}
\label{commc}
& & x z = z x \hbox{ , }x w = w x \hbox{ , } y z = z y\nonumber\\
& & y w = w y \hbox{ , }x y = y x \hbox{ , } z w = w z
\end{eqnarray}
and the matrix equation
\begin{equation}
\label{eqc} 
x y = w z 
\end{equation}
which is just the conifold equation. 
All the mesons commute (\ref{commc}) and the $N\times N$ matrices $x,y,z,w$ can be simultaneous diagonalized. 
The eigenvalues are required to satisfy the conifold equation
(\ref{eqc}) and therefore the moduli space is given by the symmetrized
product of $N$ copies of the conifold Sym$^N ( \mathcal{C} )$, as expected. 

In the remaining sections of this chapter we will use the two examples we have just introduced: $\mathbb{C}^3$, and the conifold $\mathcal{C}$, 
to illustrate some generic features of gauge theories on D3 branes at singularities. In section \ref{a1} 
we explain the link between D3 branes at singularities and the $AdS/CFT$ correspondence. In section \ref{b1} 
we give some basic notions of toric geometry, and in section \ref{c1} we summarize the generic features of 
the gauge theories living on D3 branes at toric CY singularities. In section \ref{d1} we illustrate some aspects of 
the mesonic moduli space and of the mesonic chiral ring. We introduce the Hilbert series, 
as interesting tools to study the chiral ring and the moduli space, 
and we define and give the basic properties of the plethystic exponential PE.

\section{Branes at Singularities and the AdS$/$CFT Correspondence}\label{a1}

The setup of branes at singularity is very useful for many theoretical and phenomenological applications: 
going from model building to the study of strongly coupled gauge theories. 
In particular with the AdS/CFT tools we can rewrite the gauge theory on the N D3 branes as a string theory 
in certain ten dimensional space time background. In principle it is possible to rewrite the non perturbative 
dynamics of the gauge theory as the dynamics of classical supergravity, and the non perturbative quantum dynamics 
of string theory as the weak coupling perturbative dynamics of the gauge theory. As we have already explained we are 
interested in singularities $\cX$ that are conical singularities, this means that the variety $\cX$ can be 
written as a real cone over a five dimensional base $H$ i.e. $\cX = C(H)$. If the cone is a CY the base $H$ is called 
a Sasaki-Einstein space. The low energy field theory is a quiver gauge theory superconformal invariant. 
In this setup the AdS/CFT correspondence makes an incredible prediction: this flat gauge theory is ``dual'' to type 
$IIB$ string theory propagating in the background $AdS_5 \times H$ with N units of five form flux turned on: 
$\int_H F_5 = N$. 
In the $\mathbb{C}^3$, and $ \mathcal{C}$ case the horizon manifolds are respectively $S^5$ and $T^{1,1}$.

Recently, there has been renewed interest in generalizing
the $AdS/CFT$  correspondence to generic Sasaki-Einstein manifolds $H$. 
This interest has been initially motivated by the discovery of new infinite classes of non compact 
$CY$ metrics \cite{Gauntlett:2004zh,Gauntlett:2004yd,Gauntlett:2004hh,Cvetic:2005ft,Cvetic:2005vk,Martelli:2005wy} and 
the construction of their dual  $\mathcal{N}=1$ supersymmetric $CFT$ \cite{Benvenuti:2004dy,Benvenuti:2005ja,Butti:2005sw,Franco:2005sm}. 

The $AdS_5 \times H$ geometry is obtained by taking the near horizon limit of the stack of  $N$ $D3$ at the tip of the cone. 
We call this correspondence a duality because it relates the strong coupling regime of the gauge theory to 
the weak classical regime of the string and vice versa. 
The duality of the correspondence is easily illustrated in the example of $\mathcal{N}=4$ gauge theory. 
In the field theory side there exist two parameters: the gauge coupling $g_{YM}^2$ and the number of colors $N$. 
In the string theory side the two parameters are the string coupling $g_s$ and the string length $\alpha '$ 
parameterizing respectively the quantum corrections to the classical string and the stringy corrections to the supergravity. 
If we define the 't Hooft coupling $\lambda = g_{YM}^2 N$, the matching between the parameters of the two sides of the 
correspondence is the following:
\begin{eqnarray}
4\pi g_s = \frac{\lambda}{N} &,& \frac{R^2}{\alpha '} = \sqrt{\lambda} 
\end{eqnarray}
The large $N$ limit in the field theory side: $N \rightarrow \infty$ with $\lambda$ fixed, correspond to the classical limit $g_s \rightarrow 0$ in the string theory side. The strong coupling limit $\lambda \rightarrow \infty $ in the field theory correspond to the $\alpha ' \rightarrow 0 $ limit of the string theory. Namely, according to the AdS/CFT correspondence, the strong coupling limit of the gauge theory is mapped to classical gravity. This kind of correspondence among 
the parameters is going to repeat, in a more difficult setup, in all the gauge gravity duality examples. 
We can rephrase the correspondence saying that in the strong coupling limit the correct degrees of 
freedom of a superconformal field theory are the classical strings of the type IIB string theory on 
some specific vacuum. The correspondence predicts the match between the spectrum of operators of the SCFT and the states 
of the string theory. Namely to every gauge invariant operator in the field theory corresponds a 
state in the string theory side. This fact will be very important for us because we will be able to 
check properties of the BPS spectrum in the weak coupling regime, using the field theory, 
and in the strong coupling regime using the string theory.

Our main interest will be to study some supersymmetric properties of these gauge theories, obtained as 
low energy limit of N D3 brane at conical CY singularities, and their string dual
realization: namely their moduli space and their BPS spectrum. 

\section{Toric CY Singularities}\label{b1}

The number of non compact conical CY singularities is infinite, and the complete understanding of the relations 
between singularities and gauge theories is far from being achieved. There exist a particular sub class of
 singularities that is very well understood: namely the toric singularities. 
Roughly speaking a six dimensional manifold is toric 
if it has at least $U(1)^3$ isometry. This is for sure the case of $\mathbb{C}^3$, and the conifold we have just analyzed. 
The family of six dimensional toric CY conical singularities 
is infinite and contains as easy examples all the abelian orbifolds of $\mathbb{C}^3$. 

\begin{figure}[t]
\begin{center}
\includegraphics[scale=0.6]{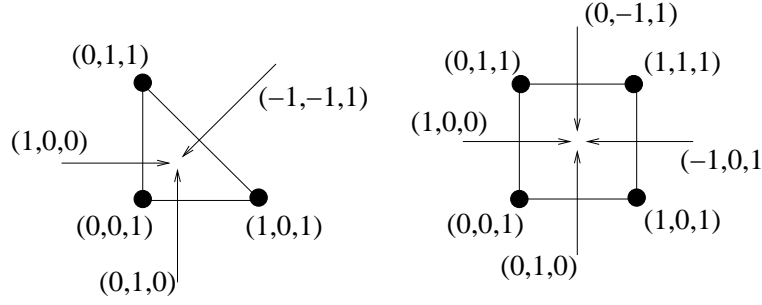} 
\caption{\small The toric diagram and the vectors generating the dual cone for a) for $\mathbb{C}^3$ and b) for the conifold.}
\label{n4conc}
\end{center}
\end{figure}

>From the algebraic-geometric 
point of view  the data of a conical toric Calabi-Yau are encoded in a 
rational polyhedral cone  $\sigma$ in $\mathbb{Z}^3$ defined by a set of
vectors $V_{i}$, $i=1,...,d$. For a CY cone, using an $SL(3, \mathbb{Z})$ transformation, 
it is always possible to carry these vectors in the form $V_{i}=(x_{i},y_{i},1)$. 
In this way the toric diagram can be drawn in the $x,y$ plane (see for example 
Figure \ref{n4conc}).
In the flat $\mathbb{C}^3$ case and in the conifold case $\mathcal{C}$ the defining vectors are respectively:
\begin{equation}
V_1=(0,0,1) \hbox{ , } V_2=(0,1,1) \hbox{ , } V_3=(1,0,1)
\end{equation} 
and
\begin{equation}
V_1=(0,0,1) \hbox{ , } V_2=(1,0,1) \hbox{ , } V_3=(1,1,1) \hbox{ , } V_4=(0,1,1)
\end{equation}
 
The CY equations can be reconstructed from this set of combinatorial data using the dual cone $\sigma^*$. 
This is defined by: 
\beq
\label{cone}
{\cal \sigma}^* = \{ {\rm y}\in \mathbb{R}^3 | l_{i} ( {\rm y}) = V^{j}_{i} y_j
\ge 0, \, i = 1 \ldots d \} \eeq 
where $V_{i}$ are the inward pointing vectors orthogonal to the facets of the polyhedral cone $\sigma^*$.  
The two cones are related as follow. The geometric generators for 
the cone $\sigma^*$, which are vectors aligned along the edges of $\sigma^*$, are the perpendicular 
vectors to the facets of $\sigma$.

To give an algebraic-geometric description of the CY, we 
need to consider the cone $\sigma^*$ as a semi-group and find 
its generators over the integer numbers. The primitive vectors pointing along 
the edges generate the cone over the real numbers but we generically need 
to add other vectors to obtain a basis over the integers. We denote by 
$W_j$ with $j=1,...,k$ a set of generators of $\sigma^*$ over the 
integers.

\begin{equation}
\label{intdual}
 \sigma^* \cap \mathbb{Z} = Z_{\geq 0 } W_1 + ... + Z_{\geq 0 } W_k
\end{equation}

The vectors generating the dual cone in the flat $\mathbb{C}^3$ case and in the conifold case $\mathcal{C}$ 
are respectively ( see Figure \ref{n4conc}):
\begin{equation}\label{c3}
W_1=(1,0,0) \hbox{ , } W_2=(0,1,0) \hbox{ , } W_3=(-1,-1,1)
\end{equation} 

and
\begin{equation}\label{con}
W_1=(1,0,0) \hbox{ , } W_2=(0,1,0) \hbox{ , } W_3=(-1,0,1) \hbox{ , } W_4=(0,-1,1)
\end{equation} 

The $k$ vectors $W_j$ generating the dual cone in $\mathbb{Z}^3$ are clearly linearly dependent 
for $k > 3$, and they satisfy the additive relations 

\begin{equation}
\label{reldual}
\sum _{j=1}^k p_{s,j} W_j = 0
\end{equation}

The three vectors in (\ref{c3}) are linearly independent, while the four vectors in (\ref{con}) satisfy the linear relation:
\begin{equation}\label{cond}
W_1 + W_3 - W_2 - W_4 =0
\end{equation}

To every vector $W_j$ it is possible to associate a coordinate $x_j$ in some 
ambient space $\mathbb{C}^k$.
The linear relations (\ref{reldual}) translate into a set of multiplicative relations among the coordinates $x_j$. 

\begin{equation}
\label{cartacoord}
 \cX = \{ (x_1, ...,x_k) \in \mathbb{C}^k | x_1^{p_{s,1}} x_2^{p_{s,2}}... x_k^{p_{s,k}} = 1\text{ for } \forall s \}  
\end{equation}

These are the algebraic equations defining the six-dimensional CY cone.

In the $\mathbb{C}^3$ case the three coordinates do not satisfy any constrains and they describe a copy of $\mathbb{C}^3$, while in the conifold they reproduce the conifold's defining equation: $x_1 x_3 = x_2 x_4 $. 

The class of toric singularities is interesting for two main reasons: It is large enough to make general 
predictions and it is easy enough to make complete computations. 
Indeed the correspondence between quiver gauge theories (considering equivalent two
theories that flow to the same IR fixed point) and singularities is well understood in the toric case \cite{Hanany:2005ve,Franco:2005rj,Hanany:2005ss,Feng:2005gw}. Namely give the geometric singularity 
it is known how to obtain the gauge theory (modulo Seiberg dualities ) and vice versa, 
from the gauge theory is is possible to reproduce the geometric singularity.

\section{Toric Quiver Gauge Theories}\label{c1}

As we have already explained a gauge theory obtained as low energy dynamics of N D3 branes at a 
singularity is of quiver type. It means that the possible gauge groups and chiral matter is highly 
restricted and can be easily described by a graph called the quiver diagram, see Figure \ref{Ccon}. 
Roughly speaking these graphs give the structure of fractional branes and open strings at the singularity. 
To complete the description of the gauge theory we must add to the graph the informations regarding the superpotential. 
This is an extra information and it is not included in the quiver graph. 

In the particular case of toric CY singularities the relation between the geometry and the gauge theory is 
well under control \cite{Benvenuti:2005ja,Franco:2005sm,Bertolini:2004xf,Hanany:2005ve,Franco:2005rj,Hanany:2005hq,Martelli:2005tp,Benvenuti:2005cz,Butti:2005ps,Butti:2005vn,Hanany:2005ss,Feng:2005gw}. 
The non toric case is still less understood: 
there exist studies on generalized conifolds \cite{Gubser:1998ia,Lopez:1998zf}, del Pezzo series \cite{Hanany:2001py,Wijnholt:2002qz,Franco:2004rt}, 
and more recently there was a proposal to construct new non toric examples \cite{Butti:2006nk}.

\begin{figure}[h]
\begin{center}
\includegraphics[scale=0.6]{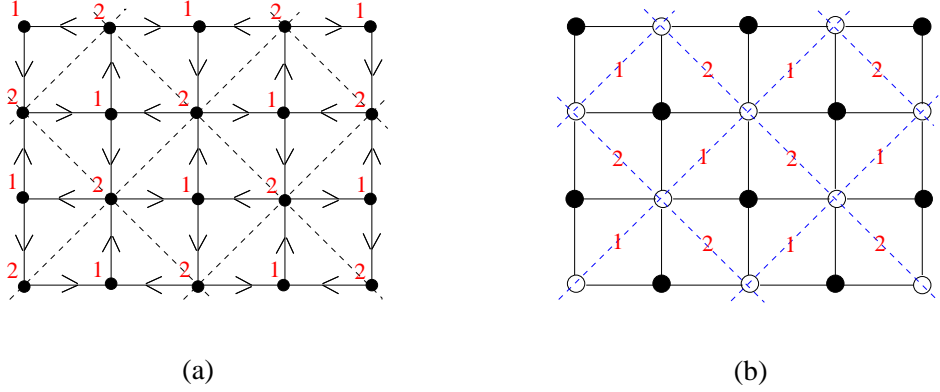} 
\caption{\small a) The periodic quiver and b) the brane tiling for the conifold. The dotted lines give the $T^2$ identifications.}
\label{boh2}
\end{center}
\end{figure}

In the toric case what make the things easier is that it is possible to include the informations 
regarding the superpotential in an graph called the periodic quiver or in its dual version called 
the brane tiling. See \cite{Kennaway:2007tq, Yamazaki:2008bt} for good reviews. Indeed in the toric case the gauge theory is completely identified by the
\emph{periodic quiver}, a diagram drawn on $T^2$ (it is the ``lift'' 
of the usual
quiver to the torus): nodes represent $SU(N)$ gauge groups, oriented
links represent chiral bifundamental multiplets and faces represent
the superpotential: the trace of the
product of chiral fields of a face gives a superpotential 
term ( see Figure \ref{boh2}). 
Equivalently the gauge theory is described by the
\emph{dimer configuration}, or \emph{brane tiling},
the dual graph of the periodic quiver, drawn also on a torus $T^2$.
In the dimer the role of faces and vertices is exchanged: 
faces are gauge groups and vertices are superpotential terms.
The dimer is a bipartite graph: it has an equal number of white and 
black vertices (superpotential terms with sign + or - respectively)  
and links connect only vertices of different colors ( see Figure \ref{boh2}).
The Lagrangian of the quiver gauge theories describe the UV degrees of freedom. 
These theories generically flow in the IR to superconformal field theories. It is well known 
that different gauge theories in the UV can flow to the same fixed point: a famous example are the electric and 
magnetic SQCD. These theories are very different in the UV but in the IR they are exactly the same, or better they are dual to each other: 
when one theory is weakly coupled the other is strongly coupled and vice versa. The 
field theory transformation that map one theory to its dual is called Seiberg duality. In the $SU(N_c)$ SQCD with $N_c$ 
colors and $N_f$ flavors $Q_i$ without superpotential and $ N_c+1 <N_f< 3N_c$ the Seiberg duality maps this 
theory to a dual SQCD with $SU(N_f-N_c)$ gauge group, 
$N_f$ dual quarks $q_i$, with reverse chirality with respect to the $Q_j$, and 
a new elementary mesonic field $T^{ij}$, that in the original theory is the composite field $Q_i \tilde{Q}_j$, and a
superpotential $\tilde{q}_iT^{ij}q_j$. The two theory are exactly equivalent in the IR and they are dual to each other. It turns out that some generalization of this Seiberg dual map acts in the quiver gauge theory as well. It 
is exactly the Seiberg dual map just described for the SQCD applied to a single $SU(N_j)$ factor in the 
$\prod_{i=1}^g SU(N_i)$ gauge group of the quiver theory.  
Indeed by applying Seiberg dualities to a quiver gauge theory we can obtain
different quivers that flow in the IR to the same CFT: to a toric
diagram we can associate different quivers/dimers describing the same IR
physics. It turns out that in the toric case one can always find phases where all the 
gauge groups have the same number of colors: $\prod_{i=1}^g SU(N_i) \rightarrow SU(N)^{g}$; these are called 
\emph{toric phases}.
Seiberg dualities keep constant the number of
gauge groups $F$, but may change the number of fields $E$, and
therefore the number of superpotential terms $V=E-F$ (since the dimer is on a torus we have: $V-E+F=0$.). Applying
Seiberg dualities it is possible to obtain a toric phase with the smallest possible number of chiral fields: these
are called minimal toric phases and all along this thesis we will mainly use these phases to describe the 
quivers gauge theories.\\
The pattern of global symmetries of the gauge theories and their relation to the geometric 
properties of the singularities will be very important for us. 
In particular non anomalous $U(1)$ symmetries play a very important role in the
gauge theory. 
For smooth horizons $H$ we expect $d$ global non anomalous symmetries (we include the R symmetry in the counting of the global symmetries), 
where $d$ is the number of sides of the toric diagram in the dual theory.
We can count these symmetries from the 
number of massless vectors in the $AdS$ dual gravity background. Indeed the AdS/CFT maps the global current operators in the SCFT 
to the gauge fields in $AdS_5$. Since the manifold H is toric, 
the metric has three $U(1)$ isometries and reducing the metric over H gives three gauge vector fields in $AdS_5$.
One of these (generated by the Reeb vector: the vector fields under which the killing spinor 
of the $\mathcal{N}=1$ supersymmetric geometry is not invariant ) corresponds to the
R-symmetry while the other two give two global flavor symmetries 
in the gauge theory. There are other $g-1$ global baryonic symmetries in the gauge theory. As already
explained these come from the $U(1)$ gauge factors that decouple in the IR. Of these $g-1$ global symmetries 
just $d-3$ have geometric origin and are non anomalous in the field theory. Indeed other gauge fields in $AdS$ come 
from the reduction of the RR four form on the non-trivial three-cycles
in the horizon manifold $H$, and there are $d-3$ three-cycles in
homology \cite{Franco:2005sm} when $H$ is smooth.
On the field theory side, these gauge fields correspond to baryonic 
symmetries.
Summarizing, the global non anomalous symmetries are:
\begin{equation}
U(1)^d=U(1)_R \times U(1)^2_F \times U(1)^{d-3}_B
\label{count}
\end{equation}
If the horizon $H$ is not smooth (that is the toric diagram has integer
points lying on the edges), equation (\ref{count}) is still true with $d$
equal to the perimeter of the toric diagram in the sense of toric
geometry (d = number of vertices of toric diagram + number of integer
points along edges). 

The $\mathcal{N}=4$ gauge theory has an $SU(4)$ global symmetry coming from 
the isometry of the $S^5$ sphere in $\mathbb{C}^3=C(S^5)$. The non anomalous abelian symmetries are the $T^3$ abelian torus contained in $SU(4)$. 
The three homology of $S^5$ is trivial (no three cycle over which reduce the RR four form), implying that the theory do not have baryonic symmetry.

In the conifold case the global symmetry of the gauge theory is easily readable from the quotient structure of $H$, indeed $T^{1,1}=\frac{SU(2) \times SU(2)}{U(1)_D}$. The field theory has a global symmetry group containing the factor $ SU(2) \times SU(2)\times U(1)$ coming from the isometry of $T^{1,1}$. The conifold theory has also a baryonic symmetry coming from the non trivial topological three cycles structure of $T^{1,1}$. Indeed topologically $T^{1,1} \sim S^2 \times S^3$ and the RR four form reduced on $S^3$ gives another vector field in $AdS_5$ corresponding to the baryonic symmetry in the conformal field theory. The abelian symmetries of the conifold correspond to the abelian $T^3$ torus contained in the isometry group of the $T^{1,1}$ times the $U(1)_B$ baryonic symmetry.

This way of proceeding is generic: the reduction of the metric feels the $U(1)^3$ isometry of the toric variety, while the reduction of the four form RR field feels the topological structure of $H$ that can be shown to be topologically a connected sum of $d-3$ $S^2 \times S^3$ spaces, namely: $H \sim \#^{d-3}( S^2 \times S^3 ) $.

In the following we will use the fact that these $d$ global non anomalous charges ( including the R charge) can be parametrized by $d$ parameters $a_1, a_2, \ldots ,a_d$ \cite{Butti:2005ps,Butti:2005vn}, each ones associated 
with a vertex of the toric diagram. There exist a simple algorithm to compute the charge of a
generic link in the dimer in function of the parameters $a_i$ in a way in which this parametrization 
automatically solve the conditions of vanishing of gauge and superpotential beta functions 
\cite{Butti:2005ps,Butti:2005vn}, and the $d-1$ independent free quantities $a_i$ parametrize
the $d-1$ global abelian non R symmetries of the gauge theory that can mix with the
R-symmetry. The value of $a_i$ at the fixed point can be found by using 
a-maximization \cite{Intriligator:2003jj}.

%\begin{figure}[h!]
%\centering
%\includegraphics[scale=0.5]{ziza.eps}
%\caption{(i) On the left: part of a toric diagram $P$ with the charge
%  distribution (a trial charge $a_i$ for every vertex $V_i$ ) and two
%  vectors $v_i$ and $v_j$ of the $(p,q)$-web. (ii) On the right:
%  the corresponding zig-zag paths and the link with charge $a_{i+1} +
%  a_{i+2} + a_{i+3}+a_{i+4}+a_j$ according to the recipe explained in
%  the main text.}\label{ziza} 
%\end{figure}

\section{The Mesonic Chiral Ring}\label{d1}

There are two main lessons we can learn from what we said till now: 
\begin{itemize}
\item{the gauge theories on D3 branes at singularities have two geometrically different abelian symmetries: the global ones plus the R symmetry, 
that we can summarize as mesonic symmetries, and the baryonic symmetries;}
\item{a sub set of the moduli space of one D3 brane at the singularity $\cX$ must reproduce the geometric 
singularity itself, and the moduli space of a gauge theory can be expressed as set equations 
among gauge invariant operators.}
\end{itemize}

Putting together these informations and the example of $\mathbb{C}^3$ and of the conifold already discussed, 
we can guess that the geometric singularity $\cX$ is reproduced in the field theory language as the constrains among 
the gauge invariant mesonics operators for $N=1$ D3 brane, and the N times symmetric product of the singularity Sym$^N(\cX)$ 
is the mesonic moduli space for generic N number of D3 branes.

We know that the gauge theory has other symmetries than the mesonic ones, and other gauge invariant operators charged under these symmetries. 
For example in the case of the conifold the operators $\det ( A_i ) $, $\det ( B_j )$ do not take vev over the N times 
symmetric product of the conifold. For these reasons we can guess that the mesonic moduli space can be obtained 
as some kind of quotient of the complete moduli space: it is the subset of the complete moduli space 
where all the operators charged under the baryonic symmetries get zero vacuum expectation values. We will 
explore the complete moduli space and the structure of the baryonic chiral ring all along this thesis. 
For the moment it is worthwhile to make some comments regarding the mesonic moduli space and its chiral ring. 
We will find again this structure as a very particular case of the general discussion about the complete moduli 
space of the gauge theory and the complete chiral ring.

There exist a very straightforward correspondence between the chiral
ring of mesonic operators in the SCFT and 
the semigroup of integer points in $\sigma^*$, the dual cone of the fan $\sigma$ \cite{Hanany:2006nm,Benvenuti:2005cz,Butti:2006hc}. 
The idea of the correspondence is simple: part of the moduli space of a 
superconformal quiver gauge theory in the abelian case $N=1$, with a single 
regular brane, is the toric CY cone $\cX$
where the D3-brane can move in the string theory set up. 
This toric cone is described by a convex rational
polyhedral cone in $\mathbb R^3$, the fan $\sigma$.
Mesonic operators are closed oriented loops in field theory and can be 
considered as well defined functions on the toric cone: the value
of the function in every point of the moduli space is the vev of 
the mesonic operator in that vacuum and such functions are the same for F-term equivalent operators.
In toric geometry the ring of algebraic functions on the toric cone 
is in one to one correspondence with the semigroup of integer points 
in the dual cone $\sigma^*$. We may then expect a one to one correspondence between the chiral 
ring of mesonic operators, equivalent up to F-terms, and the 
integer points in $\sigma^*$. This is indeed the case, as it will come out from the 
general discussion in the following chapters. 
We will see that the three integer numbers that are associated 
to the points in $\sigma^*$ are (related to) the charges of the mesons 
under the three mesonic field theory $U(1)$ (flavors and R symmetry ), coming from isometries 
of the geometry. As a consequence of this guessed correspondence to every mesons
 in the chiral ring modulo F-term equations it is associated an integer points in the dual cone $\sigma^*$.
 These integer points are nothing else that the holomorphic functions over the algebraic variety 
$\cX$. They satisfy the constrains induced by the algebraic relations defining the variety. Hence all 
the relations between points in the dual cone become relations among mesons
in the field theory. In fact,
using toric geometry and dimer technology, it is possible to show that there 
exists a one to one correspondence between the integer points inside
$\sigma^*$ and the mesonic operators in the dual field theory, modulo 
F-term constraints \cite{Martelli:2006yb,Benvenuti:2005cz}. 
To every integer point $m_j$ in  $\sigma^*$ we indeed associate a meson $M_{m_j}$ 
in the gauge theory with $U(1)^3$ charge
$m_j$.  In particular, the mesons are uniquely determined by their charge
under $U(1)^3$. The first two coordinates $Q^{m_j}=(m_j^1,m_j^2) $ 
of the vector $m_j$ are the charges of the meson under the two flavor $U(1)$ symmetries. 
Since the cone $\sigma^*$ is generated as a semi-group by the vectors $W_j$
the generic meson will be obtained as a product of basic mesons $M_{W_j}$,
and we can restrict to these generators for all our purposes.
The multiplicative relations satisfied by the coordinates $x_j$ become
a set of multiplicative 
relations among the mesonic operators $M_{W_j}$ inside the chiral ring of the gauge theory. 
It is possible to prove that these relations are a consequence
of the F-term constraints of the gauge theory.
The abelian version of this set of relations is just the
set of algebraic equations defining the CY variety as embedded in 
$\mathbb{C}^k$. 

The examples of ${\cal N}=4$ gauge theory and the conifold are shown in Figure \ref{n4conc} and were 
already discussed at the beginning of this Section. In the case of ${\cal N}=4$, the three basic mesons 
$X$, $Y$, $Z$ correspond to independent charge vectors in $\sigma  ^*$ and to the three independent coordinates 
$x$, $y$, $z$, in $\mathbb{C}^3$. In the conifold case, the four mesons $x$, $y$, $w$, $z$ correspond 
to the four vectors with one linear relation generating the dual cone $\sigma^*$ for the conifold,
and to the four embedding coordinates $x$, $y$, $w$, $z$ of the conifold in $\mathbb{C}^4$. The four basic mesonic
operators satisfy the relations induced by vector generating $\sigma^*$ and we obtain the description 
of the conifold as a quadric $xy=zw$ in $\mathbb{C}^4$ from the constrains on the mesonic chiral ring.

Mesons correspond to closed loops in the quiver 
and, as shown in \cite{Hanany:2006nm,Butti:2006hc}, for any meson
there is an F-term equivalent meson that passes for a given gauge group.
We can therefore assume that all meson loops have a base point at a 
specific gauge group and consider them as $N\times N$ matrices 
${\cal M}_\alpha^\beta$. The F-term equations imply that all mesons commute and can be 
simultaneously diagonalized. 
The additional F-term constraints require that the mesons, and therefore
all their eigenvalues, satisfy the algebraic equations defining the Calabi-Yau.
This gives a moduli space which is the $N$-fold symmetrized
product of the Calabi-Yau. This has been explicitly verified in \cite{Grant:2007ze} 
for the case of the quiver theories \cite{Benvenuti:2004dy,Benvenuti:2005ja,Butti:2005sw,Franco:2005sm} corresponding to the $L^{pqr}$ manifolds. 

\subsection{The Coordinate Ring}

Behind the previous discussion there is a more algebraic approach to the equations 
describing the geometric singularity. The modern geometry has the tendency to describe the geometric 
objects using the set of functions well defined over them. For example continuous functions for topological spaces, 
differentiable functions for differential geometry, holomorphic functions for complex manifolds. 
In the case of algebraic geometry ( namely the branch of the geometry studying the zero locus of polynomial equations ) the well defined functions are the algebraic functions: complex polynomials. 
Indeed there exist a one to one correspondence between the ring of the algebraic functions over an algebraic variety 
and the algebraic variety itself. For this reason many of the geometrical problems can be translated into 
a purely algebraic language. If we are able to get all the informations about the ring of functions of a given variety 
( the coordinate ring ) we are able to obtain all the properties of the variety itself. 
Indeed the algebraic descriptions of a variety defined as zero locus of polynomials goes in the following way: 
take an ambient space $\mathbb{C}^k$ over which the polynomials defining the variety take values, 
giving an embedding of the variety in $\mathbb{C}^k$. Over $\mathbb{C}^k$ there exist a well define ring 
$\mathbb{C}[x_1,...,x_k]$ given by all the possible complex combinations of monomials in the $x_j$ complex coordinates
of $\mathbb{C}^k$. The zero locus defining the variety is given as a set of equations over this ring of function: 
$V=\mathbb{V}(p_1,...,p_l)$, where $p_i$ are polynomials in $x_j$, and the expression $\mathbb{V}(...)$ means the 
zero locus of the polynomials into the parenthesis. There exists a one to one correspondence between the variety 
$V$ and the ideal of functions $\mathcal{I}(V)$ vanishing on $V$. The ring of functions vanishing over the variety 
$V$ is finite generated by the set of polynomials defining the the variety as a zero set in $\mathbb{C}^k$: 
$\mathcal{I}(V)= (p_1,...,p_l)$. The set of functions well defined over $V$ is given by the restrictions to the variety $V$ of 
the ring of all the polynomial functions $\mathbb{C}[x_1,...,x_k]$ over $\mathbb{C}^k$. 
This is the coordinate ring of the variety $V$ and it is given by:
\begin{equation}
\mathbb{C}[V]=\frac{ \mathbb{C}[x_1,...,x_k]}{(p_1,...,p_l)} 
\end{equation}
Now to make short a long story, once we have defined the ring of functions, the algebraic geometry tells
 us that the ``correct'' way to define algebraically a variety is to define it as the set of particular ideals, 
called the prime ideals, of the coordinate ring $\mathbb{C}[V]$. This set is called the Spec of a ring, and the 
algebraic definition of an affine variety $V$ is:
\begin{equation}\label{V}
V=\hbox{Spec} \hbox{ }\mathbb{C}[V] 
\end{equation}
Intuitively we can translate the equation (\ref{V}) as: the variety $V$ is the minimal set of points 
over which the ring of functions $\mathbb{C}[V]$ is well defined. 

In the field theory language we are interested in the moduli space of the theory, that is by definition 
an algebraic variety $\CM$, and in the chiral ring of the theory ( all the $1/2$ BPS operators), 
that are the set of functions $\mathbb{C}[\CM]$ well defined over the moduli space.
 
In the case of $\mathbb{C}^3$ and of the conifold, the equation (\ref{V}) become:
\begin{equation}\label{specc3con}
\mathbb{C}^3=\hbox{Spec} \hbox{ }\mathbb{C}[x,y,z] \hbox{  } \hbox{ , } \hbox{  } \mathcal{C}= \hbox{Spec} \hbox{ } \frac{ \mathbb{C}[x,y,w,z]}{(xy-wz)}  
\end{equation}

For the toric case all the properties of the ring of functions over $\cX$ can be obtained from the lattice 
defining the toric diagram; namely a basis of holomorphic functions over $\cX$ is given by the integer 
points of the dual cone $\sigma^*$. In algebraic geometry this fact translate into the equation:
\beq
\cX  = \mbox{Spec}[\sigma^*\cap \IZ^3] \ .
\eeq

meaning that the toric variety $\cX$ is the minimal set of points over which the ring of functions 
corresponding to the integer points in $\sigma^*$ is well defined.

\subsection{The Hilbert Series and the PE Function}\label{HSPE}

We claimed that the mesonic moduli space of the gauge theory for a single D3 brane is exactly the 
transverse space $\cX$. A Natural question is: what is the spectrum of all the mesonic $1/2$ BPS operators ? 
In the $N=1$ case the answer is pretty simple: it is the spectrum of all the holomorphic functions over $\cX$.

To collect this information in a useful way we would like to have a function $H(t; \cX)$ such that, once 
expanded in series for small values of $t$, gives the number of holomorphic functions $c_{i_1,i_2,i_3}$ over 
$\cX$ with the set of $U(1)$ charges $i_1$, $i_2$, $i_3$ under the symmetry of $\cX$:
\beq
H(t_1,t_2,t_3; \cX) = \sum_{i_1,i_2,i_3} c_{i_1,i_2,i_3} t_1^{i_1}t_2^{i_2}t_3^{i_3} \ ,
\eeq

These functions are pretty famous in mathematical literature and they are called Hilbert functions. Hilbert functions are rational functions with integer coefficients, and we will learn to appreciate their properties along this thesis.
 
We can give examples of these functions for the flat space $\mathbb{C}^3$ and the conifold.
The set of $1/2$ BPS mesonic operators can be expressed as the set of all single and multi 
trace operators subject to symmetry relations and F-terms equations. It is possible to show that the Hilbert function 
for the set of $1/2$ BPS mesonic operators, single and multi trace, in the case $N=1$ is the same of the Hilbert 
fiction for the $1/2$ BPS mesonic single trace operators in the limit $N\rightarrow \infty$ \cite{Benvenuti:2006qr}. Hence counting 
the single trace operators for $N\rightarrow \infty$ and the multi trace operators for $N=1$ gives exactly 
the same result.

In the case $N=1$ the three chiral fields of the $\mathcal{N}=4$ gauge theory parametrize a copy of $\mathbb{C}^3$. 
We can assign them the three $U(1)$ charges according their directions along the complex space. 
We now introduce three chemical potentials: $t_1$, $t_2$, $t_3$ counting the three $U(1)$ charges of the mesonic 
gauge invariant operators i.e counting the number of $X$, $Y$, $Z$ fields contained in the composite operators. 
The generic function of $\mathbb{C}^3$ is $x^i y^j z^k$ representing the $1/2$ BPS operator $X^i Y^j Z^k$. 
Hence the function counting the number of $1/2$ BPS operators is:
\beq\label{Hc31}
H(t_1,t_2,t_3; \mathbb{C}^3 ) = \sum _{i,j,k} t_1^i t_2^j t_3^k = \frac{1}{(1-t_1)(1-t_2)(1-t_3)}
\eeq
For simplicity we want to count $1/2$ BPS operators according to their dimensions. We impose $t_1=t_2=t_3=t$ and we expand
 the resulting function for small values of $t$:
\beq\label{Hc32}
H(t; \mathbb{C}^3 ) = \frac{1}{(1-t)^3} = 1 + 3 t + 6t^2 + 10t^3 + ...
\eeq
Let us read this expansion: there are 1 field of zero dimension: the identity operator $\mathbb{I}$; 
3 fields of dimension 1: $\tr(X)$, $\tr(Y)$, $\tr(Z)$; 6 fields of dimension 2: $\tr(X^2)$, $\tr(Y^2)$, $\tr(Z^2)$,
 $\tr(XY)$, $\tr(YZ)$, $\tr(XZ)$; 10 fields of dimension 3: $\tr(X^3)$, $\tr(Y^3)$, $\tr(Z^3)$,
 $\tr(X^2Y)$, $\tr(X^2Z)$, $\tr(XY^2)$, $\tr(Y^2Z)$, $\tr(XZ^2)$, $\tr(YZ^2)$,
 $\tr(XYZ)$; and so on...

For the conifold theory the situation is a bit more complicated because the ring of holomorphic functions 
is not freely generated (\ref{specc3con}). But there is another intuitive way for counting chiral operators: 
the gauge theory dual to the conifold singularity contains the global symmetry $SU(2) \times SU(2)$, and all the 
chiral mesonic operators are in representation of this symmetry. 
The elementary fields $A_i$ and $B_j$ transform respectively under the (2,1) and (1,2) representation. 
Because we consider the abelian case $N=1$ the elementary fields $A_i$ and $B_j$ are c-numbers and commute 
among themselves. Hence the chiral operators will be in the symmetric representations of $SU(2) \times SU(2)$. 
These representations have dimension $(n+1)(m+1)$. Because we are considering only mesonic operators 
we want the numbers of $A$ fields to be equal to the number of $B$ fields: the chiral mesonic operators 
transform under the symmetric representations of $SU(2) \times SU(2)$ of dimension $(n+1)^2$. 
Defining the chemical potential $t$ counting the number of elementary mesonic fields $x$, $y$, $w$, $z$ 
in the chiral mesonic ring we obtain:
\beq\label{Hcon}
H(t; \mathcal{C} ) = \sum _n (1+n)^2 t^{n} = \frac{1-t^2}{(1-t)^4}= \frac{1+t}{(1-t)^3}
\eeq

As we will see there exist a systematic way to construct such functions. The Hilbert series contains a 
lot of informations regarding the structure of the chiral ring at weak and strong coupling, 
the geometric properties of the singularity, informations regarding important matches between observables 
in the gravity and the gauge theory setup in the context of the AdS/CFT correspondence, 
and many other informations that we will learn to appreciate in the following.

As we have just seen in explicit examples the Hilbert series has the form of a rational function:
\begin{equation}
H(t;V) = \frac{ Q(t)}{(1-t)^k}
\end{equation}

where $k$ is the dimension of the embedding space, while $Q(t)$ is a polynomial of integer coefficients. 
According to the form of $Q(t)$ the moduli space of the gauge theory ( in this context the mesonic moduli space ) 
divide in three different classes: 
\begin{itemize}
\item{1. if $Q(t)=1$ the moduli space is freely generated: the generators don't have any relations 
among themselves, for example $\mathbb{C}^3$ (\ref{Hc32});}
\item{2. if $Q(t)$ is a finite product of monomials of the form $(1-t^{d_j})$: $Q(t) = \prod_j^M(1-t^{d_j})$, 
the moduli space is a complete intersection: the generators satisfy a set of algebraic relations 
such that the number of relations plus the dimension of the moduli space is equal to the number of generators; 
for example the conifold (\ref{Hcon});} 
\item{3.If $Q(t)$ cannot be written as a finite product of monomials of the form $(1-t^{d_j})$, the moduli space 
is not a complete intersection, namely the number of relations among the generators is bigger than the codimension 
of the moduli space inside the embedding space $\mathbb{C}^k$. }
\end{itemize}

The last class of models is by far the largest class. Most moduli spaces of vacua do not admit a simple 
algebraic description which truncates at some level. It is important, however, to single out and specify 
the moduli spaces which fall into the first two classes as they are substantially simpler to describe and 
to analyze. Along this thesis we will meet a lot of moduli spaces that are not complete intersections.

In the context of the counting procedure there exist a very important mathematical function called the 
Plethystic Exponential: $PE[...]$. We will see how naturally it appears in the context of counting 
BPS operators and what is its physical meaning. Its definition is the following: take a function $f(t)$ 
that vanishes at the origin, $f(0) = 0$ its Plethystic Exponential is given by\footnote{Note that in order 
to avoid an infinity the PE is defined with respect to a function that vanishes when all chemical potentials are set to zero.}:
\beq\label{defPEE}
PE \left [ f (t) \right ] := \exp\left( \sum\limits_{k=1}^\infty\frac{f (t^k)}{k}\right)  \ .
\eeq

The plethystic exponential has the role of taking a generating function for a set of operators and of counting 
all possible symmetric functions of them. It has two main roles in the counting procedure. First of all it allows to pass from the counting for one brane to the counting for generic number of branes N. 
Indeed in the case of the mesonic chiral ring the reason is easy to explain: the mesonic moduli space for 
N brane is the N times symmetric product of the mesonic moduli space for a single brane. 
Hence the chiral spectrum will be the symmetric product of the chiral spectrum for just one brane. 
Given the generating function for the mesonic chiral ring for one brane $g_1(t)$, let us define the 
chemical potential $\nu$ counting the number of D3 branes;   
we obtain the generating function $g_N(t)$ for generic $N$ applying the $\nu$-inserted\footnote{Note 
that the $\nu$ insertion satisfies the condition that the argument of PE vanishes when all chemical 
potentials are set to zero.} plethystic exponential $PE_{\nu}$: 
\beq
PE_{\nu}[ g_1(t)] = \exp\left( \sum\limits_{r=1}^\infty\frac{\nu^r g_1(t^r)}{r}\right) = \sum\limits_{N=0}^\infty g_N(t) \nu^N.
\eeq
The second role is given by its inverse function called the plethystic logarithm (PL$=$\hbox{PE$^{-1}$}). 
It is defined as:

\begin{equation}
\hbox{PE$^{-1}$}[f(t)] \equiv \sum_k^{\infty} \frac{\mu (k)}{k} \log (f(t^k)) ,
\end{equation}

where $\mu (k)$ is the M\"obius function\
\begin{equation}
\mu(k) = \left\{\begin{array}{lcl}
0 & & k \mbox{ has one or more repeated prime factors}\\
1 & & k = 1\\
(-1)^n & & k \mbox{ is a product of $n$ distinct primes}
\end{array}\right. \ .
\end{equation}

The important fact about this operator is that acting with \hbox{PE$^{-1}$} 
on a generating function we obtain the generating series for the generators and the relations in the chiral ring.
The result is generically\footnote{In the $AdS/CFT$ correspondence the moduli space of the gauge theory %$CFT$
is typically not a complete intersection variety.} an infinite series in 
which the first terms with the plus sign give the basic generators while 
the first terms with the minus sign give the relations between these basic generators. 
Then there is an infinite series of terms with plus and minus signs due to the fact 
that the moduli space of vacua is not a complete intersection and the relations in the 
chiral ring is not trivially generated by the relations between the basic invariants, 
but receives stepwise corrections at higher degrees. For the particular case of complete intersections
the series has a finite extension, with the first terms with the plus sign corresponding to the 
generators of the chiral ring, while the other terms with the minus sign corresponding to the 
relations among the generators. 
Indeed the form of this function greatly simplify in the case of the freely generated moduli space 
and in the case of complete intersection moduli spaces. 

Let us apply this function to the two easy cases we have already studied. 
$\mathbb{C}^3$ is freely generated:
\begin{equation}
PE^{-1}[H(t;\mathbb{C}^3)]=3t
\end{equation}
meaning that coordinate ring of $\mathbb{C}^3$ is generated by 3 generators: $x$, $y$, $z$ without relations.

For the conifold we have:
\begin{equation}
PE^{-1}[H(t;\mathcal{C})]=4t-t^2
\end{equation}
meaning that the coordinate ring of the conifold is generated by 4 generators of degree one: $x$, $y$, $w$, $z$, 
that satisfy a quadratic relation: $xy=wz$.

As we have just seen in the case of freely generated or complete intersection moduli space the 
expansion of $PE^{-1}$ has a finite extension, in all the other case the expansion of the $PE^{-1}$ 
will be infinite with plus and minus sign. In the following we will see some examples of this phenomenon. 

\section{Summary and Discussions}

In this chapter we introduced the subject of D3 branes at singularity $\cX$. Using the easy examples of 
$\mathbb{C}^3$ and the conifold we illustrated various properties of the IR gauge theory living on the D branes. 
We explained the general structure of toric geometry, quiver gauge theories, their moduli space and their chiral ring,
and we put this brane setup in relation with the AdS/CFT correspondence. In particular we focalized on 
the mesonic branch of the gauge theory and on the structure of the mesonic chiral ring. We tried to proceed in 
an intuitive way without paying too much attention to the mathematical rigor. In the rest of the thesis we 
will do a more formal and careful analysis of the holomorphic properties of D3 branes at singularities. 
In particular we will study the complete moduli space and the complete chiral ring, 
and the result of this chapter will be seen as a special case of this general and more rigorous analysis.

In the last part of this chapter we introduced the plethystic function PE trying to explain its 
importance in the counting procedure. Indeed the PE function acting on a Hilbert series for 
one brane is the generating function for the Hilbert series for the generic N cases, counting all the
N times symmetric products of the function counted by $N=1$ generating function. The inverse function of the
plethystic exponential PL is the generating function for the generators and the relations of the chiral ring. 
We have seen these properties in the context of the mesonic moduli space and the mesonic chiral ring,
 using the intuitive picture that the mesonic moduli space for N D3 branes is the N times symmetric product of $\cX$. 
We will be finally interested in writing partition functions for the complete spectra of BPS operators, 
including all the operators charged under the baryonic symmetries. 
We will learn that even in this much more complicated case there will be a symmetric structure 
inside the chiral counting and the PE function will be very useful for passing from the case of $N=1$ 
brane to the generic number $N$ of branes. Applying the PL function to the complete partition 
function will tell us important informations regarding the spectrum of generators of 
the chiral ring and its relations. 

We will solve the problem of counting the complete spectrum of BPS operators in the following chapters. 
This study will take us busy for long. Before entering in this very interesting problem we will analyze 
the complete moduli space $\CM$ of SCFT at singularity $\cX$. 
This space is composed by a mesonic branch, that we have just analyzed, and a baryonic branch, 
that we will analyze in the following chapter. The two branches are strictly merged and we would like to 
have a complete characterization of the total moduli space. Indeed, as we will show in the next chapter, 
it is possible to almost completely characterize the moduli space of $N=1$ brane. 
The moduli space for generic $N$ is a much harder beast. But thanks to the ability we will develop 
for counting chiral BPS operators, we will be able to extract very interesting informations 
about the moduli space for generic $N$. 
Let us start this long strip with a general analysis of the moduli spaces $\CM$ of SCFT at singularities $\cX$.

%\part{Moduli Space and Chiral Ring}

\chapter{The Master Space of $\CN=1$ Gauge Theories}\label{Master}

The vacuum moduli space $\CM$, of supersymmetric gauge theories is one
of the most fundamental quantities in modern physics.
In this chapter we want to study the full moduli space $\CM$ of the class of $\CN=1$ supersymmetric gauge 
theories obtained as the low energy limit of N D3 branes at toric conical CY singularity $\cX$. 
$\CM$ is a combination of the mesonic and baryonic branches, the former being the symmetric product of $\cX$, 
as already discussed in the previous chapter. 

In consonance with the mathematical literature, the single brane moduli space is called the master space $\f$. 
Illustrating with a host of explicit examples, we exhibit many algebro-geometric properties of the master space 
such as when $\f$ is toric Calabi-Yau, behavior of its Hilbert series, its irreducible components and its symmetries.

Once we understand the full moduli space of a class of gauge theories the next step is to study its spectra of 
$1/2$ BPS operators. 
This study will bring us in a long trip in the algebraic geometric properties of the singularity $\cX$ 
and in the AdS/CFT correspondence. We will study the chiral ring of these theories in the following chapters. 
This study will give us a lot of informations regarding the SCFT and, in the context of the moduli space $\CM$, 
will allow us to extract very interesting informations regarding the symmetries of the moduli space for the generic non abelian case with N D3 branes. We will investigate this point in chapter \ref{backtoms}.

\section{Generalities}\setall

The moduli space $\CM$ is given by the
vanishing of the scalar potential as a function of the scalar components
of the superfields of the field theory. This, in turn, is the set of
zeros, dubbed flatness, of so-called {\bf D-terms} and {\bf F-terms},
constituting a parameter, or moduli, space of solutions describing the
vacuum. The structure of this space is usually complicated, and should be best cast in the 
language of algebraic varieties.  Typically, $\CM$ consists of a union of various branches, 
such as the mesonic branch or the baryonic branch, the Coulomb branch or the Higgs branch; 
the names are chosen according to some characteristic property of the specific branch.

It is a standard fact now that $\CM$ can be phrased in a succinct
mathematical language: it is the symplectic quotient of the space of
F-flatness, by the gauge symmetries provided by the D-flatness. We will
denote the space of F-flatness by $\f$ and symmetries prescribed by
D-flatness as $G_{D^{\flat}}$, then we have
\begin{equation}\label{M}
 \CM \simeq \f // G_{D^{\flat}} \ .
\end{equation}
Using this language, we see that $\f$ is a parent space whose quotient is a moduli space. In
the mathematical literature, this parent is referred to as the {\bf
master space} \cite{master} and to this cognomen we shall
adhere.

In the context of certain string theory backgrounds, $\CM$ has an elegant geometrical realization. 
When D-branes are transverse to an affine (non-compact)
Calabi-Yau space $\cX$, a supersymmetric gauge theory exists on the world-volume of the branes. 
Of particular interest is, of course, when the  gauge theory, prescribed by D3-branes, is in four-dimensions. 
Our main interest is the IR
physics of this system, where all Abelian symmetry decouples and the gauge
symmetry is fully non-Abelian, 
typically given by products of $SU(N)$ groups. The Abelian factors
are not gauged but rather appear as global baryonic symmetries of the
gauge theory. 

Under these circumstances, the moduli space $\CM$ is a
combined mesonic branch and baryonic branch. These branches are not necessarily 
separate (irreducible) components of $\CM$ but are instead in most cases
intrinsically merged into one or more components in $\CM$. Even when mesonic
and baryonic directions are mixed, it still makes sense to talk about
the more familiar mesonic moduli space ${}^{{\rm mes}}\!{\cal M}$, as the sub-variety of $\CM$ parameterized by
mesonic operators only. Since mesonic operators have zero
baryonic charge, and thus invariant under the $U(1)$ Abelian factors,
the mesonic moduli space can be obtained as a further quotient of 
$\CM$ by the Abelian symmetries:

\begin{equation}
{}^{{\rm mes}}\!{\cal M} \simeq \CM // U(1)_{D^{\flat}} \ .
\label{mesmod}
\end{equation}

This is the subset of the full moduli space already discussed in the previous 
chapter in the case of $\mathbb{C}^3$ and of the conifold.

We are interested in the theory of physical $N$ 
branes probing the singularity; 
the gauge theory on the world-volume is then superconformal.

It is of particular interest to consider the case of a single D3-brane
transverse to the Calabi-Yau three-fold $\cX$, which will enlighten the geometrical
interpretation of the moduli space. 
Since the motion of the D-brane is parameterized by this transverse space, part of the vacuum moduli space 
$\CM$ is, {\it per constructionem}, precisely the space which the brane probes. 
The question of which part of the moduli space is going to be clarified in detail in this chapter. 
For now it is enough to specify that for a single D-brane it is the mesonic branch: 
 
\beq
\CM \supset {}^{{\rm mes}}\!{\cal M} \simeq \cX \simeq \mbox{non-compact Calabi-Yau threefold transverse to 
D3-brane.}
\eeq

In this chapter we are interested in studying the full moduli space $\CM$. 
In general, $\CM$ will be an algebraic variety of dimension greater than three. In the case of a single D3-brane, $N=1$, the
IR theory has no gauge group and the full moduli space $\CM$ is given by
the space of F-flatness $\f$. Geometrically, $\f$ is a $\IC^{dim \f -3}$ fibration over the mesonic moduli space $\cX$ given by relaxing the
$U(1)$ D-term constraints in \eref{mesmod}. Physically, $\f$ is obtained
by adding {\it baryonic} directions to the mesonic moduli space. Of course,  
we can not talk about baryons for $N=1$ but we can alternatively interpret
these directions as Fayet-Iliopoulos (FI) parameters in the stringy realization of the $N=1$ gauge theory. Indeed on the world-volume of a single D-brane there is a collection of $U(1)$ gauge groups, each giving rise to a FI parameter, which 
relax the D-term constraint in \eref{mesmod}. When these FI parameters acquire vacuum expectation values they induce non-zero values for the collection of fields in the problem and this is going to be taken to be the full moduli space  $\CM\equiv \f$. If one further restricts the moduli space to zero baryonic number we get the mesonic branch which is $\cX$, the Calabi-Yau itself.

For $N>1$ number of physical branes, the situation is more subtle. 
The mesonic moduli space,
probed by a collection of $N$ physical branes, 
is given by the symmetrized product
of $N$ copies of $\cX$ \footnote{Cf. ~\cite{Berenstein:2002ge} for a consistency analysis of this identification.}. The full moduli space $\CM$ is a bigger algebraic
variety of more difficult characterization. One of the purposes of this chapter is to elucidate this situation and to show how the properties of $\CM$
for arbitrary number of branes are encoded in the master space $\f$ for a 
single brane.
In view of the importance of the master space $\f$ for one brane even for
larger $N$, we will adopt the important convention 
that, in the rest of the chapter, the word {\bf master space}
and the symbol $\f$ will refer to the $N=1$ case, unless explicitly stated.

The case of $\cX$ being a {\bf toric Calabi-Yau
space} has been studied extensively over the last decade.
The translation between
the physics data of the D-brane world volume gauge theory and the
mathematical data of the geometry of the toric $\cX$ was initiated in
\cite{Douglas:1997de,Beasley:1999uz,Feng:2000mi1}. 
In the computational language of \cite{Feng:2000mi1,Feng:2000mi2}, the process of arriving at the toric
diagram from the quiver diagram plus superpotential was called the
forward algorithm, whereas the geometrical computation of the
gauge theory data given the toric diagram was called the inverse
algorithm. The computational intensity of these algorithms,
bottle-necked by finding dual integer cones, has been a technical
hurdle.
Only lately it is realized that the correct way to think
about toric geometry in the context of gauge theories is through the
language of {\bf dimer models} and {\bf brane tilings}
\cite{Hanany:2005ve,Franco:2005rj,Hanany:2005ss,Feng:2005gw} that we have already introduced in the previous chapter. 
Though the efficiency of this new perspective
has far superseded the traditional approach of the partial resolutions
of the inverse algorithm, the canonical toric language of the latter
is still conducive to us, particularly in studying objects beyond
$\cX$, and in particular, $\f$. We will thus make extensive use of this 
language as well as the modern one of dimers.

Recently, a so-called {\bf plethystic programme}
\cite{Benvenuti:2006qr,Feng:2007ur,Butti:2006au,Forcella:2007wk,Butti:2007jv,Forcella:2007ps} has been
advocated in counting the gauge invariant operators of supersymmetric
gauge theories, especially in the above-mentioned D-brane quiver theories.
For mesonic BPS operators, the fundamental generating function turns out to
be the Hilbert series of $\cX$ \cite{Benvenuti:2006qr,Feng:2007ur}, as already illustrated in the previous 
chapter in the case of $\mathbb{C}^3$ and of the conifold. The beautiful fact is that the full
counting \cite{Forcella:2007wk}, including baryons as well, 
is achieved
with the Hilbert series of $\f$ for one brane! Indeed, mesons have gauge-index
contractions corresponding to closed paths in the quiver diagram and the
quotienting by $G_{D^{\flat}}$ achieves this. Baryons, on the other
hand, have more general index-contractions and correspond to all paths
in the quiver; whence their counting should not involve the quotienting
and the master space should determine their counting. We will come back on this points 
in the following chapters.

In light of the discussions thus far presented, it is clear that the
master space $\f$ of gauge theories, especially those arising from toric
Calabi-Yau threefolds, is of unquestionable importance. It is therefore
the purpose of this chapter to investigate their properties in detail.
We exhibit many algebro-geometric properties of the master space $\f$
for one brane, including its decomposition into irreducible components,
its symmetry and
the remarkable property of the biggest component of being always
{\bf toric Calabi-Yau} if $\cX$ is.

We point out that even though we mainly concentrate on the master space $\f$
for one brane, we are able, using the operator counting technique, to extract important information about the 
complete moduli space $\mathcal{M}$, informations such as its symmetries for arbitrary number of branes, 
as will be shown in chapter \ref{backtoms}.

The organization of the chapter is as follows. In \ref{s:master} we introduce the concept 
of the master space $\f$ starting with various computational approaches for some orbifold singularities $\cX$, 
emphasizing on the Hilbert series. In section \ref{s:toric} we put the accent on the toric presentation of $\f$, and we will 
explain how to refine the Hilbert series. We will then explain other two very useful and more physical descriptions of $\f$: the symplectic quotient, 
known in the physics literature as the linear sigma model, and the dimers/perfect matchings realization, that represent the 
singularities as a system of fractional branes. In section \ref{s:case} we give some non orbifold examples, namely: the conifold, the Suspended Pinched Point SPP, and the complex cone over the $\mathbb{F}_0$ surface. In particular we give a very preliminary discussion on the effect of Seiberg duality on $\f$. In section \ref{s:branch} we give a 
possible interpretation of the linear branches of $\f$ as parameterizing the IR flows of the gauge theories. In section \ref{mathH} we give a proof of the 
CY condition for the biggest irreducible component of $\f$ and we illustrate 
some general mathematical properties. We conclude in section \ref{s:concM} giving a summary of the properties of
$\f$ and an illustrative table for some selected examples.

%%%%%%%%%%%%%%%%%%%%%%%%%%%%%%%%%%%%%%%%%%%%%%%%%%%
%%%%%%%%%%%%%%%%%%%%%%%%%+++++++++++++++++++++++++++++++
%%%%%%%%%%%%%%%%%%%%%%%%%%%%%%%%%%%%%%%%%%%%%%%%%%%
\section{The Master Space: the Basis and Orbifold Singularities}\label{s:master}\setall

It was realized in \cite{Beasley:1999uz} that for a single D3-brane probing a toric Calabi-Yau threefold $\cX$, 
the space of solutions to the F-terms is also a toric variety, albeit of higher dimension. In particular, for a quiver theory with $g$ nodes, it is of dimension $g+2$. Thus we have the first important property for the master space $\f$ for a toric $U(1)^g$ quiver theory:
\begin{equation}\label{dimF}
\dim(\f) = g+2 \ .
\end{equation}
This can be seen as follows. The F-term equations are invariant  under a $(\mathbb{C}^*)^{g+2}$ action,
given by the three mesonic symmetries of the toric quiver, one R and two flavor
symmetries, as well as the $g-1$ baryonic symmetries, 
including the anomalous ones.
This induces an action of $(\mathbb{C}^*)^{g+2}$ on the master space.
Moreover, the dimension of $\f$ is exactly $g+2$ as the following
field theory argument shows. We know that the mesonic moduli space has
dimension three, being isomorphic to the transverse Calabi-Yau manifold $\cX$.
As described in the introduction, the mesonic moduli space is obtained as the solution of both F-term and D-term constraints for the $U(1)^g$ quiver theory. The full master space is obtained by relaxing the $U(1)$ D-term constraints. Since an overall $U(1)$ factor is decoupled, we have $g-1$ independent baryonic parameters corresponding to the values of the $U(1)$ D-terms, which, by abuse of language, we can refer to as FI terms. As a result, the dimension of the master space is $g+2$, given by three mesonic parameters plus $g-1$ FI terms.

A second property of the Master Space is its reducibility. We will see in several examples below that it decomposes into several irreducible components, the largest of which turns out to be of the same dimension as the master space, and more importantly the largest component is a toric Calabi-Yau manifold in $g+2$ complex dimensions which is furthermore a cone over a Sasaki Einstein manifold of real dimension $2g+3$. Examples follow below, as well as a proof that it is Calabi-Yau in section \ref{mathH}.

Having learned the dimension of the master space, let us now see a few warm-up examples of what the space looks like. Let us begin with an illustrative example of perhaps the most common affine toric variety, viz., the Abelian orbifolds.

%%%%%%%%%%%%%%%%%%%%%%%%%%%%%%%%
\subsection{Warm-up: an Etude in $\f$}
The orbifold $\IC^3 / \IZ_k \times \IZ_m$ is well-studied. It is an affine toric singularity whose toric diagrams are lattice points enclosed in a right-triangle of lengths $k \times m$ (cf.~e.g., \cite{fulton}). The matter content and
superpotential for these theories can be readily obtained from
brane-box constructions \cite{Hanany:1997tb,Hanany:1998it}. We summarize as follows:
\beq\label{zkzm}
\ba{cl}
\mbox{Gauge Group Factors:} & m n; \\
\mbox{Fields:} & \mbox{bi-fundamentals} 
  \{ X_{i,j}, \ Y_{i,j}, \ Z_{i,j} \} \mbox{ from node $i$ to node $j$}
  \\
  & (i,j) \mbox{ defined modulo } (k,m) \ , \qquad
  \mbox{ total } = 3 m n; \\
\mbox{Superpotential:} & 
W = \sum\limits_{i=0}^{k-1} \sum\limits_{j=0}^{m-1} 
  X_{i,j} Y_{i+1, j} Z_{i+1, j+1} -
  Y_{ij} X_{i, j+1} Z_{i+1, j+1} \ .
\\
\ea
\eeq
We point out here the important fact that in the notation above, when
either of the factors $(k,m)$ is equal to 1, the resulting theory is
really an $\CN=2$ gauge theory since the action of $\IZ_k \times
\IZ_m$ on the $\IC^3$ is chosen so that it degenerates to have a line
of singularities when either $k$ or $m$ equals 1. In other words, if
$m=1$ in \eref{zkzm}, we would have a $(\IC^2 / \IZ_k) \times \IC$
orbifold rather than a proper $\IC^3 / \IZ_k$ one
%(in the language of \cite{su4}, this proper action would be called
%``transitive'').
We shall henceforth
be careful to distinguish these two types of 
orbifolds with this notation.

\subsubsection{Direct Computation}
Given the matter content and superpotential of an $\CN=1$ gauge theory, a convenient and algorithmic method of computation, emphasizing \eref{M}, is that of \cite{Gray:2006jb}. We can immediately compute
$\f$ as an affine algebraic variety: it is an ideal in $\IC^{3 m n}$
given by the $3mn$ equations prescribed by $\partial W = 0$. The
defining equation is also readily obtained:
it is simply the image of the ring map $\partial W$ from 
$\IC^{3 m n} \rightarrow \IC^{3 m n}$, i.e., the syzygies of the $3mn$
polynomials given by $\partial W$.
To specify $\f$ explicitly as an affine algebraic variety, let us use the notation that
\beq
\label{DDE}
(d , \delta | p) := \mbox{affine variety of dimension $d$ and degree
  $\delta$ embedded in $\IC^p$ } \ .
\eeq

Subsequently, we present what $\f$ actually is as an algebraic variety
for some low values of $(m,k)$ in Table \ref{t:fzkzm}.
We remind the reader that, of course, quotienting these above spaces by $D^{\flat}$, which in the algorithm of \cite{Gray:2006jb} is also achieved by a ring map, should all give our starting point of $\cX = \IC^3 / \IZ_k \times \IZ_m$.

\begin{table}[t]
\begin{center}
$\ba{|c||c|c|c|c|c|}\hline
m \backslash k & 1 & 2 & 3 & 4 & 5 \\ \hline \hline
1 & (3,1|3) & (4,2|6) & (5,4|9) & (6,8|12) & (7,16|15) %& (8,32|18) 
\\ \hline 
2 & (4,2|6) & (6,14|12) & (8,92|18) & (10,584|24) & (12, 3632|30)
\\ \hline
3 & (5,4|9) & (8,92|18) & (11,1620|27) & (14,26762|36) &
(17, 437038 | 45)
\\ \hline
\ea$
\end{center}
\caption{{\sf The master space $\f$ for $\IC^3 / \IZ_k \times \IZ_m$ as explicit algebraic varieties, for some low values of $k$ and $m$.}}
\label{t:fzkzm}
\end{table}

As pointed out above, the limit
of either $k$ or $m$ going to 1 in the theory prescribed in
\eref{zkzm} is really just $(\IC^2 / \IZ_k) \times \IC$ with the
$X$-fields, say, acting as adjoints. The first row and column
of Table \eref{t:fzkzm} should
thus be interpreted carefully since they are secretly $\CN=2$ theories
with adjoints. We shall study proper $\IC^3 / \IZ_k$ theories later.

%%%%%%%%%%%%%%%%%%%%%%%%
\subsubsection{Hilbert Series}\label{s:hilb1}
One of the most important quantities which characterize an algebraic variety is the Hilbert series\footnote{Note, however, that the Hilbert series is not a topological invariant and does depend on embedding.}.
In \cite{Benvenuti:2006qr,Feng:2007ur}, it was pointed out that it is also key to the problem of counting gauge invariant operators in the quiver gauge theory. We have already seen its form and its application in the simple case of $\mathbb{C}^3$ and the conifold in (\ref{Hc31},\ref{Hc32}) and (\ref{Hcon}) respectively. Let us thus calculate this quantity for $\f$.

We recall that for a variety $M$ in $\IC[x_1,...,x_k]$, the Hilbert series is the generating function for the dimension of the graded pieces:
\beq
H(t; M) = \sum\limits_{i=-\infty}^{\infty} \dim_{\IC} M_i t^i \ ,
\eeq
where $M_i$, the $i$-th graded piece of $M$ can be thought of as the number of independent degree $i$ (Laurent) polynomials on the variety $M$. The most useful property of $H(t)$ is that it is a rational function in $t$ and can be written in 2 ways:
\beq\label{hs12}
H(t; M) = \left\{
\ba{ll}
\frac{Q(t)}{(1-t)^k} \ , & \mbox{ Hilbert series of First Kind} \ ;\\
\frac{P(t)}{(1-t)^{\dim(M)}} \ , & \mbox{ Hilbert series of Second Kind} \ . 
\ea
\right.
\eeq
Importantly, both $P(t)$ and $Q(t)$ are polynomials with {\em integer}
coefficients. The powers of the denominators are such that the leading
pole captures the dimension of the manifold and the embedding space,
respectively. In particular, $P(1)$ always equals the degree of the
variety. 
For example in the equation (\ref{Hcon}) the first Hilbert series is of the First Kind: the conifold is embedded in $\mathbb{C}^4$; while the second Hilbert series is of the Second Kind: the conifold has complex dimension three.

We can also relate the Hilbert series to the Reeb vector, as is nicely explained in \cite{Martelli:2006yb}, and we will utilize this relation in the next chapter.

For now, let us present in Table \ref{hilbzkzm} the Hilbert series, in Second form, of some of the examples above in Table \ref{t:fzkzm}.
\begin{table}[t]
\begin{center}
$\ba{|c|c|c|c|}\hline
(k, m) & \f & \mbox{Hilbert Series } H(t; \f) \\ \hline
(2,2) & (6,14|12) & 
\frac{1 + 6\,t + 9\,t^2 - 5\,t^3 + 3\,t^4}{{\left( 1 - t \right) }^6}
\\ \hline
(2,3) & (8,92|18) & \frac{1 + 10\,t + 37\,t^2 + 47\,t^3 - 15\,t^4 +
  7\,t^5 + 5\,t^6}{{\left( 1 - t \right) }^8} \\ \hline
(2,4) & (10,584|24) &
\frac{1 + 14\,t + 81\,t^2 + 233\,t^3 + 263\,t^4 - 84\,t^5 + 4\,t^6 +
  71\,t^7 - 7\,t^8 + 8\,t^9}{{\left( 1 - t \right) }^{10}} \\ \hline
(3,3) &  (11,1620|27) &
\frac{1 + 16\,t + 109\,t^2 + 394\,t^3 + 715\,t^4 + 286\,t^5 - 104\,t^6
  + 253\,t^7 - 77\,t^8 + 27\,t^9}{{\left( 1 - t \right) }^{11}} \\
\hline
\ea$
\end{center}
\caption{{\sf The Hilbert series, in second form, of the master space $\f$ for $\IC^3 / \IZ_k \times \IZ_m$, for some low values of $k$ and $m$.}}
\label{hilbzkzm}
\end{table}
We see that indeed the leading pole is the dimension
of $\f$ and that the numerator evaluated at 1 (i.e., the coefficient of
the leading pole) is equal to the degree of $\f$. Furthermore, we point out that the Hilbert series thus far defined depends on a single variable $t$, we shall shortly discuss in \sref{s:hilb} how to refine this to multi-variate and how these variables should be thought of as chemical potentials.

%%%%%%%%%%%
\subsubsection{Irreducible Components and Primary Decomposition}\label{s:irred}
The variety $\f$ may not be a single irreducible piece, but rather, be
composed of various components. This is a well recognized feature in supersymmetric gauge theories. The different components are typically called {\bf branches} of the moduli space, such as Coulomb or Higgs branches, mesonic or baryonic branches. 

%Possibly the most famous case is the Seiberg-Witten solution to ${\cal N}=2$ supersymmetric gauge theories which deals mainly with the Coulomb branch but gives some attention to the other components on the moduli space which are generically called the Higgs branch. 

It is thus an interesting question to identify the different components since sometimes the massless spectrum on each component has its own unique features. We are naturally lead to a process to extract the various components which in the math literature is called {\bf primary decomposition} of the ideal corresponding to $\f$. This is an extensively studied algorithm in computational algebraic geometry (cf.~e.g.~\cite{m2}) and a convenient program which calls these routines externally but based on the more familiar Mathematica interface is \cite{stringvacua}.

\paragraph{Example of $\IC^2 / \IZ_3$: }
Let us first exemplify with the case of $(\IC^2 /
\IZ_3) \times \IC$ (i.e., $(k,m) = (1,3)$). This case, having ${\cal N}=2$ supersymmetry, is known to have a Coulomb branch and a Higgs branch which is 
a combined mesonic and  baryonic branch. The superpotential is
\begin{eqnarray}
W_{(\IC^2 / \IZ_3) \times \IC} &=& 
 X_{0,0}\,Y_{0,0}\,Z_{0,1} - 
 Y_{0,0}\,X_{0,1}\,Z_{0,1} + 
 X_{0,1}\,Y_{0,1}\,Z_{0,2} \nonumber\\
& &-Y_{0,1}\,X_{0,2}\,Z_{0,2} +
 X_{0,2}\,Y_{0,2}\,Z_{0,0} - 
 Y_{0,2}\,X_{0,0}\,Z_{0,0}  
 \ ,\nonumber\\
\end{eqnarray}
composed of a total of 9 fields, where the $X$ fields are adjoint fields in the ${\cal N}=2$ vector multiplet of the corresponding gauge group. Here, since we are dealing with a single D-brane, these fields have charge 0. Hence, $\f$ is defined, as an ideal in
$\IC^9$, by 9 quadrics:
\beq\label{f-z3}\ba{rcl}
\f_{(\IC^2 / \IZ_3) \times \IC} &=& \{
- Y_{0,2}\,Z_{0,0} + Y_{0,0}\,Z_{0,1} \ , \;
- Y_{0,0}\,Z_{0,1} + Y_{0,1}\,Z_{0,2} \ , \;
  Y_{0,2}\,Z_{0,0} - Y_{0,1}\,Z_{0,2} \ , \\
&&  \left( X_{0,0} - X_{0,1} \right) \,Z_{0,1}\ ,  \;
  \left( X_{0,1} - X_{0,2} \right) \,Z_{0,2}\ ,  \;
  \left( -X_{0,0} + X_{0,2} \right) \,Z_{0,0}\ , \\
&&  \left( -X_{0,0} + X_{0,2} \right) \,Y_{0,2}\ ,  \;
  \left( X_{0,0} - X_{0,1} \right) \,Y_{0,0}\ ,  \; 
  \left( X_{0,1} - X_{0,2} \right) \,Y_{0,1}
\} \ .
\ea\eeq
Immediately one can see that on one branch, the so-called Higgs branch, which we shall denote as 
$\f_1$, the adjoint fields $X$ do not participate. Thus it is defined by the first 3 equations in \eref{f-z3}: 3 quadrics in 6 variables. 
Furthermore, one can
see that one of the quadrics is not independent. Therefore $\f_1$ is a complete intersection of 2
quadratics in $\IC^6$, of dimension 4. To this we must form a direct
product with the Coulomb branch which is parametrized by the $X$-directions, which turns out to be one dimensional
$X_{0,0} = X_{0,1} = X_{0,2}$ (in order to satisfy the remaining 6
equations from \eref{f-z3} such that the $Y$'s and $Z$'s are
non-zero). Hence, $\f$ is 5-dimensional (as we expect from \eref{dimF} since there are $g=3$ nodes), of degree 4, and composed of 2 quadrics in $\IC^6$ crossed with $\IC$.

Now, this example may essentially be observed with ease, more
involved examples requires an algorithmic approach, as we shall see in 
many cases which ensue. 
The decoupling of
the $X$'s is indicative of the fact that 
%we have an non-transitive action and 
this is indeed just an orbifold of $\IC^2$.

\begin{figure}[t]\begin{center}
$\ba{ccc}
\ba{l} \epsfxsize = 2.5cm\epsfbox{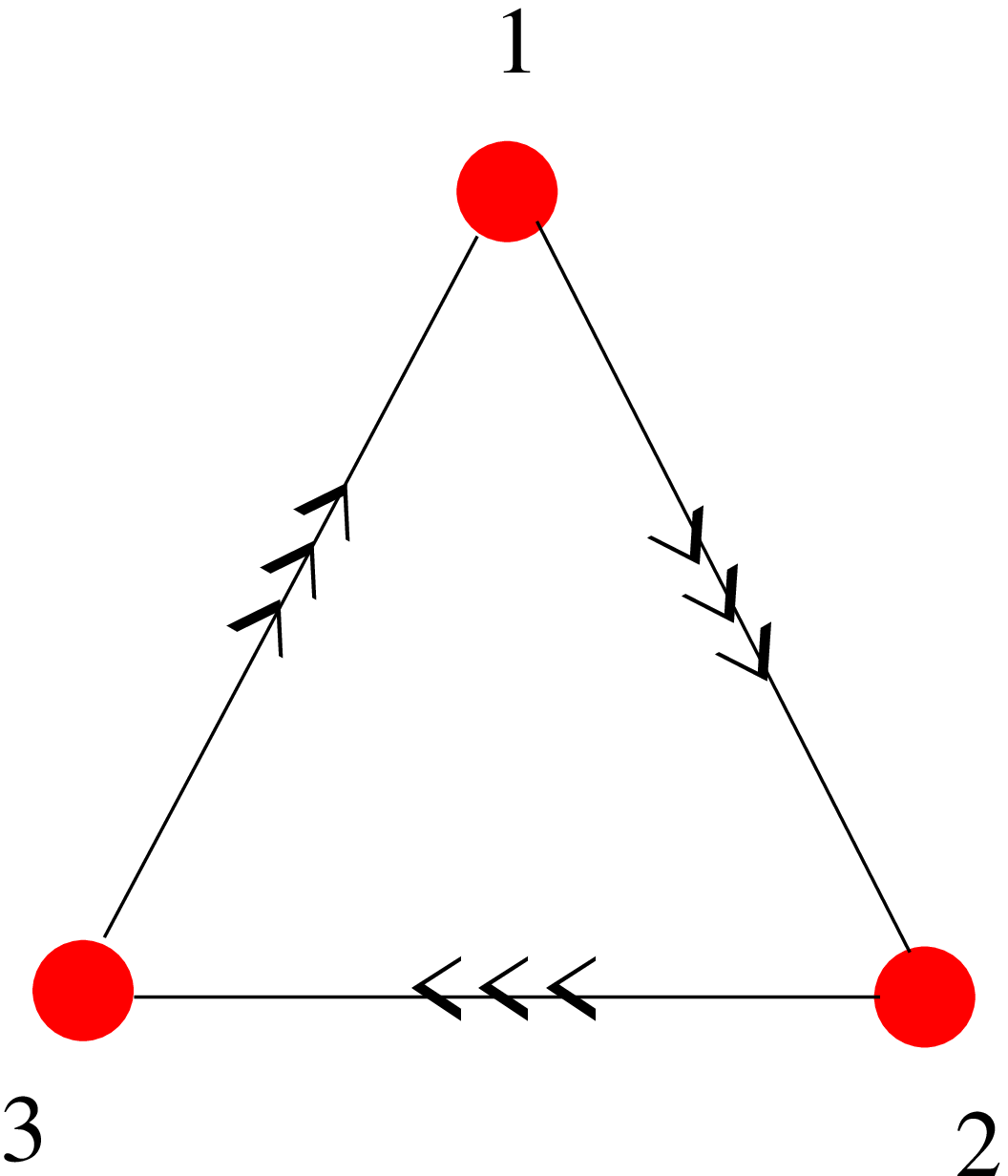} \ea
& \qquad &
\ba{l}
W_{\IC^3 / \IZ_3}=\epsilon_{i j k} X^{i}_{12}
  X^{j}_{23} X^{k}_{31} \\
\mbox{9 fields and 27 GIO's} \\
X^{i}_{12}, X^{j}_{23},
X^{k}_{31}, \hbox{  } i, j, k=1,2,3
\ea
\ea$
\caption{{\sf The quiver diagram and superpotential for $dP_0$.}}
\label{f:dP0theory}
\end{center}\end{figure}

\paragraph{Example of $dP_0 = \IC^3 / \IZ_3$: }
Let us next study a proper orbifold $\IC^3 / \IZ_3$ with a
non-trivial action, say $(1,1,1)$, on the $\IC^3$. This is also referred to in the literature as $dP_0$, 
the cone over the zeroth del Pezzo surface. In other words, this is the total space of the line bundle 
$\cO_{\IP^2}(-3)$ over $\IP^2$, once the $\IP^2$ is blown down. Here, there are 9 fields and the theory is summarized in \fref{f:dP0theory}.
Now, the F-terms are 
\beq\label{fc3z3}\ba{rcl}
\f_{\IC^3 / \IZ_3} & = & \{
-X_{2,3}^3\,X_{3,1}^2 + X_{2,3}^2\,X_{3,1}^3 \ , \;
X_{2,3}^3\,X_{3,1}^1 - X_{2,3}^1\,X_{3,1}^3\ , \;
-X_{2,3}^2\,X_{3,1}^1 + X_{2,3}^1\,X_{3,1}^2, \\
&&  X_{1,2}^3\,X_{3,1}^2 - X_{1,2}^2\,X_{3,1}^3\ , \;
-X_{1,2}^3\,X_{3,1}^1 + X_{1,2}^1\,X_{3,1}^3\ , \;
X_{1,2}^2\,X_{3,1}^1 - X_{1,2}^1\,X_{3,1}^2, \\
&&-X_{1,2}^3\,X_{2,3}^2 + X_{1,2}^2\,X_{2,3}^3\ , \;
  X_{1,2}^3\,X_{2,3}^1 - X_{1,2}^1\,X_{2,3}^3\ , \;
  -X_{1,2}^2\,X_{2,3}^1 +  X_{1,2}^1\,X_{2,3}^2
\}
\ea\eeq
and a direct primary decomposition shows that $\f$ is itself
irreducible and it is given, using the notation in (\ref{DDE}), as
\beq\label{FdP0}
\f_{\IC^3 / \IZ_3} \simeq (5,6 | 9), 
\eeq
a non-complete-intersection of 9 quadrics as given in \eref{fc3z3}, embedded in $\IC^9$. We see that the dimension is 5 since there are 3 nodes. The Hilbert
series (cf.~\cite{Butti:2007jv} and a re-derivation below) is
\begin{equation}
H(t; \f_{\IC^3 / \IZ_3}) = \frac {1+4t+t^2}{(1-t)^5} .
\label{HSdP0-1}
\end{equation}

\paragraph{Example of $\IC^3 / \IZ_2 \times \IZ_2$: }
Finally, take the case of $(k,m)=(2,2)$, or the Abelian orbifold $\IC^3 / \IZ_2 \times \IZ_2$, studied in detail by \cite{Beasley:1999uz,Feng:2000mi1}. The reader is referred to \fref{f:pmC3Z2Z2}.
% which we present in the next section.
Here, there are $2^2=4$ nodes and we expect the dimension of the master space to be 6.
Again, we can obtain $\f$ from \eref{zkzm} and perform primary decomposition
on it. We see, using \cite{m2}, that there are 4 irreducible components, three of which are merely coordinate planes and trivial. Along these directions the gauge theory admits an accidental supersymmetry enhancement to ${\cal N} = 2$ and each direction can be viewed as a Coulomb branch of the corresponding ${\cal N} = 2$ supersymmetric theory. 

The non-trivial piece of  $\f_{\IC^3 / \IZ_2 \times \IZ_2}$ is a Higgs branch and is an irreducible variety which we shall call $\firr{\IC^3 / \IZ_2 \times \IZ_2}$; it is also of dimension 6. Moreover, it is of degree 14, and is prescribed by the intersection of 15 quadrics in 12 variables. The Hilbert series for $\firr{\IC^3 / \IZ_2 \times \IZ_2}$ is given by
\begin{equation}\label{Hz2z2}
H(t;~\firr{\IC^3/\IZ_2\times \IZ_2})=\frac{1+6 t+6 t^2+t^3}{(1-t)^6} \ .
\end{equation}

%%%%%%%%%%%%%%%%
\paragraph{Summary: }q
We have therefore learned, from our few warm-up exercises, that one can compute $\f$ directly, its Hilbert series, dimension, degree, etc., using the methods of computational algebraic geometry, using, in particular, computer packages such as \cite{m2}. In general, the master space $\f$ need not be irreducible. We will see this in detail in the ensuing sections. The smaller components are typically referred to as {\bf Coulomb Branches} of the moduli space.

The largest irreducible component of the master space $\f$ will play a special 
r\^{o}le in our analysis and deserves a special notation. We will denote it
$\firr{~}$. In the toric case, it is also known as the
{\bf coherent component} of the master space.  In all our examples, this component actually has the same dimension as the full master space and, as we will see in detail example by example and in general in section \ref{mathH}, is in fact Calabi-Yau. Let us redo Table \ref{hilbzkzm}, now for the coherent component; we present the result in Table \ref{hilbzkzm-irr}.

\begin{table}[h]
\begin{center}
$\ba{|c|c|c|c|}\hline
(k, m) & \firr{~} & \mbox{Hilbert Series } H(t; \firr{~}) 
\\ \hline
(2,2) & (6,14|12) & 
\frac{1 + 6\,t + 6\,t^2 + t^3}{{\left( 1 - t \right) }^6}
\\ \hline
(2,3) & (8,92|18) & \frac{1 + 10\,t + 35\,t^2 + 35\,t^3 + 10\,t^4 +
  t^5}{{\left( 1 - t \right) }^8} \\ \hline
(2,4) & (10,584|24) &
\frac{1 + 14\,t + 78\,t^2 + 199\,t^3 + 199\,t^4 + 78\,t^5 + 14\,t^6 +
  t^7}{{\left( 1 - t \right) }^{10}} \\ \hline
(3,3) &  (11,1620|27) &
\frac{1 + 16\,t + 109\,t^2 + 382\,t^3 + 604\,t^4 + 382\,t^5 + 109\,t^6
  + 16\,t^7 + t^8}{{\left( 1 - t \right) }^{11}} \\
\hline
\ea$
\end{center}
\caption{{\sf 
The Hilbert series, in second form, of the coherent component of the master space $\f$ for $\IC^3 / \IZ_k \times \IZ_m$, for some low values of $k$ and $m$. The reader is referred to Table \ref{hilbzkzm} for contrast.}}
\label{hilbzkzm-irr}
\end{table}

We note that the degree and dimension of $\firr{~}$ is the same as that of $\f$, again suggesting that the smaller dimensional components are merely linear pieces. Nevertheless the linear pieces play a crucial r\^{o}le in the analysis of the physics of these theories since there is a rich structure of mixed Higgs and Coulomb branches; we will see this in \sref{s:branch}.
Moreover, we observe that the numerator now becomes symmetric (palindromic), a remarkable fact that will persist throughout our remaining examples; we will show why in section \ref{mathH}.

%%%%%%%%%%%%%%%%
\section{Toric Presentation}
\label{s:toric}
We have so far seen the application of computational algebraic geometry in studying the master space as an explicit algebraic variety. 
This analysis has not fully exploited the fact that $\f$ is in fact a toric variety; that is, we have been simplifying and primary decomposing 
the ideal corresponding to $\f$ without utilizing its inherent combinatorial, toric nature. 
Now, given an ideal ${\cal I}$ of a polynomial ring, when each generator of ${\cal I}$, i.e., each polynomial equation, 
is written in the form ``monomial = monomial,'' then ${\cal I}$ is known as a {\bf toric ideal} and the variety thus 
defined will be toric \cite{sturmfels}. The F-terms arising from the partials of the 
superpotential in \eref{zkzm} clearly obey this condition and this is true for all toric Calabi-Yau spaces $\cX$.

A single matrix can in fact encode all the information about the ideal of $\f$, called the {\bf $K$-matrix} in \cite{Beasley:1999uz,Feng:2000mi1}. 
For orbifolds of the form $\IC^3 / \IZ_n$ with action $(a,b,-1)$ the $K$-matrix were constructed in Eqs 4.1-3 of \cite{Douglas:1997de} 
and that of  $\IC^3 / \IZ_3 \times \IZ_3$, in \cite{Beasley:1999uz,Feng:2000mi1} (see also \cite{Muto:2001gu,Sarkar:2000iz}). 
In general, the procedure is straight-forward: solve the F-terms explicitly so that each field can be written as a 
fraction of monomials in terms of a smaller independent set. Then, translate these monomials as a matrix of exponents; 
this is the $K$-matrix.

We have seen from above that the master space $\f$ and its coherent component $\firr{~}$ of a toric $U(1)^g$ quiver gauge theory is a 
variety of dimension $g+2$. The F-terms provide $E$ equations for the $E$ fields in the quiver. 
Not all of them are algebraically independent, since the F-terms are invariant
under the $(\mathbb{C}^*)^{g+2}$ toric action. It follows that the $E$
fields can be parameterized in terms of $g+2$ independent fields.   
$K$ is therefore a matrix of dimensions $g+2$ by $E$.

%%%%%%%%%%%%
\paragraph{$\IC^3 / \IZ_3$ Revisited: }
For the $\IC^3 / \IZ_3$ example above let us illustrate the
procedure. Solving \eref{fc3z3}, we have that
\beq
{{X_{1,2}^1}= 
    {\frac{X_{1,2}^3\,X_{3,1}^1}{X_{3,1}^3}}}, \
  {{X_{1,2}^2}= 
    {\frac{X_{1,2}^3\,X_{3,1}^2}{X_{3,1}^3}}}, \ 
  {{X_{2,3}^1}= 
    {\frac{X_{2,3}^3\,X_{3,1}^1}{X_{3,1}^3}}}, \
  {{X_{2,3}^2}= 
    {\frac{X_{2,3}^3\,X_{3,1}^2}{X_{3,1}^3}}} \ .
\eeq
We see that there are 5 fields $\{X_{1, 2}^3, X_{2, 3}^3, X_{3, 1}^1,
  X_{3, 1}^2, X_{3, 1}^3\}$ which parameterize all 9 fields, signifying
  that $\f$ is 5-dimensional, as stated above.
Whence we can plot the 9 fields in terms of the 5 independent ones as:
\beq
\ba{c|ccccccccc}
& X_{1, 2}^1 & X_{1, 2}^2& X_{1, 2}^3& X_{2, 3}^1& X_{2, 3}^2& X_{2,
  3}^3&  X_{3, 1}^1& X_{3, 1}^2& X_{3, 1}^3 
\\ \hline
X_{1, 2}^3 & 1 & 1 & 1 & 0 & 0 & 0 & 0 & 0 & 0 \\
X_{2, 3}^3 & 0 & 0 & 0 & 1 & 1 & 1 & 0 & 0 & 0 \\
X_{3, 1}^1 & 1 & 0 & 0 & 1 & 0 & 0 & 1 & 0 & 0 \\
X_{3, 1}^2 & 0 & 1 & 0 & 0 & 1 & 0 & 0 & 1 & 0 \\
X_{3, 1}^3 & -1 & -1 & 0 & -1 & -1 & 0 & 0 & 0 & 1
\ea
\eeq
where we read each column as the exponent of the 5 solution
fields. This is the $K$-matrix, and captures the toric information
entirely. 
In particular, the number of rows $g+2$ of the $K$-matrix is the dimension of $\f$ and the columns of $K$ are the charge vectors of the toric action
of $(\mathbb{C}^*)^{g+2}$ on $\f$. 

The $K$-matrix gives a nice toric presentation
for the coherent component $\firr{~}$ of the master space. 
It defines an integer cone $\sigma_K^*$ in $\mathbb{Z}^{g+2}$
prescribed by the non-negative integer span of the columns of
$K$. Then, in the
language of \cite{fulton}, $\firr{~}$ as an algebraic (toric) variety of
dimension $g+2$, is given by
\beq
\firr{~} \simeq \mbox{Spec}_{\IC}[\sigma_K^* \cap \IZ^{g+2}] \ .
\eeq

This is the generalization to $g+2$ dimensions of the construction of toric varieties we have already explained in the previous chapter. 
The cone generated by the $K$ matrix is just the dual cone $\sigma^*$ in the three dimensional language, and the columns of $K$ are the multidimensional vectors corresponding to the generators of $\sigma^*$.

Now, the toric diagram of the variety is not, customarily, given by 
$\sigma^*$, but, rather, by the dual cone $\sigma$ ( this is the $g+2$ dimensional generalization of the three dimensional fan described 
in the previous chapter ). 
Let us denote the generators of $\sigma$ as the matrix $T$, then, using the
algorithm in \cite{fulton}, we can readily find that

\beq\label{TdP0}
{\scriptsize
T = \left( \begin{matrix}
   0 & 0 & 1 & 1 & 0 & 0 \cr
   0 & 0 & 1 & 0 & 1 & 0 \cr
   1 & 0 & 0 & 0 & 0 & 1 \cr
   0 & 1 & 0 & 0 & 0 & 1 \cr
   0 & 0 & 1 & 0 & 0 & 1 \cr  \end{matrix} \right) \ .
}
\eeq
$T$ is a matrix of dimensions $g+2$ by $c$, where the number of its columns, $c$, is a special combinatorial number which is specific to the particular toric phase \cite{Feng:2000mi1,Feng:2000mi2}. We recall that the dual cone consists of all lattice points which have non-negative inner product with all lattice points in the original cone. In terms of our dual matrices,
\beq\label{Pmat}
P := K^t \cdot T \ge 0 \ .
\eeq
The columns of $T$, plotted in $\IZ^5$, is then the toric diagram, and the number of vectors needed to characterize the toric diagram in $\IZ^5$ is $c$ which for our particular case is equal to 6. The $P$ matrix takes the form
\beq\label{PdP0}
P= {\tiny \left(
\begin{array}{llllll}
 1 & 0 & 0 & 1 & 0 & 0 \\
 0 & 1 & 0 & 1 & 0 & 0 \\
 0 & 0 & 1 & 1 & 0 & 0 \\
 1 & 0 & 0 & 0 & 1 & 0 \\
 0 & 1 & 0 & 0 & 1 & 0 \\
 0 & 0 & 1 & 0 & 1 & 0 \\
 1 & 0 & 0 & 0 & 0 & 1 \\
 0 & 1 & 0 & 0 & 0 & 1 \\
 0 & 0 & 1 & 0 & 0 & 1
\end{array}
\right) } .
\eeq

In fact, one can say much more about the product matrix $P$, of dimensions $E$ by $c$: it consists of only zeros and ones. In \cite{Hanany:2005ve,Franco:2005rj,Feng:2005gw,Franco:2006gc}, it was shown that this matrix, which translates between the linear sigma model fields and space-times fields, also encodes perfect matchings of the dimer model description of the toric gauge theory. This provides a more efficient construction of the master space. We will return to a description of this in \sref{s:dimer}.

%%%%%%%%%%%%%%%%%%%%%%
\paragraph{$\IC^3 / \IZ_2 \times \IZ_2$ Revisited: }
Next, let us construct the $K$-matrix for our $\IC^3 / \IZ_2 \times
\IZ_2$ example. We recall that the master space and its coherent component are of dimension 6. Using the superpotential
\eref{zkzm} to obtain the 12 F-terms, we can again readily solve the
system and obtain
\beq
{\scriptsize
K = \left( \begin{array}{cccccccccccc}
1 & 1 & 1 & 1 & 0 & 0 & 0 & 0 & 0 & 0 & 0 & 0 \cr 
   0 & 1 & 1 & 0 & 0 & 1 & 1 & 0 & 0 & 0 & 0 & 0 \cr 
   0 & -1 & 
    -1 & 0 & 1 & 0 & 0 & 1 & 0 & 0 & 0 & 0 \cr 1 & 1 & 
   0 & 0 & 0 & 0 & 0 & 0 & 1 & 1 & 0 & 0 \cr 0 & 0 & 
   0 & 0 & 1 & 0 & 1 & 0 & 1 & 0 & 1 & 0 \cr -1 & 
    -1 & 0 & 0 & -1 & 0 & -1 & 0 & 
    -1 & 0 & 0 & 1 \cr 
\end{array} \right) \ ,
}
\eeq
giving us the toric diagram with 9 vectors in 6-dimensions as
\beq\label{Tz2z2}
{\scriptsize
T = \left( \begin{array}{ccccccccc}
   0 & 0 & 1 & 0 & 0 & 1 &0 & 0 & 1   \cr 
   0 & 1& 0& 1 & 0 & 0 & 0 & 1 & 0   \cr 
0 & 1 & 0 & 1 & 0 & 0 &    0 & 0 & 1 \cr 
1 & 0 & 0 & 1 & 0 & 0 & 1 & 0 & 0   \cr 
   1 &0 & 0 & 0 & 1 & 1 & 0 & 0 & 0  
\cr 1 & 0 & 0 & 1 & 0  & 1 & 0 & 0 & 0  \cr  
\end{array} \right) \ .
}
\eeq

%%%%%%%%%%%%%%%%%%%%%%
\subsection{Computing the Refined Hilbert Series} 
\label{s:hilb}
Let us now study the Hilbert series in the language of the $K$-matrix.
We mentioned in section \ref{s:hilb1} that the Hilbert series should be refined and 
we have already seen the $\mathbb{C}^3$ example in (\ref{Hc31}). 
This is easily done and is central to the counting of gauge invariants in the plethystic programme. 
Recall that the master space $\f$ and its coherent component
$\firr{~}$ are given by a set of algebraic equations
in $\IC[X_1,...,X_E]$, where $E$ is the number of fields in the quiver.
Since we are dealing with a toric variety of dimension $g+2$ we have
an action of $(\IC^*)^{g+2}$ on $\f$ and $\firr{}$ and we can give
$g+2$ different weights to the variables $X_i$. 

What should these weights be ? Now, all information about
the toric action is encoded in the matrix $K$. Therefore, a natural weight is to simply use the columns of $K$! There are $E$ columns, each being a vector of charges of length $g+2$, as needed, and we can assign the $i$-th column to the variable $X_i$ for $i=1,\ldots,E$. Since each weight is a vector of length $g+2$, we need a $g+2$-tuple for the counting which we can denote by $\underline{t}={t_1,...,t_{g+2}}$. Because the dummy variable $\underline{t}$ keeps track of the charge, we can think of the components as chemical potentials \cite{Benvenuti:2006qr,Butti:2007jv}. With this multi-index variable (chemical potential)
we can define the {\bf Refined Hilbert Series} of $\f$ as
the generating function for the dimension of the graded pieces:
\beq\label{refineHilb}
H(\underline{t}; \f) = \sum_{{\underline \alpha}} \dim_{\IC} \f_{{\underline \alpha}}\, {\underline t}^{\underline \alpha} \ ,
\eeq
where $\f_{\underline \alpha}$, the ${\underline \alpha}$-th multi-graded piece of $\f$, can be thought of as the number of independent multi-degree ${\underline \alpha}$ Laurent monomials on the variety $\f$. A similar expression
applies to $\firr{~}$. 

The refined Hilbert series for $\f$ and $\firr{~}$ can be computed
from the defining equations of the variety, using computer algebra program
and primary decomposition, as emphasized in \cite{Gray:2006jb}. 
In addition, for the coherent component $\firr{~}$, 
there exists an efficient algorithm \cite{M2book} 
for extracting the refined Hilbert series from the matrix $K$ that can be
implemented in Macaulay2 \cite{m2}. 
The reader interested in the actual code could look at \cite{Forcella:2008bb}.

A crucial step in the above analysis seems to rely upon our ability to
explicitly solve the F-terms in terms of a smaller independent set of
variables. This may be computationally intense. For Abelian orbifolds
the solutions may be written directly using the symmetries of the
group, as was done in \cite{Douglas:1997de,Beasley:1999uz}; 
in general, however, the $K$-matrix may not be immediately obtained.
We need, therefore, to resort to a more efficient method.

%%%%%%%%%%%%%%%%%%%%%%%%%
\subsection{The Symplectic Quotient Description}\label{s:Molien}
There is an alternative and useful description of the toric 
variety $\firr{~}$ as a symplectic quotient \cite{fulton}. In the math language this is also known as the 
{\bf Cox representation} of a toric variety \cite{cox} 
and in physics language, as a {\bf linear sigma model}. We will review more carefully this ``alternative'' construction 
of toric varieties in chapter \ref{D3SE} where it will be useful to understand the structure of line bundles and divisors of a toric variety. In this chapter we will show that thanks to this representation there exist a nice way of computing the refined Hilbert series using the Molien invariant. 

Now, the $c$ generators of the dual cone $T$ are not independent in $\mathbb{Z}^{g+2}$. The kernel of the matrix $T$
\begin{equation}
T \cdot Q =0
\end{equation}
or, equivalently, the kernel of the matrix $P = K^t \cdot T$
\begin{equation}
P \cdot Q =0
\end{equation}
defines a set of charges for the symplectic action. The $c-g-2$ rows of $Q^t$
define vectors of charges for the $c$ fields in the linear sigma model 
description of $\firr{~}$ \cite{Douglas:1997de,Beasley:1999uz,fulton}:
\begin{equation}
\firr{~} = \mathbb{C}^c//Q^t \ .
\end{equation}

A crucial observation is that if the rows of $Q^t$ sum to zero, then $\firr{~}$ is Calabi-Yau. 
In the following we will see that this is the case for all the examples we
will encounter. Indeed, it is possible to show this is general; for clarity
and emphasis we will leave the proof of the fact to the section \ref{mathH} and first marvel at this fact for the detailed examples.

The refined Hilbert series for $\firr{~}$ can be computed using the
Molien formula \cite{Butti:2007jv,Pouliot:1998yv}, by projecting the trivial Hilbert series of $\mathbb{C}^c$ onto $(\mathbb{C}^{*})^{c-g-2}$ invariants. We will need
in the following the refined Hilbert series depending on some or all of the
$g+2$ chemical potentials $t_i$ and, therefore, we keep our
discussion as general as possible. The dependence on the full set of
parameters $t_i$ is given by using the Cox homogeneous coordinates for
the toric variety \cite{cox}. We introduce $c$ homogeneous variables $p_\alpha$ with chemical potentials 
$y_\alpha, \alpha=1,...,c$ acted on by $(\mathbb{C}^{*})^{c-g-2}$ with charges 
given by the rows of $Q^t$.
The Hilbert series for $\mathbb{C}^c$ is freely generated and is simply:
\begin{equation}
H(\underline{y},\IC^c) = 
H(\{y_{\alpha}\},\IC^c) = \prod_{\alpha=1}^c \frac{1}{1-y_\alpha} \ ,
\end{equation}
where we have written $\underline{y}$ as a vector, in the notation of \eref{refineHilb}, to indicate refinement, i.e., $H$ depends on all the 
$\{y_{\alpha}\}$'s.

Next, the vector of charges of $p_\alpha$ under the $(\mathbb{C}^{*})^{c-g-2}$ action is given by $\{Q_{1\alpha},...,Q_{c-g-2,\alpha}\}$. 
By introducing $c-g-2$ $U(1)$ chemical potentials $z_1,...,z_{c-g-2}$
% and the multi-index notation ${\underline z}^{\underline q}=\prod\limits_{i=1}^{c-g-2} z_i^{q_i}$ 
we can write the Molien formula, which is a localization formula of the Hilbert series from the ambient space to the embedded variety of interest, as
\begin{eqnarray}\label{Molieneq}
H(\underline{y},~\firr{~}) &=&
\int \prod_{i=1}^{c-g-2} \frac{dz_i}{z_i} H(\{y_{\alpha} \ z_1^{Q_{1\alpha}} \ldots z_{c-g-2}^{Q_{c-g-2,\alpha}} \},\IC^c) \nonumber\\
&=&
\int \prod_{i=1}^{c-g-2} \frac{dz_i}{z_i} \prod_{\alpha=1}^c \frac{1}{1-y_\alpha z_1^{Q_{1\alpha}} \ldots z_{c-g-2}^{Q_{c-g-2,\alpha}}} 
%\prod_{k-1}^{c-g-2} z_k^{Q_{i\alpha}}
 \ .
\end{eqnarray}
The effect of the integration on the $U(1)$ chemical potentials $z_i$ is
to project onto invariants of the $U(1)$'s. In this formula we integrate over the unit circles in $z_i$ and we should take $|y_\alpha|<1$.

Due to the integration on the $c-g-2$ variables $z_i$, the final result 
for the Hilbert series depends only on $g+2$ independent combinations
of the parameters $y_\alpha$, which can be set in correspondence with the
$g+2$ parameters $t_i$. We can convert the $y_\alpha$ variables to the set
of independent $g+2$ chemical potential $t_i$ for the toric action using the 
matrix $T$ as \cite{cox} 
\begin{equation}
t_i =\prod_{\alpha=1}^c y_\alpha^{T_{i\alpha}} \ .
\label{toricsympaction}
\end{equation}

Recall that the $g+2$  variables $t_i$ are the chemical potentials
of the $g+2$ elementary fields that have been chosen to parametrize $\f$. The weight of the $i$-th elementary field
$X_i$ for $i=1,...,E$ under this parameterization is given by 
the $i$-th column of the matrix $K$. Denoting with $q_i\equiv 
q_i(\underline{t})$
the chemical potential for the $i$-th field we thus have
\begin{equation}
q_i =   \prod_{\alpha=1}^c y_\alpha^{P_{i\alpha}}
\label{toricsympaction2}
\end{equation}
where we used $K^t\cdot T=P$. 

Formula (\ref{toricsympaction}), or equivalently (\ref{toricsympaction2}),
 allows us to
determine the parameters $y_i$ entering the Molien formula in terms of
the chemical potentials for the elementary fields of the quiver gauge theory.
This identification can be only done modulo an intrinsic $c-g-2$ 
dimensional ambiguity parameterized by the matrix $Q$: $y_\alpha$ 
are determined
by \eref{toricsympaction} up to vectors in the kernel of $P$.  
We will see in the next section that there is an efficient way of assigning
charges under the non-anomalous symmetries to the variables $y_\alpha$ using
perfect matchings. In particular, if we are interested in the Hilbert series
depending on a single parameter $t$, we can always assign charge $t$ to
the variables corresponding to external perfect matchings and charge one to
all the other variables. 
Let us now re-compute the refined Hilbert series for the two examples studied above, using (\ref{Molieneq}). For simplicity, we compute the Hilbert series depending on one parameter $t$, referring to \cite{Forcella:2008bb} for an example
of computation of the refined Hilbert series depending on all parameters. 

%%%%%%%%%%%%%%%%%%% 
\paragraph{Symplectic Quotient for $dP_0=\IC^3 / \IZ_3$: } 
The kernel of the matrix $T$, from \eref{TdP0},
can be easily computed to be the vector $Q$:
\begin{equation}
P \cdot Q = 0 \qquad \Rightarrow \qquad Q^t = \left(
\begin{array}{llllll}
- 1 & -1 & -1 & 1 & 1 & 1
\end{array}
\right) ,
\label{chargesdp0}
\end{equation}
which forms the vector of charges for the linear sigma model description of the master space for $dP_0$. In this description, therefore, we find that the master space, which we recall to be irreducible, is given by
\begin{equation}
\mathbb{C}^6//[-1,-1,-1,1,1,1] \ .
\label{MSdP0}
\end{equation}

We can compute the Hilbert series using the Molien formula (\ref{Molieneq}). For 
simplicity, we consider the Hilbert series depending on a single chemical
potential $t_i\equiv t$. This is obtained by assigning 
chemical potential  $t$ to all fields of the linear sigma model with 
negative charge. This assignment of charges 
is consistent with formula (\ref{toricsympaction}) and, as we will see in the next section, is equivalent to assigning $t$ to the three external perfect 
matchings and $1$ to the three internal ones. 
We introduce a new chemical potential $z$ for the $U(1)$ charge
and integrate on a contour slightly smaller than the unit circle.
Using the residue technique outlined in Section 3.2 of \cite{Forcella:2007wk},
 we find that the contribution to the integral comes from the pole at $z=t$, whence
\begin{equation}
H(t; \f_{dP_0}) = \oint \frac{dz}{ 2 \pi i z(1-t/z)^3 (1-z)^3} = \frac {1+4t+t^2}{(1-t)^5} ,
\label{HSdP0}
\end{equation}
agreeing precisely with \eref{HSdP0-1}.

%%%%%%%%%%%%%%%%%%%%%%%
\paragraph{Symplectic Quotient for $\IC^3 / \IZ_2\times \IZ_2$: } 
In this case the kernel for the matrix $T$, from \eref{Tz2z2}, is three dimensional and it is encoded by the matrix:
\beq\label{chargesZ2}
Q^t= \left(
\begin{array}{lllllllll}
 -1 & -1 & 0 & 1 & 1 & 0 & 0 & 0 & 0 \\
 -1 & 0 & -1 & 0 & 0 & 1 & 1 & 0 & 0 \\
 0 & -1 & -1 & 0 & 0 & 0 & 0 & 1 & 1
\end{array}
\right) .
\eeq
The rows of $Q^t$ induce a $(\mathbb{C}^{*})^ 3$ action on $\mathbb{C}^9$ which allows us to represent the coherent component of $\f_{\IC^3 / \IZ_2\times \IZ_2}$ as a symplectic quotient:
\begin{equation}
\firr{\IC^3 / \IZ_2\times \IZ_2} = \mathbb{C}^9//(\mathbb{C}^*)^3 \ .
\end{equation}
We compute here the Hilbert series depending on a single parameter 
$t_i\equiv t$. Formula (\ref{toricsympaction}) is consistent with assigning
chemical potential $t$ to the fields of the sigma model with negative charges 
and chemical potential $1$ to all the others. As we will see in the next section, this corresponds to  the
natural choice which assigns $t$ to the external perfect matchings and
$1$ to the internal ones. The Molien formula reads
\begin{eqnarray}
H(t,\firr{\IC^3 / \IZ_2\times \IZ_2})&=&\int \frac{dr dw ds}{r w s}\frac{1}{(1-t/r w)(1- t/r s)(1- t/w s) (1-r)^2(1-w)^2(1-s)^2} \nonumber\\
& = & \frac{1+6 t+6 t^2+t^3}{(1-t)^6} \ ,
\end{eqnarray}
which agrees with \eref{Hz2z2} exactly. The computation of the refined Hilbert
series depending on all six parameters is deferred to \cite{Forcella:2008bb}.

%%%%%%%%%%%%%%%%%%%%%%%%%%%%%%%%%%%%%%%%%%%%%%%%%%%%%%%%%%%%%%%%%%%%%%%%%%%%
%%%%%%%%%%%%%%%%%%%%%%%%%%%%%%%%
\subsection{Dimer Models and Perfect Matchings}\label{s:dimer}

\begin{figure}[t]
\begin{center}
\includegraphics[scale=0.4]{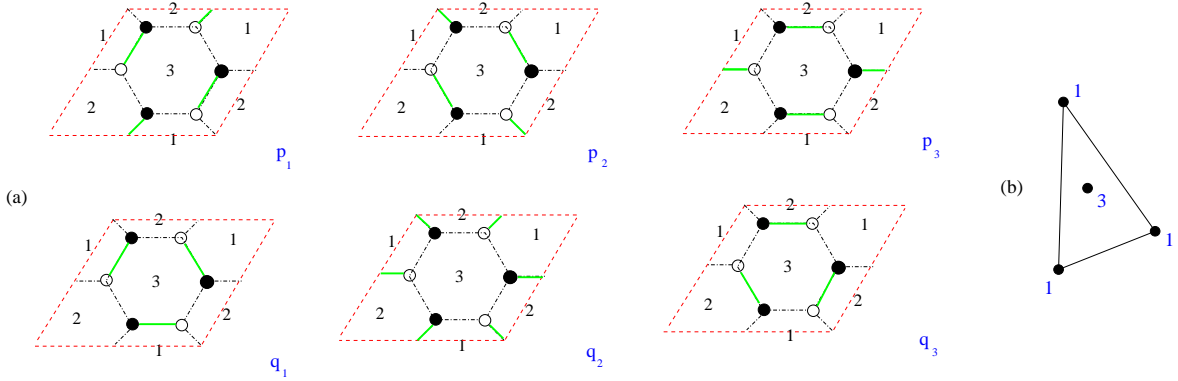} 
\caption{{\sf (a) The perfect matchings for the dimer model corresponding to $dP_0$, with $p_i$ the external matchings and $q_i$, the internal;
(b) The toric diagram, with the labeled multiplicity of GLSM fields, of $dP_0$.}}
\label{f:pmdP0}
\end{center}
\end{figure}

As we have already explained It was recently realized that the most convenient way of describing 
toric quiver gauge theories is that of dimers and brane-tilings. In this section we will see a nice application of this technology. 
Let us re-examine our above analysis using the language of dimers and perfect matchings. As we have illustrated in the previous chapter,
 a dimer or brane tiling  associated to a gauge theory living on a D3 brane at singularity is a bipartite graph, where to every face is 
associated a gauge factor, to every edge a bifundamental fields, and to every vertex a term in the superpotential 
with plus or minus sign according to the color of the vertex, see for example Figure \ref{f:pmdP0}. 
A perfect matching is a subgraph of the dimer that contains all the nodes in such a way that the set of edges don't touch among 
themselves (see Figure \ref{f:pmdP0}). The reader is referred to \cite{Hanany:2005ve,Franco:2005rj,Feng:2005gw,Franco:2006gc} and for a comprehensive introduction, especially to \cite{Kennaway:2007tq}. We will focus on perfect matchings and the matrix $P$ defined in \eref{Pmat}. Now, $K$ is of size $(g+2) \times E$ with $E$ the number of fields, and $g$ the number of gauge group factors. The matrix $T$ is of size $(g+2) \times c$, where $c$ is the number of generators of the dual cone. Thus, $P$ is a matrix of size $E \times c$. The number $c$ is, equivalently, the number of perfect matchings for the corresponding tiling (dimer model). In fact, the matrix $P$ contains entries which are either $0$ or $1$, encoding whether a field $X_i$ in the quiver is in the perfect matching $p_\alpha$:
\begin{equation}
P_{i \alpha} =
\begin{cases} 1 & \text{if $X_i \in p_\alpha$,} \\
0 &\text{if $X_i \not \in p_ \alpha $.}
\end{cases}
\end{equation}

%%%
\paragraph{Dimer Model for $dP_0$: }

Let us first discuss in detail the example of $dP_0 = \IC^3/\IZ_3$.
Using the $P$ matrix in Equation (\ref{PdP0}) we can draw the 6 different perfect matchings. They are shown in Figure \ref{f:pmdP0}.
The first three perfect matchings are identified as the external perfect matchings $p_{1,2,3}$ while the last three are the internal perfect matchings $q_{1,2,3}$ associated with the internal point in the toric diagram of $dP_0$. For reference we have also drawn the toric diagram, together with the multiplicity of the gauged linear sigma model fields associated with the nodes, which we recall from \cite{Feng:2000mi1,Feng:2000mi2}. In fact, it is this multiplicity that led to the formulation of dimers and brane tilings as originally discussed in the first reference of \cite{Hanany:2005ve,Franco:2005rj,Feng:2005gw,Franco:2006gc}. Here we find another important application of this multiplicity.

Now, from Figure \ref{f:pmdP0} we notice that the collection of all external perfect matchings cover all edges in the tiling. Similarly, the collection of all internal perfect matchings cover all edges in the tiling, giving rise to a linear relation which formally states $p_1+p_2+p_3 = q_1+q_2+q_3$, or as a homogeneous relation, $-p_1-p_2-p_3+q_1+q_2+q_3=0$. Since the $P$ matrix encodes whether an edge is in a perfect matching, the linear combination of matchings encodes whether an edge is in that linear combination. Using the homogeneous form of the relation we in fact find that the vector $(-1,-1,-1,1,1,1)$ is in the kernel of $P$ and thus forms a row of the kernel matrix $Q^t$. Since the rank of the matrix $P$ is equal to the dimension of the master space, $g+2=5$, we conclude that this is the only element in the matrix $Q$. We have thus re-obtained the result
\eref{chargesdp0}.

\begin{figure}[t]
\begin{center}
\includegraphics[scale=0.4]{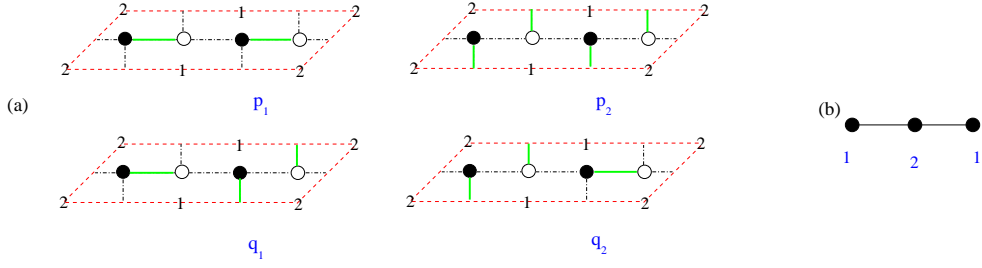} 
\caption{{\sf (a) The perfect matching for the dimer model corresponding to $\IC^2/\IZ_2$. The two upper perfect matchings are associated with the two external points in the toric diagram and the two lower perfect matchings are associated with the internal point in the toric diagram, drawn in (b).}}
\label{f:pmC2Z2}
\end{center}
\end{figure}

%%%%
\paragraph{Dimer Model for $\IC^2/\IZ_2$: }
Next, let us look at the example of $\IC^2/\IZ_2$. The toric diagram, with multiplicity 1, 2, 1, and the corresponding perfect matchings are shown in \fref{f:pmC2Z2}, denoting the two external perfect matchings by $p_{1,2}$ and those of the internal point by $q_{1,2}$. 
A quick inspection of the perfect matchings shows a linear relation $-p_1-p_2+q_1+q_2=0$, leading to a charge matrix $(-1,-1,1,1)$ for the linear sigma model description of the master space for the orbifold $\IC^2/\IZ_2$. As is computed 
in \cite{Forcella:2007wk,Butti:2007jv} and as we shall later encounter in detail in \ref{C2Z2} we find that the master space is nothing but the conifold.

\begin{figure}[t]
\begin{center}
\includegraphics[scale=0.6]{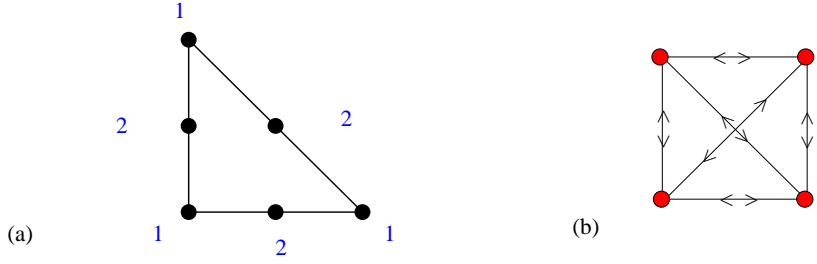} 
\caption{{\sf (a) The toric diagram for $\IC^3/\IZ_2\times \IZ_2$ together with the GLSM multiplicities/perfect matchings marked for the nodes. There is a total of 9 perfect matchings in the dimer model, which for sake of brevity we do not present here; (b) The associated quiver diagram.}}
\label{f:pmC3Z2Z2}
\end{center}
\end{figure}

%%%%%%%%%%%%%%%%%%%%%
\paragraph{Dimer Model for $\IC^3/\IZ_2\times \IZ_2$: }
The above arguments can also be used to compute the linear sigma model description of the orbifold $\IC^3/\IZ_2\times \IZ_2$. The toric diagram, shown in Figure \ref{f:pmC3Z2Z2}, consists of 6 points, 3 external with perfect matchings $p_{1,2,3}$, and 3 internal. 
The 3 internal points form a local $\IC^2/\IZ_2$ singularity and have multiplicity 2. We find 6 internal perfect matchings $q_{1,2,3,4,5,6}$. There are 3 sets of 3 points each, sitting on a line. Each such line is a $\IC^2/\IZ_2$ singularity and can therefore be used to write down a relation between the 4 perfect matchings, giving rise to 3 conifold like relations, $p_1 + p_2 = q_1+q_2, p_1+p_3 = q_3+q_4, p_2+p_3 = q_5+q_6$. Thus the master space of the orbifold $\IC^3/\IZ_2\times \IZ_2$ is an intersection of 3 conifold-like (quadric) relations in $\IC^9$, with a charge matrix
\beq\label{QC2Z2Z2}
Q^t= \left(
\begin{array}{lllllllll}
 -1 & -1 & 0 & 1 & 1 & 0 & 0 & 0 & 0 \\
 -1 & 0 & -1 & 0 & 0 & 1 & 1 & 0 & 0 \\
 0 & -1 & -1 & 0 & 0 & 0 & 0 & 1 & 1
\end{array}
\right) .
\eeq
which precisely agrees with Equation \eref{chargesZ2}.

We see that we can find a diagrammatic way, using dimers and perfect matchings, to find the charges of the matrix $Q$ and thereby the charges of the linear sigma model which describes the master space. This description is good for a relatively small number of perfect matchings and small number of fields in the quiver. When this number is large we will need to refer to the computation using the kernel of the $P$ matrix. We thus reach an important conclusion:
\begin{observation}
The coherent component of the master space of a toric quiver theory is generated by perfect matchings of the associated dimer model.
\end{observation}
%This should be a corollary of the more general Birkhoff-Von Neumann Theorem \cite{BvN}:
%\begin{theorem}
%An $n \times n$ doubly stochastic matrix (i.e., a square matrix of non-negative entries each of whose row and column sums to 1) is a convex linear combination of permutation matrices.
%\end{theorem}

We can easily make contact between the perfect matching description and the
more mathematical description of the master space outlined in the previous 
section. As shown in
\cite{Hanany:2005ve,Franco:2005rj,Feng:2005gw,Franco:2006gc}, the perfect matchings $p_\alpha$ parameterize the solutions of the
F-terms condition through the formula
\begin{equation}
X_i =\prod_{\alpha=1}^c p_\alpha^{P_{i \alpha}} \ .
\label{pmpar}
\end{equation}
This equation determines the charges of the perfect matchings (modulo
an ambiguity given by $Q^t$) in terms of the $g+2$ field theory charges.
In the previous section we introduced a homogeneous variable $y_\alpha$ for each perfect matching $p_\alpha$. We see that formula \eref{toricsympaction} for the chemical potential of the field $X_i$
\begin{equation}
q_i =   \prod_{\alpha =1}^c y_\alpha^{P_{i\alpha}} \, ,
\label{toricsympaction22}
\end{equation}
following from the
Cox description of the toric variety, nicely matches with \eref{pmpar} 
obtained from the dimer description.

Finally, there is a very simple way of determining the {\bf non-anomalous} 
charges of the perfect matchings, which is useful in computations 
based on the Molien formula.
The number of non-anomalous $U(1)$ symmetries of a toric 
quiver gauge theory is precisely the number of external perfect matchings,
or equivalently, the number $d$ of external points in the toric diagram.
This leads to a very simple parameterization for the non-anomalous 
charges \cite{Butti:2005vn,Butti:2005ps}: 
assign a different chemical potential $x_i$ for $i=1,...,d$ to each
external perfect matching and assign $1$ to the internal ones. 
It follows
from \eref{pmpar} that this prescription is equivalent to the one discussed
in \cite{Butti:2005vn,Butti:2005ps}. 
In particular, in the computation of the Hilbert series depending on just
one parameter $t$, we can assign chemical potential $t$ to all the external
matchings and $1$ to the internal ones, as we did in section \ref{s:Molien}.

%%%%%%%%%%

%%%%%%%%%%%%%%%%%%%%%%%%%%
%%%%%============================================
%%%%%%%%%%%%%%%%%%%%%%%%%%%
\section{Some non Orbifold Examples}\label{s:case}

Enriched by a conceptual grasp and armed with the computational techniques 
for describing the master space, we would like to explore some examples of 
non orbifold toric singularities. We will leave a more extensive analysis 
of other singularities to the original literature \cite{Forcella:2008bb}. The following set of 
non orbifold examples will confirm the fact that $\f$, for the $U(1)^g$ quiver 
theory, is of dimension $g+2$ and generically decomposes into several components: 
the top dimensional one of which is also a Calabi-Yau variety of dimension $g+2$, 
as well as some trivial lower-dimensional linear pieces defined by the vanishing of combinations of the coordinates.

\subsection{The Conifold}

The conifold is the easiest example of non orbifold singularity. In the case of just one brane $N=1$ it is extremely easy. Indeed the superpotential of the gauge theory is trivial in the abelian case and the four field $A_i$, $B_j$ are just c-numbers and parametrize a copy of $\mathbb{C}^4$. Hence $\f_{\mathcal{C}}$ is irreducible and trivially CY. Its Hilbert series is simply:
\begin{equation}
H(t;\f_{\mathcal{C}}) = \frac{1}{(1- t)^4}
\end{equation}

\subsection{The Suspended Pinched Point}\label{SPPmast}
\begin{figure}[t]
\begin{center}
\includegraphics[scale=0.6]{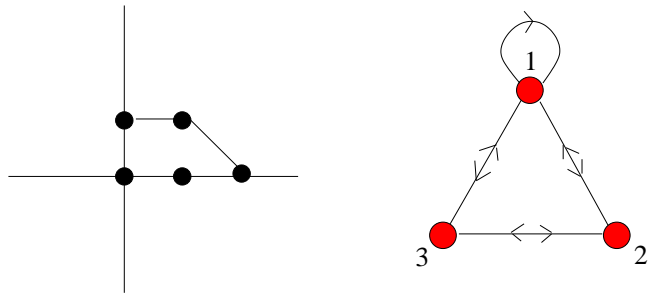} 
\caption{{\sf The toric diagram and the quiver for the $SPP$ singularity.}}
\label{f:SPPtq}
\end{center}
\end{figure}

The next easier example of non-orbifold type of singularity is the so-called suspended pinched point (SPP), first studied in \cite{Morrison:1998cs}.
The toric and the quiver diagrams are presented in Figure~\ref{f:SPPtq} 
and the superpotential is
\begin{equation}
W_{SPP} = X_{11}(X_{12}X_{21} - X_{13} X_{31}) + X_{31}X_{13}X_{32}X_{23}- X_{21}X_{12}X_{23}X_{32} \ .
\end{equation}
The SPP singularity is the easiest example of a family of singularities called the generalized conifolds. 
It is a near parent to the conifold singularity, but, as we will see in a moment, it has a very rich structure of moduli space.

The matrices $K$, $T$ and $P$ can readily found to be
\begin{equation}
K = \tmat{ 1 & 0 & 0 & 0 & 0 & 0 & 1 \cr 0 & 1 & 0 & 0 & 0 & 
   0 & 1 \cr 0 & 0 & 1 & 0 & 0 & 1 & 0 \cr 0 & 0 & 0 & 
   1 & 0 & 1 & 0 \cr 0 & 0 & 0 & 0 & 1 & -1 & 0 \cr}, \quad
T = \tmat{
0 & 0 & 0 & 0 & 0 & 1 \cr 0 & 0 & 0 & 0 & 1 & 0 \cr 
   0 & 0 & 1 & 1 & 0 & 0 \cr 1 & 1 & 0 & 0 & 0 & 0 \cr 
   0 & 1 & 0 & 1 & 0 & 0 \cr 
}, \quad
P = \tmat{
0 & 0 & 0 & 0 & 0 & 
    1 \cr 0 & 0 & 0 & 0 & 1 & 0 \cr 0 & 0 & 1 & 1 & 
   0 & 0 \cr 1 & 1 & 0 & 0 & 0 & 0 \cr 0 & 1 & 0 & 1 & 
   0 & 0 \cr 1 & 0 & 1 & 0 & 0 & 0 \cr 0 & 0 & 0 & 0 & 
   1 & 1 \cr 
} \ .
\end{equation}

In this example, we need to weight the variables appropriately by giving weight $t$ to all external points in the toric diagram, as discussed above:
\beq
\{X_{21},X_{12},X_{23},X_{32},X_{31},X_{13},X_{11} \} \rightarrow
\{1,1,1,1,1,1,2 \} \ .
\eeq
In the actual algebro-geometric computation this means that we weigh the variables of the polynomials with the above degrees and work in a weighted space.
\comment{
  \begin{equation}
    \hbox{R}(SPP)=\mathbb{C}[x21,x12,x23,x32,x31,x13,x11,
      Degrees=>{1,1,1,1,1,1,2}]
  \end{equation}
  \begin{eqnarray}
    \hbox{I}(SPP)&=&(x12*x23*x32-x12*x11,x21*x23*x32-x21*x11,\nonumber\\
    & & x21*x12*x32-x32*x31*x13,x21*x12*x23-x23*x31*x13,\nonumber\\
    & & x32*x23*x13-x13*x11,x32*x23*x31-x31*x11,x13*x31-x12*x21)\nonumber\\
  \end{eqnarray}
}
Now, we find that the moduli space is a reducible variety $\f_{SPP} = \firr{SPP} \cup L_{SPP}$ with Hilbert series:
\begin{equation}
H(t;\f_{SPP}) = \frac{1 + 2t + 2t^2 - 2t^3 + t^4}{(1- t)^4(1 -t^2)}
\end{equation}
and
\begin{eqnarray}
\firr{SPP} &=& \mathbb{V}(X_{23} X_{32} - X_{11}, X_{21} X_{12} - X_{31} X_{13}) \nonumber\\
L_{SPP} &=& \mathbb{V}(X_{13}, X_{31}, X_{12}, X_{21}) \ ,
\end{eqnarray}
where, as in the previous chapter, we have used the standard algebraic geometry notation that, given a set
$F=\{f_i\}$ of polynomials, $\IV(F)$ is the variety corresponding to the vanishing locus of $F$.
The top component $\firr{SPP}$ is a toric variety of complex dimension 5 which is the product of a conifold and a plane $\IC^2$; it has Hilbert series:
\begin{equation}
H(t;~\firr{SPP}) = \frac{1-t^2}{(1-t)^4} \frac{1}{(1-t)^2} = 
\frac{1+t}{(1-t)^5} \ ,
\end{equation}
with a palindromic numerator, as was with the \'etudes studied above.
The other component $L_{SPP}$ is a plane isomorphic $\IC^3$ with Hilbert series:
\begin{equation}
H(t;L_{SPP}) = \frac{1}{(1-t^2)(1-t)^2} \ .
\end{equation}
The two irreducible components intersect in a $\IC^2$ plane with Hilbert series:\begin{equation}
H(t;~\firr{SPP} \cap  L_{SPP}) = \frac{1}{(1-t)^2} \ .
\end{equation}
We observe that the Hilbert series of the various components satisfy the additive relation:
\begin{equation}
H(t;\f_{SPP}) = H(t;~\firr{SPP})+ H(t;L_{SPP})- H(t;~\firr{SPP} \cap  L_{SPP}) \ .
\end{equation}
This is, of course, the phenomenon of ``surgery'' discussed in \cite{Hanany:2006uc}. This ``surgery'' property is generic for all the reducible moduli space. We refer the reader to the original literature \cite{Forcella:2008bb} for many other examples of this property.

In the symplectic quotient description, we find that the kernel of the $T$-matrix is $Q^t = \left( 1, -1, -1, 1, 0, 0 \right)$ and hence $\firr{SPP} \simeq 
\IC^6 // Q^t$.
The symmetry group of the coherent component is easily found to be $SU(2)^3\times U(1)^2$, a rank 5 group as expected from the toric property of this space. The non-Abelian part is realized as the repetition of the charges, $1, -1,$ and $0$, respectively.

%%%%%%%%%%%%%
\subsection{Cone over Zeroth Hirzebruch Surface}\label{s:F0}
\begin{figure}[t]
\begin{center}
  \epsfxsize = 11cm
  \centerline{\epsfbox{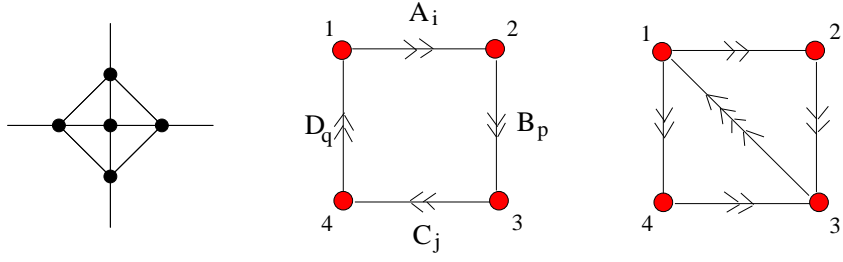}}
  \caption{{\sf The toric diagram and the quivers for phases I and II of $\mathbb{F}_0$.}}
  \label{f:F0III}
\end{center}
\end{figure}

We continue with a simple toric threefold: the affine complex cone over the zeroth Hirzebruch surface $\mathbb{F}_0$, 
which is simply $\IP^1 \times \IP^1$. The complex cone $C_{\mathbb{C}}(\mathbb{F}_0)$ is a regular $\mathbb{Z}_2$ 
orbifold of the conifold, but it presents a much more rich structure. The toric diagram is drawn in Figure \eref{f:F0III}.
As already explained in chapter \ref{braneasing} to a given CY singularity $\cX$ generically correspond more than just one field theory. 
These set of field theory are different UV descriptions of the same IR superconformal fixed manifold \cite{Feng:2000mi1,Beasley:2001zp}. 
Indeed all the SCFT associated to a given singularity are related by a chain of Seiberg dualities. 
The $\mathbb{F}_0$ theory is an example of this phenomena and has two toric/Seiberg 
(see for example \cite{Feng:2001bn, Franco:2002mu}) dual phases, $(\mathbb{F}_0)_I$ and $(\mathbb{F}_0)_{II}$, 
of the gauge theory \cite{Feng:2000mi1,Feng:2000mi2}, and the quivers and superpotentials are:
\begin{equation}\label{WF0}\ba{rcl}
W_{(\mathbb{F}_0)_I} &=& \epsilon_{ij}\epsilon_{pq} {A}_i{B}_p{C}_j{D}_q ;
\\
W_{(\mathbb{F}_0)_{II}} &=& \epsilon_{ij}\epsilon_{mn} X^i_{12}X^m_{23}X_{31}^{jn}- \epsilon_{ij}\epsilon_{mn} X^i_{14}X^m_{43}X_{31}^{jn} \ .
\ea\end{equation}
\comment{
  \begin{equation}
    R(F_0)=QQ[A1,A2,B1,B2,C1,C2,D1,D2,Degrees=>{1,1,1,1,1,1,1,1}]
  \end{equation}
  \begin{eqnarray}
    I(F_0) &=&(B1*C2*D2-B2*C2*D1,B2*C1*D1-B1*C1*D2,\nonumber\\
    & & A1*C2*D2-A2*C1*D2,A1*C2*D1-A2*C1*D1,\nonumber\\
    & & A2*B2*D1-A2*B1*D2,A1*B1*D2-A1*B2*D1,\nonumber\\
    & & A1*B2*C2-A2*B2*C1,A1*B1*C2-A2*B1*C1)
  \end{eqnarray}
}
%%%%%%%%%********
\paragraph{Toric Phase I: }
We can readily find the F-terms from $W_{(\mathbb{F}_0)_I}$ in \eref{WF0} and using the techniques outlined above, we can find the $K$-matrix to be
\begin{equation}
{\scriptsize
K=
\left(
\begin{array}{lllllllll}
    & A_1& A_2 & B_1 & B_2 & C_1 & D_1 & C_2 & D_2 \\
A_1 & 1 & 0 & 0 & 0 & 0 & 0 & -1 & 0 \\
A_2 & 0 & 1 & 0 & 0 & 0 & 0 & 1  & 0 \\
B_1 & 0 & 0 & 1 & 0 & 0 & 0 & 0  & -1 \\
B_2 & 0 & 0 & 0 & 1 & 0 & 0 & 0  &  1 \\
C_1 & 0 & 0 & 0 & 0 & 1 & 0 & 1  & 0 \\
D_1 & 0 & 0 & 0 & 0 & 0 & 1 & 0  & 1 \\
 \end{array}
\right) }
.
\end{equation}
Whence, primary decomposition gives that there are three irreducible pieces:
\beq\label{F0-I}\ba{rcl}
\f_{(\mathbb{F}_0)_I} &=& \firr{(\mathbb{F}_0)_I} \cup L^1_{(\mathbb{F}_0)_I} \cup L^2_{(\mathbb{F}_0)_I}, \\
\ea\eeq
with
\beq\label{FF0}
\ba{rcl}
\firr{(\mathbb{F}_0)_I}  &=& \mathbb{V}(B_2 D_1 - B_1 D_2, A_2 C_1 - A_1 C_2)\\
L^1_{(\mathbb{F}_0)_I} &=& \mathbb{V}(C_2, C_1, A_2, A_1)\\
L^2_{(\mathbb{F}_0)_I} &=& \mathbb{V}(D_2,D_1, B_2, B_1) \ .
\ea\eeq
%The piece $\firr{(F_0)_I}$ is what we have called $\firr{~}$ above and we emphasize here it is Calabi-Yau and will use this notation as well.
With weight $t$ to all 8 basic fields in the quiver the Hilbert series of the total space is given by
\beq
H(t;~\f_{(\mathbb{F}_0)_I}) = \frac{1 + 2t + 3t^2 - 4t^3 + 2t^4}{(1-t)^6} \ .
\eeq

The top-dimensional component of $\f_{(\mathbb{F}_0)_I}$ is toric Calabi-Yau, and here of dimension 6; 
this is consistent with the fact that the number of nodes in the quiver is 4. Specifically, $\firr{(\mathbb{F}_0)_I}$ is the product of two conifolds and it has Hilbert series (again with palindromic numerator):
\begin{equation}\label{2conifmaster}
H(t;~\firr{(\mathbb{F}_0)_I})=\frac{(1 + t)^2}{(1-t)^6} \ .
\end{equation} 
The two lower-dimensional components are simply $\IC^4$, with Hilbert series
\beq
H(t;~L^1_{(\mathbb{F}_0)_I}) = H(t;~L^2_{(\mathbb{F}_0)_I}) = \frac{1}{(1-t)^4} \ .
\eeq
These two hyperplanes are, as mentioned above, Coulomb branches of the moduli space, they intersect the 
$\firr{(\mathbb{F}_0)_I}$ along one of the two three dimensional conifolds which have Hilbert series
\begin{equation}
H(t;~\firr{(\mathbb{F}_0)_I} \cap L^1_{(\mathbb{F}_0)_I}) = H(t; ~\firr{(\mathbb{F}_0)_I} \cap L^2_{(\mathbb{F}_0)_I}) = \frac{1+t}{(1-t)^3} \ .
\end{equation}
The Hilbert series of various components again satisfy the additive surgical relation of \cite{Hanany:2006uc}:
\begin{eqnarray}
\hspace{-1cm}
H(t;~\f_{(\mathbb{F}_0)_I} )
&=&H(t;~\firr{(\mathbb{F}_0)_I})+ H(t;~L^1_{(\mathbb{F}_0)_I}) + H(t;~L^2_{(\mathbb{F}_0)_I}) \nonumber\\
& & - H(t;~\firr{(\mathbb{F}_0)_I} \cap L^1_{(\mathbb{F}_0)_I} ) - H(t;~\firr{(\mathbb{F}_0)_I} \cap L^2_{(\mathbb{F}_0)_I}) \ .
\end{eqnarray}

For reference, the dual cone $T$-matrix and the perfect matching matrix $P = K^t \cdot T$ are:
\beq
T = \tmat{
0 & 0 & 0 & 0 & 0 & 0 & 1 & 1 \cr 0 & 0 & 0 & 0 & 0 & 1 & 0 & 1 \cr 0 & 0 & 0 & 1 & 1 & 0 & 0 & 0 \cr 0 & 0 & 1 & 0 & 1 & 
   0 & 0 & 0 \cr 0 & 1 & 0 & 0 & 0 & 0 & 1 & 0 \cr 1 & 0 & 0 & 1 & 0 & 0 & 0 & 0 \cr 
} \ , \qquad
P = \tmat{
 0 & 0 & 0 & 0 & 0 & 0 & 1 & 1 \cr 0 & 0 & 0 & 0 & 0 & 1 & 0 & 1 \cr 0 & 0 & 0 & 1 & 1 & 0 & 0 & 0 \cr 0 & 0 & 1 & 0 & 1 & 0 & 0 & 0 \cr 0 & 1 & 0 & 0 & 0 & 0 & 1 & 0 \cr 1 & 0 & 0 & 1 & 0 & 0 & 0 & 
 0 \cr 0 & 1 & 0 & 0 & 0 & 1 & 0 & 0 \cr 1 & 0 & 1 & 0 & 0 & 0 & 0 & 0 \cr 
} \ .
\eeq
Subsequently, their kernel is $Q^t$, giving us
\beq
Q^t = \tmat{ 0 & 1 & 0 & 0 & 0 & -1 & -1 & 1 \cr 1 & 0 & 
    -1 & -1 & 1 & 0 & 0 & 0 \cr  } \Rightarrow
\firr{(\mathbb{F}_0)_I} \simeq \IC^8 // Q^t \ .
\eeq
The fact that the rows of $Q^t$ sum to 0 means that the toric variety is indeed Calabi-Yau.
The symmetry of the coherent component is $SU(2)^4\times U(1)^2$, suitable for a product of two conifolds. We note that the charge matrix $Q$ has 8 columns which are formed out of 4 pairs, each with two identical columns. This repetition of columns in the charge matrix is another way of determining the non-Abelian part of the symmetry group of the coherent component.

%%%%%%%%%%%%
\paragraph{Toric Phase II: }
We can perform a similar analysis for the second toric phase which is a Seiberg dual of the first. Note from \eref{WF0} that the gauge-invariant terms in $W_{(\mathbb{F}_0)_{II}}$ now have a different number of fields; correspondingly, 
we must thus assign different weights to the variables. 
The ones that are composed of Seiberg dual mesons of the fields of the first toric phase should be assigned twice the weight:
\begin{equation}
\{X^1_{12},X^1_{23},X^{22}_{31},X^2_{23},X^{21}_{31},X^2_{12},X^{12}_{31},X^{11}_{31},X^1_{14},X^1_{43},X^2_{43},X^2_{14} \} \rightarrow \{ {1,1,2,1,2,1,2,2,1,1,1,1} \} \ .
\end{equation}
In the actual algebro-geometric computation this means that we weight the variables of the polynomials with the above degrees and work in a weighted space.
\comment{
  \begin{equation}
    \hbox{R}(\mathbb{F}_0 II) = \mathbb{C}[x12,x23,yy31,y23,yx31,y12,xy31,xx31,
      x14,x43,y43,y14,Degrees=>{1,1,2,1,2,1,2,2,1,1,1,1}]
  \end{equation}
  Observe the different weights for the various fields: the ones with weights $2$ are the ones that 
are composed mesons of the fields of the first toric phase.
  \begin{eqnarray}
    \hbox{I}(\mathbb{F}_0 II)&=&(x23*yy31-y23*yx31,x12*yy31-y12*xy31,\nonumber\\
    & & x12*x23-x14*x43,x12*yx31-y12*xx31,x12*y23-x14*y43,\nonumber\\
    & & x23*xy31-y23*xx31,y12*x23-y14*x43,y12*y23-y14*y43,\nonumber\\
    & & x43*yy31-y43*yx31,x14*yy31-y14*xy31,x14*yx31-y14*xx31,\nonumber\\
    & & x43*xy31-y43*xx31) 
  \end{eqnarray}
}
Subsequently, we find that
\beq
{\tiny
K=
\left(
\begin{array}{lllllllllllll}
  & X^1_{43} & X^{11}_{31}&X^2_{14} &X^1_{23}&X^{21}_{31}& X^2_{23}& X^1_{12}& X^1_{14}& X^2_{12}& X^{12}_{31}& X^2_{43}& X^{22}_{31}\\
X^1_{43}   & 1 & 0 & 0 & 0 & 0 & 0 & 1  & 0 & 1 & 0 & 1 & 0 \\
X^{11}_{31} & 0 & 1 & 0 & 0 & 0 & 0 & 1  & 1 & 0 & 1 & 0 & 0 \\
X^2_{14}  &0 & 0 & 1 & 0 & 0 & 0 & 1  & 1 & 1 & 0 & 0 & 0 \\
X^1_{23}  &0 & 0 & 0 & 1 & 0 & 0 & -1 & 0 & -1&-1 &-1 &-1 \\
X^{21}_{31}& 0 & 0 & 0 & 0 & 1 & 0 & -1 & -1& 0 & 0 & 0 & 1 \\
X^2_{23}  & 0 & 0 & 0 & 0 & 0 & 1 & 0  & 0 & 0 & 1 & 1 & 1 \\
 \end{array}
\right)
} \ .
\eeq
The master space affords the primary decomposition
$\f_{(\mathbb{F}_0)_{II}} = \firr{(\mathbb{F}_0)_{II}} \cup L^1_{(\mathbb{F}_0)_{II}} \cup L^2_{(\mathbb{F}_0)_{II}} \cup L^3_{(\mathbb{F}_0)_{II}}$ with
\beq\begin{array}{rcl}
\firr{(\mathbb{F}_0)_{II}}  &=& \mathbb{V}(X^{12}_{31}X^1_{43} - X^{11}_{31} X^2_{43}, X^2_{23}X^1_{43} - X^1_{23}X^2_{43}, X^{22}_{31}X^1_{43} - X^{21}_{31} X^2_{43},
X^2_{12}X^1_{14} - X^1_{12}X^2_{14},\\
&&
X^{21}_{31}X^1_{14} - X^{22}_{31}X^2_{14}, X^{22}_{31}X^1_{14} - X^{12}_{31}X^2_{14}, X^{21}_{31}X^{12}_{31} -  X^{22}_{31}X^{22}_{31}, X^1_{23}X^{12}_{31} - X^2_{23}X^{11}_{31},  \\ 
&&
X^2_{23}X^2_{12} - X^2_{43}X^2_{14}, X^1_{23}X^2_{12} - X^1_{43}X^2_{14}, X^1_{12}X^{21}_{31} - X^2_{12}X^{11}_{31},X^1_{12}X^2_{23} - X^1_{14}X^2_{43}, \\
&& 
X^1_{23}X^{22}_{31} - X^2_{23}X^{21}_{31}, X^1_{12}X^{22}_{31} - X^2_{12}X^{12}_{31}, X^1_{12}X^1_{23} - X^1_{14}X^1_{43})\\
L^1_{(\mathbb{F}_0)_{II}} &=& \mathbb{V}(X^2_{14}, X^2_{43}, X^1_{43}, X^1_{14}, X^2_{12}, X^2_{23}, X^1_{23},X^1_{12})\\
L^2_{(\mathbb{F}_0)_{II}} &=& \mathbb{V}(X^2_{43}, X^1_{43}, X^{11}_{31}, X^{12}_{31}, X^{21}_{31}, X^2_{23}, X^{22}_{31}, X^1_{23})\\
L^3_{(\mathbb{F}_0)_{II}} &=& \mathbb{V}(X^2_{14}, X^1_{14},X^{11}_{31}, X^{12}_{31}, X^2_{12}, X^{21}_{31}, X^{22}_{31}, X^1_{12}) \ .
\end{array}
\eeq
The Hilbert series is
\beq
H(t;~\f_{(\mathbb{F}_0)_{II}}) = \frac{1 + 6t + 17t^2 + 24t^3 + 14t^4 - 4t^5 + 4t^7 + 2t^8}{(1- t)^2(1 - t^2)^4} \ .
\eeq
We see that $\f_{(\mathbb{F}_0)_{II}}$ is composed of four irreducible components: 
a six dimensional $\firr{(\mathbb{F}_0)_{II}}$ with the same Hilbert series of the product of two conifolds:

\begin{equation}
H(t;~\firr{(\mathbb{F}_0)_{II}}) = \frac{(1 + t)^2}{(1-t)^6} \ .
\end{equation}

$\firr{(\mathbb{F}_0)_{II}}$ is the biggest irreducible component and it has 
the same Hilbert series (\ref{2conifmaster}) of $\firr{(\mathbb{F}_0)_{I}}$: the biggest irreducible component 
of the first toric phase.

The other three irreducible components are three four dimensional complex planes each defined by the vanishing of 
8 coordinates out of the 12 total. These planes intersect only at the origin of the coordinate system 
and have the Hilbert series
\beq
H(t;~L^1_{(\mathbb{F}_0)_{II}})=\frac{1}{(1-t^2)^4}, \quad
H(t;~L^2_{(\mathbb{F}_0)_{II}})= H(t;~L^3_{(\mathbb{F}_0)_{II}})=\frac{1}{(1-t)^4} \ .
\eeq
We see that $L^1_{(\mathbb{F}_0)_{II}}$ has $t^2$ in the denominator instead of $t$ because 
it is parameterized precisely by the coordinates of twice the weight.

The three $\IC^4$ components and the $\firr{(\mathbb{F}_0)_{II}}$ component intersect in a three dimensional 
conifold variety but with different grading of the coordinates:
\begin{eqnarray}
& & H(t;~\firr{(\mathbb{F}_0)_{II}} \cap L^1_{(\mathbb{F}_0)_{II}}) = \frac{1+t^2}{(1-t^2)^3}, \nonumber\\
& & H(t;~\firr{(\mathbb{F}_0)_{II}} \cap L^2_{(\mathbb{F}_0)_{II}}) = H(t;~\firr{(\mathbb{F}_0)_{II}} \cap L^3_{(\mathbb{F}_0)_{II}}) = \frac{1+t}{(1-t)^3} \ .
\end{eqnarray}
Once again, we have a surgery relation:
\begin{eqnarray}
H(t; ~\f_{(\mathbb{F}_0)_{II}}) & = & H(t;~\firr{(\mathbb{F}_0)_{II}} )+H(t;~L^1_{(\mathbb{F}_0)_{II}})
+ H(t;~L^2_{(\mathbb{F}_0)_{II}}) + H(t;~L^3_{(\mathbb{F}_0)_{II}} )\nonumber\\
& & - H(t;~\firr{(\mathbb{F}_0)_{II}} \cap L^1_{(\mathbb{F}_0)_{II}} ) - H(t;~\firr{(\mathbb{F}_0)_{II}} \cap L^2_{(\mathbb{F}_0)_{II}} ) - H(t;~\firr{(\mathbb{F}_0)_{II}} \cap L^3_{(\mathbb{F}_0)_{II}}). \nonumber\\
\end{eqnarray}

The dual cone $T$-matrix and the perfect matching matrix $P = K^t \cdot T$ are:
\beq
T = \tmat{ 0 & 0 & 0 & 0 & 0 & 0 & 1 & 1 & 1 \cr 0 & 0 & 0 & 
   0 & 1 & 1 & 0 & 0 & 1 \cr 0 & 1 & 1 & 1 & 0 & 0 & 
   0 & 0 & 0 \cr 0 & 0 & 0 & 1 & 0 & 0 & 0 & 1 & 1 \cr
   0 & 0 & 1 & 0 & 0 & 1 & 0 & 0 & 1 \cr 1 & 0 & 0 & 
   1 & 0 & 0 & 0 & 1 & 0 \cr} \ , \qquad
P = \tmat{ 0 & 0 & 0 & 0 & 0 & 0 & 1 & 1 & 1 \cr 0 & 0 & 0 & 
   0 & 1 & 1 & 0 & 0 & 1 \cr 0 & 
    1 & 1 & 1 & 0 & 0 & 0 & 0 & 0 \cr 0 & 0 & 0 & 1 &
   0 & 0 & 0 & 1 & 1 \cr 0 & 0 & 1 & 0 & 0 & 1 & 0 & 
   0 & 1 \cr 1 & 0 & 0 & 1 & 0 & 0 & 0 & 1 & 0 \cr 0 &
   1 & 0 & 0 & 1 & 0 & 1 & 0 & 0 \cr 0 & 1 & 0 & 1 & 
   1 & 0 & 0 & 0 & 0 \cr 0 & 1 & 1 & 0 & 0 & 0 & 1 & 
   0 & 0 \cr 1 & 0 & 0 & 0 & 1 & 1 & 0 & 0 & 0 \cr 1 &
   0 & 0 & 0 & 0 & 0 & 1 & 1 & 0 \cr 1 & 0 & 1 & 0 & 
   0 & 1 & 0 & 0 & 0 \cr} \ .
\eeq
Hence, the kernel is $Q^t$ and we have
\beq
Q^t = \tmat{
 1 & 1 & 0 & -1 & 0 & -1 & 
    -1 & 0 & 1 \cr 0 & 1 & 0 & -1 & 0 & 0 & 
    -1 & 1 & 0 \cr 0 & 1 & -1 & 0 & 
    -1 & 1 & 0 & 0 & 0 \cr } \Rightarrow
\firr{(\mathbb{F}_0)_{II}} \simeq \IC^9 // Q^t \ .
\eeq
Again, the rows of $Q^t$ sum to 0 and the toric variety is Calabi-Yau.
The second and third rows are conifold like relations and the first row is a relation which 
is found for the master space of $\IC^3/\IZ_3$ and is not in the first phase of $\mathbb{F}_0$.

There are some important comments to do. We have seen that the master space for two Seiberg equivalent 
gauge theories is not the same. In particular the structure of the irreducible components and their intersections change under
Seiberg duality. A nice observation is that the Hilbert series of the coherent components of the master space 
is exactly the same.
 
We see here that a manifestation of Seiberg duality is in the fact that the coherent component of each of 
the phases have the same Hilbert series. In this section we used just the chemical potential $t$, but It is possible to check that the Hilbert series for $\firr{~}$ for the various Seiberg dual phases remain equal even if we refine them with all the non anomalous global symmetries, while if we add also the anomalous symmetries the Hilbert series are generically different \cite{Forcella:2008ng}. We will see the explicit example of $\mathbb{F}_0$ in section \ref{FOannonan}.
Indeed it is possible to verify in other examples that different toric phases of a given 
Calabi-Yau singularity $\cX$ exhibit the same Hilbert series for the coherent component $\firr{~}$ of the master space \cite{Forcella:2008ng}. 
This is going to be our conjectured relation:
\begin{conjecture}\label{seibcong}
Quivers which are toric (Seiberg) duals have coherent component of the master space with the same Hilbert series written in term
of the non anomalous charges of the gauge theories.
\end{conjecture}
It would be interesting to understand the fate of the linear components under Seiberg duality, 
and the explicit relation between the coherent components of Seiberg dual gauge theories.
For some progress along this line we refere the reader to \cite{Forcella:2008ng}.

\subsubsection{Comments on Seiberg Dual Phases}

It is important to note that our analysis of the moduli space is just the classical 
analysis of the abelian moduli space of gauge theory: namely we are considering the theory with just 
``$SU(1)$'' gauge factors and imposing the classical F-flat conditions. 
There are a couple of things that could go in the wrong direction: the Seiberg duality is a non abelian duality, 
and its application to the abelian case is not clear in principle; the classical F-term conditions 
can have some quantum corrections that lift at least part of the classical moduli space. 

About the first problem we can say that in the examples studied in literature there is a geometric 
extension of the Seiberg duality to the case $N=1$. 
Indeed in abelian case the Seiberg duality could 
be interpreted as a geometric operation called toric duality.
% and this could remain in principle a correct duality even in the abelian case. 
We observe that all the previous claims done in literature
about toric duality were related just to the mesonic branch of the complete moduli space. 
For such cases we have 2 different theories which have the same mesonic moduli space, for any N, including $N=1$. 
We now face a new situation in which the full moduli space is discussed - both mesonic and baryonic, on all of its components.
Indeed in the previous section we analyzed the $\mathbb{F}_0$ gauge theory and we discovered that:
 $(\mathbb{F}_0)_I$ and $(\mathbb{F}_0)_{II}$  have the same mesonic moduli space, coherent  
component with the same Hilbert series for non anomalous $U(1)$ charges, but the linear components of the first 
phase are contained in the linear component of the second phase. Does this qualify as Seiberg duality for $N=1$? 
Is it possible to extend the geometric operations on the full moduli space? Observe that for $N=1$ there are no quantum 
effect and no lifting. This is correct since $(\mathbb{F}_0)_I$ and $(\mathbb{F}_0)_{II}$ are not Seiberg equivalent 
in Field Theory sense. If toric duality applies, it is possible that it implies some equivalence of the
coherent components only. The $N=2$ case would be the right laboratory: here the theory is non abelian and 
Seiberg duality implies the equivalence of the full moduli space, including
linear components. But here come some computational problems. 
If we try to compute the classical Hilbert series with the help of the package Macaulay2 it get stacked and 
we are not able to obtain a result. As we will see in the following chapters 
there exist another way for computing Hilbert series: 
namely passing to the gravity dual of the $AdS/CFT$ correspondence. 
This procedure give the Hilbert series for the complete moduli space for all the values of $N$. 
We will discover that it is possible to apply this procedure only to the coherent component of the moduli space. 

A computation done in the gravity side in principle takes into account all the strong coupling corrections. 
The fact that the gravity computation is sensible only to the coherent component and that, 
in the case just analyzed, 
the coherent components of two Seiberg dual theories have the same Hilbert series, give us arguments to guess that the conjecture 
\ref{seibcong} could be correct even at quantum level. 

We hope to come back in the future on the problem of studying Seiberg dualities in these classes of theories, 
but we can try to make an heuristic analysis of the particular case of the $\mathbb{F}_0$ gauge theory. For more systematic, but still incomplete analysis, we refer the reader to \cite{Forcella:2008ng}. 
The linear components $L^2_{(\mathbb{F}_0)_{II}}$ and $L^3_{(\mathbb{F}_0)_{II}}$ of the second phase of $\mathbb{F}_0$ match well with the two linear components $L^1_{(\mathbb{F}_0)_I}$,  $L^2_{(\mathbb{F}_0)_{I}}$ 
of the first phase of $\mathbb{F}_0$: they have the same dimension, the same Hilbert series, 
the same intersection with the coherent component of the master space.
%and it seems that giving vev to their coordinate fields, the $(\mathbb{F}_0)_{II}$ theory flows
% to the Coulomb branch of the $\frac{\mathbb{C}^2}{\mathbb{Z}_2}\times \mathbb{C}$ theory 
%in four different ways as it happens for $(\mathbb{F}_0)_I$ as we will see in section \ref{s:branch}.
In the Hilbert series language these facts can be summarized by the relation:
\begin{eqnarray}
\hspace{-1cm}
H(t;~\f_{(\mathbb{F}_0)_I} )
&=&H(t;~\firr{(\mathbb{F}_0)_{II}})+ H(t;~L^2_{(\mathbb{F}_0)_{II}}) + H(t;~L^3_{(\mathbb{F}_0)_{II}}) \nonumber\\
& & - H(t;~\firr{(\mathbb{F}_0)_{II}} \cap L^2_{(\mathbb{F}_0)_{II}} ) - H(t;~\firr{(\mathbb{F}_0)_{II}} \cap L^3_{(\mathbb{F}_0)_{II}}) \ .
\end{eqnarray}
 
The question is: what is the third linear branch of $(\mathbb{F}_0)_{II}$ and why it does not exist in $(\mathbb{F}_0)_I$. 
Maybe It could be possible to see that that this kind of branch is lifted in the quantum theory for $N>1$. 
The branch $L^1_{(\mathbb{F}_0)_{II}}$ for $ N=1$ is parametrized only by the vevs of the dual mesons generated 
by Seiberg duality. One could argue that, if for $N>1$ there is a similar branch, it is lifted. 
First, generically in this branch, all or part of the other fields get masses, 
leading to no other obvious fixed point and to a theory which is still strongly coupled where we could expect 
non trivial dynamics. Second, the situation is remarkably similar to SQCD: 
we know that in the electric theory the moduli space of SQCD is exact at classical level, 
while in the magnetic theory the moduli space receives quantum corrections; 
in particular the flat direction in the magnetic theory with zero vev for the dual quarks and non zero vev 
for the dual mesons M 
(situation similar to $L^1_{(\mathbb{F}_0)_{II}}$) is partially lifted. A simple way to see it is: 
in the dual branch the F terms $q M \bar{q}$ allow flat directions with $q= \bar{q}=0$ and with 
$M$ a generic matrix $N_f$ by $ N_f$. 
However $M$ is dual to the electric meson $Q\bar{Q}$ which is a $N_f$ by $N_f$ with maximum rank 
equal to $N_c < N_f$; for this reason some components of $M$ should be lifted. 
As discussed by Seiberg \cite{Seiberg:1994pq} these are lifted because $M$ gives mass to the dual quarks 
reducing the effective number of flavor to $N_{f_{eff}} \sim N_c$ and thus generating 
non-perturbative corrections to the moduli space. Following this line of reasoning one could expect that something 
similar happens in $(L^1)_{(\mathbb{F}_0)_{II}}$. To reach this conclusion we need to make many assumptions: 
we extrapolate Seiberg duality to $N=1$ and we focus on the gauge group which undergoes 
duality ignoring all the others. 
The 2 by 2 matrix $M$ with the four mesons $X_{31}^{ij}$ along the branch
 $L^1_{(\mathbb{F}_0)_{II}}$ should have rank one: 
thus $\det M = 0$ which is the equation of the conifold; this is just the intersection of the linear 
branch with the coherent component, as discussed in the previous section. 
So $(L^1)_{(\mathbb{F}_0)_II}$ is partially lifted: 
what remains is already contained in the big branch and it is not a new branch of the theory. 
The above argument should be taken as an indication of what we may expect for $N>1$. 
It is a question whether the other linear branches are partially lifted. 

We would like to summarize the various possibilities for the ``quantum'' master space: one possible expectation is 
that the branches of the minimal phase of a CY are genuine branches also in the IR, 
but every time we perform a Seiberg duality we introduce spurious flat directions. This possibility 
has some motivations due to the above discussion in field theory. 

Another possible expectation is that just the coherent component of the moduli space is invariant under Seiberg 
duality, while all the other components are lifted by ``quantum'' correction. 
This possibility has some justifications in the dual gravity picture, where, as we will see in the following, 
the counting of D3 brane states dual to
BPS operators seems to see just the coherent component of the master space.

A third possibility, the one we will support, is that even the coherent component of the master space is not invariant under Seiberg duality, 
but only its Hilbert series written just in term of the non abelian $U(1)$ symmetries is and invariant of the duality transformation. 

Clearly a much more careful investigation must be done. In section \ref{FOannonan} we will make some other comments, and in \cite{Forcella:2008ng} we analyse some related topics.

%%%%%%%%%%%%%%%%%%%%%%%%%%%%%%%%%

\section{Linear Branches of Moduli Spaces}\label{s:branch}
\setall

>From the previous sections we learn that the master space $\f$ of an $\mathcal{N}=1$ 
quiver gauge theory on a toric singularity $\cX$ generically presents a reducible moduli space. 
This fact could appear surprising. Indeed the $\mathcal{N}=2$ supersymmetric gauge theories are 
the classical examples of theories with reducible moduli spaces. These theories present two 
well separated branches: the Higgs branch and the Coulomb branch. 
Moving along these different branches has a well defined characterization both in terms of the geometry 
of the moduli space and in terms of the VEV structure of the gauge theory. 
It would be interesting to have a similar interpretation in this more generic setup.

In the case of just $N=1$ brane on the tip of the conical singularity $\cX$, the reducibility of $\f$ 
follows easily from the toric condition. Indeed, the equations defining $\f$ are of the form 
``monomial $=$ monomial'', whereby giving us toric ideals as discussed in section \ref{s:toric}. 
Let us embed this variety into $\mathbb{C}^d$ with coordinates $\{x_1,...,x_d\}$, then its algebraic equations have the form:
\begin{equation}\label{fget}
x_1^{i_{j_1}}...x_d^{i_{j_d}}\Big( M_{1_j}(x)-M_{2_j}(x) \Big)=0 \ ,
\end{equation}
where $j=1,...,k$ runs over the number of polynomials defining the variety and the polynomials 
$M_{1_j}(x)-M_{2_j}(x)$ are irreducible. If some $i_{j_p}$ is different from zero the variety 
is reducible and is given by the union of the zero locus of $M_{1_j}(x)-M_{2_j}(x)$ together with a set 
of planes (linear components) $L^l$ given by the zero locus of the factorized part in (\ref{fget}).

Hence, the master space $\f$ could be the union of a generically complicated $g+2$ dimensional variety with 
a set of smaller dimensional varieties that are generically linear varieties 
$\{\mathbb{C}^{d_1},...,\mathbb{C}^{d_n}\}$ parameterized by combinations 
of the coordinates\footnote{The reducibility of the moduli space could persist for $N > 1$ (see section \ref{molienN} and \cite{Butti:2007jv} for some examples) and it would be interesting to have a clearer geometric picture even in these more complicated cases. In this section we will concentrate on the $N=1$ case.}; 
this is certainly true for the cases studied in the above section \ref{s:case}, and for the other examples
presented in \cite{Forcella:2008bb}. 
It would be nice to give a gauge theory interpretation to these smaller dimensional linear branches. 

We will try to give a  nice picture for these planes, looking at the first phase of $\mathbb{F}_0$. 
The reader is referred to \cite{Forcella:2008bb} for more examples. 
The lesson we will learn is that {\it these planes of the master space could parametrize flows in the gauge theory}. 
Specifically, there may be chains of flows from one irreducible component of the master space of a theory to another. 
The archetypal example which will be the terminus of these flows will be the gauge theory on a 
D3-brane probing the Calabi-Yau singularity $\mathbb{C}^2/\mathbb{Z}_2\times \mathbb{C}$, to which we alluded in section \ref{s:dimer}. Let us first briefly review this theory, continuing along the same vein as our discussions in section \ref{s:irred}.

%%%%%%%%%%%%%%%%%%%%%%%%
\subsection{The $(\mathbb{C}^2/\mathbb{Z}_2) \times \mathbb{C}$ Singularity}
\label{C2Z2}
\begin{figure}[t]
\begin{center}
  \epsfxsize = 11cm
  \centerline{\epsfbox{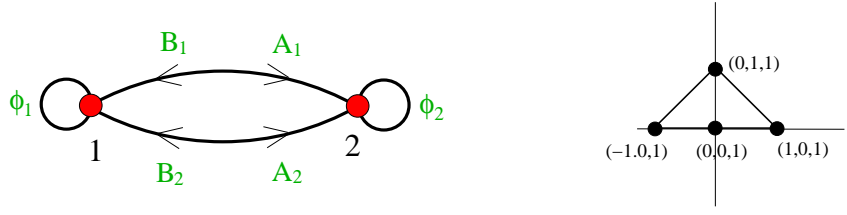}}
  \caption{{\sf Quiver and toric diagrams for $(\mathbb{C}^2/\mathbb{Z}_2) \times \mathbb{C}$.}}
  \label{f:c2z2quiver}
\end{center}
\end{figure}

The quiver gauge theory for $(\mathbb{C}^2/\mathbb{Z}_2) \times \mathbb{C}$ has ${\cal N}=2$ supersymmetry with two vector multiplets and two bi-fundamental hyper-multiplets. In ${\cal N}=1$ notation we have six chiral multiplets denoted as $\phi_1, \phi_2, A_1, A_2, B_1, B_2$, with a superpotential
\begin{equation}
W = \phi_1 (A_1 B_1 - A_2 B_2) + {\phi}_2 (B_2 A_2 - B_1 A_1)
\end{equation}
and the quiver and toric diagrams are given in \fref{f:c2z2quiver}.
The master space $\f_{(\mathbb{C}^2/\mathbb{Z}_2)\times \mathbb{C}}$ of the theory can be easily found to be%\footnote{Note that in the presece of adjoint fields $\f$ does not coincide with the moduli space $\mathcal{M}$ in the case $N=1$ because we must consider $U(1)$ groups and not $SU(1)$.}
\begin{equation}\label{N2id}
\f_{(\mathbb{C}^2/\mathbb{Z}_2)\times \mathbb{C}}=\mathbb{V}(A_1B_1-A_2B_2,\phi_1A_1-\phi_2A_1,\phi_1A_2-\phi_2A_2,\phi_1B_1-\phi_2B_1,\phi_1B_2-\phi_2B_2) \ .
\end{equation}
Now, $\f_{(\mathbb{C}^2/\mathbb{Z}_2)\times \mathbb{C}}$ is clearly reducible and decomposes into the following two irreducible components as $\f_{(\mathbb{C}^2/\mathbb{Z}_2)\times \mathbb{C}}= \firr{(\mathbb{C}^2/\mathbb{Z}_2)\times \mathbb{C}} \hbox{  } \cup \hbox{  }L_{(\mathbb{C}^2/\mathbb{Z}_2)\times \mathbb{C}}$ with
\beq\label{N2red}
\firr{(\mathbb{C}^2/\mathbb{Z}_2)\times \mathbb{C}}=\mathbb{V}(\phi_1 - \phi_2, A_1B_1 - A_2B_2) \ , \qquad
L_{(\mathbb{C}^2/\mathbb{Z}_2)\times \mathbb{C}} =\mathbb{V}(A_1, A_2, B_1, B_2) \ .
\eeq
Specifically, $\firr{(\mathbb{C}^2/\mathbb{Z}_2)\times \mathbb{C}}$ is $\mathbb{C} \times \mathcal{C}$, where the $\mathbb{C}$ is defined by $\phi_1 = \phi_2$ and the conifold singularity $\mathcal{C}$ is described by the chiral fields $\{A_1,A_2,B_1,B_2\}$ with the constraint $A_1 B_1 = A_2 B_2 $. The component $L_{(\mathbb{C}^2/\mathbb{Z}_2)\times \mathbb{C}} = \mathbb{C}^2$ is parametrized by the fields $\{\phi_1,\phi_2 \}$. These two branches meet on the complex line parametrized by $\phi_1=\phi_2$. 

The field theory interpretation of these two branches is standard: moving in $L_{(\mathbb{C}^2/\mathbb{Z}_2)\times \mathbb{C}}$ 
we are giving VEV to the scalars in the vector multiplet and hence we call $L_{(\mathbb{C}^2/\mathbb{Z}_2)\times \mathbb{C}}$ 
the Coulomb branch; while moving in $\firr{(\mathbb{C}^2/\mathbb{Z}_2)\times \mathbb{C}}$ we are giving VEV to 
the scalars in the hyper-multiplets and hence we call $\firr{(\mathbb{C}^2/\mathbb{Z}_2)\times \mathbb{C}}$ the Higgs branch.

Let us take $\mathbb{F}_0$ as our main example and let us revisit the 
reducibility its master space trying to give it a gauge theory interpretation.

%%%%%%%%%%%%%%%%%%%
%\subsubsection{Case Studies Re-examined}

%%%%%
\subsection{The Linear Branches of $\mathbb{F}_0$ and its IR Flows}
Let us start by re-examining $(\mathbb{F}_0)_I$, encountered in section \ref{s:F0}. We recall from \eref{F0-I} that 
the master space $\f_{\mathbb{F}_0}$ is the union of three branches: the biggest one is six dimensional and is the 
set product of two conifold singularities, i.e., $\mathcal{C} \times \mathcal{C}$, and the two smallest ones 
are two copies of $\mathbb{C}^4$, parametrized respectively by the VEV of 
$\{B_1, B_2, D_1, D_2\}$ and $\{A_1, A_2, C_1, C_2\}$.
\beq\ba{rcl}
L^1_{(\mathbb{F}_0)_I}&=&\{B_1, B_2, D_1, D_2\}\\
L^2_{(\mathbb{F}_0)_I}&=&\{A_1, A_2, C_1, C_2\} \ .
\ea\eeq

\begin{figure}[t]
\begin{center}
\includegraphics[scale=0.5]{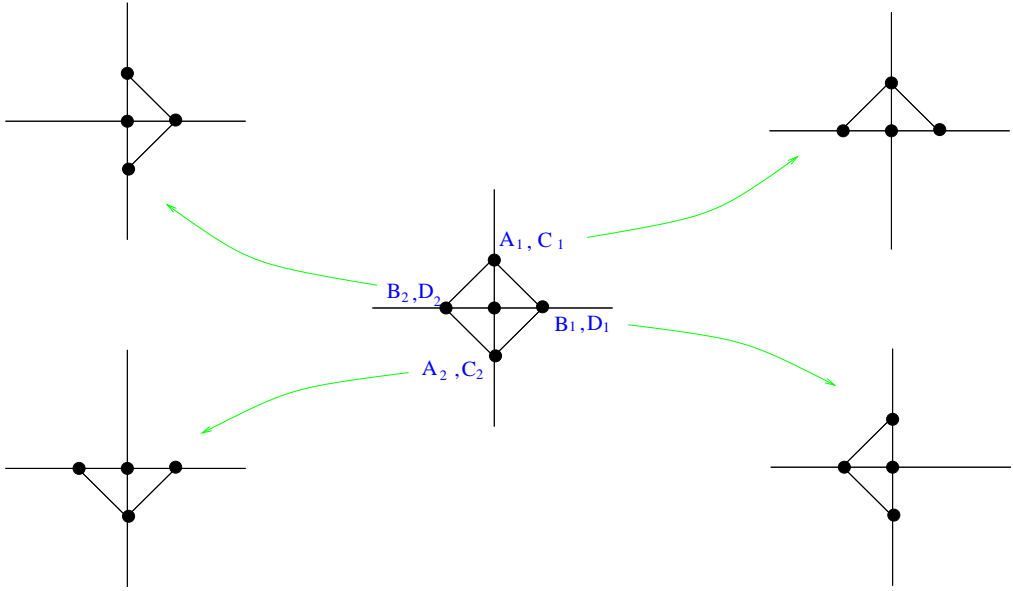} 
\caption{{\sf The four possible flows from $\mathbb{F}_0$ to $(\mathbb{C}^2/\mathbb{Z}_2) \times \mathbb{C}$ as exhibited in the toric diagrams.}}
\label{flowF0}
\end{center}
\end{figure}

Looking at the toric diagram one can understand the possible flows of the gauge theory. 
Indeed we remind the reader that the prescription is as follows \cite{Beasley:1999uz,Feng:2000mi1,Feng:2002fv}. 
If one toric diagram is derived by deleting external vertices of the other, then the associated gauge theories 
can flow one into the other. Geometrically this is blowing up the singularity, 
while in the gauge theory a FI term is turned on and as a
consequence gives VEV to some chiral fields, for a product of $U(N_i)$ gauge groups; or a baryonic VEV is 
turned on in the case of product of $SU(N_i)$ gauge groups. 
It is indeed possible to associate fields of the gauge theory to each
vertex in the toric diagram, as shown in Figure \ref{flowF0}. 
The deleted vertex can be associated with the fields that are getting a VEV. 

Using this prescription one can easily understand that the gauge theory can flow to the one associated to the 
$\mathbb{C}^2/\mathbb{Z}_2 \times \mathbb{C}$ singularity in four different ways by giving VEV to one out of the four possible sets of fields: $\{A_1,C_1\}$, $\{A_2,C_2\}$, $\{B_1,D_1\}$, $\{B_2,D_2\}$. In this way the linear components of the master space will flow among themselves according to the scheme:
\begin{eqnarray}\label{fllin}
& &\vev{{A_1,C_1}} \neq 0 \hbox{ or } \vev{{A_2,C_2}} \neq 0: \hbox{  } L^2_{(\mathbb{F}_0)_I} \rightarrow L_{(\mathbb{C}^2/\mathbb{Z}_2)\times \mathbb{C}} \nonumber\\
& &\vev{{B_1,D_1}} \neq 0 \hbox{ or } \vev{{B_2,D_2}} \neq 0: \hbox{  } L^1_{(\mathbb{F}_0)_I} \rightarrow L_{(\mathbb{C}^2/\mathbb{Z}_2)\times \mathbb{C}}
\end{eqnarray}

Once arrived at the fixed point one can move along the Coulomb branch of the IR theory giving VEV respectively to the set of fields: $\{A_2,C_2\}$, $\{A_1,C_1\}$, $\{B_2,D_2\}$, $\{B_1,D_1\}$ that parametrize the linear component $L_{(\mathbb{C}^2/\mathbb{Z}_2)\times \mathbb{C}}$ of the master space of the SCFT at $\mathbb{C}^2/\mathbb{Z}_2\times \mathbb{C}$ singularity.
Hence, along the two smaller branches of the moduli space the theory admits an accidental $\mathcal{N}=2$ supersymmetry and the theory moves along the Coulomb branch of the resulting gauge theory. Giving vev to the field belonging to tha linear components, the hyperplanes of the UV theory flow into the hyperplanes of the IR theory. This is a situation that is 
going to repeat in many other examples, and we refer the reader to \cite{Forcella:2008bb} for a more comprehensive explanation.

%%%%%%%%%%%%%%%%%%+++++++++++++++++++++++++++++++++++++++++++
%%%%%%%%%%%%% GLOBAL SYMMETRIES
%%%%%%%%%%%%%%%%%%+++++++++++++++++++++++++++++++++++++++++++

%%%%%%%%%%%%%%%
\section{The Master Space: Hilbert Series and CY Geometry}\label{mathH}

We have studied, with examples for toric theories and just one physical D3-branes 
(look \cite{Forcella:2008bb} for more examples), the algebraic variety formed by the F-flatness equations.
We have seen that in general this is a reducible variety, with a top Calabi-Yau component of dimension $g+2$.
In this section we recapitulate and deepen the properties of the coherent component 
$\firr{~}$ of the master space $\f$. We study some properties of its Hilbert series 
and we give a demonstration that $\firr{~}$ is always a $g+2$ dimensional CY.

\paragraph{As a Higher-Dimensional Calabi-Yau: }
The most suggestive property is that $\firr{~}$ is always a $g+2$
dimensional {\em Calabi-Yau manifold}. This has been explicitly checked
in all the examples we have discussed. It is simple to check the
Calabi-Yau condition in the linear sigma model description, since
it corresponds there to the fact that the vectors of charges in $Q^t$
are traceless, or equivalently, the toric diagram has all the vectors ending on the same hyper-plane. 
%This can be easily checked in equation (\ref{chargesdp0}) for 
%$\IC^3/\IC_3$, in equation (\ref{chargesZ2}) for $\IC^3/\IC_2\times\IC_2$ 
%and in all other examples by explicit computation.

There is a remarkably simple proof that $\firr{~}$ is Calabi-Yau
which emphasizes the r\^{o}le of perfect matchings. Recall from \cite{Hanany:2005ve,Franco:2005rj,Feng:2005gw,Franco:2006gc} and section \ref{s:dimer} 
that the quiver gauge theory has a description in terms of a dimer model,
this is a  bi-partite tiling
of the two torus, with $V/2$ white vertices and $V/2$ black vertices,
where $V$ is the number of superpotential terms. The elementary fields
of the quiver correspond to the edges in the tiling, each of which connects
a black vertex to a white one. Now, by definition, a perfect matching
is a choice of fields/dimers that cover each vertex
precisely once. In particular, a perfect matching contains exactly $V/2$
dimers, connecting the $V/2$ black vertices to the $V/2$ white ones.
Since the columns of the matrix $P$, of size $E \times c$,
tell us which fields/dimers occur in a given perfect matching we have
the nice identity
\begin{equation}
\underbrace{(1,1,....,1)}_{E} \cdot P = \frac{V}{2}
\underbrace{(1,1,....,1)}_{c} ,
\end{equation}
which basically counts the number of edges in each perfect matching.

By multiplying this equation on the right by the matrix
kernel $Q$ of $P$, $P\cdot Q=0$, we obtain
\begin{equation}
0 = \underbrace{(1,1,....,1)}_{c} \cdot Q
\end{equation}
and we conclude that the vector of charges of the linear sigma model,
which are the rows of the matrix $Q^t$, are traceless. This proves that
$\firr{~}$ is Calabi-Yau.

We again see that the prefect matchings description is crucial in our understanding of the properties of the master space.
As explained in section \ref{s:dimer}, the perfect matchings generates the
coherent component of the master space.

%%%

%%%%%%%%%%%%%%%
\paragraph{Palindromic Hilbert Series: }
An intriguing property of the Hilbert series for $\firr{~}$
is its symmetry. As manifest from all our examples, the numerator of
the Hilbert series (in second form) for $\firr{~}$ is a polynomial in $t$
\begin{equation}
P(t)=\sum_{k=0}^N a_k t^k
\end{equation}
with symmetric coefficients $a_{N-k}=a_k$. This means that
there is a remarkable symmetry of the Hilbert series for $\firr{~}$ under 
$t\rightarrow 1/t$,
\begin{equation}
H(1/t;\firr{~})= t^w H(t;\firr{~})
\end{equation}
where the modular weight $w$ depends on $\firr{~}$. 

A polynomial with such a symmetry between its consecutively highest and lowest coefficients $a_{N-k} \leftrightarrow a_k$ is known as a {\bf palindromic polynomial}. A beautiful theorem due to Stanley \cite{stanley} states the following
\begin{theorem}
The numerator to the Hilbert series of a graded Cohen-Macaulay domain $R$ is palindromic iff $R$ is Gorenstein.
\end{theorem}
What this means, for us, is that the coordinate ring of the affine variety 
$\firr{~}$ must be Gorenstein. However, an affine toric variety is Gorenstein
precisely if its toric diagram is co-planar, i.e., it is Calabi-Yau 
(cf.~e.g.~\S~4.2 of \cite{Morrison:1998cs}). Thus we have killed two birds with one stone: proving that $\firr{~}$ is affine toric Calabi-Yau above from perfect matchings also shows that the numerator of its Hilbert series has the palindromic symmetry.

As we will see in chapter \ref{backtoms}, this symmetry extends to the refined Hilbert series written
in terms of the R-charge parameter $t$ and chemical potentials for a global
symmetry $G$. Although $G$ has been up to now Abelian, we will see that
in some special cases of theories with hidden symmetries $G$ becomes non
Abelian. Introducing chemical potentials $z$ for Cartan sub-algebra of $G$, 
we will write the refined Hilbert series as a sum over $G$-characters
in the general form,
\begin{equation}
H(t,G) = \left (\sum_{k=0}^N \chi_k(z) t^k\right ) PE\left [\sum_{i=1}^G \chi_i(z) t^i\right ]
\end{equation} 
where PE is the plethystic exponential function already introduced in section \ref{HSPE}, 
computing symmetric products on
a set of generators of the coherent component. The refined Hilbert series
is now invariant under the the combined action of $t\rightarrow 1/t$ and
charge conjugation, $\chi_{N-k}=\chi_k^*$. 
%We already saw an explicit example
%in equation \eref{rHS} for $\IC^3/\IZ_2\times \IZ_2$ and w
We will see examples in chapter \ref{backtoms}.

%%%%%%%%%%%%%%%%

%%%%%%%%%%%%%%+++++++++++++++++++++++++++++++++++++++++++
%%%%%%%%%%%%%%%%%%
%%%%%%%%%%%%%%%%%%+++++++++++++++++++++++++++++++++++++++

%%%%%%%%%%%%%%%%%%+++++++++++++++++++++++++++++++++++++++++++
\section{Summary and Discussions}\label{s:concM}\setall
We have enjoyed a long theme and variations in $\f$, touching upon diverse motifs. 
Let us now part with a recapitulatory cadence. We have seen that for a single brane, 
the master space is the algebraic variety formed by the space of F-terms and it coincide with the moduli space 
of the gauge theory $\CM=\f$. In the case of the singularity $\cX$ which the D3-brane probes being a toric Calabi-Yau threefold, 
we have a wealth of techniques to study $\f$: direct computation, toric cones via the $K$ and $T$ matrices, 
symplectic quotients in the Cox coordinates as well as dimer models and perfect matchings. 
Using these methods we have learned that
\begin{itemize}
\item $\cX$ is the mesonic branch of the full moduli space $\CM$ of a single  D3-brane gauge theory and is the symplectic quotient of $\f$ by the Abelian D-terms;
\item For a $U(1)^g$ toric quiver theory, $\f$ is a variety of dimension $g+2$;
\item The master space $\f$ is generically reducible, its top dimensional component, called the coherent component $\firr{}$, is a Calabi-Yau variety, of the same dimension and degree as $\f$. The lower-dimensional components are generically linear pieces $L^i$, composed of coordinate hyperplanes;
\item The coherent component $\firr{~}$ can be quite explicitly characterized. 
Denoting, as usual, with $E$ the number of fields in the quiver, $g$ the number of nodes in the quiver and with $c$ the number of perfect matchings in the dimer realization of the gauge theory, we have defined three matrices:
\begin{itemize}
\item{A $(g+2) \times E$ matrix $K$ obtained by solving the F-terms 
in terms of a set of independent fields. The columns give the 
charges of the elementary fields under the $(\IC^*)^{g+2}$ action;
in a more mathematical language, they give the
semi-group generators of the dual cone $\sigma_K^*$
in the toric presentation for $\firr{~}$:
\beq
\firr{~} \simeq \mbox{Spec}_{\IC}[\sigma_K^* \cap \IZ^{g+2}]
\eeq
}
\item{A $(g+2) \times c$ matrix $T$, defined by $K^t \cdot T \ge 0$, and
representing the cone $\sigma_K$ dual to $\sigma_K^*$. The columns are 
the $c$ toric vectors of the $g+2$ dimensional variety $\firr{~}$.
We see that the number of perfect matchings in the dimer realization of the
quiver theory is the number of external points in the 
toric diagram for $\firr{~}$. This generalizes the fact that the
{\it external} perfect matchings are related to the external points of the 
toric diagram for the three dimensional transverse Calabi-Yau space $\cX$.}
\item{A $E \times c$ matrix $P=K^t \cdot T$ which defines the perfect matchings as collections of elementary fields.}
\end{itemize}

\item The variety $\firr{~}$ also has a linear sigma model, or symplectic quotient, description as
\begin{equation}
\firr{~} = \mathbb{C}^c//Q^t \ ,
\end{equation}
where $Q$ is the kernel of the matrices $P$ and $T$.

\item $\firr{}$ is generated by the perfect matchings in the dimer model (brane tiling) corresponding to the quiver theory; 
%This should follow from the Birkhoff-von Neumann theorem;
\item In the field theory, $\firr{}$ often realizes as the Higgs branch, and the hyper-planes $L^i$, the Coulomb branch of the moduli space $\CM$. The acquisition of VEVs by the fields parameterizing $L^i$ can cause one theory to flow to another via the Higgs mechanism; we have illustrated the flows of the $(\mathbb{F}_0)_I$ theory, but another archetypal example is the chain 
of $dP_n$ theories \cite{Forcella:2008bb};
\item Under Seiberg/toric duality, we conjecture that the Hilbert series for $\firr{}$, refined with all the non anomalous symmetries of the field theory, remains invariant; 
\item According to the plethystic program, the Hilbert series of $\cX$ is the generating function for the BPS mesonic operators. In order to count the full chiral BPS operators, including mesons and baryons, we need to find the refined (graded by various chemical potentials) Hilbert series of $\f$;
\item The Hilbert series of the various irreducible pieces of $\f$, obtained by primary decomposition, obey surgery relations;
\item The numerator of the Hilbert series of $\firr{}$, in second form, is palindromic; 
%This follows from the Stanley theorem;
\item The gauge theory possesses hidden global symmetries corresponding to the symmetry of $\firr{}$ which, 
though not manifest in the Lagrangian, is surprisingly encoded in the algebraic geometry of $\firr{}$. 
%In particular, we can re-write the terms of the single brane generating function, i.e., the refined Hilbert series, of $\firr{}$ in the weights of the representations of the Lie algebra of the hidden symmetry in a selected set of examples. Via the plethystic exponential, this extends to an arbitrary number $N$ of branes.
\end{itemize}

\begin{table}[t]
$
\begin{array}{|c|c|c|c|c|} \hline
\cX & \f & \firr{} & H(t;~\firr{}) & \mbox{Global Symmetry} \\ \hline \hline
\IC^3 & \IC^3 & \IC^3 & (1-t)^{-3} & U(3) \\ \hline
\cC & \IC^4 & \IC^4 & (1-t)^{-4} & U(1)_R \times SU(4)_H \\ \hline
(\IC^2 / \IZ_2) \times \IC & (4,2) & \cC \times \IC & \frac{1 + t}{(1-t)^4} &
  U(1)_R \times SU(2)_R \times U(1)_B \times SU(2)_H \\ \hline
\IC^3 / \IZ_2\times \IZ_2 & (6,14) & - & \frac{1+6 t+6 t^2+t^3}{(1-t)^6} &
  U(1)_R \times U(1)^2 \times SU(2)^3_H \\ \hline
SPP & (5,2) & \cC \times \IC^2 & \frac{1 + t}{(1-t)^5} &
  U(1)_R \times U(1)_M \times SU(2)_H^3 \\ \hline
dP_0 & (5,6) & \simeq \f & \frac {1+4t+t^2}{(1-t)^5} &
  U(1)_R \times SU(3)_M \times SU(3)_H \\ \hline
\mathbb{F}_0 & (6,4) & \cC \times \cC & \frac{(1 + t)^2}{(1-t)^6} &
  U(1)_R \times U(1)_B \times SU(2)^2_M \times SU(2)_H^2 \\ \hline
dP_1 & (6,17) & -& \frac{1 + 4 t + 7 t^2 + 4 t^3 + t^4}{(1 - t)^6(1+t)^2} &
  U(1)_R \times SU(2)_M \times U(1)^3 \times SU(2)_H \\ \hline
dP_2 & (7,44) & -& \frac{1 + 2t + 5 t^2 + 2t^3 + t^4}{(1-t)^7(1+t)^2} &
  U(1)_R \times SU(2)_H  \times U(1)^5 \\ \hline
dP_3 & (8,96) & -& \frac{1 + 4t^2 + t^4}{(1 - t)^8(1+t)^2} &
  (SU(2) \times SU(3))_H \times U(1)^5 \\ \hline
\end{array}
$
\caption{{\sf    
The master space, its coherent component and Hilbert space as well
as the global symmetry of the gauge theory. The notation $(n,d)$ denotes the
dimension and degree respectively of $\f$. For the symmetries, the subscript $R$ denotes R-symmetry, $M$ denotes the symmetry of the mesonic branch, $B$ denotes baryon charge, while $H$ denotes the hidden global symmetry. Note that the rank of the global symmetry group is equal to the dimension of $\f$.}}
\label{t:sum}
\end{table}

In Table \ref{t:sum}, we illustrate some of the above points with a select class of examples. We present the single-brane master space, its coherent component (by name if familiar), the associated Hilbert series as well as the global symmetry, standard as well as hidden. Some of the examples in the table were not computed in this thesis and we refer to the original literature \cite{Forcella:2008bb}.

For a general number $N$ of D3-branes, the situation is more subtle. The moduli space is now the variety of 
F-flatness quotiented by the special unitary factors of the gauge group, 
so that when quotiented by the $U(1)$ factors as a symplectic quotient we once more arrive at the mesonic branch, 
which here is the $N$-th symmeterized product of the Calabi-Yau space $\cX$. 
A direct attack to the problem of generic $N$ seems very difficult, however, in the following chapter we will learn 
to construct the Hilbert series of the moduli space $\CM_N$ for generic $N$ number of branes, and to extract 
from these generating functions very important informations. In particular we will learn how 
to encode the representation of the hidden symmetries, observed in the $N=1$ case for the coherent component  $\firr{}$ of 
the master space, in the plethystic exponential. Indeed, we will learn to re-write the terms of the single 
brane generating function, i.e., the refined Hilbert series, of $\firr{}$ in the weights of 
the representations of the Lie algebra of the hidden symmetry in a selected set of examples. Via the plethystic exponential, 
this extends to an arbitrary number $N$ of branes. This fact will give us the possibility to extend the analysis 
of symmetries of the moduli space to the generic non abelian case. 

Before coming back to study properties of the moduli space for generic $N$ we need to learn how to count the complete spectrum
 of BPS operators. This counting problem will take us busy for the next three chapters, after them we 
will be able to come back to the problem of studying the moduli space $\CM_N$ for generic $N$ number of branes.

%%%%%%%%%%%%%%%%%%%%%%%%%%%%%%%%%%%%%%%%%%%%%%%%%%%%%%%%%%%%%

\newpage 

%%%%%%%%%%%%%%%%%%%%
%%%%%%%

\chapter{Counting BPS D3 Branes and BPS Baryonic Operators }\label{D3SE}

In the previous chapter we developed a good understanding of the moduli space $\CM$ 
of a D3 brane sitting at the singularity $\cX$. 
If we really want to understand the supersymmetric properties of these SCFT we need to do two other major steps: 
study the spectra of $1/2$ BPS operators and study the non abelian case: namely generic number of branes $N$. 
It will turn out that the properties of the moduli space $\CM_N$ for generic $N$ are very well described by the 
counting of $1/2$ BPS operators. 

The set of $1/2$ BPS operators are nothing else that the coordinate ring of the algebraic variety $\CM$. 
The algebraic geometry tells that there exist a one to one correspondence between an algebraic variety and 
its coordinate ring. This means that if we will be able to get informations regarding the $1/2$ BPS operators for generic $N$, 
then we will be able to get informations regarding the moduli space $\CM_N$ for generic $N$. 
The study of the counting procedure will take us busy for long, but in the meanwhile we will learn how 
the geometric informations of the singularity $\cX$ are encoded in the spectrum of the SCFT and  
how to map strong coupling counting to weak coupling counting using the AdS/CFT correspondence. 

Our analysis of the counting procedure will be spread over the following three chapters. 
The first and the present one have a more geometric approach, the second one will have a more field theory approach, 
while in the last one we will try to give a synthesis of the two approaches to formulate a general recipe for counting $1/2$ 
BPS gauge invariant operators for generic number of branes $N$ and for generic toric conical CY singularity $\cX$.

In this chapter we start this long trip studying supersymmetric D3 brane configurations wrapping internal cycles
of type IIB backgrounds $AdS_5\times H$ for a generic Sasaki-Einstein
manifold H. We will show that these configurations correspond to BPS baryonic operators
in the dual quiver gauge theory using the AdS/CFT conjecture. In each sector with given baryonic charge,
we write explicit partition functions counting all the BPS operators according to their flavor and R-charge. This result 
will be completely general and will reduce to the mesonic counting in the particular case of zero baryonic charges, showing that the 
results we have intuitively explained in the first chapter are consistent.

We also show how to extract geometrical information about H from the partition functions; in particular, we give general
formulae for computing the volume of H and of all non trivial three cycles in H. 
>From these volumes one can obtain the strong coupling value of the central charge and of all the R charges of the SCFT.

\section{Generalities}

The study of $BPS$ states in quantum field theory and in string theory is clearly a very important topic. 
These states are generically protected against quantum corrections and contain information regarding the 
strong coupling behavior of supersymmetric field theories and superstring theories. 
In the past years they were especially important in the study of strong weak dualities, like the AdS/CFT conjecture \cite{Maldacena:1997re} 
which gives a connection between the $BPS$ operators in conformal field  theories and $BPS$ states in string theory. 

In this chapter we discuss the set of one half $BPS$ states in string theory realized as $D3$ branes wrapped on 
(generically non trivial) three cycles in the supergravity background $AdS_5 \times H$, where $H$ is a Sasaki-Einstein 
manifold \cite{Acharya:1998db,Morrison:1998cs,Klebanov:1998hh}.
These states are holographically dual to baryonic $BPS$ operators in $\mathcal{N}=1$ four dimensional 
$CFT$s \cite{Gubser:1998fp}, which are quiver gauge theories. 

Recently there has been some interest in counting $BPS$ states in the $CFT$s dual to CY singularities 
\cite{Romelsberger:2005eg,Kinney:2005ej,Biswas:2006tj,Mandal:2006tk,Benvenuti:2006qr,Martelli:2006vh}. 
As we have already anticipated in chapter \ref{braneasing}, and we will better appreciate in the present chapter,
the partition function counting mesonic $BPS$ gauge invariant operators according to their flavor quantum  
numbers contains a lot of information regarding the geometry of the $CY$ 
\cite{Benvenuti:2006qr,Martelli:2006yb}, including the algebraic equations of the singularity. 
Quite interestingly, it also provides a formula for the volume of $H$
\cite{Martelli:2006yb}. This geometrical information has a direct counterpart in field theory, since, according to  the $AdS/CFT$ correspondence, the 
volume of the total space and of the three cycles are duals to the central 
charge and the $R$ charges of the baryonic operators respectively \cite{Gubser:1998fp,Gubser:1998vd}.

The countings in the mesonic gauge invariant sector of the $CFT$ is not too difficult and basically 
correspond of counting the spectra of holomorphic function over the cone $\cX$. 
Geometrically these corresponds to consider giant graviton configurations \cite{McGreevy:2000cw} 
corresponding to $BPS$ $D3$ branes wrapped on trivial three cycles in $H$. 
%From the partition 
%function of these string theory objects one can compute the volumes of the total space $H$ but is unable to gain any information 
%regarding the volume of the three cycles in $H$. 
In this chapter we analyze the generic baryonic
$BPS$ operators, corresponding to $D3$ branes wrapped on generically non trivial three cycles inside $H$. 
We succeed in counting $BPS$ states charged under the baryonic charges of the field theory and we write explicit 
partition functions at fixed baryonic charges. This is the first step to write down the 
complete partition functions and it reduce to the mesonic case once we fix all the baryonic charges equal to zero. 
We investigate in details their geometrical properties.
In particular we will show how to extract 
from the baryonic partition functions a formula for 
the volume of the three cycles inside $H$ and for the volume of $H$ itself. 
We will mostly concentrate on the toric case but our procedure seems adaptable to the 
non toric case as well.
The results of this chapter will be fundamental for the following ones in which we will manage to construct the complete partition functions.

The chapter is organized as follows.
In Section \ref{toric} we review some basic elements of toric geometry related to line bundles and divisors that will be very 
useful in the following discussion. In Section \ref{susyD3}
we formulate the general problem of describing and quantizing the $BPS$ 
$D3$ brane configurations.  
We will use homomorphic surfaces to parameterize the supersymmetric $BPS$ configurations of $D3$ brane wrapped in $H$, following 
results in \cite{Mikhailov:2000ya,Beasley:2002xv}. 
In the case where $\cX$ is a toric variety we have globally defined homogeneous 
coordinates $x_i$ which are charged under the baryonic charges of the theory and which we can use to parametrize these surfaces. 
We will quantize configurations of D3 branes wrapped on these surfaces and we will find the Hilbert space of $BPS$ states 
using a prescription found by Beasley \cite{Beasley:2002xv}. The complete $BPS$ Hilbert space factorizes in sectors with 
definite baryonic charges. Using toric geometry tools, 
we can assign to each sector a convex polyhedron $P$. 
The BPS operators in a given sector are in one-to-one correspondence with 
symmetrized products of N (number of colors) integer points in $P$.
In Section \ref{bar} we discuss the assignment of charges and we set the general counting problem. 
In Section \ref{fieldtheo} we make some comparison with the field theory side. 
%These states are the symmetric product of $N$ constituents and these are put in one to one relation with integer points of $P_D$. 
%The corresponding $BPS$ operators in field theory are constructed by taking the basic baryons and replacing the $N$ 
%constituents elementary chiral superfields with suitable 
%sequences of chiral superfields which are associated to points in $P$. 
In Section \ref{counting}, we will write a partition function $Z_D$ counting the integer points in $P_D$ and a partition function $Z_{D,N}$ 
counting the integer points in the symmetric product of $P_D$. 
Here we will see how the plethystic function $PE[...]$ appears as a fundamental step in the counting problems. We will take the conifold
as an easy explanatory example and we will give for it a geometric prescription to compute the partition function for 
all the scalar BPS operators. 

%The most interesting fact is that $Z_{D,N}$ counts the $BPS$ states with a 
%given set of baryonic and global charges and hence it is the full baryonic partition function. 
>From $Z_D$, taking a suitable limit, 
we will be able to compute the volume of $H$ and of all the three cycles in $H$, and we will put in relation these results 
with the central charge and the R charges of the field theory, as described in Section \ref{volc3}. 
Although we mainly focus on the toric case the general prescriptions can be applied to every conical CY singularity,
and the formula for the computation of the volume of the three cycles, and of $H$ are valid in general. 
The extension to the non toric case is more a technical problem than a conceptual one.
% filling also the second missing points: find a general procedure to extract the volume of the cycles for the non-toric case.  
   
>From the knowledge of $Z_{D,N}$ we can 
reconstruct the complete partition function for the chiral ring of quiver gauge theories. This is a quite hard problem in
field theory, since we need to count gauge invariant 
operators modulo F-term relations and to take into account the finite number of colors $N$ which induces relations 
among traces and determinants. 
The geometrical computation of $Z_{D,N}$ allow to by-pass these problems as we will explicitly see in the next chapters.
 
In this chapter we will mainly focus on the geometrical properties of the partition functions $Z_D$, although some preliminary 
comparison with the dual gauge theory  is made in Section \ref{fieldtheo}. In forthcoming
chapters, we will show how to compute the complete partition function for selected examples and how to compare with field theory 
expectations \cite{Forcella:2007wk,Butti:2007jv}.

\section{Divisors and Line Bundles in Toric Geometry}\label{toric} 

Before starting our discussion about D3 brane embedding we need to review some basic topics of 
toric geometry. As often happens for a single phenomena there exist different descriptions, 
and some descriptions are more comfortable for the understanding of certain properties of the phenomena. 
Toric geometry admits different descriptions, we will not interested in exploring all the possible descriptions, and we refer to the literature for extensive explanations \cite{fulton,cox}. Two description of toric geometry are very useful for us: 
the first one presents a toric variety as an intersection in some ambient space and find it physical realizations 
for example in the study of the moduli space of CFT $\cX$, $\f$. 
We have already explained its general structure and its applications in chapters \ref{braneasing}; 
the second one describe a toric variety as a symplectic quotient. 
We have already seen this description in the study of the coherent component of the moduli space $\firr{~}$ 
and its refined Hilbert series in section \ref{s:Molien}.

In this section we plan to study a bit deeply its properties and in particular 
how to describe divisors and line bundles on toric varieties. 
As we will see this second description is very well suited for the counting problem. 

\begin{figure}[h!!!]
\begin{center}
\includegraphics[scale=0.6]{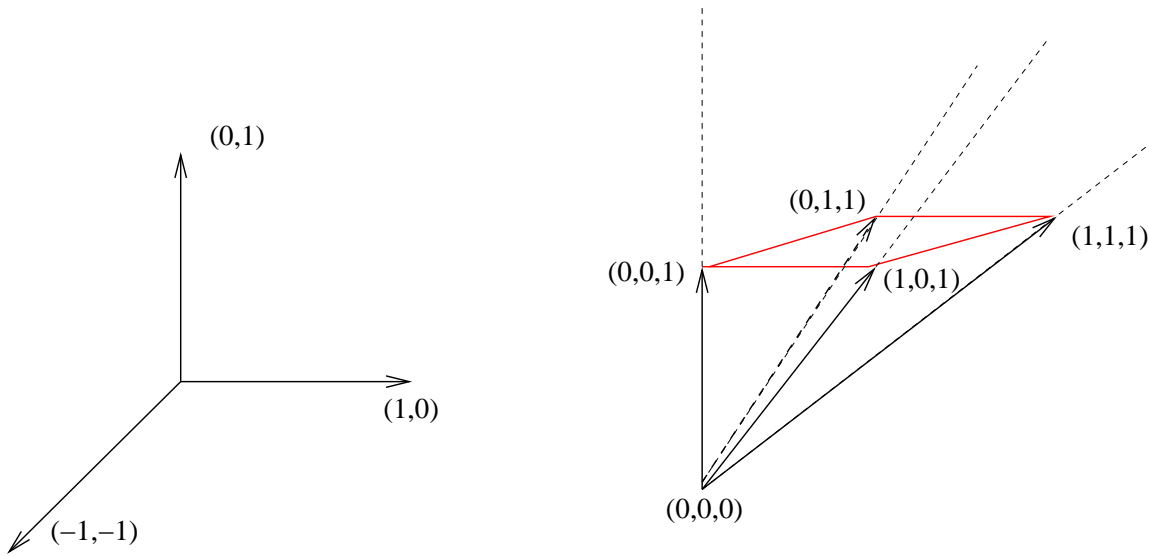} 
\caption{On the left: the fan for $\mathbb{P}^2$ with three maximal cones
of dimension two
which fill completely $\mathbb{R}^2$; there are three one dimensional cones
in $\Sigma(1)$ with generators $\{(1,0),(0,1),(-1,-1)\}$. On the right: the
fan for the conifold with a single maximal cone of dimension three; there are
four one dimensional cones  in $\Sigma(1)$ with generators $\{ (0,0,1),(1,0,1),(1,1,1),(0,1,1)\}$.}\label{fan}
\end{center}
\end{figure}

As we have already explained in the first chapter a toric variety $V_{\Sigma}$ 
is defined by a fan $\Sigma$: a collection of strongly  convex rational polyhedral cones in the 
real vector space $N_\mathbb{R} = N \otimes _ {\mathbb{Z}} \mathbb{R}$ ($N$ is an $n$ dimensional lattice $N \simeq \mathbb{Z}^n$). 
Some examples are presented in Figure \ref{fan}.

Starting from these data we can construct a toric variety as a complex intersection following the procedure we have already explained 
in previous chapters, or as a symplectic quotient, as presented in \ref{s:Molien} and explained in the following. 
The two procedures are equivalent, but the symplectic quotient construction is more useful for what we are going to discuss in this chapter.
We define the variety $V_{\Sigma}$ as a symplectic quotient \cite{fulton,cox}. Consider the one dimensional cones of 
$\Sigma $ and a minimal integer generator $V_i$ of each of them. Call the set of one dimensional cones $\Sigma (1) $. 
Assign a ``homogeneous coordinate'' $x_i$ to each $V_i \in \Sigma (1) $.
If $d=$ dim$\Sigma (1) $, $x_i $ span $\mathbb{C}^d$. 
Consider the group
\begin{equation}
G=\{(\mu_1,...,\mu_d\}\in (\mathbb{C}^*)^d | \prod_{i=1}^d \mu_i^{<m,V_i>} =1\, ,\,\,\,\, m\in \mathbb{Z}^3\}\, ,
\label{Ggroup}\end{equation}
which acts on $x_i$ as
\begin{equation}\label{Ggroupa}
(x_1,...,x_d) \rightarrow (\mu_1 x_1,...,\mu_d x_d)\, .
\end{equation}
 $G$ is isomorphic, in general, to $(\mathbb{C}^*)^{d-n}$ times a discrete group. The continuous part $(\mathbb{C}^*)^{d-n}$ 
can be described as follows.
Since $d \geqslant n $ the $V_i $ are not linearly independent. They determine $d-n$ linear  relations:
\begin{equation}\label{symplq}
\sum_{i=1}^{d} Q^{(a)}_i V_i = 0  
\end{equation}
with $a=1,...,d-n$ and $Q_i^{(a)}$ generate a $(\mathbb{C}^*)^{d-n}$ action on $\mathbb{C}^d$:
\begin{equation}(x_1,...,x_d) \rightarrow (\mu^{Q_1^{(a)}}x_1,...,\mu^{Q_d^{(a)}}x_d)\label{resc}\end{equation}
where $\mu \in \mathbb{C}^*$. 

For each maximal cone $\sigma \in \Sigma $ define the function $f_{\sigma}= \prod_{n_i \notin \sigma} x_i$ and the locus 
$S$ as the intersection of all the hypersurfaces $f_{\sigma}=0$. Then the toric variety is defined as: 
$$V_{\Sigma}= ( \mathbb{C}^d - S )/ G$$
There is a 
residual $(\mathbb{C}^*)^{n}$ complex torus action acting on $V_{\Sigma}$, from
which the name {\it toric variety}. In the following, we will denote with $T^n\equiv U(1)^n$ the real torus contained in $(\mathbb{C}^*)^{n}$.

In all the examples we will consider in this chapter $G=(\mathbb{C}^*)^{d-n}$ and the previous quotient is interpreted as a symplectic reduction. 
The case where $G$ contains a discrete part includes further orbifold quotients. These cases can be handled similarly to the ones 
discussed in this chapter as we will see some examples in the following chapters. 

Using these rules to construct the toric variety, it is easy to recover
 the usual representation for $\mathbb{P}^n $:
$$\mathbb{P}^n = (\mathbb{C}^{n+1} - \{ 0 \})/ \{ x \sim \mu x \}$$
where the the minimal integer generators of $\Sigma (1)$ are $n_i = \{ e_1 ,..., e_n, -\sum_{k=1}^n e_k \}$, $d=n+1$ and 
$Q = (1,...,1)$ (see Figure \ref{fan} for the case $n=2$).

We will be interested in affine toric varieties, where the fan is a single cone $\Sigma$ = $\sigma$. 
In this case $S$ is always the null set. It is easy, for example, to find the symplectic quotient representation of the conifold:
$$C(T^{1,1}) = \mathbb{C}^4 / (1,-1,1,-1)$$ 
where $d=4$, $n=3$, $n_1=(0,0,1)$, $n_2=(1,0,1)$, $n_3=(1,1,1)$, $n_4=(0,1,1)$ and we have written $(1,-1,1,-1)$ for 
the action of $\mathbb{C}^*$ with charges $Q= (1,-1,1,-1)$. 

This type of description of a toric variety is the easiest one to study divisors and line bundles. Each $V_i \in \Sigma (1)$ 
determines a $T$-invariant divisor $D_i$ corresponding to the zero locus $\{ x_i = 0 \}$ in $V_{\Sigma}$. $T$-invariant means 
that $D_i$ is mapped to itself by the torus action $(\mathbb{C}^*)^{n}$ (for simplicity we will call them simply divisors from now on). 
The $d$ divisors $D_i$ are not independent but satisfy the $n$ basic equivalence relations:
\begin{equation}\label{eqrel}
\sum _{i=1}^d < e_k , V_i > D_i = 0
\end{equation}
where $e_k$ with $k= 1,...,n$ is the orthonormal basis of the dual lattice $M \sim \mathbb{Z}^n $ with the natural paring: 
for $n\in N$, $m\in M$ $<n,m> = \sum _{i=1}^n n_i m_i$ $\in \mathbb{Z}$.
Given the basic divisors $D_i$ the generic divisor $D$ is given by the formal
sum $D = \sum_{i=1}^d c_i D_i$ with $c_i \in \mathbb{Z}$. Every divisor $D$
determines a line bundle $\mathcal{O}(D)$ \footnote{The generic divisor $D$ on an
affine cone is a Weyl divisor and not a Cartier divisor \cite{fulton}; for this reason the map between divisors 
and line bundles is more subtle, but it can be easily generalized using the homogeneous coordinate ring of the toric 
variety $V_{\Sigma}$ \cite{cox} in a way that we will explain. 
%The net result is a generalization of the the concept of line bundle, however we 
With an abuse of language, we will continue to call the sheaf $\mathcal{O}(D)$ the line bundle associated with the divisor $D$.}.

There exists a simple recipe to find the holomorphic sections of
 the line bundle $\mathcal{O}(D)$.
Given the $c_i$, the global sections of $\mathcal{O}(D)$ can be determined by looking at the polytope (a convex rational polyhedron in $M_{\mathbb{R}}$):
\begin{equation}\label{poly}
P_D = \{ u \in M_{\mathbb{R}}| <u,V_i >\hbox{  }\geq \hbox{  } - c_i \hbox{  }, \hbox{  }\forall i \in \Sigma(1) \}
\end{equation} 
where $ M_{\mathbb{R}}= M \otimes _ {\mathbb{Z}} \mathbb{R}$.
Using the homogeneous coordinate $x_i$ it is easy to associate a section $\chi^m$ to every point $m$ in $P_D$:
\begin{equation}\label{sect}
\chi ^m = \prod _{i=1}^d x_i ^{< m , V_i > + c_i} .
\end{equation} 
Notice that the exponent is equal or bigger than zero.
Hence the global sections of the line bundle $\mathcal{O}(D)$ over $V_{\Sigma}$ are:
\begin{equation}\label{sectgen}
H^0 ( V_{\Sigma}, \mathcal{O}_{V_{\Sigma}}(D))= \bigoplus _{m \in P_D \cap M } \mathbb{C} \cdot \chi ^m 
\end{equation} 
At this point it is important to make the following observation: all monomials $\chi ^m$ have the same charges under the $(\mathbb{C}^*)^{d-n}$ 
described at the beginning of this Section (in the following these charges will be identified with the baryonic charges of the dual gauge theory). 
Indeed, under the $(\mathbb{C}^*)^{d-n}$ action we have:
\begin{equation}\label{baryonicaction}
\chi ^m \rightarrow \prod _{i=1}^d(\mu ^{<m, Q_i^{(a)}V_i>+Q_i^{(a)}c_i})x_i ^{< m , V_i > + c_i} = \mu ^{\sum_{i=1}^d Q_i^{(a)} c_i} \chi ^m 
\end{equation} 
where we have used equation (\ref{symplq}). Similarly, all the sections have 
the same charge under the discrete part of the group $G$. 
This fact has an important consequence. The generic polynomial 
$$f = \sum a_m \chi^m \in H^0 ( V_{\Sigma}, \mathcal{O}_{V_{\Sigma}}(D))$$
is not a {\it function} on $V_{\Sigma}$, since it is not invariant under the $(\mathbb{C}^*)^{d-n}$ action (and under possible discrete orbifold actions).
However, it makes perfectly sense to consider the zero locus of $f$. Since all monomials in $f$ have the same charge under  $(\mathbb{C}^*)^{d-n}$, 
the equation $f=0$ is well defined on $V_{\Sigma}$ and defines a divisor \footnote{In this way, we can set a map between linearly equivalent divisors 
and sections of the sheaf $\mathcal{O}_{V_{\Sigma}}(D)$ generalizing the usual
map in the case of standard line bundles.}.

\subsection{A Simple Example}
After this general discussion, let us discuss an example to clarify 
the previous definitions. 

Consider the toric variety $\mathbb{P}^2$. 
The fan $\Sigma $ for $\mathbb{P}^2$ is generated by:
\begin{equation}
V_1 = e_1 \hbox{   } \hbox{ , } \hbox{  } V_2 = e_2 \hbox{   }\hbox{  ,  } \hbox{  } V_3 = - e_1 - e_2 
\end{equation}
The three basic divisors $D_i$ correspond to $\{ x_1 = 0 \}  $, $\{ x_2 = 0 \}  $, $\{ x_3 = 0 \}  $, and they satisfy the following relations 
(see equation (\ref{eqrel})):
\begin{eqnarray}
D_1 - D_3 = 0 \nonumber \\
D_2 - D_3 = 0 \nonumber
\end{eqnarray}
and hence $D_1 \sim D_2 \sim D_3\sim D$. All line bundles on $\mathbb{P}^2$
are then of the form ${\cal O}(n D)$ with an integer $n$, and are usually
denoted as ${\cal O}(n)\rightarrow \mathbb{P}^2$. It is well known that  
the space of global holomorphic sections of ${\cal O}(n)\rightarrow \mathbb{P}^2$ is
given by the homogeneous polynomial of degree $n$ for $n\ge 0$, while
it is empty for negative $n$. We can verify this statement using the 
general construction with polytopes.

\begin{figure}[h!!!]
\begin{center}
\includegraphics[scale=0.6]{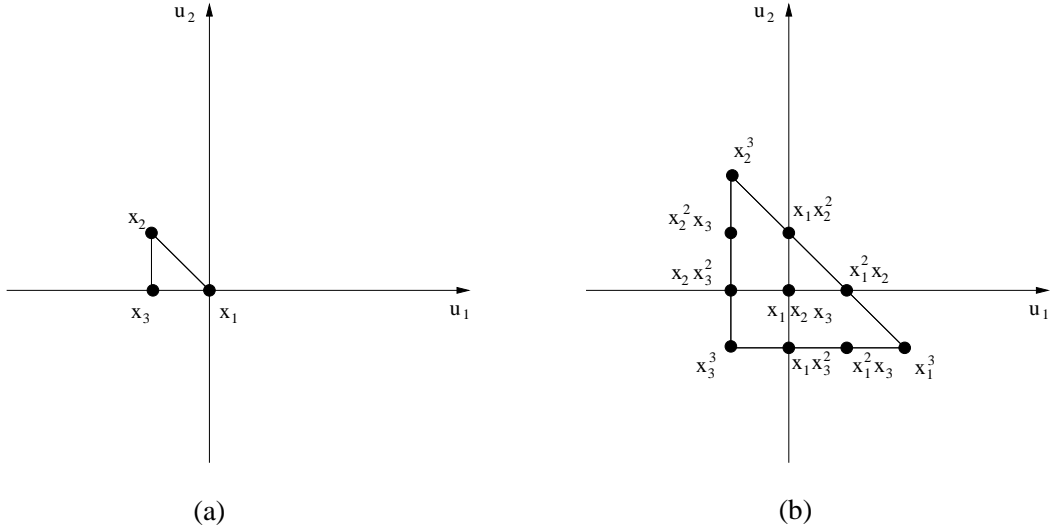} 
\caption{(a) The polytope associated to the line bundle $\mathcal{O}(1)
  \rightarrow \mathbb{P}^2$. (b) The polytope associated to the line bundle $\mathcal{O}(3) \rightarrow \mathbb{P}^2 $.}\label{p2o1o3}
\end{center}
\end{figure}

Consider the line bundle $\mathcal{O}(D_1)$ associated with the divisor $D_1$.
In order to construct its global sections we must first determine the polytope $P_{D_1}$ ($c_1 = 1, c_2 = c_3 = 0$):
\begin{equation}
P_{D_1} = \{ u_1 \geqslant -1 ,\hbox{  } u_2 \geqslant 0 ,\hbox{  } u_1 + u_2 \leqslant 0\}
\end{equation}
Then, using (\ref{sect}), it easy to find the corresponding sections:
\begin{equation}
\{ x_1,\hbox{  } x_2,\hbox{  } x_3 \}
\end{equation}
These are the homogeneous monomials of order one over $\mathbb{P}^2$. Indeed we have just constructed the line bundle 
$\mathcal{O}(1) \rightarrow \mathbb{P}^2$ (see Figure \ref{p2o1o3}). 

Consider as a second example the line bundle $\mathcal{O}(D_1 + D_2 + D_3)$.
In this case the associated polytope is:
\begin{equation}
P_{D_1+D_2+D_3 } = \{ u_1 \geqslant -1 ,\hbox{  } u_2 \geqslant -1 ,\hbox{  } u_1 + u_2 \leqslant 1\}
\end{equation}
Using (\ref{sect}) it is easy to find the corresponding sections:
\begin{equation}
\{ x_1^3,\hbox{  } x_1^2 x_2,\hbox{  } x_1x_2x_3, ...\}
\end{equation}
These are all the homogeneous monomials of degree $3$ over $\mathbb{P}^2$; we have indeed constructed the line bundle 
$\mathcal{O}(3) \rightarrow \mathbb{P}^2$ (see Figure \ref{p2o1o3}).

The examples of polytopes and line bundles presented in this Section are analogous to the ones that we will use in the 
following to characterize the $BPS$ baryonic operators. The only difference (due to the fact that  we are going to consider 
affine toric varieties) is that the polytope $P_D$ will be a non-compact rational convex polyhedron, and the space of 
sections will be infinite dimensional.\\  

\section{BPS D3 Brane Configurations}\label{susyD3}

We could in principle attack the problem of counting the BPS operators for generic $N$ directly in field theory, 
following the line of the previous chapter. 
The mathematical branch that take care of this problem is the invariant theory and essentially it consist 
in finding the set of invariant under the action of some group acting on the  elements of a ring of polynomials 
subject to a set of constrains on its generators. The constraints are given by the F-flat equations of the gauge 
theory while the group action is just the gauge group action of the field theory. In sections \ref{invmolien}, \ref{molienN}
 we will see some applications of invariant theory. This way to construct Hilbert series meets some difficult computational problems 
already at small rank of the gauge group. It is essentially very difficult to go over $N=3$. 
We will like to have a generic structure for the Hilbert series for arbitrary $N$. In section \ref{HSPE}
we learnt that to compute Hilbert series of the mesonic moduli space for generic $N$ it is enough to count the 
holomorphic functions over the $N$ times symmetric product of the transverse space $\cX$ to the system of D3 brane. 
As we will see in the following \cite{Mikhailov:2000ya} the set of all the holomorphic function of the variety $\cX$ 
parametrize the set of all possible trivial D3 brane embedding in $\cX$. 
These objects are in one to one correspondence to all the possible supersymmetric D3 branes wrapping a trivial three cycle in 
$H$ and spinning like a relativistic particle in the $AdS$ space. These are all the set of giant gravitons in 
$AdS\times H$ and according to the AdS/CFT correspondence they are dual to the set of all mesonic operators in the field theory. 

>From this specific example and from the computational difficulties of the field theoretic approach we will look to the dual geometry, 
and to the set of states dual to all the $1/2$ BPS operators of the gauge theory. 
Looking at the gravity dual of the gauge theory we will reach two main results: a generic formulation of the Hilbert series 
counting the BPS spectra for generic $N$; and the structure of the chiral ring of the gauge theory at strong coupling.

In this Section we discuss the prescription \cite{Beasley:2002xv} 
for determining the BPS 
Hilbert space corresponding to supersymmetric D3 brane configurations. We 
generalize the example of the conifold presented in \cite{Beasley:2002xv}
to the case of a generic toric Calabi-Yau cone.

\subsection{Motivations}\label{motivations}
%Let us move now to the physical problem. 
Consider the supersymmetric background of type $IIB$ supergravity $AdS_5 \times H$ with $H$ a Sasaki-Einstein manifold. This geometry is 
obtained by taking the near horizon geometry of a stack of  $N$ $D3$ on the isolated Goreinstein singularity of a local Calabi-Yau three-fold 
given by the real cone $C(H)$ over the base $H$. The D3 branes fill the four dimensional Minkowski space-time $M_4$ in $M_4 \times C(H)$. 

The dual superconformal field theory is a quiver gauge theory with gauge group $\prod_{i=1}^g SU(N_i)$.
% an $\mathcal{N}=1 $ supersymmetric quantum field theory with gauge group $SU(N_1) \times ...SU(N_k)$ and chiral superfields that transform under the fundamental of a gauge group and the anti-fundamental of another gauge group. 
Due to the presence of $SU(N)$ type groups these theories have generically baryonic 
like operators inside their spectrum and these are the 
objects we are interested in. 

Let us take the field theory dual to the conifold singularity as a basic example. We are already familiar with this
theory from the previous chapters. The theory has gauge group $SU(N) \times SU(N)$ and 
chiral superfields $A_1$, $A_2$ that transform under the fundamental of the first gauge group and under the anti-fundamental of the second one, 
and $B_1$, $B_2$ that transform under the conjugate representation. There exists also a non-abelian global symmetry $SU(2) \times SU(2)$ under which 
the $A$ fields transform as $(2,1)$ and the $B$ as $(1,2)$.  The superpotential
is $W=\epsilon_{ij}\epsilon_{pq} A_i B_p A_j B_q$.  
It is known that this theory has one baryonic charge and that the $A_i$ fields have charge one under this symmetry and the $B_i$ fields have 
charge minus one. Hence one can build the two basic baryonic operators:
\begin{eqnarray}
\label{exbarBS}
\epsilon^1 _{p_1,...,p_N}  \epsilon_2 ^{k_1,...,k_N} (A_{i_1})^{p_1}_{k_1}... (A_{i_N})^{p_N}_{k_N} = (\det A)_{ ( i_1,...,i_N ) } \nonumber \\
\epsilon^1 _{p_1,...,p_N}  \epsilon_2 ^{k_1,...,k_N} (B_{i_1})^{p_1}_{k_1}... (B_{i_N})^{p_N}_{k_N} = (\det B)_{ ( i_1,...,i_N ) }
\end{eqnarray}
These operators are clearly symmetric in the exchange of the $A_i$ and $B_i$ respectively, and transform under $(N+1,1)$ and $(1,N+1)$ 
representation of $SU(2) \times SU(2)$. The important observation is that these are the baryonic operators with the smallest possible dimension: 
$\Delta _{\det A ,\det B} = N \Delta _{A,B}$.    
One can clearly construct operator charged under the baryonic symmetry with bigger dimension in the following way. Defining the operators 
\cite{Berenstein:2002ke,Beasley:2002xv} \footnote{which are totally symmetric in the $SU(2)\times SU(2)$ indices due to the F-term 
relations $A_i B_p A_j=A_j B_p A_i$, $B_p A_i B_q = B_q A_i B_p$.}
\begin{equation}\label{AAAnB}
A_{I;J}= A_{i_1}B_{j_1}...A_{i_m}B_{j_m}A_{i_{m+1}}
\end{equation}
the generic type $A$ baryonic operator is:
\begin{equation}
\label{genbarABNS}
\epsilon^1 _{p_1,...,p_N}  \epsilon_2 ^{k_1,...,k_N} (A_{I_{1};J_{1}})^{p_1}_{k_1}... (A_{I_{N};J_{N}})^{p_N}_{k_N}\, . 
\end{equation}
One can clearly do the same with the type $B$ operators.

Using the tensor relation
\begin{equation}
\label{epsilonss}
\epsilon_{\alpha_1...\alpha_N}\epsilon^{\beta_1...\beta_N}=\delta^{\beta_1}_{[\alpha_1}...\delta^{\beta_N}_{\alpha_N]}\, ,
\end{equation} 
depending on the symmetry of (\ref{genbarABNS}), one can sometimes factorize the operator in a basic baryon  times operators 
that are neutral under the baryonic charge \cite{Berenstein:2002ke,Beasley:2002xv}. 

It is a notorious fact that the $AdS/CFT$ correspondence maps the basic baryonic operators (\ref{exbarBS}) to static $D3$ branes 
wrapping specific three cycles of $T^{1,1}$ and minimizing their volumes. The volumes of the $D3$ branes are proportional to 
the dimension of the dual operators in $CFT$. Intuitively, the geometric dual of an operator (\ref{genbarABNS}) is a fat brane 
wrapping a three cycle, not necessarily of minimal volume, and moving in the $T^{1,1}$ geometry (we will give more rigorous arguments below). 
If we accept this picture the factorisable operators in field theory can be interpreted in the geometric side as the product of gravitons/giant 
gravitons states with a static $D3$ brane wrapped on some cycle, and the non-factorisable ones are interpreted as excitation states of the 
basic $D3$ branes or non-trivial brane configurations. 

We would like to generalize this picture to a generic conical $CY$ singularity $\cX$. Using a nice parametrization 
of the possible $D3$ brane $BPS$ configurations in the geometry found in \cite{Mikhailov:2000ya,Beasley:2002xv}, 
we will explain how it is possible to characterize all the baryonic operators in the dual $SCFT$, count them according to their 
charges and extract geometric information regarding the cycles.

\subsection{Supersymmetric D3 Branes}\label{supD3}

Consider supersymmetric $D3$ branes wrapping three-cycles 
in $H$. There exists a general characterization of these types of configurations \cite{Mikhailov:2000ya,Beasley:2002xv} that 
relates the $D3$-branes wrapped on $H$ to holomorphic four cycles in $C(H)$. The argument goes as follows. Consider the euclidean 
theory on $\mathbb{R}^4 \times C(H)$. 
It is well known that one $D3$-brane wrapping a holomorphic surface $S$ in $C(H)$ preserves supersymmetry. If we put $N$ $D3$-branes 
on the tip of the cone $C(H)$ and take the near horizon limit the supergravity background becomes $Y_5 \times H$ where $Y_5$ is the 
euclidean version of $AdS_5$. We assume that $S$ intersects $H$ in some three-dimensional cycle $C_3$. 
The $BPS$ $D3$ brane wrapped on $S$ looks like a point in $\mathbb{R}^4$ and like a line in $Y_5$: it becomes a brane wrapped on a 
four-dimensional manifold in $\gamma \times H$ where $\gamma$ is the geodesic in $Y_5$ obtained from the radial direction in $C(H)$. 
Using the $SO(5,1)$ global symmetry of $Y_5$ we can rotate $\gamma$ into any other geodesic in $Y_5$. For this reason when we make the 
Wick rotation to return to Minkowski signature (this procedure preserves supersymmetry) we may assume that $\gamma $ becomes a time-like 
geodesic in $AdS_5$ spacetime. 
In this way we have produced a supersymmetric $D3$ brane wrapped on a three cycle in $H$ which moves along $\gamma$ in $AdS_5$. 
Using the same argument in the opposite direction, we realize also that any supersymmetric $D3$ brane wrapped on $H$ can be 
lifted to a holomorphic surface $S$ in $C(H)$.

 Due to this characterization, we can easily parametrize the classical phase space $\mathcal{P}_{cl}$ of supersymmetric 
$D3$ brane using the space of holomorphic surfaces in $C(H)$ without knowing the explicit metric on the Sasaki-Einstein space $H$ 
(which is generically unknown!).

The previous construction 
characterizes all kind of supersymmetric configurations
of wrapped D3 branes. These include branes wrapping trivial cycles and
stabilized by the combined action of the rotation and the RR flux, which
are called giant gravitons in the literature \cite{McGreevy:2000cw}. Except
for a brief comment on the relation between giant gravitons and dual giant 
gravitons, we will be mostly interested in D3 branes wrapping non trivial
cycles. These correspond to states with non zero baryonic charges in the
dual field theory. The corresponding surface $D$ in $C(H)$ is then a non
trivial divisor, which, modulo subtleties in the definition of the sheaf
${\cal O}(D)$, can be written as the zero locus of a section of ${\cal O}(D)$
\begin{equation}
\chi =0\, \qquad\qquad\qquad \chi\in H^0(\cX,{\cal O}(D))\label{gen}
\end{equation}
%Except that for a brief commets on the relation between giant gravitons and dual giant gravitons, 
\subsubsection{The Toric Case}
\begin{figure}
\begin{center}
\includegraphics[scale=0.6]{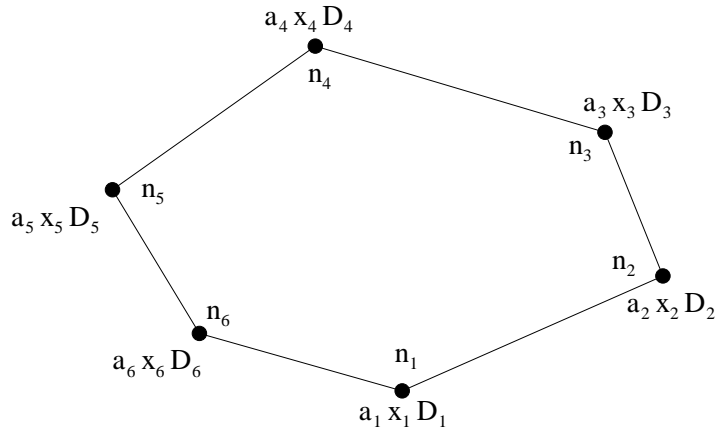} 
\caption{A generic toric diagram with the associated trial charges $a_i$, homogeneous coordinates $x_i$ and divisors $D_i$.}\label{hom}
\end{center}
\end{figure}

The previous discussion was general for arbitrary Calabi-Yau cones $C(H)$.
>From now on we will mostly restrict to the case of an affine toric Calabi-Yau cone $C(H)$. For this type of toric manifolds 
the fan $\Sigma$ described in Section \ref{toric} is just a single cone $\sigma$, due to the fact that we are considering a singular affine variety. 
Moreover, the Calabi-Yau nature of the singularity requires that all the generators of the one dimensional cone in $\Sigma(1)$
lie on a plane; this is the case, for example, of the conifold
pictured in Figure \ref{fan}. 
We can then characterize the variety with the convex hull 
of a fixed number of integer points in the plane: the toric diagram (Figure \ref{hom}). 
For toric varieties, the equation for the D3 brane configuration can be written quite explicitly using homogeneous coordinates. 
As explained in Section \ref{toric}, we can associate to every vertex of the toric diagram a global homogeneous coordinate $x_i$. 
Consider a divisor $D=\sum c_i D_i$. All the
supersymmetric configurations of D3 branes corresponding to surfaces linearly
equivalent to $D$ can be written as the zero locus of the generic section
of $H^0 ( V_{\Sigma}, \mathcal{O}_{V_{\Sigma}}(D))$
\begin{equation}\label{holsur}
P(x_1,x_2,...,x_d)\equiv\sum_{m\in P_D \cap M} h_m \chi^m = 0
\end{equation}
As discussed in Section 2, the sections take the form of the
monomials (\ref{sect}) 
$$\chi ^m = \prod _{i=1}^d x_i ^{< m , V_i > + c_i} $$
and there is one such monomial for each
integer point $m\in M$ in the polytope $P_D$ associated with $D$ as 
in equation (\ref{poly})
$$ \{ u \in M_{\mathbb{R}}| <u,V_i >\hbox{  }\geq \hbox{  } - c_i \hbox{  }, \hbox{  }\forall i \in \Sigma(1) \}$$
%We can equivalently associated to a set of baryonic charges polytope $P_D$, as explained in Section \ref{toric}. Using the recipe equation (\ref{sect}) we can construct the global sections $\chi ^m$  and the generic holomorphic global section of the line bundle $\mathcal{O}(D)$ (see (\ref{sectgen})) as a linear combination of $\chi ^m$:
%\begin{equation}\label{holsect}
%H^0 ( C(H), \mathcal{O}(D))= \sum _{m \in P_D \cap M } h_m \chi ^m
%\end{equation}
%\begin{equation}\label{holsur}
%P(x_1, x_2,...,x_d)= h + h_i x_i + h_{i,j}x_ix_j + h_{i,j,k}x_i x_j x_k + ... = \sum_I h_I x_I
%\end{equation}
%where we sum over repeated indices, we have introduced the multi-index $I$ and $h_I$ are complex constant parameters. 
As already noticed, the $x_i$ are only
defined up to the rescaling (\ref{resc}) but the equation
$P(x_1,...x_d)=0$ makes sense since all monomials have the same charge
under $(\mathbb{C}^*)^{d-3}$ (and under possible discrete orbifold actions). 
Equation (\ref{holsur}) generalizes the familiar
description of hypersurfaces in projective spaces $\mathbb{P}^n$ as zero
locus of homogeneous polynomials. In our case, since we are considering affine varieties, 
the polytope $P_D$ is non-compact and the space of holomorphic global sections is infinite dimensional.
 
We are interested in characterizing the generic supersymmetric $D3$ brane configuration with a fixed baryonic charge. 
We must therefore understand the relation between divisors and baryonic charges: it turns out that there is
a one-to one correspondence between baryonic charges and classes of divisors
modulo the equivalence relation (\ref{eqrel}) \footnote{In a fancy mathematical way, we could say that the 
baryonic charges of a D3 brane configuration are
given by an element of the Chow group $A_2(C(H))$.}. We will understand
this point by analyzing in more detail the $(\mathbb{C}^*)^{d-3}$ action  defined in Section \ref{toric}.  

%This can be done understanding that $P(x_1,...,x_d)$ is the generic element of the homogeneous coordinate ring over $C(H)$ and this one can be decomposed in direct sum of spaces having the same charges under the $(\mathbb{C}^*)^{d-n}$ action. 
\subsubsection{The Assignment of Charges}\label{charges}
To understand the relation between divisors and baryonic charges, 
we must make a digression and recall how one can assign $U(1)$ global charges to the homogeneous coordinates associated 
to a given toric diagram \cite{Benvenuti:2005ja,Franco:2005sm,Butti:2005ps,Butti:2005vn}.

Non-anomalous $U(1)$ symmetries play a very important role in the dual 
gauge theory and it turns out that we can easily parametrize these global symmetries directly from the toric diagram. 
In a sense, we can associate field theory charges directly to the homogeneous coordinates.

As explained in chapter \ref{braneasing} for a background with  horizon $H$, 
we expect $d-1$ global non-anomalous symmetries, 
where $d$ is number of vertices of the toric diagram \footnote{More precisely,
$d$ is the number of integer points along the perimeter of the toric diagram.
Smooth horizons have no integer points along the sides of the toric diagram
except the vertices, and $d$ coincides with the number of vertices. 
Non smooth horizons have sides passing through integer points and these must be
counted in the number $d$.}.
We can count these symmetries by looking at the number of massless vectors in the $AdS$ dual. And in chapter \ref{braneasing}
we obtained that the global abelian non anomalous symmtries are $U(1)^d = U(1)_R \times U(1)^2_F \times U(1)^{d-3}_B$, where the 
first three $U(1)$s come from the reduction of the metric along the isometries of $\cX$, while the other factors come from the
the reduction of the RR four form fields over the non trivial three cycles in H.

In this paper we use the fact that the $d-1$ non-R symmetry global non-anomalous charges can be parametrized by $d$
parameters $a_1, a_2, \ldots ,a_d$, each associated 
with a vertex of the toric diagram (or a point along an edge), 
satisfying the constraint:
\begin{equation}
\sum_{i=1}^d a_i = 0
\label{sum}
\end{equation}
The $d-3$ baryonic charges are those satisfying the further
constraint \cite{Franco:2005sm}:
\begin{equation}
\sum_{i=1}^d a_i V_i=0
\label{bari}
\end{equation}
where the vectors of the fan $V_i$ have the form $V_i=(y_i,z_i,1)$ with
$(y_i,z_i)$ the coordinates of integer points along the perimeter of
the toric diagram.
 
The R-symmetries are parametrized with the $a_i$
in a similar way of the other non-baryonic global symmetry, but they satisfy the different constraint
\begin{equation}
\sum_{i=1}^d a_i = 2
\label{sumr}
\end{equation} 
due to the fact that the terms in the superpotential must have
$R$-charges equal to two.

Now that we have assigned trial charges to the vertices of a toric diagram and hence to the homogeneous coordinates $x_i$, 
we can return to the main problem of identifying supersymmetric D3 branes with fixed baryonic charge. 
Comparing equation (\ref{symplq}) with equation (\ref{bari}), we realize that the baryonic charges $a_i$ in 
the dual field theory are the charges $Q^{(a)}_i$ of the action of $(\mathbb{C}^*)^{d-3}$ on the homogeneous coordinates $x_i$. 
We can now assign a baryonic charge to each monomials made with the homogeneous coordinates $x_i$. 
All terms in the equation (\ref{holsur}) corresponding to a D3 brane wrapped on $D$ are global sections $\chi ^m$ of 
the line bundle $\mathcal{O}(\sum_{i=1}^d c_i D_i)$ and they have all the same $d-3$ baryonic charges $B^a=\sum_{i=1}^d Q_i^{(a)} c_i$; 
these are determined only in terms of the $c_i$ defining the corresponding line bundle (see equation (\ref{baryonicaction})). 

Using this fact we can associate a divisor in $C(H)$ to every set of $d-3$ baryonic charges. The procedure is as follows. 
Once we have chosen a specific set of baryonic charges $B^a$ we determine the corresponding $c_i$ using the
relation $B^a=\sum_{i=1}^d c_i Q_i^{(a)}$.  These coefficients define a divisor $D=\sum_{i=1}^d c_i D_i$.  
It is important to observe that, due to equation (\ref{symplq}), the $c_i$ are defined only modulo the equivalence relation 
$c_i \sim c_i + <m,V_i>$ corresponding to the fact that the line bundle $\mathcal{O}(\sum_{i=1}^d c_i D_i)$ is identified only 
modulo the equivalence relations (\ref{eqrel}) $D \sim D + \sum_{i=1}^d <m,V_i> D_i $. We conclude that baryonic charges are 
in one-to-one relation with divisors modulo linear equivalence.

%Summarizing: once we have chosen a set of baryonic charges and determined the corresponding $c_i$'s, all the possible supersymmetric $D3$ brane configurations with that given baryonic charges are given by the set of all possible holomorphic surfaces $D$ with that specific charges under the $(\mathbb{C}^*)^{d-3}$ action. 
%These ones are the zero locus of sections of $H^0 ( C(H), \mathcal{O}(D))$
%as in equation (\ref{holsur}). Since the space of sections is infinite
%dimensional, we allow the integers $m\in P_D\cap M$ to run  
%over an infinite set. 

For simplicity, we only considered continuous baryonic charges. In the case
of varieties which are orbifolds, the group $G$ in equation (\ref{Ggroup})
contains a discrete part. 
In the orbifold case, the quantum field theory contains 
baryonic operators with discrete charges. This case can be easily
incorporated in our formalism as we will see in chapter \ref{chiralcount}.

\subsubsection{The Quantization Procedure}

Now we want to quantize the classical phase space $\mathcal{P}_{cl}$ using geometric quantization \cite{Wood} following \cite{Beasley:2002xv}.

Considering that $P$ and $\lambda P$ with $\lambda \in \mathbb{C}^*$ vanish on the same locus in $C(H)$, 
it easy to understand that the various distinct surfaces in $C(H)$ correspond to a specific set of $h_m$ 
with $m \in  P_D \cap M $ modulo the equivalence relation $h_m \sim \lambda h_m $: this is a point in the 
infinite dimensional space $\mathbb{CP}^{\infty}$ in which the $h_m$ are the homogeneous coordinates. 
Thus we identify the classical configurations space $\mathcal{P}_{cl}$ of supersymmetric $D3$ brane associated 
to a specific line bundle $\mathcal{O}(D)$ as $\mathbb{CP}^{\infty}$ with homogeneous coordinates $h_m$.
%Thanks to the fact that we are using homogeneous coordinates the holomorphic surface are divided in two classes. The generic surface with at least one non-vanishing baryonic charge represent a $D3$ brane wrapped on homologically non-trivial three cycles and we may call these configurations the baryonic ones. The other class of surfaces are the one with all the baryonic charges equal to zero, these are associated to $D3$ branes wrapped on trivial three cycles that in literature are called giant gravitons \cite{McGreevy:2000cw}.\\
A heuristic way to understand the geometric quantization is the following.
We can think of the $D3$ brane as a particle moving in $\mathcal{P}_{cl}$ and we can associate to it a wave 
function $\Psi$ taking values in some holomorphic line bundle $\mathcal{L}$ over $\mathbb{CP}^{\infty}$.  
The reader should not confuse the line bundle ${\cal L}$, over the classical
phase space $\mathcal{P}_{cl}$ of wrapped D3 branes, with the lines bundles ${\cal O}(D)$, which are defined on $C(H)$.  
Since all the line bundles ${\cal L}$ over a projective space are determined by their degree 
(i.e. they are of the form $\mathcal{O}(\alpha)$) we have only to find the value of $\alpha$. 
This corresponds to the phase picked up by the wave function $\Psi$ when the $D3$ brane goes around a closed path in $\mathcal{P}_{cl}$. 
Moving in the phase space $\mathcal{P}_{cl}$ corresponds to moving
 the $D3$ brane in $H$. Remembering that a $D3$ brane couples to the four form field $C_4$ of the supergravity and that 
the backgrounds we are considering are such that $\int_H F_5 =N$, it was argued in \cite{Beasley:2002xv} that the wave function 
$\Psi$ picks up the phase $e^{2\pi i N}$ and $\alpha = N$. 
For this reason $\mathcal{L}=\mathcal{O}(N)$ and the global holomorphic sections of this line bundle over 
$\mathbb{CP}^{\infty}$ are the degree $N$ polynomials in the homogeneous coordinates $h_m$. 

Since the $BPS$ wave functions are the global holomorphic sections of $\mathcal{L}$, we have that the $BPS$ 
Hilbert space $\mathcal{H}_D$ is spanned by the states: 
\begin{equation}
| h_{m_1}, h_{m_2},...,h_{m_N} > 
\label{bpsstateH}
\end{equation} 
This is Beasley's prescription. 

We will make a correspondence 
in the following between $h_m$ and certain operators in the field theory
with one (or more) couple of free gauge indices
\begin{equation}
h_m\, ,\qquad\qquad\qquad (O_m)_{\alpha\beta}\,\, (\rm{or}\,(O_m)_{\alpha_1\beta_1...\alpha_k\beta_k})
\end{equation}
where the $O_m$ are operators with fixed baryonic charge. The generic state
in this sector $| h_{m_1}, h_{m_2},...,h_{m_N} >$ will be identify with a
gauge invariant operator obtained by contracting the $O_m$ with one
(or more) epsilon symbols. The explicit example of the conifold is
discussed in details in \cite{Beasley:2002xv}: the homogeneous coordinates 
with charges $(1,-1,1,-1)$ can be put in one-to-one correspondence with the elementary fields $(A_1,B_1,A_2,B_2)$ 
which have indeed baryonic charge $(1,-1,1,-1)$. The use of the divisor $D_1$ (modulo linear equivalence) allows to
study the BPS states with baryonic charge $+1$. 
It is easy to recognize that the operators $A_{I,J}$ in equation (\ref{AAA}) have baryonic charge $+1$ and are in 
one-to-one correspondence with the sections of ${\cal O}(D_1)$
$$\sum_{m\in P_{D_1}\cap M} h_m \chi^m =h_1 x_1+ h_3 x_3+...$$
The BPS states $|h_{m_1}, h_{m_2},...,h_{m_N}>$
are then realized as all the possible determinants, as  in equation 
(\ref{genbarABNS}).
 
\subsection{Giant Gravitons and Dual Giant Gravitons}
%As we have previously explained, given a set of baryonic charges and 
%the corresponding $c_i$ modulo the identification $c_i \sim c_i + <m,n_i>$, the generic $D3$ brane classically supersymmetric configurations with that baryonic charges is given by the zero locus of the generic section \ref{holsect}. The quantization procedure assigns to every set of $N$ elements $h_m$ in $\mathbb{CP}^{\infty}$ a wavefunction \ref{bpsstate} i.e. degree $N$ holomorphic polynomial in the coordinates $h_{m_i}$. 

Among the $\mathcal{O}(D)$ line bundles there is a special one: the bundle of holomorphic functions $\mathcal{O}$ 
\footnote{Clearly there are other line bundles equivalent to $\mathcal{O}$. For example, since we are considering Calabi-Yau spaces, 
the canonical divisor $K= - \sum_{i=1}^d D_i$ is always trivial and $\mathcal{O} \sim \mathcal{O}(\sum_{i=1}^d D_i)$.}. 
It corresponds to the supersymmetric $D3$ brane configurations wrapped on homologically trivial three cycles $C_3$ in $H$ 
(also called giant gravitons \cite{McGreevy:2000cw}). Indeed we now understand that the counting procedure for the mesonic spectra of
the field theory, intuitively explained in chapter \ref{braneasing}, is indeed the counting of all the possible 
giant gravitons in the dual string theory background $AdS \times H$.

When discussing trivial cycles, we can parameterize holomorphic surfaces just by using the embedding 
coordinates \footnote{These can be also expressed as specific polynomials in the homogeneous coordinates 
such as that their total baryonic charges are zero.}. Our discussion here can be completely general and not 
restricted to the toric case.
Consider the general Calabi-Yau algebraic affine varieties $\cX$ 
that are cone over some compact base $H$ (they admit at least a $\mathbb{C}^* $ action $\cX=C(H)$). 
These varieties are the zero locus of a collection of polynomials in some $\mathbb{C}^k$ space. 
We will call the coordinates of the $\mathbb{C}^k$ the embedding coordinates $z_j$, with $j=1,...,k$. 
The coordinate rings $\mathbb{C}[\cX]$ of the varieties $\cX$ are the restriction of the polynomials in $\mathbb{C}^k$ 
of arbitrary degree on the variety $\cX$:   
\begin{equation}
\mathbb{C}[\cX]= \frac{\mathbb{C}[z_1,...z_k]}{(p_1,...,p_l)}= \frac{\mathbb{C}[z_1,...z_k]}{\mathcal{I}(\cX)}
\label{corring}
\end{equation} 
where $\mathbb{C}[z_1,...,z_k]$ is the $\mathbb{C}$-algebra of polynomials in $k$ variables and $p_j$ are 
the defining equations of the variety $\cX$.
We are going to consider the completion of the coordinate ring (potentially infinity polynomials) 
whose generic element can be written as the (infinite) polynomial in $\mathbb{C}^k$
\begin{equation}
P(z_1,...z_k) = c + c_i z_i + c_{ij}z_iz_j+...= \sum _I c_I z_I
\label{polCk}
\end{equation} 
restricted by the algebraic relations $\{ p_1=0,...,p_l=0\}$\footnote{ There exist a difference 
between the generic baryonic surface and the mesonic one: the constant $c$. The presence of this 
constant term is necessary to represent the giant gravitons, but if we take for example the constant 
polynomial $P=c$ this of course does not intersect the base $H$ and does not represent a supersymmetric $D3$ brane. 
However it seems that this is not a difficulty for the quantization procedure \cite{Beasley:2002xv,Biswas:2006tj}}. 

At this point Beasley's prescription says that the $BPS$ Hilbert space of the 
giant gravitons $\mathcal{H}_g$ is spanned by the states  
\begin{equation}
| c_{I_1}, c_{I_2},...,c_{I_N} > 
\label{bpsstategiant}
\end{equation} 
These states are holomorphic polynomials of degree $N$ over $\mathcal{P}_{cl}$ and are obviously 
symmetric in the $c_{I_i}$. For this reason we may represent (\ref{bpsstategiant}) as the symmetric product:
\begin{equation}
Sym ( | c_{I_1}> \otimes | c_{I_2}> \otimes ... \otimes |c_{I_N} >) 
\label{bpsstategiantsym}
\end{equation} 
Every element $|c_{I_i} > $ of the symmetric product is a state that 
represents a holomorphic function over the variety $C(H)$. This is easy to understand if one takes the polynomial 
$P(z_1,...z_k)$ and consider the relations among the $c_I$ induced by the radical ideal $\mathcal{I}(\cX)$ (in a sense 
one has to quotient by the relations generated by  $\{ p_1=0,...,p_l=0\}$). For this reason the Hilbert space of giant gravitons is:
\begin{equation}
\mathcal{H}_g =  \bigotimes ^{N \hbox{   } Sym} \mathcal{O}_{\cX} 
\label{bpsstategiants}
\end{equation} 
Obviously, we could have obtained the same result in the toric case 
by applying the techniques discussed in this Section. Indeed if we put all the $c_i$ equal to zero the 
polytope $P_D$ reduces to the dual cone $\sigma^*$ of the toric diagram, whose integer points corresponds 
to holomorphic functions on $\cX$ \cite{Martelli:2006yb}. Indeed in the toric case we reproduce the structure of the mesonic chiral ring we claimed in 
the chapter \ref{braneasing}. Namely the mesonic chiral ring is the set of all $N$ times symmetric holomorphic 
function over $\cX$. Given the expression \ref{bpsstategiants} we can claim that the mesonic moduli space of the 
field theory is $N$ times the symmetric product of the geometric singularity $\cX$. We have obtained this result 
in the dual gravity description of the gauge theory and indeed this result is a strong coupling result. 
To make this statement completely rigorous we need to have a clearer matching between the giant gravitons 
in string theory and the mesonic operators in the gauge theory. This match will be done in section \ref{fieldtheo}.

There exist another set of D3 brane configurations in $AdS_5 \times H$ that can be considered dual to the mesonic operators 
in the field theory. These configurations are called dual giant gravitons and consist of D3 branes wrapped on a three-sphere in 
$AdS_5$ and spinning along the Reeb direction. In \cite{Martelli:2006vh} it was shown that the Hilbert space of a dual giant graviton
$\mathcal{H}_{dg}$ in the background space $AdS_5 \times H$, where $H$ is a generic Sasaki-Einstein manifold, is the space of 
holomorphic functions $\mathcal{O}_{\cX}$ over the cone $\cX$. A single dual giant graviton reproduce just the abelian structure of the field theory, and to 
have informations regarding the non abelian case we need to construct the gran canonical ensemble of N dual giant gravitons. 
At the end of the day what come out is that the Hilbert space of N dual giant gravitons $\mathcal{H}_{dg}^N$ 
is the N times symmetric product of holomorphic functions over $\cX$:
\begin{equation} 
\mathcal{H}_{dg}^N = \bigotimes ^{N \hbox{   } Sym} \mathcal{O}_{\cX}
\end{equation}
In the discussion in literature there is the observations that giant gravitons and dual giant gravitons are in some sense 
dual to each other. At this point it is easy to understand in complete generality this duality:
\begin{equation}
\mathcal{H}_g =  \mathcal{H}_{dg}^N 
\label{bpsstategiantsh}
\end{equation} 
and indeed this is the reason why the counting of $1/2$ $BPS$ giant graviton states and 
dual giant graviton states give the same result\cite{Biswas:2006tj,Mandal:2006tk}: 
the counting of $1/2$ $BPS$ mesonic state in field theory. 

This section was a bit technical, it is therefore worthwhile to make the point

\paragraph{Summary: } we have learned that every line bundle over $\cX$ individuate a topological embedding for a supersymmetric
D3 brane in H. Every section of the given line bundle distinguishes the actual BPS embedding inside the given topological class.
A quantum state of a D3 brane is given by the N time symmetric product of sections of the particular line bundle. These states 
are the ones to be matched with the BPS operators. If the line bundle is non trivial the three cycle wrapped in H by the D3 brane
is non trivial and the operator dual to the D3 brane states will be a baryonic operator, otherwise the three cycle will be trivial 
and the dual operator will be a mesonic operator.

\section{Flavor Charges of the BPS Baryons}\label{bar}
In the previous Section, we discussed supersymmetric $D3$ brane configurations with specific baryonic charge. 
Now we would like to count, in a sector with given baryonic charge, the states with a given 
set of flavor charges $U(1)\times U(1) $ and $R$-charge $U(1)_R$. The generic state of the $BPS$ Hilbert 
space (\ref{bpsstateH}) is, by construction, a symmetric product of the single states $|h_m>$. 
These are in a one to one correspondence with the integer points in the polytope $P_D$, which 
correspond to sections $\chi^m$. As familiar in toric geometry \cite{fulton,cox}, a integer point $m\in M$ contains information about
the charges of the $T^3$ torus action, or, in quantum field theory language,
 about the flavor and $R$ charges.

Now it is important to realize that, as already explained in Section \ref{susyD3}, the charges $a_i$ 
that we can assign to the homogeneous coordinates $x_i$ contain information about the baryonic charges 
(we have already taken care of them) but also about the flavor and $R$ charges in the dual field theory. 
%In fact one can understand the homogeneous coordinates as an equivalence class of fields in the $CFT$: each homogeneous coordinate corresponds to a set of 
%elementary fields with given baryonic, flavor and $R$ charges \cite{tomorrow,aZequiv}. 
If we call $f^k_i$ with $k=1,2$ the two flavor charges and $R_i$ the $R$-charge, 
the section $\chi^m$ has flavor charges (compare equation (\ref{sect})):
\begin{equation}
f^k_{m} = \sum_{i=1}^d (<m,V_i> + c_i)f^k_i
\label{flav}
\end{equation}  
and $R$-charge:
\begin{equation}
R_{m} = \sum_{i=1}^d (<m,V_i> + c_i)R_i
\label{erre}
\end{equation} 

All the singularity $\cX$ have a special vector $K$ called the Reeb vector. It is constant in norm over all $\cX$ and parametrizes the 
isometry of $\cX$ dual to the R symmetry in the field theory. It is the vector field that move the killing spinor $\psi$ of the CY $\cX$: $\mathcal{L}_{K} \psi \ne 0$, and it can be expanded over a basis $\partial_{\phi_i}$ of the $T^3$ toric fibration: $K= \sum_{i=1}^3 b_i \partial_{\phi_i}$.

Making use of the Reeb vector it is possible to refine the last formula (\ref{erre}). Indeed 
the $R_i$, which  are the R-charges of a set of elementary fields of the gauge theory
\cite{Franco:2005sm,Butti:2005ps} \footnote{The generic elementary field in the gauge
theory has an R-charge which is a linear combination of the $R_i$ \cite{Butti:2005ps,Butti:2005vn}.},
%Indeed a static D3 brane wrapped on  the three cycles $C_3^i$ corresponding to the elementary divisors $D_i$ restricted at $r=1$ gives rise to a di-baryonic
%operator $\det X^{(i)}$ in the gauge theory. The elementary field $X^{(i)}$
%has R-charge $R_i$. 
%One can also parametrizes the $R_i$ in terms of the volume of the Sasaki-Einstein space $H$ and of the volume of $C_3^i$ \cite{Gubser:1998fp}:
%\begin{equation}
%R_i = \frac{\pi \hbox{Vol}(C_3^i)}{3 \hbox{Vol}(H)}
%\label{errevol}
%\end{equation} 
%Here $\hbox{Vol}(C_3^i)$ and $\hbox{Vol}(H)$ 
are completely determined by the Reeb vector of $H$ and the vectors $V_i$ defining the toric diagram \cite{Martelli:2005tp}. 
%Using this parametrization 
Moreover, it is possible to show that $\sum_{i=1}^d V_i R_i = \frac{2}{3}b $ \cite{Butti:2005ps,Butti:2005vn}, where $b$ specifies how the Reeb vector 
lies inside the  $T^3$ toric fibration $b=(b_1,b_2,b_3)$. Hence:
\begin{equation}
R_{m} = \frac{2}{3} <m,b> + \sum_{i=1}^dc_iR_i
\label{errebb}
\end{equation} 
This formula generalizes an analogous one for mesonic operators \cite{Butti:2006nk}. 
%This $R$ charge is really what we expect from an operator in field theory that is given by an elementary field with some baryonic charges 
%and a tower of mesons over it \cite{Butti:2006nk}. 
Indeed if we put all the $c_i$ equal to zero the polytope $P_D$ reduces to the 
dual cone $\sigma^*$ of the toric diagram \cite{Martelli:2006yb}. We know  that the elements of the 
mesonic chiral ring of the $CFT$ correspond to integer points in this cone and they have $R$-charge 
equal to $\frac{2}{3}<m,b>$. 
In the case of generic $c_i$, the right most factor of (\ref{errebb}) is in a sense the background $R$ charge: 
the $R$ charge associated to the fields carrying the non-trivial baryonic charges. In the simple example of 
the conifold discussed in subsection \ref{motivations}, formula (\ref{errebb}) applies to the operators (\ref{AAAnB}) 
where the presence of an extra factor of $A$ takes into account the background charge. In general the R charge (\ref{errebb})
is really what we expect from an operator in field theory that is given by elementary fields with some baryonic charges 
dressed by ``mesonic insertions''.

The generic baryonic configuration is constructed by specifying $N$ integer points $m_{\rho}$ in the polytope $P_D$. 
 Its $R$ charge $R_B$ is
\begin{equation}
R_{B} = \frac{2}{3} \sum_{\rho = 1}^N <m_{\rho},b> + N \sum_{i=1}^dc_iR_i
\label{errebar}
\end{equation} 
This baryon has $N$ times the baryonic charges of the associated polytope. 
Recalling that at the superconformal fixed point dimension $\Delta$ and $R$-charge of a chiral superfield are related 
by $R = 2 \Delta / 3$, it is easy to realize that the equation (\ref{errebar}) is really what is expected for a baryonic 
object in the dual superconformal field theory. Indeed if we put all the $m_{\rho} $ equal to zero we have 
(this means that we are putting to zero all the mesonic insertions)
\begin{equation}
\Delta _{B} = N \sum_{i=1}^dc_i \Delta _i
\label{basicbar}
\end{equation} 
This formula can be interpreted as follows.
The elementary divisor $D_i$ can be associated with (typically more than one) elementary field in the gauge theory, with R charge $R_i$. 
By taking just one of the $c_i$ different from zero in formula (\ref{basicbar}), we obtain the dimension of a 
baryonic operator in the dual field theory: take a fixed field, compute its determinant and the dimension of 
the operator is $N$ times the dimension of the individual fields. These field operators correspond to $D3$ branes 
wrapped on the basic divisors $D_i$ and are static branes in the $AdS_5 \times H$ background \footnote{The generic configuration of a $D3$ brane wrapped on a three cycle $C_3$ in $H$ is given by a holomorphic section of $\mathcal{O}(D)$ that is a 
non-homogeneous polynomial under the $R$-charge action. For this reason, and holomorphicity, 
it moves around the orbits of the Reeb vector \cite{Mikhailov:2000ya,Beasley:2002xv}. 
Instead the configuration corresponding to the basic baryons is given by the zero locus of a 
homogeneous monomial (therefore, as surface, invariant under the $R$-charge action), and for this reason it is static.}. 
They wrap the three cycles $C_3^i$ obtained by restricting the elementary divisors $D_i$  at $r=1$. 
%gives rise to a di-baryonic
%operator $\det X^{(i)}$ in the gauge theory. The elementary field $X^{(i)}$
%has R-charge $R_i$. 
One can also write the $R_i$ in terms of the volume of the Sasaki-Einstein space $H$ and of the volume of $C_3^i$ \cite{Gubser:1998fp}:
\begin{equation}
R_i = \frac{\pi \hbox{Vol}(C_3^i)}{3 \hbox{Vol}(H)}
\label{errevol}
\end{equation} 
%Here $\hbox{Vol}(C_3^i)$ and $\hbox{Vol}(H)$ are completely determined by the Reeb vector of $H$ and the vectors $n_i$ defining the toric diagram \cite{MSY}.
Configurations with
more than one non-zero $c_i$ in equation (\ref{basicbar}) correspond to
basic baryons made with elementary fields whose R-charge is a linear combination of the $R_i$ (see \cite{Butti:2005ps,Butti:2005vn}) 
or just the product of basic baryons. 

The generic baryonic configuration has $N$ times the $R$-charge and the global charges of the basic baryons 
(static branes which minimize the volume in a given homology class) plus the charges given by the 
fattening  and the motion of the three cycle inside the geometry (the mesonic fluctuations on the $D3$ 
brane or ``mesonic insertions'' in the basic baryonic operators in field theory). It is important to notice 
that the BPS operators do not necessarily factorize in a product of basic baryons times mesons 
\footnote{In the simple case of the conifold this is due to the presence of two fields $A_i$ with the same gauge 
indices; only baryons symmetrized in the indices $i$ factorize \cite{Berenstein:2002ke,Beasley:2002xv}. 
In more general toric quiver gauge theories it is possible to find different strings of elementary fields with the same
baryonic charge connecting a given pair of gauge groups; their existence 
prevents the generic baryons from being factorisable.}.

\subsection{Setting the Counting Problem}
In Section \ref{counting} we will count the baryonic states of the theory with given baryonic charges (polytope $P_D$) 
according to their $R$ and flavor charges. 
Right now we understand the space of classical supersymmetric $D3$ brane configurations $\mathcal{N}$ as a 
direct sum of holomorphic line bundles over the variety $\cX$:
\begin{equation}
\mathcal{N} = \bigoplus_{c_i \sim c_i + <m,V_i> } \mathcal{O}\Big(\sum_i^d c_i D_i \Big)
\label{dirs}
\end{equation} 
where the $c_i$ specify the baryonic charges. We have just decomposed the
space $\mathcal{N}$ into sectors according to the grading given by the baryonic
symmetry. Geometrically, this is just the decomposition of the homogeneous
coordinate ring of the toric variety under the grading given by the action
of $(\mathbb{C}^*)^{d-3}$. Now, we want to introduce a further grading.
Inside every line bundle there are configurations with different flavor and $R$ charges corresponding to 
different sections of the same line bundle. 

Once specified the baryonic charges, the Hilbert space of BPS
operators is the $N$ order symmetric product of the corresponding line bundle. 
Hence the $1/2$ $BPS$ Hilbert space is also decomposed as:
\begin{equation}
\mathcal{H} = \bigoplus_{c_i \sim c_i + <m,V_i> } \mathcal{H}_{D}
\label{dirsH}
\end{equation} 
We would like to count the baryonic operators of the dual $SCFT$ with a given set of flavor and $R$ charges. 
We can divide this procedure into three steps:
\begin{itemize}
\item{find a way to count the global sections of a given holomorphic line bundle (a baryonic partition function $Z_D$);}
\item{write the total partition function for the $N$-times symmetric product of the polytope $P_D$ (the partition function $Z_{D,N}$). 
This corresponds to find how  many combinations there are with the same global charges $a_B^k$ 
(with $k=1,2$ for the flavor charges and $k=3$ for the $R$ charge ) for a given baryonic state: the possible set of $m_{\rho}$ such that: 
\begin{equation}
a_B^k - N \sum_{i=1}^d c_i a_i^k = \sum_{m_{\rho} \in P_D \cap M}\sum_{i=1}^d <m_{\rho},V_i> a_i^k\, .
\label{flavBN}
\end{equation}
}  
\item{write the complete $BPS$ partition function of the field theory by summing over
all sectors with different baryonic charges. 
Eventually
we would also like 
%to translate these considerations on the homogeneous coordinates $x_i$ on a counting on the fields of the $SCFT$, and eventually 
to write the complete $BPS$ partition function of the field theory including all the $d$ charges at a time: 
$d-3$ baryonic, $2$ flavor and $1$ $R$ charges \cite{Forcella:2007wk,Butti:2007jv}.}
\end{itemize} 
In the following Sections, we will solve completely the first two steps. 
The third step is complicated by various facts. First of all 
the correspondence between the
homogeneous coordinates and fields carrying the same $U(1)$ charges is not one
to one. From the gravity side of the $AdS/CFT$ correspondence one can explain
this fact as follow \cite{Franco:2005sm}. The open strings attached to a $D3$
brane wrapped on the non-trivial three cycles corresponding to the basic
baryons in the dual field theory have in general many supersymmetric vacuum
states. This multiplicity of vacua corresponds to the fact that generically
the first homotopy group of the three cycles $\pi_1(C_3)$ is non-trivial and
one can turn on a locally flat connection with non-trivial Wilson lines. The
different Wilson lines give the different open string vacua and these are
associated with different elementary fields $X_{ij}$ (giving rise to basic
baryons $\det X_{ij}$ with the same global charges). One has then to include
non trivial multiplicities for the $Z_{D,N}$ when computing the complete BPS 
partition function. Moreover one should pay particular attention to the
sectors with higher baryonic charge. All these issues and the
determination of the partition function depending on all
the $d$ charges are discussed in \cite{Forcella:2007wk,Butti:2007jv}, and will be review them in the following two chapters.

\section{A First Comparison with the Field Theory Side}\label{fieldtheo} 
At this point it is probably worthwhile to make a more straight contact with 
the field theory. This is possible at least for all toric CY because the
 dual quiver
gauge theory is known \cite{Hanany:2005ve,Franco:2005rj,Hanany:2005ss,Feng:2005gw}. In this chapter we will
mainly focus on the partition function $Z_D$ for supersymmetric D3 brane 
configurations. In forthcoming chapters we will show how to
compute the partition function for the chiral ring and how to compare with
the full set of BPS gauge invariant operators. In this Section we show that,
for a selected class of polytopes $P_D$, there is a simple correspondence
between sections of the line bundle $D$ and operators in the gauge theory.
  
\begin{figure}
\begin{center}
\includegraphics[scale=0.5]{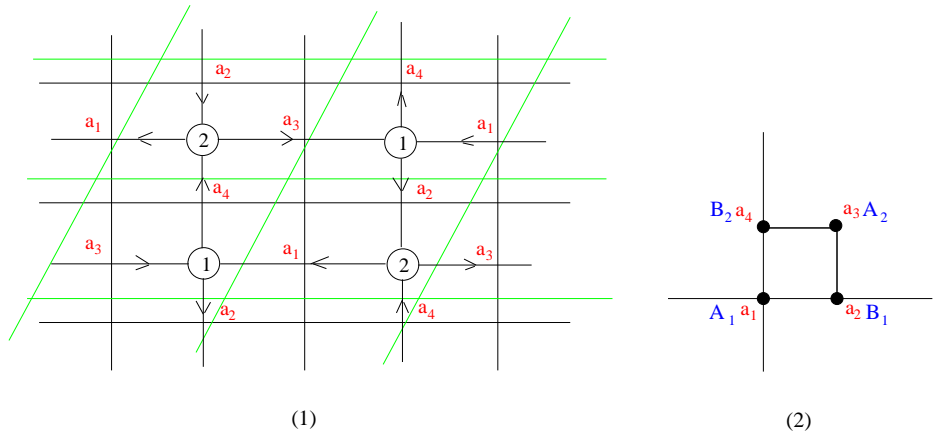} 
\caption{(1) Dimer configuration for the field theory dual to $C(T^{1,1})$
  with a given assignment of charges $a_i$ and the orientation given by the
  arrows linking the gauge groups. We have drawn in green the bounds of the
  basic cell. (2) Toric diagram for the singularity $C(T^{1,1})$.}\label{baryt}
\end{center}
\end{figure}

%Before doing this we very briefly review some results about the AdS/CFT
%correspondence for the class of toric geometries.
As previously explained the gauge theory dual to a given toric singularity is completely identified by the \emph{dimer configuration}, 
or \emph{brane tiling} 
%(Figure \ref{barydp}) 
\cite{Hanany:2005ve,Franco:2005rj,Hanany:2005ss,Feng:2005gw}.
There is a general recipe for assigning baryonic, flavors and R charges to
the elementary fields for a minimal toric phase of the $CFT$ \cite{Benvenuti:2005ja,Franco:2005sm,Butti:2005ps,Butti:2005vn}.
As described in Section (\ref{charges}), we can parameterize all charges
with $d$ numbers $a_i$, $i=1,...,d$ associated with the 
vertices of the toric diagram subject to the constraint (\ref{sum})
in case of global symmetries and (\ref{sumr}) in case of R symmetries.
Every elementary field can be associated with a brane wrapping a
particular divisor 
\begin{equation}
D_{i+1}+D_{i+2}+ \ldots D_{j}
\end{equation}
and has charge $a_{i+1}+a_{i+2}+ \ldots a_{j}$ \footnote{Call $C$ the set of all the unordered pairs of vectors in the $(p,q)$ web (the $(p,q)$ web 
is the set of vectors $v_i$ perpendicular to the edges of the toric diagram and with the same length as the corresponding edge); 
label an element of $C$ with the ordered indexes $(i,j)$, with the convention that  
the vector $v_i$ can be rotated to $v_j$ in the counter-clockwise direction
with an angle $\leq 180^o$. With our conventions $|\langle v_i, v_j \rangle|=\langle v_i, v_j \rangle$, 
where with $\langle \hbox{ },\hbox{ } \rangle$ we mean the determinant of the $2 \times 2$ matrix. One can associate with any element of $C$ the divisor 
$D_{i+1}+D_{i+2}+ \ldots D_{j}$ 
and a type of chiral field in the field theory with multiplicity $\langle v_i, v_j \rangle$ 
and global charge equal to $a_{i+1}+a_{i+2}+ \ldots a_{j}$ \cite{Butti:2005ps,Butti:2005vn}. 
The indexes $i$, $j$ are always understood to be defined modulo $d$.}. Various fields have the same
charge; as mentioned in the last part of the Section \ref{bar}, this multiplicity is due to the non-trivial homotopy of the corresponding cycles.
% the baryonic ground state of the $D3$ branes in $H$ has multiplicity and this fact gives 
%in the $SCFT$ many fields with the same charges. 
Explicit methods for computing the charge of each link in the dimer are
given 
% and the multiplicity of the elementary
in \cite{Franco:2005sm,Butti:2005ps} and the reader is refer to these papers for an exhaustive list of examples. In Figure \ref{baryt} we report the easiest example of the conifold.

%As explained in Section \ref{toric} one can 
%associate with each vector $n_i$ a chiral field with trial global charge $a_i$. 

With this machinery in our hands we can analyze the the field theory operators 
corresponding to the $D3$ brane states analyzed in the previous Sections. 
The first thing to understand is the map between the section $\chi^m$ of the 
line bundle we are considering and the field theory operators. The case of 
the trivial line bundle is well known: the corresponding polytope
is the cone of holomorphic functions which are in one-to-one correspondence
with mesonic operators. The latter are just closed loops in the quiver.
It is possible to construct a map between closed loops in the quiver 
and points in the cone of holomorphic functions; it can be
shown that closed loops mapped to the same point in the cone correspond
to mesons that are $F$-term equivalent \cite{Hanany:2006nm,Butti:2006hc}. This map give us an in principle one to 
one relation between the points in the dual cone, and hence the giant gravitons states, and the mesonic operators. It is an 
example of states/operators map in AdS/CFT, and justifies our discussion about the mesonic BPS operators in chapter \ref{braneasing}. 
  
In this section we construct the same map with generically open paths in the dimer, and it will reduce to the before 
mentioned map in the special case of closed paths. In particular
we would like to associate to every point in the polytope $P_D$ a sequence of 
contractions of elementary fields modulo $F$-term equivalence. This is indeed
possible for a particular class of polytopes which we now describe. This general map of course 
will reduce to the mesonic map in the case of trivial divisors. 

Let us start by studying open paths in the dimer.
Take two gauge group $U$, $V$ in the dimer and draw an oriented path $P$ connecting them 
(an oriented path in the dimer is a sequence of chiral fields oriented 
in the same way and with the gauge indices contracted along the path). 
The global charges of $P$ is the sum of the charges of the fields contained in $P$ and can be schematically written as:
\begin{equation}\label{P}
\sum_{i=1}^d c_i a_i
\end{equation}
with some integers $c_i$. Draw another oriented path $Q$ connecting the same gauge groups. 
Consider now the closed non-oriented path $Q-P$; as explained in \cite{Butti:2006hc} the charges 
for a generic non-oriented closed path can be written as \footnote{For non-oriented paths one has 
to sum the charges of the fields with the same orientation and subtract the charges 
of the fields with the opposite orientation.}:
\begin{equation}\label{P-Qa}
\sum_{i=1}^d <m,V_i> a_i 
\end{equation}
with $m$ a three dimensional integer vector.
Hence the charges for a generic path $Q$ connecting two gauge groups are:
\begin{equation}\label{P-Qb}
\sum_{i=1}^d (<m,V_i> + c_i) a_i\, .
\end{equation}
 Because the path $Q$ is oriented we have just added the charges along it and the coefficients of the $a_i$ are all positive:
\begin{equation}\label{P-Qc}
<m,V_i> + c_i \geqslant 0\, .
\end{equation}
We have the freedom to change $P$; this means that the $c_i$ are only 
defined up to the equivalence relation $c_i \sim c_i + <m,V_i>$.
Observe that (\ref{P-Qc}) is the same condition on the exponents for the 
homogeneous coordinates $x_i$ in the global sections $\chi^m$ (\ref{poly}), 
and the equivalence relation $c_i \sim c_i + <m,V_i>$ corresponds to the equivalence relation 
on the divisors $D \sim D + \sum_{i=1}^d <m,V_i> D_i$. 
Hence we realize that to every path connecting a pair of gauge groups we can assign 
a point in a polytope associated with the divisor  $\sum_i c_i D_i$  modulo 
linear equivalence. In particular, all operators associated with open paths
between two gauge groups $(U,V)$ have the same baryonic charge, as it can
be independently checked.
 
Now that we have a concrete map between the paths in the dimer and the integer 
points in the polytope we have to show that this map is well defined. Namely 
we have to show that we map $F$-term equivalent operators to the same point in the polytope and 
that to a point in the polytope corresponds only one operator in field theory modulo $F$ terms relations 
(the injectivity of the map). 
The first step is easy to demonstrate: paths that are $F$-term equivalents have the 
same set of $U(1)$ charges and are mapped to the same point $m$ in the polytope $P_D$. 
Conversely if paths connecting two gauge groups are mapped to the same point $m$ it means 
that they have the same global charges. The path $P-P'$ is then
a closed unoriented path with charge 0. As shown in \cite{Hanany:2006nm,Butti:2006hc}
$P$ and $P'$ are then homotopically equivalent \footnote{ Indeed it is possible to 
show that $m_1$ and $m_2$ of a closed path are its homotopy numbers around the dimer. }.  
%with $m$ different from zero and hence the path $P-P'$ is homotopically non-trivial. This implies that $P$ is not homotopically equivalent to $P'$. This means that two paths mapped to the same point in $P_D$ have the same homotopy and $U(1)$ charges. 
Now we can use the Lemma $5.3.1$ in \cite{Hanany:2006nm} that says: ``in a consistent tiling, 
paths with the same $R$-charge are $F$-term equivalent if and only if they are homotopic'' 
to conclude that paths mapped to the same point $m$ in $P_D$ are $F$-term equivalent. 
Surjectivity of the map is more difficult to prove, exactly as in the case
of closed loops \cite{Hanany:2006nm,Butti:2006hc}, but it is expected
to hold in all relevant cases.

%Moreover if we suppose that the map is surjective on $P_D$, than we conclude that to 
%every point in $P_D$ we have only one field theory operator in the chiral ring of the 
%$CFT$ \footnote{This uniqueness is only up to the multiplicity of the fields with the 
%same charges ( i.e paths with the same charges linking different gauge groups ) 
%due to the fact that the paths are open.}.

In particular we will apply the previous discussion to the case of neighbouring
gauge groups $(U,V)$ connected by one elementary field. If the charge of
the field is $a_{i+1}+...+a_j$ we are dealing with the sections of the line
bundle ${\cal O}(D_{i+1}+...+D_j)$
$$\sum_{m\in P_{D_{i+1}+...+D_j}} h_m \chi^m =h\,  x_{i+1}...x_j+...$$
The section $x_{i+1}...x_j$ will correspond to the elementary field itself
while all other sections $\chi^{m_j}$ will correspond to operators with
two free gauge indices $(O_m)_{\alpha\beta}$ under $U$ and $V$ which correspond to open paths from $U$ to $V$.  
%When we can find $c_i$ such that the only possible paths with that baryonic charges are single connected paths in the dimer connecting different gauge groups 
The proposal for finding the gauge invariant operator dual to the $BPS$ state $|h_{m_1},...,h_{m_N}>$ 
is then the following\footnote{ This is just a simple generalization of the one in \cite{Beasley:2002xv}.}. 
We associate to every $h_{m_j}$, with section $\chi^{m_j}$, 
an operator in field theory with two free gauge indices 
in the way we have just described (from now on we will call these paths the building blocks of the baryons).  
%Starting from the $BPS$ state $|h_{m_1},...,h_{m_N}>$  in the gravity side
%with the multiplicity given by the multiplicity  $\#_D$ associated to the shortest paths with those baryonic charges. 
%We assume that in this construction we encountered only single connected paths
%in the dimer connecting different gauge groups; the general case will be discussed in section \ref{comments}. 
%The building blocks are then operators with
%just two free gauge indices $(O_m)_{\alpha\beta}$. 
%Then we take all the operators connecting the same gauge groups and 
Then we construct a gauge invariant operator by contracting all the $N$ free indices of one gauge group with its epsilon tensor and all the $N$ free indices of the other gauge group with its own epsilon. 
The field theory operator we have just constructed has clearly the same global charges 
of the corresponding state in the string theory side and due to the epsilon contractions 
is symmetric in the permutation of the field theory building blocks like the string theory state. 
This generalizes Equation (\ref{genbarABNS}) to the case of
a generic field in a toric quiver.  
%The previous set of BPS states correspond to all the baryonic 
%gauge invariant operators that we can construct by contracting with
%two epsilon operators $(O_m)_{\alpha\beta}$ with free gauge indices under
%the gauge groups $U$ and $V$. 
By abuse of language, we can say that
we have considered all the single determinants that we can make with
indices in $(U,V)$.

As already mentioned, D3 branes wrapped
on three cycles in $H$ come with a multiplicity which is given by the
non trivial homotopy of the three cycle. On the field theory side, this
corresponds to the fact there is a multiplicity of elementary fields with
the same charge. Therefore a polytope $P_D$
is generically associated to various 
different pairs of gauge groups $(U^a,V^a)$, $a=1,...\#_D$. 
For this reason we say that the polytope $P_D$ has a multiplicity $\#_D$. 
This implies that there is an isomorphism between the set of 
open paths (modulo $F$-terms) connecting the different pairs $(U^a,V^a)$. 
Similarly, the single determinant baryonic operators
constructed as above from different pairs $(U^a,V^a)$ come isomorphically
from the point of view of the counting problem.

%The number of baryonic operators (modulo $F$-term relations) with flavors and $R$ charges 
%associated with $m$ can be obtained by multiplying in all possible ways
%$N$ sections $\chi^{m_i}$ with $\sum_{i=1}^N m_i=m$ and by multiplying
%for the multiplicity $\#_D$ of the polytope.     

%Summarizing we have shown that, once we have chosen a pair of gauge groups $(U,V)$, 
%all the paths $P$ linking the gauge groups $U$ and $V$ can be mapped to points in a polytope $P_D$ 
%defined by the charge $\sum_{i=1}^d c_i a_i$ of the shortest path ($m=0$) connecting the two groups. 
%Clearly in general there will be many pairs of gauge groups with the same $c_i$ and hence the same polytope $P_D$. 
%For this reason we say that the polytope $P_D$ has a multiplicity $\#_D$. 
%The important point is that for every single polytope in the family of $P_D$ the map between 
%the points and the paths/operators is one to one modulo $F$-term relations.

%Now we can ask how many baryonic operators with that set of $c_i$ and given values of the flavors and $R$ charges we can have modulo the $F$-term relations. We will answer to this question in the following Sections where we will construct the partition functions counting the various operators. Finally,
%the number of baryonic operators with given $T^3$ charge predicted by the partition function must be multiplied by the multiplicity of the polytope.

Obviously, the baryonic operators we have constructed are just a subset
of the chiral ring of the toric quiver gauge theory. They correspond to
possible single determinants that we can construct. In the case of greater
baryonic charge (products of determinants) the relation between points
in the polytope $P_D$ and operators is less manifest and a preliminary discussion is given in the following
section. 

For an explicit example of this construction the reader is referred to \cite{Butti:2006au} where the baryonic building blocks 
associated with a line bundle over the Calabi-Yau cone $C(Y^{2,1})$ are discussed.

\subsection{Comments on the General Correspondence}\label{comments}

In the case of a generic polytope $P_D$, not associated with elementary
fields, the correspondence between sections of the line bundle ${\cal O}(D)$
and operators is less manifest. The reason is that
we are dealing with higher baryonic charges and the corresponding gauge
invariant operators are generically products of determinants.

%In the generic case there are various subtleties related to the construction of gauge invariant operators and the multiplicity
%of the polytope. 
%In the previous Sections we have shown that to every pair of gauge groups we can associate certain $c_i$ and hence a polytope
% with a multiplicity in the geometric side, and for every point in the polytope there are as many field operators as the multiplicity of the polytope modulo $F$-term relations.
%The point is that we would like to do the contrary: decide the baryonic charges compatible with the geometry $B^{(a)}=\sum_{i=1}^d Q_i^{(a)}c_i$ find a set of corresponding $c_i$, construct the polytop, find the multiplicity of the polytop, count the numbers of building blocks with given $U(1)$ charges, construct the baryonic gauge invariant operators, count the number of baryons with given $U(1)$ charges.\\
%We will explain the counting procedure in the next Sections. 
%The difficult part of the above procedure is finding the multiplicity of a given polytope and construct the gauge invariant baryonic operators. 

Let us consider as an example the case of the conifold. Suppose we want to 
study the polytope $P_{2 D_1}$ which corresponds to 
classify the $BPS$ operators with baryonic charges equal to $2N$.
In field theory we certainly have baryonic operators with charge $2N$, 
for example $\det A_1\cdot \det A_1 $. 
%Following our recipe we must find the pairs of gauge groups with, for example, $c_1=2$ and hence charge $2a_1$. 
Clearly all the products of two baryonic operators 
with baryon number $N$ give a baryonic operator with baryon number $2N$. 
In a sense in the conifold all the operators in sectors with baryonic charge 
with absolute value bigger than one are factorized \cite{Berenstein:2002ke,Beasley:2002xv}. However we cannot find a simple prescription 
for relating sections  of $P_{2D_1}$ to paths in the dimer. Certainly
we can not find a single path in the dimer (Figure \ref{baryt}) 
connecting the two gauge groups with  charge $2 a_1$. 
%Therefore
%the prescription given  . 

%It would be interesting to see if one can make a more detailed correspondence
%between sections of the line bundles and operators with free gauge indices.
One could speculate that the prescription valid for basic polytopes
has to be generalized by allowing the use of paths and multipaths.
%For every set of $c_i$ we determine all the possible paths and multi-paths with the corresponding charges.
%With this generalized prescription it is easy to reproduce the missing operators. 
For example, to the section $\chi^m$ in the polytope $P_{2D_1}$ for the conifold 
we could assign two paths connecting the two gauge groups with charges $\sum_{i=1}^d <m^{(1)},V_i> + a_1$ 
and  $\sum_{i=1}^d <m^{(2)},V_i> + a_1$ with $m=m^{(1)}+m^{(2)}$ and therefore a building block
consisting of two operators $(O_{m^{(1)}})_{\alpha_1\beta_1}$, $(O_{m^{(2)}})_{\alpha_2\beta_2}$. We should now construct the related gauge invariant operators. 
Out of these building blocks we cannot construct a single determinant because 
we don't have an epsilon symbol with $2N$ indices, but we can easily construct 
a product operator using four epsilons. 
We expect, based on Beasley's prescription, a one to one correspondence 
between the points in the $N$ times symmetric product of the polytope $P_{2 D_1}$ 
and the baryonic operators with baryonic charge $+2$ in field theory. 
Naively, it would seem that, with the procedure described above, we have found many more operators. 
Indeed the procedure was plagued by two ambiguities: in the construction of the 
building blocks, it is possible to find more than a pair of paths corresponding
to the same $m$ (and thus the same $U(1)$ charges) that are not $F$-term equivalent; 
in the construction of the gauge invariants we have the ambiguity on how to distribute 
the operators between the two determinants. 
The interesting fact is that these ambiguities seem to disappear when we
consider the final results for gauge invariant operators, due to the $F$-term relations and the properties of the epsilon symbol. 
One can indeed verify, at least in the case of $P_{2 D_1}$ and for 
various values of $N$, 
there is exactly a one to one correspondence between the points in the $N$ times symmetric 
product of the polytope and the baryonic operators in field theory. 
%( this correspondence is better understood once we are able to construct the partition functions associated to polytops and their symmetric products as we will explain in the next Sections).
% but fully understanding the
%multiplicity of a polytope is of fundamental importance if one wants to write
%the complete partition function of a given $SCFT$ \cite{ADAZ}. 
%This is clearly only a preliminary discussion. 
In the following chapters we would try to make this kind of prescription more rigorous.

The ambiguity in making a correspondence between sections of the polytope
and operators is expected and it is not particularly problematic.
The correct correspondence is between the states $|h_{m_1},...,h_{m_N}>$ and 
baryonic gauge invariant operators. 
The sections in the geometry are not states 
of the string theory and the paths/operators are not gauge invariant operators. What the $AdS/CFT$ 
correspondence tells us is that there exists a 
one to one relation between states in string theory 
and gauge invariant operators in field theory, and this is a one to one 
relation between 
the points in the $N$-fold symmetric product of a given polytope and 
the full set of gauge invariant operators with given baryonic charge.

The comparison with field theory should be then done as follows. One
computes the partition functions $Z_{D,N}$ of the $N$-fold symmetrized
product of the polytope $P_D$ and compares it with all the gauge invariant 
operators in the sector of the Hilbert space with given baryonic charge.
We have explicitly done it for the conifold for the first few values of $N$
and the operators with lower dimension. In forthcoming chapters 
we will actually re-sum the partition functions $Z_{D,N}$ and we will write
the complete partition function for the chiral ring of the conifold and other
selected examples; we will compare the result with the dual field theory
finding perfect agreement.

The issues of multiplicities that we already found in the case of polytopes
associated with elementary fields persists for generic polytopes. Its
complete understanding is of utmost importance for writing a full partition
function for the chiral ring \cite{Forcella:2007wk,Butti:2007jv}.

\section{Counting BPS Baryonic D3 Branes}\label{counting}
 
In this Section, as promised, we count the number of BPS baryonic operators corresponding to the BPS D3 branes
in the sector of the Hilbert space $\cal{H}_D$, associated with a divisor $D$.
All operators in $\cal{H}_D$ have fixed baryonic charges. Their number
is obviously infinite, but, as we will show, the number of operators with
given charge $m\in T^3$ under the torus action is finite. It thus 
makes sense to write a partition function $Z_{D,N}$ for the BPS baryonic operators 
weighted by a $T^3$ charge $q=(q_1,q_2,q_3)$. $Z_{D,N}$ will be a polynomial in the $q_i$ 
such that to every monomial $n \hbox{ } q_1^{m_1} q_2^{m_2} q_3^{m_3}$ we associate $n$ $BPS$ 
D3 brane states with the R-charge and the two flavor charges
parametrized by $\sum_{i=1}^d (<m,V_i> a_i + N c_i a_i)$. 

The computation of the weighted partition function is done in two steps.
We first compute a weighted partition function $Z_D$, 
or character, counting the sections of ${\cal O}(D)$; these correspond to the $h_m$ which are 
the elementary constituents of the baryons. 
In a second time, we determine the total partition function $Z_{D,N}$ for
the states $|h_{m_1}...h_{m_N}>$ in $\cal{H}_D$.

\subsection{The Character $Z_D$}\label{riemannroch}
We want to re-sum the character, or weighted partition function, 
\begin{equation}
Z_D = \rm{Tr} \{ q | H^0(\cX,{\cal O}(D)\} = \sum_{m\in P_D  \cap  M} q^m
\label{character}
\end{equation}   
counting the integer points in the polytope $P_D$ weighted with their
charge under the $T^3$ torus action. 

In the trivial case ${\cal O}(D)\sim {\cal O}$, $Z_D$ is
just the partition function for holomorphic functions discussed in
\cite{Martelli:2006yb,Benvenuti:2006qr}, which can be computed using the Atyah-Singer index
theorem \cite{Martelli:2006yb}. Here we show how to extend this method to the
computation of $Z_D$ for a generic divisor $D$. The results of this section of course reduce to the 
ones of \cite{Martelli:2006yb,Benvenuti:2006qr} once we consider the trivial line bundle. 

Suppose that we
have a smooth variety and a line bundle ${\cal O}(D)$ with a holomorphic action of $T^k$ (with $k=1,2,3$ and $k=3$ is the toric case). 
Suppose also that the higher dimensional cohomology of the line bundle vanishes, $H^{i}(\cX,{\cal O}(D))=0$, for $i\ge1$. 
The character (\ref{character}) then coincides with the Leftschetz number
\begin{equation}
\chi(q,D) = \sum_{p=0}^3 (-1)^p  \rm{Tr} \{ q | H^{p}(\cX,{\cal O}(D))\}
\end{equation}
which can be computed using the index theorem: 
we can indeed write $\chi(q,D)$
as a sum of integrals of characteristic classes over
the fixed locus of the $T^k$ action. In this thesis, we will only consider 
cases where $T^k$ has isolated fixed points $P_I$. The general case can be 
handled in a similar way.
In the case of isolated fixed points,
the general cohomological formula\footnote{Equivariant Riemann-Roch, or the Lefschetz fixed point formula, reads
\begin{equation}
\chi(q,D) = \sum_{F_i} \int_{F_i} \frac {{\rm Todd}(F_i) Ch^q (D)}{\prod_{\lambda}(1-q^{m_\lambda^i} e^{-x_\lambda})}
\end{equation}
where $F_i$ are the set of points, lines and surfaces which are fixed by
the action of $q\in T^k$, ${\rm Todd}(F)$ is the Todd class ${\rm Todd}(F) =1 + c_1(F)+...$ and, on a fixed locus, $Ch^q(D)= q^{m^0}e^{c_1(D)}$ where $m^0$ is the weight of the $T^k$ action. The normal bundle $N_i$ of each fixed submanifold $F_i$ has been splitted in
line bundles; $x_\lambda$ are the basic characters and $m_\lambda^i$ the weights of the $q$ action on the line bundles.} considerably simplifies and can be 
computed by linearizing the $T^k$ action near the fixed points. The
linearized action can be analyzed as follows.  
Since $P_I$ is a fixed point, the
group $T^k$ acts linearly on the normal (=tangent) space at $P_I$,
$TX_{P_I}\sim \mathbb{C}^3$. 
%and it makes
%sense to consider the determinant of the action of $q$, 
%$\det(1-q)|_{TX_{P_I}}$. An abelian group has only dimension one 
%representations and 
The tangent space will split into three one dimensional
representations $TX_{P_I}=\sum_{\lambda=1}^3 L^\lambda$ of the abelian group 
$T^k$. We denote the
corresponding weights for the $q$ action with $m_I^\lambda, \lambda=1,2,3$.
Denote also with $m_I^0$ the weight of the action of $q$ on the $\mathbb{C}$ 
fiber of the line bundle ${\cal O}(D)$ over $P_I$. The equivariant Riemann-Roch   formula expresses the Leftschetz number as a sum over the
fixed points
\begin{equation}
\chi(q,D) = \sum_{P_I} \frac{q^{m_I^0}}{\prod_{\lambda=1}^3 (1-q^{m_I^\lambda})}
\end{equation}

We would like to apply the index theorem to our CY cone $\cX$.
Unfortunately, $\cX=C(H)$ is not smooth and a generic element of $T^k$ has a fixed point at the apex of the cone, which is exactly the singular point.
To use Riemann-Roch we need to resolve the cone $\cX$ to a smooth variety 
$\tilde \cX$
and to find a line bundle ${\cal O}(\tilde D)$ on it with the following two
properties: i) it has the same space of sections, $H^0 (\tilde \cX,{\cal O}(\tilde D)) = H^0(\cX,{\cal O}(D))$, ii) it has vanishing higher cohomology
$H^{i}(\tilde \cX,{\cal O}(\tilde D))=0, i\ge 1$.

Notice that the previous discussion was general and apply to all Sasaki-Einstein manifolds $H$. It gives 
a possible prescription for computing $Z_D$ even in the non toric case.
In the following we will consider the case of toric cones where the
resolution $\tilde \cX$ and the divisor $\tilde D$ can be explicitly found.

\begin{figure}
\begin{center}
\includegraphics[scale=0.6]{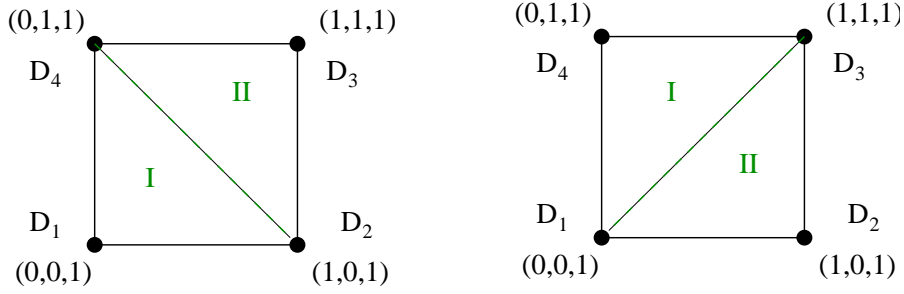} 
\caption{The two resolutions for the conifold. 
No internal points have been blown up. In each case, one line have been added to the original fan; there are two maximal cones and corresponding fixed points, denoted I,II.}
\label{conifoldC}
\end{center}
\end{figure}

Toric Calabi-Yau cones have a pretty standard resolution by triangulation
of the toric diagram, see Figure \ref{conifoldC}. In chapter \ref{chiralcount} we will see more 
involved examples in which the complete resolution of the singularity imply the introductions of new divisors.

The fan of the
original variety $\cX$ consists of a single maximal cone, with a set of
edges, or one-dimensional cones, $\Sigma(1)$ whose generators $V_i$ are not
linearly independent in $\mathbb{Z}^3$. The 
resolutions of $\cX$ consist of all the possible subdivisions of the fan in
smaller three dimensional cones $\sigma_I$. The new variety $\tilde \cX$
is still a Calabi-Yau if all the minimal generators $V_i$ of the 
one-dimensional cones lie on a plane. This process looks like 
a triangulation of the toric diagram. If each three-dimensional cone is 
generated by linearly independent primitive vectors, the variety is smooth.
The smooth Calabi-Yau resolutions of $\cX$ thus consist of all the triangulation
of the toric diagram which cannot be further subdivided. 
Each three dimensional cone $\sigma_I$ is now a copy of $\mathbb{C}^3$ and
the smooth variety $\tilde \cX$ is obtained by gluing all these $\mathbb{C}^3$ 
according to the rules of the fan.
$T^3$ acts on each $\sigma_I$ in a simple way: the three
weights of the $T^3$ action on a copy of $\mathbb{C}^3$ are just given by
the primitive inward normal vectors $m_I^\lambda$ 
to the three faces of $\sigma_I$.
Notice that each $\sigma_I$ contains exactly one fixed point of $T^3$ (the origin in the copy of $\mathbb{C}^3$) with weights given by the vectors 
$m_I^\lambda$.

The line bundles on $\tilde \cX$ are given by $\tilde D=\sum_i c_i D_i$
where the index $i$ runs on the 
set of one-dimensional cones $\tilde\Sigma(1)$, which is typically bigger than
the original $\Sigma(1)$. Indeed, each integer internal point of the toric 
diagram gives rise in the resolution $\tilde \cX$ to a new divisor. The space of sections of $\tilde D$ are still determined by the 
integral points of the polytope
\begin{equation}\label{poly2}
\tilde P_D = \{ u \in M_{\mathbb{R}}| <u,V_i >\,  \ge\, - c_i\, , \,\,  \forall i\in \tilde 
\Sigma(1) \}
\end{equation} 
It is important for our purposes that each maximal cone $\sigma_I$ determines a integral point $m_I^0\in M$ as the
solution of this set of three equations:
\begin{equation}\label{charge}
< m_I^0, V_i > = - c_i, \,\,\, V_i\in \sigma_I,  
\end{equation}
In a smooth resolution $\tilde X$ this equation has always integer solution 
since the three generators $V_i$ of $\sigma_I$ are a basis for $\mathbb{Z}^3$.
As shown in \cite{fulton}, $m_I^0$ is the charge of the local equation for
the divisor $\tilde D$ in the local patch $\sigma_I$ and it is equal to zero in the case of 
trivial divisor. It is therefore the
weight of the $T^3$ action on the fiber of ${\cal O}(D)$ over the fixed
point contained in $\sigma_I$.

The strategy for computing $Z_D$ is therefore the following. We smoothly
resolve $\cX$ and find a divisor $\tilde D= \sum_i c_i D_i$ by assigning
values $c_i$ to the new one-dimensional cones in $\tilde \Sigma(1)$
that satisfies the two conditions
\begin{itemize}
\item{It has the same space of sections, $H^0(\tilde \cX,{\cal O}(\tilde D)) = H^0(\cX,{\cal O}(D))$. Equivalently, the polytope $\tilde P_D$ has the same integer points of $P_D$.}
\item{It has vanishing higher cohomology
$H^{i}(\tilde \cX,{\cal O}(\tilde D))=0, i\ge 1$. As shown in
\cite{fulton} this is the case if there exist integer 
points $m_I^0\in M$ that satisfy the convexity condition
\footnote{The $m_I^0$s determine a continuous piecewise linear function $\psi_D$ on
the fan as follows: in each maximal cone $\sigma_I$ the function $\psi_D$
is given by $<m_I^0, v>$, $v\in \sigma_I$. As shown in \cite{fulton},
the higher dimensional cohomology vanishes, $H^i(\tilde \cX, {\cal O}(D))=0, \, i\ge 1$, whenever the function $\psi_D$ is upper convex.}
\begin{eqnarray}
< m_I^0, V_i > &=& - c_i, \,\,\, V_i\in \sigma_I \nonumber\\
< m_I^0, V_i > &\ge& - c_i, \,\,\, V_i\notin \sigma_I\label{convex}
\end{eqnarray}
These equations are of course trivially satisfied in the case of trivial divisors: namely $c_i=0$ and $m_I^0=0$. 
}
\end{itemize}

There are many different smooth resolution of $\cX$, corresponding to
the possible complete triangulation of the toric diagram. It is possible to show \cite{Butti:2006au} 
that we can always find a compatible resolution $\tilde \cX$
and a minimal choice of $c_i$ that satisfy the two given conditions.

The function $Z_D$ is then given as
\begin{equation}
Z_D = \sum_{P_I} \frac{q^{m_I^0}}{\prod_{\lambda=1}^3 (1-q^{m_I^\lambda})}
\label{sumzd}
\end{equation}
where in the toric case for every fixed point $P_I$ there is a maximal cone $\sigma_I$, $m_I^\lambda$ are the three inward primitive normal vectors of 
$\sigma_I$ and $m_I^0$ are determined by equation (\ref{convex}). 
This formula can be conveniently generalized to the case where the fixed points are not isolated but there are curves or surfaces fixed by the torus action, and reduce to:
\begin{equation}
Z_{\mathcal{O}}= \sum_{P_I} \frac{1}{\prod_{\lambda=1}^3 (1-q^{m_I^\lambda})}
\label{sumzdt}
\end{equation}
in the case of trivial divisors.

We finish this Section with two comments. The first is a word of caution. Note
that if we change representative for a divisor in its equivalence class
($c_i\sim c_i+<M,V_i>$) the partition function $Z_D$ is not invariant, 
however it is just rescaled by a factor $q^M$.

The second comment concerns toric cones. For toric CY cones there is an
alternative way of computing the partition functions $Z_D$ by expanding
the homogeneous coordinate ring of the variety according to the decomposition
(\ref{dirs}). Since the homogeneous coordinate ring is freely generated by the $x_i$, 
its generating function is simply given by 
$$\frac{1}{\prod_{i=1}^d (1-x_i)}\, .$$
By expanding this function according to the grading given
by the $(\mathbb{C}^*)^{d-3}$ torus action we can extract all the $Z_D$.
This approach will be discussed in detail in the following chapters and in section \ref{geometricamusement}. 

\subsection{A Simple Example: the Conifold}\label{conifoldexample}

We now want to use the right now familiar example of the conifold to give explicit computations and results 
of the general theory just explained. The reader is referred to \cite{Butti:2006au} for a more extensive list of examples.

The four primitive generators for the one dimensional cones of the conifold 
are $\{(0,0,1),(1,0,1),\\(1,1,1),(0,1,1)\}$ and we call the associated divisors $D_1,D_2,D_3$ and
$D_4$ respectively. They satisfy the equivalence relations $D_1\sim D_3\sim -D_2\sim -D_4$. There is only one baryonic symmetry under which
the four homogeneous coordinates transform as
\begin{equation}\label{conact}
(x_1,x_2,x_3,x_4)\sim (x_1 \mu ,x_2/\mu,x_3 \mu ,x_4/\mu )
\end{equation}
where $\mu \in \mathbb{C}^*$.
The conifold case is extremely simple in that the chiral fields of the 
dual gauge theory are in one-to-one correspondence with the homogeneous
coordinates: $(x_1,x_2,x_3,x_4)\sim (A_1,B_1,A_2,B_2)$. Recall that
the gauge theory is $SU(N)\times SU(N)$ with chiral fields $A_i$ and
$B_p$ transforming as $(N,\bar{N})$ and $(\bar{N},N)$ and as $(2,1)$ and $(1,2)$ under the enhanced $SU(2)^2$ global flavor symmetry.

The two possible resolutions for the conifold are presented in Figure \ref{conifold}.
We would like first to compute the partition function for the trivial $0$ divisor.
For the trivial line bundle $\mathcal{O}$ all $c_i=0$ and $m_I^0=0$ and the localization formula (\ref{sumzdt}) is independent of the particular resolution chosen. In the conifold case there are two possible complete resolutions ( see Figure \ref{conifoldC}). Regions $I$ and $II$ correspond
to the two maximal cones in the resolution and, therefore, to the  two fixed
points of the $T^3$ action. Denote also $q=(q_1,q_2,q_3)$. Using the prescriptions given above, we compute the three primitive inward 
normals to each cone.

For the first resolution we have the vectors:
\begin{eqnarray}
{\rm Region\, I}\qquad &m_I^\lambda = \{ ((1,0,0),(0,1,0),(-1,-1,1)\} \nonumber\\
{\rm Region\, II}\qquad & m_{II}^\lambda = \{(-1,0,1),(0,-1,1),(1,1,-1)\}
\end{eqnarray}

and the generating function is:
\begin{eqnarray}
Z_{\mathcal{O}}^a &=& \frac{1}{(1 - q_1)(1 - q_2)(1 - \frac{q_3}{q_1 q_2})} + \frac{1}{(1 - \frac{q_3}{q_1})(1 - \frac{q_3}{q_2})(1 - \frac{q_1 q_2}{q_3})}\nonumber \\
&=& \frac{q_1 q_2 (1 - q_3)}{(1- q_1)(1 - q_2) (q_1 - q_3)(q_2 - q_3)}
\end{eqnarray}

For the second resolution we have the vectors:
\begin{eqnarray}
{\rm Region\, I}\qquad &m_I^\lambda = \{ ((1,0,0),(0,-1,1),(-1,1,0)\} \nonumber\\
{\rm Region\, II}\qquad & m_{II}^\lambda = \{(0,1,0),(-1,0,1),(1,-1,0)\}
\end{eqnarray}

and the generating function is:
\begin{eqnarray}
Z_{\mathcal{O}}^b &=& \frac{1}{(1 - q_1)(1 - \frac{q_3}{q_2})(1 - \frac{q_2}{q_1})} + \frac{1}{(1 - q_2 )(1 - \frac{q_3}{q_1})(1 - \frac{q_1}{q_2})}\nonumber \\
&=& \frac{q_1 q_2 (1 - q_3)}{(1- q_1)(1 - q_2) (q_1 - q_3)(q_2 - q_3)}
\end{eqnarray}

And indeed we have $Z_{\mathcal{O}}^a=Z_{\mathcal{O}}^b=Z_{\mathcal{O}}$. Hence we conclude that the localization formula for the conifold for the trivial line bundle $\mathcal{O}$ is invariant under flop transitions. For a complex three dimensional toric variety $\cX$ any complete resolutions can be reach from any other by a sequence of flop transitions. In toric geometry the toric flop transitions are local operations and are essentially the flop transition for the conifold that we just discussed. Because we have shown that the conifold partition function is invariant under flop transitions, we conclude that the partition function $Z_{\mathcal{O}}$ for any toric three dimensional variety $\cX$ is independent of the particular resolution. 

We can consider the partition function just along the Reeb vector $(3/2,3/2,3)$ by putting $q_1=q_2=q,q_3=q^2$:
\begin{equation}\label{mesZ}
Z_{\mathcal{O}}= \frac{1+q}{(1 - q)^3}
\end{equation}
This is exactly the equation \ref{Hcon} of chapter \ref{braneasing} with $q=t$. Hence in the particular case of trivial line bundle our general procedure provides a systematic way to reproduce all the results we claimed in chapter \ref{braneasing}. 
We can now expand (\ref{mesZ}) in powers of $q$:
\begin{equation}\label{mesZex}
Z_{\mathcal{O}} = 1 + 4q + 9q^2 +...
\end{equation}
This counting match the mesonic counting of operators in field theory. Indeed it counts operators in the field theory according to their R charges in the $(n+1)^2$ symmetric representation of the $SU(2)\times SU(2)$ global symmetry group. The expansion (\ref{mesZex}) reproduce the $N=1$ multitrace counting:
\begin{equation}
1\, , \qquad A_iB_p\, , \qquad A_i B_p A_j B_q\, , \qquad ....
\end{equation}  
or equivalently the $N \rightarrow \infty $ single trace counting:
\begin{equation}
1\, , \qquad Tr(A_iB_p)\, , \qquad Tr(A_i B_p A_j B_q)\, , \qquad ....
\end{equation}  

We can now pass to compute partition functions for non trivial divisor $D$.
We first compute the partition function for the divisor $D_1$ using the
resolution on the left hand side of the figure.  Using the prescriptions given above, we compute the three primitive inward 
normals to each cone  and the weight of
the $T^3$ action on the fiber. It is manifest that the
conditions required in equation (\ref{convex}) are satisfied.
\begin{eqnarray}
{\rm Region\, I}\qquad &m_I^\lambda = \{ ((1,0,0),(0,1,0),(-1,-1,1)\}\qquad & m_I^0=(1,1,-1)\nonumber\\
{\rm Region\, II}\qquad & m_{II}^\lambda = \{
(0,-1,1),(-1,0,1),(1,1,-1)\}\qquad & m_{II}^0=(0,0,0)\nonumber
\end{eqnarray}
\vskip -0.5truecm
\begin{equation}
 Z_{D_1} = \frac{q_1(q_2-q_3)+q_3-q_2 q_3}{(1-q_1)(1-q_2)(1-q_3/q_1)(1-q_3/q_2)q_3}\label{conD1}
\end{equation}
Once again, for simplicity, let us expand $Z_{D_1}$ along the direction of  the Reeb vector $(3/2,3/2,3)$ by putting $q_1=q_2=q,q_3=q^2$. This corresponds to count
mesonic excitations 
according to their R-charge, forgetting about the two $U(1)^2$ flavor indices. 
\begin{equation}
Z_{D_1}=\frac{2}{(1-q)^3}=\sum_{n=0}^\infty (n+1)(n+2) q^n = 2 + 6 q + 12 q^2 + ... 
\end{equation}
This counting perfectly matches the list of operators in the gauge 
theory. In the sector of Hilbert space with  baryonic charge $+1$ we find
the operators (\ref{AAA})
\begin{equation}
A_i\, , \qquad A_i B_p A_j\, , \qquad A_i B_p A_j B_q A_k\, , \qquad ....
\end{equation}
The F-term equations $A_i B_p A_j=A_j B_p A_i$, $B_p A_i B_q = B_q A_i B_p$
guarantee that the $SU(2)\times SU(2)$ indices are totally symmetric. The
generic operator is then of the form $A(BA)^n$ transforming in the 
$(n+2,n+1)$ representation of $SU(2)\times SU(2)$ thus 
exactly matching the $q^n$ term in $Z_{D_1}$.  
The R-charge of the operators in $Z_{D_1}$ 
is accounted by the exponent of $q$ by adding the factor $q^{\sum c_i R_i}=q^{1/2}$ which is common to all the operators in this sector (cfr. equation (\ref{erre})). The result perfectly matches with the operators  $A(BA)^n$ since the 
exact R-charge of $A_i$ and $B_i$ is $1/2$.
We could easily include the $SU(2)^2$ charges in this counting.

Analogously, we obtain for $Z_{D_3}$
%\begin{eqnarray}
%{\rm Region\, I}\qquad &m_I^\lambda = \{ ((1,0,0),(0,1,0),(-1,-1,1)\}\qquad &m_I^0=(0,0,0)\nonumber\\
%{\rm Region\, II}\qquad & m_{II}^\lambda = \{ (0,-1,1),(-1,0,1),(1,1,-1)\}\qquad &m_I^0=(-1,-1,1)\nonumber\\
\begin{equation}
 Z_{D_3} = \frac{q_1(q_2-q_3)+q_3-q_2 q_3}{(1-q_1)(1-q_2)(q_1-q_3)(q_2-q_3)}
= q_3 Z_{D_1}/(q_2 q_1)\label{conD3}
\end{equation}
Since $D_1\sim D_3$ the polytope $P_{D_3}$ is obtained by $P_{D_1}$ by a
translation and the the two partition functions $Z_{D_1}$ and $Z_{D_3}$ are
proportional. Finally, the partition functions for $D_2$ and $D_4$ are obtained by choosing the resolution in the right hand side of figure \ref{conifold}, for which is possible to satisfy the convexity condition (\ref{convex})
\begin{eqnarray}
Z_{D_2} &=& \frac{q_2 (q_1 + q_2 - q_1 q_2 -q_3)}{(1 - q_1)(1 - q_2)(q_1 - q_3)(q_2 -q_3)}\nonumber\\
Z_{D_4} &=& \frac{q_1 (q_1 + q_2 - q_1 q_2 -q_3)}{(1 - q_1)(1 - q_2)(q_1 - q_3)(q_2 -q_3)} = q_1 Z_{D_2} /q_2
\end{eqnarray}

\subsection{The $N=1$ Generating Function}
\label{geometricamusement}

Our major objective is to obtain partition functions counting all the operators in the SCFT according to their set of charges under the complete global abelian symmetry group. This means that we need to learn how to sum up the partition functions for fixed baryonic number. We do not have enough elements yet to give a general recipe, and the reader is referred to chapter \ref{chiralcount} for the 
general procedure. But it is worthwhile to give right now the general geometric procedure for the conifold in the case of $N=1$.
Indeed we have all the elements we need for this particular case and we will get a result from a purely 
geometric procedure that can be compared with the one obtained from the field theory point of view in the next chapter, with perfect 
agreement. 

Using the homogeneous coordinates the conifold can be described as the algebraic quotient of $\mathbb{C}^4$ 
by the $\mathbb{C}^*$ action as in equation (\ref{conact}). In this description all the four abelian symmetries of the
gauge theory are manifest as the four isometries of $\mathbb{C}^4$, which 
descend to the R symmetry, the two flavor symmetries and one baryonic symmetry of the 
theory on the conifold. In $\mathbb{C}^4$ the symmetries act 
directly on the homogeneous coordinates. We will use notations where
$x_i$ denotes both the homogeneous coordinates and the chemical potentials
for the standard basis of the four $U(1)$ symmetries of $\mathbb{C}^4$.
Because in the case of the
conifold there is just a one-to-one correspondence between the four
homogeneous coordinates $(x_1,x_2,x_3,x_4)$ and the four elementary
fields $(A_1,B_1,A_2,B_2)$, looking at the transformation rules of the fields corresponding to the various homogeneous coordinates 
we can set the correspondence:

\begin{equation}\label{eqcon}
(x_1,\, x_2,\, x_3,\, x_4)\, =\, (t b x, \frac{t y}{b}, \frac{t b}{x}, \frac{t}{b y}) .
\end{equation}

where $b$ is the field theory baryonic charge, $t$ the R charge, and x, y are the weights under the two $SU(2)$ global symmetries.
Note in particular that the rescaling symmetry of the $x_i$, by which
we mod out to obtain the conifold, can be identified with the baryonic 
symmetry. 
\comment{More generally, a CY with a toric diagram with $d$ vertices
is the algebraic quotient of $\mathbb{C}^d$ by the $d-3$ $\mathbb{C}^*$ actions
given by the $\mu_i$ obeying (\ref{bartoric}) which correspond to the $d-3$
baryonic symmetry of the dual gauge theory.
}

%Now we are ready for discussing the geometric interpretation of $g_{1,B}$.
As we will see in the following chapter the generating function of 
homogeneous monomials $x^\alpha=\prod_{i=1}^4 x_i^{\alpha_i}$ on the conifold:
\begin{equation}\label{congeom1}
\frac{1}{\prod_{i=1}^4 (1-x_i)}=\frac{1}{(1-t b x)(1-\frac{t y}{b})(1- \frac{t b}{x})(1-\frac{t}{b y})}
\end{equation}
 is just the $N=1$ generating function $g_1(b,t,x,y; {\cal C})$ for the SCFT counting the strings of fields according 
to their charges. Indeed in the case $N=1$ the superpotential of the conifold is trivial and the 
moduli space is just $\mathbb{C}^4$ with coordinates the four elementary fields of the conifold, with the same $U(1)$ charges
of the homogeneous coordinates $x_i$. We can indeed assign a degree $B$ to a monomial by
looking at the scaling behavior under the $\mathbb{C}^*$ action in equation
(\ref{eqcon}): $x^\alpha\rightarrow \lambda^B x^\alpha$.
Let us define the functions $g_{1,B}(t,x,y; {\cal C})$ counting the subset
of monomials of degree $B$. 
\comment{This agrees with our previous definition
of $g_{1,B}(t,x,y; {\cal C})$ as the generating function for operators
of baryonic charge $B$ since the elementary fields are in correspondence with
the homogeneous coordinates and the $\mathbb{C}^*$ action 
coincides with the baryonic symmetry. 
}
%$g_{1,B}(t,x,y; {\cal C})$ is thus the generating functiowe denote as ${\cal O}(B)$n for
%homogeneous polynomials of degree $B$ on the conifold. 
Homogeneous polynomials
 are not functions on the conifold in a strict sense, as they transform non trivially under a rescaling of the $x_i$, but are rather sections of a line bundle.
Line bundles on the conifold are labeled by an integer B. For simplicity,
we denote them as ${\cal O}(B)$. The holomorphic sections of 
${\cal O}(B)$ are just the polynomials of degree $B$. 
The construction exactly  parallels the familiar case of $\mathbb{P}^k$ where 
the homogeneous polynomials of degree $n$ are the sections of
the line bundle ${\cal O}(n)$. 
\comment{In toric geometry, sections of line bundles, when arranged according to
their charge under the torus action, fill a convex polytope in $\mathbb{Z}^3$.
For each integer point $P$ in the polytope there is exactly one section with charges given by the integer entries of the point $P\in\mathbb{Z}^3$. For the
conifold, these polytopes have the shape of  integral conical pyramids and
will be discussed in details in the next Section.}

It is a general result of toric geometry
that the homogeneous coordinate ring of a toric variety 
decomposes as a sum over non-trivial
line bundles \cite{cox}, which is equivalent to the statement that
polynomials can be graded by their degree. In the case of the conifold we have

\begin{equation}
\mathbb{C}[x_1,x_2,x_3,x_4] = \sum_{B=-\infty}^{\infty} {\rm H}^0 ({\cal C}, {\cal O}(B)) .
\end{equation}

The equation
\begin{equation}
g_1(t,b,x,y; {\cal C}) = \sum_{B=-\infty}^\infty b^B g_{1,B}(t,x,y; {\cal C}),
%\label{g1coni}
\end{equation}
translates the previous mathematical identity at the level of generating
functions. In this context 
$g_{1,B}(t,x,y; {\cal C})$ is interpreted as a
 character under the action of the three abelian isometries of the conifold
with chemical potential $x,y,t$,
\begin{equation}
g_{1,B}(t,x,y; {\cal C}) = {\mbox Ch}\, {\rm H}^0({\cal C}, {\cal O}(B))\equiv
{\rm Tr}\{x,y,t| {\rm H}^0({\cal C}, {\cal O}(B))\} .
\label{charac}
\end{equation}

and it is the kind of partition functions we learnt to compute in this chapter. 
Indeed for generic $B$ we can expresses
the result as a sum over the fixed points $P_I$ of the torus $T^3$ action in a
smooth resolution of the conifold

\begin{equation}
g_{1,B}(q;  {\cal C}) = q^{n_B}\sum_{P_I} \frac{q^{m^{(I)}_B}}{\prod_{i=1}^3 (1-q^{m^{(I)}_i})} ,
\label{locC}
\end{equation}
where the index $I$ denotes the set of isolated fixed points
and the four vectors
$m^{(I)}_i,\, i=1,2,3$, $m^{(I)}_B$ in $\mathbb{Z}^3$ are the weights of the linearized action of $T^3$ on ${\cal C}$  and the fiber of the line bundle, respectively.

\begin{figure}[h!!!!!]
\begin{center}
\includegraphics[scale=0.5]{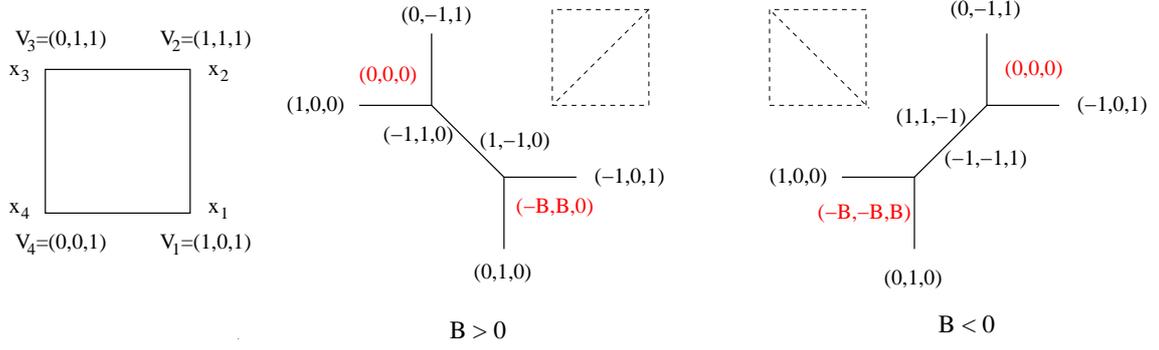} 
\caption{Localization data for the $N=1$ baryonic partition functions. The
vertices $V_i$ are in correspondence with homogeneous coordinates $x_i$ and
with a basis of  divisors $D_i$. Two different resolutions, related by
a flop, should be used for positive and negative $B$, respectively. 
Each resolution has two fixed points,
corresponding to the vertices of the $(p,q)$ webs; the weights $m^{(I)}_i,\, i=1,2,3$
and $m^{(I)}_B$ at the fixed points are indicated in black and red, respectively.}
\label{coooC}
\end{center}
\end{figure}

We notice that to perform a 
computation using the index theorem we need to choose a correctly normalized
basis $q\equiv(q_1,q_2,q_3)$ for the $T^3$ torus action on the manifold, the same used in the previous 
section for the elementary line bundle $\mathcal{O}(-1)$, $\mathcal{O}$, $\mathcal{O}(1)$.
This implies taking square roots of the variables $(x,y,t)$ (that have
been chosen for notational simplicity). The relation between variables
is easy computed by using
\begin{equation}
q_k=\prod_{i=1}^4 x_i^{\langle e_k,V_i\rangle}\,\qquad\qquad k=1,2,3, 
\label{rel}
\end{equation} 
where $e_k$ are the basis vectors of $\mathbb{Z}^3$ and $V_i$ the vertices of
the toric diagram (notice that all dependence on baryonic charges drops from
the right hand side by equation (\ref{bari})).
We thus have

\begin{eqnarray}
\nonumber
q_1&=&x_1 x_2 = t^2 x y,\\ \nonumber
q_2&=&x_2 x_3 = \frac{t^2 y}{x},\\ 
q_3&=&x_1 x_2 x_3 x_4 = t^4 .
\label{GLr}
\end{eqnarray}
 
%The two possible resolutions of the conifold are shown in  Figure \ref{co}.
%The fixed points of the 
%torus action are in correspondence with the vertices of the $(p,q)$ web
%or, equivalently with the triangles in the subdivision of the toric diagram. 
%The vectors  $m^{(I)}_i,\, i=1,2,3$ in the denominator of
%formula (\ref{loc}) are computed as the three primitive inward normal vectors
%of the cone $\sigma_I$ in $\mathbb{Z}^3$ made with the three vertices $V_{i}$ of
%the I-th triangle. 
For computing  the character $g_{1,0}$ of holomorphic 
functions on the conifold we use formula (\ref{locC}) with $m_B^{(I)}=n_B=0$ \cite{Martelli:2006yb}; both resolutions can be used and give the same result.   
For computing the character $g_{1,B}$ for sections of a line bundle of degree
$B$ we need to choose a convenient resolution and compute the vectors 
$m_B^{(I)}$ and $n_B$. This can be done as already explained in the previous sections. 
\comment{
A generic line bundle
can be associated with a linear combination $\sum_{i=1}^d c_i D_i$ of the basic
divisors. There is a basic divisor $D_i$ for each vertex and we can assign
to it the same charges of the corresponding $x_i$; two divisors are linearly
equivalent if and only if they have the same baryonic charge. Given a divisor
$\sum_{i=1}^d c_i D_i$ of degree $B$ we can assign numbers $c_i$ to the
vertices $V_i$ of the toric diagram. Each fixed point $I$ determines
a vector $m_B^{(I)}$ as the integer solution of the linear system of three
equations   

\begin{equation}\label{charge}
\langle m_B^{(I)}, V_i \rangle = - c_i, \,\,\, V_i\in \sigma_I,  
\end{equation}
where the $V_i$ are the vertices of the $I$-th triangle.
A resolution gives the correct result for $g_{1,B}$ whenever all the vectors $m_B^{(I)}$ satisfy the convexity condition
$\langle m_B^{(I)}, V_i \rangle \ge - c_i$ for the vertices $V_i$ not belonging to
the $I$-th triangle}
The prefactor $q^{n_B}$ in formula (\ref{locC}) is just given
by the $T^3$ charge of the monomial $\prod_{i=1}^d x_i^{c_i}$ to correctly sum up all the generating functions for fixed $B$.

In the case at hand, the partition functions $g_{1,B}$ can be
computed using the divisor $B D_1$ and the resolution
on the left in Figure (\ref{coooC}) for $B>0$
and the divisor $|B| D_4$ and the resolution on the right 
for $B<0$. It is interesting to note that in passing from positive to 
negative baryonic charges we need to perform a flop on the resolved 
conifold. 
%We will say more about the flop transition in the next Section. 
The weights are reported in Figure \ref{coooC}. Formula (\ref{locC}) gives

\begin{eqnarray}
g_{1,B\ge0}(q; {\cal C}) &=& {q_3^{\frac{B}{4}} (\frac{q_1}{q_2})^{\frac{B}{2}}}
\left(\frac{1 } { (1 - q_1) (1-\frac{q_3}{q_2})  (1-\frac{q_2}{q_1}) }+ \frac{( \frac{q_2}{q_1})^B} { (1 - q_2) (1-\frac{q_3} {q_1})  (1-\frac{q_1}{q_2}) }\right) , \nonumber \\ \nonumber
g_{1,B\le0}(q; {\cal C}) &=& \frac{(q_1 q_2)^{\frac{B}{2}}}{q_3^{\frac{3B}{4}}}
\left( \frac{(\frac{q_3}{q_1 q_2 })^B } { (1-q_1) (1-q_2)  (1-\frac{q_3}{q_1 q_2}) }+ \frac{1} { (1-\frac{q_3}{q_2}) (1-\frac{q_3} {q_1})  (1-\frac{q_1 q_2}{q_3}) }\right ) .
\end{eqnarray}

This formula, when expressed in terms of $(x,y,t)$, using Equation (\ref{eqcon}), becomes

\begin{eqnarray}\label{g1Bgeom}
g_{1,B\ge0}(t,x,y; {\cal C}) &=& \frac{t^B x^{B} } { (1 - \frac{1}{x^2}) (1-t^2 x y)  (1-\frac{t^2 x}{y}) }+ \frac{t^B x^{-B}} { (1 - x^2) (1-\frac{t^2 y} {x})  (1-\frac{t^2}{x y}) } , \nonumber \\
g_{1,B\le0}(t,x,y; {\cal C}) &=& \frac{t^{-B} y^{-B} } { (1-\frac{1}{y^2}) (1-t^2 x y)  (1-\frac{t^2 y}{x}) }+ \frac{t^{-B} y^{B}} { (1-y^2) (1-\frac{t^2 x} {y})  (1-\frac{t^2}{y x}) }. \nonumber\\ 
\end{eqnarray}

In the next chapter we will see how it is possible to extract the complete generating function for $N=1$ directly from the field theory. There we will learn that the most efficient way of computing the $g_{1,B}(t,x,y; {\cal C})$, once you know $g_1$, is to expand the generating function of the coordinate ring $g_1$ in a generalized Laurent series. Indeed we will see that the formulae (\ref{g1Bgeom}), computed here just using the geometric properties of $\cX$, can be equivalently obtained from a completely field theory computation. Indeed this is just a particular example of a general fact: there exist two different geometric  ways 
of computing the characters $g_{1,B}(\{t_i\}; CY)$ 
for general CY manifold $\cX$ : by Laurent expansion of the generating function for
homogeneous coordinates or by using the index theorem.

\subsection{The Partition Function for BPS Baryonic Operators}\label{geomBZ}

Let us now pass to shortly analyze the non abelian generic $N > 1$ case.
The BPS baryonic states in a sector of the Hilbert space associated
with the divisor $D$ are obtained from the $h_m$ by considering
the N-fold symmetrized combinations $|h_{m_1},....,h_{m_N}>$. The 
partition function $Z_{D,N}$ for BPS baryon is obtained from $Z_D$ by
solving a combinatorial problem \cite{Kinney:2005ej,Benvenuti:2006qr}. 

Given $Z_D$ as sum of integer points in the polytope $P_D$
\begin{equation}
Z_D(q) = \sum_{m\in P_D \cap M} q^m
\end{equation}
the generating function $G_D(\nu,q)$ for symmetrized products of elements in $Z_D$ is
given by
\begin{equation}
G_D(\nu,q)=\prod_{m\in P_D \cap M} \frac{1}{1- \nu \hbox{ }q^m} =\sum_{N=0}^\infty \nu^N Z_{D,N}(q)
\label{multi}\end{equation}
This formula is easy to understand: if we expand $(1- \nu \hbox{ } q^m)^{-1}$ in 
geometric series, the coefficient of the term $\nu^k$ is given by 
all possible products of $k$ elements in $P_D$, and this is clearly a $k$-symmetric product.

It is easy to derive the following relation between $Z_D(q)$ and $G_D(\nu,q)$
\begin{equation}\label{expG}
G_D(\nu,q)=e^{\sum_{k=1}^\infty \frac{\nu^k}{k} Z_D(q^k)}
\end{equation}
$G_D$ is exactly the Plethystic function $PE_{\nu}[...]$ applied to the function $Z_D(q)$. Indeed we have already introduced this function in chapter \ref{braneasing} and we remind the reader that $PE_{\nu}[g(t)]=e^{\sum_{k=1}^\infty \frac{\nu^k}{k} g(t^k)}$. We will use the properties of this function in the following chapters. Here we see as the $PE$ function naturally appears even in the counting problem of baryonic operators. It will reveal a fundamental tool for studying the counting problem for the complete chiral ring.

In the case we have computed $Z_D(q)$ in terms of  the fixed point data of a compatible resolution as in equation  (\ref{sumzd}) 
$$Z_D(q)=\sum_I \frac{q^{m_I^0}}{\prod_{\lambda=1}^3(1-q^{m_I^\lambda})}=\sum_I\sum_{s_I^1,s_I^2,s_I^3} q^{m_I^0}q^{\sum_{\lambda=1}^3 s_I^\lambda m_I^\lambda}$$
formula (\ref{expG}) allows, 
with few algebraic manipulation, to write the generating
function  as follows
\begin{equation}
G_D(\nu,q)=\prod_{P_I}\prod_{s_I^\lambda=0}^\infty \frac{1}{1- \nu \hbox{ } q^{m_I^0} q^{\sum_{\lambda=1}^3 s_I^\lambda m_I^\lambda}}\end{equation}

 We are eventually interested
in the case of BPS baryonic operators associated with the symmetrized
elements $|h_{m_1},....,h_{m_N}>$, and thus to the $N$-fold symmetric partition function:
\begin{equation}\label{sympar}
Z_{D,N}(q)\equiv \frac{1}{N!}\frac{\partial ^N G_D(\nu,q)}{\partial \nu^N } \Big| _{\nu=0}
\end{equation}
Thanks to $G_D(0,q)=1$ (see eq. (\ref{expG})) we can easily write $Z_{D,N}$ in function of $Z_D$.
For example we have:
$$Z_{D,1}(q)= Z_D(q)$$
$$Z_{D,2}(q) = \frac{1}{2} (Z_D(q^2)+Z_D^2(q))$$
$$Z_{D,3}(q) = \frac{1}{6} (2 Z_D(q^3)+ 3 Z_D(q^2) Z_D(q) + Z_D^3(q))$$
Once we know $Z_D$ for a particular baryonic sector of the $BPS$ Hilbert space it is easy to write down the complete partition function $Z_{D,N}$.

\section{Volumes and Field Theory Charges}\label{volc3}
One of the predictions of the AdS/CFT correspondence for the
background $AdS_5\times H$ is that the volume of $H$ is related
to the central charge $a$ of the CFT, and the volumes of the three cycles
wrapped by the D3-branes are  related to the R-charges of the corresponding
baryonic operators \cite{Gubser:1998fp,Gubser:1998vd}:
\begin{equation}\label{aV}
a=\frac{\pi^3}{4 \hbox{ Vol(H)}} \qquad R_i = \frac{\pi \hbox{Vol}(C_3^i)}{3 \hbox{Vol}(H)}
\end{equation}
To many purposes, it is useful to consider the volumes
as functions of the Reeb Vector $b$.
Recall that each K\"ahler metric on the cone, or equivalently a Sasakian structure on the base $H$,  determines a Reeb vector $b=(b_1,b_2,b_3)$ and
that the knowledge of $b$ is sufficient to compute all volumes in $H$ \cite{Martelli:2005tp}. 
Denote with $\rm{Vol}_H(b)$ the volume of the base of a K\"ahler cone with Reeb vector $b$.
The Calabi-Yau condition $c_1(\cX)$ requires $b_3=3$ \cite{Martelli:2005tp}. As shown in \cite{Martelli:2005tp,Martelli:2006yb}, the Reeb vector $\bar b$ 
associated with the Calabi-Yau metric can be obtained by minimizing the function $\rm{Vol}_H(b)$ with respect to $b=(b_1,b_2,3)$. This volume minimization is the geometrical 
counterpart of a-maximization in field theory \cite{Intriligator:2003jj}; the equivalence of a-maximization and volume minimization has been explicitly proven for all toric cones in \cite{Butti:2005ps,Butti:2005vn} and for a class a non toric cones in \cite{Butti:2006nk}. For each Reeb vector $b=(b_1,b_2,b_3)$ we can also define the
volume of the three cycle obtained by restricting a divisor $D$ to the base,
$\rm{Vol}_{D}(b)$. We can related the value $\rm{Vol}_{D}(\bar b)$ at the 
minimum to the exact R-charge of the lowest dimension baryonic operator associated with the divisor $D$ \cite{Gubser:1998fp,Gubser:1998vd} as in formula (\ref{errevol}).

All the geometrical informations about volumes can be
extracted from the partition functions.
The relation between the character $Z_{\cal{O}}(q)$ 
for holomorphic functions on $C(H)$ and the volume of $H$ was suggested
in \cite{Bergman:2001qi} and proved for all K\"ahler cones in \cite{Martelli:2006yb}. If we
define $q=(e^{-b_1 t},e^{-b_2 t},e^{-b_3 t})$, we have \cite{Bergman:2001qi,Martelli:2006yb}
\begin{equation}
\rm{Vol}_H (b) =\pi^3 \lim_{t\rightarrow 0} t^3 Z_{\cal{O}} (e^{-b t})\label{vol}
\end{equation}
This formula can be interpreted as follows: the partition function $Z_{\cal{O}}(q)$ has a pole for $q\rightarrow 1$, and the order of the pole - three - reflects the complex dimension of $C(H)$ while
the coefficient is related to the volume of $H$. 
We can write a general formula for the volume of $H$ in term of the fixed point data:
 \begin{equation}
 \rm{Vol}_H(b)= \sum_{P_I} \frac{\pi ^3}{\prod_{\lambda = 1}^3 (m_I^\lambda, b)}
\label{volsum}
\end{equation}

For the conifold example we have:
\begin{equation}
\lim_{t\rightarrow 0} t^3 Z_{\cal{O}} (e^{-b t})= \frac{b_3}{b_1 b_2(b_1 - b_3)(b_2 - b_3)}
\end{equation}
and this correctly reproduce the result of \cite{Martelli:2005tp} for the volume.

In a similar way  the partition functions $Z_D$ contain the information about the three-cycle volumes $\rm{Vol}_{D}(b)$. Indeed we suggest that,
for small $t$ \footnote{And a convenient choice of $D$ in its equivalence class.},
\begin{equation}
\frac{Z_D(e^{-b t})}{Z_{\cal{O}}(e^{-b t})} \sim 1 + t \frac{\pi  \rm{Vol}_D(b)}{2 \rm{Vol}_H(b)}+...\label{voldiv}
\end{equation}
Notice that the leading behavior for all partition functions $Z_D$ is the same and proportional to the volume of $H$; for $q\rightarrow 1$ the main contribution comes from states with arbitrarily large dimension and it seems that
states factorized in a minimal determinant times gravitons dominate the
partition function. The proportionality to $\rm{Vol}_H$ is then expected
by the analogy with giant gravitons probing the volume of $H$. The subleading
term of order $1/t^2$ in $Z_D$ then contains information about the dimension 
two complex divisors. 
Physically it is easy to understand that $Z_D$ contains the information about the volumes of the divisors. 
We can think at $Z_D$ as a semiclassical parametrization of the holomorphic non-trivial surfaces in $\cX$, 
with a particular set of charges related to $D$; while $Z_{\mathcal{O}}$ parametrizes the set of trivial 
surfaces in $\cX$. Thinking in this way it is clear that both know about the volume of the compact space, 
but only $Z_D$ has information on the volumes of the non-trivial three cycles. 

For divisors $D$ associated with elementary fields we can rewrite 
equation (\ref{voldiv}) in a simple and suggestive way in terms 
of the R-charge, or dimension,  of the elementary field (see equation (\ref{errevol}))
\begin{equation}
\frac{Z_D(e^{-b t})}{Z_{\cal{O}}(e^{-b t})} \sim 1+ t \frac{3 R_D(b)}{2}+...=1+ t \Delta(b)+...
\end{equation}
 
 As a check of formula (\ref{voldiv}), we can expand the partition functions
for the conifold computed in the previous Section
\begin{equation}
\frac{Z_{D_i}}{Z_{\cal{O}}}\sim(1+\frac{(b_1-b_3)(b_2-b_3)t}{b_3},1+\frac{(b_1 b_3-b_1 b_2)t}{b_3},1+\frac{b_1 b_2 t}{b_3},1+\frac{(b_2 b_3 -b_1 b_2)t}{b_3})
\nonumber
\end{equation}
 and compare it with the formulae in \cite{Martelli:2005tp}
\begin{equation}
\rm{Vol}_{D_i}(b)=\frac{2\pi^2 \det \{n_{i-1},n_i,n_{i+1}\}}{\det\{b,n_{i-1},n_i\} \det \{b,n_i,n_{i+1}\}}\, , \qquad \rm{Vol}_H(b)=\frac{\pi}{6}\sum_{i=1}^d \rm{Vol}_{D_i}(b)\label{tor}
\end{equation}
One can perform similar checks for some $Y^{p,q}$, $L^{p,q,r}$ and $dP_n$ singularities, with perfect agreement. A sketch of a general 
proof for formula (\ref{voldiv}) is given in \cite{Butti:2006au}.

We would like to notice that, by expanding equation (\ref{sumzd}) for $q=e^{-b t}\rightarrow 1$ and comparing with formula (\ref{voldiv}), we are able to write
a simple formula for the volumes of divisors in terms of the fixed point
data of a compatible resolution
 \begin{equation}
 \rm{Vol}_D(b)=2 \pi^2 \sum_{P_I} \frac{(-m_I^0,b)}{\prod_{\lambda = 1}^3 (m_I^\lambda, b)}
\label{volsum}
\end{equation}
This formula can be conveniently generalized to the case where the fixed points are not isolated but there are curves or surfaces fixed by the torus action.

The previous formula is not specific to toric varieties. It can be used  whenever we are able to resolve the cone $C(H)$ and to reduce the computation of $Z_D$ to a sum over isolated fixed points (and it can be generalized to the case 
where there are fixed submanifolds). As such, it 
applies also to non toric cones. The relation between
volumes and characters may give a way for computing volumes of
divisors in the general non toric case, where explicit formulae like 
(\ref{tor}) are not known. 

In this section we start to appreciate how the geometric informations of $\cX$ are encoded in partition functions like $Z_D$. This 
is just the starting point. In the following chapters we will see that a huge quantity of informations regarding the chiral ring
of the gauge theory is contained in the partition function we are going to compute.  

\section{Summary and Discussions}

In this chapter we analyzed a general procedure to construct partition functions counting both 
baryonic and non baryonic $BPS$ operators of a field theory dual to a toric geometry $\cX$. These partition functions 
contains the structure of the BPS chiral ring of the gauge theory and, because we computed them 
counting D3 brane states in the dual gravity background,
 in principle they are valid at strong coupling.  
The partition functions contain a lot of informations regarding the geometry of $\cX$. 
Indeed we explained how one can extract the volume of the 
horizon manifold H and the volumes of all the elementary three cycles from the partition functions. 
These geometrical volumes encode the values of the central charge and of 
all the R charges of the gauge theory at the IR fixed point. 

Some observations are mandatory.

As we said our computation is done on the supergravity side, and it is therefore valid
at strong coupling and large N. In the following chapter we will see that, similarly to 
the partition function for BPS mesonic operators \cite{Kinney:2005ej,Biswas:2006tj,Benvenuti:2006qr}, 
we will be able to extrapolate the result to finite value for the coupling and for small N. 
This fact suggest some non renormalization properties of the spectra, 
meaning that the number of BPS operators with a given set of global charges 
is invariant under the RG flow. 

It would be interesting to understand better the counting of multiplicity, and to find a way 
to write down a complete partition function for the $BPS$ gauge invariant scalar operators. We will analyze this issue in the 
following chapters.

It would be also interesting to understand better the non toric case. Most
of the discussions in this chapter apply to this case as well. The classical 
configurations of BPS D3 branes wrapping a divisor $D$ are still 
parameterized by the generic section of $H^0(\cX,{\cal O}(D))$ and Beasley's 
prescription for constructing the BPS Hilbert space is unchanged. The partition
function $Z_D(q)$ can be still defined, with the only difference that 
$q\in T^k$ with $k$ strictly less than three. $Z_D(q)$ can be still
computed by using the index theorem as explained in Section \ref{counting}
and the relation between $Z_D(q)$ and the three cycles volumes should be
still valid. In particular, when $\cX$ has a completely smooth resolution
with only isolated fixed points for the action of $T^k$, formulae
 (\ref{sumzd}) and (\ref{volsum}) should allow to compute the partition functions and 
the volumes. What is missing in the non-toric case is an analogous of the
homogeneous coordinates, the polytopes and the existence of a canonical
smooth resolution of the cone. But this seems to be just a technical problem.

\vspace{2em}

\chapter{Baryonic Generating Functions}\label{barcon}

In this chapter we will develop some techniques to write down complete partition functions 
for quiver gauge theories. We will show how it is possible to use the plethystic function $PE$, 
in order to compute baryonic generating functions that count BPS operators in the chiral ring of 
quiver gauge theories living on the world volume of D branes probing the non compact CY manifold $\cX$. 
These baryonic generating functions count the mesonic and baryonic chiral spectrum of operators according to 
their baryonic charges and for generic number of colours $N$. Compare to the previous chapter, we will follow a 
more field theory approach, but we will be able to reproduce all the result already obtained form the more 
geometric approach just explained, and we will go far beyond. We will give special attention to the conifold theory, 
where exact expressions for generating functions are given in detail, and we will leave a complete discussion for 
generic $\cX$ toric singularity in the next chapter. In this chapter we do the first step towards the solution of the long 
standing problem for the combinatorics of quiver gauge theories with baryonic moduli spaces. Quite surprisingly, 
looking to the conifold example, the baryonic charge turns out to be the quantized K\"ahler modulus of the geometry. 
In the next chapter we will see that the relation between the baryonic charges and the K\"ahler moduli of the geometry $\cX$ 
is the key point for solving the multiplicity problem and construct complete partition 
functions for all the SCFT dual to toric singularity $\cX$. 

\section{Generalities}
\setall

Enriched by our experience in the previous chapters we understand that counting problems in supersymmetric gauge theories 
reveals a rich field theory and geometrical structure: the generators and the relations in the chiral ring, as we have already 
seen in the easy examples of the $N=1$ mesonic moduli space for $\mathbb{C}^3$ and the conifold; and the volume of H and of all
its non trivial three cycles, has we have just seen in the previous sections. Using the generating functions for 
counting BPS operators in the chiral ring, one can get further information about the dimension of the moduli space 
of vacua and the effective number of degrees of freedom in the system. %Such counting problems were done in recent publications
The computation of such generating functions is generically a very hard problem. 
There are, however, special cases in which the problem simplifies considerably and one can get an exact answer for the generating function.

As we have already seen there is an amazing simplification of the problem of counting BPS operators in the chiral 
ring when the gauge theory is living on $N$ D brane in Type II superstring theory \cite{Benvenuti:2006qr,Butti:2006au,Feng:2007ur}. 
For such a case one is applying the plethystic program which is introduced in \cite{Benvenuti:2006qr} and discussed in further d
etail in \cite{Feng:2007ur}. The main point in this program is the use of the plethystic exponential and its inverse, 
the plethystic logarithm, in a way which allows the computation of finite $N$ generating functions in terms of 
only one quantity -- the generating function for a single D brane. The generating function for one D brane is calculated 
in turn by using geometric properties of the moduli space of vacua for this gauge theory, typically the moduli space 
being a CY manifold. In many cases the generating function for one D brane is computed even without the detailed knowledge 
of the quiver gauge theory and bypasses this data by just using the geometric properties of the moduli space. 
The PE essentially generate the functions counting symmetric product of functions counted by the Hilbert series 
over which the PE operates. Hence the importance of the PE is very easy understandable for the mesonic spectra 
where the moduli space for N brane is the N times symmetric product of the geometric singularity $\cX$. 
The importance of the PE function for the counting of operators with non trivial baryonic number is may be less intuitive; 
but it is a fact, coming from the geometric quantization procedure, as reviewed in the previous chapter, 
that the PE naturally appears even in the baryonic counting. We will be more explicit in the following. 

The plethystic program brings about an important distinction between different types of moduli spaces 
-- we have three such types: 1. There are moduli spaces which are freely generated 
-- that is the set of generators obey no relations. 2. There are moduli spaces which are complete intersections 
-- that is there are generators that satisfy relations such that the number of relations plus the dimension of 
the moduli space is equal to the number of generators. 3. The rest are moduli spaces which are not complete intersections. 
The last class of models is by far the largest class. Most moduli spaces of vacua do not admit a simple algebraic description 
which truncates at some level. It is important, however, to single out and specify the moduli spaces which fall into the first 
two classes as they are substantially simpler to describe and to analyze. 
In this chapter we will analyze the moduli space of the conifold for different values of $N$ 
and we will see how the PE function and its inverse function are very useful for studying these kind of properties.

Counting problems in supersymmetric gauge theories have been a subject of recent and 
not so recent interest. The first paper possibly dates back to 1998, \cite{Pouliot:1998yv} 
and is independently reproduced in \cite{Romelsberger:2005eg}. 
Other works include \cite{Kinney:2005ej, Biswas:2006tj, Mandal:2006tk,Martelli:2006vh,Basu:2006id}, 
this subject is of increased interest in recent times. Mesonic generating functions were given in \cite{Benvenuti:2006qr}, 
mixed branches were covered using ``surgery'' methods \cite{Hanany:2006uc}. 
In all these works miss a general and powerful enough procedure for counting the complete spectrum 
including the operators charged under the various baryonic symmetries, that are by far the most generic operators. 
In the previous chapter we explain how to construct baryonic generating functions for divisors in toric geometry. 
This procedure is at the basis of other counting procedure and reduce to them in particular limits. 
The combination of these problems brings about the next interesting problem -- that of counting baryonic operators on the moduli space.

\comment{
Baryons have been considered in detail in supersymmetric gauge theories and were subject of intensive study in the context of the AdS/CFT correspondence \cite{Gubser:1998fp}. Of particular mention are the works by \cite{Berenstein:2002ke} were baryons were discussed in detail from a gauge theory point of view; \cite{Mikhailov:2000ya} with the relation of giant gravitons to holomorphic curves in the geometry; and \cite{Beasley:2002xv} with the quantization of the system, leading to the right way to treat counting of baryons in quiver gauge theories.
}

In this chapter we compute generating functions for baryonic BPS operators. 
The crucial methods which are used are the successful combination of the plethystic program 
\cite{Benvenuti:2006qr, Feng:2007ur}, introduced in the chapter \ref{braneasing},
together with the baryonic generating functions for 
divisors in the geometry \cite{Butti:2006au}, explained in the previous chapter. 
Together they form an exact result that computes BPS operators on baryonic branches.

The chapter is organized as follows. In Section \ref{generalCY} we introduce the plethystic program for baryonic generating functions 
for theories with one baryonic charge. In Section \ref{conifoldB} we look in detail on baryonic generating functions for the conifold theory for $N=1$ branes. 
We discuss in detail the generating function for unfixed and for fixed baryonic charges. 
The relation to the geometry is elaborated and diagrammatic rules are composed for computing generating functions for $N=1$ D brane 
and fixed baryonic charge. The relation to tilings of the two dimensional plane is discussed and the lattice of BPS charges reveals 
itself beautifully as the (p,q) web of the toric diagram. In section \ref{conifoldNN} we then proceed to discuss the general 
$N$ case and write the relations between the different generating functions, f
or fixed and unfixed D brane charge and for fixed and unfixed baryonic charges. 
The case of $N=2$ and small number of branes is computed and discussed in detail. 
Aspects of the plethystic exponential and the plethystic 
logarithm are then discussed in detail in section \ref{PEquant} and in section \ref{PLlog}
and lead to the understanding of the generators and relations for the chiral ring
of the conifolds for different values of $N$ in \ref{PLlog}. In section \ref{Toys} we turn to some toy models, 
termed ``half the conifold'' and the ``$\frac{3}{4}$ the conifold''. 
For the first example we demonstrate how the plethystic program for baryons computes the generating function exactly, 
leading to a freely generated moduli space for fixed number of branes, $N$. 
A comparison with the Molien invariant is then pursued. 
The other example is discussed in detail and some explicit formulae are given.

\section{Generating Functions for CY Manifolds with One Baryonic Charge}
\label{generalCY}
\setall

In this section we will give general prescriptions on the computation of generating functions for BPS operators 
in the chiral ring of a supersymmetric gauge theory that lives on a D brane which probes a generic non-compact 
CY manifold $\cX$ which has a single baryonic charge. A class of such manifolds includes the $Y^{p,q}$ theories \cite{Gauntlett:2004zh,Gauntlett:2004yd,Gauntlett:2004hh}, 
or more generally it includes the $L^{abc}$ theories \cite{Cvetic:2005ft,Cvetic:2005vk,Benvenuti:2005ja,Butti:2005sw}. 
A challenge problem is to compute the generating functions for baryonic operators. 
This chapter will be devoted to the study of this problem, with success for a selected set of theories. 
The problem of computing generating functions for a general CY manifold is still very difficult to solve but with the methods of this section we will be able to reduce the problem of computing the generating function for a generic number of D branes, 
$N$ and generic baryonic number $B$, to a much simpler problem. This reduction is done using the plethystic program \cite{Benvenuti:2006qr} (See further details in \cite{Feng:2007ur}).

Recall that, as explained in chapter \ref{braneasing} and derived in the previous chapter,
 in the case of baryon number $B=0$, namely for mesonic generating functions 
\cite{Benvenuti:2006qr}, the knowledge of the generating function for $N=1$ is enough to compute the generating function for any $N$. 
This is essentially due to the fact that the operators for finite $N$ are symmetric functions of the operators for $N=1$ 
and this is precisely the role which is played by the plethystic exponential -- 
to take a generating function for a set of operators and count all possible symmetric functions of it.

For the present case, in which the baryon number is non-zero, it turns out that the procedure is not too different 
than the mesonic case. Indeed as shown for some particular cases in the previous chapter, one needs to have the 
knowledge of a single generating function, $g_{1,B}$, for one D brane, $N=1$ and for a fixed baryon number $B$, 
and this information is enough to generate all generating functions for any number of D branes and for a fixed baryonic 
number \cite{Butti:2006au}.

Given a ${\cal N}=1$ supersymmetric gauge theory with a collection of $U(1)$ global symmetries, $\prod_{i=1}^r U(1)_i$, 
we will have a set of $r$ chemical potentials $\{t_i\}_{i=1}^r$. The generating function for a gauge theory living on a 
D brane probing a generic non-compact CY manifold is going to depend on the set of parameters, $t_i$. 
There is always at least one such $U(1)$ global symmetry and one such chemical potential $t$, corresponding to the $U(1)_R$ symmetry, and in 
the case in which $\cX$ is a toric variety $r=g+2$, where $g$ is the number of gauge factors and we are including the anomalous 
$U(1)$ symmetries.
For a given CY manifold $\cX$ we will denote the generating function for a fixed number of D branes, $N$ and a fixed baryonic charge 
$N B$ by $g_{N,N B}(\{t_i\}; \cX).$

Let us introduce a chemical potential $\nu$ for the number of D branes. 
Then the generating function for any number of D branes and for a fixed baryonic charge $B$ is given by

\begin{eqnarray}
\nonumber
g_B(\nu; \{t_i\}; \cX) &=& \hbox{PE$_{\nu}$}[g_{1,B}(\{t_i\}; \cX)] \equiv \exp\biggl(\sum_{k=1}^\infty \frac{\nu^k}{k}g_{1,B}(\{t_i^k\}; \cX) \biggr)\\
&=& \sum_{N=0}^\infty \nu^N g_{N,N B}(\{t_i\}; \cX) .
\label{g1pletT}
\end{eqnarray}

This formula can be inverted to give the generating function for fixed baryonic charge and fixed number of branes, in case the generating function for fixed baryonic charge is known,

\begin{equation}
\nonumber
g_{N,N B}(\{t_i\}; \cX) = \frac{1}{2\pi i} \oint \frac{d\nu}{\nu^{N+1}} g_B(\nu; \{t_i\}; \cX) .
%\label{g1coni}
\end{equation}

We can further introduce a chemical potential $b$ for the baryon number and sum over all possible baryon numbers go get the generating function for an unfixed baryonic number,

\begin{eqnarray}
g(\nu; \{t_i\}; \cX) = \sum_{B=-\infty}^\infty b^{N B} \exp\biggl(\sum_{k=1}^\infty \frac{\nu^k}{k}g_{1,B}(\{t_i^k\}; \cX) \biggr) .
%\label{g1coni}
\end{eqnarray}

This formula is very schematic, however sufficient to the purposes
of this chapter. The correct general formula would accommodate
multiple baryonic charges, not necessarily running over all integers, and
multiplicities for the components $g_B$. 
From the previous formula we can extract the generating function for a fixed number of D branes $N$,

\begin{eqnarray}
g(\nu; \{t_i\}; \cX) = \sum_{N=0}^\infty \nu^N g_N(\{t_i\}; \cX).
%\label{grandcancon}
\end{eqnarray}

This set of equations form the basis of this chapter 
and allow for the computation of $g_{N,N B}$ for any $N$ and $B$, once $g_1$ is known. 
As a result, the problem of computing generating functions for baryonic operators greatly 
simplifies and amounts to figuring out the much simpler case of one D brane, $g_1$.

From the previous chapter we learnt that the generating function $g_{1,B}$ can be given a geometric interpretation. 
$g_{1,0}$ is the character for holomorphic functions on the CY manifold \cite{Martelli:2006yb}, $g_{1,B}$, for generic $B$, 
is the character for holomorphic sections of suitable line bundles \cite{Butti:2006au}. 
%This is essentially due to the fact that we can use holomorphic surfaces to parameterize supersymmetric configurations of D3 branes wrapped on non trivial cycles on the CY horizon \cite{Mikhailov:2000ya,Beasley:2002xv}. In this construction the baryonic charge in gauge theory is related to the homology of the cycles in geometry and $g_{1,B}$ gets contribution from surfaces (zero loci of sections) belonging to a given homology class. 
Indeed formula (\ref{g1pletT}) can be alternatively interpreted as giving 
the BPS Hilbert space of states with baryonic charge $B$ and number of branes $N$ as the geometric
quantization of the classical configuration space of supersymmetric D3
branes wrapped on cycles of homology $B$, as already explained in the previous chapter.

To demonstrate this method of computation we will now turn to few simple examples 
in which one can compute the generating functions and perform some consistency checks of this proposal.

\section{Baryonic Generating Functions for the Conifold $N=1$}
\label{conifoldB}
\setall
In this chapter the conifold will be our main example.
Let us start analyzing how we can encode the global symmetries of the conifold theory in a generating function.
The gauge theory on the conifold has a global symmetry $SU(2)_1\times SU(2)_2\times U(1)_R\times U(1)_B$. It has 4 basic fields $A_{1,2}$ and $B_{1,2}$ that transform under these symmetries according to the following table:

\begin{table}[htdp]
\caption{Global charges for the basic fields of the quiver gauge theory on the D brane probing the Conifold.}
\begin{center}
\begin{tabular}{|c||c c|c c|c|c||c|}
%\begin{tabular}{|1||1 1||1 1||1|1||1|}
\hline
 & \multicolumn{2} {c|} {$SU(2)_1$} & \multicolumn{2} {c|} {$SU(2)_2$} & $U(1)_R$ & $U(1)_B$ & monomial\\ \hline
\cline{2-5}
& $j_1$ & $m_1$ & $j_2$ & $m_2$&&&\\ \hline \hline
$A_1$ & $\frac{1}{2}$ & $+\frac{1}{2}$ & 0 & 0 & $\frac{1}{2}$ & 1& $t_1 x$\\ \hline
$A_2$ & $\frac{1}{2}$ & $-\frac{1}{2}$ & 0 &0 & $\frac{1}{2}$ & 1& $\frac{t_1}{x}$ \\ \hline
$B_1$ & 0 & 0 & $\frac{1}{2}$ & $+\frac{1}{2}$ & $\frac{1}{2}$ & -1& $t_2 y$ \\ \hline
$B_2$ & 0 & 0 & $\frac{1}{2}$ & $-\frac{1}{2}$ & $\frac{1}{2}$ & -1& $\frac{t_2}{y}$ \\ \hline
\end{tabular}
\end{center}
\label{globalconifold}
\end{table}

The last column represents the corresponding monomial in the generating function for BPS operators in the chiral ring. $t_1$ is the chemical potential for the number of $A$ fields, $t_2$ is the chemical potential for the number of $B$ fields, $x$ is the chemical potential for the Cartan generator of $SU(2)_1$ and $y$ is the chemical potential for the Cartan generator of $SU(2)_2$. A generic operator that carries a spin $j_1$ with weight $m_1$ under $SU(2)_1$, spin $j_2$ with weight $m_2$ under $SU(2)_2$ has $2j_1$ $A$'s, $2j_2$ $B$'s and therefore will be represented by the monomial

\begin{equation}
t_1^{2j_1} x^{2m_1} t_2^{2j_2} y^{2m_2}.
\label{monom}
\end{equation}

We can also keep track of the R-charge and the baryonic charge $B$ by introducing chemical potentials $t$ and $b$, respectively. With this notation we have $t_1=t b$ and $t_2 = \frac{t}{b}$, and a generic operator represented by the monomial (\ref{monom}) is given by

\begin{equation}
t^{2j_1+2j_2} x^{2m_1} b^{2j_1-2j_2} y^{2m_2},
%\label{monom}
\end{equation}
and carries $R=j_1+j_2$ and $B=2(j_1-j_2)$.
With this notation we can proceed and write generating functions which count BPS operators in the chiral ring of the conifold theory.

%\subsection{$N=1$ for the Conifold}
%\label{N1}

Since the superpotential for the $N=1$ theory is $W=0$, it is natural to expect that the generating function is freely generated by the 4 basic fields of the conifold gauge theory and it takes the form

\begin{equation}
g_1(t_1,t_2,x,y; {\cal C}) = \frac{1}{(1-t_1 x) (1-\frac{t_1}{x})(1-t_2 y) (1-\frac{t_2}{y})}.
\label{g1coniC}
\end{equation}

\begin{figure}[h!!!]
\begin{center}
\includegraphics[scale=0.55]{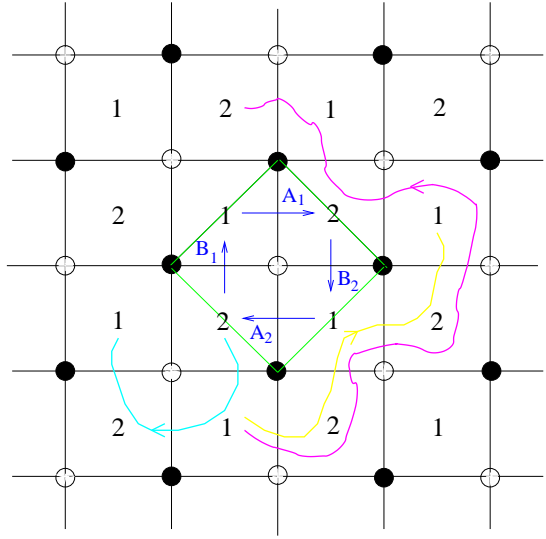} 
\caption{The dimer representation for the conifold. The fundamental domain is depicted in green. We have drawn some paths representing BPS operators in the chiral ring in the case $N=1$: in cyan $B_2A_2B_1$; in yellow $A_1B_1A_1B_1$, in magenta $A_1B_1A_1B_1A_2B_1A_2$.}
\label{coni}
\end{center}
\end{figure}

This generating function has a simple interpretation in the dimer representation \cite{Hanany:2005ve} for the conifold (see Figure \ref{coni}). Recall that the tiling for the conifold gets the form of a chessboard array \cite{Franco:2005rj} with two tiles in the fundamental domain and with horizontal arrows $A_{1,2}$ going right and left respectively, and vertical arrows $B_{1,2}$ going up and down respectively. A generic BPS operator in the chiral ring for $N=1$ will be represented by an open path or possibly by many open paths in the tiling and the function (\ref{g1coniC}) represents all such open paths.

The dimension of the moduli space for the $N=1$ theory is 4, as can be seen from the behavior near $t_i =1$, where one finds the generating function for $\mathbb{C}^4$: 

\begin{equation}
g_1(t_1=t,t_2=t,x=1,y=1; {\cal C}) = \frac{1}{(1-t)^4}.
%\label{g1coni}
\end{equation}

This is natural to expect, as this theory has 4 fundamental fields, a global symmetry of rank 4, and a vanishing superpotential, leading to a freely generated chiral ring. We will see other examples in which these conditions are not met and, as a result, the behavior is different.

It is useful to rewrite equation (\ref{g1coniC}) in terms of the baryonic and R-charge chemical potentials.

\begin{equation}
g_1(t,b,x,y; {\cal C}) = \frac{1}{(1-t b x) (1-\frac{t b}{x})(1- \frac {t y}{b}) (1-\frac{t}{b y})},
\label{g1conitbxy}
\end{equation}
and to expand $g_1$ in a  ``generalized'' Laurent expansion

\begin{equation}
g_1(t,b,x,y; {\cal C}) = \sum_{B=-\infty}^\infty b^B g_{1,B}(t,x,y; {\cal C}),
%\label{g1coni}
\end{equation}
where $g_{1,B}(t,x,y; {\cal C})$ is the generating function for BPS operators at fixed number 
of branes with $N=1$ and fixed baryonic charge $B$. The expression (\ref{g1conitbxy}) is exactly the generating function 
(\ref{congeom1}) we wrote in the previous chapter, using the homogeneous coordinates of the conifold. $g_{1,B}(t,x,y; {\cal C})$ 
can be computed using the inversion formula

\begin{equation}
g_{1,B}(t,x,y; {\cal C}) = \frac{1}{2\pi i} \oint \frac {db} {b^{B+1}} g_1(t,b,x,y; {\cal C})\equiv \frac{1}{2\pi i} \oint db I ,
\label{res1C}
\end{equation}
with a careful evaluation of the contour integral for positive and negative values of the baryonic charge $B$. We have denoted for simplicity the integrand to be $I$. For $B\ge0$ the contribution of the contour integral comes from the positive powers of the poles for $b$ and we end up evaluating the two residues 

\begin{equation}
B\ge0 : - res|_{b=\frac{1}{t x}} I - res|_{b=\frac{x}{t}} I, \nonumber
\label{res2}
\end{equation}

For $B\le0$ the contribution of the contour integral comes from the negative powers of the poles for $b$ and we end up evaluating the two residues 

\begin{equation}
B\le0 : res|_{b=\frac{t}{y}} I + res|_{b= t y} I .
\label{res3}
\end{equation}

Collecting these expressions together we get the generating functions for fixed baryonic number

\begin{eqnarray}
g_{1,B\ge0}(t,x,y; {\cal C}) &=& \frac{t^B x^{B} } { (1 - \frac{1}{x^2}) (1-t^2 x y)  (1-\frac{t^2 x}{y}) }+ \frac{t^B x^{-B}} { (1 - x^2) (1-\frac{t^2 y} {x})  (1-\frac{t^2}{x y}) } , \nonumber \\ \nonumber
g_{1,B\le0}(t,x,y; {\cal C}) &=& \frac{t^{-B} y^{-B} } { (1-\frac{1}{y^2}) (1-t^2 x y)  (1-\frac{t^2 y}{x}) }+ \frac{t^{-B} y^{B}} { (1 - y^2) (1-\frac{t^2 x} {y})  (1-\frac{t^2}{y x}) } .\\
\label{resconF}
\end{eqnarray}

Observe that these are exactly the formulae (\ref{g1Bgeom}) we already obtained in section \ref{geometricamusement} 
just applying the index theorems to the generic line bundle associated to the baryonic charge $B$.

Each of these equations reflects the Weyl symmetry of each $SU(2)$ global symmetry, 
acting as $x\leftrightarrow\frac{1}{x}$, $y\leftrightarrow\frac{1}{y}$, respectively. 
Furthermore, under the map of $B\leftrightarrow-B$ and $x\leftrightarrow y$ these equations are invariant, 
reflecting the fact that the two $SU(2)$ global symmetries are exchanged when the baryon number reverses sign.

\begin{figure}[h!!!]
\begin{center}
\includegraphics[scale=0.65]{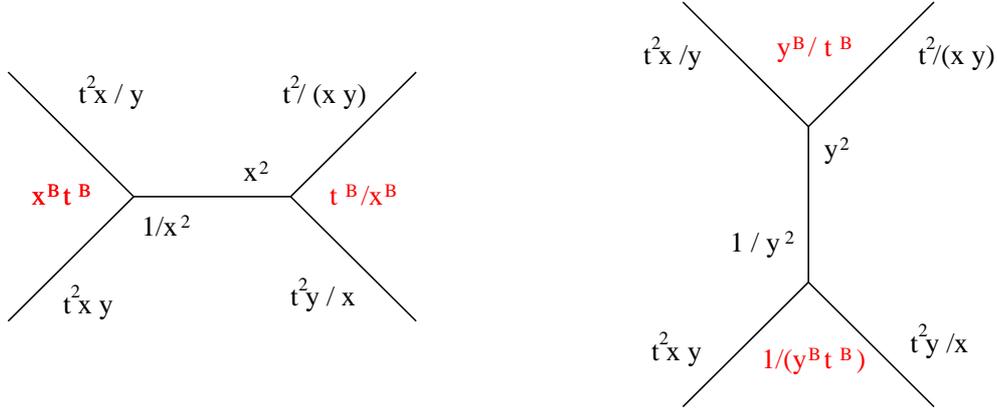} 
\caption{The $(p,q)$ webs for the two possible resolution of the conifold. 
The left figure corresponds to positive baryon number, $B>0$, 
while the right figure corresponds to negative baryon number, $B<0$.
The vertices correspond to the fixed points of the torus action on the 
resolution.}
\label{conifoldlocalization2}
\end{center}
\end{figure}
The equations for $g_{1,B}$ take the general form of equation (\ref{sumzd}) in the previous chapter. It is therefore suggestive that the formulas for $g_{1,B}$ will always be of this form and will come from localization, giving rise to contributions from the vertices of the $(p,q)$ web which are the fixed points of the torus action. Figure \ref{conifoldlocalization2} shows the contribution of each leg and each vertex in the $(p,q)$ web of the conifold theory. The left (right) vertex gives rise to the first (second) contribution for $g_{1,B\ge0}$, respectively, while the bottom (top) vertex gives rise to the first (second) contribution for $g_{1,B\le0}$, respectively. 
%A more detailed discussion of the localization formula for the conifold is given in Section \ref{geometricamusement}.

We can ignore the $SU(2)$ quantum numbers in Equations (\ref{resconF}) by setting $x=y=1$ and get

\begin{equation}
g_{1,B}(t,1,1; {\cal C}) = \frac{t^{ |B| } (1+t^2)} { (1- t^2)^3}+ \frac{ |B| t^{ |B| } } { (1- t^2 )^2 } ,
\label{rescon11}
\end{equation}
which indeed for the mesonic generating function, $B=0$, coincides with Equation (\ref{Hcon}):
\begin{equation}
g_{1,0}(t,1,1; {\cal C}) = \frac{ (1+t^2)} { (1- t^2)^3} ,
\label{rescon11}
\end{equation}
when $t^2 \rightarrow t $, and generalizes it to any baryonic number.

\subsection{Relation to Dimers and Integer Lattices}\label{pyramid}

It is instructive to draw the lattice of charges that the generating functions for fixed baryon numbers represent. Taking a power expansion of the form

\begin{equation}
g_{1,B}(t,x,y; {\cal C}) = \sum_{R, m_1, m_2} d_{R, B, m_1, m_2}t^{2R}x^{2m_1}y^{2m_2} ,
%\label{g1coni}
\end{equation}
we find that the coefficients $d_{R, B, m_1, m_2}$ are either $0$ or $1$. As such they form some kind 
of a fermionic condition on an occupation of the lattice point given by the three integer coordinates 
$2R, 2m_1, 2m_2$ (recall that $R, m_1, m_2$ admit half integral values and therefore twice their value is an integer). 
This phenomenon of fermionic exclusion was already observed for mesonic BPS gauge invariant operators 
in \cite{Benvenuti:2006qr} and persists for the case of baryonic BPS operators  \cite{Butti:2006au}. It is expected to persist 
for any singular toric non-compact CY manifold $\cX$. In contrast, for the non-toric case, although there exist a 
lattice structure related to the generating function, the fermionic conditions seems to be lost \cite{Butti:2006nk}. 

Furthermore this lattice of points forms an ``integral conical pyramid" 
which is given by the intersection of two objects: 
A) a conical pyramid with a rectangular base, and B) the three dimensional integral lattice. 
Let us see this in some detail for several cases. Let us set $q=t^2$.
With this definition the power of $q$ gives the $R$ charge of the corresponding operator. 
We first look at $B=0$ and expand equation (\ref{resconF}) up to order $q^4$. 
We then write down a matrix which represents the two dimensional lattice in the $x,y$ coordinates.

\begin{equation}
\left(
\begin{array}{lllllllll}
 q^4 & 0 & q^4 & 0 & q^4 & 0 & q^4 & 0 & q^4 \\
 0 & q^3 & 0 & q^3 & 0 & q^3 & 0 & q^3 & 0 \\
 q^4 & 0 & q^4+q^2 & 0 & q^4+q^2 & 0 & q^4+q^2 & 0 & q^4
   \\
 0 & q^3 & 0 & q^3+q & 0 & q^3+q & 0 & q^3 & 0 \\
 q^4 & 0 & q^4+q^2 & 0 & q^4+q^2+1 & 0 & q^4+q^2 & 0 & q^4
   \\
 0 & q^3 & 0 & q^3+q & 0 & q^3+q & 0 & q^3 & 0 \\
 q^4 & 0 & q^4+q^2 & 0 & q^4+q^2 & 0 & q^4+q^2 & 0 & q^4
   \\
 0 & q^3 & 0 & q^3 & 0 & q^3 & 0 & q^3 & 0 \\
 q^4 & 0 & q^4 & 0 & q^4 & 0 & q^4 & 0 & q^4
\end{array}
\right) .
\end{equation}

\begin{figure}[h!!!]
\begin{center}
\includegraphics[scale=0.45]{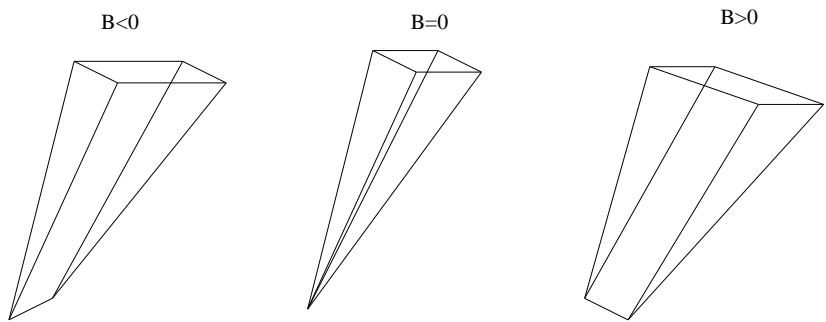} 
\caption{Examples of the ``integral conical pyramid''s for the conifold in the cases $B<0$, $B=0$, $B>0$.}\label{bariflop}
\end{center}
\end{figure}

The point in the middle contains $1=q^0$ and represents the identity operator. 
The 4 points around it contain a single power of $q$ and represent the 4 operators, 
$A_i B_j, i=1,2, j=1,2$, corresponding to the spin $(\frac{1}{2},\frac{1}{2})$ 
representation of $SU(2)_1\times SU(2)_2$. These points form an integral square of size 1. 
Next there are 9 points with $q^2$ that form a bigger integral square of size 2. These 9 
points represent operators of the form $ABAB$ where we omit the indices but due to F term 
constraints we have 9 instead of 16 and transform now in the spin $(1,1)$ representation. 
Next 16 points in $q^3$ represent the operators $(AB)^3$ forming an integral square of size 3, etc. 
The generic case of $q^n$ will have $n^2$ points, representing the operators $(AB)^n$ which transform in the spin
$(\frac{n}{2},\frac{n}{2})$ representation, forming an integral square of size $n$. 
In fact this lattice is the weight lattice of the $SU(2)_1\times SU(2)_2$ irreducible representations of equal spin. 
We thus see that the lattice points are in correspondence with chiral operators in the $N=1, B=0$ theory and form a 
three dimensional lattice in the $q,x,y$ plane. Note that there are 4 diagonal lines which form the boundaries 
of this three dimensional lattice. These diagonals coincide with the $(p,q)$ web of the conifold theory, 
at the point where the size of the two cycle shrinks to zero, as in the middle of Figure \ref{bariflop}. 
We will soon identify the size of the two-cycle with the baryon number, 
or more precisely its absolute value. It is curious to note that this size is quantized, 
consistent with observations in \cite{Gopakumar:1998ki, Iqbal:2003ds}.

Let us turn to the case of baryonic charge $B=1$. We expand equation (\ref{resconF}) to order $q^4$ and write the matrix for the two dimensional lattice in $x,y$,

\begin{equation}
q^{\frac{1}{2}}
\left(
\begin{array}{lllllllllll}
 q^4 & 0 & q^4 & 0 & q^4 & 0 & q^4 & 0 & q^4 & 0 & q^4 \\
 0 & q^3 & 0 & q^3 & 0 & q^3 & 0 & q^3 & 0 & q^3 & 0 \\
 q^4 & 0 & q^4+q^2 & 0 & q^4+q^2 & 0 & q^4+q^2 & 0 &
   q^4+q^2 & 0 & q^4 \\
 0 & q^3 & 0 & q^3+q & 0 & q^3+q & 0 & q^3+q & 0 & q^3 & 0
   \\
 q^4 & 0 & q^4+q^2 & 0 & q^4+q^2+1 & 0 & q^4+q^2+1 & 0 &
   q^4+q^2 & 0 & q^4 \\
 0 & q^3 & 0 & q^3+q & 0 & q^3+q & 0 & q^3+q & 0 & q^3 & 0
   \\
 q^4 & 0 & q^4+q^2 & 0 & q^4+q^2 & 0 & q^4+q^2 & 0 &
   q^4+q^2 & 0 & q^4 \\
 0 & q^3 & 0 & q^3 & 0 & q^3 & 0 & q^3 & 0 & q^3 & 0 \\
 q^4 & 0 & q^4 & 0 & q^4 & 0 & q^4 & 0 & q^4 & 0 & q^4
\end{array}
\right) .
\end{equation}

Note the factor of $q^{\frac{1}{2}}$ in front of this expression, indicating that there is one more $A$ field, carrying an $R$ charge $\frac{1}{2}$. For this case there are two lowest order lattice points with
$q^{\frac{1}{2}}$, corresponding to the operators $A_1, A_2$. These two points transform in the spin $(\frac{1}{2},0)$ representation and form an integral rectangle of sizes $2\times1$. At order $q^{\frac{3}{2}}$ we find 6 lattice points transforming in the spin $(1,\frac{1}{2})$ representation and forming an integral  rectangle of sizes $3\times2$, etc. At order $q^{n+\frac{1}{2}}$ we find $(n+2)(n+1)$ lattice points transforming in the spin $(\frac{n+1}{2},\frac{n}{2})$ and forming an integral rectangle of sizes $(n+2)\times(n+1)$. In summary the lattice we get is the weight lattice of all $SU(2)_1\times SU(2)_2$ representations with a difference of spin $\frac{1}{2}$. The 4 diagonal lines which form the boundaries of this three dimensional lattice now coincide with the $(p,q)$ web of the conifold theory, at a point where the size of the two cycle is non-zero, as in the right of Figure \ref{bariflop}.

As a last example let us turn to the case of baryonic charge $B=-2$. We expand equation (\ref{resconF}) to order $q^4$ and write the matrix for the two dimensional lattice in $x,y$,

\begin{equation}
\left(
\begin{array}{lllllll}
 q^4 & 0 & q^4 & 0 & q^4 & 0 & q^4 \\
 0 & q^3 & 0 & q^3 & 0 & q^3 & 0 \\
 q^4 & 0 & q^4+q^2 & 0 & q^4+q^2 & 0 & q^4 \\
 0 & q^3 & 0 & q^3+q & 0 & q^3 & 0 \\
 q^4 & 0 & q^4+q^2 & 0 & q^4+q^2 & 0 & q^4 \\
 0 & q^3 & 0 & q^3+q & 0 & q^3 & 0 \\
 q^4 & 0 & q^4+q^2 & 0 & q^4+q^2 & 0 & q^4 \\
 0 & q^3 & 0 & q^3+q & 0 & q^3 & 0 \\
 q^4 & 0 & q^4+q^2 & 0 & q^4+q^2 & 0 & q^4 \\
 0 & q^3 & 0 & q^3 & 0 & q^3 & 0 \\
 q^4 & 0 & q^4 & 0 & q^4 & 0 & q^4
\end{array}
\right) .
\end{equation}

In this case there are two more $B$ fields than $A$ fields and hence there are three lowest order lattice points with $q^{{1}}$, corresponding to the operators $B_1B_1, B_1B_2, B_2B_2$. These three points transform in the spin $(0,1)$ representation and form an integral rectangle of sizes $1\times3$. At order $q^{{2}}$ we find 8 lattice points transforming in the spin $(\frac{1}{2},\frac{3}{2})$ representation and forming an integral  rectangle of sizes $2\times4$, etc. At order $q^{n+1}$ we find $(n+1)(n+3)$ lattice points transforming in the spin $(\frac{n}{2},\frac{n}{2}+1)$ and forming an integral rectangle of sizes $(n+1)\times(n+3)$. In summary the lattice we get is the weight lattice of all $SU(2)_1\times SU(2)_2$ representations with a negative difference of spin $1$. The 4 diagonal lines which form the boundaries of this three dimensional lattice now coincide with the $(p,q)$ web of the conifold theory after a flop transition, at a point where the size of the two cy!
 cle is non-zero, as in the left of Figure \ref{bariflop}.

%Examples of this ``integral conical pyramid" are given in Figure \ref{bariflop}.

We summarize this discussion by the observation that the boundaries of the conical pyramid are identified with the $(p,q)$ web of the conifold. This $(p,q)$ web can be resolved in two phases, separated by a flop transition. Each of these phases is characterized by the sign of the baryonic charge $B$ and the flop transition appears at $B=0$. In this sense one can identify the baryonic charge with the K\"ahler modulus for the conifold.

The relation to dimers then becomes natural. Let us fix a baryonic charge $B$. Take the tiling of the conifold, fix a reference tile and consider all open paths that begin with this tile, consists of $n_1$ arrows of type $A$ (horizontal) and $n_2$ arrows of type $B$ (vertical) such that the numbers $n_1$ and $n_2$ satisfy the condition $n_1-n_2=B$. For a given endpoint there may be more than one path which connects it to the reference tile. All such paths are equal by imposing the F-term conditions. See \cite{Hanany:2006nm} for a detailed explanation and \cite{Butti:2006hc,Butti:2006au} for the applications to the mesonic and baryonic BPS operators. The collection of endpoints then forms the two dimensional projection of the three dimensional ``conical pyramid'' lattice discussed above. What we have just seen in this section 
can be seen as the dimer realization of the geometric localization techniques used in the previous chapter.

\section{General Number of Branes, $N$, for the Conifold}\label{conifoldNN}

In the previous section we saw the generating function for the case of $N=1$ for the conifold. 
In this section we will develop the general $N$ case.
This is done using the Plethystic Exponential. Recall that in the case of baryon number $B=0$, 
namely for mesonic generating functions, \cite{Benvenuti:2006qr} the knowledge of the generating 
function for $N=1$ is enough to compute the generating function for any $N$. This is essentially 
due to the fact that the operators for finite $N$ are symmetric functions of the operators for 
$N=1$ and this is precisely the role which is played by the plethystic exponential -- to take a 
generating function for a set of operators and count all possible symmetric functions of it.

For the present case, in which the baryon number is non-zero, it turns out that the procedure 
is not too different than the mesonic case. One needs to have the knowledge of a single generating 
function, $g_{1,B}$, for one D brane, $N=1$ and for a fixed baryon number $B$, 
and this information is enough to compute all generating functions for any number of D 
branes and for a fixed baryonic number \cite{Butti:2006au} (see section \ref{geomBZ}).
Indeed the field theory operators with fixed baryonic numbers, when the rank of the gauge group factor is N,
are obtained contracting N strings of elementary fields of the $N=1$ case with two epsilons, 
and hence they are symmetric in the $N=1$ operators.

We can summarize this information by writing an expression for the generating 
function of any number of branes and any baryonic number,

\begin{eqnarray}
\nonumber
g(\nu; t,b,x,y; {\cal C})
&=& \sum_{N=0}^\infty \nu^N g_N(t, b, x, y; {\cal C}) \\
\label{grandcancon}
%g(\nu; t,b,x,y; {\cal C})
&=& \sum_{B=-\infty}^\infty b^B \exp\biggl(\sum_{k=1}^\infty \frac{\nu^k}{k}g_{1,B}(t^k, x^k, y^k; {\cal C}) \biggr) \\
%g(\nu; t,b,x,y; {\cal C}) 
&=& \sum_{B=-\infty}^\infty b^B \exp\biggl(\sum_{k=1}^\infty \frac{\nu^k}{k}\frac{1}{2\pi i} \oint \frac {db'} {b'^{B+1}} g_1(t^k,b'^k,x^k,y^k; {\cal C}) \biggr) ,
\nonumber
\label{PEcon}
\end{eqnarray}
where the first equality indicates that this is a generating function for fixed number of branes, the second equality indicates that we are summing over all contributions of fixed baryonic numbers and that each contribution is the plethystic exponential of the generating function for one D brane and for fixed baryonic charge. The third equality expresses the generating function in terms of a multi-contour integral of the generating function for one D brane.
For completeness, using Equation (\ref{g1conitbxy}) and Equation (\ref{resconF}) we write down the generating function as explicit as possible,

\begin{eqnarray}
& &g(\nu; t,b,x,y; {\cal C}) = \\
\nonumber
\label{grandcanconex}
&=& \sum_{B=-\infty}^\infty b^B \exp\biggl(\sum_{k=1}^\infty \frac{\nu^k}{k}\frac{1}{2\pi i} \oint \frac {db'} {b'^{B+1}} \frac{1}{(1-t^k b'^k x^k) (1-\frac{t^k b'^k}{x^k})(1- \frac {t^k y^k}{b'^k}) (1-\frac{t^k}{b'^k y^k})}  \biggr) \\
\nonumber
&=& \sum_{B=0}^\infty b^B \exp\biggl(\sum_{k=1}^\infty
\frac{\nu^k t^{kB} x^{kB} } {k (1 - \frac{1}{x^{2k}}) (1-t^{2k} x^k y^k)  (1-\frac{t^{2k} x^k}{y^k}) }+ \frac{\nu^k t^{kB} x^{-kB}} { k (1 - x^{2k}) (1-\frac{t^{2k} y^k} {x^k})  (1-\frac{t^{2k}}{x^k y^k}) } \biggr) \\ \nonumber 
&+&\sum_{B=1}^{\infty} b^{-B} \exp \biggl ( \sum_{k=1}^\infty \frac{\nu^k t^{k B} y^{ k B} } { k (1-\frac{1}{y^{2k}}) (1-t^{2k} x^k y^k)  (1-\frac{t^{2k} y^k}{x^k}) }+ \frac{ \nu^k t^{k B} y^{- k B}} { k (1 - y^{2 k}) (1-\frac{t^{2 k} x^k} {y^k})  (1-\frac{t^{2 k}}{y^k x^k}) } \biggr) .
\nonumber
\end{eqnarray}

If we further ignore the $SU(2)$ quantum numbers and set $x=y=1$ we get

\begin{eqnarray}
g(\nu; t,b,1,1; {\cal C})
\label{grandcancon11}
&=& \exp\biggl(\sum_{k=1}^\infty
\frac{ \nu^k (1+t^{2k})} { k (1- t^{2k})^3} \biggr) \\ &+& \sum_{B=1}^\infty ( b^B + b^{-B} )\exp\biggl(\sum_{k=1}^\infty
\frac{ \nu^k t^{ k B } (1+t^{2k})} { k (1- t^{2k})^3}+ \frac{ B \nu^k t^{ k B } } { k (1- t^{2 k} )^2 }\biggr) .
\nonumber
\end{eqnarray}

The exponential terms have an equivalent representation as infinite products and take the form

\begin{eqnarray}
g(\nu; t,b,1,1; {\cal C})
%\label{grandcancon11}
= \prod_{n=0}^\infty \frac{1}{(1-\nu t^{2n})^{(n+1)^2}}
+ \sum_{B=1}^\infty ( b^B + b^{-B} ) \prod_{n=0}^\infty \frac {1} {(1-\nu t^{2n+B})^{(n+1)(n+B+1)} } .
\nonumber
\end{eqnarray}

\subsection{$N=2$ for the Conifold}
\label{Nis2}

Taking the expansion of Equation (\ref{PEcon}) to second order in the chemical potential $\nu$ for the number of D branes, we find

\begin{eqnarray}
g_2(t, b, x, y; {\cal C}) &=& \sum_{B=-\infty}^\infty b^B \frac{1}{2}\biggl[\frac{1}{2\pi i} \oint \frac {db'} {b'^{B+1}} g_1(t,b',x,y; {\cal C}) \biggr]^2 \\ \nonumber
&+& \sum_{B=-\infty}^\infty b^B \frac{1}{2}\frac{1}{2\pi i} \oint \frac {db'} {b'^{B+1}} g_1(t^2,b'^2,x^2,y^2; {\cal C}) = F_1 + F_2 ,
%\label{g1coni}
\end{eqnarray}
and we divided the contribution into two terms, $F_1$ and $F_2$. We next use the identity

\begin{eqnarray}
\sum_{B=-\infty}^\infty b^B \frac{1}{2\pi i} \oint \frac {db'} {b'^{B+1}} = \delta(b-b').
%\label{g1coni}
\end{eqnarray}
and evaluate $F_2$ to be

\begin{eqnarray}
F_2=\frac{1}{2}g_1(t^2,b^2,x^2,y^2; {\cal C}) .
%\label{g1coni}
\end{eqnarray}

$F_1$ is slightly more involved since it contains two contour integrals but still is relatively easy to evaluate,

\begin{eqnarray}
\nonumber
F_1 &=& \sum_{B=-\infty}^\infty b^B \frac{1}{2}\frac{1}{2\pi i} \oint \frac {db'} {b'^{B+1}} g_1(t,b',x,y; {\cal C}) \frac{1}{2\pi i} \oint \frac {db''} {b''^{B+1}} g_1(t,b'',x,y; {\cal C}) \\ \nonumber
&=& \sum_{B=-\infty}^\infty b^B \frac{1}{2}\frac{1}{(2\pi i)^2} \oint \frac {db'} {b'^{B+1}} \oint \frac {db''} {b''^{B+1}} g_1(t,b',x,y; {\cal C})  g_1(t,b'',x,y; {\cal C}) \\ \nonumber
&=& \sum_{B=-\infty}^\infty b^B \frac{1}{2}\frac{1}{(2\pi i)^2} \oint \oint \frac {ds'} {s'^{B+1}} \frac {ds} {s} g_1(t,s s',x,y; {\cal C})  g_1(t,\frac{s'}{s},x,y; {\cal C}) \\ \nonumber
&=& \frac{1}{2}\frac{1}{(2\pi i)} \oint \frac {ds} {s} g_1(t,b s,x,y; {\cal C})  g_1(t,\frac{b}{s},x,y; {\cal C}) . \\ \nonumber
%\label{g1coni}
\end{eqnarray}

We can now collect all the terms and get an expression for $N=2$,
%which is in fact correct for any non-compact CY manifold with one baryonic charge and not just the conifold.

\begin{equation}
g_2(t, b, x, y; {\cal C}) = \frac{1}{2}g_1(t^2,b^2,x^2,y^2; {\cal C})
+ \frac{1}{2}\frac{1}{(2\pi i)} \oint \frac {ds} {s} g_1(t,b s,x,y; {\cal C})  g_1(t,\frac{b}{s},x,y; {\cal C}) .
\label{g2coniC}
\end{equation}

We can recall the chemical potentials for counting the $A$ and $B$ fields and rewrite

\begin{equation}
g_2(t_1, t_2, x, y; {\cal C}) = \frac{1}{2}g_1(t_1^2, t_2^2, x^2, y^2; {\cal C})
+ \frac{1}{2}\frac{1}{(2\pi i)} \oint \frac {ds} {s} g_1(t_1 s,\frac{t_2}{s}, x, y; {\cal C})
g_1(\frac{t_1}{s},t_2 s, x, y; {\cal C}) .
%\label{g1coni}
\end{equation}

Let us evaluate the second term and write it explicitly using equation (\ref{g1coniC}).

\begin{equation}
\frac{1}{2}\frac{1}{(2\pi i)} \oint \frac {ds} {s}
\frac{1}{(1-t_1s x) (1-\frac{t_1s}{x})(1-\frac{t_2 y}{s}) (1-\frac{t_2}{s y})}
\frac{1}{(1-\frac{t_1 x}{s}) (1-\frac{t_1}{s x})(1-t_2 s y) (1-\frac{t_2 s}{y})} .
%\label{g1coni}
\end{equation}

The residue integral now gets contributions from 4 different points at

\begin{equation}
s=t_1 x, \qquad s=\frac{t_1}{x}, \qquad s=t_2 y, \qquad s=\frac{t_2}{y}.
%\label{g1coni}
\end{equation}

The computations are a bit lengthy but after some work we get an expression for the generating function for BPS operators on N=2 D branes probing the conifold. If we ignore the $SU(2)$ weights by setting $x=y=1$, we obtain:

\comment{
$$
g_2(t_1, t_2, x, y; {\cal C}) =\,\,\qquad\qquad\qquad\qquad\qquad\qquad\qquad\qquad\qquad\qquad\qquad\qquad\qquad\qquad$$
$$
\frac
{1-t_1^3t_2^3(1-t_1^2)(1-t_2)^2\chi_{\frac{1}{2}}(x)\chi_{\frac{1}{2}}(y)-t_1^2t_2^2[(1-t_1^2)t_2^2\chi_1(x)+(1-t_2^2)t_1^2\chi_1(y)]+t_1^4t_2^4(1-t_1^2-t_2^2)}{(1-t_1^2)(1-t_2^2)( 1 - t_1^2 x^2 )( 1 - t_2^2 y^2 )(1-\frac{t_1^2}{x^2})(1-\frac{t_2^2}{y^2})(1-t_1t_2xy)(1-\frac{t_1t_2x}{y})(1-\frac{t_1t_2y}{x})(1-\frac{t_1t_2}{xy})} . \nonumber
%\label{g1coni}
$$

%\begin{eqnarray}
%&&g_2(t_1, t_2, x, y; {\cal C})  \nonumber\\ 
%&=&\frac
%{1-t_1^3t_2^3(1-t_1^2)(1-t_2)^2\chi_{\frac{1}{2}}(x)\chi_{\frac{1}{2}}(y)-t_1^2t_2^2[(1-t_1^2)t_2^2\chi_1(x)+(1-t_2^2)t_1^2\chi_1(y)]+t_1^4t_2^4(1-t_1^2-t_2^2)}{(1-t_1^2)(1-t_2^2)( 1 - t_1^2 x^2 )( 1 - t_2^2 y^2 )(1-\frac{t_1^2}{x^2})(1-\frac{t_2^2}{y^2})(1-t_1t_2xy)(1-\frac{t_1t_2x}{y})(1-\frac{t_1t_2y}{x})(1-\frac{t_1t_2}{xy})} . \nonumber
%\label{g1coni}
%\end{eqnarray}
}

%We can ignore the $SU(2)$ weights by setting $x=y=1$ and by using $\chi_j(1) = 2j+1$,

\begin{equation}\label{N2con}
g_2(t_1, t_2, 1, 1; {\cal C}) = \frac{1+t_1t_2+t_1^2t_2^2[1-3(t_1^2+t_2^2)]+t_1^3t_2^3(t_1^2+t_2^2-3)+4t_1^4t_2^4}
{(1-t_1^2)^3(1-t_1t_2)^3(1-t_2^2)^3} .
%\label{g1coni}
\end{equation}

Similar expressions can be obtained for $N=3$ 

\begin{equation}\label{conifoldN3}
g_3(t_1,t_2;{\cal C})=\frac{F(t_1,t_2)}{(1-t_1^3)^4(1 - t_1 t_2)^3 (1-t_1^2 t_2^2)^3(1- t_2^3)^4} ,
\end{equation}
\begin{eqnarray}\label{conifoldN3F}
 F(t_1,t_2)&=& 1 +  t_1^{15}
    t_2^9 + + 3 t_1^{14} t_2^8 (-1 + 2 t_2^3) + t_1 (t_2 + 2 t_2^4)
 + t_1^2(7 t_2^2 - 4 t_2^5) +
        t_1^3(7 t_2^3 - 10 t_2^6) \nonumber\\
& +&  3 t_1^{13} t_2^7(-1 + 2
t_2^6)
+
 t_1^{12} t_2^6(9 - 34 t_2^3 + 22 t_2^6)
  +  t_1^7 t_2^4(-22 + 35 t_2^3 + 8 t_2^6 - 3 t_2^9)
\nonumber\\
& -& t_1^8
    t_2^5(4 - 35 t_2^3 + 10 t_2^6 + 3 t_2^9)
+ t_1^6
          t_2^3(-10 - t_2^3 + 4 t_2^6 + 9 t_2^9) + t_1^{10}(6
        t_2^4 + 8 t_2^7 - 26 t_2^{10})
\nonumber\\
 &+&
        t_1^9 t_2^6(4 + 31 t_2^3 - 34 t_2^6 +
        t_2^9)
+ 2 t_1^{11} t_2^5(3 - 5 t_2^3 -
    7 t_2^6 + 3 t_2^9)
\nonumber\\ &+& 2
        t_1^4(t_2 + t_2^4 - 11 t_2^7 + 3 t_2^{10}) + 2 t_1^5 t_2^2(-2 - 5 t_2^3 - 2
          t_2^6 + 3 t_2^9)
\end{eqnarray}
and greater values of $N$. 

The properties of these generating functions for different values of $N$ are 
discussed in Section \ref{PLlog}.

\section{The PE, the Baryonic Chiral Ring and the Geometric Quantization}\label{PEquant}

It should be clear from the previous sections that the knowledge 
of the generating function $g_{1,B}(q;\mathcal{C})$ for $N=1$ and fixed baryonic number 
$B$ is enough to compute the generating function for any 
$N$ and $B$. Intuitively this is essentially due to the fact that the 
chiral BPS operators for finite $N$ are symmetric functions of the operators for $N=1$.
 The plethystic exponential has the role of taking a generating function 
for a set of operators and of counting all possible symmetric functions of them, 
therefore it allows to pass from $g_{1,B}(q;\mathcal{C})$ to $g_{N, N B}(q;\mathcal{C})$. 
In this section we want to explain in detail how this procedure works for the simple case of 
the conifold for baryonic charge $N$, and the geometric realization of this procedure.

%The gauge theory dual to the conifold singularity has four elementary fields $(A_1,B_1,A_2,B_2)$ which have baryonic charge $(1,-1,1,-1)$ and gauge group $SU(N) \times SU(N)$. 
In section \ref{motivations} we have seen that the generic baryonic operators with baryonic charge N is obtained 
taking N strings of the elementary fields $(A_1,B_1,A_2,B_2)$, each one having baryonic charge one: 
\begin{equation}\label{AAA}
A_{I;J}= A_{i_1}B_{j_1}...A_{i_m}B_{j_m}A_{i_{m+1}},
\end{equation}
and contract them with two epsilon symbols:

\begin{equation}
\label{genbarA}
\epsilon^1 _{p_1,...,p_N}  \epsilon_2 ^{k_1,...,k_N} (A_{I_{1};J_{1}})^{p_1}_{k_1}... (A_{I_{N};J_{N}})^{p_N}_{k_N}\, .\end{equation}

Thanks to the $\epsilon$ symbols these operators are completely symmetric in the exchange of the $A_{I;J}$'s. Now we want to understand the role of the plethystic exponential in the problem of counting all the operators of the form (\ref{genbarA}). 
%Once we have constructed the generating function $g_{1,1}(q;\mathcal{C})$ counting all the possible chiral operators of the form (\ref{AAA}) looking at (\ref{genbarA}) 
It is clear that the generating function counting all the possible BPS operators with baryonic charge N 
is the one that counts all the possible symmetric products of N elements of the form (\ref{AAA}). The BPS operators (\ref{AAA}) are in correspondence with points inside an integral conical pyramid (see Section \ref{pyramid}), that we denote $P_{B=1}$. Therefore counting the operators (\ref{AAA}) is the same as counting the integer points $m=(m_1,m_2,m_3) \in \mathbb{Z}^3$ inside $P_{B=1}$, and the counting procedure for generic $N$ is simply achieved by:
\begin{equation}\label{PLnu}
\prod_{m \in P_{B=1}} \frac{1}{1-\nu q^m}= 
\sum_{N = 0}^{\infty}\nu^N \hbox{ (all symmetric products of $N$ elements in $P_{B=1}$) } ,
\end{equation}
where we have introduced the chemical potential $\nu$ for the number of branes, and $q=(q_1,q_2,q_3)$ are the chemical potentials for the $T^3$ toric action.
%Then we have Taylor expanded the infinite product for small $\nu$ and weighed the symmetric products according to the number of basic chiral fields used to construct them. 
>From this expansion it is clear that the LHS of Equation (\ref{PLnu}) is the generating function for all the possible symmetric products of $N$ elements inside $g_{1,1}(q;\mathcal{C})$. 
The RHS of Equation (\ref{PLnu}) is what we write as $\sum_{N=0}^{\infty}\nu^N g_{N,N}(q;\mathcal{C})$, and it is easy to show that:

\begin{equation}\label{PLnufr}
\prod_{m \in P_{B=1}} \frac{1}{1-\nu q^m}= \exp{\Big(\sum_{k=1}^{\infty} \frac{\nu ^{k}}{k} g_{1,1}(q^k;\mathcal{C})}\Big)
\end{equation}

The RHS of Equation (\ref{PLnufr}) is the definition of the plethystic exponential, hence we have the relation:

\begin{equation}\label{PLnuex}
\hbox{PE$_{\nu}$} [ g_{1,1}(q;\mathcal{C})]= \sum_{N=0}^{\infty}\nu^N g_{N,N}(q;\mathcal{C}) .
\end{equation}

This is the physical justification of the use of the plethystic exponential in the baryonic counting problem: 
the generic BPS operator of the gauge theory in the presence of N D3 branes, 
Equation (\ref{genbarA}), is a symmetric function of the chiral operators in 
the case of $N=1$, Equation (\ref{AAA}). Hence once we know $g_{1,B}(q;\mathcal{C})$ 
we obtain $g_{N,NB}(q;\mathcal{C})$ by counting all possible symmetric functions of 
the operators in $g_{1,B}(q;\mathcal{C})$, and this is exactly the role of the plethystic exponential.

It is also possible to relate the plethystic exponential to the result of the geometric quantization 
procedure of a system of D3 branes wrapped on three cycles inside $T^{1,1}$. 
It is known that the gauge theory we have discussed so far is dual to Type IIB string theory on the 
$AdS_5 \times T^{1,1}$ supergravity background. In the geometric side the baryonic operators correspond 
to D3 branes wrapped over non trivial three cycles in $T^{1,1}$. The supersymmetric D3 brane
 wrapping three-cycles in $T^{1,1}$ are in one to one correspondence with the holomorphic surfaces 
$S$ in the real cone $C(T^{1,1})$ over $T^{1,1}$ \cite{Mikhailov:2000ya,Beasley:2002xv}. 
As explained in section \ref{geometricamusement}, we can associate to every vertex of 
the toric diagram a global homogeneous coordinate $x_i$. 
In the conifold case the homogeneous coordinates $(x_1,x_2,x_3,x_4)$ with charges $(1,-1,1,-1)$
 can be put in one-to-one correspondence with the elementary fields $(A_1,B_1,A_2,B_2)$  
which have indeed the same charges. Hence all the supersymmetric configurations of 
D3 branes wrapped in $T^{1,1}$ with $B=1$ are zero loci of homogeneous polynomials of degree one:  
\begin{eqnarray}
P_{B=1}(x_1,x_2,x_3,x_4) &\equiv&  h_1 x_1 + h_3 x_3 + \nonumber \\ 
                              & &  h_{11;2} x_1^2 x_2 + h_{13;2} x_1 x_3 x_2 + h_{33;2} x_3^2 x_2 +  \nonumber \\ 
                              & &  h_{11;4} x_1^2 x_4 + h_{13;4} x_1 x_3 x_4 + h_{33;4} x_3^2 x_4 + ...   
\end{eqnarray}
where $h_{m;n}$'s are complex numbers and parametrize the supersymmetric D3 brane embeddings.
This is a particular case of the general procedure explained in the previous chapter, and we know that the phase space 
$\mathcal{P}_{cl}$ of this classical system of D3 branes is $\mathbb{CP}^{\infty}$ 
and the $h_{m;n}$'s are its homogeneous coordinate \cite{Beasley:2002xv}; and the BPS Hilbert space, obtained
using geometric quantization procedures \cite{Wood,Beasley:2002xv} over $\mathcal{P}_{cl}$, is spanned by the degree 
$N$ polynomials in the homogeneous coordinates $h_{m;n}$, 
where $N$ is the value of the integral of the five form $F_5$ over $T^{1,1}$, 
and it correspond to the number of colors in the dual gauge theory. 
These polynomials are clearly completely symmetric in the exchange of the $h_{m;n}$'s and one can write them as the symmetric states: 
\begin{equation}
| h_{m_1;n_1}, h_{m_2;n_2},...,h_{m_N;n_N} \rangle 
\label{bpsstate}
\end{equation} 
In this geometric language every $h_{m;n}$ correspond to a section of the line bundle $\mathcal{O}(B=1)$ and the problem of counting all these sections is explained in detail in section \ref{counting}. 
Let us denote the generating function for this counting problem $Z_{1,1}(q;\mathcal{C})$, meaning that 
we are counting the sections of the line bundle $\mathcal{O}(D)=\mathcal{O}(B=1)$ over the conifold: we use the 
notation explained in the previous chapter. The geometric quantization procedure prescribes that (\ref{bpsstate}) are a basis of the Hilbert space for the BPS wrapped D3 branes, and hence of all the possible symmetric products of $N$ of the sections of $\mathcal{O}(B=1)$. As explained above, it is by now clear that the role of the plethystic exponential is to pass from the generating function of the global sections of $\mathcal{O}(B=1)$ to the generating function of their $N$-times symmetric products:
\begin{equation}
\hbox{PE$_{\nu}$}[Z_{1,1}(q;\mathcal{C})]= \sum_{N=0}^{\infty} \nu^N Z_{N,N}(q;\mathcal{C})
\end{equation}
Hence in the geometric picture the appearance of the plethystic exponential is the realization of the geometric quantization procedure over the phase space of a system of wrapped D3 branes in the internal geometry.

Clearly, there exists a precise relation between the two counting problems\cite{Beasley:2002xv}. From the identification of the elementary fields of the conifold with the homogeneous coordinates $x_i$, one has the obvious association between the homogeneous coordinates $h_{m;n}$ over the phase space and the chiral operators with $B=1$:

\begin{equation} 
h_{i_1,...,i_{m+1};j_1,...,j_m} \leftrightarrow A_{i_1}B_{j_1}...A_{i_m}B_{j_m}A_{i_{m+1}}
\end{equation}
and hence the correspondence between the BPS states in $AdS_5\times T^{1,1}$ and the baryonic operators in the dual field theory \cite{Beasley:2002xv}:  
\begin{equation}
| h_{I_1;J_1},...,h_{I_N;J_N} \rangle  \leftrightarrow \epsilon^1\epsilon_2 (A_{I_{1};J_{1}},...,A_{I_{N};J_{N}}) 
\end{equation}
Therefore the interpretation of the plethystic exponential in these counting problems is two-fold: in the gauge theory side it is the realization that the generic BPS baryonic operator is a symmetric product of some basic building blocks; in the geometric side it is the direct outcome of the geometric quantization procedure of a system of D3 branes. The two approaches are related by the $AdS/CFT$ correspondence\footnote{ In the generic toric case there exist a relation between the two counting problem, but it is more subtle, mainly due to the fact that the correspondence between the homogeneous coordinates $x_i$ and the elementary fields of the dual gauge theory is not one to one.}:
\begin{equation}\label{relgg}
g_{1,1}(q;\mathcal{C}) = Z_{1,1}(q;\mathcal{C})  \hbox{             }  \hbox{           }  \hbox{ , } \hbox{             } \hbox{             } g_{N,N}(q;\mathcal{C}) = Z_{N,N}(q;\mathcal{C})
\end{equation}

Now we understand a bit better the correspondence between the geometric and the field theory counting. 
The correspondence for general toric variety $\cX$ and general line bundle is much more complicated and we will 
explain it in the following chapter. But, for notation simplicity, we will leave from on the symbol $Z_{D,N}$ for the 
generating function and we will use just the symbol $g_{N,NB}$ both for the field theory and the geometric computations.

\section{The Structure of the Chiral Ring for the Conifold}\label{PLlog}

Now that we have under control the generating functions for the conifold for generic baryonic number 
$B$ and number of branes $N$, it is interesting to understand the structure of 
the chiral ring for different values of $N$.

For simplicity we consider the generating functions for the conifold 
$g_N(t_1,t_2;\cal{C})$ with $x=1$, $y=1$. In this way it counts only the number of fields 
$A_i$ ($t_1$) and $B_i$ ($t_2$) in the gauge invariant operators without taking into account 
their weights under the global $SU(2)$ symmetries.
Our procedure will be to take the plethystic logarithm (\hbox{PE$^{-1}$}, 
or equivalently \hbox{PL}) of the generating function $g_N(t_1,t_2;\cal{C})$. 
This operation is the inverse function of the plethystic exponential and it was defined in section \ref{HSPE}:

\begin{equation}
\hbox{PE$^{-1}$}[g_N(t_1,t_2;{\cal C})] \equiv \sum_k^{\infty} \frac{\mu (k)}{k} \log (g_N(t_1^k,t_2^k;\cal{C})) ,
\end{equation}
where $\mu (k)$ is the M\"obius function
\comment{
\footnote{\begin{equation}
\mu(k) = \left\{\begin{array}{lcl}
0 & & k \mbox{ has one or more repeated prime factors}\\
1 & & k = 1\\
(-1)^n & & k \mbox{ is a product of $n$ distinct primes}
\end{array}\right. \ .
\end{equation}}.

}
We remind the reader that acting with \hbox{PE$^{-1}$} on a generating function gives back the generating series for the generators and the relations in the chiral ring.
We are going to see that the result is generically an infinite series in which the first terms with the plus sign give the basic generators while the first terms with the minus sign give the relations between these basic generators. Then there is an infinite series of terms with plus and minus signs due to the fact that the moduli space of vacua is not a complete intersection. In the introductory section \ref{HSPE} we
have just seen examples of \hbox{PE$^{-1}$} that has a finite extension: namely free generated rings and 
complete intersections. In this section we will see examples of the generic 
infinite extension of the \hbox{PE$^{-1}$} for non complete intersections.

Let us start with the simplest case $N=1$. In this case the generating function for the conifold is simply:
\begin{equation}
g_1(t_1,t_2;{\cal C})= \frac{1}{(1 - t_1)^2(1 - t_2)^2} .
\end{equation} 
Taking the plethystic logarithm we obtain:

\begin{equation}
\hbox{PE$^{-1}$} [g_1(t_1,t_2;{\cal C})]= 2 t_1 + 2 t_2 .
\end{equation}

This means that in the case $N=1$ the chiral ring is freely generated by $A_1$, $A_2$ and $B_1$, $B_2$ without any relations. Indeed in this case the gauge theory is Abelian and we have no matrix relations, the superpotential is zero and we have no $F$-term relations.

We can now pass to the more interesting case of $N=2$. The generating function is given in Equation (\ref{N2con}):

\begin{equation}
g_2(t_1,t_2;{\cal C})= \frac{1 + t_1 t_2 + t_1^2 t_2^2 - 3 t_1^4 t_2^2 - 3 t_1^2 t_2^4 + t_1^5 t_2^3 + t_1^3 t_2^5  - 3 t_1^3 t_2^3 + 4 t_1^4 t_2^4}{(1 - t_1^2)^3(1 - t_1 t_2)^3 (1 - t_2^2)^3} .
\end{equation} 

The first terms of its  plethystic logarithm are:

\begin{equation}\label{PLg2}
\hbox{PE$^{-1}$}[g_2(t_1,t_2;{\cal C})] = 3t_1^2 + 4t_1 t_2 + 3t_2^2 - 3 t_1^4 t_2^2 - 4t_1^3 t_2^3 -3t_1^2 t_2^4 + ...
\end{equation} 

As explained above the interpretation of equation (\ref{PLg2}) is: the first three monomials are the basic generators of the chiral ring and we can recognize them in the following gauge invariant operators in the quiver theory
%$CFT$
:

\begin{eqnarray}\label{basicgen}
3 t_1^2 & \rightarrow & \epsilon \epsilon A_1 A_1 \hbox{ , } \epsilon \epsilon A_1 A_2 \hbox{ , } \epsilon \epsilon A_2 A_2 \nonumber \\
3 t_2^2 & \rightarrow & \epsilon \epsilon B_1 B_1 \hbox{ , } \epsilon \epsilon B_1 B_2 \hbox{ , } \epsilon \epsilon B_2 B_2 \nonumber \\
4t_1 t_2 & \rightarrow & \hbox{Tr}(A_1 B_1) \hbox{ , } \hbox{Tr}(A_1 B_2) \hbox{ , } \hbox{Tr}(A_2 B_1) \hbox{ , } \hbox{Tr}(A_2 B_2) 
\end{eqnarray}
where the indices contraction between fields and epsilon symbols is implicit. These operators transform under spin (1,0), (0,1), and $(\frac{1}{2},\frac{1}{2})$ of $SU(2)_1\times SU(2)_2$, respectively.

The second three monomials give the quantum number of the relations between the basic generators of the chiral ring. The presence of these relations means that the chiral ring in the case $N=2$ is not freely generated. It is possible to understand this fact looking at the higher degree gauge invariant fields in the chiral ring. At order $t_1^4 t_2^2$ using the ten basic generators we can write the operators:
\begin{equation}\label{relb}
(\epsilon \epsilon A_i A_j)^2 (\epsilon \epsilon B_k B_l) \hbox{ , } (\hbox{Tr}(A_i B_j))^2 (\epsilon \epsilon A_k A_l) \hbox{ :          18 + 30 operators}\end{equation}
 Using the tensor relation
\begin{equation}\label{tensrel}
\epsilon_{\alpha_1...\alpha_N}\epsilon^{\beta_1...\beta_N} = \delta_{[ \alpha_1}^{\beta_1}...\delta_{\alpha_N]}^{\beta_N}
\end{equation}
and some tensor algebra we can rewrite the gauge invariants in equation (\ref{relb}) in terms of the operators:
\begin{equation}\label{relcC}
\epsilon \epsilon (A_i B_j A_k)(A_l B_m A_n) \hbox{ , } \epsilon \epsilon (A_i B_j A_k B_l A_m)(A_n) \hbox{ :          21 + 24 operators}
\end{equation}
Hence it is possible to write the $48$ operators in equation (\ref{relb}) in terms of the $45$ operators in equation (\ref{relcC}). This means that there exist 
at least $3$ relations with quantum numbers $t_1^4t_2^2$ between the ten basic generators in Equation (\ref{basicgen}). One can check that the relations are exactly three and these are precisely the ones predicted by the term $-3t_1^4 t_2^2$ in Equation (\ref{PLg2}).
To justify the term  $-3t_1^2 t_2^4$ in (\ref{PLg2}) one works in the same way as before exchanging the $A$ fields with the $B$ fields.
In a similar way we can justify the term $-4t_1^3 t_2^3$ in (\ref{PLg2}). Using the ten basic generators we can write the following $56$ operators at level $t_1^3 t_2^3$:
\begin{equation}\label{reld}
(\epsilon \epsilon A_i A_j)(\epsilon \epsilon B_k B_l)\hbox{Tr}(A_m B_n) \hbox{ , } (\hbox{Tr}(A_i B_j))^3 \hbox{ :          36 + 20 operators}
\end{equation}
these can be written in terms of the $52$ operators:
\begin{equation}\label{rele}
\epsilon \epsilon (A_i B_j A_k B_l)(A_m B_n) \hbox{ , } \epsilon \epsilon (A_i B_j A_k B_l A_m B_n)(1) \hbox{ :           36 + 16 operators}
\end{equation}
where $1$ is the identity operator.

Hence we see that in field theory there exist $4$ relations with the quantum numbers $t_1^3t_2^3$, and these are the ones predicted by the  plethystic logarithm in equation (\ref{PLg2}).

The higher monomials in equation (\ref{PLg2}) mean that the moduli space of the conifold at $N=2$ is not a complete intersection.

At $N=2$ one may expect the presence of the mesonic operators $\hbox{Tr}(ABAB)$ among the basic generators of the chiral ring, but it is easy to show that:

\begin{equation}\label{abab}
(\epsilon \epsilon A_i A_j )(\epsilon \epsilon B_k B_l ) = (\hbox{Tr}(A_iB_j))^2 - \hbox{Tr}(A_iB_kA_jB_l) ,
\end{equation}
and hence we conclude that $\hbox{Tr}(ABAB)$ do not appear independently, but are generated by the ten basic gauge invariant operators in (\ref{basicgen}) as predicted by the plethystic logarithm in Equation (\ref{PLg2}).

To check the consistency of the discussion about the basic generators and their relations one can expand the generating function $g_2(t_1,t_2;{\cal C})$ at the first few orders:
\begin{eqnarray}\label{g2ffew}
g_2(t_1,t_2;{\cal C})&=& 1 + 3t_1^2 + 4t_1 t_2+ 3t_2^2  + 6 t_1^4  + 12t_1^3t_2 + 19t_1^2t_2^2 +  12t_1 t_2^3 + 6t_2^4 + \nonumber\\
& & 10t_1^6+ 24t_1^5t_2 + 45t_1^4t_2^2  + 52 t_1^3t_2^3  + 45 t_1^2 t_2^4 + 24t_1t_2^5 + 10t_2^6 + ... \nonumber\\
\end{eqnarray}
and try to construct the gauge invariant operators associated with each monomial.
It is convenient to form this expansion into a two dimensional lattice in the $t_1, t_2$ coordinates,
\begin{equation}
\left(
\begin{array}{lllllllll}
 1 & 0 & 3 & 0 & 6 & 0 & 10 & 0 & 15 \\
 0 & 4 & 0 & 12 & 0 & 24 & 0 & 40 & 0 \\
 3 & 0 & 19 & 0 & 45 & 0 & 81 & 0 & 127 \\
 0 & 12 & 0 & 52 & 0 & 112 & 0 & 192 & 0 \\
 6 & 0 & 45 & 0 & 134 & 0 & 258 & 0 & 417 \\
 0 & 24 & 0 & 112 & 0 & 280 & 0 & 504 & 0 \\
 10 & 0 & 81 & 0 & 258 & 0 & 554 & 0 & 934 \\
 0 & 40 & 0 & 192 & 0 & 504 & 0 & 984 & 0 \\
 15 & 0 & 127 & 0 & 417 & 0 & 934 & 0 & 1679
\end{array}
\right) .
\end{equation}
The $1$ is the identity operator, the quadratic monomials are the basic generators already discussed, at the quartic order we begin to have composite operators made out of the 10 generators. Let us count them taking into account their symmetry properties:
\begin{eqnarray}\label{quart}
6 t_1^4 & \rightarrow & (\epsilon \epsilon A_i A_j)^2  \hbox{ :         6 operators} \nonumber \\
12 t_1^3t_2 & \rightarrow & (\epsilon \epsilon A_i A_j) \hbox{Tr}(A_k B_l) \hbox{ :        12 operators} \nonumber \\
19 t_1^2 t_2^2 & \rightarrow & (\epsilon \epsilon A_i A_j)  (\epsilon \epsilon B_k B_l) \hbox{ , } (\hbox{Tr}(A_i B_j))^2\hbox{ :         9 + 10 operators} \nonumber\\
12 t_1t_2^3 & \rightarrow & (\epsilon \epsilon B_i B_j) \hbox{Tr}(A_k B_l) \hbox{ :         12 operators} \nonumber \\
6 t_2^4 & \rightarrow & (\epsilon \epsilon B_i B_j)^2 \hbox{ :         6 operators} 
\end{eqnarray}
Thus, the counting as encoded in the generating function and the %$CFT$
explicit counting in the gauge theory nicely agree.
It is interesting to do the same analysis at dimension six, because at this level the relations among the basic generators enter into the game (see equation (\ref{PLg2})):
\begin{eqnarray}\label{sixth}
10t_1^6& \rightarrow & (\epsilon \epsilon A_i A_j)^3  \nonumber \\ & & \hbox{          10 operators} \nonumber \\
24t_1^5t_2 & \rightarrow & (\epsilon \epsilon A_i A_j)^2  \hbox{Tr}(A_k B_l) \nonumber \\ & & \hbox{         24 operators} \nonumber \\
45t_1^4 t_2^2 & \rightarrow & (\epsilon \epsilon A_i A_j)^2  (\epsilon \epsilon B_k B_l) \hbox{ , } (\hbox{Tr}(A_i B_j))^2(\epsilon \epsilon A_n A_m) \nonumber \\ & & \hbox{        18 + 30 operators - 3 relations = 45 operators } \nonumber\\
52 t_1^3t_2^3 & \rightarrow & (\epsilon \epsilon A_i A_j)  (\epsilon \epsilon B_k B_l) \hbox{Tr}(A_n B_m) \hbox{ , } (\hbox{Tr}(A_i B_j))^3  \nonumber \\ & & \hbox{        36 + 20 operators - 4 relations = 52 operators } \nonumber\\
45t_1^2 t_2^4 & \rightarrow & (\epsilon \epsilon B_i B_j)^2  (\epsilon \epsilon A_k A_l) \hbox{ , } (\hbox{Tr}(A_i B_j))^2 (\epsilon \epsilon B_n B_m)  \nonumber \\ & & \hbox{        18 + 30 operators - 3 relations = 45 operators } \nonumber\\
24t_1t_2^5 & \rightarrow & (\epsilon \epsilon B_i B_j)^2 \hbox{Tr}(A_k B_l)  \nonumber \\ & & \hbox{         24 operators} \nonumber \\
10t_2^6& \rightarrow & (\epsilon \epsilon B_i B_j)^3   \nonumber \\ & & \hbox{          10 operators} 
\end{eqnarray}
Also at this level the computation with the generating function and with the conformal field theory techniques, taking into account the relations between the basic generators previously explained, nicely agree.

It is now interesting to give a look at the case of $N=3$. This is the first level where non factorisable baryons appear \cite{Berenstein:2002ke,Beasley:2002xv}. 
These types of baryons are the ones that cannot be written as a product of mesonic operators (traces of fields) and the basic baryons (for example $ \epsilon \epsilon A_i A_j A_k$ at the level $N=3$ is a basic baryon).
The generating function for $N=3$ has the form:
\begin{equation}\label{g3con}
g_3(t_1,t_2;{\cal C})=\frac{F(t_1,t_2)}{(1-t_1^3)^4(1 - t_1 t_2)^3 (1-t_1^2 t_2^2)^3(1- t_2^3)^4} ,
\end{equation}
where $F(t_1,t_2)$ is a polynomial in $t_1$, $t_2$ give in equation (\ref{conifoldN3F}).
The first few terms in the $t_1, t_2$ lattice look like

\begin{equation}
\left(
\begin{array}{lllllllllll}
 1 & 0 & 0 & 4 & 0 & 0 & 10 & 0 & 0 & 20 & 0 \\
 0 & 4 & 0 & 0 & 18 & 0 & 0 & 48 & 0 & 0 & 100 \\
 0 & 0 & 19 & 0 & 0 & 78 & 0 & 0 & 198 & 0 & 0 \\
 4 & 0 & 0 & 72 & 0 & 0 & 260 & 0 & 0 & 624 & 0 \\
 0 & 18 & 0 & 0 & 224 & 0 & 0 & 738 & 0 & 0 & 1686 \\
 0 & 0 & 78 & 0 & 0 & 620 & 0 & 0 & 1854 & 0 & 0 \\
 10 & 0 & 0 & 260 & 0 & 0 & 1545 & 0 & 0 & 4246 & 0 \\
 0 & 48 & 0 & 0 & 738 & 0 & 0 & 3508 & 0 & 0 & 8958 \\
 0 & 0 & 198 & 0 & 0 & 1854 & 0 & 0 & 7414 & 0 & 0 \\
 20 & 0 & 0 & 624 & 0 & 0 & 4246 & 0 & 0 & 14728 & 0 \\
 0 & 100 & 0 & 0 & 1686 & 0 & 0 & 8958 & 0 & 0 & 27729
\end{array}
\right) .
\end{equation}

We now take the plethystic logarithm to understand the structure of basic generators at level $N=3$:

\begin{equation}\label{PLg3con}
\hbox{PE$^{-1}$} [g_3(t_1,t_2;{\cal C})] = 4t_1^3 + 2 t_1^4 t_2  + 2t_1t_2^4 + 4t_2^3 + 4 t_1 t_2  + 9t_1^2 t_2^2+...
\end{equation}

with the dots meaning the presence of relations among the basic generators at higher orders in $t_1$, $t_2$. On the $t_1,t_2$ lattice this expression looks like this

\begin{equation}
\left(
\begin{array}{lllllllllll}
 0 & 0 & 0 & 4 & 0 & 0 & 0 & 0 & 0 & 0 & 0 \\
 0 & 4 & 0 & 0 & 2 & 0 & 0 & 0 & 0 & 0 & 0 \\
 0 & 0 & 9 & 0 & 0 & -6 & 0 & 0 & -3 & 0 & 0 \\
 4 & 0 & 0 & 0 & 0 & 0 & -18 & 0 & 0 & 12 & 0 \\
 0 & 2 & 0 & 0 & -26 & 0 & 0 & 24 & 0 & 0 & 27 \\
 0 & 0 & -6 & 0 & 0 & -16 & 0 & 0 & 136 & 0 & 0 \\
 0 & 0 & 0 & -18 & 0 & 0 & 133 & 0 & 0 & -134 & 0 \\
 0 & 0 & 0 & 0 & 24 & 0 & 0 & 144 & 0 & 0 & -1064 \\
 0 & 0 & -3 & 0 & 0 & 136 & 0 & 0 & -871 & 0 & 0 \\
 0 & 0 & 0 & 12 & 0 & 0 & -134 & 0 & 0 & -1292 & 0 \\
 0 & 0 & 0 & 0 & 27 & 0 & 0 & -1064 & 0 & 0 & 6072
\end{array}
\right)
\end{equation}

Now we would like to match the predictions of equation (\ref{PLg3con}) with the gauge invariant operators in quiver theory
%$CFT$.
\begin{eqnarray}\label{n3}
4t_1^3 & \rightarrow & \epsilon \epsilon A_i A_j A_k  \hbox{ :         4 operators} \nonumber \\
2t_1^4t_2 & \rightarrow & ... \hbox{ :        2 non factorizable baryons } \nonumber \\
2t_1 t_2^4 & \rightarrow & ... \hbox{ :       2 non factorizable baryons }\nonumber\\
4t_2^3 & \rightarrow & \epsilon \epsilon B_i B_j B_k  \hbox{ :       4 operators } \nonumber\\
4 t_1 t_2 & \rightarrow &  \hbox{Tr}(A_i B_j) \hbox{ :       4 operators } \nonumber\\
9t_1^2t_2^2 & \rightarrow & \hbox{Tr}(A_i B_j A_k B_l) \hbox{ :        9  operators} \nonumber \\
\end{eqnarray}
As at the level $N=2$ the operators of the type $\hbox{Tr}(ABABAB)$ do not appear independently but are generated by the basic operators in (\ref{n3}). This can be understood in the same way as the case $N=2$ using the tensor relation (\ref{tensrel}). 

The most interesting basic generators are the ones related to the monomials $2t_1^4 t_2$, $2t_1t_2^4$: these are the first cases of non factorizable baryons. Let us analyze the monomial $2t_1^4 t_2$ (the discussion for the other monomial is the same if one exchanges the $A$ fields with the $B$ fields and $t_1$ with $t_2$). At this level, if all the baryonic operators were factorizable, the only gauge invariant operators we could construct with the right quantum numbers would be of the type $(\epsilon \epsilon A_i A_j A_k)(\hbox{Tr} A_l B_m)$ and there are just $16$ of them. If instead we follow the general proposal \cite{Beasley:2002xv,Butti:2006au} to relate monomials in the homogeneous coordinates to baryonic gauge invariant operators, at the level $t_1^4 t_2$ we would write the operators

\begin{equation}\label{nonfact}
\epsilon \epsilon (A_lB_mA_i) (A_j) (A_k),
\end{equation}
where the epsilon terms are contracted once with each index in the brackets.
If we count how many operators exist of this form we discover that they are $18$. This suggests that among the operators in (\ref{nonfact}) there are $2$ that are not factorizable and must be included as generators of the ring. An explicit
computation shows that there are indeed two non factorizable determinants.
These two ``special'' baryons in the spectra of the field theory are the ones that make the generating function $g_3(t_1,t_2;{\cal C})$ non trivial, and they are precisely the ones required by the plethystic logarithm (\ref{PLg3con}) as basic generators at the level $t_1^4 t_2$. The existence of non factorizable baryons is related
to the flavor indices of the $A_i$ \cite{Berenstein:2002ke}; 
only those baryons where the $A_i$ are
suitably symmetrized can be factorized using the relation (\ref{tensrel}).

Next, one can go up with the values of $N$ and try to extract the general pattern of basic generators for the chiral ring of the gauge theory
%$CFT$
dual to the conifold singularity.

We use a notation where $( . , . )$ denotes the weights of the gauge invariant operators under $t_1$, $t_2$. %and $(j_1, j_2)$ is the spin representation under $SU(2)_1\times SU(2)_2$.
The basic generators for the first few values of $N$ are: 
\begin{eqnarray}\label{NNN}
N=1 &\rightarrow& 2 (1,0) + 2 (0,1) \hbox{ :   4 operators} \nonumber\\ 
N=2 &\rightarrow& 3 (2,0) + 4 (1,1) + 3 (0,2)   \hbox{ :  10 operators} \nonumber\\  
N=3 &\rightarrow& 4 (3,0) + 2 (4,1) + 4 (1,1) + 9 (2,2) +  \nonumber\\ & &  2 (1,4) + 4 (0,3)   \hbox{ :  25 operators} \nonumber\\     
N=4 &\rightarrow& 5 (4,0) + 4 (5,1) + 4 (1,1) + 9 (2,2) + 16 (3,3) +  \nonumber\\ & & 4 (1,5) + 5(0,4)   \hbox{ :  47 operators} \nonumber\\ 
N=5 &\rightarrow& 6 (5,0) + 6 (6,1) + 6 (7,2) + 4 (1,1) + 9 (2,2) + 16 (3,3) + 25 (4,4) + \nonumber\\ & & 6 (2,7)+ 6 (1,6)+6 (0,5)   \hbox{ :  90 operators} 
\end{eqnarray}
From (\ref{NNN}) it seems that the pattern of the basic generators for generic $N$ is the following (we assume reflection symmetry under the exchange of $t_1$ and $t_2$):
\begin{itemize}
\item{The mesonic generators of the chiral ring are all the mesons (single traces) starting from weight (1,1) up to weight $(N-1,N-1)$. There are $(n+1)^2$ generators with weight $(n,n)$, transforming in the spin $(\frac{n}{2},\frac{n}{2})$ representation, leading to a total of $\frac{(N-1)(2N^2+5N+6)}{6}$ generators;}
\item{The baryonic generators of the chiral ring have a jump of 2 in the level:
\begin{itemize}
\item{$N+1$ generators of weight $(N,0)$;}
\item{$2(N-2)$ generators of weight $(N,1)$ starting at level $N=3$, which are exactly the non factorizable baryons we have already discussed;}
\item{generators of weight $(N,2)$ starting at level $N=5$, which are new non factorizable baryons;}
\item{etc.}
\end{itemize}}\end{itemize}
The non factorizable baryons appear for the first time at level $N=3$, and going up with the number of branes the number and type of non trivial baryons increase.

\section{Some Toy Models}\label{Toys}

In this section we are going to look at some toy models to give more explicit formulae
and a more complete comparison with the field theory BPS spectrum.

\subsection{Half the Conifold}

\begin{figure}[h!!!!!!!!!!!!!!!!!]
\begin{center}
\includegraphics[scale=0.5]{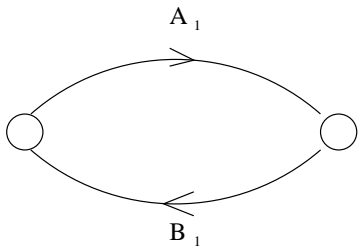} 
\caption{Quiver for half the conifold.}
\label{half}
\end{center}
\end{figure}

A nice toy model consists of half the matter content of the conifold. 
This model is particularly simple as it has no F term relations and all baryons are factorized. 
The gauge theory data can be encoded in the quiver in Figure \ref{half} with $W=0$.

The chiral multiplets are assigned charges according to Table \ref{globalhalfconifold} which is a reduction of Table \ref{globalconifold}.

\begin{table}[htdp]
\caption{Global charges for the basic fields of the quiver gauge theory for ``Half the Conifold."}
\begin{center}
\begin{tabular}{|c||c|c||c|}
%\begin{tabular}{|1||1 1||1 1||1|1||1|}
\hline
& $U(1)_R$ & $U(1)_B$ & monomial\\ \hline
$A_1$ & $\frac{1}{2}$ & 1& $t_1 = t b$\\ \hline
$B_1$ & $\frac{1}{2}$ & -1& $t_2 = \frac{t}{b}$ \\ \hline
\end{tabular}
\end{center}
\label{globalhalfconifold}
\end{table}%

With this toy model we are counting the subset of BPS operators of the conifold
with no occurrences of $A_2$ and $B_2$.

As for the conifold case, the generating function for one D brane and any baryon number is freely generated by the two chiral multiplets $A_1$ and $B_1$ and takes the form

\begin{equation}
g_1(t_1,t_2; \frac{1}{2}{\cal C}) = \frac{1}{(1-t_1)(1-t_2)} = \frac{1}{(1-t b) (1- \frac {t}{b})}.
\label{g1halfconi}
\end{equation}

>From which it is easy to extract the generating function for fixed baryonic charge $B$ and one D brane, $N=1$,

\begin{eqnarray}
g_{1,B\ge0}(t_1, t_2; \frac{1}{2}{\cal C}) &=& \frac{t_1^B} {1-t_1t_2} = \sum_{n=0}^\infty t_1^{n+B} t_2^n, \nonumber \\ \nonumber
g_{1,B\le0}(t_1, t_2; \frac{1}{2}{\cal C}) &=& \frac{t_2^{-B}} {1-t_1t_2} = \sum_{n=0}^\infty t_1^{n} t_2^{n-B}.
\label{reshalfcon}
\end{eqnarray}
which indeed reflects the symmetry of taking $B\leftrightarrow -B$, $t_1\leftrightarrow t_2$ simultaneously.
Setting $\nu$ to be the chemical potential for the number of D branes, $N$, and taking the $\nu$-inserted plethystic exponential of these expressions we can easily extract the the generating function for fixed number of baryons

\begin{eqnarray}
g_{B}(\nu; t_1, t_2; \frac{1}{2}{\cal C}) &=& \exp\biggl(\sum_{k=1}^\infty \frac{\nu^k}{k}g_{1,B}(t_1^k, t_2^k; \frac{1}{2}{\cal C}) \biggr), \nonumber \\ \nonumber
g_{B\ge0}(\nu; t_1, t_2; \frac{1}{2}{\cal C}) &=& \exp\biggl(\sum_{k=1}^\infty \frac{\nu^k t_1^{B k}}{k (1-t_1^k t_2^k)} \biggr), \nonumber \\ \nonumber
g_{B\le0}(\nu; t_1, t_2; \frac{1}{2}{\cal C}) &=& \exp\biggl(\sum_{k=1}^\infty \frac{\nu^k t_2^{ - B k}}{k (1-t_1^k t_2^k)} \biggr).
%\label{reshalfcon}
\end{eqnarray}
We can compare this generating function to the generating function of the complex line, taken from \cite{Benvenuti:2006qr},

\begin{equation}
g(\mu ; t ; \mathbb{C}) = \exp\biggl(\sum_{k=1}^\infty \frac{\mu^k}{k(1-t^k)}\biggr) = \sum_{N=0}^\infty \mu^N \prod_{i=1}^N \frac{1}{(1-t^i)} = \prod_{k=0}^\infty \frac{1}{(1-\mu t^k)},
\end{equation}
which upon substitution $\mu=\nu t_1^B$ and $t=t_1 t_2$ for positive baryonic charge and similar expressions for negative baryonic charge, leads to the generating function
for fixed number of D branes $N$ and fixed baryonic charge $N B$,

\begin{eqnarray}
g_{N,N B\ge0}(t_1, t_2; \frac{1}{2}{\cal C}) &=& t_1^{NB}\prod_{i=1}^{N}\frac{1} {(1-t_1^it_2^i)}, \nonumber \\ \nonumber
g_{N,N B\le0}(t_1, t_2; \frac{1}{2}{\cal C}) &=& t_2^{-NB}\prod_{i=1}^{N}\frac{1} {(1-t_1^it_2^i)}.
%\label{reshalfcon}
\end{eqnarray}

Note the relation  $g_{N,N B\ge0}=t_1^{NB}g_{N,0}$ (and a similar expression for $B<0$) which implies that the partition function at baryonic number $N B$ is
proportional to the mesonic partition function. This can be understood
by observing that, since there are no flavor indices, using equation (\ref{tensrel}), we can factorize all baryons into a product of the basic 
baryons $\det (A_1)$ and $\det (B_1)$ times mesons. 
%The fact that the $g_{N,N B}$ is actually 
%proportional to $g_{N,0}$ follows from the absence of F term relations among
%fields.  

Summing over all baryonic charges we can get the generating function for fixed number of D3-branes, $N$

\begin{equation}
g_N(t_1,t_2; \frac{1}{2}{\cal C}) = \frac{1}{(1-t_1^N)(1-t_2^N)} \prod_{i=1}^{N-1}\frac{1} {(1-t_1^it_2^i)}.
\label{gNhalfconi}
\end{equation}

Which turns out to be freely generated. The corresponding chiral ring of BPS operators is generated by the $N+1$ operators, $\det (A_1)$, $\det (B_1)$, and Tr $(A_1B_1)^i, i=1\ldots N-1$.

\subsubsection{Comparing with the Molien Invariant}\label{invmolien}

The generating functions for cases in which the superpotential is zero can be computed using a different method -- that of the Molien Invariant \cite{Pouliot:1998yv,Romelsberger:2005eg}. In the next chapter we will see a generalization of this techniques to include the case in which the superpotential is non trivial. 

In the zero super potential case the method of the Molien Invariant uses a multi-contour integral formula and goes as follows. Given a supersymmetric ${\cal N} = 1$ gauge theory with a gauge group $G$ of rank $r$ and Weyl group of order $|W|$, and a set of chiral multiplets transforming in representations, $R_k$, we set a chemical potential $t_k$ for each representation and compute the generating function for BPS operators in the chiral ring

\begin{equation}\label{molienSUN}
g(\{t_k\}; G) = \frac{1}{|W|} \prod_{j=1}^r \oint \frac{dw_j}{2 \pi i w_j} \frac{\prod_\alpha(1-w^{h(\alpha)})}{\prod_k \prod_{\lambda_k}(1- t_k w^{h(\lambda_k)})} ,
%\label{g1halfconi}
\end{equation}
where $h(\alpha)$ are weights of the adjoint representation of $G$, while $h(\lambda_k)$ are weights of the representation $R_k$.

This function is also used in the Commutative Algebra literature (see for example \cite{Djokovic:1981bh}) and is called there the Molien--Weyl invariant.

For the case of a gauge group $SU(N)$ and $n$ chiral multiplets in the fundamental representation we introduce $n$ chemical potentials, $t_k, k=1\ldots n$ and this multi-contour integral can be extended to an $N$-dimensional contour integral of the form

\begin{equation}
g(\{t_k\}; SU(N) ) = \frac{1}{N!} \prod_{j=1}^N \oint \frac{dw_j}{2 \pi i w_j} \frac{\prod_{i<j} (w_i-w_j)^2}{1-\prod_{j=1}^N w_j} \frac {1}{\prod_{k=1}^n \prod_{j=1}^N (1- t_k w_j)} .
%\label{g1halfconi}
\end{equation}

This expression counts all gauge invariant operators which contain fields in the fundamental representation but not fields in the anti-fundamental representation.
If we would like to count BPS operators which are constructed out of $n_1$ chiral multiplets in the fundamental representation and $n_2$ chiral multiplets in the anti-fundamental representation we introduce chemical potentials $t_{k_1}, k_1=1\ldots n_1$ and $q_{k_2}, k_2=1\ldots n_2$ and write
the integral

\begin{eqnarray}
& & g(\{t_{k_1}\}, \{q_{k_2}\}; SU(N) ) = \\ \nonumber
& & \frac{1}{N!} \prod_{j=1}^N \oint \frac{dw_j}{2 \pi i w_j}
\frac{\prod_{i<j} (w_i-w_j)^2}{1-\prod_{j=1}^N w_j}
\frac {1}{\prod_{k_1=1}^{n_1} \prod_{j=1}^N (1- t_{k_1} w_j)}
\frac {1}{\prod_{k_2=1}^{n_2} \prod_{j=1}^N (1- t_{k_2} w_j^{-1})}.
%\label{g1halfconi}
\end{eqnarray}

We are now ready to present the formula for the case of discussion in this section, half the conifold. We have two gauge groups, $SU(N)\times SU(N)$ for which we introduce 2 sets of $N$ variables each, $w_i, i=1\ldots N$, and $v_i, i=1\ldots N$. There is one chiral multiplet $A_1$ with a chemical potential $t_1$ and one chiral multiplet $B_1$ with 

\begin{eqnarray}
& & g_N(t_1, t_2; \frac{1}{2}{\cal C}) = \\ \nonumber
& & \frac{1}{(N!)^2}
\prod_{r=1}^N \oint \frac{dw_r}{2 \pi i w_r}
\prod_{s=1}^N \oint \frac{dv_s}{2 \pi i v_s}
\frac{\prod_{i<j} (w_i-w_j)^2 (v_i-v_j)^2}{(1-\prod_{j=1}^N w_j) (1-\prod_{j=1}^N v_j)}
\frac {1}{\prod_{r,s=1}^N (1- \frac{t_1 w_r}{v_s})(1- \frac{t_2 v_s}{w_r})}.
%\label{g1halfconi}
\end{eqnarray}

The case $N=2$ was computed explicitly and the result coincides with Equation (\ref{gNhalfconi}), while other cases were not checked due the heavy computations they involve but definitely can be checked for higher values of $N$.

\subsection{$\frac{3}{4}$ Conifold}

Here we are considering another example of counting BPS operators which are a subset of the states in the conifold theory. We are going to restrict to all the operators that include the fields $A_1, A_2, B_1$ in Table \ref{globalconifold} but not $B_2$. Since these are 3 out of 4 of the fundamental fields of the conifold theory we use the suggestive name ``$\frac{3}{4}$ Conifold''. It is amusing to see that some of the expressions can be derived exactly and for any number of branes $N$ and for any baryonic number $B$, even though the full answer can be represented implicitly but no explicit expression is known.
The first point to note is that the field $B_1$ has no flavor index and therefore all baryons with negative baryonic charge factorize. This allows us to sum exactly all the generating functions with negative baryonic charge and fixed $N$. The F term relations are indeed relevant for this problem but they turn out to affect the mesons only ($B=0$). For simplicity of computations we are going to set $x=1$ while the chemical potential $y$ for the second $SU(2)$ is not needed.

The $N=1$ generating function is again freely generated and is given by equation (\ref{g1coniC}) with the factor corresponding to $B_2$ removed. It takes the form 

\begin{equation}
g_1(t_1,t_2; \frac{3}{4}{\cal C}) = \frac{1}{(1-t_1)^2(1-t_2)} = \frac{1}{(1-t b)^2 (1- \frac {t}{b})}.
\label{g134coni}
\end{equation}

Using techniques which by now are standard (see equations (\ref{res1C}, \ref{res2}, \ref{res3})) we have

\begin{eqnarray}
g_{1,B\ge0}(t_1, t_2; \frac{3}{4}{\cal C}) &=& \frac{t_1^B (1+B - B t_1 t_2)} {(1-t_1t_2)^2} = \sum_{n=0}^\infty (n+B+1) t_1^{n+B} t_2^n, \nonumber \\
g_{1,B\le0}(t_1, t_2; \frac{3}{4}{\cal C}) &=& \frac{t_2^{-B}} {(1-t_1t_2)^2} = \sum_{n=0}^\infty (n+1) t_1^{n} t_2^{n-B}.
%\label{reshalfcon}
\end{eqnarray}
and applying the plethystic exponential we find the generating function for fixed $B$.

\begin{eqnarray}
\nonumber
g_{B\ge0}(\nu; t_1, t_2; \frac{3}{4}{\cal C}) &=& \exp\biggl(\sum_{k=1}^\infty \frac{\nu^k t_1^{ B k}}{k (1-t_1^k t_2^k)^2} \biggr) \biggl[ \exp\biggl(\sum_{k=1}^\infty \frac{\nu^k t_1^{B k}}{k (1-t_1^k t_2^k)} \biggr) \biggr]^B \\ \nonumber
g_{B\le0}(\nu; t_1, t_2; \frac{3}{4}{\cal C}) &=& \exp\biggl(\sum_{k=1}^\infty \frac{\nu^k t_2^{ - B k}}{k (1-t_1^k t_2^k)^2} \biggr) = \prod_{n=0}^\infty \frac{1}{(1 - \nu t_1^n t_2^{n-B})^{n+1}}
\end{eqnarray}

We notice that, as in the half-conifold case, 

\begin{equation}
g_{B\le0}(\nu; t_1, t_2; \frac{3}{4}{\cal C}) = g_{B=0}(\nu t_2^{-B}; t_1, t_2; \frac{3}{4}{\cal C}) ,
\end{equation}

This implies an equality satisfied by generating functions with fixed number of branes and fixed negative baryonic charge by taking the power expansion in $\nu$ on both sides,

\begin{equation}
g_{N,N B\le0}(t_1, t_2; \frac{3}{4}{\cal C}) = t_2^{- N B} g_{N,0}(t_1, t_2; \frac{3}{4}{\cal C}) ,
\end{equation}

Next we sum over all negative baryonic charges and get

\begin{equation}
\sum_{B=0}^\infty t_2^{N B} g_{N,0} (t_1, t_2; \frac{3}{4}{\cal C}) = \frac{g_{N,0}(t_1, t_2; \frac{3}{4}{\cal C})}{1-t_2^N}
\end{equation}
so that just one free generator ($\det (B_1)$) is added to the mesonic ones.
In agreement with the fact that baryons are factorized and relations only
apply to mesons.

Notice that the partition function for mesons, $g_{B=0}$ with $t_1 t_2=t$
is the mesonic partition function for $\mathbb{C}^2$, namely

\begin{equation}
\nonumber
g_{B=0}(\nu; t_1, t_2; \frac{3}{4}{\cal C}) = \exp\biggl(\sum_{k=1}^\infty \frac{\nu^k}{k (1-t_1^k t_2^k)^2} \biggr) = g(\nu; t_1 t_2 ; \mathbb{C}^2) 
\end{equation}

This follows from the fact that the two matrices $A_1 B_1$ and $A_2 B_1$ are adjoint valued and
commute due to F-term relations. As a result their moduli space is a copy of $\mathbb{C}^2$.

The other half of the spectrum with $\det (A_i)$ is more difficult to analyze:
there are non factorizable baryons and non trivial relations. To get some expressions for the positive case we write 

\begin{eqnarray}
\nonumber
g_{N B\ge0}(\nu; t_1, t_2; \frac{3}{4}{\cal C}) &=& \biggl[ \sum_{N=0}^\infty \nu^N t_1^{N B} g_{N,0}(t_1, t_2; \frac{3}{4}{\cal C})\biggr] \biggl[ \sum_{N=0}^\infty \nu^N t_1^{N B} g_N(t_1 t_2 ; \mathbb{C}) \biggr]^B \end{eqnarray}

The baryon number dependence is indeed isolated but does not seem to simplify to be summed over.

\subsubsection{$N=2$ for $\frac{3}{4}$ the Conifold}

Getting a general expression for any $N$ seems to be too hard. We can nevertheless use the techniques described in Section \ref{Nis2} in order to get an explicit expression for $N=2$. Using equations (\ref{g2coniC}) and (\ref{g134coni}), we find

\begin{eqnarray}
\nonumber
g_2(t_1, t_2; \frac{3}{4} {\cal C}) &=& \frac{1}{2}g_1(t_1^2,t_2^2; \frac{3}{4} {\cal C})
+ \frac{1}{2}\frac{1}{(2\pi i)} \oint \frac {ds} {s} g_1(t_1 s,\frac{t_2} {s}; \frac{3}{4} {\cal C})  g_1(\frac{t_1}{s},t_2 s; \frac{3}{4} {\cal C}) \\
&=& \frac{1-t_1^4 t_2^2}{(1-t_1^2)^3(1-t_1 t_2)^2(1-t_2^2)},
\label{g234coni}
\end{eqnarray}
implying that the moduli space of vacua for this theory is a complete intersection and is generated by the 6 operators, the spin 1 baryons, $\det (A_1), \epsilon \epsilon A_1 A_2, \det (A_2)$, the spin $\frac{1}{2}$ mesons $Tr (A_1 B_1), Tr (A_2 B_1)$, and the spin 0 baryon $\det (B_1)$. The moduli space is five dimensional, and is given by a degree (4,2) (4 $A$'s, 2 $B$'s) relation in these 6 generators.

\section{Summary and Discussions}

The results of this chapter were presented in an almost purely field theory language.
Looking at the previous chapter we can review the success of the present one as the nice marriage
of two basic principles to construct the 
exact generating function which counts baryonic and mesonic BPS 
operators in the chiral ring of quiver gauge theories. We gave 
explicit formulas for the conifold theory and some subsets of the fields in it 
(these are termed ``half the conifold'' and ``$\frac{3}{4}$ the conifold'', respectively). 
The first principle relies on the plethystic program \cite{Benvenuti:2006qr,Feng:2007ur} 
that consist of a set techniques to generate partition functions for generic number of branes once the 
$N=1$ case is known. The second principle \cite{Butti:2006au} uses some index theorems 
to handle baryonic generating functions with specific divisors in the geometry. 
This is translated into generating functions for fixed baryonic numbers and hence 
leads to the full generating function for any number of D branes, N, and any baryonic charge, $B$, 
as well as the other charges in the system such as $R$ charges and flavor charges.

It is interesting to remark that the results in this chapter can be also interpreted as 
based on the semiclassical quantization of wrapped supersymmetric branes in an $AdS_5$ 
background dual to a strongly coupled gauged theory at large $N$ \cite{Butti:2006au}. 
We compared this strong coupling computations with 
the spectrum of operators obtained directly in field theory, for some specific values of 
$N$ and some set of operators. The match seems one to one at least for the specific case described in 
the chapter. This fact induce us to guess that such countings 
are valid also for weak coupling constant and small values of $N$, 
as similarly observed in the case of $\mathbb{C}^3$ \cite{Biswas:2006tj}. 
We will see some other examples of this kind of interpolation between strong 
coupling large N and weak coupling small N in the following chapter.   

An important outcome, that will be fundamental in the following chapter, 
is the identification of the baryon number $B$ with the K\"ahler modulus. 
This identification is very explicit in the conifold theory and it is expected 
to be true for any CY manifold $\cX$. In the next chapter we will see in great 
detail how the K\"ahler modulus of the geometry $\cX$ and the baryonic charges 
of the gauge theory are related. 

We can make some short speculation: the fact that computation of generating functions 
reveals an important connection between gauge theories and their CY moduli spaces, may suggest that 
the baryon number in field theory may be interpreted as the discrete area of two cycles in $\cX$. 
Pushing on this correspondence we can guess that the discrete values of the baryonic charges are 
the quantum volume of the two and four cycles in a regime in which the string theory 
is strongly coupled and the gauge theory description reveals an underlying discrete structure. 
We expect this correspondence could produce a new line of research for strongly coupled string theories, and we will 
present more data in favor of this correspondence in the next chapter.

\chapter{Partition Functions for General Quiver Gauge Theories}\label{chiralcount}

In the previous two chapters we accumulated enough informations to attack the problem of counting 
BPS gauge invariant operators in the chiral ring of quiver gauge theories living on D-branes 
probing generic toric $CY$ singularities $\cX$. 
This is in principle an extremely difficult task. 
In this chapter we will try to give a synthesis of what we learnt both from 
the geometry and the field theory to develope powerful techniques to solve the general problem.

In the previous chapter, in some short comments, 
we claimed some kind of relations between baryonic charges and K\"ahler structure of $\cX$. 
In this chapter we will discover 
that the computation of generating functions that 
include counting of baryonic operators is strongly related to a possibly 
deep connection between the baryonic charges in field theory and the K\"ahler moduli of $\cX$. 

A study of the interplay between gauge theory and geometry shows that given
geometrical sectors appear more than once in the field theory, leading to a notion of
``multiplicities''.
We explain in detail how to decompose the generating function for 
one D-brane into different sectors and how to compute their relevant 
multiplicities by introducing geometric and anomalous baryonic charges. 
The Plethystic Exponential remains a major tool for passing from one 
D-brane to arbitrary number $N$ of D-branes.
We will illustrate a general procedure valid for all the toric CY variety $\cX$.
 
To make the discussion more easy to follow we will revisit 
the analysis the conifold and we will study in detail the explicit 
examples of $\mathbb{C}^3/\mathbb{Z}_3$ and $\mathbb{F}_0$.

\section{Generalities}

This chapter is devoted to the study of the (baryonic and mesonic) generating function for the chiral ring in
quiver gauge theories. Extending the results of the previous chapter, we compute the
generating functions including baryonic degrees of freedom for various theories. We first study
in detail the generating function for one D-brane and we decompose it into sectors with definite
baryonic charges. This decomposition is closely related to the geometry and to the generating
functions for holomorphic curves obtained by localization in the CY manifold $\cX$.
We conjecture that  the generating function for a number $N$ of D-branes is completely determined by
the generating function for a single D-brane and it is obtained by applying the plethystic
exponential to each sector \cite{Benvenuti:2006qr, Butti:2006au, Hanany:2006uc, Feng:2007ur,
Forcella:2007wk}. This conjecture, which can be proved in the case of mesonic operators, is
inspired by the geometrical quantization of the classical D3-brane configurations in the
gravitational dual. We explicitly compute the generating functions for a selected set of
singularities, including $\mathbb{C}^3/\mathbb{Z}_3$, $\mathbb{F}_0$ and we make 
various checks in the dual field theory. We refer the reader to \cite{Butti:2007jv} for other examples.

In chapter \ref{barcon} we studied the simple and elegant
case of the conifold.
A new feature, which
arises for more involved singularities, like for example $\mathbb{C}^3/\mathbb{Z}_3$ and 
$\mathbb{F}_0$, is the existence of multiplicities, namely the fact
that geometrical sectors appear more than once in field theory.
As we go over these examples in detail, we find that multiplicities have a geometrical
interpretation and can be resolved, with a construction that ties together
in a fascinating way the algebraic geometry of the CY and the combinatorics
of quiver data.

The chapter is organized as follows. In section \ref{a}, we discuss the basics of generating functions
and plethystics. In section \ref{decomposition}, we apply these tools and we explain in detail the nice interplay between the 
geometric properties of the singularity $\cX$ and the graph properties of the field theory, that conspire to give the 
general recipe to obtain the complete generating function for every toric CY singularities. We give
a detailed discussion of the partition functions from both
the field theory and the geometry perspectives. The GKZ decomposition is introduced and the
auxiliary GKZ partition function is defined. Section \ref{examples} contains detailed examples: 
we revise the conifold and we study in detail the $\mathbb{C}^3/\mathbb{Z}_3$ 
and the $\mathbb{F}_0$. For these examples we explicitly
compute generating functions both for $N=1$ and for small $N>1$. Section \ref{molienN} deals with 
$N>1$ D-branes and gives a systematic 
approach to the field theory computation by means of the Molien formula. 
In this chapter we will mainly interested in regular horizon, the reader interested 
in the singular horizon case, is referred to \cite{Hanany:2006nm}, where he can find some preliminary
discussions and observations.

\section{General Structure of Generating Functions for BPS Operators}\label{a}

In this section we will give general prescriptions on the computation of generating functions
for BPS operators in the chiral ring of a supersymmetric gauge theory that lives on a D-brane
which probes a generic non-compact CY manifold $\cX$. These will be of course compatible with 
the simpler cases of the conifold that 
we have already discussed in detail 
in the previous chapter.

Given an ${\cal N}=1$ supersymmetric gauge theory with a collection of $U(1)$ global symmetries,
$\prod_{i=1}^r U(1)_i$, we have a set of $r$ chemical potentials $\{t_i\}_{i=1}^r$. The
generating function for a gauge theory living on a D-brane probing a generic non-compact CY
manifold $\cX$ depends on the set of parameters, $t_i$. There is always at least one such $U(1)$
global symmetry and one such chemical potential $t$, corresponding to the $U(1)_R$ symmetry.

The global charges are divided into classes: {\bf baryonic charges}, and {\bf flavor charges}
(by abuse of language, we will include the $R$-symmetry in this latter class). The number of
non-anomalous flavor symmetries, related to the isometries of $\cX$, is less than three while
the number of non-anomalous baryonic symmetries, related to the group of divisors in $\cX$, can
be quite large. In certain cases, we can also have baryonic discrete symmetries. As is
demonstrated below, in addition to the non-anomalous baryonic charges we need to consider the
{\bf anomalous baryonic charges}. We will only consider the case of toric CY where the number of
flavor symmetries is three. When it will be necessary to make distinctions, we will denote with
$x,y$ or $q_i$ the flavor chemical potentials and with $b_i$ the non-anomalous baryonic chemical potentials. 
Chemical potentials for anomalous charges are denoted by $a_i$.

For a given CY manifold $\cX$, we denote the generating function for $N$ D-branes by $g_{N}(\{t_i\}; \
\cX)$. The generating function for $N=1$ is simple to compute by using field theory arguments.
%Recall that , which can be
%determined in the toric case using dimer technology \cite{Hanany:2005ve,
%Franco:2005rj},\footnote{See \cite{ Hanany:2005ss,Feng:2005gw, Hanany:2006nm, Garcia-Etxebarria:2006aq, Brini:2006ej, Butti:2006hc, Benvenuti:2005ja, Benvenuti:2005cz,Franco:2005sm,Butti:2005sw, Butti:2005ps,Franco:2006gc, Benvenuti:2006xg,Lee:2006ru, Imamura:2006ub,Ueda:2006wy,Butti:2006nk,Imamura:2006ie,Oota:2006eg,Kato:2006vx,Imamura:2007dc} for a rich set of subsequent developments.} 
Indeed the quiver gauge theory living on the world-volume of the D3-branes consists of a gauge group $SU(N)^g$, adjoint or bi-fundamental chiral
fields 
%\footnote{Henceforth we denote fields by bold characters to distinguish them from global quantum numbers.} 
${\bf X}^J$, which can be considered as $N\times N$ matrices, and a
superpotential $W({\bf X}^J)$.

For $N=1$ the matrices ${\bf X}^J$ reduce to numbers and the F-term conditions become polynomial
equations in the commuting numbers ${\bf X}^J$. We can consider the polynomial ring
$\mathbb{C}[{\bf X}^J]$ to be graded by the weights $t_i$. Since the gauge group is acting
trivially for $N=1$, the ring of gauge invariants is just the quotient ring
$${\cal R}_{N=1}^{inv}=\mathbb{C}[{\bf X}^J]/{\cal I}$$
where $\cal I$ is the set of F-term constraints $dW({\bf X}^J)/d{\bf X}^J$. The generating function
for polynomials is the right now familiar Hilbert series which we got used in chapter \ref{Master} to study 
the master space $\f$; indeed ${\cal R}_{N=1}^{inv}= \mathbb{C}[\f]$. It can be computed using the methods explained in chapter \ref{Master} and it coincide with 
the Hilbert series of the master space $g_1(\{t_i\})=H(\{t_i\};\f)$. In particular
there is an algorithmic way to compute it using computer algebra programs like Macaulay2
\cite{M2book}. 
We can therefore assume that the generating function $g_1(\{t_i\})$ for ${\cal R}_{N=1}^{inv}$ is
known.

We proceed to the determination of $g_N$ with a general conjecture:

\begin{itemize}
\item{
{\it For the class of theories considered here (D-branes probing non-compact CY which are any of
toric, orbifolds or complete intersections), the knowledge of the generating function for $N=1$
is enough to compute the generating function for any $N$.}}
\end{itemize}

>From the last two chapters this fact must be the right now familiar statement that the mesonic and the 
baryonic operators for finite $N$ are symmetric functions of the mesonic and baryonic operators for $N=1$.

The general construction is as follows. There exists a decomposition of the $N=1$ ring of
invariants ${\cal R}_{N=1}^{inv}$, and consequently of its generating function, into sectors ${\cal S}$ of
definite baryonic charges

\begin{equation}
g_1(\{t_i\}; \cX)=\sum_{{\cal S}} g_{1,\cal S}(\{t_i\}; \cX)
\end{equation}
where $g_{1,\cal S}$ is the generating function for the subsector ${\cal S}\subset {\cal
R}_{N=1}^{inv}$. All elements in ${\cal S}$ have the same baryonic charges, and, except for a
multiplicative factor, $g_{1,\cal S}$ only depends on the flavor charges $q_i$. In simple cases,
like the conifold, ${\cal S}$ is just a label running over all the possible values of the
non-anomalous baryonic charge. The understanding of the precise decomposition of ${\cal
R}_{N=1}^{inv}$ into subsectors in the general case is a nontrivial task and is one of the
subjects of this chapter.

The generating function for $N$ branes is then obtained by taking $N$-fold symmetric products of
elements in each given sector $\cal S$. This is precisely the role which is played by the Plethystic
Exponential (PE) -- to take a generating function for a set of operators and count all possible
symmetric functions of it. If we introduce a chemical potential $\nu$ for the number of
D-branes, the generating function for any number of D-branes is given by
\begin{eqnarray}
\nonumber
g(\nu; \{t_i\}; \cX)
&= &\sum_{\cal S} \hbox{PE$_{\nu}$}[g_{1,\cal S}(\{t_i\}; \cX)] \equiv \sum_{\cal S} \exp\biggl(\sum_{k=1}^\infty \frac{\nu^k}{k}g_{1,\cal S}(\{t_i^k\}; \cX) \biggr)\nonumber \\
&\equiv& \sum_{N=0}^\infty g_{N}(\{t_i\};\cX) \nu^N
\label{g1plet}
\end{eqnarray}

This the generalization of the prescription we gave in section \ref{generalCY} to a generic toric CY $\cX$.  
The detailed description of the decomposition into sectors $\cal S$ is given
in the rest of this chapter, but it is important to notice from the very
beginning that such a decomposition is not unique. As already mentioned above,
gauge invariants in the same sector have the same baryonic
charges. One can take these baryonic charges to be non-anomalous. 
This however, turns out to be not enough and we seem to need a finer decomposition of the ring of invariants
which is obtained by enlarging the set of non-anomalous baryonic charges to a
larger set. There are two basic extensions, one related to an expansion
in a full set of baryonic charges, anomalous and non-anomalous, and the other extension 
is related to a full set of discretized K\"ahler moduli on the CY resolutions. We thus have two
complementary points of view:

\begin{itemize}
\item{{\bf Quantum field theory perspective}:
the most general decomposition of the generating function $g_1(\{t_i\})$ is into the full set of
baryonic charges. Let us extend the set of chemical potentials $t_i$ to all the baryonic
charges, including the anomalous ones, denoted by $a_i$. There are $g-1$ independent baryonic
charges, where $g$ is the number of gauge groups. We can thus decompose ${\cal R}_{N=1}^{inv}$ into sectors with definite charges under
$U(1)^{g-1}$. $g_{1}(\{t_i\})$ will decompose into a formal Laurent series in the baryonic
chemical potentials $b_i$ and $a_i$. The ${\cal R}_{N}^{inv}$ rings of invariants for number of colors
$N$ will similarly decompose into sectors of definite baryonic charge.
%The
%generating function $g_N(t_i)$ can be computed according to equation
%(\ref{g1plet}).
We can formally extend the gauge group $SU(N)^g$ to $U(N)^g/U(1)$ by gauging the baryonic
symmetries.\footnote{The theory will of course be anomalous. The overall $U(1)$ is discarded
since it acts trivially.} From this perspective, the decomposition of the ring of $SU(N)^g$
invariants into Abelian representations of the extended group $U(N)^g/U(1)$ is sometimes called
an expansion in {\it covariants} and is extremely natural from the point of view of invariant
theory. All sectors ${\cal S}$ appear with multiplicity one in the decomposition of Equation
(\ref{g1plet}).}
\item{{\bf The dual geometrical perspective}: the full set of BPS states
of the dual gauge theory can be obtained by quantizing the classical configuration space of
supersymmetric D3-branes wrapped on the horizon. This problem can be equivalently rephrased in
terms of holomorphic surfaces in the CY with $g_1$ as generating function \cite{Mikhailov:2000ya,Beasley:2002xv}.
Quite remarkably, $g_1$ has a decomposition

\begin{equation}
g_1(\{t_i\}; \cX)=\sum_{\cal S} m ( {\cal S} ) \ g_{1,\cal S}(\{t_i\}; \cX),
\end{equation}

where the parameters $\cal S$ can be identified with a complete set of {\it discretized  K\"ahler
moduli} and the integers $m ({\cal S})$ are multiplicities. We will call it the {\bf GKZ decomposition}, from the known description of the K\"ahler
cone in terms of a secondary fan given by the GKZ construction. The functions $g_{1,\cal S}$
can be explicitly determined with the computation of a character using the equivariant index
theorem. This geometrical decomposition has multiplicities $m({\cal S})$ which will be interpreted in the
following sections. The result for finite $N$ is generated by the following function

\begin{equation}
g(\nu;\{t_i\}; \cX)=\sum_{\cal S} m ( {\cal S} ) {\rm PE}_\nu [g_{1,\cal S}(\{t_i\}; \cX)] ,
\end{equation}

and can be interpreted as the result of quantizing the classical BPS D3-brane
configuration in each sector $\cal S$.}
\end{itemize}

The two decompositions of the $N=1$ generating function are different and complementary. For a
toric CY manifold that has a toric diagram with $d$ external vertices and $I$ internal integral
points, the number of non-anomalous baryonic symmetries is $d-3$, the number of anomalous baryonic symmetries is $2 I$ and the dimension of the K\"ahler moduli space is $d-3+I$. The field theory expansion is thus based on a lattice $\Gamma_{(b,a)}$ of dimension $d-3+2 I$ consisting of all baryonic
charges, anomalous or not, while the geometrical expansion is based on a lattice $\Gamma_{GKZ}$
of dimension $d-3+I$. The two sets have a nontrivial intersection $\Gamma_b$, consisting of
non-anomalous baryonic charges.
% but the latter is not strictly contained in the first one.

At the end,  we will be interested in the generating function for BPS operators
with chemical potential with respect to the {\it non-anomalous} charges.
To this purpose, we must project any of the two lattices on their
intersection, which is the $d-3$ lattice of non-anomalous baryonic symmetries
$\Gamma_b$
$$g_1(\{t_i\})=\sum_{k \in \Gamma_b} m ( k )\ g_{1,k}(\{t_i\})$$
and multiplicities will generically appear.

On the other hand, we could even imagine to enlarge our lattices. Adding the anomalous baryonic
charges to the GKZ fan we obtain a lattice of dimension $d-3+3 I$. The points give hollow polygons
over the GKZ fan. All these issues will be discussed in detail in the rest of the chapter.

%\newpage

\section{Expanding the $N=1$ Generating Function}
\label{decomposition}

As we have seen in the previous chapters our decomposition of the ring of invariants of the gauge theory is a decomposition into
different types of {\it determinants}. Let us review its structure and fix the notations. For simplicity, we will use the following notation: given
a pair of gauge groups $(\alpha,\beta),\ \alpha,\beta=1,...,g$, we call {\bf determinant of type $(\alpha,\beta)$} a gauge
invariant of the form

$$\epsilon^{i_1,...,i_N} ({\bf X}_{I_1}^{(\alpha,\beta)})_{i_1}^{j_1} .... ({\bf X}_{I_N}^{(\alpha,\beta)})_{i_N}^{j_N}\epsilon_{j_1,...,j_N}$$
where $({\bf X}_I^{(\alpha,\beta)})_{i}^{j}$ denotes a string of elementary fields with all gauge indices
contracted except two indices, $i$ and $j$, corresponding to the gauge groups $(\alpha,\beta)$. 
The index $I$ runs over all possible strings of elementary fields
with these properties.
The full
set of invariants is obtained by arbitrary products of these determinants. Using the tensor
relation
$$\epsilon^{i_1,...,i_N}\epsilon_{j_1,...,j_N} = \delta^{i_1}_{[j_1}\cdots \delta^{i_N}_{j_N]} $$
some of these products of determinants are equivalent and some of these are actually equivalent
to mesonic operators made only with traces. In particular, mesons are included in the above
description as determinants of type $(\alpha,\alpha)$.

We can decompose the ring of invariants according to the baryonic charges,
which indeed distinguish between different types of determinants, or baryons.
This decomposition is natural in field theory and it also has a simple interpretation in the dual gravity theory.

In fact, in theories obtained from D3-branes at CY singularities, baryons can be identified with
branes wrapped on nontrivial cycles on the base $H$ of the CY as explained in detail in chapter \ref{D3SE}. 
The non-anomalous symmetries can
be clearly identified in the dual theory. In particular, states with the same non-anomalous
baryonic charges can be continuously deformed into each other: indeed we can relate the set of
non-anomalous baryonic charges to the group of three-cycles in $H$. The homology
$H^3(H,\mathbb{Z})=\mathbb{Z}^{d-3}\times \Gamma$ determines $d-3$ continuous baryonic charges
($d$ is the number of vertices of the toric diagram) and possibly a set of discrete baryonic
charges from the torsion part $\Gamma$.

Let us remind what happens in the particular case of the conifold, previously analyzed.
In the conifold there is one non-anomalous baryonic charge (since $d=4$) which is
related to the single three-cycle in the base $T^{1,1}$. There are only two gauge groups and two
types of determinants: $(1,2)$ and $(2,1)$. The invariants decompose according to the baryonic
charge:

\begin{enumerate}
\item $B=0$ corresponds to the mesons (D3-branes wrapping trivial cycles, giant
gravitons \cite{McGreevy:2000cw}),
\item $B>0$ corresponds to the sector containing the determinants $(\det {\bf
A})^B$ and all possible mesonic excitations (D3-branes wrapping $B$ times the 3-cycle),
\item finally, $B<0$ corresponds to the sector containing the determinants $(\det {\bf B})^{|B|}$
and all possible mesonic excitations (D3-branes wrapping $|B|$ times the 3-cycle with opposite
orientation).
\end{enumerate}

The conifold picture is nice and in many ways elegant. However, a simple look at any other
quiver gauge theory reveals that this simple picture is too naive. % (the conifold strikes again as a prototypical example for nothing..).
Consider, for~example, the case of the orbifold $\mathbb{C}^3/\mathbb{Z}_3$ (see chapter \ref{Master} and
\fref{dP0quiver}), that already reveals all types of oddities:

\begin{itemize}
\item{Since $d=3$, there is no continuous non-anomalous baryonic symmetry.
However, $H^3(S^5/\mathbb{Z}_3,\mathbb{Z})=\mathbb{Z}_3$ and there is a discrete baryonic
symmetry. We can indeed construct determinants
%\footnote{for simplicity in this chapter we will follow the notation: $X_{1,2}^i=U_i$, $X_{2,3}^i=V_i$, $X_{3,1}^i=W_i$}
for example, using the fields ${\bf X}_{12},{\bf X}_{23}$
and ${\bf X}_{31}$ with $\mathbb{Z}_3$ charge $+1$. These do not carry any continuous conserved
charge since the product $\det {\bf X}_{12} \det {\bf X}_{23} \det {\bf X}_{31}$ can be rewritten as a meson in
terms of traces; for example, using $\epsilon^{i_1,...,i_N}\epsilon_{j_1,...,j_N} = \delta^{i_1}_{[j_1}\cdots \delta^{i_N}_{j_N]} $ we can write,
$$ \det {\bf X}_{12}^1\det{\bf X}_{23}^1\det{\bf X}_{31}^1 =\det ( {\bf X}_{12}^1{\bf X}_{23}^1 {\bf X}_{31}^1) = \tr\left(\left( {\bf X}_{12}^1{\bf X}_{23}^1{\bf X}_{31}^1\right)^N \right) +... \pm \left(\tr({\bf X}_{12}^1{\bf X}_{23}^1{\bf X}_{31}^1) \right)^N $$
On the other hand, $(\det {\bf X}_{12}^1)^3$ cannot be reduced to traces simply because there are no
gauge invariant traces we can make with ${\bf X}_{12}^1$ alone:  we have actually an infinite number
of products of determinants ($(\det {\bf X}_{12})^n$ for $n=1,2,...$ for example) that cannot be
rewritten in terms of mesons. All these operators correspond in the ring  of invariants to
sectors that cannot be distinguished by the discrete baryonic charge.}
\item{The BPS D3-brane configurations wrap divisors in the CY: for
$\mathbb{C}^3/\mathbb{Z}_3$ we have just a single divisor $D$ satisfying $3 D=0$ and this agrees
with the homology of the base $S^5/\mathbb{Z}_3$. However, we also have a vanishing compact
four-cycle which is represented in toric geometry by the integer internal point of the toric
diagram. The size of this cycle becomes finite when we blow up the orbifold. It is conceivable
that the inclusion of compact four-cycles such as this one will affect the description of the classical
configuration space of D3-branes. This will add a new parameter, related to the group of
divisors on the CY resolution, which has dimension one.}
\item{We could distinguish among elementary fields and types of determinants
by using all the possible baryonic charges, including the anomalous ones.
For $\mathbb{C}^3/\mathbb{Z}_3$ this would lead to the inclusion of the
two existing anomalous baryonic charges. As we will explain,
this set of charges is different and complementary with respect to that
related to the group of divisors on the resolution. }
\end{itemize}

Having this example in mind, we now discuss the two possible expansions
of the ring of invariants. Explicit examples of the decompositions will be
presented in Section \ref{examples}. We encourage the reader to jump forth
and back with Section \ref{examples} for a better understanding of the
material.

\subsection{Expanding in a Complete Set of Baryonic Charges}\label{full}

The most general decomposition of the $g_1(\{t_i\})$ generating function is according to the
full set of baryonic charges, including the anomalous ones, denoted by $a_i$. The sectors ${\cal
S}$ in this case correspond to sectors with definite anomalous and non-anomalous baryonic
charges.

There are $g-1$ independent baryonic charges, where $g$ is the number of gauge
groups. By gauging the baryonic symmetries, we would obtain a quiver theory
with the same fields and superpotential, and a gauge group
$$\prod_{i=1}^g U(N)\, /\, U(1) ,$$
where we factor out the overall decoupled $U(1)$. Some of the $U(1)$ factors will be anomalous,
of course. The baryonic charges have a very natural description: they  correspond to the $U(1)$
factors in $U(N)^g/U(1)$. In this way, different elementary fields have the same baryonic
charges if and only if they are charged under the same gauge groups. This allows to efficiently
distinguish between invariants belonging to different sectors. Notice that non-anomalous
baryonic symmetries alone would not distinguish all inequivalent possibilities. For example, in
$\mathbb{C}^3/\mathbb{Z}_3$, the mesonic operator $\det {\bf X}_{12}^1\det {\bf X}_{23}^1\det {\bf X}_{31}^1$ and the
determinant $(\det {\bf X}_{12}^1)^3$ have the same charge under $\mathbb{Z}_3$, but different charges
under the two anomalous baryonic symmetries.

Let us thus extend the set of chemical potentials $t_i$ to all the baryonic charges, including
the anomalous ones.  We can therefore decompose ${\cal R}_{N=1}^{inv}$ into sectors with definite
charges under $U(1)^{g-1}$.

The $N=1$ generating function $g_{1}(\{t_i\})$ will decompose into a formal Laurent series in
the baryonic chemical potentials $b_i$ and $a_i$. The explicit decomposition of $g_1$ into a
formal Laurent series can be done by repeatedly applying the residue theorem, in a similar way to what we explained in the 
previous chapter; the computation
however quickly becomes involved, since the order of integration becomes crucial and divides the
result into many different cases. 

The ring of invariants ${\cal R}_{N}^{inv}$, will similarly decompose
into sectors of definite baryonic charges. The generating function $g_N(\{t_i\})$ can then be
computed according to Equation (\ref{g1plet}).

We can understand this decomposition in terms of representation theory.
%We can formally extend the gauge group $SU(N)^G$ to
%$U(N)^G/U(1)$  by gauging the baryonic symmetries
%(and discarding the trivial $U(1)$).
>From this perspective, we have decomposed the ring of $SU(N)^g$ invariants into Abelian
representations of the extended group $U(N)^g/U(1)$. This is sometimes called an expansion in
{\it covariants} and is extremely natural from the point of view of invariant theory. From our
point of view, covariants are just the possible set of independent {\it determinants}.   Each
sector $\cal S$ in ${\cal R}_N^{inv}$ will be specified by a certain number of gauge group pairs $(\alpha_i,\beta_i)$
and is associated to the subsector of the ring of gauge invariants made with products of the
{\it determinants} of type $(\alpha_i,\beta_i)$.

To make connection with the toric quiver gauge theories we notice that we
have the relation \cite{Feng:2002zw}:
\begin{equation}
g-1=2I+(d-3)
\end{equation}
where $d$ in the number of vertices of the toric diagram and $I$ the number of
integer internal points.
Only $d-3$ of these baryonic symmetries are not anomalous.

We expect all sectors ${\cal S}$
to appear with multiplicity one in the decomposition of
Equation (\ref{g1plet}). In the following we will show some examples of this decomposition.

\subsection{Supersymmetric D3-Branes and the GKZ Decomposition}\label{D3sup}
As explained in chapter \ref{D3SE} the full set of BPS states of the dual gauge theory can be obtained by quantizing the classical
configuration space of supersymmetric D3-branes wrapped on the horizon H. D3-brane configurations are in 
one-to-one correspondence with holomorphic four-cycles in the CY variety $\cX=C(H)$.
%This is clear for static D3-branes wrapping a three-cycle in the horizon: the
%corresponding divisor is the cone over the three-cycle. For more general configurations of
%excited and rotating D3-branes we obtain a four-cycle by a Euclidean continuation: we can
%replace time with the radial coordinate using the isometries of (Euclidean) $AdS_5$. 
Indeed our problem
can be equivalently rephrased in terms of holomorphic surfaces in $\cX$ with $g_1$ as a
generating function.

>From this perspective, we have an obvious decomposition into sectors ${\cal S}$ corresponding to
Euclidean D3-branes that can be continuously deformed into each other in $\cX$. Such D3-branes
have the same non-anomalous baryonic numbers; indeed, geometrically, the non-anomalous baryonic
charges are identified with the group of divisors modulo linear equivalence. 
%Let us discuss this
%point in detail for toric varieties since it will be crucial in the following.

Recall from section \ref{toric} that our conical CY $\cX$ is specified by a toric diagram in the plane with $d$ vertices having
integer coordinates $V_i$. 
%By embedding the plane in three dimensions we have a toric cone with
%edges $V_i=(n_i,1)\in \mathbb{Z}^3$ (the toric fan of our conical CY). Call the set of edges
%$\Sigma (1) $. Assign a ``homogeneous coordinate'' $x_i$ to each $V_i \in \Sigma (1) $; 
To each vertices $V_i$ we can associate an homogeneous coordinate $x_i$, spanning $\mathbb{C}^d$, in term of which $\cX$ 
is defined by the symplectic quotient: $\cX = (\mathbb{C}^d \ \backslash \ S)/ G$, where S is some set, that in our affine case will be always the null set, and the group G is defined in equation (\ref{Ggroup}), (\ref{Ggroupa}).
%Consider the group
%\begin{equation}
%K=\left\{(\mu_1,...,\mu_d)\in (\mathbb{C}^*)^d \left | \prod_{i=1}^d \mu_i^{\langle m,V_i\rangle } =1, m\in \mathbb{Z}^3 \right. \right\} ,
%\label{Kgroup}\end{equation}
%which acts on $x_i$ as
%$$(x_1,...,x_d) \rightarrow (\mu_1 x_1,...,\mu_d x_d)\, .$$
$G$ is isomorphic, in general, to $(\mathbb{C}^*)^{d-3}$ times a discrete group.
%Then, 
%where $\Delta$ is a subset fixed by the action of $K$. 
%Geometrically, the $x_i$ can be
%interpreted as homogeneous coordinates on the CY, just like the familiar coordinates for
%projective spaces. 
The residual $(\mathbb{C}^*)^{3}$ complex torus action acting on $\cX$ is
dual to the flavor symmetry group in the gauge theory, while the group $G$ is dual to the
non-anomalous baryonic symmetry group. Notice that the flavor and baryonic symmetries nicely
combine in the full group of $d$ non-anomalous charges which act naturally on the $d$
homogeneous coordinates $x_i$ as $(\mathbb{C}^*)^d$ acts on $\mathbb{C}^d$. 
%In the tiling
%construction, the $x_i$ are used to assign the non-anomalous charges to each field in the
%quiver \cite{Franco:2005sm,Benvenuti:2005cz,Butti:2005vn}.

In toric geometry each edge $V_i$ determines a (not necessarily compact)
four-cycle $D_i$ in $\cX$, and a generic four-cycle $D$ is given by a linear combination of basic
divisors $D=\sum_{i=1}^d c_i D_i$ where $c_i$ are integer coefficients. The divisors are subject to the three linear 
equivalence conditions $\sum_{i=1}^d \langle e_k, V_i\rangle  D_i = 0$, $k=1,2,3 $,
where $e_k$ is a basis for $\mathbb{Z}^3$. The group of four-cycles
modulo linear equivalence is isomorphic to the baryonic group $G$. 
It follows that the
non-anomalous baryonic symmetry distinguishes 
deformation equivalence classes of Euclidean
D3-branes.

However this is not the end of the story. The decomposition into non-anomalous baryonic charges
is not fine enough. A D3-brane state with baryonic charge $B$ can form a sort of bound state which
distinguishes it from a set of $B$ D3-brane states with baryonic charge one. This typically
happens in theories where there are elementary fields with multiple non-anomalous baryonic
charges. By going over examples, it easy to convince
oneself that the classical
D3 brane configurations obtained from divisors on the singular CY $\cX$ do not
exhaust all possible sectors of the dual gauge theory. However,
as already mentioned, we have a plethora of compact vanishing four-cycles that are
expected to enter in the description of the set of supersymmetric D3-branes and solve these
ambiguities. We have exactly $I$ compact vanishing four-cycles, one for each integer internal
point in the toric diagram. These cycles become of finite size in the smooth resolutions $\tilde\cX$ of $\cX$. We will see that with the addition of these divisors
we can give a convenient description of all sectors in the dual gauge theory.
It would be interesting to understand the necessity for the inclusions of
these divisors directly from the point of view of the geometric quantization
of classical supersymmetric branes living on the horizon.

We are led to enlarge the set of basic divisors of size $d-3$ to a larger set of size
$d-3+I$ by adding a divisor $D_i$ for each internal point of the
toric diagram. We now have a larger group of divisors which strictly
contains the baryonic symmetry group. The larger set of divisors
immediately leads us to the description of the K\"ahler moduli space $\cX$, of dimension $d-3+I$.
This moduli space is still a toric variety described by the so-called secondary fan,
or GKZ fan, and it is indeed parameterized by the divisors $D_i$ in the larger set.

\subsubsection{The GKZ Decomposition}\label{dec}

It is well known that there are many different smooth resolutions of the CY corresponding to the
possible complete triangulations of the toric diagram. Different resolutions are connected by
flops. The number of K\"ahler moduli of the CY is $I+d-3$ where $I$ is the number of internal
points; this is the same as the number of geometrical FI terms that appear in the symplectic
quotient description of the resolved manifold. This number can be greater than the number of
non-anomalous baryonic symmetries in field theory, which is $d-3$.

There is an efficient description of the K\"ahler moduli space in terms of divisors \cite{Cox:2000vi}. Take a
complete resolution of the variety and consider the set of all effective divisors

\begin{equation}
\left\{ \sum_{i=1}^{d+I} c_i D_i, \ \mbox{such that} \ c_i\ge 0 \hbox{ , }\forall i \right\}
\end{equation}

modulo the three linear equivalence conditions given by $\sum_{i=1}^{d+I} \langle e_k,V_i\rangle  D_i=0$ where $e_k$ is the standard basis for $\mathbb{Z}^3$ and $V_i$ are the vertices of the toric
diagram, including the internal integer points. The $c_i\in \mathbb{R}^+$ give a parametrization of the $d+I-3$ dimensional K\"ahler
moduli provided we impose a further condition: to have all cycles of positive volumes we must
consider only {\bf convex} divisors. The convexity conditions can be expressed as follows.
Assign a number $c_i$ to the $i$-th point in the toric diagram. To each triangle $\sigma$ in
the triangulation of the toric diagram we assign a vector $m_\sigma \in \mathbb{Z}^3$ which is the integer solution of the system of three linear equations\footnote{ Actually, these equations can be solved for all simplicial resolutions, corresponding to not necessarily maximal triangulations of the toric diagram. If we allow triangles with area greater than one, we have resolved varieties which still have orbifold singularities. For completely smooth resolutions, the vertices of all triangles $\sigma$ are
primitive vectors in $\mathbb{Z}^3$ and Equation (\ref{weights}) has integer solutions. Observe that, to avoid ambiguities, we have changed a bit the notation from the one we used in chapter \ref{D3SE}.},

\begin{equation}
\langle  m_\sigma, V_i \rangle  = -c_i,\,\,\,   i\in \sigma
\label{weights}
\end{equation}

 and impose the inequalities

\begin{equation}
\langle  m_\sigma, V_i \rangle \,  \ge \,  -c_i,\,\,\,   i\not\in \sigma
\label{convexity}
\end{equation}

The set of inequalities (\ref{convexity}), as $\sigma$ runs over all the triangles, determines
the convexity condition for the divisor. For a given resolution, the set of convex divisors forms a 
cone in the $\mathbb{R}^{d+I-3}$ vector space, that parameterizes the K\"ahler moduli of the
resolution. The boundary of this cone corresponds to the vanishing of some cycle. If
we can perform a flop, we enter a new region in the moduli space corresponding to a different
resolution. Indeed, the cones constructed via the convexity condition for the various
possible resolutions of the toric diagram form regions in the $\mathbb{R}^{d+I-3}$ vector space
that are adjacent; altogether these reconstruct a collection of adjacent cones
(a fan in toric language) in $\mathbb{R}^{d+I-3}$. The toric variety
constructed from this fan in $\mathbb{R}^{d+I-3}$ is the
K\"ahler moduli space of the CY.
This is known as the GKZ fan, or secondary fan \cite{OdaTadao:19910900,Gelf}.
We move from a cone in the
GKZ fan to another by performing flops (or in case we also consider orbifold resolutions
by flops or further subdivisions of the toric diagram). It is
 sufficient for us to consider smooth varieties and we thus reserve
the name {\bf GKZ fan} to the collections of cones corresponding to smooth
resolutions.

The GKZ fans %for the conifold, $\mathbb{C}^3/\mathbb{Z}_3$,
for the conifold and $\mathbb{F}_0$ are given in
Figures \ref{GKZconifold},
%\ref{dP0quiver},
\ref{FF}, respectively.

We form a lattice by considering the integer points in the GKZ fan.
We claim that the $N=1$ generating function, or in other words the Hilbert series for the Master space that we computed in chapter \ref{Master}, has an expansion in sectors
corresponding to the integer points of the GKZ lattice. Denote by $P$ an integer point in the GKZ lattice, then

\begin{equation}
g_1(\{t_i\})= \sum_{P\in GKZ} m(P) g_{1,P}(\{t_i\}) , \label{GKZex}
\end{equation}

where $m(P)$ is the multiplicity of the point $P$. Furthermore, we conjecture that the finite $N$ generating function can be obtained as

\begin{equation}
\sum_N g_N(\{ t_i\}) \nu^N =\sum_{P\in GKZ} m(P) {\rm PE}_\nu [g_{1,P}(\{t_i\})]
\end{equation}

\subsubsection{GKZ and Field Theory Content}
\label{auxiliary}

At the heart of the previous formulae, there is a remarkable connection between the GKZ
decomposition and the quiver gauge theory. To fully appreciate it we suggest to the reader to
read this and the following subsections in close parallel with section \ref{examples} where explicit examples
are given.

The integer points in the GKZ fan correspond to
sectors in the quantum field theory Hilbert space made out of determinants. Recall that mesons in
the quiver gauge theories correspond to closed paths in the quiver. We want to associate
similarly the other independent sectors made out of determinants with equivalence classes of open
paths in the dimers. The open paths fall into equivalence classes $A$
specified by the choice of ending points on the dimer.
The open path in a given class can be reinterpreted in the gauge
theory as strings of elementary fields with all gauge indices contracted
except two corresponding to a choice of a specific pair of gauge groups;
let us call these {\it composite fields}.
Baryonic operators are written as  ``$\det
{ A}$'', which is a schematic expression for two epsilons contracted with $N$ composite fields
freely chosen among the representatives of the class $A$. Generic sectors are made with
arbitrary products $\det A \det B ....$etc. Whenever open paths $A$ and $B$ can be composed to
give the open path $C$, there is at least one choice of representatives for $A$ and $B$ such that
we can write $\det A \det B=\det C$ and we want to consider the two sectors $\det A\det B$ and
$\det C$ equivalent. This can be enforced as follows. Denote with letters $a,b,c...$ the
equivalence classes of arrows in the quiver connecting different gauge groups. By decomposing
open paths in strings of letters, we can associate a sector with a string of letters. We should
however take into account the fact that if, for example, the arrows $a,b,c$ make a closed loop,
the operator $$(\det a)( \det b)( \det c)=\det abc$$ is a meson. We take into account this fact
by imposing the constraint $abc=0$. Moreover, composite fields connecting the same pairs of
gauge groups as an elementary fields do not determine the existence of new independent
determinants; to avoid overcountings, the corresponding string of letters should be set to zero.
Analogously, whenever two different strings of letters correspond to open paths
with the same endpoints, these strings should be identified.
We call ${\cal I}$ the set of
constraints obtained in this way and construct the ring
$${\cal R}_{GKZ}=\mathbb{C}[a,b,c...]/{\cal I}$$

Quite remarkably, the monomials in the ring ${\cal R}_{GKZ}$ are in correspondence with the integer
points in the GKZ fan. More precisely, we can grade the ring with $d-3+I$ charges in such a way
that the generating function of ${\cal R}_{GKZ}$, which we call the {\bf auxiliary GKZ partition
function}, and is denoted by $Z_{\rm aux}$, counts the integer points in the GKZ fan. Moreover, any integer point $P$ comes with
a {\bf multiplicity} $m(P)$ which is the one in Equation (\ref{GKZex}).

We have explicitly verified the above statement in all the examples we studied and we conjecture
that it is a general result for all toric diagrams: the auxiliary partition function counting
open paths in the quiver modulo equivalence coincides with the generating function for the GKZ
lattice dressed with field theory multiplicities. Just another remarkable connection between
apparently different objects: combinatorics on the tiling, geometry of $\cX$ and gauge theory!

%Even more fascinating, as we will discuss in detail in Section \ref{empty}, we can enlarge the
%grading of ${\cal R}_{GKZ}$ to the anomalous baryonic charges thus obtaining the partition function for
%a hollow cone over the GKZ fan where all points have now multiplicity one.

As another example of this fascinating correspondence, we will see that it is
possible to eliminate multiplicities
by refining the GKZ lattice. This is
done by enlarging the set of charges. In particular, we have found that,
if we refine the GKZ lattice by adding the anomalous baryonic charges,
we obtain a ``hollow cone'' in $\mathbb{Z}^{d-3+3 I}$ with no multiplicities.

As previously explained in a generic toric quiver gauge theory there are
 $2I$ anomalous baryonic symmetries, where $I$ is the number of internal
integer points. The variables in the auxiliary GKZ ring ${\cal R}_{GKZ}$ corresponds to arrows
in the quiver and therefore can be assigned a definite charge under the baryonic symmetries that
are just the ungauged $U(1)$ gauge group factors. We can therefore grade the auxiliary ring
${\cal R}_{GKZ}$ with a set of $d-3+3 I$ weights: the original $d-3 +I$ discretized K\"ahler
parameters $(\beta_1,...,\beta_{d-3+I})$ of the GKZ lattice plus the $2 I$ anomalous baryonic
weights $a_i$. The power series expansion of the auxiliary partition function
$Z_{aux}(\beta_i,a_i)$ will draw a lattice in $\mathbb{Z}^{d-3+3 I}$ which has the shape of an
hollow cone over the GKZ fan: over each point of the GKZ fan there is a hollow polygon $C(P)$
whose shape is related to the pq-web of the toric geometry. Quite remarkably, all points in
the lattice come with multiplicity one. Examples for $\mathbb{C}^3/Z_3$ and $\mathbb{F}_0$
are presented in Figure \ref{emptyC3} and \ref{emptyF0}.
%, \ref{emptydP1_1} and \ref{emptydP1_2}.

\subsubsection{Computing $g_{1,P}$ for one D Brane in a Sector $P$ Using Localization}\label{localization}

To compute the partition functions $g_{1,P}$ we need to apply the procedure explained in section \ref{counting} to 
every integer point $P$ in the GKZ fan. Indeed to every $P$ is associated a smooth resolution of $\cX$ and a
particular divisor $\sum c_i D_i$ on it.
Extending our interpretation of the BPS states in terms of
supersymmetric D3-branes wrapping holomorphic cycles in the singular CY to its resolution, 
we have a natural definition for the function $g_{1,P}$:
it  should count all
the sections of the line bundle ${\cal O} ( \sum c_i D_i)$ corresponding to holomorphic surfaces
in the given linear equivalence class. Therefore $g_{1,P}$ is just the character
${\rm Tr} \left\{ H^0\left(\cX,{\cal O}\left(\sum c_i D_i\right)\right) \bigr| q\right\}$
under the action of the element $q\in T^3$ of the torus of
flavor symmetries. All elements in
 $ H^0(\cX,{\cal O}(\sum a_i D_i))$ have the same baryonic charges. As we have explained section \ref{counting} 
the higher cohomology groups
of a convex line bundle vanish and the character then coincides with the
holomorphic Lefschetz number that can be computed with the equivariant index
theorem. Following section \ref{counting} we can expresses
the result as a sum over the fixed points $P_I$ of the torus $T^3$ action on
the particular smooth resolution of $\cX$ corresponding to the point $P$
in the GKZ lattice: 
\begin{equation}
g_{1,P}(\{t_i\}; \cX) = t^{n_P}\sum_{P_I} \frac{q^{m^{(I)}_P}}{\prod_{i=1}^3 (1-q^{m^{(I)}_i})} ,
\label{loc}
\end{equation}
where the index $I$ denotes the set of isolated fixed points
and the four vectors
$m^{(I)}_i,\, i=1,2,3$, $m^{(I)}_P$ in $\mathbb{Z}^3$ are the weights of the linearized action of $T^3$ on the resolved CY and the fiber of the line bundle
${\cal O} (\sum c_i D_i)$, respectively. The fixed points of the
torus action are in correspondence
with the triangles in the subdivision of the toric diagram.
The vectors  $m^{(I)}_i,\, i=1,2,3$ in the denominator of
Equation (\ref{loc}) are computed as the three primitive inward normal vectors
of the cone $\sigma_I$ in $\mathbb{Z}^3$ made with the three vertices $V_{i}$ of
the I-th triangle. The vector $m^{(I)}_0$ in the numerator is instead computed as in Equation (\ref{weights}): $\langle  m^{(I)}_0, V_i \rangle  = -c_i,\,\,\,   i\in \sigma_I $. The prefactor $t^{n_P}$ in Equation (\ref{loc}) is just the charge
of the divisor $\sum c_i D_i$. The full dependence on baryonic charges
in encoded in this prefactor.

In explicit computations, some care should be paid to the choice of charges. There is a natural
geometric basis for the non-anomalous charges of the gauge theory. In fact, the homogeneous
coordinates $x_i$ that are used to define $\cX$ as a symplectic quotient are extremely useful 
to assign a full set of $d$ (flavor+baryonic) charges to the
elementary fields in the quiver; this is done using zig-zag paths and standard dimers techniques
\cite{Butti:2005vn,Butti:2005ps}. All the elementary fields have charge which is given by a
product of the $x_i$. We can also assign charge $x_i$ to the divisor $D_i$ of the singular cone.
In all the examples we have considered there is a natural way to assign charges to the enlarged
set of divisors entering the GKZ decomposition. This allows to compute the prefactor $t^{n_P}$.
The $x_i$ decompose into three flavor charges and $d-3$ baryonic charges. The splitting of the
charges $x_i$ into flavor and baryonic charges is not unique in general; flavor charges can be
always redefined by adding a linear combination of the baryonic charges.
However, a toric diagram comes with a specific basis for the flavor $T^3$ action
which is determined by the equation
\begin{equation}
q_k=\prod_{i=1}^d x_i^{\langle e_k,V_i\rangle}\,\qquad\qquad k=1,2,3,
\label{rel}
\end{equation}
where $e_k$ are the basis vectors of $\mathbb{Z}^3$ and $V_i$ the vertices of
the toric diagram. Notice that all dependence on baryonic charges drops from
the right hand side. This is the $T^3$ basis
that should be used in the localization formula (\ref{loc}).

\subsubsection{Checks with all Charges: GKZ Approach vs. Field Theory}\label{checks}

%%%%%Now that we have understood how to resolve the multiplicities in the GKZ cones we would like to check our computations against the one done directly in field theory. Eventually we would like to write down generating functions counting operators without multiplicities: for each set of charges there is just one operator. To reach this task we will use all the $U(1)$ symmetries of our theories: anomalous and non-anomalous.\\
%As previously explained in a generic quiver toric gauge theory there are 2 flavor symmetries, 1 $R$ symmetry, $d-3$ non-anomalous baryonic symmetries, and $2I$ anomalous baryonic symmetries. Where $d$ is the number of points in the perimiter of the toric diagram and $I$ is the number of internal integer points.\\
%The baryonic symmetries are related to the possible formal $FI$ terms we can turn on in the field theory. If the theory has $G$ gauge groups then the numbers of $FI$ terms, and hence $U(1)$ baryonic symmetries, are $G-1$, because the diagonal $U(1)$ always decouples. Using toric geometry we have the relation:
%\begin{equation}
%G-1=2I+(d-3)
%\end{equation}
%In this short section we show the generating functions for $N=1$ with all the charges for the theories previously analized, namely: $\mathbb{C}^3/\mathbb{Z}_3$, $\mathbb{F}_0$, $dP1$. We computed the generating functions in two different ways and the two results nicely match.\\

Having understood how to compute and resolve the multiplicities in the GKZ cones and 
to compute the partition functions we can also refine our decomposition
of the $N=1$ generating function by adding the anomalous baryonic charges.

Using the equivariant index theorem we compute all the generating functions $g_{1,P}$ for all
points $P$ in the GKZ lattice. We can use $d-3+I$ coordinates for the GKZ cone
$\beta=(\beta_1,...,\beta_{d-3+I})$. Denote also with $B(\beta)$ the non-anomalous baryonic
charge corresponding to the point $P$ of the GKZ lattice. As discussed in Section
\ref{localization}, the generating functions depend on the baryonic charges only by a
multiplicative factor: $g_{1,P}=b^{B(\beta)} g_{1,\beta}(q)$ and all the other dependence is on
the flavor charges $q_i$. Thanks to the auxiliary generating functions we were able to find
expressions for the multiplicities $m(\beta)$ of the fields over each point of the GKZ
lattice. These functions sum up to the complete generating functions with $N=1$ and with all the
non-anomalous charges:
\begin{equation}
g_1(q,b)=\sum_{P \hbox{ }\in \hbox{ }GKZ} m(P) \hbox{ }g_{1,\beta}(q) \hbox{ }b^{B(\beta)}
\end{equation}
where $b$ are the chemical potentials for the non-anomalous baryonic charges.
To resolve the multiplicities $m(\beta)$ we construct the hollow cone by
adding the anomalous baryonic charges. Over each point in the GKZ lattice there
is a hollow polygon $C(\beta)$ which can be parametrized
in terms of the set of $2I$ anomalous charges $a_j$
with $j=1,...,2I$, such that:

\begin{equation}
\sum_{K_j \hbox{ } \in \hbox{ } C(\beta)} a_1^{K_1}...\hbox{  }a_{2I}^{K_{2I}} \Big|_{(a_1=1,\hbox{  }...,\hbox{ }a_{2I}=1)} = m(P)
\end{equation}

Using these resolutions we obtain the resolved generating functions for $N=1$ with all the charges, anomalous and non-anomalous:

\begin{equation}\label{allgkz}
g_1(q,b,a)=\sum_{\beta \hbox{ }\in \hbox{ }GKZ}\sum_{K_j \hbox{ }\in\hbox{ } C(\beta)} a_1^{K_1}...\hbox{ }a_{2I}^{K_{2I}} \hbox{   } b^{B(\beta)}\hbox{  } g_{1,\beta}(q)
\end{equation}

The non flavor charges do not appear in the basic generating functions $g_{1,\beta}(q)$, but
they are multiplicative factors over which one has to sum up, in the same way one does for the
usual non-anomalous baryonic charges.

We would like to stress that Equation (\ref{allgkz}) points to a remarkable connection between the
geometry of $\cX$, which is used to compute the right hand side, and the
field theory, that can be used to determine the left hand side (the
$N=1$ generating function).
In other words, we have two different ways of computing the $N=1$ generating
$g_1(q,b,a)$ function which nicely match:

\begin{itemize}
\item{In the first case we use the GKZ geometric picture explained in the previous sections. We first compute the generating functions $g_{1,\beta}(q)$ for each point in the GKZ lattice which depend on the flavor charges.
We next sum over all the points of the hollow cone by dressing
$g_{1,\beta}(q)$ with the appropriate weight under the baryonic symmetries.}
\item{In the second case we use the field theory picture. We can take
the fields of the gauge theory as basic variables
and we assign to them all the possible charges anomalous and non-anomalous.
This means that we
construct the ring generated by the fundamental fields and we grade it with all the charges. Then we construct the quotient ring by modding out the  ring of elementary fields by the ideal
generated by $F$-term equations. Using Macaulay2 we compute the Hilbert series of the quotient
ring obtaining the completed resolved generating function with all the charges of the field
theory: $g_1(q,b,a)$.}
\end{itemize}
As we will explicitly demonstrate on the examples in Section \ref{examples},
the two computations completely agree.

As a final remark, we notice that Equation (\ref{allgkz}) is the most general decomposition of
the $N=1$ generating function that we can write. We can close our circular discussion and go
back to the initial point. Equation (\ref{allgkz}) has been obtained by enlarging the GKZ
lattice in order to eliminate multiplicities. The hollow cone is a lattice in $d-3+3I$
dimensions. The corresponding $d-3+3I$ charges contain, as a subset, all the anomalous and
non-anomalous baryonic charges that are in number $d-3+2I$. Notice that the terms in series in
Equation (\ref{allgkz}) depend on the extra $I$ GKZ parameters only through the factor
$b^{B(\beta)}$. By projecting the hollow cone on the  $d-3+2I$ dimensional space of baryonic
charges we obtain the explicit expansion of the $N=1$ generating function $g_1(q,b,a)$ in a
complete set of baryonic charges which was discussed in Section \ref{full}. One can compare this
expansion with the one obtained by performing repeated contour integrations. As one can check
explicitly, the points in the baryonic charge lattice have still multiplicity one.

\section{Examples}\label{examples}

In this section we explicitly compute the $N=1$ generating function for certain toric CY manifolds $\cX$ 
and decompose it. Then we apply the PE function to every sector and we obtain the generating
functions for $N>1$. We start by revisiting the example of the conifold, and then we present 
the example of $\mathbb{C}^3/\mathbb{Z}_3$ and $\mathbb{F}_0$.

\subsection{The Conifold Revisited}\label{conif}

To demonstrate our general discussion above and to prepare for more involved cases we start by
reviewing the generating function for the conifold.
The generating function for the conifold can be written as \ref{g1coniC}:

\begin{equation}
g_1(t_1,t_2,x,y; {\cal C}) = \frac{1}{(1-t_1 x) (1-\frac{t_1}{x})(1-t_2 y) (1-\frac{t_2}{y})}.
\label{g1coniCCC}
\end{equation}

Let us start setting $x=y=1$ for simplicity. 
%General formulae including the $SU(2)$ chemical potentials can be found in the previous chapter and in Section \ref{examples}.

$g_1$ decomposes into sectors with fixed baryonic charge $B$, each with multiplicity
one:
\begin{eqnarray}
g_1(t_1,t_2; {\cal C}) &=& \sum_{B=-\infty}^\infty g_{1,B}(t_1,t_2; {\cal C}),
\nonumber\\
g_{1,B>0}(t_1,t_2; {\cal C}) &=&\sum_{n=0}^\infty (n+1+B)(n+1) t_1^{n+B} t_2^{n}\nonumber\\
 g_{1,B<0}(t_1,t_2; {\cal C}) &=&\sum_{n=0}^\infty (n+1)(n+1+|B|) t_1^{n} t_2^{n+|B|} \nonumber\\
 g_{1,0}(t_1,t_2; {\cal C}) &=&\sum_{n=0}^\infty (n+1)^2 t_1^{n} t_2^{n}
\label{g1conit}
\end{eqnarray}

\begin{figure}[h!!!]
\begin{center}
\includegraphics[scale=0.65]{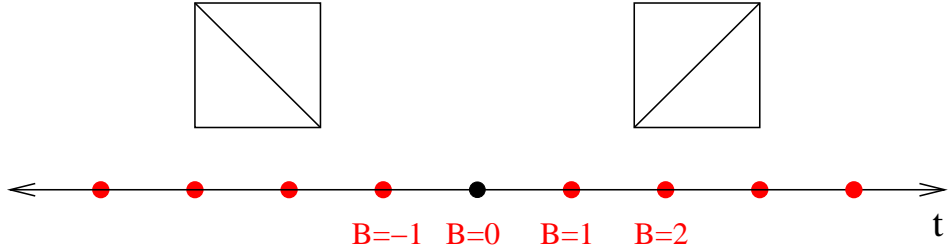}
\caption{The GKZ decomposition for the K\"ahler moduli space of the conifold, consisting of two
one-dimensional cones connected by a flop. The coordinate $t$ on the moduli space is associated
with the volume of the two-cycle in the resolution of the conifold. When $t$ goes to zero, the
cycle vanishes and we can perform a flop on the variety by inflating a different two-cycle. A
natural discretization of the GKZ fan is in correspondence with the decomposition of the $g_1$
generating function. }
\label{GKZconifold}
\end{center}
\end{figure}

It is manifest that each term in $g_{1,B}$ has a monomial $b^B$ corresponding to a baryonic
charge $B$. Observe that the decomposition in equation (\ref{g1conit}) shows explicitly that every sector with
fixed baryonic charge $|B|$ decompose in symmetric representations of dimension $(n+1+|B|)(n+1)$ of the 
global symmetry group $SU(2)_1 \times SU(2)_2$. As we explained in the previous two chapters the decomposition into each baryonic 
charge can be computed by expanding $g_1(t,b;{\cal C})$ in a  formal Laurent series in 
$b$ or by determining the functions $g_{1,B}$ by
resolving $\cX$, see Figure \ref{GKZconifold}, and using the equivariant index theorem. 
% Both
% computations have been discussed in detail in \cite{Forcella:2007wk} and will be reviewed in the
% following Sections.

It is right now clear that the decomposition has an interpretation in terms of supersymmetric D3-branes states:
$g_{1,B}$ counts the supersymmetric D3-branes wrapping cycles of homology $B$ thus corresponding
to states with baryonic number $B$. As already observed in the previous chapter quite remarkably, 
the structure of the integer lattices with
generating function $g_{1,B}$ and the explicit computation with the index theorem, strongly
suggests a relation between $B$ and a discretized K\"ahler modulus of the resolved CY
\cite{Forcella:2007wk}.

The result for generic $N$ is obtained as follows
\begin{equation}
g(\nu; t_1,t_2; {\cal C}) = \sum_{B=-\infty}^\infty {\rm PE}_\nu [g_{1,B}(t_1,t_2; {\cal C})] = \sum_{N=0}^\infty \nu^N g_{N}(t_1,t_2; {\cal C}) 
\end{equation}
and a list of generating functions for small values of $N$ is given in section \ref{Nis2}.

The conifold has only one baryonic charge, not anomalous, which can be
used to parametrize the K\"ahler moduli space. The two expansions,
one in baryonic charges, the other according to the GKZ lattice, coincide.

\subsubsection{Baryonic Charge Expansion}
We first expand the $N=1$ generating function, Equation (\ref{g1coniCCC}) for the
conifold according to the baryonic charge

\begin{equation}
g_1(t,b,x,y; {\cal C}) = \sum_{B=-\infty}^\infty b^B g_{1,B}(t,x,y; {\cal C}),
%\label{g1coni}
\end{equation}
$g_{1,B}(t,x,y; {\cal C})$ can be computed using the inversion formula

\begin{equation}
g_{1,B}(t,x,y; {\cal C}) = \frac{1}{2\pi i} \oint \frac {db} {b^{B+1}} g_1(t,b,x,y; {\cal C}) ,
\label{res1}
\end{equation}
with a careful evaluation of the contour integral for positive and negative values of the baryonic charge $B$ (see the previous
chapter for a detailed explanation): 
%For $B\ge0$ the contribution of the contour integral comes from the positive powers of the poles for $b$ ($b=x/t,1/(xt)$)
%while for $B\le0$ the contribution of the contour integral comes from the negative powers of the poles for $b$ ($b=t y,t/y$)
\begin{eqnarray}
g_{1,B\ge0}(t,x,y; {\cal C}) &=& \frac{t^B x^{B} } { (1 - \frac{1}{x^2}) (1-t^2 x y)  (1-\frac{t^2 x}{y}) }+ \frac{t^B x^{-B}} { (1 - x^2) (1-\frac{t^2 y} {x})  (1-\frac{t^2}{x y}) } , \nonumber \\ \nonumber
g_{1,B\le0}(t,x,y; {\cal C}) &=& \frac{t^{-B} y^{-B} } { (1-\frac{1}{y^2}) (1-t^2 x y)  (1-\frac{t^2 y}{x}) }+ \frac{t^{-B} y^{B}} { (1 - y^2) (1-\frac{t^2 x} {y})  (1-\frac{t^2}{y x}) } .\\
\label{rescon}
\end{eqnarray}
By setting $x=y=1$ and $t_1=b t,t_2=t/b$ we recover expansion (\ref{g1conit}).

\subsubsection{GKZ Decomposition}

We can similarly perform a GKZ decomposition of the $N=1$ generating function.
In Figure \ref{coooC} the toric diagram and the two resolutions of the conifold
are reported. There are four divisors $D_i$ subject to three relations that
leave an independent divisor $D$,
$D_1=D_3=-D_2=-D_4\equiv D$.
Consider the cone of effective divisors $\sum c_i D_i, \, c_i\ge 0$
modulo linear equivalence in $\mathbb{R}$
$$\sum_{i=1}^4 c_i D_i \equiv (c_1+c_3-c_2-c_4) D \equiv B D$$
where we defined $B=c_1+c_3-c_2-c_4$. For each resolution, we solve Equation (\ref{weights}) for
the two triangles in the resolution, or, equivalently, the two vertices of the pq-web; the
resulting vectors $m_i^{(I)}$ and $m_B^{(I)}$ are reported in black and red respectively in Figure \ref{coooC}.
%\begin{figure}[h!!!!!]
%\begin{center}
%\includegraphics[scale=0.5]{co2.eps}
%\caption{Localization data for the $N=1$ baryonic generating functions. The vertices $V_i$ are in
%correspondence with homogeneous coordinates $x_i$ and with a basis of  divisors $D_i$. Two
%different resolutions, related by a flop, should be used for positive and negative $B$,
%respectively. Each resolution has two fixed points, corresponding to the vertices of the
%pq-webs; the weights $m^{(I)}_i,\, i=1,2,3$ and $m^{(I)}_B$ at the fixed points are indicated in
%black and red, respectively.} \label{co2}
%\end{center}
%\end{figure}
The convexity condition, Equation (\ref{convexity}), then tells us that the resolution on the
left corresponds to $B>0$ and the resolution on the right to $B<0$. Altogether we obtain two
half lines (cones) in $\mathbb{R}$ that form the GKZ fan as in Figure \ref{GKZconifold}. The
point of intersection $B=0$ of the two cones corresponds to the singular conifold and the two
cones in the fan are related by a flop transition.

We now compute the generating functions $g_{1,B}$ using localization. As mentioned in the
previous section, we must pay attention to the normalization of charges. The homogeneous
coordinates for the conifold are extremely simple: $({\bf A}_1\, ,\, {\bf B}_1\, ,\, {\bf A}_2\, ,\, {\bf B}_2\,) \longrightarrow (x_1\, ,\, x_2\, ,x_3\, ,\, x_4)$, 
which can be easily translated in the notations of Section \ref{conif}.
The natural flavor $T^3$ basis is then given by Equation (\ref{rel})

\begin{equation}
q_1=x_1 x_2 = t^2 x y \hbox{  } \hbox{ , } \hbox{  } q_2=x_2 x_3 = \frac{t^2 y}{x} \hbox{  } \hbox{ , } \hbox{  } q_3=x_1 x_2 x_3 x_4 = t^4 .
\label{GLr}
\end{equation}

We are ready to apply the localization formula. Each point in the GKZ fan is associated with a
resolution and a divisor: for $B>0$ we use the resolution on the left in Figure \ref{coooC} and $B
D_1$, while for $B<0$ the resolution on the right and the divisor $|B| D_4$. The weights are
reported in Figure \ref{coooC}. Equation (\ref{loc}) and Equation (\ref{GLr}) give

\begin{eqnarray}
g_{1,B\ge0}(t,x,y; {\cal C}) &=& \frac{t^B x^{B} } { (1 - \frac{1}{x^2}) (1-t^2 x y)  (1-\frac{t^2 x}{y}) }+ \frac{t^B x^{-B}} { (1 - x^2) (1-\frac{t^2 y} {x})  (1-\frac{t^2}{x y}) } , \nonumber \\ \nonumber
g_{1,B\le0}(t,x,y; {\cal C}) &=& \frac{t^{-B} y^{-B} } { (1-\frac{1}{y^2}) (1-t^2 x y)  (1-\frac{t^2 y}{x}) }+ \frac{t^{-B} y^{B}} { (1-y^2) (1-\frac{t^2 x} {y})  (1-\frac{t^2}{y x}) } .
\end{eqnarray}
which coincides with the result previously obtained in Equation (\ref{rescon}).

\subsubsection{Conifold: Multiplicities in the GKZ Decomposition}

The multiplicities in the GKZ decomposition of the $N=1$ generating function for the conifold are trivial since they are all equal to 1. Nevertheless it is instructive to follow the procedure which is outlined in Section \ref{auxiliary} in order to compute the multiplicities using the auxiliary GKZ partition function $Z_{\rm aux}(t)$ which counts independent sectors in the ring of invariants. As explained in Section \ref{auxiliary}, we assign a letter $a,b$ to the two types of arrows in the quiver ${\bf A}_i,{\bf B}_i$. There is only one relation $ab=0$ corresponding to the closed loop in the quiver. The polynomial ring for the GKZ decomposition of the conifold is therefore

\begin{equation}
\label{ZGKZconifold}
{\cal R}_{GKZ} ( {\cal C} ) = \mathbb{C}[a,b]/(ab),
\end{equation}

We thus compute the generating function for the polynomial ring (which can be easily computed by observing it is a complete intersection), with chemical potential $t_1$ to $a$ and $t_2$ to $b$ we find

\begin{equation}
\label{Zauxconifold}
Z_{\rm aux}(t_1, t_2; {\cal C}) = \frac{1-t_1 t_2}{(1-t_1) (1-t_2)} = 1+ \sum_{B = 1}^\infty t_1^B
+ \sum_{B = 1}^{\infty} t_2^{B}
\end{equation}

By expanding this auxiliary partition function we find multiplicity $1$ for the integer points $B>0$, multiplicity 1 for the integer points $B<0$, and multiplicity 1 for the point $B=0$, reproducing the lattice depicted in Figure \ref{GKZconifold}.
We finally have

\begin{equation}
g_1(t_1,t_2) = g_{1,0}(t_1,t_2) + \sum_{B=1}^\infty g_{1,B}(t_1,t_2) + \sum_{B=-\infty}^{-1} g_{1,B}(t_1,t_2) ,
\end{equation}
which appears to be a trivial observation as a Laurent series in the baryonic chemical potential
but in fact turns out to be nontrivial for more involved singularities.

Now that we have rephrased the conifold example in the general setup of the GKZ decomposition, we can apply this general
method to more involved singularities $\cX$.
 
\subsection{Generating Functions for $\mathbb{C}^3/\mathbb{Z}_3$}
\label{N1C3Z3}

\begin{figure}[ht]
\begin{center}
  \epsfxsize = 10cm
  \centerline{\epsfbox{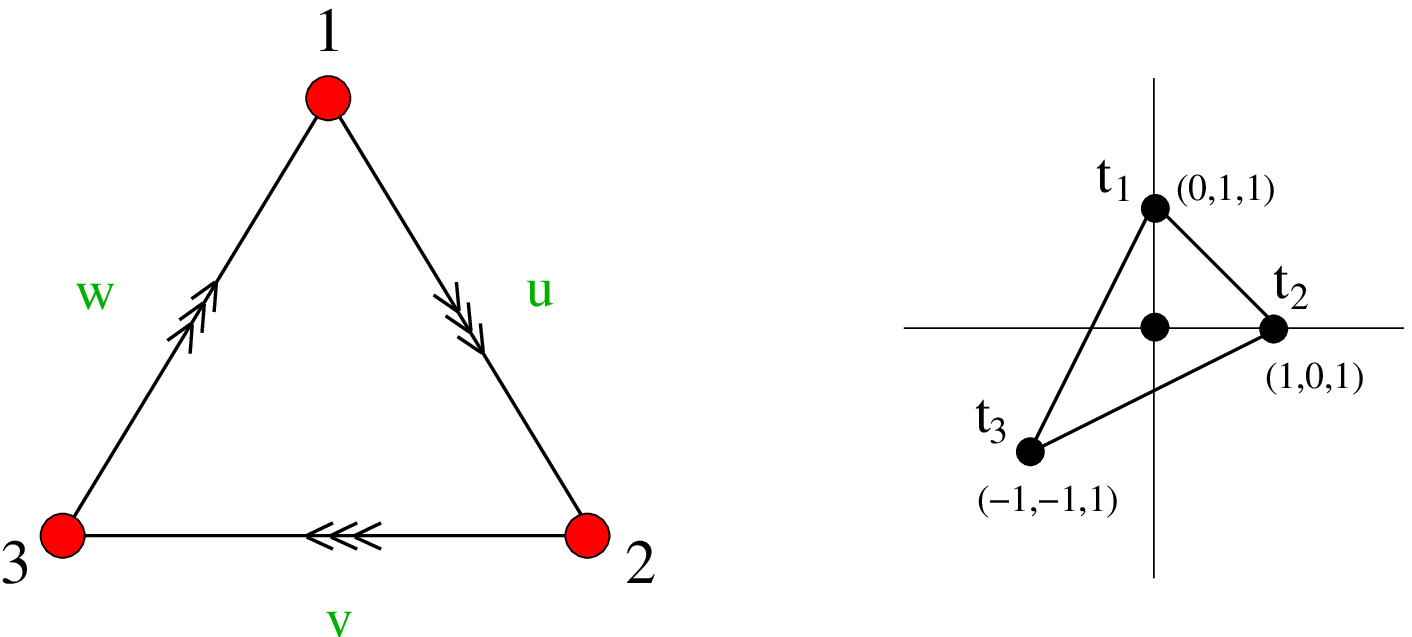}}
  \caption{Quiver and toric diagram for $\IC^3/\IZ_3$.}
  \label{dP0quiver}
\end{center}
\end{figure}

We have already analyzed the master space of the $\mathbb{C}^3/\mathbb{Z}_3$ in chapter \ref{Master}
: it is a five dimensional irreducible toric CY variety.  
For visual simplicity we report here in Figure \ref{dP0quiver} its quiver and toric diagram. 
The gauge theory has three sets of bifundamental fields ${\bf X}_{12}^{i},{\bf X}_{23}^{j},{\bf X}_{31}^{k}$ with $i, j, k=1,2,3$ 
and a superpotential $\epsilon_{i j k }{\bf X}_{12}^{i}{\bf X}_{23}^{j}{\bf X}_{31}^{k}$.

\paragraph{Symmetries and Geometry}

In chapter \ref{Master} we discovered that the master space $\f_{\mathbb{C}^3/\mathbb{Z}_3}$ has 
the symmetries $U(1)_R\times SU(3)_M \times SU(3)_H$, where the last factor is an ``hidden symmetry'' not manifest in the UV 
Lagrangian. In this section we will just consider the UV symmetries and we will come back to the hidden ones in the 
following chapter.

The UV global flavor symmetry is $SU(3)\times U(1)_R$. All the fields have R-charge $2/3$ 
and each set of three fields transform in the fundamental representation of $SU(3)$. 
One can also define two anomalous baryonic $U(1)$ charges which can be chosen to be

\bean
&A_1:& ({\bf X}_{12},{\bf X}_{23},{\bf X}_{31}) \mapsto (a_1 {\bf X}_{12},\ a_1^{-1}{\bf X}_{23},\ {\bf X}_{31}) \\
&A_2:& ({\bf X}_{12},{\bf X}_{23},{\bf X}_{31}) \mapsto (a_2 {\bf X}_{12},\ a_2 {\bf X}_{23},\ a_2^{-2} {\bf X}_{31})
\eean

There are some non-anomalous discrete symmetries \cite{Gukov:1998kn} acting on the fields as follows,
\bean
&A:& ({\bf X}_{12},{\bf X}_{23},{\bf X}_{31}) \mapsto ({\bf X}_{31},{\bf X}_{12},{\bf X}_{23}) \\
&B:& ({\bf X}_{12},{\bf X}_{23},{\bf X}_{31}) \mapsto (b{\bf X}_{12},\ b^{-1}{\bf X}_{23},\ {\bf X}_{31}) \quad \mbox{where } b^3 = 1 \\
&C:& ({\bf X}_{12},{\bf X}_{23},{\bf X}_{31}) \mapsto (c{\bf X}_{12},\ c{\bf X}_{23},\ c^{-2}{\bf X}_{31}) = (c{\bf X}_{12},\ c{\bf X}_{23},\ c {\bf X}_{31}) \quad \mbox{where } c^3 = 1
\eean
We see that $B$ is a subgroup of $A_1$ %On the other hand,
and $C$ is subgroup of $A_2$. $C$ is related to the torsion homology group for three cycles and, in a sense,
is a discrete baryonic charge.
%and the R-charge  (DON'T UNDERSTAND - C IS NOT AN R SYMMMETRY...??).
%The discrete transformations form the Heisenberg group with relations
%\be
%  AB=BAC\ , \ \
%  AC=CA\ , \ \
%  BC=CB\ , \ \
%  A^3=B^3=C^3=1
%\ee
%On the gravity side, we have $AdS_5 \times X_5$ where the Sasaki-Einstein manifold is now $X_5 =
%S^5/ \IZ_3$ with homology groups
%\bean
%  H_1(S^5 / \IZ_3, \IZ)=\IZ_3 \\
%  H_3(S^5 / \IZ_3, \IZ)=\IZ_3.
%\eean
%Hence, we can wrap fundamental strings, D1- and D3-branes on $S^1 / \IZ_3$ and $S^3 / \IZ_3$
%cycles in $S^5 / \IZ_3$. The number of such wrapped objects is conserved modulo 3 and measured
%by $A, B$ and $C$, respectively.\footnote{Since $A$ and $B$ do not commute, one cannot determine
%the number of F- and D-strings at the same time. This is an interesting consequence of RR-flux
%studied in detail in \cite{Gukov:1998kn}.}

%In this paper, we do not consider the operator $A$, because wrapped fundamental strings can be
%seen in the $1/N$ expansion and here we study baryons. We characterize the states by the charges
%under $B$ and $C$.

%\begin{figure}[ht]
%\begin{center}
%  \epsfxsize = 4cm
%  \centerline{\epsfbox{Z3}}
%  \caption{Toric diagram for $\IC^3/\IZ_3$. (SHOULD WE PUT THIS TOGETHER WITH
%THE QUIVER IN FIGURE 3?)}
%  \label{Z3orbifold}
%\end{center}
%\end{figure}

The symplectic quotient description of $\IC^3/\IZ_3$ is as follow.
Since $d=3$ we introduce three homogeneous
coordinates $x_i$. The group $G$ of baryonic charges is defined by the Equations (\ref{Ggroup}), (\ref{Ggroupa}):
$$\prod_{j=1}^3 \mu_j^{\langle e_i,V_j\rangle }=1$$ which implies
$\mu_j=c$ with $c^3=1$.
The symplectic quotient description of the CY just reduces to the orbifold
description $\mathbb{C}[x_1,x_2,x_3]/\mathbb{Z}_3$, as expected.

As already discussed, the homogeneous coordinates can be used to give a full set of weights for
non-anomalous symmetries to the elementary fields. We can write  $x_i= c t_i$ in terms of the
discrete baryonic charge $c$ and the flavors $t_i$. The assignment of charges to the fields is
done using standard dimer techniques  and is reported in Table \ref{chargesC3Z3}. One can notice
that $t_i$ are the charges of the original ${\cal N}=4$ SYM. To keep track of the coordinates on
the two dimensional projection in Figure \ref{dP0quiver} we introduce three chemical potentials
$t, x, y$ which count the $R$-charge, and the $(x,y)$ integral positions, respectively and read
from the figure $t_1=t y, t_2 = t x, t_3 = \frac{t}{x y}$.

The full set of continuous charges, anomalous or not, is summarized in the Table \ref{chargesC3Z3}, as well as the chemical potentials and the assignment of homogeneous coordinates to the fields.

%%%%%We notice that the homogeneous coordinates basis for charges is the most
%natural one from a geometric perspective and it exists for all type of toric
%CY.
%As usual, we denote with $b$ the chemical potentials for
%the non-anomalous baryonic charges and $a_i$ the chemical potentials for the anomalous baryonic
%charges. \
\begin{table}
\begin{center}
\begin{tabular}{|c|c|c|c|c||c||c|}
\hline
\ {\bf field} \ &   {\bf $SU(3)$}    & \ \  {\bf $R$} \ \  & \ \  {\bf $A_1$} \ \  & \ \  {\bf $A_2$}  \ \ &
\  chemical    \ & \  homogeneous \\
   & & & & & potentials & coordinates \\
\hline\hline $({\bf X}_{12}^1\, ,{\bf X}_{12}^2\, , {\bf X}_{12}^3) $  & ${\bf 3}$    & $\frac{2}{3}$  &   1 & $1$ & $(t a_1 a_2 y\, , t a_1 a_2 x\, ,t a_1 a_2 / x y)$  &
$(c t_1\, , c t_2\, , c t_3)$ \\
$({\bf X}_{23}^1\, ,{\bf X}_{23}^2\, , {\bf X}_{23}^3) $  & ${\bf 3}$    & $\frac{2}{3}$       & $-1$ & 1 & $(t a_2 y / a_1 \, , t a_2 x / a_1\, ,t a_2 / a_1 x y ) $  &
$(c t_1\, , c t_2\, , c t_3)$ \\
 $({\bf X}_{31}^1\, ,{\bf X}_{31}^2\, , {\bf X}_{31}^3) $ & ${\bf 3}$    & $\frac{2}{3}$ & 0 & $-2$  & $(t y / a_2^{2} \, , t x /a_2^{2} \, ,t / a_2^{2} x y)$  &
$(c t_1\, , c t_2\, , c t_3)$ \\
\hline
\end{tabular}
\end{center}
\caption{Global charges for the basic fields of the quiver gauge theory
living on the D-brane probing the orbifold $\IC^3 / \IZ_3$. The $x$ and $y$ chemical potentials count $SU(3)$ weights, while $A_1$ and $A_2$ count anomalous baryonic charges.}
\label{chargesC3Z3}
\end{table}

\subsubsection{The $N=1$ Generating Function}
The $N=1$ generating function, or Hilbert series for the master space $\f_{\IC^3/\IZ_3}$,
was computed in section \ref{s:master} with respect to just the R charge with the result:
% s generated by the elementary fields ${\bf U}_i,{\bf V}_j,{\bf
%W}_k$ modulo nine F-term relations which can be expressed through the ideal
%\begin{eqnarray}
% {\cal I}= ({\bf V}_2 {\bf W}_3-{\bf V}_3 {\bf W}_2,\ {\bf V}_1 {\bf W}_3-{\bf V}_3 {\bf
%W}_1, \ {\bf V}_1 {\bf W}_2-{\bf V}_2 {\bf W}_1, \nonumber \\
%{\bf U}_2 {\bf W}_3-{\bf U}_3 {\bf W}_2, \ {\bf U}_1 {\bf W}_3-{\bf U}_3 {\bf W}_1,\ {\bf U}_1
%{\bf W}_2-{\bf U}_2 {\bf W}_1,\nonumber \\ {\bf V}_2 {\bf U}_3-{\bf V}_3 {\bf U}_2,\ \ \ {\bf
%V}_1 {\bf U}_3-{\bf V}_3 {\bf U}_1,\ \ \ {\bf V}_1 {\bf U}_2-{\bf V}_2 {\bf U}_1) . \ \nonumber
%\end{eqnarray}
%Each field carries R-charge $\frac{2}{3}$ and therefore we can give to all the same weight for the chemical potential, $t$.
%Computing the Hilbert series of the polynomial ring
%\begin{equation}
%\label{ringC3Z3}
%{\cal R}_{N=1} ( \IC^3/\IZ_3 ) = \mathbb{C}[\{ {\bf U}_i\},\{{\bf V}_j\},\{ {\bf W}_k\}]/{\cal I}
%\end{equation}
%with Macaulay2 we obtain
\begin{equation}
g_1(t; \IC^3/\IZ_3)=\frac{1+4 t+t^2}{(1-t)^5}\label{N1} .
\end{equation}
%For simplicity we give weight $t$ to all variables.
%The dimension of an  irreducible algebraic variety $V$ can be computed from its Hilbert series
%$g(t)$  by looking at the order of the pole for $t\rightarrow 1$
%\begin{equation}
%g_1(t) \sim   \frac{A}{(1-t)^{{\rm dim} V}},
%\label{dim}
%\end{equation}
%with the residue $A$ a measure for the volume of this variety. The $N=1$ moduli space of vacua
%for $\mathbb{C}^3/\mathbb{Z}_3$ has dimension five, as can be seen from
%$$g_1(t)\sim \frac{6}{(1-t)^5}$$
%This can be understood as three mesonic directions describing the CY plus two independent
%baryonic parameters that correspond to the gauge theory FI terms. As usual, the mesonic moduli
%space for $N=1$ is isomorphic to the CY geometry. The additional baryonic parameters come from
%the fact that the gauge group is $SU(N)^G$ and not $U(N)^G/U(1)$. Since we do not have to impose
%the $U(1)$ D-term conditions, this leaves $G-1$ additional free parameters that can be
%identified with the FI terms in the gauge theory. Since in general the mesonic flat directions
%are given by the symmetrized product of $N$ CY's for $N$ D-branes, giving $3N$ parameters, the
%dimension of the moduli space for generic $N$ and $G$ is expected to be $3 N+G-1$.

%The $N=1$ generating function can be decomposed in two different ways: according to the two anomalous baryonic charges and according to the one-dimensional GKZ lattice.

\subsubsection{The GKZ Decomposition}

In the singular CY there is only one independent divisor $D_i\equiv D$ with $3 D=0$. This reflects the
$\mathbb{Z}_3$ discrete charge. However, on the smooth resolution of the orbifold there is a new
divisor $D_4$ corresponding to the internal point. $D_1=D_2=D_3=D$ is still true but now $3 D$
is non-zero, but equal instead to $-D_4$. The cone of effective divisors in $\mathbb{R}$ is
given by
$$\sum_{i=1}^4 c_i D_i = (c_1+c_2+c_3-3 c_4) D \equiv \beta D\, , \qquad \qquad c_i\ge 0$$
and the convexity condition, Equation (\ref{convexity}), requires $\beta\ge 0$. The GKZ fan is
thus a half-line in $\mathbb{R}$. The integer parameter $\beta$ turns out to be 
the discrete K\"ahler modulus of the resolution of $\mathbb{C}^3/\mathbb{Z}_3$, measuring the discrete area of the two cycle.

The right basis for localization is given
by $$q_i=\prod_{j=1}^3 x_j^{\langle e_i,V_j \rangle}$$
and we compute (notice that the discrete baryonic charge $c$ correctly drops out from this formula)
 $q_1=\frac{t_2}{t_3}, q_2=\frac{t_1}{t_3}, q_3=t_1 t_2 t_3$. We thus obtain\footnote{
These partition functions reduce for $\beta=0,1,2$ to the three independent partition functions
for nontrivial divisors on the singular cone.}

\begin{eqnarray}
\label{g1bC3Z3}
g_{1,\beta} (t_1,t_2,t_3) &=& \frac{t_1^\beta}{(1-t_1^3)(1-\frac{t_2}{t_1})(1-t_3/t_1)} \\
&+&\frac{t_2^\beta}{(1-t_1/t_2)(1-t_2^3)(1-t_3/t_2)}+\frac{t_3^\beta}{(1-t_1/t_3)(1-t_2/t_3)(1-t_3^3)}.
\nonumber
\end{eqnarray}

\vskip 0.3cm $g_{1,0}$ is identified with the Molien invariant for the discrete group
$\mathbb{Z}_3$ and indeed computes the mesonic generating function as explained in detail in
\cite{Benvenuti:2006qr}. In the limit $t_i=t$ we find

\begin{eqnarray}
\label{g1geomZ3}
g_{1,\beta} (t,t,t) = t^\beta \left ( \frac{1+7t^3+t^6}{(1-t^3)^3} + \frac{3\beta(1+t^3)}{2(1-t^3)^2} + \frac{\beta^2}{2(1-t^3)} \right ) .
\end{eqnarray}

\subsubsection{Multiplicities}

The multiplicities in the GKZ decomposition of the $N=1$ generating function can be computed
using $Z_{\rm aux}(t)$, the auxiliary GKZ partition function counting independent sectors in the
ring of invariants. As explained in Section \ref{auxiliary}, we assign a letter $u,v,w$ to the
three types of arrows in the quiver ${\bf X}_{12}^{i},{\bf X}_{23}^{j},{\bf X}_{31}^{k}$. There is only one relation
$uvw=0$ corresponding to the closed loop in the quiver. We thus get the polynomial ring

\begin{equation}
\label{ZGKZC3Z3}
{\cal R}_{GKZ} ( \IC^3/\IZ_3 ) =\mathbb{C}[u,v,w]/(uvw),
\end{equation}
and compute the generating function (which can be easily computed by assuming it is a complete intersection), with charge $t$ to all letters obtaining

\begin{equation}
\label{ZauxC3Z3}
Z_{\rm aux}(t ; \IC^3/\IZ_3 ) = \frac{1-t^3}{(1-t)^3} = 1+ \sum_{\beta = 1}^\infty 3\beta t^\beta .
\end{equation}

By expanding this auxiliary partition function we find
multiplicity $3 \beta$ for the point $\beta>0$ and multiplicity 1 for the point $\beta=0$. This is easily understood:
the independent sectors contain determinants of the form
$(\det {\bf X}_{12})^n (\det {\bf X}_{23})^m$ with $n+m=\beta$ or similar with ${\bf X}_{12},{\bf X}_{23},{\bf X}_{31}$ permuted;
there are $3 \beta$ such sectors. This point will be further elaborated below.

We finally have

\begin{equation}
g_1(t_1,t_2,t_3) = g_{1,0}(t_1,t_2,t_3) + \sum_{\beta=1}^\infty 3 \beta g_{1,\beta}(t_1,t_2,t_3) .
\label{expa}
\end{equation}

This can be summed easily using Equations (\ref{g1bC3Z3}) and (\ref{ZauxC3Z3}) and gives

\begin{eqnarray}
\label{g1C3Z3}
g_{1} (t_1,t_2,t_3; \IC^3/\IZ_3) &=& \frac{1}{(1-t_1)^3 (1-\frac{t_2}{t_1})(1-t_3/t_1)} \\
&+&\frac{1}{(1-t_1/t_2)(1-t_2)^3(1-t_3/t_2)}+\frac{1}{(1-t_1/t_3)(1-t_2/t_3)(1-t_3)^3}.
\nonumber
\end{eqnarray}

For the special case $t_1=t_2=t_3=t$ we can take the limit or resum, using Equation (\ref{g1geomZ3}),

$$g_1(t,t,t; \IC^3/\IZ_3)=\frac{1+4 t+t^2}{(1-t)^5}$$

which is exactly Equation (\ref{N1}).

\subsubsection{Refining the GKZ Decomposition}

Using Equation (\ref{ZauxC3Z3}) we summarize the multiplicities
\begin{equation}
  m(\beta) =
  \left\{ \begin{array}{ll}
 1  & \textrm{for} \ \beta=0 \\
 3\beta & \textrm{for} \ \beta > 0.
  \end{array}
  \right.
\end{equation}

For a dibaryon, the AdS/CFT dual object is a D3-brane that wraps a $C_3 = S^3 / \IZ_3$
cycle\footnote{Generically, $C_3$ is a Lens space.} in $S^5 / \IZ_3$. The homology tells us
that the wrapping number is characterized by an integer, modulo 3. By resolving the singular
Calabi-Yau, this gets promoted to a (non-negative) integer which is just the coordinate $\beta$ in the
GKZ cone. The GKZ fan does not take into account the possible topologically nontrivial flat
connections on the wrapped D3-brane. To avoid multiplicities, we include the U(1) extensions of
all the discrete charges. The R-charge is already a coordinate in the GKZ fan and 
its corresponding GKZ auxiliary generating function is given in Equation (\ref{ZauxC3Z3}). The
remaining charges are the anomalous charges $A_1$ and $A_2$, as given in Table \ref{chargesC3Z3}, which we now add to the lattice as
extra coordinates. The points in the resulting lattice form a ``hollow cone'' and have no
multiplicities.

%In the singular case, the charges under discrete subgroups of the anomalous U(1)'s determine the
%(torsion) D1-charge of the wrapped D3-brane.

%\begin{eqnarray}\label{auxC3}
%\nonumber
%& &Z_{\rm aux}(t, a_1, a_2; \IC^3/\IZ_3 ) =
%\frac{1-t^3}{(1-ta_1/a_2^2)(1-ta_2/a_1)(1-ta_2)} = \\ &=& 1 + (a_1a_2^{-2} + a_2 + a_2a_1^{-1})t + \\ \nonumber
%&+& (a_1^2 a_2^{-4} + a_2^{-1} + a_1 a_2^{-1} + a_2^2 + a_2^2 a_1^{-2} + a_2^2 a_1) t^2+ \\
%&+& (a_1^{-1}+a_1+a_1^3 a_2^{-6} + a_1 a_2^{-3} + a_1^2 a_2^{-3} + a_2^3 + a_2^3 a_1^{-3} + a_2^3 a_1^{-2} + a_2^3 a_1^{-1}) t^3 + \ldots \nonumber
%\end{eqnarray}

The dressed auxiliary GKZ partition function which now also contains the anomalous charges can
be computed using the assumption that the polynomial ring, Equation (\ref{ZGKZC3Z3}), is a
complete intersection,
\begin{eqnarray}
\label{auxC3}
\nonumber
& &Z_{\rm aux}(t, a_1, a_2; \IC^3/\IZ_3 ) =
\frac{1-t^3}{(1-ta_1 a_2)(1-ta_2/a_1)(1-t/a_2^2)} = \\ &=& 1 +
(a_2^{-2} + a_1^{-1} a_2 + a_1 a_2)t + \\ \nonumber
&+& ( a_2^{-4} + a_1^{-1}a_2^{-1} + a_1 a_2^{-1} + a_2^2 + a_1^{-2}
a_2^2 + a_1^2 a_2^2 ) t^2+ \\
&+& (a_1^{-2}+a_1^2+a_2^{-6} + a_1^{-1} a_2^{-3} + a_1 a_2^{-3} +
a_1^{-3} a_2^3 + a_1^{-1}a_2^3
+ a_1 a_2^3 + a_1^3 a_2^3 ) t^3 + \ldots \nonumber
\end{eqnarray}

\begin{figure}[ht]
\begin{center}
  \epsfxsize = 8cm
  \centerline{\epsfbox{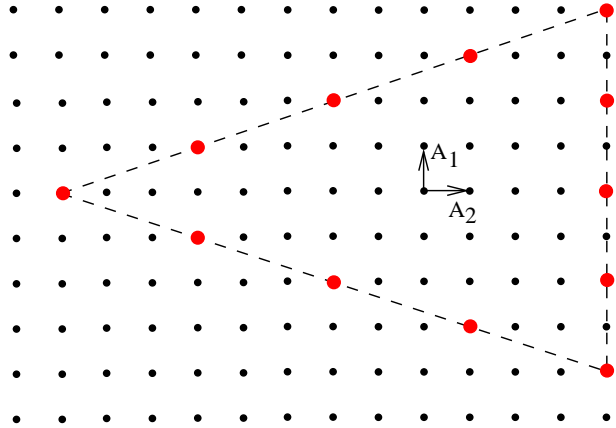}}
  \caption{The hollow triangle $C(4)$ above $R=4$, i.~e. the terms containing $t^4$. It gives the multiplicity $4\times 3 = 12$.}
  \label{emptyC3}
\end{center}
\end{figure}

\begin{figure}[ht]
\begin{center}
  \epsfxsize = 3cm
  \centerline{\epsfbox{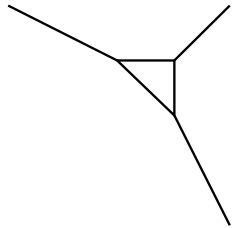}}
  \caption{The pq-web of $\IC^3 / \IZ_3$. The blown-up four-cycle is associated to the triangle in the middle.}
  \label{C3_Z3_pq}
\end{center}
\end{figure}

By drawing the lattice points in the $(A_1, A_2)$ lattice one can see that there is a ``hollow triangle'' $C(\beta)$ above each point $\beta$
in the 1d GKZ cone (\fref{emptyC3}), with edge length measured by the R-charge. This gives the $1,3,6,9,\ldots$ multiplicities. The same triangle appears in the pq-web (\fref{C3_Z3_pq}) of the geometry. This is a general feature as we will see in other examples. The polygon in the fiber parameterized by the anomalous charges nicely matches the shape of the blown-up cycle in the pq-web.

We can now refine the decomposition (\ref{expa}) by introducing the anomalous
charges. We first write the expansion (\ref{auxC3}) in the form

$$Z_{\rm aux}(t, a_1, a_2; \IC^3/\IZ_3 ) = \sum_{\beta=0}^\infty  \left(\sum_{K\in C(\beta)} a_1^{K_1} a_2^{K_2} \right) t^\beta $$
where the two-dimensional index $K = (K_1, K_2)$ runs over the points of the hollow triangle
$C(\beta)$.

We can then refine the decomposition (\ref{expa}) by
replacing the multiplicity $m(\beta)$ with $\sum_{K\in C(\beta)} a_1^{K_1} a_2^{K_2} $,

\begin{equation}
g_1(t_1,t_2,t_3,a_1,a_2) = \sum_{\beta=0}^\infty \left(\sum_{K\in C(\beta)} a_1^{K_1} a_2^{K_2}\right) g_{1,\beta}(t_1,t_2,t_3) .
\label{expa2}
\end{equation}

%\paragraph{The $N=1$ Generating Function with 5 Chemical Potentials}
%In the gauge theory related to the $\mathbb{C}^3/\mathbb{Z}_3$ variety there exist three flavor charges with chemical potentials $t_1, t_2, t_3,$ no non-anomalous baryonic symmetries, and two anomalous baryonic symmetries with chemical potentials $a_1,a_2$.

By explicit computation we can resum the previous series and compare with
the expected field theory result, finding perfect agreement.
The left hand side of formula (\ref{expa2}) is indeed
the $N=1$ generating function
depending on all the five chemical potentials, which can be computed as
the Hilbert series for the polynomial ring of the F-term equations
using the grading in Table \ref{chargesC3Z3},
\begin{eqnarray}
& & g_1(t_1,t_2,t_3,a_1,a_2;\mathbb{C}^3/\mathbb{Z}_3) = \nonumber\\
& & \frac{P(t_1,t_2,t_3,a_1,a_2)}{(1 - \frac{t_1}{a_2^2})(1 - \frac{a_2 t_1}{a_1})(1 - a_1 a_2 t_1)(1 - \frac{t_2}{a_2^2})(1 - \frac{a_2 t_2}{a_1})(1 - a_1 a_2 t_2)(1 -\frac{t_3}{a_2^2})(1 - \frac{a_2 t_3}{a_1})(1 - a_1 a_2 t_3)}\nonumber\\
\label{N1C3}
\end{eqnarray}
where $P(t_1,t_2,t_3,a_1,a_2)$ is a polynomial in the gauge theory chemical potentials

\begin{eqnarray}
& & P(t_1,t_2,t_3,a_1,a_2)=1 - \frac{t_1 t_2}{a_1 a_2} - \frac{a_1 t_1 t_2}{a_2} - a_2^2 t_1 t_2 + t_1^2 t_2 + t_1 t_2^2 - \frac{t_1 t_3}{a_1 a_2} - \frac{a_1 t_1 t_3}{a_2} -
    a_2^2 t_1 t_3 + \nonumber\\
& & t_1^2 t_3 - \frac{t_2 t_3}{a_1 a_2} - \frac{a_1 t_2 t_3}{a_2} - a_2^2 t_2 t_3 + 4 t_1 t_2 t_3 + \frac{t_1 t_2 t_3}{a_1^2} + a_1^2 t_1 t_2 t_3 + \frac{t_1 t_2 t_3}{a_1 a_2^3} + \frac{a_1 t_1 t_2 t_3}{a_2^3} + \frac{a_2^3 t_1 t_2 t_3}{a_1} + \nonumber\\
& & a_1 a_2^3 t_1 t_2 t_3 - \frac{t_1^2 t_2 t_3}{a_2^2} - \frac{a_2 t_1^2 t_2 t_3}{a_1} -
     a_1 a_2 t_1^2 t_2 t_3 + t_2^2 t_3 - \frac{t_1 t_2^2 t_3}{a_2^2} - \frac{a_2 t_1 t_2^2 t_3}{a_1} - a_1 a_2 t_1 t_2^2 t_3 + t_1 t_3^2 + \nonumber\\
& &  t_2 t_3^2 - \frac{t_1 t_2 t_3^2}{a_2^2} - \frac{a_2 t_1 t_2 t_3^2}{a_1} - a_1 a_2 t_1 t_2 t_3^2 + t_1^2 t_2^2 t_3^2
\end{eqnarray}

%$(((b c^3 ((b^4 t1 t2 t3 + c^6 t1 t2 t3 + b^3 ((t1 t2 t3 + c^3 t1 t2 t3 - c t1 t2 t3 ((t1 + t2 + t3)) - c^2 ((t1 t2 +((t1 + t2)) t3)%))) + b c^3 ((t1 t2 t3 + c^3 t1 t2 t3 - c t1 t2 t3 ((t1 + t2 + t3)) - c^2 ((t1 t2 + ((t1 + t2)) t3)))) + b^2 c^2 (((-t1) t2 - ((t1 +% t2)) t3 - c^2 t1 t2 t3 ((t1 + t2 + t3)) + c ((1 + t1 t2 ((t1 + t2)) + ((t1^2 + 4 t1 t2 + t2^2)) t3 + ((t1 + t2 + t1^2 t2^2)) t3^2))%)))))$

We would like to stress that decomposition (\ref{expa2}) is highly nontrivial. The right hand
side has been computed from the geometrical localization formulae and the refined GKZ auxiliary
generating function. It is then remarkable that the sum on the right hand side coincides with
the field theory $N=1$ generating function.

Using Equations (\ref{g1bC3Z3}) and (\ref{auxC3}) we get the following simpler expression

\begin{eqnarray}
\label{g1a1a2C3Z3}
g_1(t_1,t_2,t_3,a_1,a_2; \mathbb{C}^3/\mathbb{Z}_3)
&=& \frac{1}{(1-t_1 a_1 a_2)(1- \frac{t_1 a_2}{a_1})(1- \frac{t_1}{a_2^2})(1-\frac{t_2}{t_1})(1-t_3/t_1)} \\ \nonumber
&+&\frac{1}{(1-t_1/t_2)(1-t_2 a_1 a_2)(1-t_2 a_2/a_1)(1-t_2 / a_2^2)(1-t_3/t_2)} \\ \nonumber
&+&\frac{1}{(1-t_1/t_3)(1-t_2/t_3)(1-t_3 a_1 a_2)(1-t_3 a_2/a_1)(1-t_3 / a_2^2)}.
\end{eqnarray}

Formula (\ref{g1a1a2C3Z3}) has the general structure of equation (\ref{sumzdt}) and  suggests the existence of a localization formula for the
holomorphic functions on the $N=1$ moduli space, which is a five-dimensional
variety with an action of five $U(1)$ symmetries, three flavor plus two
baryonic. 
In the following section we will see that $g_1$ for the $\mathbb{F}_0$ gauge theory has the same localization structure. It is possible to verify that the $g_1$ generating function has the general structure of equation (\ref{sumzdt}) for other examples \cite{Butti:2007jv}. This is a consistency check that our counting procedure really reproduce the moduli space for $N=1$ brane. Indeed we know from
chapter \ref{Master} that the moduli space for just one brane is a $g+2$ dimensional toric variety and it is 
in some sense reasonable that the BPS spectrum over the moduli space can be counted 
directly applying an index theorem as explained in section \ref{counting}. 
We will add more comments about this phenomena in the following section when we will understand 
better the correspondence between the field theory and the geometric computation. 

By projecting the refined GKZ expansion on the plane $(A_1,A_2)$ we would
get the expansion of the $N=1$ generating functions into sectors with definite
baryonic charge. The same result can be obtained by expanding $g_1$ in a
Laurent series by means of the residue theorem. It is easy to check that the
multiplicity of each sector is one.

%\subsubsection{Expansion in anomalous baryonic charges.}

%\

%(CAN WE WRITE SUCH EXPANSION?)

\subsubsection{Generating Functions for $N>1$}
The generating function $g_N$ is now obtained from the general formula (\ref{g1plet}) starting
from {\bf any} decomposition of the $N=1$ generating function, the GKZ decomposition
(\ref{expa}), the refined GKZ decomposition (\ref{expa2}) or the anomalous baryonic charge
decomposition. Since we are interested in writing generating functions depending on the
non-anomalous charges, at the end of the computation $a_i$ should be set to one.

The more economical way of obtaining $g_N$ is to start from decomposition (\ref{expa}). The
generating function for $N$ D-branes is now given by the plethystic exponentiation

\begin{equation}
\sum_{N=0}^\infty g_N(t_1,t_2,t_3) \nu^N = \hbox{PE}_\nu[g_{1,0}(t_1,t_2,t_3)] + \sum_{\beta=1}^\infty 3 \beta \hbox {PE}_\nu [g_{1,\beta}(t_1,t_2,t_3)]
\end{equation}

The cases $N=2$ and $N=3$ (with only one charge $t$) are given by

\begin{eqnarray}
& & g_2(t,t,t) =\nonumber \\
& & \frac{1+t+13 t^2+20 t^3+53 t^4+92 t^5+137 t^6+134 t^7+146 t^8+103 t^9+55 t^{10}+19 t^{11}+9 t^{12}}{(1-t)^8(1+t)^6(1+t+t^2)^3}
\nonumber
\end{eqnarray}

\begin{eqnarray}
& & g_3(t,t,t) =\nonumber \\
& &\frac{1+32t^3+394 t^6+2365t^9+7343t^{12}+12946t^{15}+13201t^{18}+7709 t^{21}+2314t^{24}+276t^{27}+3t^{30}}{(1-t^3)^{11}(1+t^3)^3}
\nonumber
\end{eqnarray}

Taking the Plethystic Logarithm for these expressions we find 9 generators for $N=1$, 18 baryonic and 10 mesonic for $N=2$, 30 baryonic and 10 mesonic for $N=3$, 45 baryonic, 10 mesonic of $R$ charge 2, 28 mesonic of $R$ charge 4 for $N=4$, etc. By taking the order of the pole at $t=1$ we find the dimension of the moduli space is $3N+2=3N+g-1$: 3N dimensions for the mesonic moduli space and 2 dimensions 
for the baryonic fibers over it. All of this agrees with the field theory expectations.

\subsection{Generating Functions for $\mathbb{F}_0$}\label{FOannonan}

\begin{figure}[ht]
\begin{center}
  \epsfxsize = 10cm
  \centerline{\epsfbox{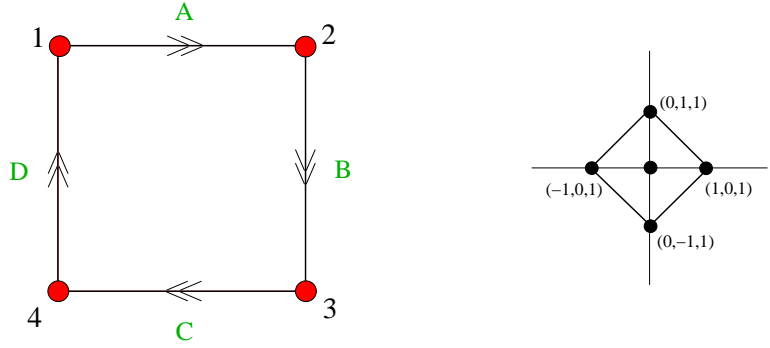}}
  \caption{Quiver and toric diagram for $\mathbb{F}_0$.}
  \label{F0quiver}
\end{center}
\end{figure}

$\mathbb{F}_0$ is a $\mathbb{Z}_2$ freely acting orbifold of the conifold. We report here for simplicity the toric diagram 
and the quiver of its first phase Figure \ref{F0quiver}. We will analyze here just the first toric phase of $\mathbb{F}_0$ and 
from now on $(\mathbb{F}_0)_I=\mathbb{F}_0$.  The quiver gauge theory has four types of fields ${\bf A},{\bf B},{\bf C},{\bf D}$.
%transforming under an $SU(2)^2$ global symmetry (see \fref{F0quiver} for the quiver).
and the superpotential is $\epsilon_{ij}\epsilon_{pq} {\bf A}_i{\bf B}_p{\bf C}_j{\bf D}_q$.

\paragraph{Symmetries and Geometry}

%The non-anomalous baryonic charge is
%\be
%B: ({\bf A, B, C, D}) \rightarrow (b {\bf A},\ b^{-1} {\bf B},\ b {\bf C},\ b^{-1} {\bf D})
%\ee

%The anomalous charges are defined as
%\bea
%A_1: ({\bf A, B, C, D}) \rightarrow (a_1 {\bf A},\ {\bf B},\ a_1^{-1} {\bf C},\ {\bf D}) \\
%A_2: ({\bf A, B, C, D}) \rightarrow ({\bf A},\ a_2 {\bf B},\ {\bf C},\ a_2^{-1} {\bf D})
%\label{f0_anom}
%\eea
In section \ref{s:F0} we have shown that the master space of $\mathbb{F}_0$ is a toric six dimensional
variety that is reducible in the union of a six dimensional CY cone: the product of two conifold $\firr{\mathbb{F}_0}= 
\mathcal{C} \times \mathcal{C}$ and two four dimensional complex planes: $\mathbb{C}^4$. The symmetry of $\firr{\mathbb{F}_0}$
is $U(1)_R \times U(1)_B \times SU(2)_M^2 \times SU(2)_H^2$. The last factor is an ``hidden symmetry''
that is not manifest in the UV Lagrangian. In this section we will analyze just the UV symmetries and we will come back
to the hidden symmetries in the next chapter.

Including flavor charges, we find a rank six global UV symmetry denoted by $SU(2)_1\times
SU(2)_2\times U(1)_R\times U(1)_B\times U(1)_{A_1}\times U(1)_{A_2}$. The basic fields have
transformation rules under the global symmetry which are summarized in Table \ref{globalF0}.

\begin{table}[htdp]
\begin{center}
\begin{tabular}{|c|c|c|c|c|c|c|c|c|}
\hline
\ {\bf field} \ & \ \  {\bf $F_1$}  \ \ & \ \  {\bf $F_2$}  \ \ & \ \  {\bf $R$} \ \ & \ \  {\bf $B$}  \ \ & \ \  {\bf $A_1$}  \ \ & \ \  {\bf $A_2$}  \ \ & \ \  chemical  \ \ & non-anomalous \\
 &  &  &  &  & &  & \ \  potentials \ \  & chemical potentials \\
\hline \hline
${\bf A}_1$  & $\frac{1}{2}$  & $0$      & $\frac{1}{2}$ & 1 & 1 & 0 & $t b x a_1$    & $t_1x =t b x$\\
${\bf A}_2$  & $-\frac{1}{2}$ & $0$       & $\frac{1}{2}$ & 1 & 1 & 0 & $\frac{t b a_1}{x}$ & $\frac{t_1}{x}=\frac {t b}{x}$ \\
${\bf B}_1$  & $0$          & $\frac{1}{2}$ & $\frac{1}{2}$ &$-1$ & 0 & 1 & $\frac{t y a_2}{b}$ & $t_2 y= \frac{ t y}{b}$   \\
${\bf B}_2$  & $0$          & $-\frac{1}{2}$ & $\frac{1}{2}$    &$-1$ & 0 & 1 & $\frac{t a_2}{b y}$ & $\frac{t_2}{y} =\frac{t}{b y}$  \\
${\bf C}_1$  & $\frac{1}{2}$    & $0$       & $\frac{1}{2}$ & 1 &$-1$ & 0 & $\frac{t b x}{a_1}$ & $t_1 x = t b x$  \\
${\bf C}_2$  & $-\frac{1}{2}$   & $0$       & $\frac{1}{2}$ & 1 &$-1$ & 0 & $\frac{t b}{x a_1}$  & $\frac{t_1}{x}=\frac{t b}{x}$ \\
${\bf D}_1$  & $0$ & $\frac{1}{2}$ & $\frac{1}{2}$ &$-1$ & 0 &$-1$ & $\frac{t y}{b a_2}$  & $t_2 y =\frac{t y}{b}$ \\
${\bf D}_2$  & $0$          & $-\frac{1}{2}$    & $\frac{1}{2}$ &$-1$ & 0 &$-1$ & $\frac{t}{b y a_2}$  & $\frac{t_2}{y}=\frac{t}{b y}$ \\
\hline
\end{tabular}
\end{center}
\caption{Global charges for the basic fields of the quiver gauge theory
living on the D-brane probing the CY with $\mathbb{F}_0$ base.}
\label{globalF0}
\end{table}

We can explicitly examine the geometry of $\mathbb{F}_0$ in the symplectic quotient description.
To this purpose, since $d=4$, we  introduce four homogeneous coordinates $x_i$ in $\mathbb{C}^4$ and
 we define the group of rescaling
$\prod_{j=1}^4 \mu_j^{\langle e_i,V_j\rangle }=1$ which consists of a
continuous charge acting as $(b,1/b,b,1/b)$ on the $x_i$ and of a discrete one $(1,e,1,e)$ with
$e^2=1$. This implies that the manifold, as we know, is a $\mathbb{Z}_2$ quotient of the conifold

\begin{equation}
{\cal R} ( {\cal C} ) = \mathbb{C}[x_1, x_2, x_3, x_4]/(x_1x_2-x_3x_4); \qquad {\cal R} ( \mathbb{F}_0 ) = {\cal R} ( {\cal C} ) / \mathbb{Z}_2
\end{equation}

The homogeneous charges $x_i$ can be represented by chemical potentials as $$(x_1,x_2,x_3,x_4) \longrightarrow (t_1 x,e t_2 y,t_1/x,e t_2/y)$$ in terms of the discrete baryonic charge $e$ and the chemical potentials $t_i$ and $x,y$;
notations are inherited from original conifold theory.
For future reference, we notice that
the right basis for localization is given by $q_i=\prod_{j=1}^4 x_j^{\langle e_i,V_j\rangle }$ and we compute
 $q_1=x^2,q_2=y^2,q_3=t_1^2 t_2^2$.

\subsubsection{The $N=1$ Generating Function}

As we have already seen in section \ref{s:F0} the master space $\f_{\mathbb{F}_0}$ is reducible. 
Here we decide to study only the irreducible component of the moduli space which contains the generic
point with all fields different from zero: $\firr{\mathbb{F}_0}$.  Algebraically, this is obtained by taking the closure
of the open set ${\bf A},{\bf B},{\bf C},{\bf D}\ne 0$ and give the product of two conifold.
 We will see that the geometry of the CY: the complex cone over $\mathbb{F}_0$: $\cX=C_{\mathbb{C}}(\mathbb{F}_0)$,
nicely captures this branch of the moduli space. The other branches can be added by performing
surgeries as in we have shown in section \ref{s:F0}. The $N=1$ generating function for the generic branch is given
in equation (\ref{2conifmaster}). Introducing just the two field theory chemical potentials $t_1$, $t_2$ in Table
\ref{globalF0} for the fields $A_i$, $C_j$ and $B_p$, $D_q$ respectively, we can distinguish between the two 
conifold in $\firr{\mathbb{F}_0}$, and we obtain:   

\begin{equation}
g_1(t_1, t_2; \mathbb{F}_0 ) = \frac{(1-t_1^2)(1-t_2^2)}{(1-t_1)^4(1-t_2)^4}
\label{N11}
\end{equation}
%By taking order of the pole at $t_1=t_2=t=1$ we find the dimension of the moduli space to be 6; this can be easily understood by having three mesonic directions (parameterizing the CY) plus three baryonic directions given by the gauge theory FI terms, consistent with a dimension formula of $3N+G-1$.

\subsubsection{The GKZ Decomposition}

\begin{figure}[ht]
\begin{center}
  \epsfxsize = 6cm
  \centerline{\epsfbox{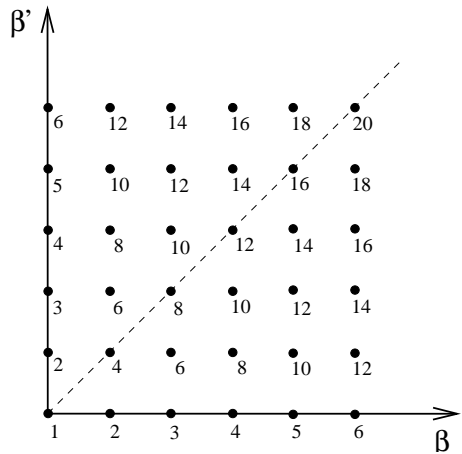}}
  \caption{GKZ decomposition for $\mathbb{F}_0$ with multiplicities.}
  \label{FF}
\end{center}
\end{figure}

On the singular cone, there is just one independent divisor $D_1=D_3=-D_2=-D_4$ as for the conifold. 
On the resolution, there is a new divisor $D_5$ corresponding to the internal point.
$D_1=D_3$ and $D_2=D_4$ are still true but now $D_1$ and $-D_2$ are different.
We can parametrize the GKZ fan in $\mathbb{R}^2$ with
$\beta D_1 +\beta^\prime D_2$. The integer parameters $\beta$ and $\beta^\prime$ have the interpretation as the discrete K\"ahler parameters of $\mathbb{F}_0$, namely the discrete areas of the two $\mathbb{P}^1$'s.
In this case, the convexity condition requires $\beta,\beta^\prime\ge 0$. The GKZ cone is depicted in Figure \ref{FF} and the multiplicities are presented in Equation (\ref{dictfig}). Notice that the QFT baryonic charge is given by $B=\beta-\beta^\prime$, so the sector $\beta^\prime < \beta$ corresponds to the positive baryonic charges and  the sector $\beta^\prime > \beta$ corresponds to the negative ones.

%\begin{figure}[h!!!]
%\begin{center}
%\includegraphics[scale=0.6]{F0.eps}
%\caption{Toric diagram, quiver and GKZ decomposition for $\mathbb{F}_0$.}
%\label{FF}
%\end{center}
%\end{figure}

It is interesting to note that the line $\beta^\prime = \beta$ contains the sectors with zero
baryonic charge. It is not however true that operators with zero baryonic charges are made with
traces; this is only true for $\beta=\beta^\prime=0$. The other sectors on the line
$\beta^\prime = \beta$ correspond to determinants of the form  $(\det {\bf A} \det {\bf B})^n$.

Localization now gives
(using the complete set of charges $(t_1x,t_2y,t_1/x,t_2/y)$)

\begin{eqnarray}
\label{ZBBF0}
g_{1,\beta,\beta^\prime} (t_1, t_2 , x, y; \mathbb{F}_0 ) &=& \frac{t_1^{\beta}t_2^{\beta^\prime} x^{-\beta} y^{-\beta^\prime}}{(1-x^2)(1- \frac{t_1^2 t_2^2}{x^2 y^2})(1-y^2)}+\frac{t_1^{\beta}t_2^{\beta^\prime} x^{\beta} y^{-\beta^\prime}}{(1-1/x^2)(1-t_1^2t_2^2x^2/y^2)(1-y^2)} \\ \nonumber
&+&\frac{t_1^{\beta}t_2^{\beta^\prime} x^{-\beta} y^{\beta^\prime}}{(1-x^2)(1-t_1^2t_2^2y^2/x^2)(1-1/y^2)}+\frac{t_1^{\beta}t_2^{\beta^\prime} x^{\beta} y^{\beta^\prime}}{(1-1/x^2)(1-t_1^2t_2^2x^2y^2)(1-1/y^2)}
\end{eqnarray}

The dependence on the baryonic charge can be obtained by replacing
$t_1\rightarrow tb$ and $t_2\rightarrow t/b$ and, as expected, is given
by a multiplicative factor
$$g_{1,\beta,\beta^\prime} (t_1, t_2 , x, y; \mathbb{F}_0 ) = b^{\beta -\beta^\prime} \hat g_{1,\beta,\beta^\prime} (t , x, y; \mathbb{F}_0 ) . $$
The generating function for $x,y=1$ can be nicely written as

\begin{equation}
g_{1,\beta,\beta^\prime} (t_1, t_2; \mathbb{F}_0 ) = \sum_{n=0}^\infty
(2 n+1+\beta)(2 n+1+\beta^\prime) t_1^{2 n+\beta} t_2^{2 n+\beta^\prime}
\label{seri}
\end{equation}

It is then obvious that, for example, the mesonic partition function $g_{1,0,0}$ can be obtained from the mesonic partition function for the conifold
\ref{g1conit}
by projecting on the $\mathbb{Z}_2$ invariant part ($t_{1,2}\rightarrow -t_{1,2}$).

\subsubsection{Multiplicities and Seiberg Duality}
To extract multiplicities we use the auxiliary partition function for the GKZ cone. We introduce
letters $a,b,c,d$ for the four possible classes of arrows. The only relation that they form is
related to the closed loop $abcd$. The generating function is the Hilbert series of the polynomial ring

\begin{equation}
{\cal R}_{\rm GKZ} ( \mathbb{F}_0 ) = \mathbb{C}[a,b,c,d]/(abcd)
\label{ringGKZF0}
\end{equation}

By assigning chemical potential $t_1$ to $a,c$ and $t_2$ to $b,d$ we obtain the
auxiliary GKZ partition function for multiplicities:

\begin{eqnarray}
\label{ZauxF0}
Z_{\rm aux}(t_1,t_2; \mathbb{F}_0 ) &=& \frac{1-t_1^2 t_2^2}{(1-t_1)^2 (1-t_2)^2} \\ \nonumber
&=& 1+ \sum_{\beta = 1}^\infty ( \beta + 1 ) t_1^\beta + \sum_{\beta' = 1}^\infty (\beta' + 1) t_2^{\beta'} + \sum_{\beta = 1}^\infty \sum_{\beta' = 1}^\infty 2 ( \beta + \beta' ) t_1^\beta t_2^{\beta'}
\end{eqnarray}

>From which we can extract the following multiplicities,
$2(\beta+\beta^\prime)$ for $\beta,\beta^\prime\ge1$, $\beta+1$ for $\beta^\prime=0$, and
$\beta^\prime+1$ for $\beta=0$.

We thus have

\begin{equation}
 g_1(\{t_i\}; \mathbb{F}_0 ) = g_{1,0,0} + \sum_{\beta=1}^\infty (\beta+1)g_{1,\beta,0}
+ \sum_{\beta^\prime=1}^\infty (\beta^\prime+1)g_{1,0,\beta^\prime}
+ \sum_{\beta,\beta^\prime=1}^\infty 2 (\beta+\beta^\prime) g_{1,\beta,\beta^\prime}
\label{F0multiplicities}
\end{equation}
and one computes, using Equation (\ref{seri}),

\begin{equation}\label{g1F0}
g_1(t_1, t_2; \mathbb{F}_0 )=\frac{(1-t_1^2)(1-t_2^2)}{(1-t_1)^4(1-t_2)^4} 
\end{equation}

which is exactly Equation (\ref{N11}). 

This is a rather important point. We have just seen that the 
geometrical counting procedure reproduce just the coherent component $\firr{\mathbb{F}_0}$ of the moduli space. 
This is a quite interesting result. Indeed in section \ref{s:F0} we accumulated some evidences, from the field theory point of view, that the 
two coherent components $\firr{~}$ of two Seiberg equivalent CFT have the same Hilbert series written just in term of the non anomalous 
symmetries. From the geometric computation we have just presented, this fact means that the GKZ construction must be invariant under Seiberg duality: namely the auxiliary GKZ partition function restricted to the non anomalous charges is the same in the two phases of $\mathbb{F}_0$.
Indeed for the second phase of $\mathbb{F}_0$  we have the quiver in figure \ref{f:F0III}. If we call $e$ the arrows along the diagonal and $a$, $b$, $c$, $d$, the four arrows along the square, the relations they satisfy are: $abe=0$, $cde=0$, $ab-cd=0$. The generating function for the 
second phase is the Hilbert series of the polynomial ring\footnote{To do not make too heavy the notation in all this chapter we will call $\mathbb{F}_0$ the first toric phase of the $\mathbb{F}_0$ gauge theory, and  $(\mathbb{F}_0)_{II}$ the second toric phase.}

\begin{equation}
{\cal R}_{\rm GKZ} ( (\mathbb{F}_0)_{II} ) = \mathbb{C}[a,b,c,d,e]/(abe,cde,ab-cd)
\label{ringGKZF0II}
\end{equation}

By assigning chemical potential $t_1$ to $a,c$, $t_2$ to $b,d$ and $t_1 t_2$ to $e$ we obtain the
auxiliary GKZ partition function for multiplicities for the second phase of $\mathbb{F}_0$:

\begin{equation}
\label{ZauxF0II}
Z_{\rm aux}(t_1,t_2; (\mathbb{F}_0)_{II} ) = \frac{1-t_1^2 t_2^2}{(1-t_1)^2 (1-t_2)^2} 
\end{equation}

This is exactly the auxiliary GKZ partition function of the phase I (\ref{ZauxF0}), meaning that the GKZ construction is invariant under
Seiberg duality. We have previously observed that the geometric reconstruction of the BPS spectrum is sensitive just to the 
coherent component of the moduli space. Indeed the invariance under Seiberg duality of the GKZ partition function is the 
geometric dual of the invariance of the Hilbert series, refined by all the chemical potentials associated to the non anomalous U(1) symmetries, 
of the coherent component of the master space.

It seems that this fact, that we have shown just for the $\mathbb{F}_0$ theory, is going to repeat for other 
examples as well. For a detailed analysis of the $\mathbb{F}_0$ case, and more general examples, we refer the reader to \cite{Forcella:2008ng}.
We conclude that both from the field theory and from the geometry we can guess that the Hilbert series 
for non anomalous charges of the coherent component of the master space is invariant under Seiberg duality. 
This fact suggests the presence of some ``quantum'' corrections to the classical moduli space of the field theory.  
We could in principle obtain some informations regarding the ``quantum'' corrected moduli space from the dual gravity setup (observe that we are here taking the non 
trivial extrapolation from $N \rightarrow \infty $ to $N=1$ !). 

In this section we have seen that 
the Hilbert series for $\firr{\mathbb{F}_0}$ is exactly reproduced by the counting of supesymmetric D3 branes in $\cX$. 
It would be interesting to understand if and how the linear components of $\f$ are reproduced by the geometry of $\cX$, 
and if the Hilbert series obtained from the geometric setup could give some hints to understand the possible deformations to the classical field theory moduli spaces and their behavior under Seiberg dualities.

\subsubsection{Refining the GKZ Decomposition and Seiberg Duality}

The auxiliary partition function in Equation (\ref{ZauxF0}) is derived by computing the Hilbert series of the GKZ ring in Equation (\ref{ringGKZF0}). By expanding this partition function we get multiplicities in the $(\beta, \beta^\prime)$ lattice given by the following infinite matrix,

\begin{equation}
\left(
\begin{array}{ccccccc}
  1 & 2 & 3 & 4 & 5 & 6 & \\
  2 & 4 & 6 & 8 & 10 & 12 & \\
  3 & 6 & 8 & 10 & 12 & 14 & \ldots \\
  4 & 8 & 10 & 12 & 14 & 16 &   \\
  5 & 10 & 12 & 14 & 16 & 18 & \\
  6 & 12 & 14 & 16 & 18 & 20 & \\
  & & \vdots &  & & & \ddots
\end{array}
\right) .
\label{dictfig}
\end{equation}

In order to avoid getting multiplicities, let us introduce the chemical potentials for the anomalous charges given in Table \ref{globalF0}. Using Macaulay2, or the fact that we are dealing with a complete
intersection, we can write the auxiliary partition function dressed with these new charges as

\begin{equation}
Z_{\rm aux}(t_1,t_2, a_1, a_2; \mathbb{F}_0 ) =  \frac{1-t_1^2 t_2^{2}}{(1-t_1 a_1)(1-t_1/a_1)(1-t_2 a_2)(1-t_2/a_2)}
\label{f0gen}
\end{equation}

\begin{figure}[ht]
\begin{center}
  \epsfxsize = 6cm
  \centerline{\epsfbox{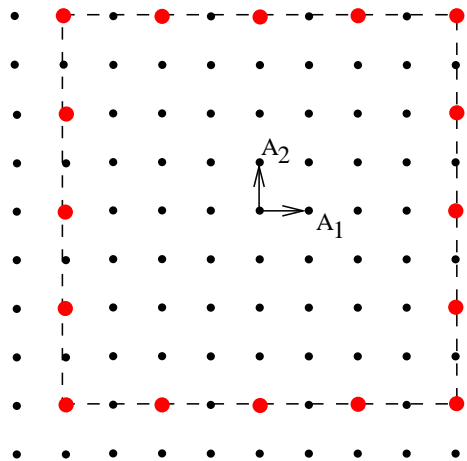}}
  \caption{The hollow rectangle $C(4,4)$ above $(\beta, \beta^\prime)=(4,4)$. It gives the multiplicity 16.}
  \label{emptyF0}
\end{center}
\end{figure}

\begin{figure}[ht]
\begin{center}
  \epsfxsize = 2.5cm
  \centerline{\epsfbox{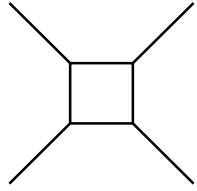}}
  \caption{The pq-web of $\mathbb{F}_0$. The blown-up four-cycle is associated to the square in the middle.}
  \label{F0_pq}
\end{center}
\end{figure}

By expanding this function, we see that above each point in the GKZ fan, parametrized by $(\beta, \beta^\prime)$, there is a rectangle $C(\beta,\beta^\prime)$ in the $(A_1, A_2)$ lattice as in \fref{emptyF0}. The related rectangle in the pq-web is shown in \fref{F0_pq}.

We can thus refine our decomposition of the $N=1$ partition function.
Equation (\ref{F0multiplicities}) can be replaced by the following formula
where all multiplicities are lifted:

\begin{equation}
g_1(t_1,t_2,a_1,a_2) = \sum_{\beta=0,\beta^\prime=0}^\infty \left(\sum_{K\in C(\beta,\beta^\prime)} a_1^{K_1} a_2^{K_2} \right) g_{1,\beta,\beta^\prime}(t_1,t_2) .
\label{expa2F0}
\end{equation}

One can explicitly resum the right hand side, computed from geometrical
data and the auxiliary GKZ generating functions, and compare it with
the $N=1$ generating function as computed by field theory.
By resumming the series in  Equation (\ref{expa2F0}) we indeed recover the
%\paragraph{The $N=1$ Generating Function with 6 Chemical Potentials}
%In the gauge theory related to the singular cone over the $\mathbb{F}_0$ surface there exist two flavor chemical potentials $x,y$, one for the $R$ charge, $t$, one non-anomalous baryonic
%chemical potential $b$, and two anomalous baryonic chemical potentials $a_1,a_2$. They give grading to the different fields using Table (\ref{globalF0}).
generating function for $N=1$ with all anomalous and non-anomalous charges; this is given by
the Hilbert series for the polynomial ring of $\firr{(\mathbb{F}_0)_I}$ in section \ref{s:F0} and is easily computed
using the fact that we deal with a complete intersection,

\begin{eqnarray}\label{F0uno}
& & g_1(x,y,t,b,a_1,a_2;\mathbb{F}_0)= \\
& & \frac{ (1 - \frac{t^2}{b^2}) (1 - b^2 t^2) }{(1 - \frac{b t}{a_1 x})(1 - \frac{a_1 b t}{x})(1 - \frac{b t x}{a_1})(1 - a_1 b t x)(1 - \frac{t}{a_2 b y})(1 - \frac{a_2 t}{b y})(1 - \frac{t y}{a_2 b})(1 - \frac{a_2 t y}{b})}\nonumber
\end{eqnarray}

For completeness we can rewrite this expression by summing using Equations (\ref{ZBBF0}) and (\ref{f0gen}),

\begin{eqnarray}\label{F0loc}
%\label{ZBBF0}
g_1(x,y,t_1,t_2,a_1,a_2;\mathbb{F}_0)
&=& \frac{1}{(1-x^2)(1-t_1 a_1/x)(1- \frac{t_1}{x a_1})(1-t_2 a_2/y)(1- \frac{t_2}{y a_2})(1-y^2)}  \nonumber \\
&+& \frac{1}{(1-1/x^2)(1-t_1 x a_1)(1-t_1 x/a_1)(1-t_2 a_2 / y)(1-t_2/ y a_2)(1-y^2)} \nonumber \\
&+& \frac{1}{(1-x^2)(1-t_1 a_1 / x)(1-t_1/ x a_1)(1-t_2 y a_2)(1-t_2 y /a_2)(1-1/y^2)} \nonumber \\
&+& \frac{1}{(1- \frac{1}{x^2})(1-t_1 x a_1)(1-t_1 x /a_1)(1-t_2 y a_2)(1-t_2 y /a_2)(1-1/y^2)} . \nonumber \\ 
\end{eqnarray}

As in the case of $\mathbb{C}^3/\mathbb{Z}_3$ in section \ref{N1C3Z3}
this formula suggests that some localization is at work in the
field theory $N=1$ moduli space $\CM=\f$, which is a six dimensional toric variety with the
action of six $U(1)$ flavor and baryonic symmetries as described in chapter \ref{Master}. From equation (\ref{g1F0}) we can refine our observation by guessing that the counting of sections of line bundles with multiplicities over $\cX$ is equivalent to the counting of holomorphic functions over $\firr{~}$.

Indeed we can count the complete chiral spectra of the gauge theory in two different geometrical way: in the first one we start from the three dimensional singularity $\cX$ and we count all the sections of all the divisors according to the GKZ decomposition taking into account the field theory multiplicities:
\begin{equation}
g_1(q,b,a; \cX)= \sum _{\vec{\beta}} \sum_{\vec{K}\in C(\vec{\beta})}\vec{a}^{\vec{K}}{\rm Tr} \left\{ H^0\left(\cX _{\vec{\beta}},{\cal O}_{\vec{\beta}}\left(\sum a_i D_i\right)\right) \bigr| q\right\}
\end{equation}
this is the way we are following in this chapter and it reveals very useful for extending the computation to generic $N$;
the second way is more natural from the point of view of the study of the complete moduli space $\CM$ of the field theory and it fit very 
well with the discussion in chapter \ref{Master}. The trivial observation is that the set all the $1/2$ BPS operators are exactly the set of all holomorphic 
functions over the complete moduli space $\CM$ ( i.e. sections of the trivial bundle $\mathcal{O}$ over $\mathcal{M}$ in the language of the section \ref{counting}). In chapter \ref{Master} we have shown that the moduli space $\CM$ for $N=1$ brane, namely the master space $\f$, is a toric variety composed by a CY toric cone of dimension $g+2$ called the coherent component $\firr{~}$, and a set of lower dimensional linear spaces $L^i$. Hence if we want to count the holomorphic functions over $\firr{~}$ we just need to extend the formalism for trivial bundle developed in section \ref{counting} from three to $g+2$ dimensions:
\begin{equation}\label{charg2}
g_1(q,b,a; \cX )={\rm Tr} \left\{ H^0\left(\firr{~} , \mathcal{O}\right) \bigr| q,b,a \right\}
\end{equation}
this character can be computed applying index theorem on the trivial line bundle $\mathcal{O}$ over $\firr{~}$
and it takes the form (\ref{F0loc}), suggesting that: 
\begin{equation}
{\rm Tr} \left\{ H^0\left(\firr{~} ,\mathcal{O}\right) \bigr| q,b,a \right\} = \sum_{\vec{\beta}}\sum_{\vec{K}\in C(\vec{\beta})}\vec{a}^{\vec{K}}{\rm Tr} \left\{ H^0\left(\cX _{\vec{\beta}},{\cal O}_{\vec{\beta}}\left(\sum a_i D_i\right)\right) \bigr| q\right\}
\end{equation}

Let us conclude with a couple of observations about the second phase of $\mathbb{F}_0$, for more details we refer the reader to \cite{Forcella:2008ng}. 
It is possible to show that the two phases of $\mathbb{F}_0$ have different completely refined Hilbert series for the coherent component of the master space. If we define the charges for the second phase as in Table \ref{globalF0II}, with as usual $t_1=t b$, $t_2=t/b$
\begin{table}[h]
\begin{center}
\begin{tabular}{|c|c|c|c|c|c|c|c|c|}
\hline
\ {\bf Phase II} \ & \ \  {\bf $F_1$}  \ \ & \ \  {\bf $F_2$}  \ \ & \ \  {\bf $R$} \ \ & \ \  {\bf $B$}  \ \ & \ \  {\bf $A_1$}  \ \ & \ \  {\bf $A_2$}  \ \ & \ \  fugacities   \\
\hline \hline
${\bf X^1_{12}}$ & $\frac{1}{2}$ & 0 & $\frac{1}{2}$ & 1 & $1 $ & $0 $ & $ t b x a_1$ \\
${\bf X^2_{12}}$ & $-\frac{1}{2}$ & 0 & $\frac{1}{2}$ & 1 & $1 $ & $0 $ & $ \frac{t b a_1}{ x} $ \\
${\bf X^1_{23}}$ & 0 & $\frac{1}{2}$ & $\frac{1}{2}$ & $-1$ & $0 $ & $1 $ & $ \frac{t y a_2}{b}$ \\ 
${\bf X^2_{23}}$ & 0 & $-\frac{1}{2}$ & $\frac{1}{2}$ & $-1$ & $0 $ & $1 $ & $ \frac{t a_2}{b y}$ \\  
${\bf X^{11}_{31}}$ & $\frac{1}{2}$ & $\frac{1}{2}$ & $1$ & $0$ & $-1 $ & $-1 $ & $ \frac{t^2 x y}{a_1 a_2} $ \\ 
${\bf X^{12}_{31}}$ & $\frac{1}{2}$ & $-\frac{1}{2}$ & $1$ & $0$ & $-1 $ & $-1 $ & $ \frac{ t^2 x}{y a_1 a_2}$ \\ 
${\bf X^{21}_{31}}$ & $-\frac{1}{2}$ & $\frac{1}{2}$ & $1$ & $0$ & $-1 $ & $-1 $ & $ \frac{t^2 y}{x a_1 a_2}$ \\ 
${\bf X^{22}_{31}}$ & $-\frac{1}{2}$ & $-\frac{1}{2}$ & $1$ & $0$ & $-1 $ & $-1 $ & $ \frac{t^2}{x y a_1 a_2}$ \\  
${\bf X^2_{14}}$ & $0$ & $-\frac{1}{2}$ & $\frac{1}{2}$ & $-1$ & $0$ & $1$ & $ \frac{t a_2}{b y}$ \\  
${\bf X^1_{14}}$ & $0$ & $\frac{1}{2}$ & $\frac{1}{2}$ & $-1$ & $0$ & $1$ & $\frac{t y a_2}{b}$ \\  
${\bf X^2_{43}}$ & $-\frac{1}{2}$ & $0$ & $\frac{1}{2}$ & $1$ & $1$ & $0$ & $\frac{t b a_1}{x}$ \\
${\bf X^1_{43}}$ & $\frac{1}{2}$ & $0$ & $\frac{1}{2}$ & $1$ & $1$ & $0$ & $t b x a_1$ \\  
\hline
\end{tabular}
\end{center}
\caption{Global charges for the basic fields of the second phase of $\mathbb{F}_0$.}
\label{globalF0II}
\end{table}
we obtain:
\begin{eqnarray}\label{HSF0}
& & g_1(x,y,t,b,a_1,a_2;(\mathbb{F}_0)_{II}) =   \frac{ P(x,y,t,b,a_1,a_2)} {(1-\frac{t^2 x y}{a_1 a_2})(1- \frac{ t^2 x}{y a_1 a_2})(1- \frac{t^2 y}{x a_1 a_2})(1- \frac{t^2}{x y a_1 a_2})} \times \nonumber \\ 
& &  \frac{1}{(1-  t a_1 b x)(1-\frac{t b a_1}{x})(1-\frac{t y a_2}{b})(1-  \frac{t  a_2}{b y})(1- t b x a_1)(1- \frac{t b a_1}{ x})(1-\frac{t y a_2}{b})(1-\frac{t a_2}{b y}) } \nonumber
\end{eqnarray}
with $P(x,y,t,b,a_1,a_2)$ a polynomial in the fugacities. One can check that the Hilbert Series for the first phase in \ref{F0uno}, and the one for the second phase in \ref{HSF0} are really different: $g_1(x,y,t,b,a_1,a_2;\mathbb{F}_0) \ne g_1(x,y,t,b,a_1,a_2;(\mathbb{F}_0)_{II})$ and that they become exactly equal for $a_1=a_2=1$, in agreement with our conjecture in chapter \ref{Master}:
\begin{eqnarray}\label{F0uno}
& & g_1(x,y,t,b;\mathbb{F}_0)= g_1(x,y,t,b;(\mathbb{F}_0)_{II}) = 
\frac{ (1 - \frac{t^2}{b^2}) (1 - b^2 t^2) }{(1 - \frac{b t}{x})^2(1 - b t x)^2(1 - \frac{t}{b y})^2(1 - \frac{t y}{b})^2}\nonumber
\end{eqnarray}
 
Moreover the the refined 
GKZ auxiliary partition functions are also different. In phase II the GKZ auxiliary partition function is:
\begin{equation}\label{ZauxgkzII}
Z_{\rm aux}(t_1,t_2, a_1, a_2; (\mathbb{F}_0)_{II} ) = \frac{(1- t_1 t_2 a_1 a_2)(1- t_1^2 t_2^2)}{(1-t_1 a_1)^2(1-t_2 a_2)^2(1-  t_1 t_2/ a_1 a_2)}
\end{equation}
and it is explicitly different from (\ref{f0gen}). The two partition functions become the same (\ref{ZauxF0},\ref{ZauxF0II}) only for
$a_1=a_2=1$. The nice observation is that $g_1$ for both the two toric phases is decomposable in terms of the elementary partition 
functions $g_{1,\beta,\beta^\prime}$ (\ref{ZBBF0}), and the formula (\ref{expa2F0}) holds for both the phases, meaning that 
$g_{1,\beta,\beta^\prime}$ combined with $Z_{\rm aux}$ of phase I and II  reconstruct $g_1$ for the phase I and II respectively. 
All this means that the computation of BPS operators for $N>1$ give the same result
in the two phases, when expressed in terms of non anomalous symmetries $t_1,t_2,x,y$, while the dependence on the anomalous charges
$a_1$ and $a_2$ is different. However this agrees with Seiberg duality, because dual operators have the same non anomalous charges but 
in general different charges under anomalous symmetries. It agrees also with the strong coupling gravity computations, because the 
anomalous symmetries are not really symmetries of the theory and are invisible in the dual gravity background.

\subsubsection{Generating Functions for $N>1$}

As before the higher $N$ generating function $g_N$ is given by the plethystic exponentiation using formula (\ref{g1plet}). We can start from {\bf any} decomposition of the $N=1$ generating function, the GKZ decomposition (\ref{F0multiplicities}), the refined GKZ decomposition (\ref{expa2F0}) or the expansion in baryonic charges. Using the GKZ decomposition we have

\begin{eqnarray}
\sum_{N=0}^\infty g_N( \{t_i\}; \mathbb{F}_0 ) \nu^N &=& \hbox{PE}_\nu[g_{1,0,0}]
+ \sum_{\beta=1}^\infty(\beta + 1) \hbox {PE}_\nu [g_{1,\beta,0}] \\
&+& \sum_{\beta^\prime=1}^\infty(\beta^\prime + 1) \hbox {PE}_\nu [g_{1,0,\beta^\prime}]
+ \sum_{\beta=1}^\infty \sum_{\beta^\prime=1}^\infty 2(\beta + \beta^\prime ) \hbox {PE}_\nu [g_{1,\beta,\beta^\prime}] \nonumber
\end{eqnarray}

>From which we can compute the generating function for $N=2$

\begin{eqnarray}
g_2(t_1,t_2) &=& \frac{F(t_1,t_2)}{(1-t_1^2)^4(1-t_1^2t_2^2)^3(1-t_2^2)^4}\nonumber\\
F(t_1,t_2) &=& 1+2t_2^2+t_1^8t_2^4(-6+8t_2^2+17t_2^4+2t_2^6)+t_1^2(2+14t_2^2+8t_2^4-3 t_2^6) \nonumber\\
& & -t_1^6t_2^2(3+20 t_2^2+10t_2^4-8t_2^6-t_2^8) +t_1^{10}(t_2^6+2 t_2^8)-t_1^4(-8 t_2^2+6 t_2^4+20 t_2^6+6 t_2^8)\nonumber
\end{eqnarray}

Taking the plethystic logarithm of the generating functions for $\mathbb F_0$,
it is  possible to find the generators of the moduli space. For $N=1$ the
plethystic logarithm of (\ref{N11}) is very simple:
\[
\left(4 t_1 + 4 t_2 \right)  - \left( t_1^2 + t_2^2 \right)
\]
and it correctly reproduces the 8 chiral field generators and
the two relations among them.
For $N=1$ the moduli space is a complete intersection, and this is reflected
by the fact that the plethystic logarithm has a finite number of terms.
In the $N=2$ case there are 12 baryonic generators and 13 mesonic ones and the
moduli space is no more a complete intersection.
By looking at the order of the pole for the computed generating functions
near $t_1=t_2=t=1$, we checked that the dimension of the moduli space is $3 N
+ g - 1$ as expected.

\subsection{Counting in $\frac{1}{2} \mathbb{F}_0$ and in $\frac{3}{4} \mathbb{F}_0$}

We can also consider, as was done for the conifold in section \ref{Toys}, fictitious
theories counting subsets of the BPS chiral operators.

The half $\mathbb{F}_0$ corresponds to considering only the BPS operators
containing ${\bf A}_1,{\bf B}_1,{\bf C}_1,{\bf D}_1$ and no occurrences of
the other four elementary fields. This is done by truncating
Equation (\ref{seri}) for $g_{1,\beta,\beta^\prime}$

\begin{equation}
g_{1,\beta,\beta^\prime} = \sum_{n=0}^\infty
 t_1^{2 n+\beta} t_2^{2 n+\beta^\prime}
\end{equation}

We easily obtain the generating functions for $N=1,2,$

\begin{eqnarray}
g_1(t_1,t_2; \frac{1}{2} \mathbb{F}_0) &=& \frac{1}{(1-t_1)^2(1-t_2)^2}\nonumber\\
g_2(t_1,t_2; \frac{1}{2} \mathbb{F}_0) &=&\frac{1}{(1-t_1^2)^2(1-t_1^2t_2^2)(1-t_2^2)^2}\end{eqnarray}

Moreover, by generalizing arguments given in section \ref{Toys} for the $1/2$ conifold, it
is easy to write a formula for generic $N$

\begin{equation}
g_N(t_1,t_2; \frac{1}{2} \mathbb{F}_0) = \frac{1}{(1-t_1^N)^2(1-t_2^N)^2}\prod_{i=1}^{N-1}\frac{1}{1-t_1^{2 i} t_2^{2 i}}
\end{equation}
which is interpreted as the fact that the ring of invariants is freely generated by four
determinants $\det {\bf A}_1,\det {\bf B}_1,\det {\bf C}_1,\det {\bf D}_1$ and $N-1$ mesons
${\rm Tr} ({\bf A}_1{\bf B}_1{\bf C}_1{\bf D}_1)^i$ with $i=1,...,N-1$. This can be understood
easily from the absence of nontrivial F term conditions and the fact that all baryons factorize
into an elementary determinant times mesons.

The $\frac{3}{4} \mathbb{F}_0$ corresponds to considering only the BPS operators
containing ${\bf A}_1,{\bf B}_1,{\bf C}_1,{\bf D}_1$ and ${\bf A}_2,{\bf C}_2$
and no occurrences of the other two elementary fields. This is done by taking

\begin{equation}
g_{1,\beta,\beta^\prime} = \sum_{n=0}^\infty (2n+1+\beta) t_1^{2 n+\beta} t_2^{2 n+\beta^\prime} .
\end{equation}

Using the same multiplicities as in Equation (\ref{F0multiplicities}) we obtain

\begin{eqnarray}
g_1(t_1,t_2; \frac{3}{4} \mathbb{F}_0) &=& \frac{1+t_1}{(1-t_1)^3(1-t_2)^2}\nonumber\\
g_2(t_1,t_2; \frac{3}{4} \mathbb{F}_0) &=&\frac{(1+t_1^2t_2^2)(1+2 t_1^2+2 t_1^4t_2^2+t_1^6t_2^2)}{(1-t_1^2)^4(1-t_1^2t_2^2)^2(1-t_2^2)^2}
\label{threequarter}
\end{eqnarray}

\section {The Molien Formula: Checks for $N>1$ }\label{molienN}

All along this thesis we have made some consistency checks of the geometric computations we proposed against a direct field theory construction. Till now we focused on some particular cases, namely the abelian $N=1$ case, some lower dimensional operators in the chiral ring, and some peculiar field theories with zero superpotential. In this section we would like to give a general method, valid for every quiver field theory, for computing generating function for generic $N$ in the field theory side.   
The baryonic generating functions found in the previous sections can indeed be checked against an
explicit field theory computation, at least for small values of $N$. The method we present here is the extension of the Molien Invariant formulae 
discussed in section \ref{Toys} to
the case in which the superpotential is an arbitrary non trivial functions of the elementary chiral superfields. 

The problem of finding
invariants under the action of a continuous group is the hearth of invariant theory and goes
back to the nineteenth  century, as most of the concepts necessary for its solution, like
syzygies and free resolutions, all amenable to Hilbert. Modern advances, such as the discovery
of Groebner basis, gave an algorithmic way of solving such problems and the advent of computer
algebra programs made some computations really doable.

The generating functions for a fixed number of colors $N$ can be reduced to a problem for polynomial rings as follows. Consider an $\mathcal{N}=1$ supersymmetric gauge theory with $F$ elementary fields $X$ and a gauge group ${\cal G}$. Since we are discussing the chiral ring we can replace ${\cal G}$ with its complexification ${\cal G}_c$.
For quiver theories, the elementary fields consist of $N\times N$ matrices.
Consider a polynomial ring in $F N^2$ variables $\mathbb{C}[X_{ij}]$
made with the entries of these matrices. The F-terms give matrix relations
whose entries are polynomial equations. We can collect all the polynomial
F-term equations in an ideal $\cal I$ and define the quotient ring
\begin{equation}
{\cal R}[X_{ij}]=\mathbb{C}[X_{ij}]/\cal I
\end{equation}
The gauge group ${\cal G}_c$ and the global symmetry group act naturally on the ring and we can grade
the elements of ${\cal R}$ with gauge and global charges. Denoting with $t_i$ the global Abelian
charges and with $z_i$ the charges under the Cartan subgroup of the gauge group ${\cal G}_c$, we can
write the generating function, or Hilbert series, of the graded ring ${\cal R}$,
\begin{equation}
H_{\cal R}(t;z)=\sum_{nm} a_{nm} z^n t^m \label{ser}
\end{equation}
which can be arranged to be a power series in the global charges $t$ and a
Laurent expansion in the gauge charges $z$. The full gauge group ${\cal G}_c$
acts on the quotient ring ${\cal R}$ and, since the gauge symmetry commutes with
the global symmetry, all the elements of ${\cal R}$ with given charge $t^m$ form
a (not necessarily irreducible) representation of ${\cal G}_c$. Therefore,
the coefficient of $t^n$ in Equation (\ref{ser}) is the
character of a ${\cal G}_c$ representation,

\begin{equation}
H_{\cal R}(t;z)=\sum_{m=0}^\infty \chi^m(z) t^m =\sum_{m=0}^\infty \left(\sum_i a^m_i \chi^{(i)}(z)\right) t^m
\end{equation}

Here we have denoted with $\chi^{(i)}$ the irreducible representations of
${\cal G}_c$ and decomposed the representation on the elements of charge $t^m$ into
irreducible ones. The generating function for invariants is given by the
projection onto the trivial representation with character $\chi^{(0)}=1$,

\begin{equation}
H_{\cal R}^{inv}(t)=\sum_{m=0}^\infty  a^m_0  t^m
\end{equation}

The projection can be easily done by averaging $H(t;z)$ on the gauge group
with the Weyl measure. The latter has indeed the property to keep only
the contribution of the trivial representation

\begin{equation}
\int d\mu(z) \chi^{(i)}(z) =\delta_{i,0}
\end{equation}

For a given group $ \cal G$ with rank $r$ we can explicitly write the measure as
a multi-contour integral

\begin{equation}
\frac{1}{|W|}\prod_{j=1}^r \int_{|z_j|=1} \frac{dz_j}{2\pi i z_j} (1-z^{h(\alpha)})
\end{equation}
where $h(\alpha)$ are the weights of the adjoint representation and $|W|$ is
the order of the Weyl group.
We finally get the Molien formula:
\begin{equation}
H_{\cal R}^{inv}(t)= \frac{1}{|W|}\prod_{j=1}^r \int_{|z_j|=1} \frac{dz_j}{2\pi i z_j} (1-z^{h(\alpha)}) H_{\cal R}(t;z)
\label{Molien}
\end{equation}
This is the generalization of the formula (\ref{molienSUN})  to the case in which the superpotential is non zero.
Since the multi-contour integrals can be evaluated with the residue theorem, the real problem in
using Equation (\ref{Molien}) is the determination of the integrand, that is the Hilbert series
of the quotient ring ${\cal R}$. Fortunately, this is the kind of problems that modern commutative
algebra made algorithmic and that can be easily solved with computer algebra programs. For
example Macaulay2 naturally deals with polynomial rings and it has a build-in command {\it
hilbertSeries}. For moderate values of $F N^2$, the computation takes fractions of second, but
it can become too hard with a standard computer already at $N=3$ and a number of fields $F$ of order 10. In these cases, one can still truncate the computation at a maximum degree in t and get a sensible
result.

It is worth mentioning that special care has to be taken when the moduli space is not irreducible and at certain points in moduli space new branches are opening up. This is the case for example for ${\cal N}=2$ theories that have additional Coulomb branches and was treated in \cite{Hanany:2006uc}. As we saw in chapter \ref{Master} the reducibility of the moduli space $\CM$ is quite generic and understanding in detail how to compute Hilbert series on this space is a fundamental task. 
The general case that we want to address is the following. Suppose that we have an $N=1$ F-term equation where one of the elementary fields can be factorized: $X_0 F(X)=0$ where $F(X)$ is a polynomial not containing $X_0$. Considering the $N=1$ moduli space as a fibration over the line parametrized by $X_0$, we see that the dimension of the fiber increases by one unit over $X_0=0$: indeed, for $X_0\ne 0$ we can impose the further constraint $F(X)=0$ which reduces by one the dimension of the fiber. This means that a new branch opens up at $X_0$ and the full moduli
space is reducible. This is for example the case for $\mathbb{F}_0$ as discussed in chapter \ref{s:F0}. 
We may want to determine the generating function
for a given irreducible component of the moduli space, in particular
the closure of the open set $X_0\ne 0$, or, for $N>1$, of $\det(X_0)\ne 0$.
This is done with a standard trick. Add a new element $q$ to the ring and
a new equation, $q \det(X_0)-1$ to the ideal $\cal I$. Clearly, the new equation
prevents $\det(X_0)$ from being zero. The irreducible component of the
moduli space we are interested in is obtained by projecting the variety
defined by the new ideal $\tilde{\cal I}=({\cal I},q\det(X_0)-1)$ on the space parameterized
by the $X,X_0$ and taking the closure. This can be done by eliminating $q$
from the ideal $\tilde{\cal I}$. This defines the elimination ideal $\cal J$ that
can be computed with the Macaulay2 command {\it eliminate}.
If we define

\begin{equation}
{\cal R}[X_{ij}]=\mathbb{C}[X_{ij}]/\cal J
\end{equation}
we can now proceed as before, compute the Hilbert series of this ring
and project it onto gauge invariants with the Molien formula. This would
give us the generating function for the particular irreducible component
of the moduli space.

We now present some explicit examples based on the conifold and $\mathbb{F}_0$. Other cases can be checked similarly, 
at least for small values of number of fields,
$F$, and number of D-branes, $N$. When $N$ increases it is necessary to truncate the series to a
maximum degree.

\subsection{$N=2$ for the Conifold}

The generating function of the conifold for $N=2$ was explicitly computed in section \ref{Nis2} 
and is given in Equation (\ref{N2con}). For $N=2$ we have four
fields ${\bf A}_i,{\bf B}_i$ that are two-by-two matrices, whose entries we denote by
$a_i^{pq},b_i^{pq}$. The four matrix F-term equations
$${\bf A}_1 {\bf B}_i {\bf A}_2={\bf A}_2 {\bf B}_i {\bf A}_1\, ,\qquad\qquad\qquad {\bf B}_1 {\bf A}_i {\bf B}_2={\bf B}_2 {\bf A}_i {\bf B}_1$$
give rises to sixteen polynomial equations for the $a_i^{p,q},b_i^{p,q}$
which generates an ideal $\cal I$ in the polynomial ring $\mathbb{C}[a_i^{p,q},b_i^{p,q}]$.
The element $(g,\bar g)$ of the
complexified gauge group $SL(2)\times SL(2)$ acts on the matrices
as ${\bf A}_i\rightarrow g {\bf A}_i \bar g^{-1}$ and ${\bf B}_i\rightarrow \bar g {\bf B}_i g^{-1}$.
All the entries $a_i^{p,q}$ and $b_i^{p,q}$ transform with a definite
charge under the Cartan subgroup

\begin{eqnarray}
g &=& \left(\begin{array}{ll} z & 0 \\ 0 & 1/z \end{array}\right ) \nonumber\\
\bar g &= & \left(\begin{array}{ll} w & 0 \\ 0 & 1/w \end{array}\right )
\end{eqnarray}

We further assign chemical potential $t_1$ to the eight fields  $a_i^{p,q}$
and chemical potential $t_2$ to the eight fields  $b_i^{p,q}$. The sixteen F-term
constraints generating the ideal $\cal I$
transform homogeneously under the gauge and global charges.
We can thus grade the quotient ring

\begin{equation}
{\cal R}_{N=2} = \mathbb{C}[a_i^{p,q},b_i^{p,q}]/\cal I
\end{equation}
with four charges, corresponding to chemical potentials, two gauge $z,w$ and two global
$t_1,t_2$. The Hilbert series of ${\cal R}_{N=2}$ can be computed using Macaulay2

\begin{eqnarray}
H_{\cal R}(t_1, t_2; z, w)=\frac{P(t_1,t_2;z,w)}{(1-t_1 zw)(1-t_1 \frac{z}{w})(1-t_1 \frac{w}{z})(1-\frac{t_1}{zw})(1-t_2 zw)(1-t_2 \frac{z}{w})(1-t_2 \frac{w}{z})(1-\frac{t_2}{zw})} \nonumber \\
\nonumber\\
\end{eqnarray}

\begin{eqnarray}
P(t_1,t_2;z,w)&=&1+4 t_1^3 t_2+4 t_1 t_2^3+ 6 t_1^2 t_2^2 + t_1^4 t_2^4-2(t_2 t_1^2+t_1 t_2^2+t_2^2 t_1^3+t_1^2 t_2^3) \left(w + \frac{1}{w}\right) \left(z + \frac{1}{z}\right) % (zw+\frac{z}{w}+\frac{w}{z}+\frac{1}{zw})
\nonumber\\
&+&t_1^2 t_2^2 \left(w + \frac{1}{w}\right)^2 \left(z + \frac{1}{z}\right)^2
%(10+2 w^2+\frac{2}{w^2}+2 z^2+\frac{2}{z^2}+\frac{z^2}{w^2}+\frac{w^2}{z^2}+w^2 z^2+\frac{1}{z^2 w^2})
\end{eqnarray}

The molien formula now reads

\begin{equation}
g_2(t_1,t_2; {\cal C})=\int_{|w|=1}\frac{dw (1-w^2)}{2\pi i w} \int_{|z|=1}\frac{dz (1-z^2)}{2\pi i z}  H_{\cal R}(t_1,t_2;z,w)
\end{equation}

Some attention should be paid in performing the contour integrals. Recall that $H_{\cal R}$ gives the
generating function for the ring ${\cal R}$ when expanded in power series in $t_1$ and in $t_2$ which are supposed to be complex numbers of modulus less than one. This should be taken into account
when performing the contour integrals on the unit circles $|z|=|w|=1$. For example, the first
contour integration in $z$ takes contribution only from the residues in the points $t_1 w, t_1/w,
t_2 w, t_2/w$ lying inside the unit circle $|z|=1$ (we take $|t_i|<1,|w|=1$). Similar arguments
apply to the second integration. After performing the two integrals we obtain
\begin{equation}
g_2(t_1,t_2)= \frac{1 + t_1 t_2 + t_1^2 t_2^2 - 3 t_1^4 t_2^2 - 3 t_1^2 t_2^4 + t_1^5 t_2^3 + t_1^3 t_2^5  - 3 t_1^3 t_2^3 + 4 t_1^4 t_2^4}{(1 - t_1^2)^3(1 - t_1 t_2)^3 (1 - t_2^2)^3} ,
\end{equation}
which perfectly coincides with Equation (\ref{N2con}).

\subsection{$N=1$ and $N=2$ for $\frac{3}{4}\mathbb{F}_0$: the Reducibility of the Moduli Space}

We now consider an example where the moduli space is not irreducible.
We consider the $\frac{3}{4} \mathbb{F}_0$ case in order to limit the number
of equations involved. The following discussion can be applied to the other cases of reducible moduli space as well.
For $N=1$ we consider the polynomial ring

\begin{equation}
{\cal R}%_{N=1} \left( \frac{3}{4} \mathbb{F}_0 \right)
 = \mathbb{C}[{\bf A}_i,{\bf B}_1,{\bf C}_i,{\bf D}_1] / \cal I
\end{equation}
with six variables. There are two F-term equations
$$ {\cal I}=( \, {\bf A}_1 {\bf B}_1 {\bf C}_2 = {\bf A}_2 {\bf B}_1 {\bf C}_1 \,\,\,  ,  {\bf C}_1 {\bf D}_1 {\bf A}_2 = {\bf C}_2 {\bf D}_1 {\bf A}_1 \, ) $$
The Hilbert series for this polynomial ring is

\begin{equation}
H_{\cal R}(t_1,t_2)=\frac{1-2 t_1^2 t_2+t_1^2 t_2^2}{(1-t_1)^4(1-t_2)^2} .
\end{equation}

As already discussed, the variety defined by $\cal I$ is not irreducible:
we are interested in the closure of the open set ${\bf B}_1,{\bf D}_1\ne 0$.
We then define a new ideal by adding two new variables $q_1,q_2$ to ${\cal R}$,

\begin{equation}
\tilde {\cal R}= \mathbb{C}[{\bf A}_i,{\bf B}_1,{\bf C}_i,{\bf D}_1, q_1,q_2] / \tilde{\cal I}
\end{equation}
and two new generators to the ideal $\cal I$

$$ \tilde{\cal I} =({\cal I}, q_1 {\bf B}_1-1, q_2 {\bf D}_1-1)$$
The closure of the open set ${\bf B}_1,{\bf D}_1\ne 0$ is obtained
by eliminating $q_1$ and $q_2$. This can be done in a polynomial
way by using the Groebner basis and the algorithm
 is implemented in Macaulay2 in the command {\it eliminate}. In our
case the elimination ideal is just
$${\cal J}=(\, {\bf A}_1 {\bf C}_2 - {\bf A}_2 {\bf C}_1 \, ) $$
and the Hilbert series of

\begin{equation}
{\cal R}^\prime = \mathbb{C}[{\bf A}_i,{\bf B}_1,{\bf C}_i,{\bf D}_1] / \cal J
\end{equation}
is

$$H_{\cal R^\prime}(t_1,t_2)= \frac{1+t_1}{(1-t_1)^3(1-t_2)^2}$$
which indeed coincides with the $g_1(t_i, \frac{3}{4} \mathbb{F}_0)$ generating
function given in Equation (\ref{threequarter}). The $N=2$ generating function
should be computed in a similar way. The fields ${\bf A}_i,{\bf B}_1,{\bf C}_i,{\bf D}_1$ are now two-by-two matrices, for a total of $24$ independent entries.
The ideal $\cal I$ now contains $8$ polynomial equations, given by
$$ \tilde{\cal I} =({\cal I}, q_1 \det {\bf B}_1 -1, q_2 \det {\bf D}_1-1) ,$$
and the elimination ideal $\cal J$ is obtained by {\it eliminating} $q_1$ and $q_2$. The Hilbert
series of $\cal J$ graded with the gauge charges is a rational function $H_{\cal
R}(t_1,t_2;z_1,z_2,z_3,z_4)$ whose expression is too long to be reported here. Four integrations
using the residue theorem finally give the $N=2$ generating function given in
(\ref{threequarter}).

%\section{Including all the charges}\label{empty}

%In Section \ref{examples} we saw in examples how multiplicities arise in the GKZ fan. The basic object which generates these multiplicities is called the {\bf auxiliary GKZ partition
%function} and is given in Equations (\ref{ZauxC3Z3},\ref{ZauxF0}). In this section we will add extra coordinates to the GKZ lattice to remove these multiplicities and present a lattice for which every point has multiplicity 1.

% \cite{Franco:2005sm} \cite{Burrington:2006uu}

\section{Summary and Discussions}

In this chapter we have performed a further step in the understanding and the computation of the
generating functions for the chiral ring of the superconformal gauge theories living on branes
at CY singularities. We have reinforced the conjecture that the generating functions for $N$
colors can be computed simply in terms of the $N=1$ generating functions through the plethystic
program. We tested this conjecture, suggested by a computation with D3-branes in the dual
background, in field theory for small values of $N$. It would be interesting to perform checks
for large $N$ as well as to investigate the statistical properties of the resulting generating
functions.

In particular we have made an explicit investigation of the properties of the complete $N=1$
generating function and we have compared the result with a geometrical computation. The emerging
structure reveals once more the deep interplay between the quiver gauge theory and the algebraic
geometry of the CY. We reinforce the relations between the K\"ahler
moduli space of the CY an the baryonic charges in the field theory we observed in the previous chapter. Indeed in this chapter we
found an intriguing relation between the decomposition of
the $N=1$ generating function in sector of given baryonic charge and the discretized K\"ahler
moduli space of the CY.

The geometrical structure of the complete moduli space for $N$ colors, which is
obtained by the $N$-fold symmetrized product of the CY by adding the baryonic
directions, is still poorly understood. In chapter \ref{Master} we have seen that already for $N=1$
the moduli space is rich and interesting. 

In this chapter we have done the nice step forward to give a general recipe, valid for
every singularity $\cX$, to compute the generating function $g_N$ for arbitrary values of N.
In the next chapter we will apply the techniques we have developed for the finite N counting
to the study of the complete moduli space $\CM_N$ for generic values of $N$.

\chapter{Back to the Moduli Space}\label{backtoms}

Enriched by the discussions of the last three chapters we can now go back to the problem of studying the moduli space $\CM$ of SCFT at singularity $\cX$. Now we know how to compute partition functions for generic $N$ number of branes and we can use the paradigm of algebraic geometry: an algebraic variety is completely described by its coordinate ring. Namely if we want to study the moduli space $\CM_N$ for a SCFT with 
$N$ number of colors, it is enough to study the spectrum of holomorphic function $\mathbb{C}[\CM_N]$ over it, or, in a field theory language, the 
spectrum of 1/2 BPS operators. We will see that the Hilbert series of $\CM_N$ for generic $N$ contain a lot of interesting informations 
about the moduli space $\CM_N$. In particular, using the PE function and the GKZ decomposition, we would be able to extend the 
hidden symmetries we have found in the Master space $\f$ for $N=1$ brane, to the complete moduli space $\CM_N$ for generic number of
branes N.

\section{Generalities}

In chapter \ref{Master} we made an extensive study of the moduli space $\CM=\f$ of the gauge theory living on a single D3 brane probing the three-fold singularity $\cX$. We would be definitively interested to obtain as many informations as possible regarding the moduli space $\CM_N$ for generic number of brane $N$. This seems in principle a very difficult problem, if we want to attack it directly as we did in the case of just one brane. However thanks to the techniques we developed to study the BPS spectrum of these kind of gauge theories, we will be able to extract non trivial informations regarding the moduli space $\CM_N$ in a rather indirect way.

In the two following subsections we review what we know about the moduli space $\CM_N$ and the PE function.
 
\subsection{The non Abelian Moduli Space}

Let us start summarizing what we can say about $\CM_N$ looking directly at the defining equations.
The case for an arbitrary number $N$ of D3-branes is much more subtle and less understood in the mathematical literature than the $N=1$ case, 
even though it is clear from the gauge theory perspective. We know that the world-volume theory for $N$ D3-branes is a quiver theory with product $U(N_i)$ gauge groups and in the IR, the $U(1)$ factors decouple since only the special unitary groups exhibit asymptotic freedom and are strongly coupled in the IR. Thus the moduli space  of interest is the space of solutions to the F-flatness, quotiented out by a non-Abelian gauge group 
%${\cal G}_N = \prod_i SU(N_i)$,
\beq
\CM_N = \f_N / (SU(N_1) \times \ldots \times SU(N_g)).
\eeq 
where the index $N$ recalls that we are dealing with $N$ branes. The moduli 
space $\CM_N$ is of difficult characterization since the quotient is
fully non-Abelian and it can not be described by toric methods, as in the $N=1$
case. This means that passing from $N=1$ to $N > 1$ we lose lot of the powerful tolls of toric 
geometry we have extensively used in the $N=1$ case.  

The more familiar mesonic moduli space is obtained by performing a
further quotient by the Abelian symmetries. Even for $N$ branes, the Abelian group will be constituted of the decoupled $U(1)$ factors, and hence will be the same as in the toric, $N=1$ case. 
Once again, we expect to have some symplectic quotient structure, in analogy with \eref{mesmod}, for this mesonic moduli space:
\beq\label{symM-N}
{}^{{\rm mes}}\!{\cal M}_N    \simeq \CM_N // U(1)^{g-1} \ .
\eeq
Hence, a toric symplectic quotient still persists for \eref{symM-N}, even though the moduli space in question is not necessarily toric.

Moreover, our plethystic techniques, which we will shortly review, will illuminate us with much physical intuition. First, the mesonic moduli space for $N$ branes is the $N$-th symmetrized product of that of $N=1$:
\begin{equation}\label{symM}
{}^{{\rm mes}}\!{\cal M}_N \simeq {\rm Sym}^N \cX := \cX^N / \Sigma_N \ ,
\end{equation}
where $\Sigma_N$ is the symmetric group on $N$ elements.
%Once again, we expect to have some symplectic quotient structure, in analogy with \eref{M}, for this mesonic moduli space:
%\beq\label{symM-N}
%\cX_N \simeq \CM_N // G_N \ .
%\eeq
We see that the mesonic moduli space, for $\cX$ a Calabi-Yau threefold, is of dimension $3N$ by \eref{symM}.
% what is not clear is what the moduli space $\CM_N$ and the group $G_N$ should be\footnote{We thank Balazs Szendroi for discussions on this point.}. 
%For $N=1$, it is clear, as was in the cases discussed above, that $\CM_1 = \f$ is simply the space of F-flatness and $G_1$ is the toric $U(1)^{g-1}$ action. Even though mathematically it may be difficult to construe, physically, there are natual candidates for $\CM_N$ and $G_N$. We know that the world-volume theory for $N$ D3-branes is a quiver theory with product $U(N_i)$ gauge groups and in the IR, the $U(1)$ factors decouple since only the special unitary groups exhibit asymptotic freedom and is strongly coupled in the IR. Therefore even for $N$ branes, $G_N$ will be constituted of the decoupled $U(1)$ factors, and hence will be the same as in the toric, $N=1$ case. Hence, a toric symplectic quotient still persists for \eref{symM-N}, even though $\CM_N$ is not necessarily toric. The moduli space, however, will be a little more involved: $\CM_N$ is the space of solutions to the F-flatness, quotiented out by the non-Abelian gauge group ${\cal G}_N = \prod_i SU(N_i)$. In su!
% mmary, gauge theory dictates that the mesonic moduli space is
%\beq
%\cX_N \simeq \left(\f / (SU(N_1) \times \ldots \times SU(N_g))\right) // U(1)^{g-1}
%\eeq
%and the combined baryonic and mesonic moduli space is the quotient 
%\beq
%\CM_N = \f / (SU(N_1) \times \ldots \times SU(N_g)).
%\eeq 
The dimension of the moduli space $\CM_N$ is thus $3N + g - 1$ for general $N$.

An important observation is that the symmetric product of a toric CY variety is no more toric and generically non CY. This mesonic moduli space is the base of the $(\mathbb{C}^*)^{g-1}$ baryonic fibrations whose total space is the complete moduli space $\CM_N$.
Combining our knowledge on the complete moduli space for just one brane $\CM=\f$ and our skill in counting BPS operators for generic $N$ we will study the properties of $\CM_N$.

\subsection{The Plethystic Programme Reviewed}\label{s:plet}

As we have already realized the PE function is a fundamental tool in the study of general $N$. Here we review/summarize some essential facts, 
that we learnt to appreciate in the previous chapters and that will be very useful in the present analysis. 

The realization in \cite{Benvenuti:2006qr,Feng:2007ur} is that the mesonic generating function, with all baryonic numbers fixed to be zero, $g_{1,0}(t;~\cX) = f_\infty(t;~\cX)$, for the single-trace mesonic operators for D3-branes probing a Calabi-Yau threefold $\cX$ at $N \rightarrow \infty$, is the Hilbert series of $\cX$. The {\bf Plethystic Exponential} of a multi-variable function $g(t_1, \ldots, t_n)$ that vanishes at the origin, $g(0, \ldots, 0) = 0,$ is defined as:

\beq\label{defPE}
PE \left [ g (t_1, \ldots, t_n) \right ] := \exp\left( \sum\limits_{k=1}^\infty\frac{g (t_1^k, \ldots, t_n^k)}{k}\right)  \ .
\eeq

Then the multi-trace mesonic operators at $N \rightarrow \infty$ are counted by the plethystic exponential
%\footnote{Note that in order to avoid an infinity the PE is defined with respect to a function that vanishes when all chemical potentials are set to zero.}
\beq\label{defPE1}
g_{\infty,0}(t;~\cX) = PE[f_\infty(t;~\cX)-1] := \exp\left( \sum\limits_{r=1}^\infty\frac{f_\infty (t^r)-1}{r}\right)  \ .
\eeq
The inverse, $f_1(t;~\cX) = PE^{-1}[f_\infty(t;~\cX)]$, is counting objects in the defining equation (syzygy) of the threefold $\cX$. The mesonic multi-trace generating function $g_N$ at finite $N$ is found by the series expansion of the $\nu$-inserted plethystic exponential
%\footnote{Note that the $\nu$ insertion satisfies the condition that the argument of PE vanishes when all chemical potentials are set to zero. 
%Any attempt to subtract something from this function leads to incorrect results.}
 as 

\beq
PE_{\nu}[ f_\infty(t;~\cX)] = \exp\left( \sum\limits_{r=1}^\infty\frac{\nu^r f_\infty(t^r)}{r}\right) = \sum\limits_{N=0}^\infty g_N(t) \nu^N.
\eeq

In general (chapter \ref{Master}) for the combined mesonic and baryonic branches of the moduli space\footnote{To be Strict, we should say here the mesonic branch together with given FI-parameters, since at $N=1$, there are no baryons. Nevertheless, we can still generate the counting for the baryons for $N>1$ using PE of the $N=1$ case.}, the $N=1$ operators are counted by the Hilbert series of the master space $\f$. The plethystic program can be efficiently applied to the study
of the coherent component of the moduli space ( chapter \ref{chiralcount} ).
%$g_1(t;~\cX)\equiv H(t; ~\firr)$ which is the Hilbert series of  the 
%master space as described repeatedly in the examples above. 
With the generating function for the coherent component of the master space,
which we denote $g_1(t;~\cX)\equiv H(t; ~\firr{~}) $, we can proceed with the
plethystic program and find the result for $g_N(t;~\cX)$, counting the combined baryonic and mesonic gauge invariant operators at finite $N$. 

As we explained in the previous chapter the implementation of the plethystic program requires a decomposition
of the $g_1(t;~\cX)$ generating function in sectors of definite baryonic 
charges, to which the plethystic exponential is applied. There exist an interesting connection of this decomposition with K\"ahler moduli. 
%This also enables  a different computation of $g_1(t;~\cX)$. Though our current techniques compute this quantity using the Hilbert series of the master space, we will later check that this indeed agrees with the formalism of \cite{Butti:2007jv}.
The decomposition of $g_1(t;~\cX)$ requires the knowledge
of two sets of partition functions, the geometrical ones, obtained by 
localization, and the auxiliary one, obtained from dimer combinatorics.
Specifically, we realized that 
\begin{equation}\label{GKZ1}
g_1(t;~\cX) = \sum\limits_{P \in GKZ} m(P) g_{1,P}(t;~\cX)
\end{equation} 
where the summation is extended over the lattice points $P$, of multiplicity $m(P)$, of the so-called {\bf GKZ (or secondary) fan} of the Calabi-Yau threefold and $g_{1,P}$ is a much more manageable object obtained from a localization computation, as given in Eq (\ref{loc}).
%(4.18) of \cite{Butti:2007jv}. 
The GKZ fan, to remind the reader, is the fan of an auxiliary toric variety, which is the space of K\"ahler parameters of the original toric threefold $\cX$. This space is of dimension $I - 3 + d$, where $I$ is the number of internal points and $d$, the number of vertices, of the toric diagram of $\cX$. 

The multiplicity $m(P)$ of points in this GKZ lattice fan is counted by an {\bf auxiliary partition function}, so-named $Z_{\rm aux}$. This is simply the (refined) Hilbert series of the following space: take the simpler quiver than the original by neglecting any repeated arrows and then form the space of open but not closed loops in this simplified quiver. The expansion of $Z_{\rm aux}$ can then be used to compute the generating function for one D3-brane. 

As a brief reminder to the reader, the procedure to determine the refined generating function $g_1$ in \eref{GKZ1} is to (1) obtain the generating function $g_{1,\beta_1, ... \beta_K}$ in terms of a set of K\"ahler parameters $\beta_1, ..., \beta_K$ using the localisation formula (\ref{loc});
%(4.18) of \cite{Butti:2007jv}; 
(2) obtain $Z_{\rm aux}(t_1,...,t_K)$ as above, and (3) replace a term $t_1^{\beta_1} ...  t_K^{\beta_K}$ in $Z_{\rm aux}$ by an expression for $g_{1,\beta_1,...\beta_K}$. 
%We will not enter in the details of this construction and we refer the reader
%to the chapter \ref{ } \cite{Butti:2007jv}.
The important point for our ensuing
discussions is that the plethystic program
can be applied to the $N=1$ partition functions at each point of the GKZ fan 
in order to obtain the finite  $N$ generating function 
\begin{equation}\label{GKZN}
g(t;~\cX) := \sum_{N=0}^\infty \nu^N g_N(t;~\cX) = \sum\limits_{P \in GKZ} m(P) PE_{\nu} \left [  g_{1,P}(t;~\cX)\right ] \ .
\end{equation}

The organization of the rest of the chapter is as follows. In section \ref{s:hidden} we introduce and elaborate on 
the remarkable feature of ``hidden symmetries'' of gauge theories from D3 branes at singularities. In chapter \ref{Master} we have already 
seen that the coherent component $\firr{~}$ of the master space for one brane has generically some hidden symmetries 
not manifest in the UV Lagrangian describing the field theory. 
In this section we will see that the symmetries of $\firr{~}$ can manifest themselves as hidden global symmetries of the gauge theory,
and indeed they can sometime be promoted to symmetries of the complete moduli space $\CM_N$ for generic value of N. 
We examine how such symmetries beautifully exhibit themselves in the plethystics of the Hilbert series 
of the master space by explicitly arranging themselves 
into representations of the associated Lie algebra. 
In sections \ref{dP0rev} and \ref{s:F0rev} we describe in details how to study the moduli 
space of gauge theories for generic N in the $dP_0$ and $\mathbb{F}_0$ examples respectively.

\section{Hidden Global Symmetries}\label{s:hidden}\setall

In chapter \ref{Master} we realized that the master space $\firr{~}=\CM$ of gauge theories at singularities $\cX$, in the case $N=1$,  
has generically more symmetries than the Lagrangian describing UV field theory. A natural question is: 
is it an accident of the $N=1$ case or it propagates to generic N ? 
May we learn something about the moduli space for generic $N$ looking at the symmetries of the N=1 moduli space ?
Or in the language of the chapter \ref{Master}: Is it possible that the master space $\firr{~}$ contains important informations regarding the complete 
moduli space $\CM_N$ ? 
In the rest of this chapter we will try to give an affirmative answer to these questions.

 All along the history, the physics made huge progress looking to the symmetries of our world. 
Indeed the idea will move our following studies will be: SYMMETRIES. 
The moduli space of a field theory may possess symmetries beyond gauge symmetry or reparametrisation. 
Searching for hidden symmetries of a given supersymmetric field theory often leads to insight of the 
structure of the theory. For D-brane quiver theories, an underlying symmetry is ever-present: 
the symmetry of the Calabi-Yau space $\cX$ is visible in the UV as a global flavor symmetry in the Lagrangian, 
while the symmetry of the full moduli space $\CM$ can reveal a new {\bf hidden global symmetry} which develops as the theory flows to the IR. 

In particular, in \cite{Franco:2004rt}, the basic fields of the quiver for $dP_n$ theories were reordered into multiplets of a proposed $E_n$ symmetry and consequently the superpotential terms into singlets of this symmetry. 
The exceptional Lie group $E_n$ acts geometrically on the divisors of the del Pezzo 
surfaces and is realized in the quantum field theory as a hidden symmetry
enhancing the $n$ non-anomalous baryonic $U(1)$'s of the $dP_n$ quiver.

Do the symmetries of the master space, whose geometrical significance we have learned to appreciate in the foregoing discussions, 
manifest themselves in the full moduli space $\CM_N$ of the gauge theory? 
Phrased another way, do these symmetries survive the symplectic quotient of \eref{M} and manifest themselves also
at finite $N$? This is indubitably a natural question. 

In this section, 
equipped with the new notion of the master space we can revisit this problem and recast all operators into irreducible representations of the symmetry of $\firr{~}$.
We will show that these symmetries are encoded in a subtle and beautiful way by the fundamental invariant of the plethystic programme for $\CM_N$:
 the Hilbert series.
Moreover, we will demonstrate in some examples
how the hidden symmetries of the moduli space come from enhancing anomalous and 
non-anomalous Abelian symmetries of the Lagrangian describing the UV field theory.

Since we will always deal with the coherent component of the moduli space in the following and there is no source for ambiguity, we will adopt a simplified notation for the Hilbert series:
\begin{equation}
g_1(t;\cX) \equiv H(t; \firr{\cX})
\end{equation}

%%%%%%%%%%%%%%%%%%%%%%%%%%%%%%%%%%%%%%%%
\subsection{Character Expansion: a Warm-up with $\mathbb{C}^3$}
The ${\cal N}=4$ supersymmetric gauge theory does not have a baryonic branch and therefore the master space $\f$ coincides with the Calabi-Yau manifold $\cX = \mathbb{C}^3$.
We therefore do not expect any new symmetries but instead can use this example as a warm-up example for expanding in terms of characters of the global symmetry. For a single D3-brane, $\f \simeq \CM \simeq \cX \simeq \IC^3$ and there is a $U(3)$ symmetry.
Now, there is an $SU(4)_R$ symmetry for which $U(3)$ is a maximal subgroup, however, we shall see below that the slightly smaller $U(3)$ suffices to keep the structure of the BPS operators.

The generating function for $\mathbb{C}^3$ is well known and was computed in various places (cf.~e.g.~\cite{Benvenuti:2006qr,Feng:2007ur}). It takes the form
\begin{equation}\label{C3}
g(\nu; t_1, t_2, t_3; \mathbb{C}^3) = \prod_{n_1=0}^\infty \prod_{n_2=0}^\infty \prod_{n_3=0}^\infty \frac{1}{1-\nu t_1^{n_1} t_2^{n_2} t_3^{n_3}} \ ,
\end{equation}
which coincides with the grand canonical partition function of the three dimensional harmonic oscillator.
This form is perhaps the simplest one can write down for the exact answer and from this extract the generating function for any fixed $N$.
In this subsection, we will rewrite it in terms of characters of the $U(3)$ global symmetry.
The expansion demonstrates how one can explicitly represent this function in terms of characters.
This will help in analyzing the next few examples in which expansion in terms of characters are done but for more complicated cases.

Equation (\ref{C3}) admits a plethystic exponential form,
\begin{equation}\label{C3PE}
g(\nu; t_1, t_2, t_3; \mathbb{C}^3) = PE_{\nu}[ g_1 ] , \quad 
g_1(t_1, t_2, t_3; \mathbb{C}^3) = \frac{1}{(1-t_1)(1-t_2)(1-t_3)} = 
PE[t_1 + t_2 + t_3] \ .
\end{equation}
We recall that $g_1$, the generating function for a single D3-brane, $N=1$, is none other than the refined Hilbert series for $\IC^3$, itself being the $PE$ of the defining equations (syzygies) for $\IC^3$, here just the 3 variables. Furthermore, $PE_{\nu}[ g_1 ]$ encodes all the generators for the multi-trace operators (symmetric product) at general number $N$ of D-branes.

We can now introduce $SU(3)$ weights $f_1, f_2$ which reflect the fact that the chemical potentials $t_1, t_2, t_3$ are in the fundamental representation of $SU(3)$, and a chemical potential $t$ for the $U(1)_R$ charge,
\begin{equation}
(t_1, t_2, t_3) = t \left (f_1, \frac{f_2}{f_1}, \frac{1}{f_2} \right ) \ .
\end{equation}
We can define the character of the fundamental representation with the symbol
\begin{equation}
[1, 0 ] = f_1+ \frac{f_2}{f_1}+ \frac{1}{f_2} \ ,
\end{equation}
and get
\begin{equation}
g_1(t_1, t_2, t_3; \mathbb{C}^3) = PE \left [ [1, 0] t  \right ] = \sum_{n=0}^\infty [n,0] t^n \ .
\end{equation}
The second equality follows from the basic property of the plethystic exponential which produces all possible symmetric products of the function on which it acts. The full generating function is now rewritten as
\begin{equation}\label{C3PE-1}
g(\nu; t_1, t_2, t_3; \mathbb{C}^3) = PE_{\nu} \left [ \sum_{n=0}^\infty [n,0] t^n \right ] ,
\end{equation}
giving an explicit representation as characters of $SU(3)$. This equation is very important, and shows explicitly that the complete spectrum
of BPS operators for the $\mathcal{N}=4$ theory, for generic $N$, reorganize in representation of $U(3)$. H
ence the complete moduli space $\CM_N$ has at least 
an $U(3)$ symmetry. This is clearly an observation that is expected and can be obtained in many other ways, but a similar fact will be rather non 
trivial for other singularity $\cX$: If the Hilbert series for generic N can be explicitly written in term of character of certain group 
G, than the moduli space $\CM_N$ for generic N will have the group G as symmetry group.
 
We can expand $g(\nu; t_1, t_2, t_3; \mathbb{C}^3) = \sum\limits_{N=0}^\infty g_N(t_1, t_2, t_3; \mathbb{C}^3) \nu^N$ and find, for example,
\begin{equation}\label{C3N2}
g_2(t_1, t_2, t_3; \mathbb{C}^3) = \left ( 1 - [0,2] t^4 + [1,1] t^6 - [0,1] t^8 \right ) PE \left [ [1,0] t + [2,0] t^2 \right ] \ .
\end{equation}
Alternatively we can write down an explicit power expansion for $g_2$ as
\begin{equation}\label{C3N2ser}
g_2(t_1, t_2, t_3; \mathbb{C}^3) = \sum_{n=0}^\infty \sum_{k=0}^{\lfloor \frac{n}{4}\rfloor + \lfloor \frac{n+1}{4}\rfloor} m(n,k)[n-2k, k] t^{n} , 
\qquad
m(n,k) 
= 
\left\{ \ba{lcl}
\lfloor \frac{n}{2}\rfloor -k+1 && n \mbox{ odd} \\ 
\lfloor \frac{n}{2}\rfloor - 2 \lfloor \frac{k+1}{2}\rfloor +1 &&
n \mbox{ even} 
\ea
\right. 
\end{equation}

Note that $g_2$ is not palindromic, indicating that the moduli space of 2 D-branes on $\mathbb{C}^3$, as expected, is not Calabi-Yau. 
%It is worth pursuing an analysis along these lines and find the various irreducible components. We will not pursue this further but 
Armed with this character expansion let us now turn to more involved cases where there is a baryonic branch.

%%%%%%%%%%%%%%%%%%%%%%%%%%%
\subsection{Conifold Revisited}\label{s:coni}
Having warmed up with $\IC^3$, let us
begin with our most familiar example, the conifold $\cX = \cC$.
In chapter \ref{Master} we learnt that the master space for the conifold is simply $\f = \mathbb{C}^4$. The symmetry of this space is $SU(4)\times U(1)$ where the $U(1)$ is the R-symmetry while the $SU(4)$ symmetry is not visible at the level of the Lagrangian and therefore will be called ``hidden''. One should stress that at the IR the two $U(1)$ gauge fields become free and decouple and one is left with 4 non-interacting fields, which obviously transform as fundamental representation of this $SU(4)$ global symmetry. Now, there is a mesonic $SU(2)\times SU(2)$ symmetry and a baryonic $U(1)_B$ symmetry, this $SU(4)$ {\bf Hidden Symmetry} is an enhancement of both. 
%
%\begin{figure}[t]\begin{center}
%$\begin{array}{cc}
%\begin{array}{c}
%\epsfxsize = 8cm \epsfbox{conifold.eps}
%\end{array}
%&
%W_{\cC} =  \tr(\epsilon_{il} \epsilon_{jk} A_i B_j A_l B_k)
%\end{array}$
%\caption{{\sf The quiver and toric diagrams, as well as the
%    superpotential for the conifold $\cC$.}}
%\label{f:coni}
%\end{center}\end{figure}
In section \ref{conifoldB} we learnt how to assign the UV
%To start let us recall the theory in \fref{f:coni}.
%Indeed, we see that when the number of branes $N=1$, we have a $U(1)^2$ theory with $W = 0$. The vanishing of the superpotential in this case means that we have four free variables $A_{1,2}$ and $B_{1,2}$ and the master space should be $\IC^4$. 
%The gauge theory has an explicit global symmetry 
$SU(2)_1\times SU(2)_2\times U(1)_R\times U(1)_B$ global symmetry charges to the $A_{i}$ and $B_{j}$ fields that are the coordinates of the master
space $\f = \mathbb{C}^4$. We obtained indeed the N=1 generating function:
%parametrizing and the four fields transform under these symmetries according to \fref{weightsConi}.
%\begin{table}[t]\begin{center}
%\begin{tabular}{|c||c|c|c|c||c|}
%\hline
% & $SU(2)_1$ $(j_1,m_1)$& $SU(2)_2$ $(j_2,m_2)$& $U(1)_R$ & $U(1)_B$ & monomial\\ \hline \hline
%$A_1$ & $(\frac{1}{2}, +\frac{1}{2})$ & $(0,0)$ & $\frac{1}{2}$ & 1& $t_1 x$\\ \hline
%$A_2$ & $(\frac{1}{2},-\frac{1}{2})$ & $(0,0)$ & $\frac{1}{2}$ & 1& $\frac{t_1}{x}$ \\ \hline
%$B_1$ & $(0,0)$& $(\frac{1}{2},+\frac{1}{2})$ & $\frac{1}{2}$ & -1& $t_2 y$ \\ \hline
%$B_2$ & $(0,0)$ & $(\frac{1}{2},-\frac{1}{2})$ & $\frac{1}{2}$ & -1& $\frac{t_2}{y}$ \\ \hline
%\end{tabular}
%\caption{{\sf The transformation, under the explicit global symmetry group 
%$SU(2)_1\times SU(2)_2\times U(1)_R\times U(1)_B$, of the 4 fields in the conifold theory. The monomials indicate the associated chemical potentials in the Plethystic programme.}}
%\label{weightsConi}
%\end{center}\end{table}
%We have marked the monomials for the counting of baryonic operators whose
%generating function for $N=1$, in the notation of the plethystic programme, was given in Equation (3.3) of \cite{Forcella:2007wk}; this is also simply the refined Hilbert series for the master space $\f_{\cC}$:
\begin{equation}\label{g1coni}
\ba{rcl}
g_1(t_1,t_2,x,y; {\cal C}) &=& H(t_1,t_2,x,y; \f_{\cC} = \IC^4)
= \frac{1}{(1-t_1 x) (1-\frac{t_1}{x})(1-t_2 y) (1-\frac{t_2}{y})} \\
&=& PE[t_1 x + \frac{t_1}{x} + t_2 y + \frac{t_2}{y}] \ .
\ea
\end{equation}
In the above, we recall from section \ref{s:plet} that the Hilbert series is itself the PE of the defining equations. 
%Moreover, as in \cite{Forcella:2007wk} we can define $b$ which counts (i.e., is the chemical potential associated to) baryon number and $t$ which counts the total R-charge; then $t_1 = t b$ and $(t_2 = t/b)$ would respectively count the number of $A$ and $B$ fields appearing in the baryonic operator. Furthermore, $x$ and $y$ keep track of the first and second $SU(2)$ weights respectively. Indeed, if we unrefined by setting $t_1=t,t_2=t,x=1,y=1$, we would obtain the familiar Hilbert series for $\IC^4$ which is $g_1(t;~\IC^4) = (1-t)^{-4}$.

Now, the fields $A_{i}$ and $B_{j}$ are in the representations of the explicit $SU(2)_1\times SU(2)_2\times U(1)_R \times U(1)_B$ symmetry, 
while our hidden symmetry is $SU(4) \times U(1)$. We will therefore introduce $SU(4)$ weights, $h_1, h_2, h_3$ and map them to the three weights of the $SU(2)_1\times SU(2)_2\times U(1)_B$ global symmetry. This is done simply by first taking the four weights of the fundamental representation of $SU(4)$, whose character we will denote as $[1,0,0]$, and which we recall can be written in terms of the weights as
\begin{equation}
[1,0,0] = h_1 + \frac{h_2}{h_1} + \frac{h_3}{h_2} + \frac{1}{h_3} \ .
\end{equation}
Then we multiply by $t$ to obtain a $SU(4) \times U(1)$ representation, 
which should be mapped them to the four weights of $SU(2)_1\times SU(2)_2\times U(1)_R \times U(1)_B$:
% in the rightmost column of \eref{weightsConi} above:
\begin{equation}
t \left ( h_1, \frac{h_2}{h_1}, \frac{h_3}{h_2}, \frac{1}{h_3} \right ) = \left (t_1 x, \frac{t_1}{x}, t_2 y, \frac{t_2}{y} \right ) \ .
\end{equation}
%This has a solution $t = \sqrt{t_1 t_2}, h_1 = b x, h_2 = b^2, h_3 = b y$.

In analogy with \eref{C3PE}, we can now write \eref{g1coni} in terms of a plethystic exponential:
\begin{equation}\label{g1conif}
g_1(t, h_1, h_2, h_3; {\cal C}) = PE \left [ [1,0,0] t \right ] = \sum_{n=0}^\infty [n, 0, 0] t^n \ ,
\end{equation}
where $[n, 0, 0]$ is the completely symmetric tensor of rank $n$ and dimension $n+3 \choose 3$. The first equality writes the 4 generators of $\IC^4$, viz., $t_1 x, \frac{t_1}{x}, t_2 y, \frac{t_2}{y}$, in the weights of $SU(4) \times U(1)$ and the second equality follows from the definition of $PE$ in \eref{defPE} and the fact that expansion of the plethystic exponential in power series of $t$ will compose the $n$-th symmetrized product. Equation \eref{g1conif} is a trivial and obvious demonstration that the $N=1$ generating function is decomposed into irreducible representations of $SU(4)$, precisely one copy of the irreducible representation $[n, 0, 0]$ at R-charge $n$. To be precise we are taking here the R-charge to be $n$ times that of the basic field $A$.

%%%%%%%%%%%%%%%%%%%%%%%%%
\subsubsection{Hidden Symmetries for Higher Plethystics}
We have now seen that the basic invariant, i.e., $g_1$, the Hilbert series, of $\f_{\cC}$, can be written explicitly as the plethystic exponential of the fundamental representation of $SU(4)$. Now, the hidden symmetry mixes baryon number with meson number and therefore we do not expect this symmetry to hold for general number $N$ of D3-branes. For the case of $N=2$, however, baryons and mesons have the same R-charge and therefore we may expect the global symmetry to be enhanced. 

Actually for $N=2$, $SU(4)$ becomes a symmetry of the Lagrangian of the conifold theory, which is an $SO(4)$ theory with four flavors in the vector
representation and an $SU(4)$ invariant superpotential. Writing the 4 flavors in the vector representation of $SO(4)$ as a $4\times4$ matrix $Q$ we find that the superpotential is $W = \det Q $. Using the formalism which we developed to count gauge invariant operators for the conifold we find that the $N=2$ generating function does indeed decompose into characters of irreducible representations of $SU(4)$.

The generating function for $N=2$ was computed in section \ref{conifoldNN}.
%Equation (3.47) and its predecessor in \cite{Forcella:2007wk}. 
This generating function could be given as a function of all 4 chemical potentials and is quite lengthy. 
Here, we will take this expression and recast it into characters of the global symmetry $SU(4)$. 
The first point to note is that the generators form the $[2, 0, 0]$ representation of $SU(4)$. 
It is natural to expect this since this representation is the second rank symmetric product of the generators for $N=1$. 
The other terms are less obvious and need explicit computation. A short computation yields
\begin{equation}\label{g2coni}
g_2(t, h_1, h_2, h_3; \cC) = \left (1 - [0,0,2] t^6 + [1,0,1] t^8 - [0,1,0] t^{10} \right ) PE \left [ [2,0,0] t^2 \right ] \ .
\end{equation}
This can be seen if we write the explicit expressions for the characters of the irreducible representations of $SU(4)$ which we can write in terms of the weights as
\begin{equation}\ba{rcl}
[2,0,0] &=& h_1^2+\frac{h_3
   h_1}{h_2}+\frac{h_1}{h_3}+\frac
   {h_3^2}{h_2^2}+h_2+\frac{1}{h_2
   }+\frac{1}{h_3^2}+\frac{h_3}{h_1}+\frac{h_2}{h_3
   h_1}+\frac{h_2^2}{h_1^2}  \\
\left [0,0,2\right ] &=& \frac{h_1^2}{h_2^2}+\frac{h_3
   h_1}{h_2}+\frac{h_1}{h_3}+h_3^2+h_2+\frac{1}{h_2}+\frac{h_2^2}
   {h_3^2}+\frac{h_3}{h_1}+\frac{h_2}{h_3 h_1}+\frac{1}{h_1^2}  \\
\left [1,0,1 \right ] &=& \frac{h_1^2}{h_2}+\frac{h_3
   h_1}{h_2^2}+h_3
   h_1+\frac{h_2
   h_1}{h_3}+\frac{h_1}{h_2
   h_3}+\frac{h_3^2}{h_2}+\frac{h_2}{h_3^2}+3+\frac{h_2
   h_3}{h_1}
   +\frac{h_3}{h_2
   h_1}+\frac{h_2^2}{h_3
   h_1}+\frac{1}{h_3
   h_1}+\frac{h_2}{h_1^2}  \\
\left [0,1,0 \right ] &=& \frac{h_3
   h_1}{h_2}+\frac{h_1}{h_3}+h_2+\frac{1}{h_2}+\frac{h_3}{h_1}+\frac{h_2}{h_3 h_1} \ .
\ea
\end{equation}
As a check, \eref{g2coni} can be expanded to first few orders in $t$,
\[\ba{rcl}
g_2(t, h_1, h_2, h_3; \cal C) &=&
1 + [2,0,0] t^2 + \left ( [4, 0, 0] + [0,2,0] \right ) t^4
+ \left ( [6,0,0] + [2,2,0] \right ) t^6  \\
&&+ \left ( [8,0,0] + [4,2,0] + [0,4,0] \right ) t^8 
+ \left ( [10,0,0] + [6,2,0] + [2,4,0] \right ) t^{10} + \ldots
\ea\]
and in series form it is
\begin{equation}
g_2(t, h_1, h_2, h_3; {\cal C}) = \sum_{n=0}^\infty \left ( \sum_{k=0}^{\lfloor \frac {n}{2} \rfloor } [2n -4k, 2k, 0] \right ) t^{ 2 n } .
\end{equation}

Now, using the formula for the dimension of a generic irreducible representation of $SU(4)$,
\begin{equation}
{\rm dim} [n_1, n_2, n_3] = \frac{ (n_1+n_2+n_3+3) (n_1+n_2+2) (n_2+n_3+2) (n_1+1) (n_2+1) (n_3+1) } {12} ,
\end{equation}
we find that the unrefined generating function for $N=2$ sums to
\begin{equation}
g_2(t, 1, 1, 1; {\cal C}) = \frac{ 1 + 3 t^2 +6 t^4 } { (1-t^2 )^7} ,
\end{equation}
as expected and in agreement with the Equations in section \ref{conifoldNN}. 
Taking the Plethystic Logarithm we find that the 7 dimensional manifold is generated by 10 operators of order 2 transforming in the $[2, 0, 0]$ representation of $SU(4)$ subject to 10 cubic relations transforming in the $[0, 0, 2]$ representation of $SU(4)$. 
Since $g_2$ is not palindromic we expect the moduli space of 2 D-branes on the conifold to be not Calabi-Yau.

This confirms the expansion in terms of characters of $SU(4)$ for the case of $N=2$. Unfortunately this symmetry does not extend to $N=3$ and is not a symmetry for higher values of $N$. 

In the next example we are going to see how a hidden symmetry extends to all values of $N$, simply because the hidden symmetry does not mix baryonic numbers with mesonic numbers as it does for the conifold. The symmetry structure then persists to all orders in $N$.

%==========================================
%%%%%%%%%%%%%%%%%%%

\section{$dP_0$ Revisited}\label{dP0rev}
The master space for $dP_0$ is calculated above in section \ref{s:toric} and is found to be irreducible.
\comment{
In fact, it can also be computed using dimer technology. 
Now, for the case of $dP_0$ the matrix $P$, defined in \eref{Pmat}, takes the form
{\tiny
\[ P= \left(
\begin{array}{llllll}
 1 & 0 & 0 & 1 & 0 & 0 \\
 0 & 1 & 0 & 1 & 0 & 0 \\
 0 & 0 & 1 & 1 & 0 & 0 \\
 1 & 0 & 0 & 0 & 1 & 0 \\
 0 & 1 & 0 & 0 & 1 & 0 \\
 0 & 0 & 1 & 0 & 1 & 0 \\
 1 & 0 & 0 & 0 & 0 & 1 \\
 0 & 1 & 0 & 0 & 0 & 1 \\
 0 & 0 & 1 & 0 & 0 & 1
\end{array}
\right) . \] 
}
The rank of this matrix should be the dimension of the master space (as encode in the cone $K$), i.e., it should be $G+2$ where $G$ is the number of gauge groups for this model. Hence, the rank is 5 and since there are 6 columns we expect a 1 dimensional kernel for $P$. This kernel can be easily computed to be the vector $Q$,
\begin{equation}
P \cdot Q = 0 \qquad \Rightarrow \qquad Q^t = \left(
\begin{array}{llllll}
 1 & 1 & 1 & -1 & -1 & -1
\end{array}
\right) ,
\end{equation}
which forms the vector of charges for the linear sigma model description of the master space for $dP_0$. In this description, therefore, we find that the master space is 
}
Its symplectic quotient description is
$\IC^6//(-1,-1,-1,1,1,1)$.
We note that the sum of charges is zero, implying that this 5-dimensional variety is Calabi-Yau, in agreement with \eref{FdP0}. This space is a natural generalization of the conifold and has the description of a cone over a 9 real dimensional Sasaki Einstein manifold given by a circle bundle over $\IP^2\times \IP^2$. This structure reveals a symmetry of the form $SU(3)\times SU(3)\times U(1)$. 

Indeed, we note that this construction is toric and therefore we would expect a symmetry which is at least $U(1)^5$ since the master space has dimension 5. However, due to the special symmetries of this space the symmetry is larger. The $U(1)$ is the R-symmetry and the first $SU(3)$ is the natural one acting on the mesonic moduli space $\cX = dP_0 = \IC^3 / \IZ_3$. The second $SU(3)$ symmetry is a ``hidden'' symmetry, and is related to the two anomalous baryonic $U(1)$ symmetries that play a r\^{o}le as the Cartan subgroup of this symmetry. We can use the full symmetry to compute the refined Hilbert series for this space. 

%%%%%%%%%%%%%%%%%%%%%%%%%%%%%%%%%%%%%%%%
%%%%%\subsubsection{Hilbert Series for the Master Space}

%Let us first re-compute the Hilbert series for just one charge, the R-charge. This can be done in several ways and here we will use the Molien Invariant. It takes into account the gauge charges in \eref{MSdP0} and assigns weight $t/w$ to the 3 external perfect matchings and weight $w$ to the 3 internal perfect matchings. We use the residue technique outlined in Section 3.2 of \cite{Forcella:2007wk} and find that the contribution to the integral comes from the pole at $w=t$, whence

The Hilbert series for just one charge was computed with the Molien formula in 
\eref{HSdP0}
\begin{equation}
g_1(t; dP_0) = \oint \frac{dw}{ 2 \pi i w(1-t/w)^3 (1-w)^3} = \frac {1+4t+t^2}{(1-t)^5} .
\label{g1dP0sec}
\end{equation}
Taking the plethystic logarithm of this expression we find 9 generators at order $t$ subject to 9 relations at order $t^2$,
\begin{equation}
PE^{-1}[ g_1(t; dP_0) ] = 9 t - 9 t^2 + \dots
\end{equation}
This agrees exactly with the content of \eref{FdP0} which says that $\f_{dP_0}$ should be the incomplete (since the plethystic logarithm does not terminate) intersection of 9 quadrics in 9 variables.

Now, we would like to refine the Hilbert series to include all the 5 global charges. This can be done using the Molien formula or any other of the methods discussed in section \ref{s:toric}. Here we find a shorter way of determining it.
To do this we recognize the 9 quiver fields as transforming in the $[1,0]\times [0,1]$ representation of $SU(3)\times SU(3)$. For short we will write an irreducible representation of this group as a collection of 4 non-negative integer numbers, here $[1,0,0,1]$ and with obvious extension to other representations. These are all the generators of the variety \eref{MSdP0}. The relations are derived from a superpotential of weight $t^3$ so we expect 9 relations at order $t^2$ transforming in the conjugate representation to the generators, $[0,1,1,0]$. To get this into effect we rewrite \eref{g1dP0sec} into a form which allows generalization to include characters (multiplying top and bottom by $(1-t)^4$):
\begin{equation}
g_1(t; dP_0) = \frac {1 - 9 t^2 + 16 t^3 - 9 t^4 + t^6 }{(1-t)^9} \ .
\end{equation}

The coefficient of the $t^2$ in the numerator can now be identified with the 9 relations, whose transformation rules are already determined to be $[0,1,1,0]$. Being irreducible we expect the Hilbert series to be palindromic. This gives as the $t^4$ term as $[1,0,0,1]$. The same property implies that the coefficient of the $t^3$ term is self conjugate and hence uniquely becomes the adjoint representation for $SU(3)\times SU(3)$, $[1,1,0,0] + [0,0,1,1]$. Finally, the denominator can be simply expressed as a plethystic exponential of the representation for the generators, $[1,0,0,1]$. In other words, $(1-t)^{-9} = PE[9 t]$. In summary, we end up with the refinement of the Hilbert series for $\f_{dP_0}$ as:
\begin{eqnarray}\label{refg1dP0}
%\ba{rcl}
&&g_1(t, f_1, f_2, a_1, a_2; dP_0) \nonumber \\
&=& \left(1 - [0,1,1,0] t^2 + ( [1,1,0,0] + [0,0,1,1] ) t^3 
 - [1,0,0,1] t^4 + t^6 \right) PE \left [ [1,0,0,1] t \right ] .\nonumber\\
%\ea
\end{eqnarray}

For completeness, we list here the explicit expressions for the characters of the representations, using weights $f_1, f_2$ for the first, mesonic $SU(3)$ and $a_1, a_2$ for the second, hidden $SU(3)$:
\begin{equation}\ba{lll}
[1, 0, 0, 1] &=& \left ( f_1 + \frac{f_2}{f_1} + \frac{1}{f_2} \right ) \left ( \frac{1}{a_1} + \frac{a_1}{a_2} + a_2 \right ),  \\
\left [ 0, 1, 1, 0 \right ] &=& \left ( \frac{1}{f_1} + \frac{f_1}{f_2} + f_2 \right ) \left ( a_1 + \frac{a_2}{a_1} + \frac{1}{a_2} \right ) ,  \\
\left [ 1, 1, 0, 0 \right ] &=& \frac{f_1^2}{f_2}+f_1 f_2+\frac{f_1}{f_2^2}+2+\frac{f_2^2}{f_1}+\frac{1}{f_1 f_2}+\frac{f_2}{f_1^2} ,  \\
\left [ 0, 0, 1, 1 \right ] &=& \frac{a_1^2}{a_2}+a_1 a_2+\frac{a_1}{a_2^2}+2+\frac{a_2^2}{a_1}+\frac{1}{a_1 a_2}+\frac{a_2}{a_1^2} .
\ea\eeq
In terms of the above weights, being generated by the representation $[1, 0, 0, 1]$, the Hilbert series, \eref{refg1dP0}, for the case of $N=1$ D3-brane admits a simple and natural series expansion of the form
\begin{equation}
\label{g1dP0-2}
g_1 \left (t, f_1, f_2, a_1, a_2; dP_0 \right ) = \sum_{n=0}^\infty \left [n, 0, 0, n \right ] t^{ n } \ .
\end{equation}

%%%%%%%%%%%%%%%%%%%%%
\subsubsection{Higher Number of Branes}
Using the formalism of the chapter \ref{chiralcount} and summarized in section \ref{s:plet}, we can compute the $N=2$ generating function in terms of characters of the global symmetry. The computation is somewhat lengthy but the result is relatively simple:
\beq\label{g2dP0}
\ba{rcl}
g_2 \left (t, f_1, f_2, a_1, a_2;~dP_0 \right ) &=& \sum\limits_{n=0}^\infty \sum\limits_{k=0}^{\lfloor \frac {n} {2} \rfloor }\left [2 n - 4 k, 2k, 0, n \right ] t^{ 2 n } +\\
&& + \sum\limits_{n_2=0}^\infty \sum\limits_{n_3=1}^\infty \sum\limits_{k=0}^{n_2} \left [2 n_2 + 3n_3 - 2 k, k, 0, n_2 \right ] t^{ 2 n_2 + 3 n_3 } ,
\ea\eeq
with coefficient 1 for each representation which appears in the expansion. It is important to identify the generators of this expression and a quick computation reveals the order $t^2$ and order $t^3$ generators to be the representations, $[2, 0, 0, 1]$, and $[3, 0, 0, 0]$, respectively. We can therefore sum the series and obtain
\begin{equation}
g_2 \left (t, f_1, f_2, a_1, a_2;~dP_0 \right ) =
A_2 PE \left [ \left [2, 0, 0 , 1 \right ] t^{ 2 } + \left [ 3, 0, 0, 0 \right ] t^{ 3} \right ] ,
\end{equation}
where $A_2$ is a complicated polynomial of order 58 in $t$ which has the first 10 terms:
\beq\ba{lll}
A_2 \left (t, f_1, f_2, a_1, a_2\right )
&=& 1- [2, 1, 1, 0] t^4 - [1, 2, 0, 1] t^5 + \\
&+& ( [4, 1, 0, 0] + [1, 1, 0, 0] + [3, 0, 1, 1] + [0, 3, 1, 1] - [0, 0, 0, 3] ) t^6  \\
&+& ([3, 2, 1, 0] + [2, 1, 1, 0] + [1, 3, 1, 0] + [1, 0, 1, 0] + [0, 2, 0, 2] + [0, 2, 1, 0]) t^7  \\
&-& ([5, 0, 0, 1] - [2, 0, 1, 2] - [2, 0, 0, 1] + [1, 2, 1, 2]) t^8  \\
&-& \left ([5, 2, 0, 0] + [3, 3, 0, 0] + [2, 2, 0, 0] + [1, 1, 0, 0] + [0, 3, 0, 0] + 2 [3, 0, 0, 0] \right .  \\
&& - \left . \,\, [3, 0, 0, 3] + [0, 0, 0, 3] + [2, 2, 1, 1] + [1, 1, 1, 1] + [0, 0, 1, 1] \right ) t^9 + \ldots
\ea\eeq

Similarly, we can obtain generic expressions for any $N$.
Recall from section \ref{N1C3Z3} that the generating function for a fixed integral K\"ahler modulus $\beta$ is equal to
\begin{equation}
g_{1, \beta} (t) = \sum_{n=0}^\infty {3 n + \beta + 2 \choose 2} t^{3 n + \beta} ;
\end{equation}
this can be easily written in terms of representations of the global symmetry as
\begin{equation}\label{g1betadP0}
g_{1, \beta} \left (t, f_1, f_2, a_1, a_2; dP_0 \right ) = \sum_{n=0}^\infty [3 n + \beta, 0, 0, 0] t^{3 n + \beta} \ .
\end{equation}

The auxiliary partition function also admits an expression in terms of representations of the global symmetry:
\begin{equation}
Z_{\rm aux}(t, a_1, a_2; dP_0 ) = (1 - t^3) PE \left [ [0, 0, 0, 1] t \right ] ,\end{equation}
which has an expansion as
\begin{equation}\label{ZauxdP0}
Z_{\rm aux}(t, a_1, a_2; dP_0 ) = \sum_{\beta = 0}^\infty [0, 0, 0, \beta ] t^\beta - \sum_{ \beta = 3 }^\infty  [0, 0, 0, \beta - 3 ] t^\beta \ .
\end{equation}
Following chapter \ref{chiralcount} or the recipe we gave in section \ref{s:plet}, we can use can use \eref{g1betadP0} and \eref{ZauxdP0} 
to compute the generating function for one D3-brane:
\begin{equation}\ba{rcl}
& &g_{1} \left (t, f_1, f_2, a_1, a_2; dP_0 \right ) = \sum\limits_{\beta = 0}^\infty \sum\limits_{n=0}^\infty [3 n + \beta, 0, 0, \beta ] t^{3 n + \beta} - \sum\limits_{\beta = 3}^\infty \sum\limits_{n=0}^\infty [3 n + \beta, 0, 0, \beta -3 ] t^{3 n + \beta}   \\
&=& \sum\limits_{\beta = 0}^\infty \sum\limits_{n=0}^\infty [3 n + \beta, 0, 0, \beta ] t^{3 n + \beta} - \sum\limits_{\beta = 0}^\infty \sum\limits_{n=1}^\infty [3 n + \beta, 0, 0, \beta ] t^{3 n + \beta} =  \sum\limits_{\beta = 0}^\infty [\beta, 0, 0, \beta ] t^{\beta} \ .
\ea\eeq
In the second equality we shifted the index $\beta$ by 3 and an opposite shift of the index $n$ by 1 in the second term. This allows us, in the third equality, to cancel all terms for $n$ except for the term in $n=0$.
Thus we reproduce \eref{g1dP0-2} in a remarkable cancellation that leaves only positive coefficients.

Subsequently, we can obtain the expression $g_N$, by computing the $\nu$-inserted plethystic exponential and series expansion:
\begin{equation}\label{gfulldP0}\ba{rcl}
&& 
\sum\limits_{N=0}^\infty \nu^N g_N\left ( \nu; t, f_1, f_2, a_1, a_2; dP_0 \right ) = 
g \left ( \nu; t, f_1, f_2, a_1, a_2; dP_0 \right )
\\ 
&=& \sum\limits_{\beta = 0}^\infty [0, 0, 0, \beta ] \left ( PE_{\nu} \left [ \sum\limits_{n=0}^\infty [3 n + \beta, 0, 0, 0 ] t^{3 n + \beta} \right ] - PE_{\nu} \left [ \sum\limits_{n=1}^\infty [3 n + \beta, 0, 0, 0] t^{3 n + 2 \beta} \right ] \right ) .
\ea\end{equation}
Note that the first $PE$ contains all the terms in the second $PE$ and hence all the coefficients in the expansion are positive.
As a check, we can expand \eref{gfulldP0} to second order in $\nu$:
\beq
g_2 \left ( t, f_1, f_2, a_1, a_2; dP_0 \right ) = \sum_{\beta = 0}^\infty \sum_{k = 0}^{\lfloor \frac{\beta}{2} \rfloor} \left [ 2 \beta - 4k, 2k, 0, \beta \right ] t^{ 2 \beta} \\ 
+ \sum_{\beta = 0}^\infty [\beta, 0, 0, \beta ] \sum_{n=1}^\infty [3 n + \beta, 0, 0, 0] t^{3 n + 2 \beta}
\eeq
which indeed agrees with \eref{g2dP0}.

Equation (\ref{gfulldP0}) is extremely important: it shows that all the BPS operators of the gauge theory for 
$dP_0$, for every value of N, reorganize in representations of the $SU(3)\times SU(3)\times U(1)$ group, where 
the second $SU(3)$ factor is an hidden symmetry, namely 
it is not visible in the UV Lagrangian describing the gauge theory. 
Because the BPS operators of the gauge theory are the holomorphic functions of the moduli space $\CM_N$, this result imply
that the moduli space $\CM_N$, for every values of N, has the symmetry $SU(3)\times SU(3)\times U(1)$. This is a very interesting 
result that come from the counting of chiral operators !

Because the symmetry $SU(3)\times SU(3)\times U(1)$ persists for all the values of N, we can interpolate this result till $N \rightarrow \infty$.
In this way we could make the prediction that the dual type IIB string on the $AdS_5\times S^5/\mathbb{Z}_3$ background 
has a sector in which there exist an hidden $SU(3)$ symmetry.
This is just a first guess and we clearly need much more studies.

\section{$\mathbb{F}_0$ Revisited}\label{s:F0rev}

Let us move on to the $\mathbb{F}_0$ theory and focus on the first toric phase, whose master space we studied in section \ref{s:F0}, 
to see another clear example of the extension of the symmetries of the coherent component of the master space $\firr{~}$ 
to the the complete moduli space $\CM_N$ and to the complete
chiral ring of the theory for every value of N.

We recall that the master space for a single D3-brane, $N=1$, is a six dimensional reducible variety composed by a set of coordinate planes and an irreducible six dimensional Calabi-Yau piece. This top piece, being toric, admits a symmetry group which is at least $U(1)^6$. Re-examining \eref{FF0} we see that it is actually the set product of two three dimensional conifold singularities:
\begin{equation}\label{dopcon}
B_2 D_1 - B_1 D_2 = 0 \ , \  A_2 C_1 - A_1 C_2 = 0 \ ;
\end{equation}
hence the group of symmetries is $SU(2)^4 \times U(1)^2$, twice the isometry group of the conifold.
%\sref{s:coni}. 
The first $SU(2)^2$ is the non-Abelian symmetry group of the variety $\mathbb{F}_0 \sim \mathbb{P}^1 \times \mathbb{P}^1$, 
and the second $SU(2)^2$ is a hidden symmetry related to the two anomalous baryonic symmetries of the gauge theory. The chiral spectrum of the theory is summarized by the Hilbert series:
\begin{equation}\label{g1f0}
g_1(t_1,t_2; F_0) = \frac {(1 - t_1^2)(1-t_2^2)}{(1-t_1)^4(1-t_2)^4} 
= PE[4 t_1 - t_1^2 + 4 t_2 - t_2^2]
\ ,
\end{equation}
where the chemical potential $t_1$ counts the fields $A_i$, $C_j$ while $t_2$ counts the fields $B_p$, $D_q$. 

Let us define the representation $[n]\times[m]\times[p]\times[q]=[n,m,p,q]$ for the $SU(2)^4$ group. Equation \eref{dopcon} and Table (\ref{globalF0})
then implies that $A,C$ are in the $[1,0,1,0]$ representation and $B,D$ are in the $[0,1,0,1]$ representation. The refinement for the Hilbert series (\ref{g1f0}) is then
\begin{equation}\label{g1f0s}
g_1(t_1,t_2,x,y,a_1,a_2; F_0) = (1 - t_1^2)(1-t_2^2) PE \left [ [1,0,1,0]t_1 + [0,1,0,1] t_2 \right ] .
\end{equation}
Amazingly, the group $SU(2)^4 \times U(1)^2$ is the symmetry for the chiral ring for generic $N$, not just for $N=1$, and hence the moduli space of the non-Abelian theory on $N$ D3-branes has this group as a symmetry group. 

The generating function for finite $N$ can be computed using the plethystic
exponential in each sector of the GKZ decomposition of the $N=1$ partition
function. The procedure is explained in chapter \ref{chiralcount} and summarized in section \ref{s:plet}.
In particular, the implementation of the plethystic program goes through the
formulae \eref{GKZ1} and \eref{GKZN} and requires the computation of
the generating functions for fixed K\"ahler moduli and the auxiliary partition
function.

Recall from equation (\ref{seri}) that the generating function for fixed integral K\"ahler moduli, 
which in this case can be parameterized by two integers  $\beta$, $\beta'$,  is equal to
\begin{equation}
g_{1, \beta, \beta ' } (t_1,t_2; F_0) = \sum_{n=0}^\infty (2 n + \beta +1 )(2 n + \beta ' + 1 ) t_1^{2 n + \beta}t_2^{2 n + \beta '} \ .
\end{equation}
This can be easily refined in terms of representations of the global symmetry as
\begin{equation}\label{g1betaF0}
g_{1, \beta, \beta '} \left (t_1,t_2, x, y, a_1, a_2; F_0 \right ) = \sum_{n=0}^\infty [2 n + \beta, 2 n + \beta ', 0, 0] t_1^{2 n + \beta}t_2^{2 n + \beta '}
\ .
\end{equation}
The auxiliary partition function in equation (\ref{f0gen}) also admits an expression in terms of representations of the global symmetry,
\begin{equation}
Z_{\rm aux}(t_1, t_2, a_1, a_2; F_0 ) = (1 - t_1^2 t_2^2) PE \left [ [0, 0, 1, 0] t_1 + [0, 0, 0, 1] t_2 \right ] ,
\end{equation}
which has an expansion as
\begin{equation}
Z_{\rm aux}(t_1, t_2, a_1, a_2; F_0 ) = \sum_{\beta, \beta ' = 0}^\infty [0, 0, \beta , \beta ' ] t_1 ^\beta t_2^{\beta '} - \sum_{ \beta, \beta' = 2 }^\infty  [0, 0, \beta -2 , \beta' - 2 ] t_1^\beta t_2^{\beta ' } \ .
\end{equation}

Recalling from section \ref{s:plet}, once we have $g_{1, \beta, \beta '}$ and $Z_{\rm aux}$ we can do the replacement $t_1^\beta t_2^{\beta'}$ in $Z_{\rm aux}$ by $g_{1,\beta,\beta'}$ (as done in line 2 below and using the fact that by our definition $[a,b,0,0] \times [0,0,c,d] = [a,b,c,d]$) to obtain the generating function for a single brane:
\begin{equation}
\hspace{-0.5in}
\ba{rcl}
&&g_{1} \left (t_1,t_2, x, y, a_1, a_2; F_0 \right )\\
&=& \sum\limits_{\beta, \beta' = 0}^\infty \sum\limits_{n=0}^\infty [2 n +\beta, 2n + \beta', \beta, \beta' ] t_1^{2 n + \beta} t_2^{2 n + \beta'} - \sum\limits_{\beta, \beta' = 2}^\infty \sum\limits_{n=0}^\infty [2 n + \beta,2n + \beta', \beta -2 , \beta' -2 ] t_1^{2 n + \beta}t_2^{2 n + \beta'}   \\
&=& \sum\limits_{\beta, \beta' = 0}^\infty \sum\limits_{n=0}^\infty [2 n + \beta, 2n + \beta', \beta, \beta' ] t_1^{2 n + \beta}t_2^{2n+\beta'} - \sum\limits_{\beta, \beta' = 0}^\infty \sum\limits_{n=1}^\infty [2 n + \beta, 2n + \beta', \beta, \beta' ] t_1^{2 n + \beta}t_2^{2n+ \beta'} \\
&=&  \sum\limits_{\beta, \beta' = 0}^\infty [\beta, \beta' , \beta, \beta' ] t_1^{\beta} t_2^{\beta'} ,
\ea\label{charg1F0}\end{equation}
leaving only positive coefficients and meaning that $SU(2)^4 \times U(1)^2$ is indeed a symmetry of the $N=1$ moduli space. The second equality follows by shifting $n$ by $-1$ and the $\beta$'s by $2$. The third equality is the remaining $n=0$ term from both contributions. It is important to note that Equation \eref{charg1F0} factorizes into two conifold generating functions,
\begin{equation}\ba{rcl}
g_{1} \left (t_1,t_2, x, y, a_1, a_2; F_0 \right ) &=& \left ( \sum\limits_{\beta = 0}^\infty [\beta, 0 , \beta, 0 ] t_1^{\beta} \right ) \left ( \sum\limits_{\beta' = 0}^\infty [0, \beta' , 0, \beta' ] t_2^{\beta'} \right ) \\
&=& \left( (1-t_1^2) \frac{}{} PE[ [1,0,1,0] t_1] \right)
\left( (1-t_2^2) \frac{}{} PE[ [0,1,0,1] t_2] \right) ,
\ea\end{equation}
which can be easily checked to equal Equation \eref{g1f0s}.

Next, using \eref{GKZN}, we can write a generic expression for any $N$.
\begin{equation}\ba{rcl}
g \left ( \nu; t_1,t_2, x, y, a_1, a_2; F_0 \right ) &= &
\sum\limits_{\beta,\beta' = 0}^\infty [0, 0, \beta, \beta'] \left ( PE_{\nu} \left [ \sum\limits_{n=0}^\infty [2 n + \beta, 2 n + \beta', 0, 0 ] t_1^{2 n + \beta} t_2^{2 n +\beta'} \right ] - \right.\\
&& \left.
 - PE_{\nu} \left [ \sum\limits_{n=1}^\infty [2 n + \beta, 2n +\beta', 0, 0] t_1^{2 n +  \beta} t_2^{2 n + \beta'} \right ] \right ) .
\ea\end{equation}
Note that the first $PE$ contains all the terms in the second $PE$ and hence all the coefficients in the expansion are positive. This is the explicit demonstration that for generic $N$ the chiral spectrum organizes into representations of $SU(2)^4 \times U(1)^2$ and hence the moduli space of the non-Abelian theory with generic rank $N$ has symmetry $SU(2)^4 \times U(1)^2$. 
As in the previous section we could infer that type IIB string propagating on the back ground $AdS_5$ times the circle bundle over 
$\mathbb{P}^1\times \mathbb{P}^1$ has a sector in which there exist an hidden symmetry $SU(2)^2$.It would be nice to study this phenomenon. 

%==========================================
%%%%%%%%%%%%%%%%%%%

\section{Summary and Discussions}\label{s:conc}\setall

In this chapter we tried to study the non abelian moduli space of brane at singularity using the properties of the chiral spectra of the gauge theory.
The situation for general number $N$ of D3-branes is much more subtle than the $N=1$ case. The moduli space is now the variety of F-flatness quotiented by the special unitary factors of the gauge group, so that when quotiented by the $U(1)$ factors as a symplectic quotient we once more arrive at the mesonic branch, which here is the $N$-th symmeterized product of the Calabi-Yau space $\cX$. In the fully non abelian case we lose the toricity properties and a direct attack to the moduli space seems quite complicated. However, the plethystic programme persists through and we can still readily extract the generating functions for any $N$. The crucial observation was that the gauge theory possesses, at least in the $N=1$ case, hidden global symmetries corresponding to the symmetry of $\firr{~}$ which, though not manifest in the Lagrangian, is surprisingly encoded in the algebraic geometry of $\firr{~}$. In particular, we can re-write the terms of the single brane generating funct!
 ion, i.e., the refined Hilbert series of $\firr{}$, in the weights of the representations of the Lie algebra of the hidden symmetry in a selected set of examples. Using the plethystic programme we were able to extend this analysis to an arbitrary number $N$ of branes. Indeed we showed that, for a selected set of examples, we are still able to reorganize the generating function for generic $N$ in representation of the hidden symmetries. This fact allowed us to study the symmetry of the moduli space $\CM_N$. In some cases the symmetries of $\firr{~}$ become symmetries of the complete moduli space $\CM_N$ and of the complete chiral ring for generic $N$. This fact gives us the 
possibility to make some predictions about the symmetry of the dual type IIB string on the $AdS \times H$ background.

We are at the portal to a vast subject. The algebraic geometry of the master space at $N > 1$ number of branes deserves a detailed study as we have done for the single-brane example; we have given its form on physical grounds and mathematically the structure is expected to be complicated.

%The top-dimensional irreducible component of the master space is seen to be an important object and we have shown a few of its properties for $N=1$. We have conjectured that Seiberg duality preserves this with our example, it would be important to prove this in general, for higher number of branes, and indeed for generic $\CN=1$ gauge theories as well. 

A very important observation is that our systematic analysis should apply to not only D-brane theories but to supersymmetric gauge theories in general; the master space and its associated physical insight need to be thus investigated panoramically. The full symphony based on our motif in $\f$ awaits to be composed.

%%%%%%%%%%%%%%%%%%%%%%%%%%%%%%%%%%%%%%%%%%%%%%%%%%%%%%%%%%%%%

\chapter{Counting the Fermionic Operators} 

In the previous chapters we discussed properties of the moduli space of quiver gauge theories and of their BPS spectrum just looking to the bosonic degrees of
freedom. A generic gauge theory contains in its BPS spectrum also spinorial degrees of freedom. These are the gauge invariant operators containing the 
superfield $W_{\alpha}$. 
%It is the task of this chapter to give a taste of the complete partition functions including the fermionic degrees of freedom.
In this chapter we discuss a general procedure to obtain $1/2$ $BPS$ partition functions for generic $\mathcal{N}=1$ quiver gauge theories counting 
the gauge invariant operators (bosonic and fermionic), charged under all the global symmetries (mesonic and baryonic), 
in the chiral ring of a given quiver gauge theory. In particular we will explain the general recipe to include
 the fermionic degrees of freedom $W_{\alpha}$ in the previously discussed bosonic partition functions.

\section{Generalities}

%\comment{
%In the few past years there were a great effort in the study of quiver gauge theories \cite{gauntlett,Benvenuti:2005cz,Hanany:2005ve,Hanany:2005ss,Feng:2005gw,Butti:2006nk,Butti:2005vn}. 
%Understanding the structure of the spectrum of quiver gauge theories is an important topic in the AdS/CFT correspondence 
%and more generically in the study of supersymmetric gauge theories and their moduli spaces. Recently a lot of papers studied 
%generating functions counting $\frac{1}{2} BPS$ operators in $\mathcal{N}=1$ quiver $CFT$ \cite{5per5,Forcella:2007wk,Benvenuti:2006qr,Butti:2006au,Martelli:2006yb,Feng:2007ur,Romelsberger:2005eg,Kinney:2005ej,Biswas:2006tj,Mandal:2006tk,Martelli:2006vh,Basu:2006id,Hanany:2006uc,Nakayama:2007jy}. 
%}
In the previous chapter we constructed a character associated to the quiver $CFT$ at the singularity $\cX$ 
that counts the elements in the chiral ring of the theory according to their charges under the various $U(1)$ 
factors of the global symmetry group. 

The counting procedure consists in the definition of a set of chemical potentials $\{t_i \}$, $i=1,...,r$ associated 
to each $U(1)$ factor in the group $U(1)^r$ (the abelian torus of the global symmetry group of the theory).
\comment{
\footnote{The quiver gauge theories we are going to consider are obtained as near horizon limit of a system of $N$ $D3$ branes placed at the tip of a conical CY singularity.
 Their gauge group is a product of $SU(N)$ factors and the matter content are chiral bifundamentals fields. 
The global symmetries are divided in two families: the ones coming from the isometries of the compactification manifold, 
and the ones coming from the reduction of the $C_4$ form over topologically non trivial three cycles. The former are the 
``flavor symmetries'' of the field theory and contain an abelian torus $U(1)^r$, ( $r>0$, the conformal symmetry imply the 
presence of the $U(1)$ $R$ symmetry); while the second is the abelian group $U(1)^{g-1}$ of the baryonic symmetries of the field theory.}
}
We can decompose the $r$ abelian symmetries in $r=f+a$. A $U(1)^f$ comes from the isometries of $\cX$ ( flavour symmetries in CFT including the R symmetry), $f=3$ in the toric case,
 while $U(1)^a_B$ comes from the topology of H ( the baryonic symmetries in the field theory), in the toric case $a=d-3$ where d is the number of external points
in the toric diagram. Once we have defined these chemical potentials 
we construct the character $g(\{t_i\})$, such that once expanded for small values of the $t_i$:
\begin{equation}
g(\{t_i\})= \sum_{i_1,...,i_{f+a}} c_{i_1,...,i_{f+a}}t_1^{i_1}...t_{f+a}^{i_{f+a}}
\end{equation} 
give us the positive integer coefficients $c_{i_1,...,i_{f+a}}$ that tell how many independent gauge invariant operators there are 
in the chiral ring with that specific set of quantum numbers. 
\comment{
\footnote{In the paper we will call $g(\{t_i\})$ with four 
different names: partition function, generating function, character, Hilbert series. The first one is the physic literature 
name while the second, third and fourth ones are the mathematical literature names.}.
}
As we have explained in the previous chapters this enumeration problem turns out to be a very powerful toll for the study 
of the properties of the $CFT$ and the dual 
$AdS_5 \times H$ geometry \cite{Butti:2007jv,Forcella:2007wk,Benvenuti:2006qr,Butti:2006au,Martelli:2006yb,Feng:2007ur,Romelsberger:2005eg,Kinney:2005ej,Biswas:2006tj,Mandal:2006tk,Martelli:2006vh,Basu:2006id,Hanany:2006uc,Nakayama:2007jy}. The function $g(\{t_i\})$ 
contains the informations regarding the algebraic equations of the moduli space of the field theory, its dimension, 
the volume of the horizon manifold $H$ and hence the value of the central charge $a$ of the $CFT$, the volumes of all 
the non trivial three cycles inside $H$ and hence of all the $R$ charges of the chiral fields in the field theory and 
the behavior of the $BPS$ spectrum under the non perturbative quantum corrections in the strong coupling regime.

It turned out that all the geometric informations are encoded in the scalar part of the chiral ring, and indeed till now we just focused on this part of the spectrum.

%\comment{Right now we have a good understanding of the scalar part of the complete chiral ring (containing all the mesonic and baryonic 
%degrees of freedom) for a generic quiver gauge theory \cite{5per5}.\\}

The task of this chapter is to make a step further and to study the generating functions for the complete chiral ring of generic quiver gauge theories containing all the scalar degrees of freedom plus the spinorial degrees of freedom originated by the $W^i_{\alpha}$ Weyl 
spinor superfields associated to the $\mathcal{N}=1$ vector multiplets. 

The complete generating function for the chiral ring of a given quiver gauge theory gives informations about the complete 
$1/2$ $BPS$ spectrum, and hence the possibility of a statistical studies of the thermodynamical properties of a 
generic strongly coupled $CFT$.

We will show how it is possible to obtain the generating functions for the complete chiral ring starting from the generating 
functions for the scalar part. This counting procedure is divided into two parts: the $N=1$ counting, and the generic finite $N$ counting.
As usual it happens that the knowledge of the $N=1$ generating function is enough to implement the finite $N$ counting. 
We will indeed introduce a ``superfield formalism'' that allows to pass from the $N=1$ scalar generating function to the $N=1$ complete one. 
In the chiral ring there are both bosonic and fermionic degrees of freedom. Indeed the bifoundamental fields are scalar superfields and 
hence bosons, while the superfiels $W_{\alpha}^i$ are Weyl spinors and hence fermions. This different statistical behavior became 
important in the finite $N$ counting. For this reason we will introduce the notion of Generalized Plethystic Exponential $PE$\footnote{In this chapter we
will use the symbol $PE$ for the generalized plethystic exponential, while we will use $PE^{\mathcal{B}}$ for the bosonic part of the spectrum ( this is what 
we have till know simply called the PE function), and $PE^{\mathcal{F}}$ for the fermionic part.}. 
This is a simple mathematical function that implement the mixed statistic of a system of bosons and fermions. Once we have obtained 
the generating function $g_1( \{ t_i \} )$ for $N=1$, to have the generating function $g_N(\{t_i\})$ counting the chiral ring 
operators for finite values of $N$ we just need to plug $g_1( \{ t_i \} )$ in the Generalized Plethystic Exponential
\footnote{This is just a schematic expression. The more precise equations will be given in following sections.}:
\begin{equation}
\sum_{N=0}^{\infty} \nu^N g_N (\{ t_i \}) = PE_{\nu}[g_1( \{ t_i \} )]= PE_{\nu}^{\mathcal{B}}[g_1^{\mathcal{B}}( \{ t_i \} )] PE_{\nu}^{\mathcal{F}}[g_1^{\mathcal{F}}( \{ t_i \} )]
\end{equation}
where the $\mathcal{B}$, $\mathcal{F}$ are for bosonic and fermionic statistic respectively; $g_1^{\mathcal{B}}( \{ t_i \} )$ 
is the bosonic part of the $N=1$ generating function, while $g_1^{\mathcal{F}}( \{ t_i \} )$ is the fermionic part of the $N=1$ 
generating function. The function $PE_{\nu}^{\mathcal{B}}[...]$ is the usual $\nu$-inserted plethystic exponential we used all along this thesis, 
while the $PE_{\nu}^{\mathcal{F}}[...]$ is its fermionic version to be defined in the following.

In the previous chapters we developed powerful tools to compute the $N=1$ scalar generating functions 
%\cite{5per5,Forcella:2007wk,Benvenuti:2006qr,Butti:2006au,Martelli:2006yb,Feng:2007ur,Romelsberger:2005eg,Kinney:2005ej,Biswas:2006tj,Mandal:2006tk,Martelli:2006vh,Basu:2006id,Hanany:2006uc,Nakayama:2007jy} 
and the finite $N$ complete generating functions is simply obtained using the algorithmic procedure previously explained.

In this chapter we will just shortly comment on this possible extension of the techniques developed in the rest of the thesis to reach the 
complete counting of the chiral ring of quiver CFT. A more careful analysis is left for the future and the reader can find some more 
insights in \cite{Forcella:2007ps}.

The chapter is organized in the following way. In section \ref{N4f} we will start with the simplest example of the $\mathcal{N}=4$ gauge theory. 
We will construct the various generating functions looking directly at the chiral ring structure. This is a simple and explicit example 
to take in mind in the next more abstract section. In section \ref{gentheo} we will give a general discussion of the 
structure of the chiral ring of quiver gauge theories. We will introduce the concept of generalized $PE$ and we will 
explain how to compute the generating functions for the complete chiral ring in the case $N=1$ and in the generic finite $N$ case. 
In section \ref{n4reload}, we will revisit, using the general tools 
introduced in section \ref{gentheo}, the computation of the generating 
functions for the $\mathcal{N}=4$ gauge theory. In section \ref{conifo} 
we will introduce our main example: the conifold gauge theory. We will construct the $N=1$ and finite $N$ generating functions for the mesonic 
sector of the chiral ring, for the sector with baryonic charge $B=1$, and finally for the complete chiral ring with all 
the charges and all the chiral fields.

\section{The $\mathcal{N}=4$ Generating Functions}\label{N4f}

Let us start with the easy explanatory example of the $\mathcal{N}=4$, $U(N)$ gauge theory.
In this theory the basic chiral operators are the three scalars superfields:
\begin{equation}
\phi_i \hbox{  }\hbox{  }\hbox{  } i = 1,2,3
\end{equation}
and the spinor superfiels:
\begin{equation}
W_{\alpha} \hbox{  } \hbox{  }\hbox{  } \alpha = +,-
\end{equation}
>From the relation in the chiral ring we have:
\begin{equation}\label{cinrel}
\{W_{\alpha},W_{\beta}\}=0 \hbox{  } \hbox{  }\hbox{  } [W_{\alpha}, \phi _i]=0
\end{equation}
Because the fields $\phi_i$ are bosons while the fields $W_{\alpha}$ are fermions, the first will be represented by 
commuting variables while the latter by anti commuting variables.
>From the super potential we have the other relations:
\begin{equation}\label{dynnnrel}
[\phi_i, \phi _j]=0
\end{equation}
We would like to write down a generating function counting all the single trace operators in the chiral ring 
of the $\mathcal{N}=4$ gauge theory.

This is easy to obtain if we remember that the generic $1/8$ $BPS$ single trace operators are given by: 
\begin{equation}\label{chi}
\hbox{ Tr }( \phi_1^i\phi_2^j\phi_3^k )\hbox{ , }\hbox{ Tr }( W_{\alpha}\phi_1^i\phi_2^j\phi_3^k )\hbox{ , }\hbox{ Tr }( W_{\alpha} W^{\alpha} \phi_1^i\phi_2^j\phi_3^k )
\end{equation}
Let us introduce the chemical potential $q$ for the dimension of the fields:
\begin{equation}
\phi_i\rightarrow q \hbox{ , } W_{\alpha} \rightarrow q^{3/2} \hbox{ , } W_{\alpha}W^{\alpha}  \rightarrow q^3
\end{equation}
the chemical potential $w$ for the fields $W_{\alpha}$:
\begin{equation}
W_{\alpha} \rightarrow w
\end{equation}
and the chemical potential $\alpha$ for the spin of the chiral fields:
\begin{eqnarray}
W_+ \rightarrow \alpha & , & W_- \rightarrow 1/\alpha 
\end{eqnarray}
Hence the operators in (\ref{chi}) will be counted by the following generating functions\footnote{we will 
put the subscript $1$ to the generating functions because it can be shown that for the mesonic part of 
undeformed quiver gauge theories the generating functions for the multi traces, and hence of the complete 
spectrum of gauge invariant operators, in the case $N=1$ are the same as the generating functions for the single 
traces in the limit $N \rightarrow \infty$ \cite{Benvenuti:2006qr}.}:
\begin{eqnarray}\label{g1bfb}
& & g_1^0(q) = \nonumber\\
& & \sum_{i,j,k} q^{i+j+k} = \sum_{n=0}^{\infty} \frac{(n+2)(n+1)}{2}q^n = \frac{1}{(1-q)^3}\nonumber\\
& & \nonumber\\
& & g_1^w(q,w,\alpha) = \nonumber\\
& & \sum_{i,j,k} q^{3/2+i+j+k}w \Big(\alpha + \frac{1}{\alpha}\Big) = \sum_{n=0}^{\infty} \frac{(n+2)(n+1)}{2}q^{3/2+n}w\Big(\alpha + \frac{1}{\alpha}\Big)= \frac{q^{3/2} w \big(\alpha + \frac{1}{\alpha}\big)}{(1-q)^3}\nonumber\\
& & \nonumber\\
& & g_1^{w^2}(q,w) = \nonumber\\
& & \sum_{i,j,k} q^{3+i+j+k}w^2 = \sum_{n=0}^{\infty} \frac{(n+2)(n+1)}{2}q^{3+n}w^2 = \frac{ q^3w^2  }{(1-q)^3} 
\end{eqnarray}
Hence the complete generating function counting all the single trace operators (\ref{chi}) is:
\begin{equation}
g_1(q,w,\alpha)=g_1^0(q)+ g_1^w(q,w,\alpha)+ g_1^{w^2}(q,w)= \frac{1 + q^{3/2} w \big(\alpha + \frac{1}{\alpha}\big)+ w^2 q^3}{(1-q)^3}
\end{equation}

Till this moment we just counted in the limit $N \rightarrow \infty$ and in the single trace sector i.e. 
without taking into account the relations among gauge invariants coming from the fact that for finite $N$ 
one can rewrite some of the operators in terms other operators with lower dimensions.

To understand the finite $N$ counting we can look to equations (\ref{cinrel}), (\ref{dynnnrel}). These ones tell 
that one can diagonalize at the same time all the $\phi_i$ and all the $W_{\alpha}$. Hence we are left with a 
system of $3N$ bosonic eigenvalues and $2N$ fermionic eigenvalues. 
Using the Bose-Einstein and Fermi-Dirac statistic we can write down the function $g(q,w,\alpha;\nu)$ that 
generate the partition functions $g_N(q,w,\alpha)$ with fixed $N$ counting all the $1/8 BPS$ gauge 
invariant operators (single and multi traces) \cite{Kinney:2005ej}:
\begin{eqnarray}\label{bosferN}
& & g(q,w,\alpha;\nu)=\sum_{N=0}^{\infty} \nu^N g_N(q,w,\alpha)=\nonumber\\
& & \prod_{n=0}^{\infty}\frac{(1+\nu \hbox{ }w\hbox{ } \alpha\hbox{ } q^{3/2+n})^{ \frac{(n+2)(n+1)}{2}}(1+ \nu \hbox{ }w\hbox{ } \frac{1}{\alpha}\hbox{ } q^{3/2+n})^{ \frac{(n+2)(n+1)}{2}}}{(1-\nu\hbox{ } q^n)^{\frac{(n+2)(n+1)}{2}}(1- \nu \hbox{ }w^2\hbox{ } q^{3+n})^{\frac{(n+2)(n+1)}{2}}} \nonumber\\
\end{eqnarray}

In the next section we will discuss how to construct in general the generating functions for the chiral ring 
of a given quiver gauge theory. After this we will revisit the $\mathcal{N}=4$ case.

\section{The General Approach to the Complete Generating Functions}\label{gentheo}

Let us try to motivate our approach to the complete partition function for the chiral ring of a given gauge theory. 
We consider gauge theories describing the low energy dynamics of D3 branes at the CY conical singularity $\cX$. These theories have a set of flavor symmetries (we include in this class also the 
always present $R$ symmetry): $U(1)^f$, $f=3$ in the toric case; and a set of non anomalous baryonic symmetries $U(1)^{a}$, $a=d-3$ in the toric case. 
We would like to write down a generating function counting all the operators in the chiral ring according to their charges under the global symmetries 
of the theory. From the computation of the generating functions in the scalar chiral ring, explained in the previous chapters,
%\cite{Forcella:2007wk, Benvenuti:2006qr,Butti:2006au,Feng:2007ur,Hanany:2006uc,5per5} 
we know that the knowledge of the generating function for a single D3 brane is enough 
to compute the generating functions for an arbitrary number $N$ of $D3$ branes\footnote{this is due to the fact that the 
gauge invariant operators for fixed $N$ are nothing else that the $N$ times symmetric product of the ones for $N=1$}. 
The last ones are indeed obtained by the former using some combinatorial tools. For this reason one has a well defined 
notion of single particle Hilbert space ($N=1$) and multi particle Hilbert spaces ($N > 1$ ), that we will use in the following. 

A generic quiver gauge theory has the following set of chiral fields:
\begin{equation}\label{scachi}
X_{i,j}^{e_{ij}}
\end{equation}
these ones are scalar bifundamental superfields transforming in the fundamental of the $SU(N)_i$ gauge group and in 
the anti fundamental of the $SU(N)_j$ gauge group. The $e_{ij}=1,...,E_{ij}$ label the number of fields between the 
$i$-th gauge group and the $j$-th gauge group;
\begin{equation}\label{spichi}
W_{\alpha}^i
\end{equation}
these ones are the Weyl spinor superfields associated to the vector supermultiplet of the $SU(N)_i$ gauge group; 
$\alpha = +,-$ label the spin state and the label $i$ goes over all the $g$ gauge groups $i=1,...,g$.

In the chiral ring they must satisfy the ``symmetry'' relations\footnote{see for example \cite{Casero:2003gf}}:
\begin{eqnarray}\label{symrel}
& & \{ W_{\alpha}^i, W_{\beta}^i \}=0 \nonumber\\
& & W_{\alpha} ^i X_{i,j}^{e_{ij}} = X_{i,j}^{e_{ij}} W_{\alpha}^j 
\end{eqnarray}
and the dynamical relations:
\begin{equation}\label{dynrel}
\frac{\partial \mathcal{W}}{\partial  X_{i,j}^{e_{ij}}}= 0 
\end{equation}
Where $\mathcal{W}$ is the superpotential of the theory under consideration.

The gauge invariant operators constructed with scalar fields $X_{i,j}^{e_{ij}}$ define a complex ambient space. 
The dynamical equations (\ref{dynrel}) define an algebraic variety in this space and hence they give constraints 
inside the chiral ring that are usually hard to deal with. In the previous chapters we have explained how to deal with the problem 
of counting the scalar part of the chiral ring and how to implement the constraints (\ref{dynrel}) in the counting procedure.
%Thanks to the recent works \cite{Forcella:2007wk, Benvenuti:2006qr,Butti:2006au,Feng:2007ur,Hanany:2006uc,5per5} the problem of counting the scalar part of the chiral ring, and hence how to deal 
%with (\ref{dynrel}) is right now under control. 
What we want to show here is how, starting from the knowledge 
of the scalar partition function, one can write down the complete $\frac{1}{2}$ $BPS$ partition function for a generic quiver gauge theory.\\
The generic gauge invariant operator inside the chiral ring will be constructed in the following way:
given a pair of gauge groups $(x,z),\ x,z=1,...,g$, we call a gauge invariant of type $(x,z)$ a gauge invariant of the form
\begin{equation}\label{ginv}
\epsilon_{x}^{k_1,...,k_N} ({\bf O}_{I_1}^{(x,z)})_{k_1}^{l_1} .... ({\bf O}_{I_N}^{(x,z)})_{k_N}^{l_N}\epsilon^{z}_{l_1,...,l_N}
\end{equation}
where $({\bf O}_I^{(x,z)})_{k}^{l}$ denotes a string of elementary chiral fields $X_{i,j}^{e_{ij}}$, $W_{\alpha}^i$ with all gauge indices
contracted except two indices, $k$ and $l$, corresponding to the gauge groups $(x,z)$. The index $I$ runs over all possible strings of 
elementary fields
with these properties.
The full set of gauge invariant operators is obtained by arbitrary products of the operators in (\ref{ginv}). Using the tensor
relation
$$\epsilon^{k_1,...,k_N}\epsilon_{l_1,...,l_N} = \delta^{k_1}_{[l_1}\cdots \delta^{k_N}_{l_N]} $$
some of these products of determinants are equivalent and some of these are actually equivalent
to mesonic operators made only with traces.

From the first equation in (\ref{symrel}) it is easy to understand that the fields $W_{\alpha}^i$ are the fermionic degrees of freedom 
of the theory, and they can appear in the chiral operator $({\bf O}_I^{(x,z)})_{k}^{l}$ alone or at most in couple in the antisymmetric 
combination: $W_{\alpha}^i W^{\alpha}_i = W_{+}^i W_{-}^i - W_{-}^i W_{+}^i$. The second equation in (\ref{symrel}) 
tells that the position of the $W_{\alpha}^i$ spinor fields inside the chiral string $({\bf O}_I^{(x,z)})_{k}^{l}$ does not matter. 
From now on we will call the ``single particle Hilbert space'' the space spanned by the operators $({\bf O}_I^{(x,z)})_{k}^{l}$. 
This space is the total space of the gauge invariant operators in the case $N=1$.

Once we know the spectrum of the operators inside the scalar chiral ring of the single particle Hilbert space 
(i.e. all the $({\bf O}_I^{(x,z)})_{k}^{l}$ that does not contain the $W^i_{\alpha}$ field), the generic operators come in classes: 
there are the ones of the scalar chiral ring, the same ones with the insertion of one field $W_{\alpha}$, 
and the same ones of the scalar chiral ring with the insertion of the antisymmetric combination  $W_{\alpha}^i W^{\alpha}_i$. 
Thanks to the ``commutativity'' properties of the fields  $X_{i,j}^{e_{ij}}$ and $W_{\alpha}$ (\ref{symrel}), 
we do not have to specify the gauge groups $i$ the $W_{\alpha}$ fields belong to. It is enough to 
pick up a representative $W_{\alpha}$ for all the gauge groups. The result of this operation is that 
we would be able to easily write down the complete partition function for the $\frac{1}{2}$ $BPS$ 
states of a given $\mathcal{N}=1$ quiver gauge theory with gauge group given by a product of $SU(N)$ factors, 
times a diagonal overall $U(1)$ factor. The latter comes from the operators that factorize in products of the type Tr$(W_{\alpha}) (...)$. 
Even if it is possible to obtain the ``pure $SU(N)$'' counting, it is more natural to continue keeping the overall $U(1)$ factor, 
and we will indeed continue in this way in the following.

To every operators $({\bf O}_I^{(x,z)})_{k}^{l}$ we can associate a state in the single particle Hilbert space.
The single particle space of the complete chiral ring would be spanned by the states:
\begin{equation}\label{spauno}
|m_1,...,m_f,B_1,...B_a,\mathcal{S}>
\end{equation}
where $m_i$ are the charges under the $T^f$ abelian torus inside the generically non abelian 
global flavor group of the gauge theory, $B_j$ are the charges under the $U(1)^a_B$ baryonic 
symmetry of the theory, and $\mathcal{S}=\mathcal{B},\mathcal{F}$ label the statistic of the state.

Now that we have understood the structure of the chiral ring we can divide the generic one particle state (\ref{spauno}) 
in three classes according to the number of $ W_{\alpha}^i$ fields.
\begin{eqnarray}\label{symrelo}
& & |m_1,...,m_f,B_1,...B_a,0> \hbox{  } \in \hbox{  } \mathcal{H}^{\mathcal{B}}_{0,1} \nonumber\\
& & |m_1,...,m_f,B_1,...B_a,w^i> \hbox{  } \in \hbox{  } \mathcal{H}^{\mathcal{F}}_{w,1} \nonumber\\
& & |m_1,...,m_f,B_1,...B_a,w^iw^i> \hbox{  } \in \hbox{  } \mathcal{H}^{\mathcal{B}}_{w^2,1}
\end{eqnarray}
where the first and the third state are bosonic states ($\mathcal{B}$), while the second one is fermionic ($\mathcal{F}$), 
the subscripts $0,w,w^2$ label the presence of $W_{\alpha}$ fields, and the second subscript label the number of particle $n$, 
in this case ( single particle ) $n=1$.

Let us now pass to the generic $N$ case and hence to the multi particle Hilbert space. 
Thanks to the division (\ref{symrelo}) the Fock space of the chiral ring ($c$.$r$.) will be divided into three parts:
\begin{equation}\label{fockchiral}
\mathcal{F}^{c.r.}= \mathcal{F}^{\mathcal{B}}_0 \otimes  \mathcal{F}^{\mathcal{F}}_w  \otimes \mathcal{F}^{\mathcal{B}}_{w^2}
\end{equation}
where the three factors on the right hand side are the Fock spaces associated to the one particle Hilbert spaces definite in (\ref{symrelo}):
\begin{equation}\label{fock3}
\mathcal{F}^{\mathcal{B}}_0=  \bigoplus_n^{\infty}\mathcal{H}^{\mathcal{B}}_{0,n} \hbox{  } \hbox{ , } \hbox{  } \mathcal{F}^{\mathcal{F}}_w=  \bigoplus_n^{\infty}\mathcal{H}^{\mathcal{F}}_{w,n}  \hbox{  } \hbox{ , } \hbox{  } \mathcal{F}^{\mathcal{B}}_{w^2}= \bigoplus_n^{\infty}\mathcal{H}^{\mathcal{B}}_{w^2,n} 
\end{equation}
where the $\mathcal{B}$ means the Fock space of a symmetrized tensor product of the single particle states, 
while the $\mathcal{F}$ means the Fock space of an anti symmetrized tensor product of the single particle states. 
The chiral ring Fock space decomposes in the following way:
\begin{equation}\label{fockchiraldec}
\mathcal{F}^{c.r.}= \bigoplus_N^{\infty} \bigoplus_{p+q+r=N}^{\infty} \mathcal{H}^{\mathcal{B}}_{0,p} \otimes  \mathcal{H}^{\mathcal{F}}_{w,q} \otimes \mathcal{H}^{\mathcal{B}}_{w^2,r}=\bigoplus_N^{\infty} \mathcal{H}^{c.r.}_N
\end{equation}
This means that we have to deal with a multi particle space that is composed by bosons and fermions. 
If we want to count the operators at finite $N$ we must introduce mathematical functions that implement 
the bosonic and the fermionic statistic.

Let us define the one particle generating function:
\begin{equation}
g_1(q)=\sum_{n=0}^{\infty} a_n q^n
\end{equation}
counting $\frac{1}{2}BPS$ operators according for example to their dimension: the integer numbers $a_n$ 
tells us how many operators we have with dimension $n$.

For the finite $N$ counting we have to implement the right statistic.
For the bosonic part of the spectrum we have the right now familiar Plethystic function already introduced in the previous chapters
%\cite{Forcella:2007wk, Benvenuti:2006qr,Butti:2006au,Feng:2007ur,Hanany:2006uc,5per5}, 
that from now on we will call it the bosonic Plethystic function ( $PE^{\mathcal{B}}$ ): 
\begin{equation}\label{PEBB}
\prod_{n=0}^{\infty}\frac{1}{(1- \nu  q^n)^{a_n}}= PE_{\nu}^{\mathcal{B}}[g_1(q)]\equiv \exp\Big( \sum_{k=1}^{\infty} \frac{\nu^k}{k}g_1(q^k)\Big)=\sum_{N=0}^{\infty}\nu ^N g_N(q)
\end{equation}
This function takes a certain generating function $g_1(q)$ and generates the new partition functions $g_N(q)$ 
counting all the possible $N$ times symmetric products of the constituents of $g_1(q)$, implementing in 
this way the bosonic statistic, as it is possible to see from the left hand side of (\ref{PEBB}).
For the fermionic part it is useful to introduce the fermionic Plethystic function ($PE^{\mathcal{F}}$ )\cite{Feng:2007ur}:
\begin{equation}\label{PEFF}
\prod_{n=0}^{\infty}(1+ \nu  q^n)^{a_n}=PE_{\nu}^{\mathcal{F}}[g_1(q)]\equiv \exp\Big( \sum_{k=1}^{\infty}(-)^{k+1} \frac{\nu^k}{k}g_1(q^k)\Big)=\sum_{N=0}^{\infty}\nu ^N g_N(q)
\end{equation}
This function generate the partition functions $g_N(q)$ counting all the possible $N$-times anti symmetric 
products of the objects counted by $g_1(q)$, and implements in this way the fermionic statistic, as it is 
possible to see from the left hand side of (\ref{PEFF}).

Once we have defined the three basic generating functions (one for each one of the states in (\ref{symrelo})), 
the counting problem for a generic quiver gauge theories translates in counting the states in the Fock space 
defined in (\ref{fockchiraldec}).

Let us define the following chemical potentials:
\begin{itemize}
\item{$q=(q_1,...,q_f)$ labels the flavor charges;}
\item{$b=(b_1,...,b_a)$ labels the baryonic charges;}
\item{$\alpha$ labels the spin}
\item{$w$ labels the number of $W_{\alpha}$ fields.}
\end{itemize}
For each of the sectors in (\ref{symrelo}) we can associate a generating function, counting the single 
particle operators with a fixed set of baryonic charges $B=(B_1,...,B_a)$:
\begin{equation}
g_{1,B}^0(q) \hbox{  } \hbox{ , } \hbox{   } g_{1,B}^w(q,w,\alpha) \hbox{  } \hbox{ , } \hbox{  } g_{1,B}^{w^2}(q,w)
\end{equation}
With these definitions the finite $N$ counting is implemented by the following total generating function of the chiral ring:
\begin{eqnarray}\label{peppa}
& &  g^{c.r.}(q,w,\alpha,b;\nu)= \sum_{N=0}^{\infty} \nu^N g_N^{c.r.}(q,w,\alpha,b)= \nonumber\\
& &\sum_{B}b^B m(B)PE_{\nu}^{\mathcal{B}}[g^{0}_{1,B}(q)]PE_{\nu}^{\mathcal{F}}[g^{w}_{1,B}(q,w,\alpha) ]PE_{\nu}^{\mathcal{B}}[g^{w^2}_{1,B}(q,w,\alpha) ]\nonumber\\\end{eqnarray}
where the expression $b^B$ means $b_1^{B_1}...b_a^{B_a}$, and $m(B)$ is what in the previous chapters we called the multiplicities: 
the number of equal generating functions with the same set of baryonic charges and distinguished just by 
the field theory content. The meaning of (\ref{peppa}) is: with the $PE$ functions we implement 
the right statistic for the various states and then we sum over all the possible sectors with fixed baryonic 
charges taking into account the possible presence of multiplicities.

To obtain the generating function for the chiral ring with fixed number of branes $N$ one have just to take the $N$ 
times derivatives of (\ref{peppa}) with respect to the parameter $\nu$:
\begin{equation}
g_N^{c.r.}(q,w,\alpha,b) = \frac{1}{N !}\frac{\partial^N g^{c.r}(q,w,\alpha,b;\nu)}{\partial \nu^N} \Big|_{\nu = 0 }
\end{equation}

What we miss is to construct the generating functions for the two sectors of the Hilbert space containing the 
fields $W_{\alpha}$ once we know $g_{1,B}^0(q)$. This one is an easy task and can be solved in an elegant way 
introducing a superfield formalism. Let us introduce the usual set of anti commuting variables $\theta_{\alpha}$ such that:
\begin{equation}
\{ \theta _{\alpha} , \theta _{\beta} \} = 0 
\end{equation}
the dimension of the theta variables is $-3/2$ and we will label it with $q^{-3/2}$, in addition they carry a 
spin degrees of freedom that we will label with $\alpha$ for the $1/2$ spin case and $1/ \alpha$ for $-1/2$ spin case. 
As we have understood in the previous chapters 
%explained in \cite{Butti:2006au} 
the generic chiral gauge invariant operators constructed with only the scalar 
superfields of the theory is an $N$-times symmetric product of $N$ chiral fields ``building blocks'': the $({\bf O}_I^{(x,z)})_{k}^{l}$ 
fields without insertion of $W^i_{\alpha}$. For simplicity from now on we will call $\phi_B^m$ the generic scalar building block, 
where with $m$ we mean the specific set of $f$ flavor charges and with $B$ the specific set of $a$ baryonic charges of the operator. 
Given the relation inside the chiral ring and the decomposition of the Hilbert space of the chiral ring we propose a superfield 
formalism in which the super chiral fields are generically given by:   
\begin{equation}\label{superfc}
\Phi^m_B = \phi_B^m + \theta_{\alpha} (W^{\alpha}  \phi_B^m)  +  \theta_{\alpha} \theta^{\alpha}(W_{\alpha} W^{\alpha} \phi_B^m )
\end{equation}
Introducing the chemical potential $w$ counting the number of $W_{\alpha}$ fields, it is by now clear that the complete 
generating function for $N=1$ is 
\begin{equation}\label{par1B}
g_{1,B}(q,w,\alpha)=g_{1,B}^0(q)\Big( 1+q^{3/2}\hbox{ }w\hbox{ }\Big( \alpha + \frac{1}{\alpha} \Big) + q^3 \hbox{ } w^2 \hbox{ }\Big)
\end{equation} 
Once we have defined (\ref{par1B}) we can safely divide it in the fermionic part and the bosonic one:
\begin{equation}\label{div}
g_{1,B}(q,w,\alpha)=g_{1,B}^0(q) +  g_{1,B}^w(q,w,\alpha) + g_{1,B}^{w^2}(q,w^2) 
\end{equation}
The statistical behavior of the field (\ref{superfc}) will be a mixed bosonic and fermionic statistic and this 
fact is implemented by the generalized $PE$ function:
\begin{equation}\label{loca}
PE_{\nu}[g_{1,B}(q,w,\alpha)]=PE_{\nu}^{\mathcal{B}}[g_{1,B}^0(q)]PE_{\nu}^{\mathcal{F}}[g_{1,B}^w(q,w,\alpha)]PE_{\nu}^{\mathcal{B}}[g_{1,B}^{w^2}(q,w^2)]
\end{equation}
Let us now take the first derivatives of this expression and look at the form of the finite $N$ generating functions:
\begin{equation}
g_{1,B}(q,w,\alpha) = \partial _{\nu} PE_{\nu}[g_{1,B}(q,w,\alpha)] \Big|_{\nu=0}= g_{1,B}^0(q) + g_{1,B}^w(q,w,\alpha) + g_{1,B}^{w^2}(q,w)\end{equation}
This expression clearly reproduce the $N=1$ counting we started with.

The two times derivatives is more interesting: here we can start observing the mixed statistic. 
\begin{eqnarray}\label{PEexpN2}
& & g_{2,B}(q,w,\alpha) = \frac{1}{2} \partial _{\nu}^2 PE_{\nu}[g_{1,B}(q,w,\alpha)] \Big|_{\nu=0} = \nonumber\\
& & \frac{1}{2} \Big( \big( g_{1,B}^0(q^2) + [g_{1,B}^0(q)]^2 \big) + \big( g_{1,B}^{w^2}(q^2,w^2) + [g_{1,B}^{w^2}(q,w)]^2 \big) + 2 g_{1,B}^0(q) g_{1,B}^{w^2}(q,w) + \nonumber\\
& & \big( - g_{1,B}^w(q^2,w^2,\alpha^2) + [g_{1,B}^w(q,w,\alpha)]^2 \big) + 2 \big( g_{1,B}^0(q) + g_{1,B}^{w^2}(q,w) \big) g_{1,B}^w(q,w,\alpha)  \Big) \nonumber\\
\end{eqnarray}
Let us comment factor by factor the equation (\ref{PEexpN2}):
the first two factors take into account the bosonic statistic of the scalar part of the chiral ring, 
and in the usual partition functions we have written in the previous chapters that all. Here instead we have many more terms. 
The second two factors are due to the bosonic statistic of the part of the chiral ring with the insertions 
of the field $W_{\alpha}W^{\alpha}$. The third factor describe the mixing between the two bosonic sectors of the chiral ring. 
The fourth two factors is probably the most interesting one: it has the same form of the first and the second, 
but it has one minus sign more: this is due to the fermionic behavior of $g_{1,B}^w(q,w,\alpha)$. 
This factor implement the fermionic statistic of the operators in the chiral ring with the insertions 
of the field $W_{\alpha}$. The last two factors are due to the mixing between the bosonic and fermionic part of the chiral ring.

%Before passing to some checks and examples of the general proposal,
Before concluding this section we can make a comment regarding the $BPS$ mesonic branch of the chiral ring for generic quiver gauge theories. 
We learnt how to obtain the generating 
functions with $N=1$ for the scalar mesonic chiral ring of a quiver gauge theory counting the 
gauge invariant operators according to their dimensions:
\begin{equation}\label{g100}
g_{1,0}^0(q)=\sum_{n=0}^{\infty} a_n q^n
\end{equation}
Once we know the $a_n$ factors in (\ref{g100}), the general form for the function $g_0(q,t,\alpha;\nu)$ 
generating the Hilbert series for the finite $N$ mesonic counting is:
\begin{equation}\label{genNmes}
g_0(q,w,\alpha;\nu)=\sum_{\nu = 0}^{\infty} \nu ^N g_{N,0}(q,w,\alpha)=\prod_{n=0}^{\infty}\frac{(1+\nu \hbox{  }w\hbox{  } \alpha\hbox{  }
 q^{3/2+n})^{a_n}(1+ \nu \hbox{  }w \hbox{  }\frac{1}{\alpha}\hbox{  } q^{3/2+n})^{a_n}}{(1-\nu \hbox{  } q^n)^{a_n}(1- \nu \hbox{  } w^2 \hbox{  }q^{3+n})^{a_n}} \nonumber\\
\end{equation}

\section{The $\mathcal{N}=4$ Case Revisited}\label{n4reload}

In this section we will very briefly review the case $\mathcal{N}=4$ using the general technology developed in the previous section.

The basic idea is to pass from the $\delta_i^j$ invariant tensor, that does not depend on the number of colors $N$, 
to the $\epsilon_{i_1,...,i_N}$ invariant tensor that has built in the dependence on $N$. In this way the basic 
generating function counting the gauge invariant operators in the chiral ring for $N=1$ is the one counting the 
single trace in the limit $N \rightarrow \infty $ and all the counting for finite $N$ can be obtained starting 
from the single trace $N \rightarrow \infty $ generating function.

Following the prescription of the previous 
section we have just to compute the generating function $g_1^0(q)$ for $N=1$. Using the equivariant index theorem 
this is easy to compute in general, and in the particular case of $\mathcal{N}=4$ this function counts the holomorphic 
functions on $\mathbb{C}^3$ according to their degree and it is exactly the first one in (\ref{g1bfb}):
\begin{equation}
g_1^0(q) = \sum_{n=0}^{\infty}\frac{(n+1)(n+2)}{2}q^n= \frac{1}{(1-q)^3}
\end{equation}
Now we must implement the superfield prescription and write:
\begin{equation}
g_1(q,w,\alpha)=g_1^0(q)\Big( 1+q^{3/2} w \Big( \alpha + \frac{1}{\alpha} \Big) + q^3 w^2\Big)
\end{equation}
which is exactly the one obtained in section \ref{N4}.

To implement the finite $N$ counting we have just to use the generalized $PE$:
\begin{equation}\label{genN4}
g(q,w,\alpha;\nu)=PE_{\nu}[g_1(q,w,\alpha)]=PE_{\nu}^{\mathcal{B}}[g_1^0(q) + g_1^{w^2}(q,w)]PE_{\nu}^{\mathcal{F}}[g_1^{w}(q,w,\alpha)]
\end{equation}
One can easily checks that (\ref{genN4}) reproduce the equation (\ref{bosferN}). 

Now one can take the first few derivatives of (\ref{genN4}) and looking for the corresponding 
gauge invariant operators finding agreement between our counting and the field theory spectrum. 
We refer the interested reader to \cite{Forcella:2007ps}.
 
\section{The Conifold}\label{conifo}

To give an idea of the general procedure it is worthwhile to shortly study the less trivial example 
of the conifold gauge theory. This example contains almost all the properties of the generic case 
(except the multiplicity problems already discussed in chapter \ref{chiralcount}). 
In this section we just give a short review of the main steps for the construction 
of the complete partition function and we refer the reader to \cite{Forcella:2007ps} 
for more a complete discussion and some checks.

The conifold gauge theory is already familiar. It has the gauge group $SU(N)_1 \times SU(N)_2$.
The basic chiral fields are the four scalar superfields: 
\begin{equation}
A_i \hbox{ , } B_j \hbox{  }\hbox{  } i,j = 1,2
\end{equation}
The $A_i$ fields transform in the fundamental of $SU(N)_1$ and the anti fundamental of $SU(N)_2$, 
while the $B_j$ fields transform in the fundamental of $SU(N)_2$ and the anti fundamental of $SU(N)_1$; 
and the chiral spinor superfields for the two factors of the gauge group $SU(N)_1 \times SU(N)_2$:
\begin{equation}
W_{\alpha}^1 \hbox{ , } W_{\alpha}^2 \hbox{  }\hbox{  } \alpha = +,-
\end{equation}
The cinematical relations in the chiral ring are:
\begin{equation}\label{relcon}
\{W_{\alpha}^i,W_{\beta}^i\}=0 \hbox{ , } W_{\alpha}^1 A_i = A_i W_{\alpha}^2 \hbox{ , } W_{\alpha}^2 B_j = B_j W_{\alpha}^1
\end{equation}
while the dynamical ones coming from the super potential are:
\begin{equation}
B_1 A_i B_2= B_2 A_i B_1 \hbox{ , } A_1 B_j A_2 = A_2 B_j A_1
\end{equation}

\subsection{The Mesonic Chiral Ring}

Let's start from the easy case of the mesonic chiral ring.

The generating function for the scalar part of the chiral ring is\footnote{to be in line with the 
literature in the generating functions for the conifold we will use the chemical potential $q$ to 
label the $R$ charge of the chiral fields: $R_{A_i}=R_{B_i}=1/2$, $R_{W_{\alpha}}=1$.} (\ref{Hcon}):
\begin{equation}
g_{1,0}^0(q) = \sum_{n=0}^{\infty} (1+n)^2 q^n =\frac{(1+q)}{(1-q)^3}
\end{equation}
Using the prescription of section \ref{gentheo} the complete $N=1$ mesonic generating function is:
\begin{equation}\label{n1conc}
g_{1,0}(q,w,\alpha)= g_{1,0}^0(q)\Big( 1+q w \Big( \alpha + \frac{1}{\alpha} \Big) + q^2 w^2 \Big)
\end{equation}
Now we have just to apply the $PE$ formalism and we obtain the mesonic generating function for finite $N$:
\begin{eqnarray}
& & g_0(q,w,\alpha;\nu)= PE_{\nu}^{\mathcal{B}}[g_{1,0}^0(q)+g_{1,0}^{w^2}(q,w)]PE_{\nu}^{\mathcal{F}}[g_{1,0}^w(q,w,\alpha)]=\nonumber\\
& & \prod_{n=0}^{\infty}\frac{(1 + \nu \hbox{ }w\hbox{ } \alpha\hbox{ } q^{1+n})^{(n+1)^2}(1 + \nu\hbox{ } w\hbox{ } \frac{1}{\alpha} \hbox{ } q^{1+n})^{(n+1)^2}}{(1-\nu\hbox{ } q^n)^{(1+n)^2}(1- \nu\hbox{ } w^2\hbox{ } q^{2+n})^{(1+n)^2}}
\end{eqnarray}
Observe that this expression has the general form of equation (\ref{genNmes}).
Using the relations (\ref{relcon}) it is easy to understand why this procedure works. 
The equations (\ref{relcon}) shows that the single trace operators of the mesonic chiral ring satisfy:
\begin{equation}
\hbox{tr}(W_{\alpha}^1 (A B )^k )= \hbox{tr}(W_{\alpha}^2 (B A )^k) \hbox{ , } \hbox{tr}(W_{\alpha}^1W^{\alpha}_1 (A B )^k )= \hbox{tr}(W_{\alpha}^2 W^{\alpha}_2 (B A )^k)
\end{equation}
which means that we must consider only the diagonal part of the spinor superfield, namely:
\begin{equation}
W_{\alpha}\equiv W_{\alpha}^1 = W_{\alpha}^2 
\end{equation}
Hence the single trace operators of the mesonic chiral ring are:
\begin{equation}\label{gentrcon}
\hbox{tr}((A B)^k) \hbox{ , } \hbox{tr}(W_{\alpha} (A B )^k )\hbox{ , } \hbox{tr}(W_{\alpha}W^{\alpha} (A B )^k) 
\end{equation}
From (\ref{gentrcon}) it is easy to understand that whenever we know the scalar mesonic 
generating function for $N=1$, the complete one is just the one dressed as in (\ref{n1conc}).

The interested reader can now expand the generating function for the first few values of $N$ 
and check the result against the field theory \cite{Forcella:2007ps}.

Now that we have understood some basic properties about the mesonic chiral ring let us pass to 
the more interesting case of the baryonic one.

\subsection{The Baryonic Conifold's Chiral Ring}

We would like to write down the complete generating function for the chiral ring 
of the conifold theory containing all the $\frac{1}{2}$ $BPS$ degrees of freedom 
of the theory: namely the mesonic sector ($B=0$), all the operators charged under 
the $U(1)$ baryonic symmetry and of course all the possible fermionic degrees of freedom $W_{\alpha}$.

\subsubsection{The $B=1$ Baryonic Sector}\label{B1N2con}

Let us start analyzing the problem in the easy case of fixed baryonic charge $B$, 
and for simplicity we will analyze the sector $B=1$.

The prescription given in section \ref{gentheo} basically says that we just need 
to know the $g_{1,1}^0(q)$ generating function for the scalar chiral ring. 
We computed it in section \ref{conifoldexample} and it is:
\begin{equation}\label{consca}
g_{1,1}^0(q)=\frac{2 q^{1/2}}{(1-q)^3}
\end{equation}
Now we have to dress it with usual ``supermultiplet'' charges: 
\begin{equation}\label{g1n1b1}
g_{1,1}(q,w,\alpha)=g_{1,1}^0(q)\Big(1+q w \Big( \alpha +\frac{1}{\alpha} \Big) +  q^2 w^2 \Big)= \frac{2 \big(q^{1/2}+ q^{3/2}w(\alpha +\frac{1}{\alpha})+  q^{5/2} w^2 \big)}{(1-q)^3}
\end{equation}
The meaning of this procedure is easily explained in the case of the conifold. 
Using the relations (\ref{relcon}) we understand that the operators in the chiral 
ring in the case $N=1$ are\footnote{Observe in the case $N=1$ 
the operators are no more matrices but numbers.}:
\begin{equation}\label{basab}
A_{i_1}B_{j_1}A_{i_2}B_{j_2}...A_{i_n}B_{j_n}A_{i_{n+1}} ,
\end{equation}
\begin{equation}\label{wab}
W_{\alpha}A_{i_1}B_{j_1}A_{i_2}B_{j_2}...A_{i_n}B_{j_n}A_{i_{n+1}} ,
\end{equation}
\begin{equation}\label{wwab}
W_{\alpha}W^{\alpha}A_{i_1}B_{j_1}A_{i_2}B_{j_2}...A_{i_n}B_{j_n}A_{i_{n+1}} .
\end{equation}
The generating function (\ref{consca}) counts all the operators of the form (\ref{basab}), 
while the dressed one (\ref{g1n1b1}) takes into account also the operators in (\ref{wab}) and (\ref{wwab}).

Let us now apply the rules explained in section \ref{gentheo} to implement the finite $N$ counting. 
\begin{eqnarray}
& & g_{1}(q,w,\alpha;\nu)=\sum_{\nu=0}^{\infty} \nu ^N g_{N,1}(q,w,\alpha)=\nonumber\\
& & PE_{\nu}[g_{1,1}(q,w,\alpha)]=PE_{\nu}^{\mathcal{B}}[g_{1,1}^0(q)+g_{1,1}^{w^2}(q,w)]PE_{\nu}^{\mathcal{F}}[g_{1,1}^w(q,w,\alpha)]\nonumber\\
\end{eqnarray}
In the case $N=2$ we have the following generating function:
\begin{eqnarray}\label{geng12yiu}
 g_{2,1}(q,w,\alpha) =
\frac{P(q,w,\alpha)}{(1 - q)^6(1 + q)^3 \alpha^2}
\end{eqnarray}
where
\begin{eqnarray}
P(q,w,\alpha)&=& q(q w + \alpha)(1 + q w \alpha)(3 \alpha + q((3 + q(9 + q)) \alpha + \nonumber\\
& & q(3 + q (3 + q(9 + q))) w^2 \alpha + (1 + 3 q(3 + q + q^2)) w (1 + \alpha^2)))\nonumber\\
\end{eqnarray}
One can now expands the various generating functions for different values of $N$ and checks the result against the 
field theory spectrum \cite{Forcella:2007ps}.

Once again observe, as already explained in section \ref{gentheo}, that we are counting 
BPS operators in the conifold with gauge group $SU(N)_1\times SU(N)_2 \times U(1)_D$, where the $U(1)_D$ factor is the
diagonal $U(1)$ of the UV gauge group $U(N)_1\times U(N)_2$. The chiral bifundamental fields $A_i$, $B_j$, 
are not charged under this diagonal $U(1)$. Even if it is possible to obtain the ``pure $SU(N)$'' counting, 
it is more natural and economic to continue to keep the overall $U(1)$ factor, 
and we will indeed continue in this way in the following.

\subsubsection{The Complete Generating Function for the Conifold}

Now we would like to write down the complete generating function for the conifold 
containing all the baryonic charges and all the fermionic degrees of freedom.
The general procedure explained in section \ref{gentheo} tells us that the complete 
generating function $g^{c.r}(q,w,\alpha,b;\nu)$ is obtained by summing over all the 
possible baryonic charges $B$ the generalized $PE$ of the $N=1$ generating functions 
with fixed baryonic charge $g_{1,B}(q,w,\alpha)$. In the conifold case there is just 
one baryonic charge $B$ running from $-\infty$ to $+\infty$ and there are no multiplicities. 
Hence equations (\ref{peppa}) and (\ref{loca}) become: 
\begin{equation}
g^{c.r.}(q,w,\alpha,b;\nu)= \sum_{B = -\infty}^{+ \infty}b^B PE_{\nu}[g_{1,B}(q,w,\alpha)]
\end{equation}
In chapter \ref{barcon} we gave the $g_{1,B}^0(q)$ generating functions for the scalar part of the chiral ring:
\begin{equation}
g_{1,B}^0(q)= \frac{(-1 + B(-1 + q)-q) q^{B/2}}{(-1 + q)^3}\nonumber\\
\end{equation}
hence
\begin{equation}
g_{1,B}(q,w,\alpha)= g_{1,B}^0(q)\Big(1 + q w \Big( \frac{1}{\alpha} + \alpha \Big) + q^2 w^2 \Big)
\end{equation}
Let us start as usual with the $N=1$ generating function.
\begin{eqnarray}\label{g1con}
g_1^{c.r.}(q,w,\alpha,b) &=& \partial _{\nu} g^{c.r.}(q,w,\alpha,b;\nu)\Big|_{\nu=0}= \nonumber\\
& & \nonumber\\
\sum_{B = -\infty}^{+ \infty}b^B g_{1,B}(q,w,\alpha) &=& \frac{1 + q w \Big( \frac{1}{\alpha} + \alpha \Big) + q^2 w^2}{\Big( 1 - \frac{q^{1/2}}{b}\Big)^2\Big( 1 - q^{1/2}b \Big)^2}
\end{eqnarray}
This result is easily explained.
In the case $N=1$ the possible generators in the chiral ring are:
\begin{eqnarray}
A_i \hbox{ , }  B_j \hbox{ , } W_{\alpha} 
\end{eqnarray}
The theory does not have superpotential and hence the scalar chiral ring is freely generated 
and one has just to impose the relations coming from the $W^i_{\alpha}$ fields. As usual one can expand the generating function and check against the field theory spectrum \cite{Forcella:2007ps}.

Now we would like to compute the generating function $g_2^{c.r.}(q,w,\alpha,b)$ for the conifold with $N=2$, 
counting all the operators in the chiral ring.\\
Using the relations:
\begin{eqnarray}
g^{c.r.}(q,w,\alpha,b ; \nu) &=& \sum_{B = -\infty}^{B = + \infty} b^B PE_{\nu}^{\mathcal{B}} \Big[ g_{1,B}^0 (q) \Big] PE_{\nu}^{\mathcal{F}} \Big[ g_{1,B}^w (q,w,\alpha) \Big]PE_{\nu}^{\mathcal{B}} \Big[ g_{1,B}^{w^2} (q,w) \Big] \nonumber\\
&=& \sum_{\nu = 0}^{\infty} \nu^N g_N^{c.r.}(q,w,\alpha,b) 
\end{eqnarray}
One can easily obtain:
\begin{eqnarray}\label{g2t1t2}
g_2^{c.r.}(q,w,\alpha,b) &=& \sum_{B = -\infty}^{B = + \infty} b^B g_{2,B}(q,w,\alpha,b) = \nonumber\\
& &\sum_{B = -\infty}^{B = + \infty} b^B \frac{1}{2} \partial_{\nu} ^2 \Big( PE_{\nu}^{\mathcal{B}}[g_{1,B}^0(q)] PE_{\nu}^{\mathcal{F}}[ g_{1,B}^w(q,w,\alpha)] PE_{\nu}^{\mathcal{B}}[g_{1,B}^{w^2}(q,w)]\Big) = \nonumber \\
& &  \sum_{B = -\infty}^{B = + \infty} b^B \frac{1}{2}\Big( g_{1,B}^0(q^2) + [g_{1,B}^0(q)]^2 + g_{1,B}^{w^2}(q^2,w^2) + [g_{1,B}^{w^2}(q,w)]^2 \nonumber\\
& & 2 ( g_{1,B}^0(q) g_{1,B}^{w^2}(q,w)) - g_{1,B}^{w}(q^2,w^2,\alpha^2) + [g_{1,B}^{w}(q,w,\alpha)]^2 +\nonumber\\
& &  2 ( g_{1,B}^{0}(q) + g_{1,B}^{w^2}(q,w) ) g_{1,B}^{w}(q,w,\alpha) \Big) \nonumber\\
\end{eqnarray}
The easiest way to do this computation is to pass from the chemical potentials $q$ and $b$, 
counting the $R$ charge and the baryonic charge, to the chemical potentials $t_1$, $t_2$ 
counting the number of $A_i$ and $B_j$ fields, and sum over all the $SU_1(2) \times SU_2(2)$ 
symmetric representations:
%\comment{\footnote{The theory has indeed an $SU_1(2) \times SU_2(2)$ global 
%flavor symmetry under which the fields $A_i$ transform as $(2,1)$ while the fields $B_i$ transform as $(1,2)$}
%}
\begin{eqnarray}\label{g2t1t2br}
 g_2^{c.r.}(t_1,t_2,w,\alpha) =\frac{P(t_1,t_2,w,\alpha)}{(1 - t_1^2)^3(1 - t_1 t_2)^3(1 - t_2^2)^3 \alpha^2}
\end{eqnarray}
where
\begin{eqnarray}
P(t_1,t_2,w,\alpha)&=&((w + \alpha)(1 + w \alpha)( \alpha + t_1^3 t_2 (-3 + t_2^2)(w + t_2^2(1 + w^2)\alpha + w \alpha^2) - \nonumber\\
& & t_1^2(-1+ 3 t_2^2)(w + t_2^2 (1 + w^2)\alpha +  w \alpha^2) + w(w \alpha + t_2^2 (1 + \alpha^2)) + \nonumber\\
& & t_1 t_2 (\alpha + w^2 \alpha - (-4 + 3t_2^2) w(1 + \alpha^2)) + t_1^4 t_2^2(-3(1 + w^2)\alpha + \nonumber\\
& & t_2^2(w + 4(1 + w^2)\alpha + w \alpha^2)) + t_1^5 t_2^3(\alpha + w(w \alpha + t_2^2 (1 + \alpha^2))))) \nonumber
\end{eqnarray}
It is now easy to expand (\ref{g2t1t2br}) in terms of $t_1$, $t_2$ and find an expression 
to compare with the field theory result. It is easy to see that the generating function
counts the gauge invariant operators in the right way \cite{Forcella:2007ps}.

\section{Summary and Discussions}\label{conl}

In this chapter we extended the analysis of the chiral ring for gauge theory on D brane at singularities 
 explaining how it is possible to add the fermionic degrees of freedom to the bosonic chiral ring considered in
previous chapters. We presented a systematic way to construct the complete generating functions 
for any quiver gauge theories of which we are able to write down the scalar part of the 
generating functions: namely the infinite class of gauge theories dual to toric singularities, 
to quotient singularities, to complex cones over delPezzo surfaces, and many more.

We solved the problem of adding the $W_{\alpha}^i$ spinorial degrees of freedom by introducing 
a kind of superfield formalism and implementing the mixed state statistic through the introduction 
of the fermionic version of the Plethystic exponential.

Right now we have a good and global understanding of the structure of the chiral ring of a huge number of quiver 
gauge theories. Possible future developments could be a systematic study of the statistical properties 
of these gauge theories, the large quantum number behavior of the various partition functions, the phase 
structure of these theories, and maybe their application to related problems such as the holographic duals 
of these thermodynamical properties. The generating functions constructed in this chapter contain the information 
about the density distribution of the $BPS$ degrees of freedom of the $CFT$ and hence about the entropy and more 
generically the statistical and thermodynamical properties of quiver gauge theories. For this reason they could 
be a good starting point for a microscopic understanding of the entropy of the recently constructed $AdS$ 
black holes \cite{Gutowski:2004ez,Gutowski:2004yv,Sinha:2006sh,Sinha:2007ni}.

A comment is mandatory: the partition functions of quiver gauge theories we studied till now are mainly based on undeformed $CFT$. Usually $CFT$ admits a 
set of marginal deformations that leave the theory conformally invariant. It would be interesting to study how the $BPS$ spectra change when we turn on marginally operators in the field theory. 
In the next chapter we will deal with this problem for a specific class of deformations.

Another very interesting problem would be to study the BPS spectra in the non conformal case. 
This topic is leave for future investigations.

\chapter{$\beta$-Deformed Gauge Theory}

Super Conformal Field Theories usually come in families. Namely there exist a manifold of coupling constants where the theory is superconformal. Along this manifold the value of the central charge and of all the R-charges of the theory are constant, but the global symmetries can be partially broken and the moduli space can be modified: namely some of the directions can be uplifted. In the case of D3 branes at singularities the uplifted of some directions in the moduli space is due to the presence of fluxes in the background that behave as potential for the D3 branes. 

There exist a marginal deformation common to all the gauge theories on D3 branes at toric CY singularities that we have studied till now: the $\beta$-deformation. This deformation is a specific modification, depending on the complex parameter $\beta$, of the coefficients of the superpotential. As a consequence the moduli space of the gauge theory is partially uplifted and the dual geometry is no more neither CY neither complex. Indeed in this chapter we consider the class of super-conformal $\beta$-deformed $\mathcal N=1$ gauge theories dual to string theory on $AdS_5 \times H$ with fluxes, where $H$ is a deformed Sasaki-Einstein manifold. 
The supergravity backgrounds are explicit examples of Generalised Calabi-Yau manifolds:
the cone over $H$ admits an integrable generalised complex structure in terms
of which the BPS sector of the gauge theory can be described. 
We will study the moduli spaces of the deformed toric $\mathcal N=1$ gauge theories
on a number of examples. Using the dual supergravity background and the 
techniques from Generalized Complex Geometry the results in gauge theory can be explicitly checked studying the
  moduli spaces of D3 and D5 static and dual giant probes.
In this chapter we will just study the mesonic moduli space and the mesonic chiral ring of the SCFT 
from the field theory prospective, and we will leave the study of the dual gravity side to the literature \cite{Butti:2007aq}.

\section{Generalities}

The super-conformal gauge theories living on D3-branes at
singularities generally admit marginal deformations. A particularly
interesting case of marginal deformation for theories with $U(1)^3$
global symmetries is the so called $\beta$-deformation \cite{Leigh:1995ep}.  The
most famous example is the $\beta$-deformation of ${\cal N}=4$ SYM which
has been extensively studied both from the field theory point of view
and the dual gravity perspective.  In
particular, in \cite{Leigh:1995ep}, Lunin and Maldacena found the supergravity dual solution, which is
a  completely regular $AdS_5$ background.
%dual to the beta deformation of ${\cal N}=4$ SYM. 
Their construction can be generalized to the 
super-conformal theories
associated with the recently discovered Sasaki-Einstein backgrounds
$AdS_5\times L^{p,q,r}$ \cite{Gauntlett:2004zh,Gauntlett:2004yd,Gauntlett:2004hh,Cvetic:2005ft,Cvetic:2005vk,Martelli:2005wy}.  More generally, 
all toric quiver gauge theories admit $\beta$-deformations \cite{Benvenuti:2005wi}
and, have regular gravitational duals. The
resulting $\beta$-deformed theories are interesting both from the
point of view of field theory and of the gravity dual.

On the
field theory side, we deal with a gauge theory with a deformed moduli
space of vacua and a deformed spectrum of BPS operators. The case of
${\cal N}=4$ SYM has been studied in details in the literature
\cite{Berenstein:2000hy,Berenstein:2000ux,Dorey:2002pq,Dorey:2003pp,Dorey:2004xm,Benini:2004nn}. In this chapter we would like to extend this analysis to
a generic toric quiver gauge theory to see how some of the properties of the moduli space and of the BPS spectrum of the theory 
studied till now are modified in the $\beta$-deformed case. The moduli space of the $\beta$-deformed
gauge theory presents the same features as in ${\cal N}=4$ case. In particular,
its  structure depends on the value of the deformation parameter $\beta$. 
For generic $\beta$ the deformed theory admits a Coulomb branch which is given by  
a set of complex lines. For $\beta$ rational there are additional directions corresponding
to Higgs branches of the theory.

On the gravity side, the dual backgrounds can be obtained from the original Calabi-Yaus
with a continuous T-duality transformation using the general method proposed in \cite{Leigh:1995ep}.
It is possible to study the $\beta$-deformed background
even in the cases where the explicit original Calabi-Yau metric is not known. The toric structure 
of the original  background is enough. Besides the relevance for AdS/CFT, 
the $\beta$-deformed backgrounds are also interesting from the geometrical point view.
They are Generalised Calabi-Yau manifolds\cite{Grana:2004bg,Grana:2005sn}:  after the
deformation the background is no longer complex, but is still admits an integrable 
generalised complex structure.
Actually the $\beta$-deformed backgrounds represent  one of the few explicit known
examples of generalised geometry solving the equation of motions of
type II supergravity. 
\footnote{For other non compact examples see \cite{Butti:2004pk,Minasian:2006hv} and for compact ones \cite{Grana:2006kf}.}.  
The extreme simplicity of such backgrounds make it possible to explicitly apply the formalism
of Generalised Complex Geometry \cite{Hitchin:2004ut,Gualtieri:2003dx} which, provides an elegant way to
study T-duality and brane probes \cite{Martucci:2005ht,Martucci:2006ij,Koerber:2006hh}.

The connection between gravity and field theory is provided by the
study of supersymmetric D-brane probes moving on the $\beta$-deformed
background. This is basically an extension of the string states/gauge operators map we have explained in the previous chapters.
%A class of brane probes, the dual giant gravitons, is of particular
%relevance for our analysis. They are   
%branes wrapping an three-sphere in $AdS_5$ and rotating in the internal manifold, and they are
%mapped on the field theory side to 
%the mesonic operators parameterising the moduli space.
Indeed it  is possible to analyze the case of static D3 and D5
probes, as well as the case of D3 and D5 dual giant gravitons \cite{Butti:2007aq}, and study in details 
existence and moduli space of such probes.
In the $\beta$-deformed background, both 
static D3 probes and D3 dual giants can only live on a set of intersecting 
complex lines inside the
deformed Calabi-Yau, corresponding to the locus where the $T^3$ toric
fibration degenerates to $T^1$.  This is in agreement with the abelian
moduli space of the $\beta$-deformed gauge theory which indeed consists
of a set of lines.  Moreover, in the case of rational $\beta$ It is possible to demonstrate the existence of 
both static D5 probes and D5 dual giant gravitons
with a moduli space isomorphic to the original Calabi-Yau divided by a
$\mathbb{Z}_n\times \mathbb{Z}_n$ discrete symmetry.  This statement
is the gravity counterpart of the fact that, for rational $\beta$, new
branches are opening up in the moduli space of the gauge theory
\cite{Berenstein:2000hy,Berenstein:2000ux,Dorey:2002pq,Dorey:2003pp,Dorey:2004xm}. 

To do not go too much far away from the main spirit of this Thesis, in this chapter we will 
just look to the field theory side, and we refer the reader to \cite{Butti:2007aq} for a complete discussion about the gravity dual.
%Our analysis also generalises the results of \cite{MSg} where it has been shown
%that the classical phase space
% of supersymmetric D3 dual giant gravitons in the undeformed
%Calabi-Yau background is isomorphic to the Calabi-Yau variety.
\comment{
The classical way to study probe configuration is to solve the equations of motion
coming from the probe Dirac-Born-Infeld action.
Generalised Complex Geometry provides an alternative 
method to approach the problem. As we will explain, a D-brane is characterized by its generalised
tangent bundle. The dual probes in the
$\beta$-deformed geometry can be obtained from the original ones applying T-duality 
to their generalised tangent bundles. 
The approach in terms of Generalised Geometry allows also to clarify how the complex 
structure of the gauge theory is
reflected by the gravity dual, which, as we have already mentioned,
is not in general a complex manifold.

The study of brane probes we present here can be seen as consisting of two
independent and complementary sections, 
one dealing with the Born-Infeld approach and
the other one using  Generalised Complex Geometry.  
We decided to keep the two analysis independent, so that the reader not
interested in one of the two can skip the corresponding section.
}

The chapter is organized as follows. In section \ref{betadef} we introduce the structure 
of the $\beta$-deformed gauge theory for a general toric CY singularity $\cX$.   
%In this analysis, we
%will clarify how the complex structure of the gauge theory is
%reflected by the gravity dual, which is not in general a complex manifold. 
In section \ref{nonab} we explain in complete generality how to obtain the non abelian BPS conditions for the mesonic chiral ring and $\beta$-deformed mesonic moduli
space directly from the geometry of $\cX$. In section \ref{abelbet} we study the mesonic abelian moduli space, and in section \ref{razbeta} we discuss the 
particular case of rational $\beta$. In these last two sections we give some explicit examples of $\beta$-deformed mesonic 
moduli space and chiral spectrum, based on $\beta$-deformation of the gauge theories
 associated to $\mathbb{C}^3$, the conifold, the SPP, and the $PdP_4$ singularities.

\section{$\beta$-Deformed Quiver Gauge Theories}\label{betadef}

The entire class of super-conformal gauge theories living on D3-branes at 
toric conical Calabi-Yau singularities $\cX$ admits marginal deformations.
The most famous example is the $\beta$-deformation of ${\cal N}=4$ SYM
with $SU(N)$ gauge group where the original superpotential 
\begin{equation}
\label{N=4} 
X Y Z - X Z Y 
\end{equation}
is replaced by the $\beta$-deformed one
\beq
\label{N=4beta} 
e^{i\pi \beta} X Y Z - e^{-i\pi \beta} X Z Y \, .  
\eeq 
A familiar argument due to Leigh and Strassler
\cite{Leigh:1995ep} shows that the $\beta$-deformed theory is conformal for all
values of the $\beta$ parameter.

Similarly, a $\beta$-deformation can be defined for the conifold
theory. The gauge theory has gauge group $SU(N)\times SU(N)$ and
bi-fundamental fields $A_i$
%_\alpha^A$ 
and $B_j$
%_A^\alpha$ with $\alpha,A=1,...,N,i,p=1,2$ 
transforming in the representations $(2,1)$ and $(1,2)$
of the global symmetry group $SU(2)\times SU(2)$,
respectively, and superpotential 
\beq\label{conifold} 
A_1 B_1 A_2 B_2 - A_1 B_2 A_2 B_1 \, .  
\eeq 
The $\beta$-deformation corresponds to the
marginal deformation where the superpotential is replaced by
\beq\label{conifoldbeta} e^{i\pi \beta} A_1 B_1 A_2 B_2 - e^{-i\pi
  \beta} A_1 B_2 A_2 B_1 \, .  \eeq

Both theories discussed above possess a $U(1)^3$ geometric symmetry corresponding to the isometries of the
internal space $\cX$, one $U(1)$ is an R-symmetry while the other two act on the fields as flavour global 
symmetries. The $\beta$-deformation is strongly related to the existence of such
$U(1)^3$ symmetry  and has a nice and
useful interpretation in terms of non-commutativity in the internal
space \cite{Leigh:1995ep}.  The deformation is obtained by selecting in $U(1)^3$ the two flavor
symmetries $Q_i$  commuting with the supersymmetry charges
and using them to define a modified non-commutative product.  This
corresponds in field theory to replacing the standard product between
two matrix-valued elementary fields $f$ and $g$ by the star-product
\begin{equation}
\label{star}
f*g \equiv e^{i \pi \beta (Q^f \wedge \hbox{ }Q^g)} f g 
\end{equation}
where $Q^f=(Q^f_1,Q^f_2)$ and $Q^g=(Q^g_1,Q^g_2)$ are the charges of the matter fields
under the two $U(1)$ flavor symmetries and
\beq
\label{wedge} 
(Q^f \wedge Q^g)= (Q^f_1 Q^g_2 -Q^f_2 Q^g_1) \, .
\eeq

The $\beta$-deformation preserves the $U(1)^3$ geometric symmetry 
of the original gauge theory, while
other marginal deformations in general further break it.

All the superconformal quiver theories obtained from toric Calabi-Yau
singularities $\cX$ have a $U(1)^3$ symmetry
corresponding to the isometries of $\cX$ and therefore admit exactly marginal
$\beta$-deformations.  The theories have a gauge group $\prod_{i=1}^g
SU(N)$, bi-fundamental fields $X_{ij}$ and a bipartite structure which is
inherited from the dimer construction \cite{Hanany:2005ve,Franco:2005rj,Feng:2005gw,Franco:2006gc}.   The superpotential contains an
even number of terms $V$ naturally divided into $V/2$ terms weighted
by a $+1$ sign and $V/2$ terms weighted by a $-1$ sign \beq
\sum_{i=1}^{V/2} W_i(X) - \sum_{i=1}^{V/2} \tilde W_i(X) \, .  \eeq
The $\beta$-deformed superpotential is obtained by replacing the ordinary
product among fields with the star-product (\ref{star}) and 
can always be written \cite{Benvenuti:2005wi,Butti:2007aq}, after rescaling fields as 
\beq\label{defSup} 
e^{i \alpha \pi \beta} \sum_{i=1}^{V/2}
W_i(X) - e^{-i \alpha \pi \beta} \sum_{i=1}^{V/2} W_i(X)
\eeq 
where $\alpha$ is some rational number.  It is obvious how ${\cal
  N}=4$ SYM and the conifold fit in this picture.
% other examples will be given in Section \ref{gaugetheory}.

\bigskip
We will be interested in the effects of the $\beta$-deformation on the moduli space and on the 
BPS operators of the field theory. For simplicity in this chapter we will concentrated in the mesonic 
moduli space and in the mesonic chiral ring, where the effect of the marginal deformation is already remarkable. 
Indeed the $\beta$-deformation drastically reduces the 
mesonic moduli space of the theory, which, as we learnt int the previous chapters, is originally isomorphic to the $N$-fold symmetric product of
the internal Calabi-Yau. To see quickly what happens consider the case
where the $SU(N)$ groups are replaced by $U(1)$'s - by abuse of language
we can refer to this as the $N=1$ case. Physically, we are considering
a mesonic direction in the moduli space where a single D3-brane is
moved away from the singularity. In the undeformed theory the D3-brane
probes the Calabi-Yau $\cX$ while in the $\beta$-deformed theory it can only
probe a subvariety consisting of complex lines intersecting at the
origin. This can be easily seen in ${\cal N}=4$ and in the conifold
case.

For ${\cal N}=4$ SYM the
F-term equations read
\beq
\label{N4}
X Y = b Y X \hbox{ , } Y Z = b Z Y \hbox{ , } Z X = b X Z \eeq 
where here and in the following $b=e^{-2i \pi\beta}$. Since $X$, $Y$, $Z$ are
c-numbers in the $N=1$ case, these equations are trivially satisfied
for $\beta=0$, implying that the moduli space is given by three
unconstrained complex numbers $X$, $Y$, $Z$ giving a copy of
$\mathbb{C}^3$.  However, for $\beta\ne 0$ these equations can be
satisfied only on the three lines given by the equations
$X=Y=0$, or $X=Z=0$, or $Y=Z=0$. Only one field among $X$, $Y$, $Z$ is different
from zero at a time.

For the conifold the F-term equations read
\bea\label{conFterm}
B_1 A_1 B_2 &=&  b^{-1}\, B_2 A_1 B_1 \, ,\nonumber\\     
B_1 A_2 B_2 &=&  \,b\,\,  B_2 A_2 B_1 \, ,\nonumber\\  
A_1 B_1 A_2 &=&  \,b\,\,  A_2 B_1 A_1 \, ,\nonumber\\  
A_1 B_2 A_2 &=&  b^{-1}  A_2 B_2 A_1 \, .
\eea
These equations are again trivial for $\beta=0$ and $N=1$, 
the fields becoming commuting c-numbers.
The  brane moduli space is parametrized by the four gauge invariant mesons
\beq x=A_1B_1,\,\,\,\, y=A_2B_2,\,\,\,\, z=A_1B_2,\,\,\,\,  w=A_2B_1\eeq
 which are not independent but
subject to the obvious relation $xy=zw$. This is the familiar description
of the conifold as a quadric in $\mathbb{C}^4$. For
$\beta\neq 0$, the F-term constraints (\ref{conFterm}) 
are solved when exactly one field $A$ and 
one field $B$ are different from zero. This implies that only one meson
can be different from zero at a time. The moduli space thus reduces to the
four lines 
\beq
\label{conifoldlines}  y=z=w=0 \, , \qquad x=z=w=0 \, , 
\qquad x=y=z=0 \, , \qquad x=y=w=0 .
\eeq

Using the dual gravity solutions it is possible to show, in complete generality \cite{Butti:2007aq}, that 
for $\beta$-deformed toric quivers the abelian mesonic moduli space is reduced to $d$
complex lines, where $d$ is the number of vertices in the toric diagram
of the singularity. We will give a generic construction of the mesonic moduli space of the gauge theory using toric 
techniques in the following sections.

Something special happens for $\beta$ rational. New branches in the
moduli space open up. The ${\cal N}=4$ case was originally discussed
in \cite{Berenstein:2000hy,Berenstein:2000ux} and the conifold in \cite{Dasgupta:2000hn}. In all cases these
branches can be interpreted as one or more branes moving on the
quotient of the original Calabi-Yau $\cX$ by a discrete $\mathbb{Z}_n\times
\mathbb{Z}_n$ symmetry. Even in this case it is possible to describe these branes explicitly in
the gravitational duals and we refer the reader to \cite{Butti:2007aq} for a detailed discussion. 
In the following sections we will just look at the field theory
analysis of these vacua and we will show that it is possible to analyze the $\beta$ rational case in complete generality.

\section{Non Abelian BPS Conditions}\label{nonab}

In order to understand the full mesonic
 moduli space of the gauge theory we need 
to study general non-abelian solutions of the F term equations. 
Before attacking the general construction, 
we consider ${\cal N}=4$ SYM and the conifold.
 
In the ${\cal N}=4$ SYM case, we form mesons out of the three adjoint fields $X_{\alpha}^{\beta}$, $Y_{\alpha}^{\beta}$, $Z_{\alpha}^{\beta}$. The non-abelian BPS conditions for these mesonic fields are given in equation (\ref{N4}) and can be considered as equations
for three $N\times N$ matrices. In the $\beta=0$ case this equations tell us that we can diagonalize the three mesons and the eigenvalues describe an $N$ times symmetric product of $\mathbb{C}^3$.

In the conifold case, we can 
define four composite mesonic fields which transform in the adjoint 
representation of one of the two gauge groups
\begin{equation}
x=(A_1B_1)_\alpha^\beta,\,\,\,\, y=(A_2B_2)_\alpha^\beta,\,\,\,\, z=(A_1B_2)_\alpha^\beta,\,\,\,\,  w=(A_2B_1)_\alpha^\beta
\end{equation}
and consider the four mesons $x,y,z,w$ as $N\times N$ matrices. We could use 
the second gauge group without changing the results. With a simple 
computation using the F-term conditions (\ref{conFterm}) we derive the
following matrix commutation equations
\begin{eqnarray}
\label{comm}
& & x z = b^{-1} z x \hbox{ , } x w  = b w x \hbox{ , } y z = b z y\nonumber\\
& & y w = b^{-1} w y \hbox{ , } x y = y x \hbox{ , } z w = w z
\end{eqnarray}
and the matrix equation
\begin{equation}
\label{eq} 
x y = b w z 
\end{equation}
which is just the conifold equation. 
%Here and in the following $b=e^{-2 i \pi \beta}$. 
For $\beta=0$ these conditions simplify.
All the mesons commute and the $N\times N$ matrices $x,y,z,w$ can be simultaneous diagonalized. The eigenvalues are required to satisfy the conifold equation(\ref{eq}) and therefore the moduli space is given by the symmetrized
product of $N$ copies of the conifold, as expected. 

An interesting observation is that, for the ${\cal N}=4$ SYM 
and (\ref{comm}) for the conifold, the mesonic F-term conditions for 
$\beta \ne 0$ can be obtained by using the non commutative product
defined in (\ref{star}) directly on the mesonic fields.
The charges of mesons for ${\cal N}=4$ and the
conifold are shown in Figure \ref{n4con}.

The BPS conditions for the Calabi-Yau case, which require that every 
pair of mesonic fields
$f$ and $g$ commute, are replaced in the $\beta$-deformed theory by a
non commutative version 
\beq \label{starcomm} [f,g]\, =\, 0  \qquad \rightarrow \qquad [f,g]_\beta\equiv f*g-g*f \, =\, 0 \, . \eeq
It is an easy exercise, using the assignment of charges shown in Figure \ref{n4con}, to show that these modified commutation relations reproduce
equations (\ref{N4}) and (\ref{comm}).

This simple structure extends to a generic toric gauge theory. The algebraic equations of the Calabi-Yau give a set of matrix equations for mesons. In the undeformed theory, all mesons commute, while in the $\beta$-deformed theory the
original commutation properties are replaced by their non commutative version (\ref{starcomm}). 
%In order to fully appreciate these statements we need to understand the structure
%of the mesonic chiral ring for toric theories \cite{vegh,MSY2,butti,BFZ,benve,Hanany:2006uc,Feng:2007ur}.
\begin{figure}[h!!!]
\begin{center}
\includegraphics[scale=0.70]{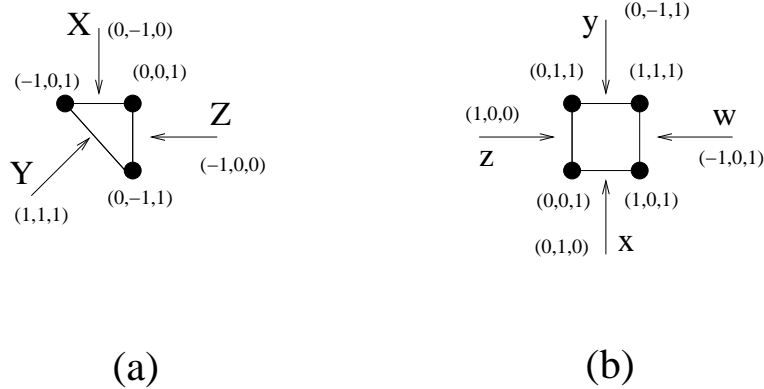} 
\caption{The toric diagram $\mathcal{C}$ and the generators of the dual cone $\mathcal{C}^*$ with the associated mesonic fields for: (a) $\mathcal{N}=4$, (b) conifold. The $U(1)^3$ charges of the mesons are explicitly indicated; the first two
entries of the charge vectors give the $U(1)^2$ global charge used to define
the non commutative product.}
\label{n4con}
\end{center}
\end{figure}

\subsection{The $\beta$-Deformed Mesonic Chiral Ring}

Let us see, in a bit more detail, how the $\beta$-deformed mesonic chiral ring is obtained for a generic CY toric singularity $\cX$.
As we explained in chapter \ref{braneasing} a toric variety is reconstructed as a set of algebraic constraints in some ambient space
looking at the linearly dependence constraints among the integer vectors generating the dual cone $\sigma^*$. The generators $W_j$, $j=1,...,k$ of the dual cone are the basis $M_{W_j}$ of the mesonic chiral ring in the gauge theory and the sum relations between points in the dual cone become multiplicative relations among mesons
in the field theory. 
 
For every integer point $m_l$ in the dual cone there exist a unique mesonic operator $M_{m_l}$ 
modulo F-term constraints.
In particular, the mesons are uniquely determined by their charge
under $U(1)^3$. The first two coordinates $Q^{m_l}=(m_l^1,m_l^2)$
of the vector $m_l$ 
are the charges of the meson under the two flavour $U(1)$ symmetries.

%We need now to understand the non abelian structure of the BPS conditions.
Mesons correspond to closed loops in the quiver 
and for any meson there is an F-term equivalent meson that passes for a given gauge group.
We can therefore assume that all meson loops have a base point at a 
specific gauge group and consider them as $N\times N$ matrices 
${\cal M}_\alpha^\beta$. In the undeformed theory,
the F-term equations imply that all mesons commute and can be 
simultaneously diagonalized. 
The additional F-term constraints require that the mesons, and therefore
all their eigenvalues, satisfy the algebraic equations defining the Calabi-Yau.
This gives a moduli space which is the $N$-fold symmetrized
product of the Calabi-Yau. 

In the $\beta$-deformed theory
 
the commutation relations among mesons are replaced by $\beta$-deformed commutators
\beq\label{mesb}
M_{m_1} M_{m_2}=e^{-2 i \pi \beta (Q^{m_1} \wedge Q^{m_2})} M_{m_2} M_{m_1}=b^{(Q^{m_1} \wedge Q^{m_2})} M_{m_2} M_{m_1} \, .
\eeq

The prescription (\ref{mesb}) will be our short-cut for computing the relevant quantities we will be interested in. 
This fact becomes computationally relevant in the generic toric case.
It is possible to show \cite{Butti:2007aq} that this procedure is equivalent to using 
the $\beta$-deformed superpotential defined in (\ref{defSup}) 
and 
deriving the constraints for the mesonic fields from the F-term relations.

Finally the mesons still satisfy a certain number of 
algebraic equations
\begin{equation}\label{modCYb}
 f({\cal M})=0 
\end{equation}
which are isomorphic to the defining equations of the original Calabi-Yau $\cX$.

\section{Abelian Moduli Space}\label{abelbet}

In this section, we give evidence, from the gauge theory 
side, that the abelian mesonic moduli space of the $\beta$-deformed theories is a set
of lines. There are exactly $d$ such lines, where $d$ is the number of vertices
in the toric diagram.
Indeed, the lines correspond to the geometric 
generators $W_j$ $j=1,...,d$ of the dual cone over the real number of the undeformed geometry, or, in other words, the edges of the polyedron $\sigma^*$ where the $T^3$ fibration degenerates
to $T^1$. Internal generators of $\sigma^*$ as a semi-group do not correspond
to additional lines in the moduli space. 
%that is the immage of the momentum 
%map of the $T^3$ action over the six-dimensional Calabi-Yau cone. 
These statements are the field theory counterpart 
of the fact that the D3 probes can move only along the edges 
of the symplectic cone.

We explained in the previous section how to obtain a set of modified 
commutation relations among mesonic fields. 
In the abelian case the mesons reduce to commuting c-numbers. 
>From the relations (\ref{mesb}) with non a trivial $b$ factor, we obtain the constraint
\begin{equation}
M_{m_1} M_{m_2}=0 \, .
\end{equation}
Adding the algebraic constraints (\ref{modCYb}) defining $\cX$, we obtain the full
set of constraints for the abelian mesonic moduli space. It is possible to show in complete generality \cite{Butti:2007aq} that the mesonic 
moduli space for a single D3 brane is:
\begin{equation}
{}^{{\rm mes}}\!{\cal M} = \bigcup _{j=1}^d \mathbb{C}_{W_j}
\end{equation}
namely the union of the $d$ complex lines associated to the generators $W_j$ of the dual cone.

We now solve the constraints in a selected set of examples, which are 
general enough to exemplify the result. We analyze ${\cal N}=4$, the conifold,
the Suspended Pinch Point ($SPP$) singularity and a more sophisticated
example, $PdP_4$, which covers the case where the generators of $\sigma^*$ 
as a semi-group are more than the geometric generators.   

\subsection{The Case of $\mathbb{C}^3$}
The $\mathcal{N}=4$ theory is simple and was already discussed in 
section \ref{betadef}. The abelian mesonic moduli space of the $\beta$-deformed theory
is done by the three lines corresponding to the geometric generators
of the dual cone as in Figure \ref{n4con}.

\subsection{The Conifold Case}
The abelian mesonic moduli space of the $\beta$-deformed conifold theory was 
already discussed in section \ref{betadef} using the elementary fields. From the 
equations (\ref{comm}) we obtain the same result: the abelian mesonic moduli space is the union of
the four lines corresponding to the external generators of the dual cone as shown in Figure \ref{n4con}. 

\subsection{The SPP Case}

The gauge theory obtained as the near horizon limit of a stack of D3-branes 
at the tip of the conical singularity
\begin{equation}\label{spp}
x y^2 = w z
\end{equation} 
is called the $SPP$ gauge theory \cite{Morrison:1998cs}. In section \ref{SPPmast} we have already studied the master space of this gauge theory.
In Figure \ref{tqspp} we give the toric diagram and the quiver of the theory.  
\begin{figure}[h!!!]
\begin{center}
\includegraphics[scale=0.55]{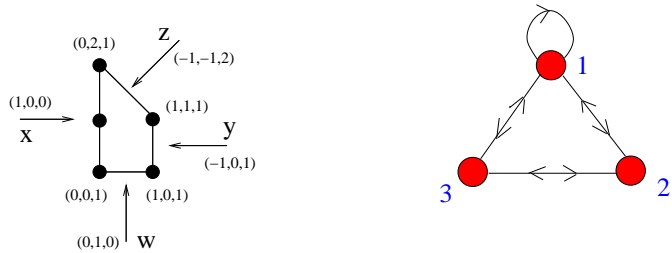} 
\caption{The toric diagram and the quiver of the $SPP$ singularity}
\label{tqspp}
\end{center}
\end{figure}
Its superpotential is
%\begin{equation}\label{superpotspp}
$W= X_{11} ( X_{13}X_{31} - X_{12}X_{21}) +X_{21}X_{12}X_{23}X_{32}  -  X_{32}X_{23}X_{31}X_{13}$ 
%\end{equation}
The generators of the mesonic chiral ring are 
\begin{eqnarray}
 & w = X_{13} X_{32} X_{21} \, , & x = X_{11} \, , \nonumber\\
 & z = X_{12} X_{23} X_{31} \, , & y = X_{12} X_{21} \, .
\end{eqnarray} 
These mesons correspond to the generators of the dual cone in Figure 
\ref{tqspp}. Their flavour charges can be read from the dual toric diagram
\begin{equation}\label{messppc}
Q_x=(1,0) \hbox{ , } Q_z=(-1,-1) \hbox{ , }  Q_y=(-1,0) \hbox{ , } Q_w=(0,1) \, .
\end{equation}
If we now $\beta$ deform the superpotential of the theory 
and we use the deformed commutation rule for mesons (\ref{mesb}) we obtain the following relations
\begin{eqnarray}\label{sppbc}
& & x w = b w x  \, , \quad z x = b x z  \, , \quad  w z = b z w \, ,\nonumber\\
& & w y = b y w  \, , \quad y z = b z y  \, . 
\end{eqnarray}
In the abelian case they reduce to
\begin{eqnarray}
& & x w = 0 \, , \quad  z x = 0  \, , \quad  w z = 0 \, ,\nonumber\\
& & w y = 0  \, , \quad  y z = 0  \, , \quad  x y^2 \sim w z \, , 
\end{eqnarray}
where the last equation is the additional F-term constraint giving the
original CY manifold.  The presence of the symbol ``$\sim$'' is due to
the fact that the original CY equation is deformed by an unimportant
power of the deformation parameter $b$, which can always be reabsorbed
by rescaling the variables.  The solutions to these equations are
\begin{eqnarray}
( x = 0 \, , \quad y = 0 \, , \quad z=0 ) \rightarrow \{ w \} \, , \nonumber\\
( x = 0 \, , \quad y = 0 \, , \quad w=0 ) \rightarrow \{ z \} \, ,\nonumber\\
( x = 0 \, , \quad z = 0 \, , \quad w=0 ) \rightarrow \{ y \} \, ,\nonumber\\
( w = 0 \, , \quad y = 0 \, , \quad z=0 ) \rightarrow \{ x \} \, , 
\end{eqnarray} 
corresponding to the four complex lines associated to the four generators of the dual cone.

\subsection{The PdP$_4$ Case}

In this section we study a bit more complicated example: the PdP$_4$ theory. 
This is the first time we meet this example in this Thesis. 
We look at this theory because this is probably the simplest example of toric singularity 
with internal generators: the perpendicular vectors to the toric diagram are enough to generate the dual
cone on the real numbers but other internal vectors are needed to
generate the cone on the integer numbers. The discussion in the dual gravity side \cite{Butti:2007aq} suggests 
that the moduli space seen by the dual giant gravitons, and hence the abelian mesonic moduli space of the gauge
theory, are exhausted by the external generators. 
We will see here evidence of this fact.

The PdP$_4$ gauge theory, \cite{Feng:2002fv}, is the theory obtained as the
near horizon limit of a stack of D3-branes at the tip of the non
complete intersection singularity defined by the set of equations
\begin{eqnarray}\label{P4F}
& & z_1 z_3 = z_2 t \hbox{ ,  } z_2 z_4 =  z_3 t \hbox{ ,  } z_3 z_5  = z_4 t \nonumber\\
& & z_2 z_5 = t^2 \hbox{ ,  } z_1 z_4 =  t^2 \, .
\end{eqnarray} 
These equations describe the complex cone over the four dimensional compact surface obtained blowing up four non generic points in $\mathbb{P}^2$.
The toric diagram and the quiver of the theory are given in Figure \ref{PdP4}.
\begin{figure}[h!!!]
\begin{center}
\includegraphics[scale=0.55]{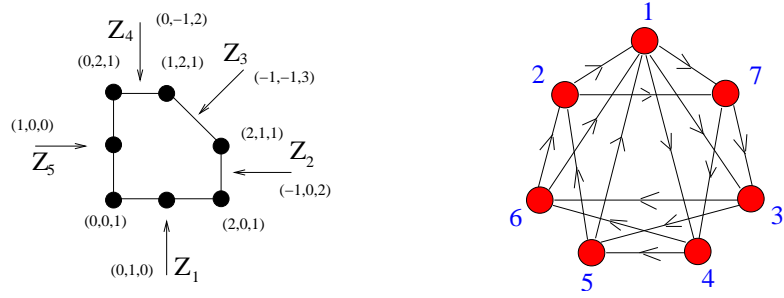} 
\caption{The toric diagram and the quiver of the $PdP_4$ singularity}
\label{PdP4}
\end{center}
\end{figure}
The superpotential of the theory is
\begin{eqnarray}\label{suppdp4}
W &=& X_{61} X_{17} X_{74} X_{46} + X_{21} X_{13} X_{35} X_{52} + X_{27} X_{73} X_{36} X_{62} + X_{14} X_{45} X_{51} \nonumber\\
&- & X_{51}X_{17} X_{73} X_{35} - X_{21} X_{14} X_{46} X_{62} - X_{27} X_{74} X_{45} X_{52} - X_{13} X_{36} X_{61} \, . 
\end{eqnarray}
The generators of the mesonic chiral ring are
\begin{eqnarray}\label{genpdp4}
& & z_1=X_{51}X_{13}X_{35} \, ,\quad  z_2=X_{51}X_{17}X_{74}X_{45} \, , 
\quad z_3=X_{21}X_{17}X_{74}X_{45}X_{52} \, ,\nonumber\\
& & z_4=X_{14}X_{45}X_{52}X_{21} \, , \quad  z_5= X_{14} X_{46} X_{61} 
\, , \quad  t=X_{13} X_{36} X_{61} \, .
\end{eqnarray}
From the toric diagram we can easily read the charges of the mesonic generators
\begin{equation}
Q_{z_1}=(0,1) \, ,\quad  Q_{z_2}=(-1,0) \, , \quad   Q_{z_3}=(-1,-1) 
\, , \quad  Q_{z_4}=(0,-1) \, , \quad  Q_{z_5}=(1,0) \, .
\end{equation}
To generate the dual cone $\sigma^*$ on the integers we need to add the internal generator $t=(0,0,1)$ with flavour charges $Q_t=(0,0)$. 
If we $\beta$-deform the superpotential (\ref{suppdp4}) and we use the rules explained in the previous section, we find that
the generators (\ref{genpdp4}) satisfy the equations (\ref{P4F}) for the PdP$_4$ singularity modified just by some irrelevant 
proportional factors given by powers of $b$.
We must then add the relations obtained from the mesonic $\beta$-deformed 
commutation rule (\ref{mesb})
\begin{eqnarray}
& &z_1 z_2 = b z_2 z_1  \, ,\quad  z_1 z_3 = b z_3 z_1  \, ,\quad
 z_5 z_1 = b z_1 z_5  \, ,\quad  z_2 z_3 = b z_3 z_2 \nonumber\\ 
& &  z_2 z_4 = b z_4 z_2  \, ,\quad  z_3 z_4 = b z_4 z_3  \, ,
\quad  z_3 z_5 = b z_5 z_3  \, ,\quad  z_4 z_5 = b z_5 z_4 \, , 
\end{eqnarray} 
that in the abelian case reduce to 
\begin{eqnarray}\label{P4cb}
& & z_1 z_2 = 0  \, ,\quad z_1 z_3 = 0  \, ,\quad z_5 z_1 = 0  \, ,\quad
 z_2 z_3 = 0 \, ,\nonumber\\
& & z_2 z_4 = 0  \, ,\quad  z_3 z_4 = 0  \, ,\quad  z_3 z_5 = 0  \, ,\quad
 z_4 z_5 = 0 \, . 
\end{eqnarray} 
The solutions to the set of equations (\ref{P4F}) and (\ref{P4cb}) are
\begin{eqnarray}
&& ( z_2 = 0  \, ,\quad  z_3 = 0  \, , \quad  z_4=0  \, ,\quad  z_5 = 0  \, ,\quad   t=0 ) 
 \rightarrow \{ z_1 \} \, , \nonumber\\
&& ( z_1 = 0  \, ,\quad  z_3 = 0  \, , \quad  z_4=0  \, ,\quad  z_5 = 0  \, ,\quad  t=0 ) 
\rightarrow \{ z_2 \} \, ,\nonumber\\
&& ( z_1 = 0 \, ,\quad  z_2 = 0  \, ,\quad  z_4=0  \, ,\quad z_5 = 0  \, ,\quad  t=0 ) 
\rightarrow \{ z_3 \} \, , \nonumber\\
&& ( z_1 = 0 \, ,\quad  z_2 = 0  \, ,\quad z_3=0  \, ,\quad z_5 = 0  \, ,\quad  t=0 ) 
\rightarrow \{ z_4 \} \, , \nonumber\\
&& ( z_1 = 0  \, ,\quad z_2 = 0  \, ,\quad  z_3=0  \, ,\quad  z_4 = 0 \, ,\quad t=0 ) 
\rightarrow \{ z_5 \} \, , 
\end{eqnarray} 
corresponding to the five external generators of the dual cone. 
We observe in particular that the complex line corresponding to the internal generators $t$ is not a solution.

\section{Non Abelian Moduli Space and Rational $\beta$}\label{razbeta}

We start to have a good understanding about the effect of the marginal $\beta$-deformation on the moduli space of the gauge theory: 
it usually drastically reduced the moduli space. In this section we will see that for some special values of $\beta$ it is possible that new branches open up in the moduli space. Let us start with a very easy and explicative example. Let us take the $\mathcal{N}=4$ theory and let us analyze a couple of BPS operators. The operator $\tr(XY)$ is not constrained by the F-term equations in the case $\beta=0$, but in the case in which $\beta \neq 0$ it is constrained to be zero on the moduli space of supersymmetric vacua. Indeed $\tr (XY) = b \tr (YX)$ using the F term equations and $b \tr (YX)= b \tr(XY) $ using the ciclicity of the trace. This imply that $\tr(XY)=0$ unless $b=1$ which imply $\beta=0$. This is an easy heuristic argument to guess that the moduli space in the deformed case is reduced, whichever is the value of $\beta \neq 0$. Let us now consider another operator: $\tr (XYXY)$. Using the F-term equations and the ciclicity of the trace we obtain $\tr (XYXY) = b^2 \tr!
  (XYXY)$. This imply that $\tr (XYXY) =0$  unless $b^2=1$. And this is possible if $b=1$ hence $\beta=0$, the undeformed case, and $b=-1$ hence $\beta=1/2$. Looking at this example we understand that for some special values of $\beta$, namely rational values, new branches of the moduli space, that are absent for generic values (irrational) of $\beta$, open up. 

Let us now try to explain the general case. The F-term equations
\begin{equation}\label{mb}
M_{m_1} M_{m_2}=e^{- 2\pi i \beta (Q^{m_1} \wedge Q^{m_2})} M_{m_2} M_{m_1}
\end{equation}
give a non commutative 't Hooft-Weyl algebra for the 
$N\times N$ matrices ${\cal M}_I$. By diagonalizing the matrix
$\theta_{m_1m_2}=(Q^{m_1} \wedge Q^{m_2})$ we can reduce the problem
to various copies of the algebra for a non commutative torus 
\beq  
M_1 M_2 = e^{2\pi i\theta} M_2 M_1 
\label{eqtheta}
\eeq
 whose representations are well known.

For generic $\beta$, corresponding to irrational values of $\theta$, the 't Hooft-Weyl algebra has no non trivial 
finite dimensional representations: we can only find solutions where
all the matrices are diagonal, and in particular equation (\ref{eqtheta}) implies $M_1 M_2= M_2 M_1=0$. The problem is thus reduced to the
abelian one and the moduli space is obtained by symmetrizing 
$N$ copies of the abelian moduli space, which consists of $d$ lines. 
This is the remaining of the original Coulomb branch
of the undeformed theory, and the mesonic moduli space is just the N-times symmetric product of the $d$ complex lines
associated to the generators of the dual cone:
\begin{equation}
{}^{{\rm mes}}\!{\cal M} = \hbox{Sym}^N \left(\bigcup _{j=1}^d \mathbb{C}_{W_j}\right)
\end{equation}

For rational $\beta= m/n$, instead, new branches are opening up in the
moduli space \cite{Berenstein:2000hy,Berenstein:2000ux,Dorey:2002pq,Dorey:2003pp,Dorey:2004xm}. Indeed, for rational $\beta$,
we can have finite dimensional representations of the 't Hooft-Weyl 
algebra  which are given by $n\times n$ matrices $(O^I)_{ij}$. 
The explicit form of the matrices $(O^I)_{ij}$ can be found in \cite{weyl},
but it is not of particular relevance for us. 
For gauge groups $SU(N)$
with $N=nM$ we can have vacua where the mesons have the form 
\begin{equation}\label{gen}
({\cal M}_I)_\alpha^\beta= {\rm Diag}({\cal M}_a) \otimes (O^I)_{ij},\quad  a=1,...,M,\quad i,j=1,...,n, \quad \alpha,\beta=1\ldots N \, .
\end{equation}
The $M$ variables ${\cal M}_a$ are further constrained by the algebraic 
equations (\ref{modCYb}) and are due to identifications
 by the action of the gauge
group. A convenient way of parameterizing the moduli space is to look at
the algebraic constraints satisfied by the elements of the center 
of the non-commutative algebra \cite{Berenstein:2000hy,Berenstein:2000ux}. 

We will give arguments showing that the center 
of the algebra of mesonic operators is the algebraic variety $ \cX /\mathbb{Z}_n \times \mathbb{Z}_n$, where $\cX$ is the original undeformed CY variety, and the two $\mathbb{Z}_n$ factors are 
abelian discrete sub-groups of the two flavours symmetries.
This statement in the field theory can be checked in the gravity side showing that the moduli space of a D5 dual 
giant graviton is the original Calabi-Yau $\cX$ 
divided by $\mathbb{Z}_n\times \mathbb{Z}_n$ \cite{Butti:2007aq}. 

The generic vacuum (\ref{gen}) corresponds to $M$ D5 dual giants moving on
the geometry. The resulting branch of the moduli space is the $M$-fold 
symmetrized product of the original Calabi-Yau divided by 
$\mathbb{Z}_n\times \mathbb{Z}_n$. Each D5 dual giant should be considered
as a fully non-abelian solution of the dual gauge theory carrying $n$ color
indices so that the total number of colors is $N=nM$. Indeed the mesonic moduli space for rational values of 
$\beta$ is:
\begin{equation}\label{mesbetra}
{}^{{\rm mes}}\!{\cal M} = \hbox{Sym}^M \left(\frac{\cX}{\mathbb{Z}_n \times \mathbb{Z}_n}\right)
\end{equation}
%We can obtain a different perspective on this branch of our gauge theory by considering
%it as the world-volume theory of D3-branes sitting at a discrete torsion
%$\mathbb{Z}_n\times \mathbb{Z}_n$ orbifold of the original singularity \cite{discrete}. In this picture,
%the D5 dual giants correspond to the physical branes surviving the orbifold
%projection. This perspective 
%has been discussed in details in the literature for ${\cal N}=4$ SYM \cite{bl} and it can be easily extended to generic toric singularities.
Let us pass to analyze some simple explicit examples, before giving the general proof of (\ref{mesbetra}). 

\subsection{The Case of $\mathbb{C}^3$}
The case of the $\beta$-deformation of $\mathcal{N}=4$ gauge theory is simple and well known \cite{Berenstein:2000hy,Berenstein:2000ux}.
 
The generators of the algebra of mesonic operators are the three elementary fields $X$, $Y$, $Z$. 
Equation (\ref{N4}) implies that it possible to write the generic element of the algebra in the ordered form
\begin{equation}\label{opC3}
\Phi_{k_1,k_2,k_3}=X^{k_1} Y^{k_2} Z^{k_3}
\end{equation}
The center of the algebra is given by the subset of operators in (\ref{opC3}) 
such that:
\begin{eqnarray}
\Phi_{k_1,k_2,k_3}\hbox{ }X = &b^{k_3 - k_2} \hbox{ }X \hbox{ }\Phi_{k_1,k_2,k_3}& = X \hbox{ }\Phi_{k_1,k_2,k_3} \, , \nonumber\\
\Phi_{k_1,k_2,k_3}\hbox{ }Y = &b^{k_1 - k_3} \hbox{ }Y \hbox{ }\Phi_{k_1,k_2,k_3}& = Y \hbox{ }\Phi_{k_1,k_2,k_3} \, ,\nonumber\\
\Phi_{k_1,k_2,k_3}\hbox{ }Z = &b^{k_2 - k_1} \hbox{ }Z \hbox{ }\Phi_{k_1,k_2,k_3}& = Z \hbox{ }\Phi_{k_1,k_2,k_3} \, .
\end{eqnarray}  
Since $b^n=1$, the center of the algebra is given by the set of $\Phi_{k_1,k_2,k_3}$ such that $k_1=k_2=k_3 \, 
{\rm mod} \, n$.

The generators of the center of the algebra are: $\Phi_{n,0,0},\Phi_{0,n,0},\Phi_{0,0,n},\Phi_{1,1,1}$. 
We call them $x,y,w,z$ respectively. They satisfy the equation
\begin{equation}\label{C3ZnZneq}
x y w = z^n
\end{equation}
which defines the variety $\mathbb{C}^3/\mathbb{Z}_n \times \mathbb{Z}_n$. 
To see this, take $\mathbb{C}^3$ with coordinate $Z^1,Z^2,Z^3$, and consider the action of the group $\mathbb{Z}_n \times \mathbb{Z}_n$ on  $\mathbb{C}^3$ 
\begin{equation}
Z^1,Z^2,Z^3 \rightarrow Z^1 \delta^{-1}\, ,\, Z^2 \delta \xi \,,\, Z^3 \xi^{-1}
\end{equation}
with $\delta^n=\xi^n=1$. The basic invariant monomials under this action 
are $x=(Z^1)^n, y=(Z^2)^n, w=(Z^3)^n, z=Z^1 Z^2 Z^3$ and they clearly satisfy 
the equation (\ref{C3ZnZneq}). 

This fact can be represented in a diagrammatic way as in Figure \ref{C3ZnZn}. This representation of the 
rational value $\beta$-deformation is valid for every toric CY singularity. 
\begin{figure}[h!!!]
\begin{center}
\includegraphics[scale=0.45]{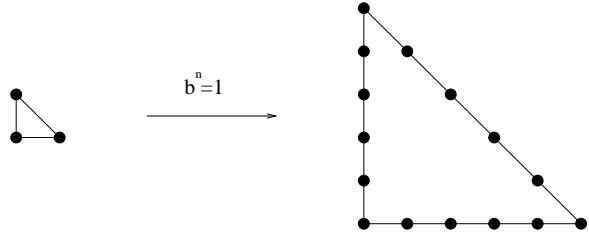} 
\caption{$\mathbb{C}^3 \rightarrow \mathbb{C}^3/\mathbb{Z}_n \times \mathbb{Z}_n$ in the toric picture, in the particular case in Figure $b^5=1$.}
\label{C3ZnZn}
\end{center}
\end{figure}

\subsection{The Conifold Case}

The case of the conifold is a bit more intricate and can be a useful example for the generic CY toric cone $\cX$. The generators of the mesonic algebra 
$x, y, z, w$ satisfy the equations (\ref{comm}). It follows that we can write 
the generic monomial element of the algebra in the ordered form
\begin{equation}\label{mescon}
\Phi_{k_1,k_2,k_3,k_4}=x^{k_1}y^{k_2}w^{k_3}z^{k_4} \, .
\end{equation}
The center of the algebra is given by the subset of the operators (\ref{mescon}) that satisfy the equations
\begin{eqnarray}
\Phi_{k_1,k_2,k_3,k_4}\hbox{ }x = &b^{k_4 - k_3} \hbox{ }x \hbox{ }\Phi_{k_1,k_2,k_3,k_4}& = x \hbox{ }\Phi_{k_1,k_2,k_3,k_4}\, , \nonumber\\
\Phi_{k_1,k_2,k_3,k_4}\hbox{ }y = &b^{k_3 - k_4} \hbox{ }y \hbox{ }\Phi_{k_1,k_2,k_3,k_4}& = y \hbox{ }\Phi_{k_1,k_2,k_3,k_4} \, ,\nonumber\\
\Phi_{k_1,k_2,k_3,k_4}\hbox{ }w = &b^{k_1 - k_2} \hbox{ }w \hbox{ }\Phi_{k_1,k_2,k_3,k_4}& = w \hbox{ }\Phi_{k_1,k_2,k_3,k_4} \, ,\nonumber\\ 
\Phi_{k_1,k_2,k_3,k_4}\hbox{ }z = &b^{k_2 - k_1} \hbox{ }z \hbox{ }\Phi_{k_1,k_2,k_3,k_4}& = z \hbox{ }\Phi_{k_1,k_2,k_3,k_4} \, . 
\end{eqnarray}  
Because $b^n=1$, the elements of the center of the algebra are the subset of the operators of the form (\ref{mescon}) 
such that $k_1=k_2$, $k_3=k_4$, mod $n$. 

The center is generated by 
$\Phi_{n,0,0,0},\Phi_{0,n,0,0},\Phi_{0,0,n,0},\Phi_{0,0,0,n},
\Phi_{1,1,0,0},\Phi_{0,0,1,1}$; we call them respectively $A,B,C,D,E,G$.
The F-term relation
\begin{equation}\label{Fconb}
x y = b w z 
\end{equation}
then implies that $E$ and $G$ are not independent: $E= b G$. 
Moreover the generators of the center of the algebra satisfy the equations
\begin{equation}\label{orbcon}
A B = C D = E^n \, .
\end{equation}

As in the previous example, it is easy to see that these are the equations of the 
$\mathbb{Z}_n \times \mathbb{Z}_n$ orbifold of the conifold.
Take indeed the coordinates $x,y,w,z$ defining the conifold as a quadric 
embedded in  $\mathbb{C}^4$. 
The action of $\mathbb{Z}_n \times \mathbb{Z}_n$ is
\begin{equation}
x, y, w, z \rightarrow x \delta \,  ,\,  y \delta^{-1}\,  ,\, w \xi^{-1}, z \xi \, ,
\end{equation}
where $\delta ^n = \xi ^n =1$.
The basic invariants of this action are $A,B,C,D,E,G$, and they are subject to the constraint (\ref{Fconb}). Hence the equations (\ref{orbcon}) define the variety $C(T^{1,1})/\mathbb{Z}_n \times \mathbb{Z}_n$. 

\begin{figure}[h!!!]
\begin{center}
\includegraphics[scale=0.45]{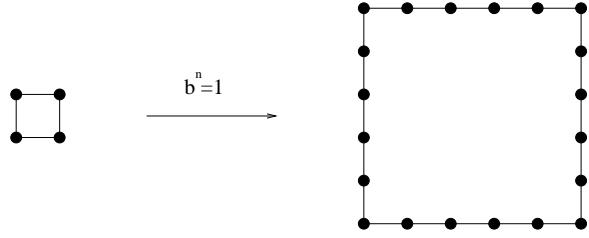} 
\caption{$C(T^{1,1})\rightarrow C(T^{1,1})/\mathbb{Z}_n \times \mathbb{Z}_n$ in the toric picture, in the particular case in Figure $b^5=1$}
\label{T11ZnZn}
\end{center}
\end{figure}

\subsection{The General Case}

Enriched by these explicit computations we now want to analyze the generic case and show that
 the center of the mesonic algebra 
for the rational $\beta$-deformed ($b^n=1$) gauge theory is the $\mathbb{Z}_n \times \mathbb{Z}_n$ 
quotient of the undeformed CY singularity $\cX$.

For a generic toric quiver gauge theory we take a set of basic mesons
$M_{W_j}$ (we will call them simply $x_j$ from now on) corresponding
to the generators $W_j$ of the cone $\sigma^*$. These are the
generators of the mesonic chiral ring of the given gauge theory.
Because they satisfy the relations (\ref{mb}) it is always possible to
write the generic monomial element of the mesonic algebra generated by
$x_j$ in the ordered form
\begin{equation}\label{genb}
\Phi_{p_1,...,p_k}= x_1^{p_1} x_2^{p_2}...x_k^{p_k} \, .
\end{equation}
We are interested in the operators that form the center of the
algebra, or, in other words, that commute with all the elements of the
algebra.  To find them it is enough to find all the operators that
commute with all the generators of the algebra, namely
$x_1,...,x_k$. The generic operator (\ref{genb}) has charge
$Q_{p_1,...,p_k}$ under the two flavor $U(1)$ symmetries, and the
generators $x_j$ have charges $Q_j$. They satisfy the following
relations
\begin{equation}\label{relc}
\Phi_{p_1,...,p_k} \hbox{ }x_j= x_j \hbox{ }\Phi_{p_1,...,p_k} \hbox{ }b^{Q_{p_1,...,p_k} \wedge Q_j } \, .
\end{equation}
This implies that the center of the algebra is formed by the set of $\Phi_{p_1,...,p_k}$ such that
\begin{equation}
Q_{p_1,...,p_k} \wedge Q_j = 0 \hbox{ mod $n$} \hbox{ , } {  j=1,...,k} \, .
\end{equation}
At this point it is important to realize that the $Q_j$ contain the
two dimensional vectors perpendicular to the edges of the two
dimensional toric diagram.  The fact that the toric diagram is convex
implies that the $Q_j$ span the $T^2$ flavor torus.  In particular
the operator $\Phi_{p_1,...,p_k}$ must commute (modulo $n$) with the
operators with charges $(1,0)$ and $(0,1)$. The first condition gives
all the operators in the algebra that are invariant under the
$\mathbb{Z}_n$ in the second $U(1)$, while the second gives all the
operators invariant under the $\mathbb{Z}_n$ contained in the first
$U(1)$.  All together the set of operators in the center of the
algebra consists of all operators $\Phi_{p_1,...,p_k}$ invariant under
the $\mathbb{Z}_n \times \mathbb{Z}_n$ discrete subgroup of the $T^2$.

The monomials made with the free $x_1,...,x_k$ coordinates of
$\mathbb{C}^k$ that are invariant under $\mathbb{Z}_n \times
\mathbb{Z}_n$, form, by definition, the quotient variety
$\mathbb{C}^k/\mathbb{Z}_n \times \mathbb{Z}_n$. The toric variety
$\cX$ is defined starting from a ring over $\mathbb{C}^k$ with relations
given by a set of polynomials $(q_1,...,q_l )$ defined by the toric
diagram
\begin{equation}
\mathbb{C}[\cX]= \frac{\mathbb{C}[x_1,...,x_k]}{(q_1,...,q_l )} \, .
\end{equation} 
Indeed the elements of the center of the algebra are the monomials made with 
the $x_j$, subject to the relations $( q_1,...,q_l )$, invariant under $\mathbb{Z}_n \times \mathbb{Z}_n$. 
This fact allows us to conclude that the center of the algebra ${}^{{\rm mes}}\!{\cal M}_{M=1}$ 
in the case $b^n=1$ is the quotient of the original $CY$ singularity $\cX$
\begin{equation}
{}^{{\rm mes}}\!{\cal M}_{M=1} = \frac{{\rm \cX}}{\mathbb{Z}_n \times \mathbb{Z}_n} \, .
\end{equation} 
The $\beta$-deformed $\mathcal{N}=4$ gauge theory and the $\beta$-deformed conifold gauge theory 
are special cases of this result and we refer the reader to \cite{Butti:2007aq} for other explicit examples.

\section{Summary and Discussions}

In this chapter we extended the analysis of the moduli space and of the chiral ring of superconformal field theories 
to the marginal deformed case. In particular we studied the marginal $\beta$-deformation, common to all the toric quiver gauge theories, and its effect on the mesonic chiral ring. We observed how the moduli space and the spectrum of the chiral ring is drastically modified by the marginal deformation and that it is possible to study in detail this effect using some kind of non-commutative generalization of the usual toric techniques. To be in line with the general philosophy of the thesis we concentrated our studies in the field theory side of the AdS/CFT correspondence, but a very careful analysis can be done even in the dual gravity side \cite{Butti:2007aq}. In particular the gravitational duals geometry is simply characterized in terms of generalised complex geometry. In this setup one can study duals states to the mesonic operators in field theory such as D3 and D5 dual giants or D3 giants configurations. It is possible to verify that the gravity computations match nicely!
  !
 with the field theory.

In this chapter we focalized on the mesonic branch of the moduli space, It would be interesting to extend this study to the baryonic spectrum 
and to the complete moduli space of the $\beta$-deformed theories.

Of course a toric quiver gauge theory has in general other marginal deformations in addition to the $\beta$-deformation. 
It would be very interesting to extend this analysis to other marginal deformations. This is a very interesting point, because the gravity 
dual of other marginal deformation is in general not explicitly known. It would be nice if the study of the moduli space and the chiral ring 
of the gauge theory could help to find the dual gravity solutions.

\chapter*{Conclusions}

In this Thesis we have introduced some new concepts to study supersymmetric gauge theories: 
the Hilbert Series, the Master Space and the Plethystic Exponential.
Using these powerful tools we have illustrated various aspects of $\mathcal{N}=1$ gauge theories. 
In particular we focalized on the study of the moduli space $\CM$ and the chiral ring 
for superconformal gauge theories with gravity duals. These theories can be engineered 
putting a stack of N D3 branes at conical Calabi Yau singularity. We were able to study 
in full generality the moduli space of the abelian theory and to get some non trivial informations
regarding the symmetries of the moduli space for generic N. We explained how to count BPS operators
and the dual stringy states, and how to get from these results a lot of informations regarding both the 
geometry and the field theory. In the last part of the thesis we generalized this approach to include
the fermionic degrees of freedom of the BPS spectrum and to study both the moduli space and the BPS spectrum for
marginal deformed gauge theories.

The Hilbert Series and the Master Space reveal very productive in the study of supersymmetric gauge theories, 
and it would be very nice to use them to study other interesting problems. 

The moduli space and the chiral ring are basic concepts, common to all the supersymmetric gauge theories,
and it would be nice to apply the tools we developed here to more phenomenological interesting theories, such as SQCD \cite{Gray:2008yu,Hanany:2008kn,Hanany:2008sb}, string compactifications \cite{Gray:2006gn,stringvacua}, or MSSM \cite{Gray:2006jb}. 
Along this line another possible step could be to study non conformal theories with gravity dual: in particular the spectrum of the cascading gauge theories along 
the renormalization group flow \cite{Klebanov:2000hb,Herzog:2004tr,Franco:2004jz,Brini:2006ej}\footnote{see also \cite{Amariti:2008es} and reference therein for the first steps towards a "geometrization" of the recently found non supersymmetric metastable vacua \cite{Intriligator:2006dd}}. 

It would be nice to start a statistical study of quiver gauge theories.
The first natural step is the analysis of the asymptotic behavior of the partition functions for large charges. See \cite{Feng:2007ur} for some related studies about the mesonic partition functions.
   
Another interesting application would be to 
extend at the quantum level the classical analysis we have done so far. 
%quantum moduli space and quantum chiral rings.
%corrections to the chiral ring and to the moduli space. 
We have in principle all the instruments: the study 
of the gravity duals must contain the quantum moduli space and the quantum chiral ring of the gauge theory, 
and along this thesis we learnt
to study the supersymmetric stringy spectrum and to map it to the gauge theory. 
Indeed in the thesis we observed that the master space 
%classical moduli space 
is not invariant under Seiberg duality. 
This fact may suggest the presence of quantum corrections. It would 
be very nice to systematically understand if there are really quantum modification 
to the moduli space and to the chiral ring, how they are done, and how they behave 
under Seiberg duality. See \cite{Forcella:2008ng} for some results.

Another interesting extension would be to study the modifications to the chiral ring and to the moduli space 
induced by general marginal deformations other than the $\beta$-deformation. It would be very nice to see if these 
kind of studies could give important informations to the dual gravity solutions. 
Indeed all the $\mathcal{N}=1$ supergravity backgrounds are described 
by the generalized complex geometry as twisted generalize Calabi Yau. It would be wonderful if, 
the techniques here developed could marry with and techniques from Generalized Complex Geometry to obtain informations about the marginal deformed supergravity backgrounds \cite{Butti:2007aq}.  

Recently there were a lot of interests in instantonic corrections to the superpotential induced by
stringy instantons. In the setup of branes at singularities the stringy instantons are euclidean D3 branes wrapped on some four cycles. 
The operators they generate are determinant like. Both the geometric embeddings and the operators they generate are very similar to the D3 brane baryonic states 
and the dual BPS operators we have considered along this thesis. 
It would be interesting to make more precise this relation \cite{Forcella:2008au}.

The AdS/CFT proposes a non perturbative definition of some string theories on some specific 
backgrounds that are asymptotically AdS. 
Recently a set of black hole solutions asymptotically AdS were found and their entropy were computed. 
In principle it could be possible to ``explain'' the microscopic origin of their entropy 
counting states in the dual field theory. 
It would be very interesting to see if the counting procedure explained in this thesis could be extended 
to this very interesting problem. The AdS black hole solutions are in general 1/4 BPS, and it would be very 
appealing to understand how to compute the spectrum of 1/4 BPS operators, and to see if this counting match with
the black hole entropy.

An even more appealing achievement would be to compute gauge theory partition functions for the 
complete spectrum, both supersymmetric and non supersymmetric.

In the recent period there is a lot of interest in the study of three dimensional SCFT that are supposed
to describe the IR dynamics of M2 branes \cite{Bagger:2006sk,Bagger:2007vi,Gustavsson:2007vu,Aharony:2008ug}. It seems that some of the  tools we developed in this 
thesis can be used also in this case \cite{Hanany:2008qc,Martelli:2008si,Hanany:2008cd,Hanany:2008fj}. It would be nice to further study this topic.  

We started counting: 1,2,3,... but there are many numbers and many things to count...the future may reserve very interesting applications 
based on the studies presented in this thesis.

\newpage

%\noindent {\LARGE {\bf Appendix: List of Notations}}

\section*{Appendix: List of notations}

\appendix

\vskip 0.5cm

\begin{tabular}{lll}
&  $\sigma$ & the direct fan for the toric diagram\\
&  $\sigma^*$ & the dual cone\\
&  $\cX$ & the six dimensional CY singularity transverse to the D branes\\
&  H & the horizon manifold\\
&  $H(t;V)$ & Hilbert series of the variety $V$ \\
&  $\f$  & the master space \\
&  $\firr{~}$ & the coherent component of the master space \\
&  $L_i$ & the linear components of the master space \\
&  $\CM$ & the moduli space of the gauge theory \\
&  ${}^{{\rm mes}}\!{\cal M}$ & the mesonic moduli space of the gauge theory \\
&  $N$ & number of D-branes \\
&  $Z_D$ & partition function for the sections of the line bundle associated to the divisor $D$ \\
&  $Z_{D,N}$ & partition function for D3 branes BPS states in the $AdS_5 \times H$ background\\ 
&  $\mathcal{P}_{cl}$ & classical phase space of BPS D3 branes on three cycles in H \\
& $g_{N,N B}(\{t_i\}; \cX)$ & the generating function for fixed number of D branes $N$ and baryonic charge $N B$ \\
& $g_B(\nu; \{t_i\}; \cX) $ & the generating function for any number of D branes and fixed baryonic charge $B$\\
& $ g_N(\{t_i\}; \cX)$ & the generating function for fixed number of D branes $N$ and any baryonic charge\\
& $g$ & number of $SU(N)$ gauge group factors in the theory \\
&  $F$ & number of fields (chiral bi-fundamental multiplets) in the theory \\
% &  $t_i$ & chemical potentials (weights) \\
&  $q_i$ & chemical potentials for the geometric $T^3$ toric action \\
&  $x, y, q_i, t_i$ & chemical potentials (weights) for flavor symmetries, R-symmetry and dimension \\
&  $b_i$ & non-anomalous baryonic weights \\
&  $a_i$ & anomalous baryonic weights \\
&  $R$ & R-charge \\
&  $B_i, B^\prime$ & baryonic charges from the gauge theory \\
&  $\beta, \beta^\prime$ & baryonic charges from the geometry\\
&  $A_i$ & anomalous baryonic charges \\
%&  ${\bf A,B,C,\ldots }$ & fields in the gauge theory \\
%&  $A,B,C$ & discrete charges generating the Heisenberg group\\
&  ${\cal R}$ & polynomial ring \\
&  ${\cal I, J}$ & ideals for a polynomial ring \\
&  $a,b,c,\ldots$ & generators of the auxiliary GKZ ring \\
&  $m(B_1, B_2, \ldots)$ & multiplicities in the GKZ fan \\
&  $I$ & number of internal (integral) points in toric diagram \\
&  $d$ & number of external points in toric diagram \\
&  $D_i$ & toric divisors assigned to the integral points in the toric diagram \\
&  $x_i$ & homogeneous coordinates \\
%&  $g_N(\{t_i\};CY)$ & baryonic generating function for $N$ D-branes probing CY \\
&  $\mbox{PE}_\nu[\ \cdot \ ]$ & plethystic exponential with weight $\nu$ for the number of
D-branes \\
&  $Z_{\beta, \beta^\prime} (\{t_i\};CY)$ & partition function from the geometry \\
&  $C(\beta)$ & hollow polygon in the fiber over the GKZ lattice \\
\end{tabular}

%\appendix

\newpage

%\bibliographystyle{plain}
%\bibliography{dodo}

\end{document}